%% file: main.tex
\DocumentMetadata{
  pdfversion = 1.7,
  pdfstandard = UA-1,
  lang        = en-US,
}

\documentclass[12pt,fleqn]{ucithesis}

\usepackage[T1]{fontenc}
\usepackage{lmodern}

\DeclareUnicodeCharacter{03C4}{\ensuremath{\tau}}
\DeclareUnicodeCharacter{2192}{\ensuremath{\rightarrow}}
\DeclareUnicodeCharacter{0302}{\textasciicircum}

\usepackage{amsmath}
\usepackage{amsthm}
\usepackage{amssymb}

\usepackage[table]{xcolor}  
\usepackage{array}
\usepackage{graphicx}
\usepackage{booktabs}      
\usepackage{caption}
\usepackage{subcaption}    
\usepackage{multirow}
\usepackage{makecell}
\usepackage{tabularx}

\usepackage[backend=biber, style=numeric-comp, natbib=true, sorting=none]{biblatex}
\usepackage{tikz}
\usepackage[compat=1.1.0]{tikz-feynman}
\usetikzlibrary{decorations.pathmorphing,decorations.markings,arrows,arrows.meta,shapes,shapes.geometric,decorations.pathreplacing,fit,patterns,patterns.meta,positioning,calc}

\usepackage{relsize}
\usepackage[titletoc]{appendix}
\usepackage{parskip}
\usepackage{enumitem}
\usepackage{float}
\usepackage{placeins}

\usepackage[
  plainpages         = false,
  hidelinks,
  unicode            = true,
  pdfdisplaydoctitle = true,
]{hyperref}

\usepackage{tagpdf}

\input{preliminaries}

\input{vita}

\let\origcurriculumvitaepage\curriculumvitaepage
\renewcommand{\curriculumvitaepage}{%
  \origcurriculumvitaepage
  \phantomsection
  \addcontentsline{toc}{chapter}{STATEMENT ON THE USE OF AI TOOLS}
  \aiusestatementpage
}

\hypersetup{
  pdftitle   = {\Thesistitle},
  pdfauthor  = {\Authorname},
  pdfsubject = {\Degreefield},
}

\usepackage{thesis-style}

\begin{document}

\preliminarypages

\include{chapter1}
\include{chapter2}

\include{chapter3}

\include{chapter4}

\include{chapter5}

\include{chapter6}

\include{chapter7}

\clearpage
\phantomsection

\printbibliography

\end{document}

%% file: preliminaries.tex
\thesistitle{A High- and Variable-Dimensional Measurement of the $Z$+jets Differential Cross Section with the ATLAS Experiment and Artificial Intelligence}

\documenttitle{Dissertation}

\degreename{Doctor of Philosophy}

\degreefield{Physics and Astronomy}

\authorname{Kevin Greif}

\committeechair{Daniel Whiteson}
\othercommitteemembers
{
  Andrew Lankford \\
  Jonathan Feng
}

\degreeyear{2026}

\copyrightdeclaration
{
  {\copyright} {\Degreeyear} \Authorname
}

\dedications
{  
  To my parents, Douglas and Anne Greif
}

\acknowledgments
{
  It takes a village to raise even a semi-effective scientist.
  I will let others judge the extent to which I qualify as one of those, but I definitely have many people to thank.
  This thesis would not be the same without every one of the people listed below, and probably many more besides.
  My heartfelt thanks to all of them, and to the reader who manages to make it all the way through.
  
  First, a few of my direct supervisors and mentors deserve special mention.
  Thank you to my advisor Daniel Whiteson, for all of the support, criticism, and motivation over these past six years.
  Working in your group has been an immense pleasure, and I am very grateful for the long leash I was given to explore my own ideas and set my own research agenda.
  Such freedom is rare in a Ph.D. program, and I believe I am a better scientist today because of it.
  Thank you also to Ben Nachman for advising on so many of my projects and being a wonderful collaborator, mentor, and host during my year at SLAC.
  I very much look forward to all of the collaboration to follow.
  To the postdocs and scientists in the Whiteson and UCI ATLAS groups who gave advice and technical support: Mike Fenton, Jonas Roemer, and Jared Sturdy.
  Thank you especially to Aishik Ghosh for being a remarkable postdoc and collaborator.
  Thank you to Andy Lankford for supervising my qualification project on the New Small Wheel, and all of the administrative support.
  It is quite likely I logged more travel reimbursements than any grad student in UCI history with your help.
  To my undergraduate research advisor Kevin Lannon, thank you for taking the time to work with a clueless sophomore who read an article about neural networks and decided this was a good reason to do particle physics.
  With your support, that hunch turned out to not be so clueless.

  My love of science was first fostered by my high school teachers Marilyn Peterson and Michelle Jungbauer.
  Thank you for planting such firm foundations.
  I doubt this thesis would look quite the same without them.

  To my many professors at Notre Dame, thank you for your commitment to education and tolerating all of my questions.
  In the physics department, special thanks to Rebecca Surman, Michael Hildreth, Morten Eskildsen, Chris Kolda, Kevin Howard, Patrick Fasano, and Sushrut Ghonge.
  In PLS, special thanks to G. Felicitas Munzel, Tarek Dika, Joseph Rosenberg, Jennifer Newsome Martin, and Henry Weinfield.
  To my professors at UCI, thank you especially to Arvind Rajaraman, Michael Ratz, and Jonathan Feng for an exceptional education in particle physics.
  Teaching can be an afterthought in academic appointments.
  I am greatly indebted to the people above who made it a priority.

  Next to my many collaborators: a great joy of particle physics is the opportunity to work with and learn from so many amazing people.
  Thank you to my co-authors on the publications presented in this thesis, many of whom did most of the work!
  First and foremost to Matthew Smith for being the other half of the $Z$+jets analysis.
  There is absolutely no way it would have come together without your help.
  Thank you also to the rest of the Carleton ATLAS group for your advice and amazing work: Dag Gilberg, Laura Miller, and Jeff Dandoy.
  To Alex Shmakov for teaching me about diffusion, spearheading the VLD work, and generally being the more interesting and knowledgeable half of the UCI deep learning satellite group at CERN.
  To many members of the Heidelberg group, including Nathan Heutsch, Javier Villadamigo, Antoine Petitjean, Victor Breso-Pla, and Sofia Palacios Schweitzer.
  Thanks especially to Tilman Plehn for several wonderful collaborations and hours of fun and productive conversations at conferences.
  To the Stanford / SLAC group, especially Jan-Lucas Uslu and Tanvi Wamorkar for the very fruitful collaborations on scaling laws and physics priors.
  Finally, thank you Vinicius Mikuni for essential advice on using Perlmutter, Omnifold, foundation models, and much more.

  Beyond direct collaborations there are many more people who provided their insights in countless discussions at UCI, CERN, SLAC, and at conferences abroad.
  These types of interactions are incredibly valuable.
  I believe the only real way to understand a scientific field and where it is going is to talk to the people who make it up.
  Thank you to all of the following for lending me a bit of your wisdom.
  The Whiteson group at UCI: Jessica Howard, Max Fieg, Makayla Vessella, Jason Baretz, Jake Rudolph, Levi Condren, Yvonne Ng, Alexis Romero, Taylor Faucett, Jacob Hollingsworth, and Ed Witkowski.
  Stanford / SLAC plus affiliates: Dennis Noll, Alkaid Cheng, Ryan Milton, Mariel Pettee, Sascha Diefenbachar, Joschka Birk, and Sam Klein.
  CERN and ATLAS: Reina Camacho Toro, Matt LeBlanc, Jennifer Roloff, Rob Les, Jad Matthieu Sardain, Jackson Barr, Fabrice Balli, Rongkun Wang, Aaron White, Bogdan Malaescu, Oleg Brandt, Frank Siegert, Chris Pollard, Michael Kagan, Jason Oliver, Julian Wollrath, Iacopo Longarini, and Dan Guest.
  Conferences and workshops: Oz Amram, Matthew Leigh, Malte Algren, Johnny Raine, Anja Butter, David Shih, Claudius Krause, François Lanusse, Huilin Qu, Nathan Suri, Louis Lyons, and Gilles Louppe.

  Thank you to my fellow members of the ATLAS ECSB.
  I am under no illusion that life in ATLAS is smooth sailing for everyone.
  My hope is that our efforts made at least a little bit of difference for the many people who bring their talent and passion to make the experiment run.
  Thank you to Christian Appelt, Petar Bokan, Hassnae El Jarrari, Hannah Arnold, Max Swiatlowski, Robin Hayes, and Rachel Hyneman.

  Thank you to my friends who I've managed to meet across these past six years.
  When considering pursuing a Ph.D., I was told by some that it would be a difficult and lonely experience.
  The people that follow proved that advice completely wrong, and my life is far richer and better because of them.
  Thank you to Mark Hayward, Chris Gardner, Tyler Smith, Cannon Vogel, Wataru Hayashi, Vidya Venkatesan, and Garrett Prechel at UCI.
  At CERN, I had the fortune of meeting many remarkable people.
  To name a few: Gabriel Matos, Punit Sharma, Andrew Smith, Elena Busch, Siddarth Singh, Sylvia Mason, Jared Burleson, Ryan Roberts, Emily Duden, Ruth del Pino Bleijerveld, Sebastian Rutherford, Patrick Dougan, Maggie Chen, Anna Mullin, Ynyr Harris, Iza Veliscek, Alessandro Ruggiero, Sid Baines, Harry Simpson, and Beth Spear.
  To my friends in Palo Alto, thank you for making me feel at home in the space of just one year: Cooper Wagner, Brendon Bullard, Laura Bullard, Prajita Bhattari, Alic Spellman, Aviv Simchony, Mirella Vassilev, Anjali Kumar, and Shubham Lochab.
  Thank you to the friends I've made by attending random conferences.
  Enjoying expensed travel and learning how to be a scientist alongside you has been a great pleasure: Rikab Ghambir, Radha Mastandrea, and Sean Benevedes.
  Thank you especially to my friends Sagar Addepalli, Eleanor Woodward, Antti Pirttikoski, Lauren Osojnak, Tom Kreße, Ki Ryeong Park, and Margot van Eijkern for being the best discoveries of my Ph.D.

  To my friends from Notre Dame who kept in touch with me through all of the years of grad school and beyond: Colin Vaughn, Saul Cortez, Sean Hullihan, David Sholar, Connor Kooistra, Taejun Kim, Hannah Oikawa, Tommy Clare, Bruce Morris, Will Neely, Jacob Naatz, John Fahrenbach, and Elizabeth Boyle.
  To the Phlounge Brew Crew, who taught me from very early on that science was better enjoyed with others than on one's own: Connor Bagwell, David Shaw, Jenna Streich, Anne Stratman, Cecilia Fasano, Joe Levano, Ryan Kim, Aaron Rea, and Ben Riordan.

  Finally very special thanks to those who traveled across an ocean to visit me in Geneva or attend my Ph.D. defense.
  To Sam Porter, thank you for the travels, friendship, and being my only friend who is a real experimentalist.
  To Rob Engelhard, high school friends who board together stay together.
  To Max Silvester, thank you for the wild adventures, good vibes, good food, and outstanding company.
  To Yao Yin, thank you for being the most exciting yet ruthlessly competent person I know.
  To Kendrick Peterson and Claire Stanecki, I love you both to death and am so happy I made you visit me together rather than separate.

  Lastly to my family.
  To Aunt Kathy and Uncle Keith for hosting me on many holidays and regular days at your home in Santa Barbara while I lived in California.
  Also to Aunt Lisa, Brianna, and Aaron for helping me feel right at home in the golden state.
  To my brothers Michael and Greg: thank you for being such remarkable people and role models.
  I hope to live with half of the courage and conviction that you do.
  To my beautiful partner Laura: thank you for your indomitable positive attitude, your deep passion for your work, and your love for me and everyone you meet.
  Meeting you was the greatest stroke of luck I think I have ever had, and I cannot wait to find out what the future holds!
  Most importantly, thank you to my wonderful parents, Douglas and Anne Greif, to whom this thesis is dedicated.
  Many people claim to have the best parents, but I have the great fortune of being correct.
  Thank you for everything.
  To steal some words from Ernest Rutherford: ``Now Doctor Greif. More your honor than mine.''

}

\newcommand{\aiusestatementpage}{%
  \begin{center}
    \textbf{\Large STATEMENT ON THE USE OF AI TOOLS}
  \end{center}
  \parskip 12pt
  \parindent 0pt

  AI tools were used frequently in the underlying research documented in this thesis, especially for writing and editing software.
  Additionally AI tools were used for the ideation and background research needed to compile this thesis, especially for Chapters 2 and 3.
  On the level of the text, AI tools were used for organizing and outlining various sections, but all of the sentences appearing in the final version of the thesis were written by hand.
  Many of the diagrams in this thesis were built with AI tools, and are labeled as such in the figure captions.

  Some overviews of the particular AI-based workflows I used when drafting this thesis are provided in the table below.
  Models used were Claude Sonnet 4.6 and 5.0, Opus 4.6 and 4.7, and GPT 5.4 and 5.5.
  Agentic tools used were Claude Code, Claude Cowork, Cursor, and OpenClaw.

  \vspace{0.5em}
  \begin{singlespace}
  \noindent\begin{tabularx}{\textwidth}{@{}
      >{\raggedright\arraybackslash}p{0.24\textwidth}
      >{\centering\arraybackslash}p{0.13\textwidth}
      X@{}}
    \toprule
    \textbf{Task} & \textbf{AI use} & \textbf{Example workflow} \\
    \midrule
    Section outlining and organization
      & Light
      & Prompt AI to translate ideas and bullet points into \LaTeX{} documents for further editing \\
    \addlinespace
    Copy editing
      & High
      & Prompt AI to read a recently written section and fix any spelling and grammar issues \\
    \addlinespace
    Diagram design
      & High
      & Prompt AI to write code for producing diagrams with Matplotlib or TikZ \\
    \addlinespace
    Literature search
      & Low
      & Prompt AI to search for relevant citations from the InspireHEP database \\
    \addlinespace
    Concept review
      & Medium
      & Prompt AI for explaining and clarifying QCD concepts \\
    \addlinespace
    \LaTeX{} formatting
      & High
      & Prompt AI to debug compilation issues, add Figures to documents, format tables, etc. \\
    \bottomrule
  \end{tabularx}
  \end{singlespace}
  \clearpage
}

\thesisabstract
{
  Proton-proton collisions at the Large Hadron Collider (LHC) offer the opportunity to observe the interactions of fundamental particles at very high energy scales. The high collision rate and large number of final state particles produced per collision imply that the datasets produced by detectors such as the ATLAS experiment are large and high dimensional. The complexity of these datasets calls for the use of novel data analysis techniques which can exploit all of the available information to illuminate known interactions and search for new ones. This thesis presents applications of artificial intelligence (AI) techniques to improve the physics results of the ATLAS experiment. It centers on a full-phase-space measurement of the Z+jets production cross section at the LHC, where the cross section is measured differential in the kinematics of every final state charged particle through the use of an AI-based unfolding algorithm. This is the first such measurement performed at the LHC, which provides a complete experimental characterization of the Z+jets production process and a proof-of-principle for the further pursuit of such measurements on a host of LHC processes.
}


%% file: vita.tex
\newcommand{\vitaentry}[3]{%
  \begin{tabular*}{\textwidth}{@{\extracolsep{\fill}}lr}
    \textbf{#1} & \textbf{#2}
  \end{tabular*}\\[-0.3em]
  \textit{#3}\vspace{0.8em}
}

\newcommand{\pubentry}[4]{\item #1. ``#2''. #3, #4.}

\curriculumvitae{

\textbf{EDUCATION}
\vspace{0.5em}

\vitaentry{Doctor of Philosophy in Physics and Astronomy}{Expected 2026}%
  {University of California, Irvine \hfill Irvine, California \\
   Advisor: Daniel Whiteson}

\vitaentry{Bachelor of Science, Magna Cum Laude}{2020}%
  {University of Notre Dame \hfill Notre Dame, Indiana \\
   Majors: Physics and Program of Liberal Studies}

\vspace{0.5em}

\textbf{RESEARCH EXPERIENCE}
\vspace{0.5em}

\vitaentry{Foundation models for particle physics}{Ongoing}%
  {Researched the application of pre-trained foundation models in particle physics. Compared their performance to equivariant neural networks, and investigated the proper composition of pre-training data to optimize scaling laws.}

\vitaentry{Artificial Intelligence based unfolding}{Ongoing}%
  {Developed novel unfolding methods based on generative models
   and contributed to community comparison studies. Studied the use of
   unbinned, highly differential cross section measurements for parameter
   estimation. Deployed Artificial Intelligence based unfolding to perform a high- and variable-dimensional
   cross section measurement of $Z$+jets production
   with the ATLAS detector.}

\vitaentry{Constituent based top tagging}{2022--2024}%
  {Benchmarked several constituent top-tagging algorithms on realistic detector
   simulation. Assessed systematic uncertainties in tagger performance.
   Prepared research-grade open data records for use by the broader particle
   physics community.}

\vitaentry{Calibration of the ATLAS New Small Wheel}{2022--2023}%
  {Calibrated the peak detector output (PDO) readout used in the ATLAS New Small
   Wheel muon detector. Participated in detector operations during Run 3.}

\vitaentry{Physics inspired neural networks}{2018--2020}%
  {Deployed a physically motivated neural network architecture on a
   classification problem in particle physics.}

\vitaentry{CMS track trigger development}{2017--2019}%
  {Improved the efficiency of cuts on track candidates in the CMS experiment's
   track trigger algorithm.}

\vspace{0.5em}

\textbf{SELECTED PUBLICATIONS}
\vspace{0.5em}

{\small Only includes papers in which I made a significant contribution,
in reverse chronological order.}
\vspace{0.3em}

\begin{enumerate}
  \pubentry{J.-L.~Uslu et al.}%
    {Towards Engineering Scaling Laws with Pretraining Data Composition}%
    {\textit{Preprint:} arXiv:2606.19781}{2026}
  \pubentry{V.~Breso-Pla et al.}%
    {Explicit or Implicit? Encoding Physics at the Precision Frontier}%
    {\textit{Preprint:} arXiv:2603.08802}{2026}
  \pubentry{A.~Petitjean et al.}%
    {Generative unfolding of jets and their substructure}%
    {\textit{Preprint:} arXiv:2510.19906}{2025}
  \pubentry{A.~Shmakov et al.}%
    {Full event particle-level unfolding with variable-length latent variational diffusion}%
    {\textit{SciPost Phys.}~18(4):117}{2025}
  \pubentry{N.~Huetsch et al.}%
    {The landscape of unfolding with machine learning}%
    {\textit{SciPost Phys.}~18(2):070}{2025}
  \pubentry{ATLAS Collaboration}%
    {Accuracy versus precision in boosted top tagging with the ATLAS detector}%
    {\textit{JINST}~19(08):P08018}{2024}
  \pubentry{A.~Shmakov et al.}%
    {End-to-end latent variational diffusion models for inverse problems in high energy physics}%
    {\textit{NeurIPS Proceedings}}{2023}
  \pubentry{K.~Greif and K.~Lannon}%
    {Physics Inspired Deep Neural Networks for Top Quark Reconstruction}%
    {\textit{EPJ Web Conf.}~245:6029}{2020}
\end{enumerate}

\vspace{0.5em}

\textbf{PRESENTATIONS}
\vspace{0.5em}

\begin{tabular*}{\textwidth}{@{\extracolsep{\fill}}p{2.2cm}p{0.78\textwidth}}
  \textbf{Jul.\ 2025} & ``Forward folding versus unfolding in the age of ML'', \textit{ML4Jets 2025} \\[0.4em]
  \textbf{Nov.\ 2024} & ``Full Event Particle-Level Unfolding with Variable Length Variational Latent Diffusion'', \textit{ML4Jets 2024} \\[0.4em]
  \textbf{Nov.\ 2023} & ``Systematic Effects in Jet Tagging Performance for the ATLAS Detector'', \textit{ML4Jets 2023} \\[0.4em]
  \textbf{Nov.\ 2023} & ``End-to-End Latent Variational Diffusion Models for Unfolding LHC Events'', \textit{ML4Jets 2023} \\[0.4em]
  \textbf{Nov.\ 2022} & ``Constituent-Based Top-Quark Tagging with the ATLAS Detector'', \textit{ML4Jets 2022} \\[0.4em]
  \textbf{Nov.\ 2019} & ``Physics Inspired Deep Neural Networks for Top Quark Reconstruction'', \textit{CHEP 2019} \\[0.4em]
\end{tabular*}

\vspace{0.5em}

\textbf{APPOINTMENTS AND MEMBERSHIPS}
\vspace{0.5em}

\vitaentry{ATLAS Early Career Scientist Board}{2024--Present}%
  {Advocate for the interests of early career scientists within the ATLAS
   collaboration. Organize educational and career development events.}

\begin{tabular*}{\textwidth}{@{\extracolsep{\fill}}lr}
  \textbf{Phi Beta Kappa Society}  & \textbf{2020--Present} \\
  \textbf{Sigma Pi Sigma Society}  & \textbf{2020--Present} \\
\end{tabular*}

\vspace{0.5em}

\textbf{AWARDS}
\vspace{0.5em}

\begin{tabular*}{\textwidth}{@{\extracolsep{\fill}}p{0.55\textwidth}r}
  \textbf{Outstanding Physics Major}  & \textbf{2020} \\
  \multicolumn{2}{p{0.95\textwidth}}{\textit{Physics Department, University of Notre Dame}} \\[0.4em]
  \textbf{Paul Chagnon Award in Physics}  & \textbf{2020} \\
  \multicolumn{2}{p{0.95\textwidth}}{\textit{Physics Department, University of Notre Dame --- recognizing service to the department and fellow students}} \\[0.4em]
  \textbf{Edward J.\ Cronin Award}  & \textbf{2020} \\
  \multicolumn{2}{p{0.95\textwidth}}{\textit{Program of Liberal Studies, University of Notre Dame --- awarded for the finest piece of written work submitted in an academic year}} \\[0.4em]
  \textbf{Hichwa Fellowship}  & \textbf{2019} \\
  \multicolumn{2}{p{0.95\textwidth}}{\textit{Physics Department, University of Notre Dame}} \\
\end{tabular*}

\vspace{0.5em}

\textbf{TEACHING EXPERIENCE}
\vspace{0.5em}

\begin{tabular*}{\textwidth}{@{\extracolsep{\fill}}lr}
  \textbf{Teaching Assistant, University of California, Irvine} & \textbf{2020--2021} \\
\end{tabular*}

\begin{itemize}
  \item Physics 3A (Mechanics), Fall 2021
  \item Physics 3LB (Electricity and Magnetism Lab), Spring 2021
  \item Physics 7LC (Mechanics Lab), Winter 2021
  \item Physics 2 (Mathematical Methods in Physics), Fall 2020
  \item Computational Methods in Physics, Spring 2020
\end{itemize}

\vspace{0.5em}

\textbf{OUTREACH}
\vspace{0.5em}

\vitaentry{ATLAS Virtual Visit Tour Guide}{2022--2025}%
  {Remote visits to the ATLAS experiment for students from around the world.}

\vitaentry{UCI Physics and Astronomy Blog}{2020--2022}%
  {Published accessible descriptions of new research from the UCI physics
   department.}

} 

%% file: chapter1.tex
\chapter{Introduction}
\label{ch:intro}



The discovery of the Higgs boson, announced on July 4th, 2012~\cite{ATLAS:2012yve,CMS:2012qbp}, is widely regarded as the most important discovery in particle physics in the 21st century.
It provided experimental confirmation of the electroweak symmetry breaking mechanism~\cite{PhysRevLett.13.321, PhysRevLett.13.508, PhysRevLett.13.585} and completed the Standard Model of particle physics~\cite{PhysRevLett.19.1264, THOOFT1972189}.
This triumph of modern science was front-page news, and is still well known in the popular imagination.
A less well known, though equally important, advancement was announced 88 days later on September 30th, 2012 when the results of the ``ImageNet Large Scale Visual Recognition Challenge'' were released.
This was a challenge within the computer vision research community to develop a machine learning model that could sort images into 1000 classes.
The training set for this challenge consisted of 1.2 million painstakingly hand-labeled images, which was unprecedented scale for labeled datasets at the time.
This scale and the recent release of Nvidia's CUDA toolkit, which allowed for optimized matrix operations on GPU hardware, allowed for a team of researchers from the University of Toronto to train a deep convolutional neural network named AlexNet that outperformed its closest competitor by 10\% in top-5 error rate\footnote{The top-5 error rate is the percentage of images that were not correctly classified into the top 5 most likely classes.}~\cite{krizhevsky2012imagenet}.
Before this moment, state-of-the-art computer vision models relied on hand-engineered features programmed by human experts, but AlexNet learned all features used to classify the images directly from the data, requiring the humans who programmed it to know essentially nothing about the images themselves.
This breakthrough became known as the ``AlexNet moment'' and marked the beginning of the modern era of machine learning.

This thesis is an attempt to understand the impact of the AlexNet moment on particle physics research in the post-Higgs era.
Both developments led directly or indirectly to a Nobel Prize in physics.
The prize was awarded to Peter Higgs and François Englert in 2013 as a direct result of the discovery of the Higgs boson, and to John Hopfield and Geoffrey Hinton in 2024 for foundational work on neural networks that enabled AlexNet.
The Nobel Prize for Higgs and Englert was the 12th Nobel Prize awarded for the development of the Standard Model\footnote{This is by my count. Some notable exclusions are the Nobel Prizes of Feynman, Schwinger, and Tomonaga for the QED theory which is a precursor of the Standard Model, and the Nobel Prizes of Kajita and McDonald for the discovery of neutrino oscillations, which are not a part of the Standard Model.}.
It was the capstone of a very fruitful and now very mature research program.
By contrast the Nobel for artificial neural networks was at least in-part an instrumentation award\footnote{An excellent parallel is Charles Wilson's development of the cloud chamber, a tool developed by a meteorologist that became very useful for fundamental physics research and eventually earned Wilson the Nobel Prize.}, significant for the other discoveries it enables.
Particle physics was one of the first scientific disciplines to make use of deep learning.
The first papers demonstrating the use of deep learning for event selection~\cite{Baldi:2014kfa} and jet tagging~\cite{Cogan:2014oua} were written in 2014, only two years after AlexNet.
Since then there has been a remarkable growth in the number of papers applying deep learning to particle physics~\cite{hepmllivingreview}.
This thesis will provide a tour of some of these applications.
Concurrently, most other sub-fields of physics have taken interest in applying deep learning tools.
Cosmology, astrophysics, gravitational wave science, condensed matter physics, plasma physics, fluid dynamics, quantum information, and optical physics have all seen fruitful applications.
There are few areas of active physics research that have not been affected by the AlexNet moment in some sense.
While these applications have yet to produce a landmark discovery or new capability like the AlphaFold protein structure prediction model that was awarded the Nobel Prize in chemistry in 2024, the breadth of these applications is a testament to the power of deep learning for physics research.

In the remainder of the Introduction, I will outline the open problems in fundamental physics\footnote{I generally refer to the fields of particle physics and cosmology as \textit{fundamental physics}, given open questions do not necessarily sort neatly into the two sub-fields.}, and very briefly summarize the last 14 years of progress in machine learning and artificial intelligence.
This review serves two purposes.
The first is to give context to the very specific research detailed in this thesis.
The second is to reflect on the fact that all open problems in particle physics are long-standing, and that the list is essentially unchanged since the 2012 discovery of the Higgs.
These problems are extremely subtle and difficult, and many years of effort by very resourceful and intelligent scientists has so far proven insufficient to solve them.
By contrast, the field of machine learning and artificial intelligence is unrecognizable compared to what it was in 2012.
The hope of this thesis is that some of this dizzying progress can finally break the current impasse facing particle physics.


\section{Open Problems in Fundamental Physics}
\label{sec:open_problems}

The most pressing open problems in fundamental physics can be understood as either the need for a better understanding of Beyond the Standard Model (BSM) physics, or evidence of Beyond $\Lambda$CDM cosmology.
BSM physics is known to exist through multiple experimental observations that cannot be explained by the particle content and forces contained in the Standard Model.
However the correct quantum field theory (or other type of theory consistent with the existing Standard Model) that explains these observations is unknown.
$\Lambda$CDM cosmology accounts for all cosmological observations very well, with the exception of some significant tensions, namely the Hubble tension.
My list of the open problems is as follows:

\begin{enumerate}
    \item{
        \textbf{Neutrino oscillations}: Neutrinos are massless under the Standard Model. However many iterations of neutrino experiments have verified that these particles ``oscillate'' between the three neutrino flavors as they propagate through space. This phenomenon is easily explained if the neutrinos carry a mass that is small enough to remain currently undetected by experiments, but the mechanism that produces this mass is unknown. This topic is not covered at all in this thesis, but current and next generation neutrino experiments are making active use of deep learning.
    }
    \item{
        \textbf{Dark matter}: Observational evidence suggests that most matter in the universe is composed of some particle that does not interact electromagnetically and so does not radiate light.
        No Standard Model particles fit this description, so the remaining matter must be composed of \textit{dark matter} that at the moment is only known to interact with the Standard Model particles through gravity.
        Huge amounts of experimental data support the existence of dark matter and much is known about how it is distributed throughout the universe, but the properties of the fundamental particles out of which it is composed are a mystery.
    }
    \item{
        \textbf{Matter--antimatter asymmetry}: The Standard Model predicts that the Big Bang should have produced nearly equal amounts of matter and antimatter, but the observable universe is composed of almost entirely matter.
        This observed asymmetry requires some mechanism for violation of the CP discrete symmetry in the early universe.
        The CP violation in the Standard Model, coming from the complex phase of the CKM quark mixing matrix and to a very small degree the strong CP phase discussed below, are too small to account for the asymmetry on their own~\cite{Sakharov:1967dj}.
        Therefore BSM physics is known to produce some amount of CP violation, but the exact mechanism remains unknown.
    }
    \item{
        \textbf{Dark energy and the Hubble tension}: Cosmological observations indicate that the universe is expanding, and that the expansion is speeding up over time.
        If the universe was composed of only the baryonic matter of the Standard Model and dark matter, this expansion would not be occurring.
        Therefore there must be an additional type of energy density, termed \textit{dark energy}, that is driving this expansion.
        A good explanation for this dark energy is a non-zero cosmological constant $\Lambda$, however this model has significant theoretical issues as discussed below.
        Additionally measurements of the expansion rate of the universe have produced inconsistencies.
        The $\Lambda$CDM model of cosmology parametrizes the expansion of the universe through the Hubble constant.
        This constant can be measured through two probes: the cosmic microwave background~\cite{Planck:2018vyg} and local distance-ladder observations~\cite{Riess:2021jrx}.
        However these two approaches yield values of the Hubble constant $H_0$ that disagree at the 4--5$\sigma$ level.
        This tension could be caused by an unaccounted for systematic effect and is under very active research.
        However if confirmed this would be a clear sign of beyond $\Lambda$CDM cosmology.
    }
    \item{
        \textbf{Gravity}: The gravitational interactions are not present within the Standard Model.
        The general theory of relativity provides an extraordinarily accurate classical theory of gravity, but none of the attempts to quantize gravity and incorporate it with the other known forces has received experimental verification.
        At the energy scales probed by the LHC gravitational effects are negligible, so this is not a direct experimental puzzle for collider physics, but our everyday experience of the gravitational force is evidence of BSM physics.
    }
    \item{
        \textbf{Origin of primordial density fluctuations}: The presence of the Baryon Acoustic Oscillation (BAO) peak in the CMB power spectrum is precisely predicted by $\Lambda$CDM cosmology, but requires the presence of primordial density perturbations to seed the oscillations.
        The mechanism that formed these perturbations in the early universe is not known, though inflation provides a compelling framework which requires the presence of a BSM field.
        Given there are competing models and experimental evidence is scarce, this is less of a clean example of BSM physics than those listed above, but it is worth mentioning.
    }
\end{enumerate}

In addition to open problems related to experimental observations, there are many more open problems related to the theory of the Standard Model and its possible extensions.
To make them more tractable I would group them into three categories.
The first, and arguably the most salient at this moment in particle physics, are related to fine-tunings of the free parameters of the Standard Model and $\Lambda$CDM.
Whether these fine-tunings are true problems or only an issue of philosophical priors is often debated, but I find them troubling enough to deserve mention.

\begin{enumerate}
    \item{
        \textbf{The hierarchy problem}: In the Standard Model, the Higgs boson mass receives quantum corrections from every field that couples to the Higgs field.
        Given it is difficult to imagine a quantum theory of gravity or generally any BSM model that does not couple to the Higgs directly or through loops, these quantum corrections should pull the Higgs mass up to the scale of the UV cutoff of the Standard Model.
        Instead the Higgs has a mass of 125~GeV.
        This requires either that the quantum corrections are cancelled by some new BSM physics near the electroweak scale, or that the scale of the corrections and the bare mass of the Higgs cancel nearly exactly despite their large size.
        This is the canonical example of a fine-tuning.
        Supersymmetry, compositeness, and extra dimensions are the most studied solutions to this problem, but none have been confirmed experimentally.
        The lack of evidence from the LHC experiments for any BSM physics near the electroweak scale has made the hierarchy problem significantly more confusing.
    }
    \item{
        \textbf{The strong CP problem}: Quantum chromodynamics (QCD) allows for a CP-violating term in the Standard Model Lagrangian.
        This term is experimentally known to be very small through limits on the neutron electric dipole moment, but there is no reason that this parameter should be small in the Standard Model.
        This is another fine-tuning problem, though less relevant to LHC physics.
    }
    \item{
        \textbf{The cosmological constant problem}: A possible explanation for the accelerating expansion of the universe mentioned above is that empty space has a non-zero energy density known as \textit{vacuum energy}.
        Naive theory estimates for this energy density predict a value $\sim 10^{120}$ times larger than the energy density which would explain observations.
        This is the most dramatic fine-tuning problem in all of physics.
        No theory which explains why the vacuum energy, if it is the mechanism that produces the observed accelerating expansion of the universe, should be so small has received experimental confirmation.
    }
\end{enumerate}

The second category consists of problems that are not fine-tunings, at least to the degree of the problems above, but are features of the Standard Model that are arbitrary and seem to beg for some deeper explanation than what is provided by the current theory.
I do not discount such feelings as irrelevant as they have produced discoveries in the past\footnote{Dirac's prediction of the positron through aesthetic considerations about negative solutions to the Dirac equation is perhaps the most relevant one. The GIM mechanism is a particle physics example of how taking a fine-tuning seriously led to new discoveries (the charm quark).}.

\begin{enumerate}
    \item{
        \textbf{The flavor puzzle}: The Standard Model contains three generations of quarks and leptons that share identical gauge quantum numbers but differ enormously in mass, with Yukawa couplings spanning nearly five orders of magnitude from the electron to the top quark.
        No symmetry principle explains why there are exactly three generations, why these Yukawa couplings span such large scales, or why the quark mixing angles (CKM matrix) are small while two of the corresponding angles in the lepton sector (PMNS matrix) are large.
        These hierarchies and mixing patterns are encoded in 13 of the 19 free parameters of the Standard Model.
        There is no theoretical motivation for their values.
        Instead they have been determined through experimental measurements, and their somewhat arbitrary structure is an odd feature of the Standard Model.
    }
    \item{
        \textbf{Accidental symmetries}: Several conservation laws of the Standard Model, for example conservation of baryon number $B$ and lepton number $L$, are not imposed by the symmetries of the theory.
        Instead, they are accidents of the Standard Model gauge group, particle content, and renormalizability constraints.
        The fact that they are present and seem to hold experimentally is a curious feature of the Standard Model.
    }
    \item{
        \textbf{V-A structure of the weak interaction}: The weak interaction couples exclusively to the left-handed fermion fields.
        Another way of stating this is that the weak interactions are maximally parity violating: the weak force would affect the dynamics of some particles but be completely turned off for their parity inverted counterparts.
        This is an experimentally established fact but there is no clear theoretical explanation for why the weak forces should single out left-handed fields.
    }
\end{enumerate}

The final category is in my opinion the most important one.
This is because it is indisputably tractable at this exact moment in particle physics, and has direct implications for how data from particle physics experiments are processed to hopefully make progress on the many problems quoted above.
Furthermore it is the only set of open questions that is directly relevant to the research presented in this thesis.
It is problems relating to our ability to calculate within the Standard Model.

\begin{enumerate}
    \item{
        \textbf{Calculation of scattering amplitudes}: The Standard Model is predictive because it can be used to calculate scattering amplitudes that are tested every day in particle physics experiments.
        These amplitudes are almost always calculated at a fixed order in some perturbative expansion, and pushing known amplitudes to ever higher orders is an important open question in particle physics.
        In particular, improved modeling of Standard Model processes, whether as signal in measurements or background in searches, depends on making progress in this direction.
    }
    \item{
        \textbf{Non-perturbative QCD}: Quantum chromodynamics is in principle a complete theory of the strong force, but the comparatively large value of the strong coupling constant $\alpha_s$ means perturbative expansions are only valid at high energies.
        At low energies, lattice QCD provides a systematic non-perturbative approach but is computationally expensive and limited in scope.
        This leaves us with no first-principles theoretical tools for modeling QCD dynamics at or below the QCD confinement scale of $\Lambda_{QCD} \sim 200$~MeV.
        Instead we have empirical models whose parameters must be tuned to match to experimental data.
        Improving these models is an active area of research, for which the cross section measurement presented in Chapter~\ref{ch:zjets} provides valuable experimental data.
        Apart from this practical problem, the fact that experimental observations like positive mass confined states cannot be understood through perturbation theory is also a deep problem of mathematical physics.
        Proving that quantum field theories like QCD produce strictly positive mass confined states is one of the Millennium Prize Problems of mathematics.
    }
\end{enumerate}

Experimental evidence is needed to guide theoretical progress on both of these problems.
It is exactly that evidence that the research in this thesis provides at a uniquely large and flexible scale.

Finally, the best way to make progress on many of these problems is unambiguously to find a new particle (necessarily one not in the Standard Model).
Some historical data is relevant here.
The current drought in new particle discoveries is currently 14 years long.
This is long compared to the ``golden era'' of particle physics between 1965 and 1985 when new particles were discovered every few years, but not incredibly long by the standards of the years preceding and following.
For example there was a 20 year gap between the discovery of the muon in 1936 and the electron neutrino in 1956.
Whatever geopolitical forces may have been at play in that time, particle physics research is clearly not always smooth sailing.
However the next particle will soon be overdue.
Some reflections on the prospects for finding it are offered in the conclusion.


\section{The Deep Learning Revolution}
\label{sec:ml_intro}

Since the AlexNet moment in 2012, the field of machine learning and artificial intelligence has undergone rapid progress.
A brief survey of this progress and how it has entered particle physics research is as follows:

In the years following AlexNet, deeper networks with many millions of parameters provided even better performance on classification tasks~\cite{Simonyan:2014vgg,He:2015resnet}.
This was an early example of how larger scale improves neural network performance.
The particle physics community adopted deep learning very soon after AlexNet showed promise in computer vision, with the first papers proposing deep-learning-based jet taggers appearing only two years after AlexNet in 2014~\cite{Cogan:2014oua,Baldi:2014kfa}.
Neural network based generative models, which produce additional samples from the underlying probability distribution of the training data, were also developed in this period.
Generative Adversarial Networks (GANs)~\cite{Goodfellow:2014gan} and Variational Autoencoders (VAEs)~\cite{Kingma:2013vae} were early generative models that were picked up by the HEP community for generative tasks like fast calorimeter simulation and anomaly detection~\cite{deOliveira:2017pjk,Paganini:2017hrr,Paganini:2017dwg,Cerri:2018anq}.
A large milestone in the development of artificial intelligence through reinforcement learning (RL) came in 2016 when AlphaGo~\cite{Silver:2016alphago} defeated the world champion Go player.

A next landmark was the introduction of the \textit{attention mechanism}~\cite{bahdanau2014neural,DBLP:journals/corr/LuongPM15} and the Transformer architecture~\cite{Vaswani:2017attention} in the years 2015--2017, which demonstrated state-of-the-art performance on machine translation tasks\footnote{It was also around this time that Google updated their Google Translate service to use neural machine translation. The development gathered enough media attention that I read about it as a first-year undergraduate. This touched off my interest in machine learning, and my decision to study physics was partially based on the fact that the CMS group was one of the few research groups utilizing deep learning at my undergraduate institution.}.
Attention's remarkable expressivity and the ability to parallelize the required computations on GPU resources meant Transformers replaced recurrent networks as the dominant sequence-processing architecture.
The first HEP papers to use attention followed a few years later~\cite{Mikuni:2020wpr}, and attention is now widely used by both ATLAS and CMS for jet classification as discussed in \Cref{sec:tagging-future}.
Many of the neural networks discussed in this thesis also utilize the attention mechanism.

Following the introduction of attention, the focus in ML/AI research was on scaling and generative modeling.
Large language models (LLMs) trained on huge amounts of text scraped from the internet began to demonstrate \textit{emergent} capabilities, which only appeared once the dataset and model sizes passed a certain scale.
For example, multi-step reasoning only began to appear once model sizes passed roughly 100 billion parameters~\cite{Wei:2022emergent}.
The CLIP model~\cite{Radford:2021clip} demonstrated the ability of neural networks to process multi-modal data, for example by developing joint embeddings for text and data simultaneously.
During this time the community of researchers applying deep learning in particle physics was broadly focused on generative models.
A lot of very rapid progress was driven by applying increasingly elaborate generative model architectures, from GANs to VAEs to normalizing flows~\cite{Durkan:2019nsf}, to diverse problems like jet generation~\cite{Butter:2019cae}, calorimeter simulation~\cite{Paganini:2017hrr}, unfolding~\cite{Datta:2018mwd, Bellagente:2020piv, Howard:2021pos}, and anomaly detection~\cite{Nachman:2020lpy, Hallin:2021wme}.
Then another landmark occurred between 2021 and 2022 with the introduction of diffusion models~\cite{Ho:2020ddpm,Song:2021score}.
In the AI community, diffusion-based image generators like DALL-E~2, Stable Diffusion, and Midjourney made headlines by producing photo realistic images conditioned on text inputs.
In HEP, diffusion models were quickly picked up and applied to the same set of generative tasks.
Diffusion model based unfolding methods are a major topic in this thesis covered in Chapter~\ref{ch:unfolding}.
Note that neural networks had gone from classifying images to generating photorealistic images of arbitrary subjects in the space of 10 years.

Another milestone occurred in late 2023 with the introduction of ChatGPT.
The introduction of a reinforcement learning based post-training~\cite{Ouyang:2022instructgpt} to LLMs allowed them to write computer code, reason through multi-step problems, and pass professional licensing examinations.
ChatGPT added 100 million users in two months, faster than any consumer product in history, and the term ``artificial intelligence'' entered colloquial use.

In the past two years, frontier LLMs have reached scales of trillions of parameters trained on many terabytes of text data.
Simultaneously, advanced RL pipelines have taught language models to use external tools, for example to edit files or search the web.
This has produced agentic AI systems that are capable of solving increasingly complicated and long-time horizon tasks autonomously.
Massive labor market displacement due to these systems is now a real concern, or more relevant to this thesis, some recent physics research papers were written almost entirely by AI systems under supervision from human experts~\cite{Schwartz:2026ekw, Guevara:2026qzd}.
Regardless of whether this is the future of scientific research, it is evidence of extraordinary progress.
I will return to some opinions on autonomous physics research in the conclusion.

Before moving on, two points about the above deserve some attention.
First, the progress described above is remarkable but it has not been evenly distributed across the entire field of machine learning.
Many foundational problems remain as unsolved as they were in 2012.
For example there is no theoretical explanation for how neural networks learn via gradient descent, making it difficult to predict or interpret the functions learned by neural networks.
Neural networks are also notoriously unreliable when applied on out-of-distribution data or subjected to adversarial attacks.
This bears some similarity to how fundamental questions in quantum mechanics were eventually set aside by the physics community because taking the formalism and running with it proved to be much more fruitful.
The lesson is that fields move forward in unexpected directions, and that progress sometimes looks like re-focusing the question rather than stubbornly pursuing a solution.
Second, the year 2022 strikes me as something of a turning point for research applying deep learning to particle physics.
Between 2012 and 2022, the main research directions of the broader AI community produced general-purpose data science tools that the HEP community could adapt relatively directly.
After 2022, the dominant direction of frontier AI research has shifted toward scaling, tool use, and the pursuit of general-purpose intelligence.
The question of how pursuit of scaling maps onto fundamental physics is a fascinating one that I will cover in \Cref{sec:tagging-future}, but generally speaking this direction does not map as cleanly onto the specific needs of fundamental physics.
Without new tools coming out of the frontier AI research community, the particle physics community has largely settled on Transformers and diffusion models as most performant deep learning tools.
The task is now to use those tools to achieve concrete physics goals.

%

\section{Thesis Roadmap}
\label{sec:roadmap}

The remainder of this thesis is organized as follows.

\Cref{ch:qcd} provides theoretical background on quantum chromodynamics (QCD), since the cross section measurement at the core of the thesis is fundamentally a measurement of QCD processes.
This chapter is highly selective.
Rather than surveying the Standard Model broadly, it focuses on the aspects of QCD that are directly relevant to jet physics and jet substructure at the LHC.
This section will also introduce some of the exotic jet substructure observables measured in \Cref{ch:zjets}.

\Cref{ch:ml} introduces the deep learning methods that are used extensively in the research chapters that follow.
The goal of this chapter is to provide the background necessary to understand the variational latent diffusion models used in the unfolding work of \Cref{ch:unfolding}, and the classifier models used for the cross section measurement in \Cref{ch:zjets}.

\Cref{ch:tagging} turns to the first of the original research contributions in the thesis.
It presents the results of an ATLAS paper which demonstrates the performance of deep-learning-based classifiers and provides a novel framework for uncertainty quantification~\cite{ATLAS:2024rua}.
Given this study was published some time ago, the chapter concludes with a summary of recent progress and future directions in jet tagging at the LHC.

\Cref{ch:unfolding} discusses the problem of \textit{unfolding} in particle physics.
It begins with a review of classical unfolding methods, specifically Iterative Bayesian Unfolding (IBU).
Several deep-learning-based unfolding methods are then described in detail.
First, the Variational Latent Diffusion approach I developed with collaborators is covered.
These results have been published in Refs.~\cite{Shmakov:2023kjj,Shmakov:2024gkd}.
Second, a method specifically for unfolding jet substructure is covered.
I advised on the development of this method, and the results have thus far been published as a preprint~\cite{Petitjean:2025tgk}.
Finally, the \Omnifold method is introduced~\cite{Andreassen:2019cjw}.

\Omnifold is then used to perform a high- and variable-dimensional cross section measurement of $Z$+jets production in \Cref{ch:zjets}.
These results have not yet been published and are currently in ATLAS internal review.
After describing the methodology, uncertainty treatment, and validation, the chapter presents results across four physics use cases that illustrate the power of unbinned and high-dimensional cross section measurements.
It concludes with future prospects and argues that unbinned and high-dimensional cross section measurements should become standard practice in particle physics experiments.

\Cref{ch:conclusion} concludes the thesis by drawing on overall themes and discussing future roles of deep learning and artificial intelligence in particle physics research.

%% file: chapter2.tex
\chapter{Quantum Chromodynamics at the Large Hadron Collider}
\label{ch:qcd}



\begin{figure}[ht]
  \centering
  \includegraphics[width=0.85\linewidth, alt={Diagram of the Standard Model elementary particles, organized by quarks, leptons, gauge bosons, and the Higgs boson, with their masses, charges, and spins labeled.}]{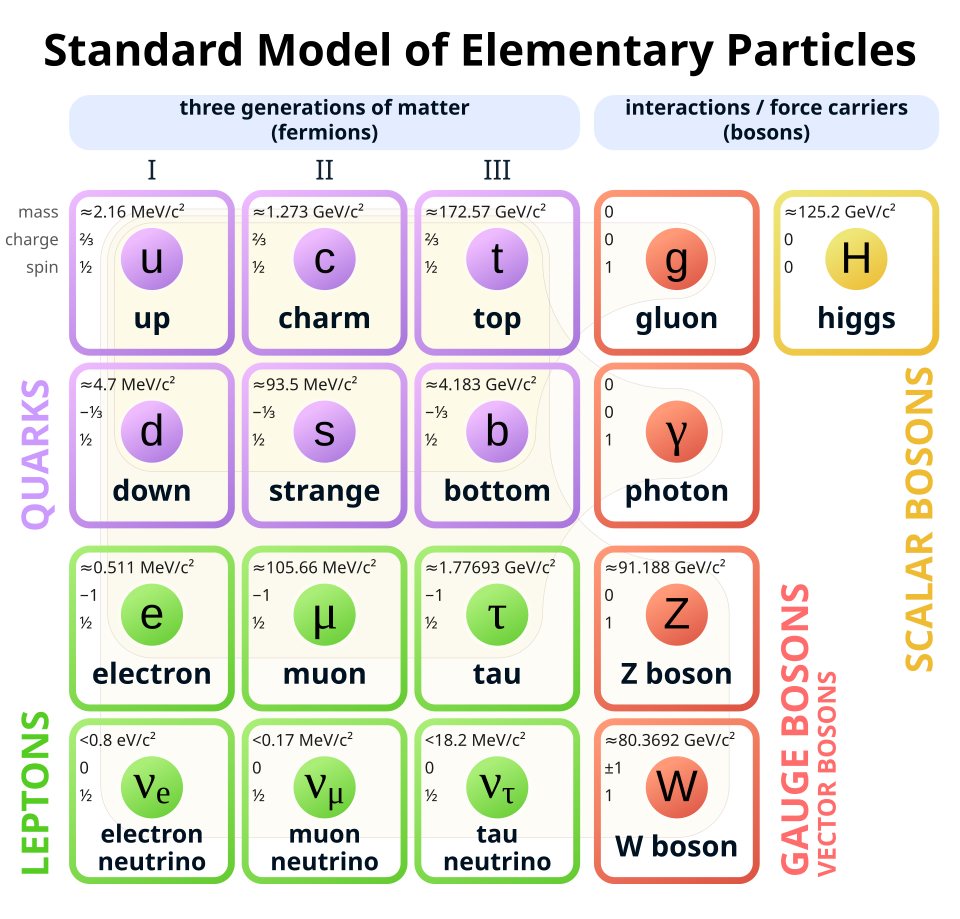}
  \caption{The elementary particles of the Standard Model. The dynamics of quarks and gluons, described by the theory of quantum chromodynamics, are most relevant to this thesis. Figure reproduced from Ref.~\cite{Cush:SMdiagram} under the CC~3.0 license.}
  \label{fig:sm-particles}
\end{figure}

\input{tab_particle_reps.tex}

This chapter provides a very high level overview of QCD, with a focus on the aspects of the theory relevant to the LHC.
It does not survey the Standard Model or QCD with any completeness.
Instead the goal is to motivate two of the observables measured in Chapter~\ref{ch:zjets} from first principles.
They will be introduced in Sections~\ref{sec:eec} and~\ref{sec:id}.

QCD is our current best theory of the strong nuclear force.
It affects the dynamics of all particles that carry color charge, meaning they transform under a non-trivial representation of the $\mathrm{SU}(3)_c$ sub-group of the Standard Model gauge group:
\begin{equation}
  \mathrm{SU}(3)_c \times \mathrm{SU}(2)_L \times \mathrm{U}(1)_Y.
  \label{eqn:sm_gauge_group}
\end{equation}
The particles of the Standard Model are shown in Figure~\ref{fig:sm-particles}.
The representations of each of the fundamental fields of the Standard Model (grouped as quarks, leptons, gauge bosons, and the Higgs doublet fields) are shown in Table~\ref{tab:sm-reps}.
All of the fundamental fields transform in the singlet representation of $\mathrm{SU}(3)_c$ except for the quarks and gluons.
These are the particles that carry color charge and participate in strong-force dynamics.
The quarks are the massive particles that make up all of the baryonic matter in the universe, and the gluons are the massless gauge bosons that mediate the strong force.
These particles are also the input particles to all LHC collisions.
The LHC is a hadron collider, meaning it collides color neutral bound states of quarks and gluons.
The measurement in Chapter~\ref{ch:zjets} is of a cross section measured in proton--proton collisions at $\sqrt{s} = 13$~TeV.
Almost all of the LHC collisions recorded by the ATLAS detector contain strong dynamics, making grappling with the complexities of this force a central challenge of LHC physics.
\textit{Jets} are the experimental signature of the production of high transverse momentum (\pt) color-charged particles.
Jets will be introduced in Section~\ref{sec:jets}, and two methods for studying them will be introduced in Sections~\ref{sec:eec} and~\ref{sec:id}.

\section{Jets and Jet Substructure at the LHC}
\label{sec:jets}

\subsection{The QCD Lagrangian and Asymptotic Freedom}

QCD dynamics are set by the Lagrangian,
\begin{equation}
  \mathcal{L}_{\mathrm{QCD}} = -\frac{1}{4} G^a_{\mu\nu} G^{\mu\nu}_a + \sum_f \bar{q}_f \left( i \gamma^\mu D_\mu - m_f \right) q_f,
  \label{eqn:qcd_lagrangian}
\end{equation}
where $G^a_{\mu\nu}$ is the gluon field strength tensor, the sum runs over the six quark flavors $f$, $D_\mu = \partial_\mu - i g_s T^a A^a_\mu$ is the covariant derivative, $g_s$ is the strong coupling constant, and $T^a$ are the generators of $\mathrm{SU}(3)$.
The gluon field strength tensor is
\begin{equation}
  G^a_{\mu\nu} = \partial_\mu A^a_\nu - \partial_\nu A^a_\mu + g_s f^{abc} A^b_\mu A^c_\nu,
  \label{eqn:field_strength}
\end{equation}
where $f^{abc}$ are the $\mathrm{SU}(3)$ structure constants and $A^a_\mu$ are the gluon fields.

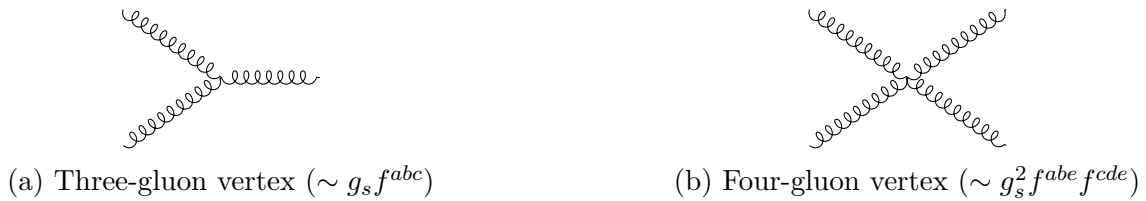
\begin{figure}[ht]
  \centering
  \begin{subfigure}[b]{0.45\textwidth}
    \centering
    \begin{tikzpicture}
      \begin{feynman}
        \vertex [particle=\(g\)] (g1) at (-1.3, 0.9);
        \vertex [particle=\(g\)] (g2) at (-1.3, -0.9);
        \vertex (v1) at (0, 0);
        \vertex [particle=\(g\)] (g3) at (1.3, 0);
        \diagram* {
          (g1) -- [gluon] (v1) -- [gluon] (g3),
          (g2) -- [gluon] (v1),
        };
      \end{feynman}
    \end{tikzpicture}
    \caption{Three-gluon vertex ($\sim g_s f^{abc}$)}
    \label{fig:3g-vertex}
  \end{subfigure}
  \hfill
  \begin{subfigure}[b]{0.45\textwidth}
    \centering
    \begin{tikzpicture}
      \begin{feynman}
        \vertex [particle=\(g\)] (g1) at (-1.3, 0.9);
        \vertex [particle=\(g\)] (g2) at (-1.3, -0.9);
        \vertex (v) at (0, 0);
        \vertex [particle=\(g\)] (g3) at (1.3, 0.9);
        \vertex [particle=\(g\)] (g4) at (1.3, -0.9);
        \diagram* {
          (g1) -- [gluon] (v) -- [gluon] (g3),
          (g2) -- [gluon] (v) -- [gluon] (g4),
        };
      \end{feynman}
    \end{tikzpicture}
    \caption{Four-gluon vertex ($\sim g_s^2 f^{abe} f^{cde}$)}
    \label{fig:4g-vertex}
  \end{subfigure}
  \caption{The gluon self-interaction vertices that arise from the non-Abelian structure of QCD.}
  \label{fig:gluon-vertices}
\end{figure}

A critical difference between QCD and QED is that the structure constants $f^{abc}$ in Equation~\ref{eqn:field_strength} are non-zero for the $\mathrm{SU}(3)$ gauge group of QCD.
The presence of non-zero structure constants introduces cubic and quartic gluon self-interaction vertices in the Feynman rules, allowing the eight gluon fields to couple to each other, as shown in Figure~\ref{fig:gluon-vertices}.
The result is that each gluon field itself carries color charge.
In QED, no analogous photon self-coupling exists and the photon does not carry electric charge.

\begin{figure}[ht]
  \centering
  \includegraphics[width=0.75\linewidth, alt={Compilation plot of measurements of the strong coupling alpha sub s as a function of the energy scale Q, showing that the coupling decreases as Q increases over a wide range of energies.}]{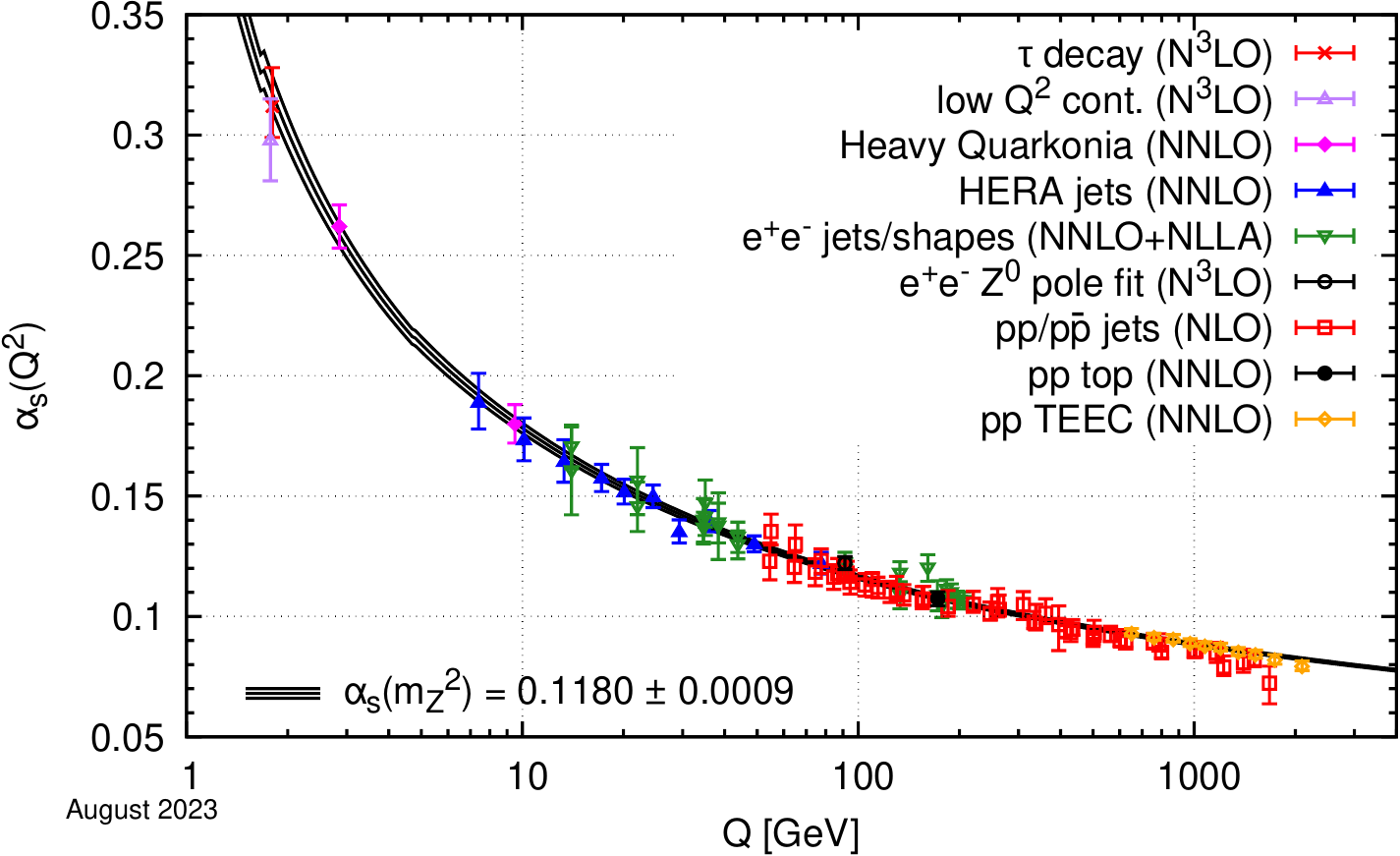}
  \caption{A compilation of experimental measurements of the strong coupling $\alpha_s$ as a function of the energy scale $Q$, illustrating asymptotic freedom through the decrease of the coupling at high energies. Figure reproduced from Ref.~\cite{Huston:2023ofk} under a CC~BY~4.0 license.}
  \label{fig:qcd_running}
\end{figure}

These self-coupling diagrams have important consequences for the running of the strong coupling constant $g_s$.
For the electromagnetic interactions, the coefficient of the one-loop beta function is positive, meaning the coupling increases with the energy scale.
For the strong interactions, the gluon self-interaction contributes a large negative term to the one-loop beta function,
\begin{equation}
  \mu \frac{dg_s}{d\mu} = -\frac{g_s^3}{16\pi^2} \left( \frac{11}{3} C_A - \frac{4}{3} T_F n_f \right) + \mathcal{O}(g_s^4),
  \label{eqn:beta_function}
\end{equation}
where $C_A = 3$ is the adjoint Casimir, $T_F = 1/2$, and $n_f$ is the number of active quark flavors.
The first term in the parentheses is the important one that arises from the gluon-gluon self interaction.
For $n_f \leq 16$ the quantity in the parentheses is positive, making the overall sign negative, so $g_s$ decreases as the energy scale $\mu$ increases.
This is in contrast to the electromagnetic interactions whose strength increases with energy scale.
The running of the QCD coupling constant, verified by many experimental measurements, is illustrated in Figure~\ref{fig:qcd_running}.
At high energy scales $Q$, the strong coupling (here plotting $\alpha_s$ rather than $g_s$) is less than 0.1.
This is the perturbative QCD regime in which a perturbative expansion in powers of the strong coupling is well defined.
Here the strong force dynamics are similar to the EM dynamics and color charged particles can freely propagate.
At low energy scales the coupling begins to diverge.
Below the QCD hadronization scale $\Lambda_{\mathrm{QCD}} \sim 200$~MeV, the strong force is so powerful that it becomes energetically favorable for color-charged particles to pull opposite color-charged particles out of the vacuum such that the overall color charge is neutralized.
This is known as \textit{hadronization}, since color charged particles like quarks and gluons become bound up in composite particles known as hadrons.
The protons and neutrons inside atomic nuclei are the most important hadrons, as they make up the vast majority of the baryonic matter in the universe.
Hadronization is an inherently non-perturbative process since the strong coupling constant is large by assumption.
The decreasing strength of the strong force with energy scale is known as asymptotic freedom~\cite{Gross:1973id, Politzer:1973fx}.
It offers an example of how an energy scale (namely $\Lambda_{\mathrm{QCD}}$) can be generated by the structure of a theory rather than being set by free parameters as is the case for many of the theoretical issues described in Chapter~\ref{ch:intro}.

\subsection{Parton Showers}

At high energies, quarks and gluons can propagate freely but this does not mean that the propagation is simple.
The QCD splitting functions, which describe the probability of a parton\footnote{Quarks and gluons are often referred to as partons, since they are the constituent parts of the hadrons we observe at low energies. At high energies, the term is still used but is less descriptive since quarks and gluons can propagate freely.} to emit a gluon, have soft and collinear divergences.
The leading-order splitting function is given schematically by
\begin{equation}
  dP = \frac{\alpha_s}{\pi} C_f \frac{d\epsilon}{\epsilon}\frac{d\theta^2}{\theta^2},
  \label{eqn:splitting}
\end{equation}
where $\epsilon = 1 - z$ is one minus the momentum fraction carried by one of the daughter partons $z$, $C_F = 4/3$ is the fundamental Casimir of $\mathrm{SU}(3)$, and $\theta$ is the emission angle between the daughter partons.
This splitting function diverges as either $\epsilon$ or $\theta$ become small, corresponding to the emission of a soft or collinear gluon by the parent parton.
These divergences exist whether the parent parton is a quark or a gluon, so gluons emitted by an initiating parton can in-turn emit more gluons.
This cascade of largely soft and collinear gluons is known as a \textit{parton shower}, which continues until all of the emitted partons reach energies near $\Lambda_{\mathrm{QCD}}$.

Any quark or gluon produced in an LHC collision immediately undergoes a parton shower before it can interact with a detector.
Given the collinear divergence of the splitting function the shower is collimated, meaning all of the resulting quarks and gluons travel roughly in the same direction.

\subsection{Jets}

These two processes, parton showering at high energies and hadronization at low energies, imply that the result of the production of a high \pt quark or gluon in an LHC collision is a collimated spray of hadrons that can interact with a detector.
This signature is known as a \textit{jet}.
Measurement of a jet requires that the four-vectors of hadrons within a jet, typically called \textit{jet constituents}, be measured by the detector.
The four-vector of the jet itself can be calculated by summing the four-vectors of the jet constituents, and then the four-vector of the jet can be used as a stand-in for the four-vector of the initiating parton when analyzing LHC collisions.
This simple prescription has several complications:

\begin{enumerate}
  \item{
    Choosing which hadrons to include in a jet is non-trivial.
    In proton--proton collisions, the presence of multiple jets, multiple parton interactions, and initial and final state radiation significantly complicate this question.
    Studying jets typically requires specifying a \textit{jet clustering algorithm}, which leads to many experimental and theoretical subtleties.
  }
  \item{
    The general purpose detectors on the LHC, ATLAS and CMS, are not sensitive to the entire four-vector for an arbitrary hadron.
    In particular the masses of individual hadrons are not constrained (see \Cref{sec:zjets-uncert-unfold}).
    Significant approximations on hadron masses are made so that the four-vector of the jet can be calculated.
  }
  \item{
    Even the parts of the hadron four-vectors that are measured are not measured with perfect resolution.
    Some smearing of the true kinematic quantities is inevitable, and can cause significant distortions in the four-vector of the jet.
  }
  \item{
    The LHC operates with many proton--proton collisions per bunch crossing.
    The presence of these \textit{pileup collisions} superimposed on top of the collision of interest can contaminate the jet, adding hadrons to it that were not produced by the shower initiated by the parton.
  }
\end{enumerate}

Each of these points adds significant complications to the experimental study of jets, even if the only goal is to infer the four-vector of the initiating parton.
However there is a huge amount of interesting and unknown physics contained in the parton shower and hadronization.
This physics can be probed through examining the underlying measurements of the jet constituents.
Studying this \textit{jet substructure} requires processing the full, high-dimensional data type that describes all hadrons within a jet.
The complexity and high dimensionality of these data types makes them natural applications for deep learning methods.
This is why the jet substructure community saw the earliest and most enthusiastic adoption of deep learning in LHC physics \cite{Larkoski:2017jix, Kogler:2018hem}.

The study of jet substructure is organized around substructure observables.
These are quantities that can be calculated using the four-vectors of the jet constituents that capture certain features of a jet.
As is the case for all physics, the study of jet substructure moves forward through an interplay between theorists and experimentalists.
Theorists make predictions of jet substructure observables within the framework of QCD, and experimentalists make measurements of these observables that can be compared to and used to refine the theory.

There are two approaches to calculating jet substructure observables within QCD.
The first approach is to directly calculate a given observable within perturbative QCD.
A complete calculation involves three ingredients~\cite{Larkoski:2017jix}: (1) a perturbative calculation at some fixed order in $\alpha_s$, (2) resummation of the large logarithms that result from the infrared and collinear divergences described above, and (3) some accounting for non-perturbative effects like hadronization, typically through an analytic shape function.
The result is prediction of the cross section of some process, typically di-jet production in electron--positron or proton--proton collisions, differential in the substructure observable~\cite{Dasgupta:2001sh}.

The second approach is to recursively apply the splitting function in Equation~\ref{eqn:splitting}.
Given the probability of a splitting at a given scale does not depend on splittings at any previous scales, the chain of splittings can be described as a Markov chain, and samples from the probability distribution over the final configuration of the parton shower can be obtained by Markov Chain Monte Carlo (MCMC) methods.
Samples provided by the MCMC are a fully exclusive partonic final state: a list of partons at or above the QCD hadronization scale $\Lambda_{\mathrm{QCD}}$.
In order to compare to experimental measurements, phenomenological models of hadronization can be applied~\cite{Marchesini:1991ch, Sjostrand:2014zea}.
Unlike the parton shower sampling, these models are not based on first-principles QCD but are models with free parameters that must be \textit{tuned} to match experimental data.
The combination of fixed-order perturbative calculations, followed by sampling of parton shower configurations through MCMC, followed by application of a hadronization model, produces an \textit{event generator}.
These event generators are capable of simulating proton--proton collisions with high accuracy~\cite{Bierlich:2022pfr,Bellm:2015jjp,Sherpa:2019gpd}.
This approach is very general-purpose, since it can provide predictions for any observable, including those that are too complicated for the analytic approach described above, or even IRC unsafe observables for which the resummation in step 2 above is not possible.
The limitation is that the accuracy of the event generator for a given observable is hard to characterize systematically, so the assigned theory uncertainties are typically less rigorous than those from resummed calculations.

An important practical distinction between these two approaches is how their outputs are presented.
Resummed calculations provide smooth analytic predictions for differential cross sections, while event generators produce discrete lists of final state hadron four-momenta.
Importantly each of these final state hadrons can be passed through a detector simulation, which simulates how the hadron interacts with a particle detector and ultimately generates experimental data~\cite{GEANT4:2002zbu, ATLAS:2010arf}.
Chaining together a fixed order calculation (either in QCD or the SM more generally), parton shower, hadronization model, and detector simulation defines a \textit{simulation chain} capable of producing simulated collision events that resemble the data collected by the experiments at the LHC.
This ability explains why event generators are used copiously in essentially every analysis of LHC data, while resummed calculations are mostly relevant within the sub-field of jet substructure and QCD.
The difference in the format of the results between the two approaches is a subtle but important motivation for the unfolding methods discussed in Chapter~\ref{ch:unfolding}.

The remainder of this chapter discusses two particular jet substructure observables which are measured in \Cref{ch:zjets}: energy correlators and the intrinsic dimensionality of jets.
The reason for exploring these observables in particular is that they are very difficult to unfold with conventional binned methods.
Detailed discussion of this will be delayed until \Cref{ch:zjets}, but some introduction to the observables from a theoretical standpoint is due first.

\section{Energy correlator observables}
\label{sec:eec}

\subsection{Context}

The $N$-point energy correlator (E$N$C) is defined as the energy-weighted $N$-point correlation function of the radiation pattern produced by a particle collision.
This mathematical object is also used in cosmology to study the CMB and in condensed matter physics to study phase transitions and criticality.
See Refs. \cite{Hofman:2008ar, Kologlu:2019mfz} for a formal treatment of these observables that make these connections explicit.
The simplest energy correlator is the two-point energy correlator (E2C or EEC).
Measuring this quantity involves building a histogram of the opening angle between objects produced by a collision, with the histogram entries weighted by the product of the energies of the two objects.
The energy weighting ensures that the observables are infrared and collinear (IRC) safe.
Either individual particles or clustered jets can be used as the input objects \cite{ATLAS:2015yaa, ATLAS:2017qir, ATLAS:2023tgo, CMS:2024mlf}.
Most studies have focused on the properties of the EEC, but higher order correlators are also of some theoretical interest.
For calculating the E$N$C, the number of $N$-tuples of objects grows combinatorially with the particle multiplicity, which makes the measurement substantially more computationally demanding~\cite{Alipour-fard:2024szj}.

The EEC was first used in particle physics as an event-shape observable in the context of electron--positron collisions \cite{Basham:1977iq, Basham:1978bw, OPAL:1991uui, SLD:1994idb} for the extraction of the strong coupling constant $\alpha_s$.
Several of the measurements shown in green in Figure~\ref{fig:qcd_running} are derived from energy correlator measurements in $e^+e^-$ collisions.
These measurements calculated the correlators between hadronic jets, rather than individual particles, so they are best understood as event shape measurements rather than jet substructure measurements.

\subsection{Energy Correlators as Jet Substructure Observables}

\begin{figure}[ht]
  \centering
  \includegraphics[width=0.65\linewidth, alt={Log-log plot of the normalized two-point energy correlator versus angular scale R sub L for CMS 2011 open data AK5 jets. The figure highlights three regions labeled Free Hadron at low R sub L, Transition at intermediate R sub L, and Quarks/Gluons at larger R sub L.}]{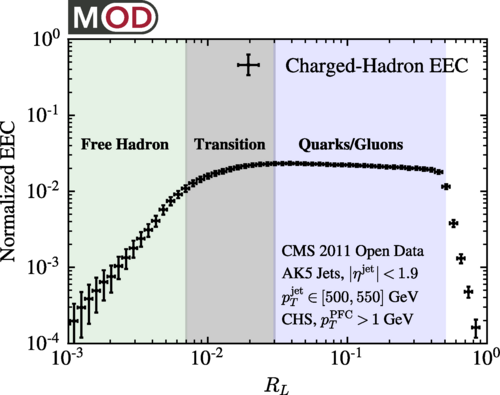}
  \caption{The two-point energy correlator (EEC) plotted using CMS open data for high \pt jets, showing distinct free-hadron, transition, and quark/gluon regimes as a function of the angular scale $R_L$ (called $z$ in the text). Figure adapted from Ref.~\cite{Komiske:2022enw} under a CC~BY~4.0 license.}
  \label{fig:cms-open-eec}
\end{figure}

Energy correlators have only recently been used as jet substructure observables.
See Ref.~\cite{Moult:2025nhu} for a recent review.
Interest has been driven by a few properties of the two-point correlator in the highly collinear limit, which is dominated by the highly collimated radiation within jets.
The two-point correlator is given as
\begin{equation}
  EEC(z) = \frac{1}{\sigma} \sum_{i,j} \int d\sigma \frac{E_i E_j}{Q^2} \delta(z - \frac{1- \cos \theta_{ij}}{2}),
  \label{eqn:eec}
\end{equation}
where $\sigma$ is the total cross section, $E_i$ is the transverse energy of the $i$-th particle, $Q$ is the sum of the transverse energies of all particles ($\sum_i E_i$), $\theta_{ij}$ is the angle between the three-momentum vectors of the $i$-th and $j$-th particles, and $z = \frac{1- \cos \theta_{ij}}{2}$.
The first property is that in perturbative QCD, the EEC is predicted to scale as a power law as a function of the opening angle $z$ between freely propagating partons.
The exponent of this power law is related to a formal field theoretical quantity known as the cusp anomalous dimension.
This provides a direct connection between a measurable observable and a fundamental property of QCD.
The larger the angular separation between particles $z$, the earlier in the jet formation process (or the higher the energy) is being probed, since larger angular separations are produced by earlier splittings in the parton shower.
This means that at moderate values of $z$, where particles are contained within a jet but last interacted with each other early in the jet formation process, a power law scaling in the EEC should be observed.

Second, deviations from this perturbative power-law scaling encode non-perturbative effects like hadronization.
As $z$ grows smaller, the EEC quantifies correlations between increasingly collinear particles, meaning they were emitted later in the jet formation process.
Therefore the very small $z$ regime should be sensitive to dynamics below the scale $\Lambda_{\mathrm{QCD}}$ where hadronization occurs.
Hadrons are color neutral and do not interact by the strong force, and so the correlation should reflect non-interacting particles.
Namely it should have a linear dependence on the angular separation.

The combination of these two properties mean that two distinct scaling regimes should be visible in the EEC: a power law corresponding to freely propagating partons at high energies at high $z$, and a linear scaling corresponding to non-interacting hadrons at low $z$.
Ref.~\cite{Komiske:2022enw} first illustrated these two distinct scaling regimes using CMS open data, as shown in \cref{fig:cms-open-eec}.
The two scaling regimes are shaded in blue and green respectively, with a transition regime corresponding to the hadronization process linking them.
The scaling regimes are very well understood from a theoretical standpoint, but the transition between them is intractable given our current theoretical tools, motivating it as a useful regime to measure experimentally.
Measuring the transition requires jets with a high energy (or \pt at a hadron collider) so that a sizeable portion of the parton shower evolution occurs above the hadronization scale.
At LEP, the available center-of-mass energy limited the maximum jet energy to around 100~GeV, which is too low to observe the perturbative scaling regime cleanly and helps explain why energy correlators were relatively unexplored as jet substructure tools in the $e^+e^-$ era.
However the large center-of-mass energies of the LHC produce the needed high \pt jets in large numbers.

The final property is that the energy correlators can be calculated to very high perturbative accuracy.
The EEC in $e^+e^-$ collisions has been computed to next-to-next-to-next-to-leading order (N$^3$LO), which is a very high level of accuracy for a jet substructure quantity~\cite{Ebert:2020sfi}.
Recent work has provided N$^3$LL accurate predictions for energy correlator observables in proton--proton collisions as well~\cite{Gao:2023ivm}.

\subsection{Experimental Status}

The renewed interest in energy correlator observables from the theory community has produced many new measurements of the observables.
I only summarize here energy correlator measurements using particles, rather than jets, since this is needed to access the collinear regime.

The most relevant existing measurement is of the EEC and E3C observables by the CMS collaboration in proton--proton collisions at the LHC.
This paper also used the measured spectra to extract $\alpha_s$~\cite{CMS:2024mlf}.
This is the only published measurement using high \pt jets from $pp$ collisions at the LHC, so it serves as a useful comparison for the EEC measurement in \Cref{sec:zjets-results-eec}.
The ALICE collaboration additionally measured the EEC in charm-tagged jets in proton--proton collisions, but at a lower center-of-mass energy and with less luminosity~\cite{ALICE:2025igw}.
Energy correlator measurements have also been carried out in heavy-ion~\cite{ALICE:2024dfl, CMS:2025ydi} and proton--ion~\cite{Liang-Gilman:2025gjl} collisions at the LHC.
These measurements feature lower \pt jets, where the observed EEC distributions are modified due to interaction with the quark-gluon plasma.
The STAR collaboration has measured the EEC in proton--proton collisions at the lower center of mass energies available at RHIC \cite{STAR:2025jut}.
Legacy analyses have reprocessed archived data from the ALEPH experiment at LEP \cite{Bossi:2024qeu}, where the EEC is measured with all tracks given the substantially lower particle multiplicity produced by electron--positron collisions.
Finally, the H1 collaboration has presented preliminary results measuring the EEC distribution in electron--proton deep inelastic scattering events using all tracks~\cite{H1:prelim}.
This result is particularly relevant because it uses the same approach to unfolding as the measurement in this thesis.
It will be discussed in detail in Chapter~\ref{ch:unfolding}.
\Cref{tab:eec-measurements} summarizes the experimental landscape.

\input{tab_eec_measurements.tex}

There are three important gaps in the existing measurements.
First, there is only one measurement from CMS which uses the high \pt jets provided by the LHC to measure the EEC well above the confinement scale~\cite{CMS:2024mlf}.
A similar measurement performed by ATLAS would be a valuable contribution.
Second, all current LHC measurements cluster particles into jets before performing the correlator measurement on the jet substructure.
This is experimentally convenient, especially for the extraction of a universal hadronization scale and for the necessary unfolding procedure discussed in \Cref{ch:unfolding}, but it introduces dependencies on the jet clustering algorithm applied and prevents measurement of the EEC over a full range of angular separations.
A measurement performed on all particles in the event, without prior jet clustering, would allow measurement of the entire perturbative scaling regime without the distribution being cut-off due to the jet radius.
Third, all existing measurements use only charged particles, since neutral particles such as photons and neutral pions are measured with substantially worse angular resolution.
Theoretical predictions for charged-particle-only measurements require the track function formalism~\cite{Chang:2013rca, Chang:2013iba}, which introduces more complexity.
Discussion of this is beyond the scope here, but work on this formalism and associated measurements is on-going~\cite{ATLAS:2025qtv} providing a promising path to calculation of energy correlators on charged particle final states.
One of the primary physics contributions of the measurement presented in this thesis is to address the first two gaps, with the third being left as a fundamental limitation of current detector technology.

\section{Intrinsic Dimensionality of Jets}
\label{sec:id}

The energy-energy correlators discussed in Section~\ref{sec:eec} are jet substructure observables that are of interest because of their connections to the underlying theory of QCD.
In contrast, the intrinsic dimensionality of jets is an observable motivated by a basic question about the structure of the data rather than ties to the underlying theory.
This is a very new observable that has only been treated, to my knowledge, in a few research papers~\cite{Komiske:2019fks, DAgnolo:2025qqr}.
There are likely interesting connections to the underlying theory.
For example the scale dependence of the intrinsic dimension (see below) could in principle encode information about the parton shower branching structure, but this remains an essentially unexplored topic.
Part of the motivation for measuring this observable is to demonstrate that measurements of it are possible, hopefully motivating further theoretical work.
The rest of the motivation is to demonstrate that deep learning methods allow us to ask questions about the data in its native, high dimensionality.

\subsection{The Manifold Hypothesis and Intrinsic Dimensionality}

The manifold hypothesis states that high-dimensional data generated by natural processes tend to cluster on lower-dimensional manifolds embedded in the high-dimensional space.
Put another way, the naive high dimensional representation we use to describe the data is redundant.
Fewer coordinates are needed to describe the data than the naive dimensionality would suggest, demonstrating that there is structure in the high dimensional space that is invisible to humans examining projections (e.g. histograms or scatter plots) but easily identifiable for deep learning models.
This observation is implicitly known in particle physics, given our data is known to be generated by relatively simple processes and respect certain symmetries, but it has not been explicitly quantified.

\begin{figure}[t]
  \centering
  \includegraphics[width=0.85\linewidth, alt={Illustration of the scale dependence of the intrinsic dimension of a dataset, showing that at large distances the data appear zero-dimensional while at smaller scales their manifold structure is revealed.}]{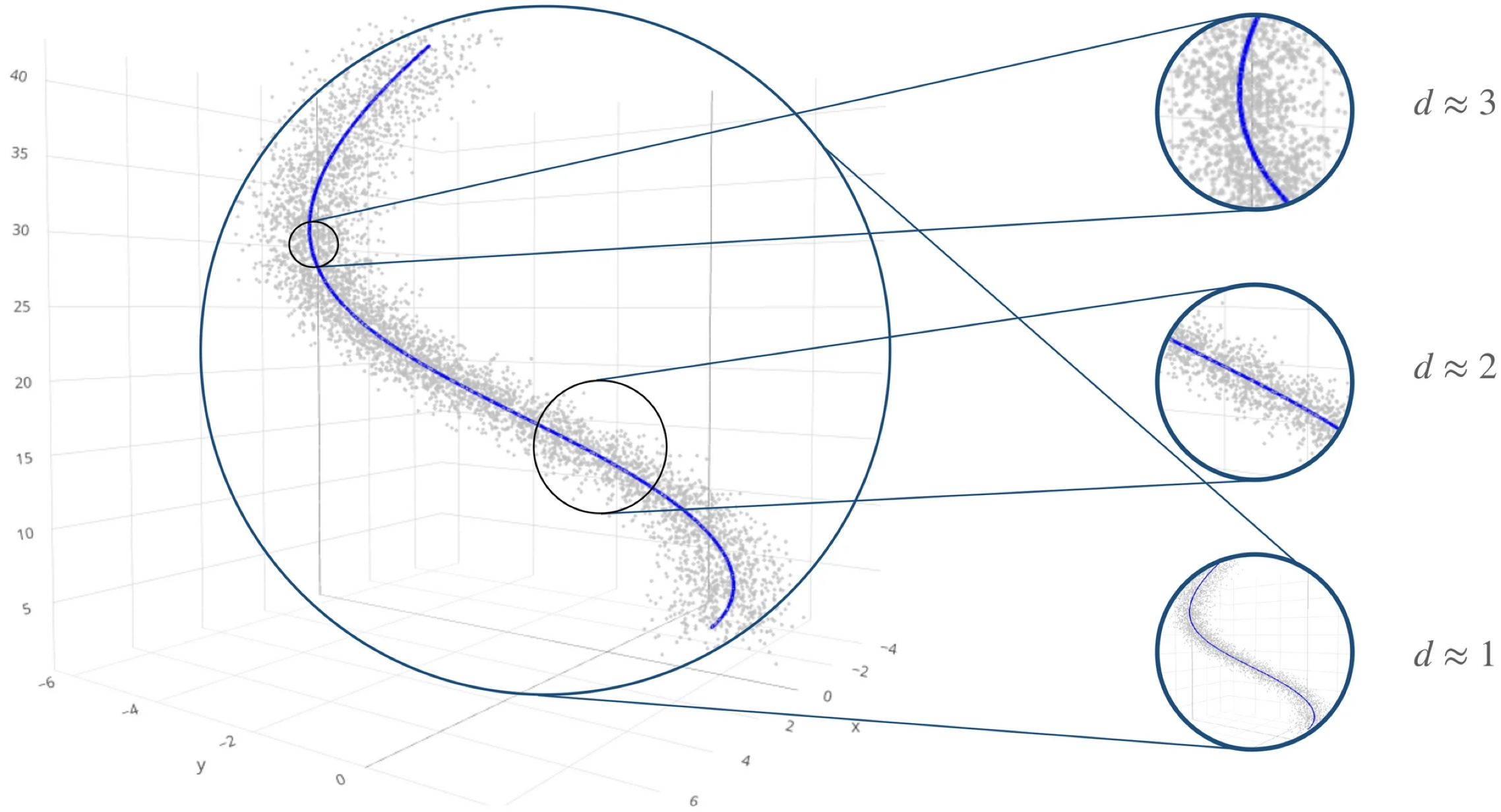}
  \caption{Illustration of the scale dependence of the intrinsic dimension. At large length scales the data appear as a single point (zero dimensions); at smaller scales the geometry of the underlying data manifold becomes visible and the estimated dimension increases. Figure adapted from Ref.~\cite{id-figure} under a CC~BY~4.0 license.}
  \label{fig:id_scale}
\end{figure}

The intrinsic dimension (ID) of a dataset is the dimension of the underlying manifold.
The manifold hypothesis states that the intrinsic dimension of a dataset is typically much less than the number of naive coordinates we use to describe the data.
An important feature of the ID is that it is \textit{scale-dependent}.
Figure~\ref{fig:id_scale} provides a visual illustration of this scale dependence.
The data cluster on a manifold embedded in the naive three-dimensional coordinates used to describe the data, but the apparent dimension of the manifold depends on the length scale.
At large distances the manifold looks like a curve that can be parametrized with one free parameter, so the dimension is one.
At smaller distances, the noise in the data gets resolved and the manifold appears first two- and then three-dimensional.
Clearly the intrinsic dimension also does not need to be an integer.
Fractal (non-integer) dimensions are expected as interpolations between the dimensions illustrated in \Cref{fig:id_scale}.

A variety of intrinsic dimension estimators have been proposed~\cite{10.1109/T-C.1971.223208, PhysRevLett.50.346, NIPS2004_74934548, CERUTI20142569, Albergante:2019ijcnn}.
A quick summary is as follows:

\begin{enumerate}
  \item{
    Principal component analysis (PCA) methods count the number of eigenvalues of the data covariance matrix that are statistically distinguishable from zero.
    They assume the data lie on a linear subspace, which is not true of curved manifolds \cite{10.1109/T-C.1971.223208}.
  }
  \item{
    Fractal and correlation dimension methods exploit how the number of data points contained within a ball of radius $r$ as $r$ scales given the dimensionality of a manifold.
    The correlation dimension~\cite{PhysRevLett.50.346} is the most widely used estimator in this family.
    This was also the first estimator to be explored for applications in particle physics~\cite{Komiske:2019fks}.
  }
  \item{
    Maximum likelihood methods write the likelihood of the observed distances between neighboring data points conditioned on the underlying dimensionality~\cite{NIPS2004_74934548}.
    This is the most broadly applicable class of estimators.
    One example was recently proposed as a test statistic for searches for physics beyond the Standard Model at the LHC \cite{DAgnolo:2025qqr}, and is measured in \Cref{sec:zjets-results-id}.
  }
  \item{
    Concentration-of-measure methods exploit the fact that vectors pointing from one data point to others cluster in particular directions in high dimensional spaces, rather than being uniformly distributed.
    These methods have not yet been applied to particle physics data~\cite{CERUTI20142569, Albergante:2019ijcnn}.
  }
\end{enumerate}

\subsection{The Energy Mover's Distance}

Most intrinsic dimension estimators require that the vector space used to describe the data be equipped with a metric so that pairwise distances between points can be calculated.
For jet data, Ref.~\cite{Komiske:2019fks} introduced the energy mover's distance (EMD) as a natural choice.
The EMD between two jets $\mathcal{E}$ and $\mathcal{E'}$ is defined as the energy cost of rearranging the substructure of one jet into the substructure of the other,
\begin{equation}
  EMD(\mathcal{E}, \mathcal{E'}) = \min_{f_{ij}} \sum_{i,j} f_{ij} \frac{\theta_{ij}}{R} + | \sum_i E_i - \sum_j E_j |,
\end{equation}
where $f_{ij}$ is the optimal transport plan moving energy from particle $i$ in jet $\mathcal{E}$ to particle $j$ in jet $\mathcal{E'}$, $\theta_{ij}$ is the angle between the three-momentum vectors of the $i$ and $j$ particles, and $R$ is the radius of the jets.
The second term accounts for differences in the overall energy normalization between the jets.
The $f_{ij}$ satisfy the following constraints:
\begin{align}
  f_{ij} \geq 0, \\
  \sum_j f_{ij} \leq E_i, \\
  \sum_i f_{ij} \leq E_j, \\
  \sum_{i,j} f_{ij} = E_{min},
\end{align}
Given the cost function for the EMD is weighted by the particle energies, the EMD is infrared and collinear safe.
To use the EMD as a substructure observable, it is useful to preprocess the jets to remove any irrelevant features, such as the location of the jet in the pseudorapidity-azimuth ($\eta$-$\phi$) plane or its rotation.
The standard approach, used throughout, is to apply a lorentz boost and rotation to the jets such that they are centered in the $\eta$-$\phi$ plane, and then rotate the jets about their axis such that the first principle component of the jet radiation pattern points toward the positive $\phi$ axis.

\subsection{The Nearest-Neighbor Intrinsic Dimension Estimator}

\begin{figure}[t]
  \centering
  \includegraphics[width=0.7\linewidth, alt={Illustration of the nearest-neighbor intrinsic dimension estimator in a two-dimensional space, showing a reference point and its nearest neighbors at two successive distance scales.}]{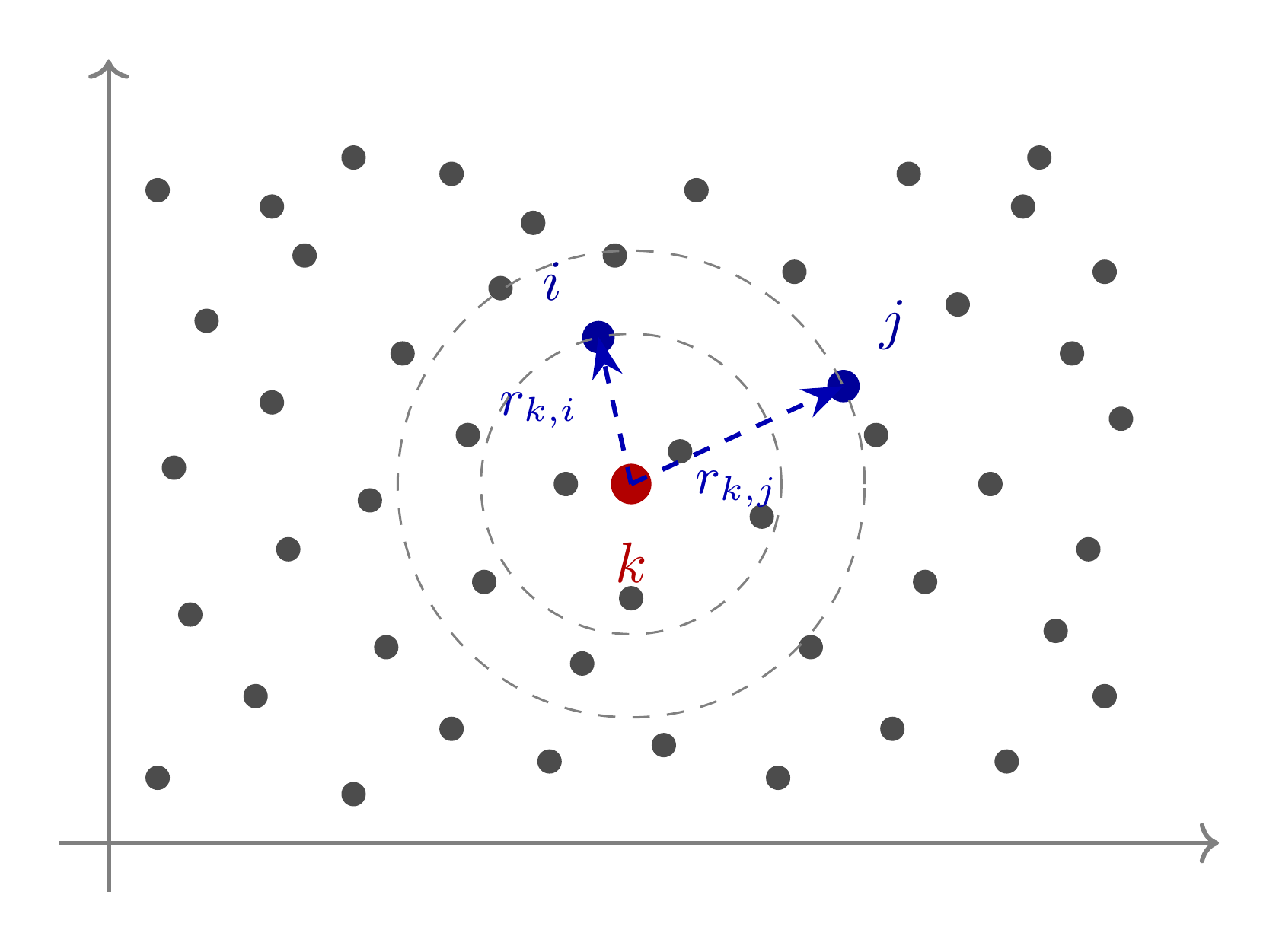}
  \caption{Illustration of the NNID estimator in a simple two-dimensional space. For a reference point (center), the distances $r_{k,i}$ to the $i$th nearest neighbor and $r_{k,j}$ to the $j$th nearest neighbor define the ratio $\mu_{k,ij}$. In this particular case $i=5$ and $j=10$. The distribution of these ratios across many reference points encodes the intrinsic dimension of the data manifold. Figure drawn by Claude Sonnet 4.6.}
  \label{fig:nnid_illustration}
\end{figure}

Given a set of $N$ jets with EMDs calculated between all pairs (the number of EMDs to compute is $N^2$), the nearest-neighbor intrinsic dimension (NNID) estimator~\cite{NIPS2004_74934548, DAgnolo:2025qqr} constructs a likelihood from the ratios of distances to successive nearest neighbors.
It is an example of a maximum likelihood method in the taxonomy above.
For a jet $k$, the distance ratio between the $j$th and $i$th nearest neighbors is given as,
\begin{equation}
  \mu_{k,ij} = \frac{\mathrm{EMD}(\mathcal{E}_k, \mathrm{NN}_j(k))}{\mathrm{EMD}(k, \mathrm{NN}_i(k))},
  \label{eqn:mu_ratio}
\end{equation}
where $\mathrm{NN}_i(k)$ denotes the $i$th nearest neighbor of jet $k$ by the EMD metric.
These distance ratios depend on the values of the unphysical indices $i$ and $j$.
In what follows, I follow Ref.~\cite{DAgnolo:2025qqr} and set $j = 2i$ throughout, reducing the number of free parameters to one.
An illustration of these ratios is provided for two dimensional data in Figure~\ref{fig:nnid_illustration}.

In a $d$-dimensional space, the ratios $\mu_{k,ij}$ follow a Pareto distribution whose shape parameter encodes the intrinsic dimensionality $d$.
The intrinsic dimension for a given choice of $i$ and $j=2i$ can then be estimated by minimizing the negative log-likelihood,
\begin{align}
 L &= -\sum_{k=1}^{N} \log f(\mu_{k,ij}; d) \notag \\
 &= -N \log d + (1 + i d)\sum_{k=1}^{N} \log \mu_{k,ij} - (j - i - 1)\sum_{k=1}^{N} \log\left(1 - \mu_{k,ij}^d\right).
 \label{eqn:likelihood}
\end{align}
$f(\mu_{k,ij}; d)$ is the Pareto probability density evaluated at the ratio for jet $k$, which has an analytic form written in the second line.
Numerically minimizing Equation~\ref{eqn:likelihood} over $d$ with all jets $k$ included in the sum yields the estimated intrinsic dimension.

Each choice of the indices $i$ and $j$ defines a new intrinsic dimensionality estimator that will probe a certain scale in the dataset.
Following fixing $j = 2i$, the choice of $i$ is the only free parameter in the calculation, but this free parameter is unphysical.
It is useful to plot the NNID estimator against a physical scale.
The median EMD between a jet and its $i$th nearest neighbor is a useful, but not unique, choice for this physical scale.
Varying the index $i$ then produces a curve that shows how the NNID estimator evolves with this scale.

Note that particle physics datasets are typically produced with event weights.
This is true for simulated datasets produced by event generators, and for the unbinned spectra that build the cross section measurement in \Cref{ch:zjets}.
The NNID calculation can be extended to accommodate event weights by scaling each jet's contribution to the likelihood in \Cref{eqn:likelihood} by its weight, and replacing the nearest neighbor assignments $i(k)$ and $j(k)$ by their weighted analogs.

\subsection{Generator-Level Results}

Given the novelty of this observable in particle physics, it is worth presenting some generator level results before moving on.
Jets clustered from particle-level charged particles are sampled from the \mgpy and \sherpa samples described in Chapter~\ref{ch:zjets}.
See \Cref{sec:zjets-results-id} for more details.
Each point in the curves corresponds to a given choice of the unphysical index $i$, which produces a NNID and median EMD estimate.
The uncertainties on the theoretical predictions are shown as shaded bands about the curve.
The results are shown in two views covering different regions of the NNID vs median EMD curve in order to resolve the fine structure.
In general the NNID versus median EMD curves agree within uncertainties between the two generators.
The observation that the intrinsic dimension of jets is low, only several dimensions, is striking given the naive representation in terms of the four-momenta of the jet constituents has several tens of dimensions.
Further the intuition about the intrinsic dimensionality of \Cref{fig:id_scale} generally applies to jet data as well.
The apparent dimensionality of the manifold decreases as the length scale increases.
There is however a small increase in the NNID between median EMD values of roughly $20$ to $30$~GeV.
This is an empirical observation that is difficult to explain from a theoretical standpoint.

\begin{figure}[t]
  \centering
  \begin{subfigure}[b]{0.48\textwidth}
    \centering
    \includegraphics[width=\linewidth, alt={NNID versus median EMD at low EMD scales, comparing MadGraph and Sherpa generators.}]{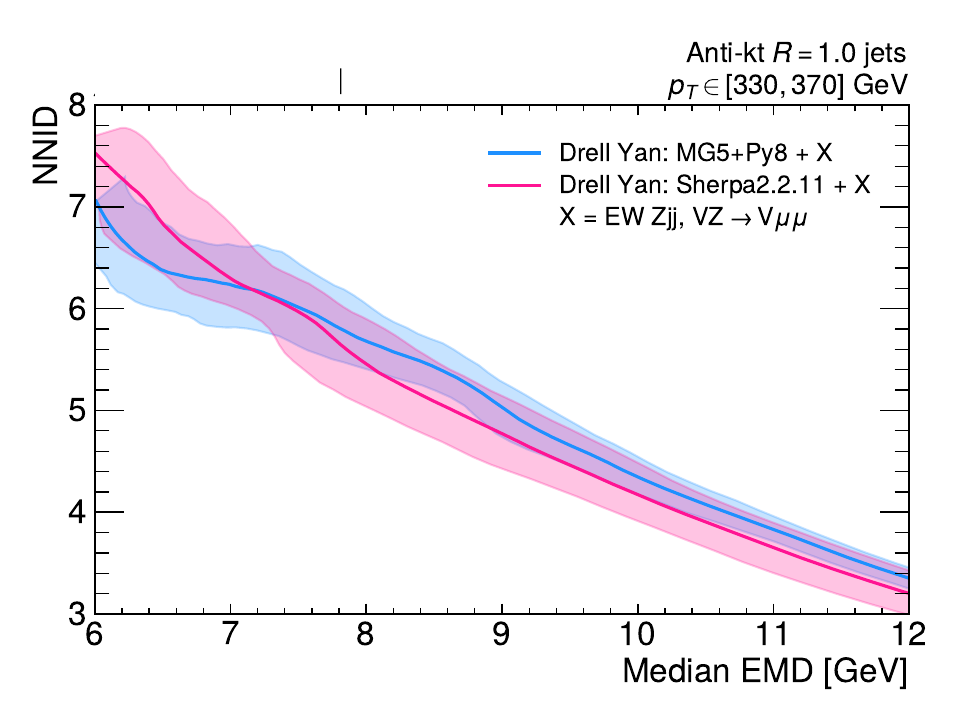}
    \label{fig:nnid_low}
  \end{subfigure}
  \hfill
  \begin{subfigure}[b]{0.48\textwidth}
    \centering
    \includegraphics[width=\linewidth, alt={NNID versus median EMD at high EMD scales, comparing MadGraph and Sherpa generators.}]{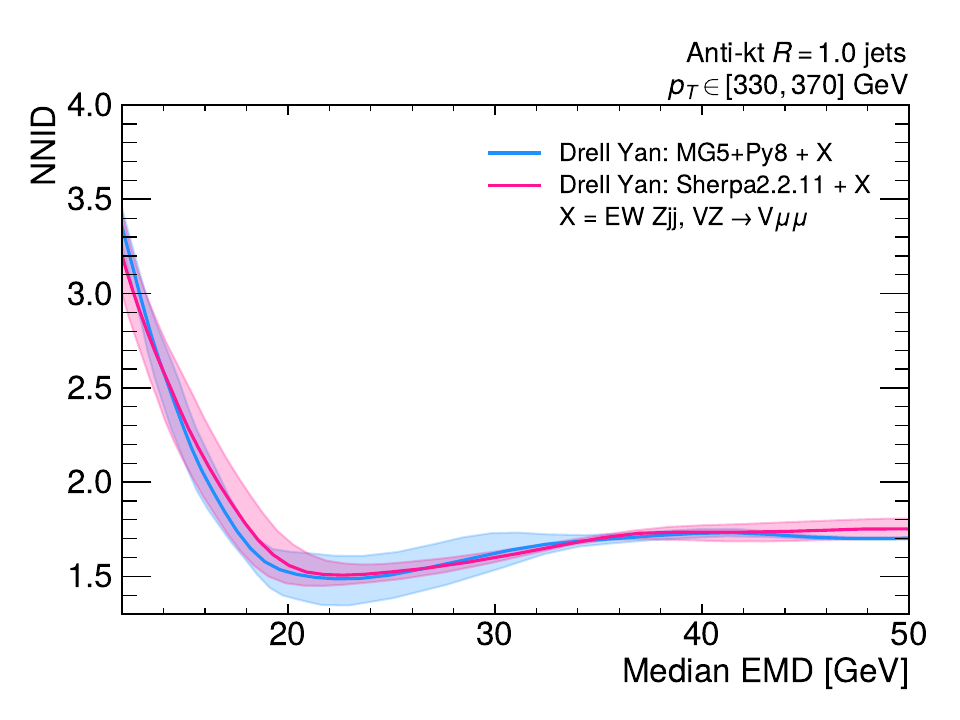}
    \label{fig:nnid_high}
  \end{subfigure}
  \caption{Nearest-neighbor intrinsic dimension (NNID) as a function of the median EMD scale, computed at truth level for the \textsc{MadGraph} (blue) and \textsc{Sherpa} (pink) event generators used in Chapter~\ref{ch:zjets}. The two views show the same curve over different ranges of the EMD axis to resolve fine structure. General agreement within uncertainties is observed.}
  \label{fig:nnid_generators}
\end{figure}

The first measurement of the intrinsic dimensionality of jet data is presented in \Cref{sec:zjets-results-id}.
It represents a somewhat different approach to jet physics, in which the focus is on asking basic questions about the experimental data ``as data'' and then evaluating the ability of QCD and the Standard Model to reproduce the answers.
This experimentally driven approach to measurements complements the typical approach of targeting observables due to their connection to the underlying theory.

%% file: tab_particle_reps.tex
\definecolor{quarkcolor}{rgb}{0.95, 0.88, 0.88}
\definecolor{leptoncolor}{rgb}{0.88, 0.92, 0.95}
\definecolor{bosoncolor}{rgb}{0.92, 0.95, 0.88}
\definecolor{higgscolor}{rgb}{0.95, 0.92, 0.82}
\definecolor{headercolor}{rgb}{0.25, 0.25, 0.35}

\begin{table}[ht!]
    \centering
    \caption{Standard Model fields and their representations under the gauge group
    $SU(3)_c \times SU(2)_L \times U(1)_Y$. Representations are written as
    $(\mathbf{d_3},\, \mathbf{d_2},\, Y)$, where $d_3$ and $d_2$ are the
    dimensions of the $SU(3)_c$ and $SU(2)_L$ representations, and $Y$ is the
    weak hypercharge under the convention $Q = T_3 + Y/2$.
    Three generations of fermions are implied; all carry identical quantum numbers.
    Table drawn by Claude Sonnet 4.6.}
    \label{tab:sm-reps}
    \renewcommand{\arraystretch}{1.4}
    \begin{tabular}{%
      >{\raggedright}p{3.6cm}
      >{\centering}p{2.2cm}
      >{\centering}p{3.2cm}
      c
    }
        \toprule
        \rowcolor{headercolor}
        \textcolor{white}{\textbf{Particle}} &
        \textcolor{white}{\textbf{Symbol}} &
        \textcolor{white}{\textbf{Representation}} &
        \textcolor{white}{\textbf{Spin}} \\
        \midrule
        
        \multicolumn{4}{l}{\textit{\textbf{Quarks (Left-handed doublets)}}} \\[2pt]
        \rowcolor{quarkcolor}
        Quark doublet & $Q_L = \binom{u}{d}_{\!L}$ & $(\mathbf{3},\,\mathbf{2},\,+\tfrac{1}{3})$ & $1/2$ \\
        
        \midrule
        \multicolumn{4}{l}{\textit{\textbf{Quarks (Right-handed singlets)}}} \\[2pt]
        \rowcolor{quarkcolor}
        Up-type quark & $u_R$ & $(\mathbf{3},\,\mathbf{1},\,+\tfrac{4}{3})$ & $1/2$ \\
        \rowcolor{quarkcolor}
        Down-type quark & $d_R$ & $(\mathbf{3},\,\mathbf{1},\,-\tfrac{2}{3})$ & $1/2$ \\
        
        \midrule
        \multicolumn{4}{l}{\textit{\textbf{Leptons (Left-handed doublets)}}} \\[2pt]
        \rowcolor{leptoncolor}
        Lepton doublet & $L_L = \binom{\nu_e}{e}_{\!L}$ & $(\mathbf{1},\,\mathbf{2},\,-1)$ & $1/2$ \\
        
        \midrule
        \multicolumn{4}{l}{\textit{\textbf{Leptons (Right-handed singlets)}}} \\[2pt]
        \rowcolor{leptoncolor}
        Charged lepton & $e_R$ & $(\mathbf{1},\,\mathbf{1},\,-2)$ & $1/2$ \\
        
        \midrule
        \multicolumn{4}{l}{\textit{\textbf{Gauge Bosons}}} \\[2pt]
        \rowcolor{bosoncolor}
        Gluons & $g^a_\mu$ & $(\mathbf{8},\,\mathbf{1},\,0)$ & $1$ \\
        \rowcolor{bosoncolor}
        $W$ bosons & $W^i_\mu$ & $(\mathbf{1},\,\mathbf{3},\,0)$ & $1$ \\
        \rowcolor{bosoncolor}
        $B$ boson & $B_\mu$ & $(\mathbf{1},\,\mathbf{1},\,0)$ & $1$ \\
        
        \midrule
        \multicolumn{4}{l}{\textit{\textbf{Higgs Sector}}} \\[2pt]
        \rowcolor{higgscolor}
        Higgs doublet & $H$ & $(\mathbf{1},\,\mathbf{2},\,+1)$ & $0$ \\
        
        \bottomrule
    \end{tabular}
\end{table}

%% file: tab_eec_measurements.tex
\begin{table}[t]
    \caption[Summary of published experimental measurements of energy-energy correlator observables.]{
      Summary of published experimental measurements of energy-energy correlator observables using particles.
      Jet transverse-momentum ranges are approximate and refer to the leading-jet or inclusive-jet selection;
      a dash (--) indicates measurements performed on all final-state particles without jet clustering.
      The asterisk denotes measurements made with charged-particle jets rather than jets that include the neutral component.
      The dagger denotes preliminary results.}
    \label{tab:eec-measurements}
    \centering
    \small
    \NoHyper
    \begin{tabularx}{\linewidth}{
      >{\hsize=1.1\hsize\centering\arraybackslash}X
      >{\hsize=0.8\hsize\centering\arraybackslash}X
      >{\hsize=0.9\hsize\centering\arraybackslash}X
      >{\hsize=1.4\hsize\centering\arraybackslash}X
      >{\hsize=0.8\hsize\centering\arraybackslash}X
    }
      \toprule \toprule
      Experiment & System & $\sqrt{s}$ & Observable & Jet $p_T$ range \\
      \midrule
      ALEPH~\cite{Bossi:2024qeu} & $e^+e^-$ & 91.2~GeV  & E2C all tracks    & -- \\
      CMS~\cite{CMS:2024mlf}           & $pp$     & 13~TeV        & E2C, E3C inside jets & 500--2500~GeV \\
      ALICE~\cite{ALICE:2025igw}       & $pp$     & 13~TeV        & E2C inside jets & 10--30~GeV* \\
      ALICE~\cite{ALICE:2024dfl}       & $pp$     & 5.02~TeV      & E2C inside jets & 20--80~GeV* \\
      STAR~\cite{STAR:2025jut}         & $pp$     & 200~GeV       & E2C inside jets & 25--55~GeV \\
      ALICE$^\dagger$~\cite{Liang-Gilman:2025gjl} & $p\text{Pb}$ and $pp$ & 5.02~TeV & E2C, E3C inside jets & 20--80~GeV* \\
      CMS~\cite{CMS:2025ydi}           & $\text{PbPb}$ & 5.02~TeV & E2C inside jets & 120--200~GeV \\
      H1$^\dagger$~\cite{H1:prelim}              & $ep$     & 320~GeV       & E2C all tracks & -- \\
      \bottomrule \bottomrule
    \end{tabularx}
    \endNoHyper
  \end{table}

%% file: chapter3.tex
\chapter{Machine Learning and Artificial Intelligence}
\label{ch:ml}


This Chapter serves as a general introduction to the deep learning methods used throughout the remainder of the thesis\footnote{Note that this can be thought of as the ``instrumentation'' chapter of this thesis, which might otherwise be dedicated to information on the ATLAS detector.}.
As in \Cref{ch:qcd}, the goal is not to be thorough.
Instead only the information needed to understand the particular deep learning methods applied to particle physics data in this thesis will be covered.

\input{chapter3_ml_overview}

\input{chapter3_deep_learning}

\input{chapter3_generative}

\input{chapter3_diffusion}

\input{chapter3_dre}

\input{chapter3_uq}

%% file: chapter3_ml_overview.tex
\section{Overview: ML as a Paradigm Shift}
\label{sec:ml-overview}

Machine learning is a set of techniques that allow rules or behaviors to be learned directly from labeled or unlabeled data by optimization algorithms.
It is often presented as the modern alternative to classical data analysis, in which a domain expert encodes their knowledge as a fixed set of rules that provide quantitative summaries of the data.
Classical data analysis is often highly effective in many scientific fields.
In particle physics specifically, many years of theoretical work have provided a rich set of relevant features that have been immensely useful for understanding both well-understood and novel features of the data.
However, this approach is rarely optimal, especially in cases where the optimal solution exists but is too complex to specify by hand.
To draw an example from \Cref{ch:tagging}, no expert-designed feature set captures the correct decision-surface for classifying jets initiated by boosted top quarks from the light quark and gluon jet background.
This is due to a lack of knowledge about certain features of the jet formation process (e.g. hadronization), but also just due to complexity.
Specifying the optimal feature by hand is probably not possible, but machine learning methods allow this complexity to be learned from the data.

Machine learning methods organize naturally into several paradigms based on the structure of the available training data:

\begin{itemize}
    \item \textit{Supervised learning} trains a model to reproduce a label or target feature that is known on some portion of the data used for training, but unknown and desired on another portion of the data.
    Canonical examples of supervised learning are classification and regression.
    Many examples of both tasks exist in particle physics, and \Cref{ch:tagging} is entirely dedicated to the jet tagging classification task.
    \item \textit{Unsupervised learning} involves learning structures in the data without relying on explicit labels.
    Examples of unsupervised tasks include clustering, compression, and generation.
    The latter task, which involves developing generative models that learn to produce new samples from $p_{\text{data}}$, is the focus of much of \cref{ch:unfolding}.
    \item \textit{Self-supervised learning} is a hybrid approach where the structure of the data themselves is exploited to provide class labels.
    Self-supervised techniques, like contrastive learning and next-token prediction, are most commonly used to learn general-purpose representations in the context of 
    \textit{foundation models} rather than solve a particular task.
    Recently self-supervised techniques such as masked particle modeling~\cite{Golling:2024abg,Birk:2024knn} have seen some preliminary application in HEP.
    \item \textit{Reinforcement learning} (RL) sets up an agent that performs actions within an environment to maximize some reward.
    The key feature of RL is that it does not require a differentiable training objective, though some methods construct a surrogate in practice.
    RL is the class of methods underlying recent high-profile advances in very difficult tasks such as game playing~\cite{Silver:2016alphago} and agentic systems~\cite{Ouyang:2022instructgpt}.
    Relatively few applications of RL within HEP have thus far been explored~\cite{Baretz:2025zsv}, but its relevance is growing with the explorations of agentic systems for automated data analysis as discussed in \cref{ch:conclusion}.
\end{itemize}

Machine learning has a long history of application in particle physics.
Boosted decision trees (BDTs) have been in use in HEP since the Tevatron~\cite{D0:2006ngk,CDF:2009itk}.\footnote{The history of BDTs in HEP is interesting. The original application of BDTs illustrated that they \textit{outperformed} artificial neural networks. Following the deep learning revolution initiated by AlexNet, BDTs were largely supplanted by neural networks, which proved far more expressive on the high-dimensional data common in HEP. More recently, BDTs have seen something of a resurgence, driven by their remarkable training speed relative to neural networks, and their robustness against noisy or irrelevant features in the density ratio estimation tasks covered in \cref{sec:dre}~\cite{Finke:2023ltw}.}
The deep learning era kicked off by AlexNet changed the scale of the datasets and models rather than introducing the concept of machine learned classifiers.
Neural networks scale better than BDTs, and so could handle the more complicated tasks that have driven much of the interest in ML in recent years.

Two key structural advantages of HEP datasets enabled the success of machine learning techniques.
The first is the ability to generate high-quality synthetic data with the Monte Carlo event generators discussed in \cref{sec:jets}.
These simulated datasets are a very close approximation to experimental data and carry \textit{truth level} information that can be used to derive effective training labels.
This is a significant advantage over other application domains where manual labeling is a significant bottleneck.
The second advantage is that the underlying data generating processes are relatively simple, are known to respect several key symmetries, and can be run in tightly controlled laboratory environments.
Recent developments in large scale ML systems are making these structural advantages less relevant as larger scale is making very complicated domains such as natural language tractable, but exploiting these advantages was a major benefit to early deep learning applications in HEP.

Deep neural networks are the dominant paradigm in modern machine learning and are used exclusively in the research of this thesis, so the remainder of this chapter provides an overview of the core network architectures and training methods used throughout.

%% file: chapter3_deep_learning.tex
\section{Neural Networks and Deep Learning}
\label{sec:dl}

\subsection{Multi-Layer Perceptrons}
\label{sec:mlp}

\begin{figure}[ht]
    \centering
    \includegraphics[width=0.75\linewidth, alt={Diagram of a multi-layer perceptron with two hidden layers. Nodes representing input features are shown on the left, connected by arrows to two columns of hidden-layer nodes in the middle, which are in turn fully connected to output nodes on the right.}]{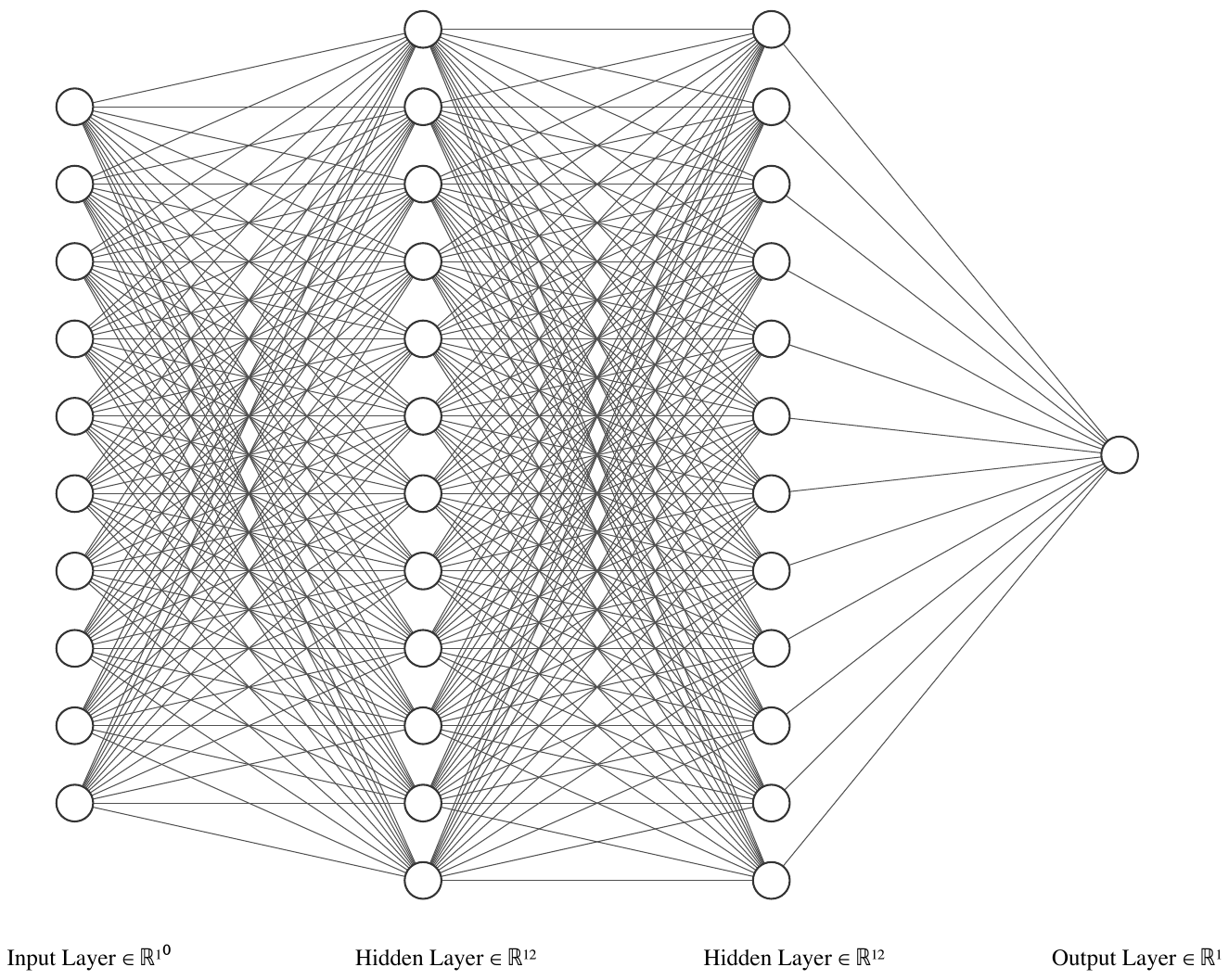}
    \caption{A diagram of a multi-layer perceptron with two hidden layers. In standard MLPs, each node in a given layer is connected to every node in the adjacent layer (fully connected). The input layer receives the feature vector $\mathbf{x}$, and the output layer produces the network's prediction. The circles represent the hidden states in between the layers of the network. The lines connecting the circles represent the weight matrices, while the biases and non-linearities are not shown.}
    \label{fig:mlp}
\end{figure}

The multi-layer perceptron (MLP) is the simplest neural network architecture.
An MLP applies a series of $L$ transformations, typically called \textit{layers}, to an input vector $\mathbf{x} \in \mathbb{R}^{d_{\text{in}}}$ to produce an output vector, not necessarily of the same length.
Each layer $\ell$ consists of an affine transformation followed by a pointwise non-linearity
\begin{equation}
    \mathbf{h}^{(\ell)} = \sigma\!\left( W^{(\ell)} \mathbf{h}^{(\ell-1)} + \mathbf{b}^{(\ell)} \right),
    \label{eqn:mlp_layer}
\end{equation}
where $\mathbf{h}^{(0)} = \mathbf{x}$, $W^{(\ell)} \in \mathbb{R}^{d_\ell \times d_{\ell-1}}$ is the weight matrix, $\mathbf{b}^{(\ell)} \in \mathbb{R}^{d_\ell}$ is the bias vector, and $\sigma$ is an elementwise nonlinear function, called the \textit{activation function}.
The final layer of the network produces the network output, and often either does not have a nonlinearity or has a task-specific one such as a softmax for classification tasks.
The weights and biases $\theta = \{W^{(\ell)}, \mathbf{b}^{(\ell)}\}_{\ell=1}^{L}$, together often referred to simply as the network \textit{weights}, are the learnable parameters of the network.
The traditional diagram of the MLP architecture is shown in \Cref{fig:mlp}.
Here $d_{\text{in}}$ is 10, $L$ is 3, $d_1$ and $d_2$ are both 12, and $d_3$ (the output dimension) is 1.
Note that the circles in the diagram correspond to the hidden states in between the layers of the network.
The lines connecting the circles represent the weight matrices, while the biases and non-linearities are not shown.

\begin{figure}[ht]
    \centering
    \includegraphics[width=0.7\linewidth, alt={Line plots of the ReLU and GELU activation functions. ReLU is zero for negative inputs and linear for positive inputs, with a sharp corner at the origin. GELU is similar but smoothly curved near zero, with a slight negative dip for small negative inputs. Figure generated by Claude Sonnet 4.6.}]{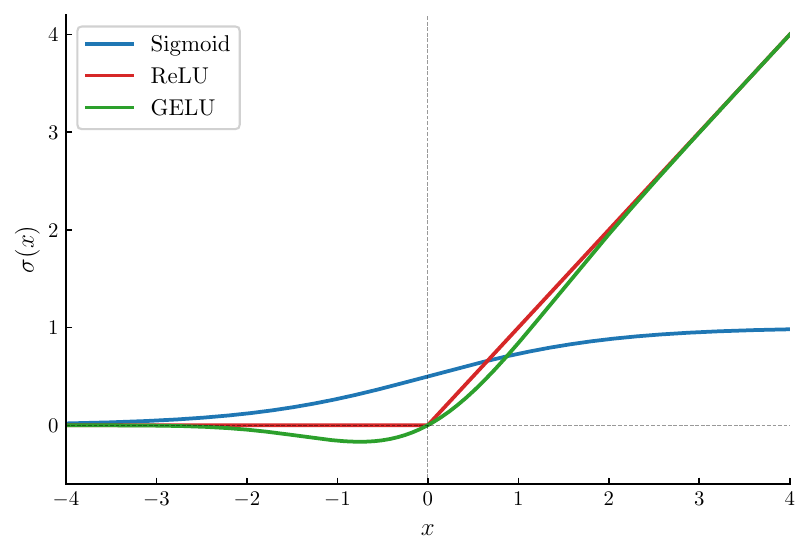}
    \caption{The ReLU and GELU functions commonly used as activations in modern neural networks, plotted against the input $x$.}
    \label{fig:activations}
\end{figure}

In principle the activation function $\sigma$ can be any non-linear function, but in practice the choice can be important for network training.
Examples of activation functions are plotted in \Cref{fig:activations}.
Early networks utilized the sigmoid or hyperbolic tangent activations, but these led to \textit{vanishing gradients} in deep networks where multiplying together gradients taken through multiple activations (see \cref{sec:training}) led to an exponential decay in the magnitude of the gradients at the first hidden layer as the network depth increased.
The \textit{rectified linear unit} (ReLU), defined as $\text{ReLU}(z) = \max(0, z)$, became the standard activation following the AlexNet moment given its gradient is either 0 or 1, meaning no exponential decay.
ReLU activations, however, introduce problems of their own, such as the \textit{dying ReLU problem} where if a neuron's value becomes very negative due to some large gradient early in training its gradients are zeroed throughout the rest of training.
ReLU-based MLPs also have a zero second derivative, which can be limiting in specific applications and prohibit the use of second-order optimizers.
Recently smoother non-linearities have become common.
The most important one is the \textit{Gaussian error linear unit} (GELU), defined as $\text{GELU}(z) = z \cdot \Phi(z)$ where $\Phi$ is the standard normal CDF.
GELU has become the preferred activation in many modern architectures, including the transformer networks used in \cref{ch:zjets}.
These functions are plotted in \Cref{fig:activations}.

The most important property of MLPs is that they are universal function approximators, meaning that an MLP with a single hidden layer of sufficient width can approximate any continuous function on a compact domain to arbitrary precision~\cite{cybenko1989approximation,hornik1989multilayer}.
The MLP is in principle flexible enough to approximate any function, even for example the very complicated functions contained in modern LLMs.
In practice this property of MLPs is of limited use because it guarantees only that a solution exists, not that it can be found by any of the gradient-based optimization strategies discussed below.
Deep MLPs and more sophisticated network architectures are used much more widely than very wide, two-layer MLPs because they are empirically observed to find much better solutions when trained with gradient descent.
There are some theoretical motivations for why this might be the case~\cite{jacot2018neural}, but there is not a rigorous theory explaining why some networks optimize better than others.
The MLP architecture on its own is little used in modern deep learning, but it is used as a component in graph and transformer architectures.

\subsection{Training Neural Networks}
\label{sec:training}

The task of training a neural network is to find parameter values $\theta$ that minimize a scalar loss function $\mathcal{L}(\theta)$ computed over a training dataset.
The canonical example of a loss function is the binary cross-entropy (BCE) loss used for binary classification tasks with inputs $\mathbf{x}_i$ and labels $y_i \in \{0, 1\}$,
\begin{equation}
    \mathcal{L}_{\text{BCE}}(\theta) = -\frac{1}{N} \sum_{i=1}^{N} \left[ y_i \log f_\theta(\mathbf{x}_i) + (1 - y_i) \log\!\left(1 - f_\theta(\mathbf{x}_i)\right) \right],
    \label{eqn:bce}
\end{equation}
where $f_\theta(\mathbf{x}_i) \in (0,1)$ is the network's predicted probability that example $i$ belongs to class 1.
Minimizing the BCE loss is equivalent to maximizing the log-likelihood of the training data under the assumption that their labels are described by a Bernoulli random variable.
Constructing loss functions that maximize the log-likelihood of the training data is a recurring pattern across machine learning.

\subsubsection{Backpropagation}

Most supervised and unsupervised machine learning methods approach minimizing $\mathcal{L}(\theta)$ by computing its gradient with respect to each parameter of the model.
The \textit{backpropagation} algorithm allows for the calculation of gradients for each parameter in a deep network by applying the chain rule of calculus layer by layer.
Concretely, the pre-activation neuron value at layer $\ell$ is $\mathbf{z}^{(\ell)} = W^{(\ell)} \mathbf{h}^{(\ell-1)} + \mathbf{b}^{(\ell)}$.
Application of the activation yields the neuron value $\mathbf{h}^{(\ell)} = \sigma(\mathbf{z}^{(\ell)})$.
Gradients must be computed for both the weights and biases of layer $\ell$.
First, the gradient of the loss with respect to the weights of layer $\ell$ is
\begin{equation}
    \frac{\partial \mathcal{L}}{\partial W^{(\ell)}} = \boldsymbol{\delta}^{(\ell)} \left(\mathbf{h}^{(\ell-1)}\right)^\top,
    \label{eqn:backprop_W}
\end{equation}
where the error signal $\boldsymbol{\delta}^{(\ell)}$ is defined recursively as
\begin{equation}
    \boldsymbol{\delta}^{(\ell)} = \left( \left(W^{(\ell+1)}\right)^\top \boldsymbol{\delta}^{(\ell+1)} \right) \odot \sigma'\!\left(\mathbf{z}^{(\ell)}\right),
    \label{eqn:backprop_delta}
\end{equation}
with $\boldsymbol{\delta}^{(L)}$ initialized from the derivative of the loss with respect to the network output and $\odot$ denoting elementwise multiplication.
Starting with $\ell = L$, $\boldsymbol{\delta}^{(\ell)}$ can be computed and then used as the starting point for the calculation of gradients at layer $\ell = L - 1$.
This process was implemented in software for early deep learning frameworks such as Theano, and later TensorFlow 1.x.
All modern deep learning frameworks (PyTorch, JAX, TensorFlow) implement automatic differentiation, which computes gradients through arbitrary compositions of differentiable tensor operations without the user needing to implement \Cref{eqn:backprop_W,eqn:backprop_delta} explicitly.\footnote{The three major frameworks differ in how they implement automatic differentiation. TensorFlow 1.x and Theano used \textit{static computation graphs} where the full computational graph is defined and compiled before any data is processed. PyTorch and TensorFlow 2.x use \textit{dynamic computation graphs} (define-by-run), where the graph is constructed on the fly during the forward pass. JAX makes automatic differentiation as a first-class function transformation (\texttt{jax.grad}) that can be composed with other transforms such as vectorization (\texttt{jax.vmap}) and JIT compilation (\texttt{jax.jit}).}

\subsubsection{Gradient Descent and Optimization}

\begin{figure}[ht]
    \centering
    \includegraphics[width=0.65\linewidth, alt={Three-dimensional surface plot of a bowl-shaped loss function over a two-dimensional parameter space. A sequence of points connected by arrows traces a path descending from a high-loss starting point down the surface toward the minimum at the bottom of the bowl.}]{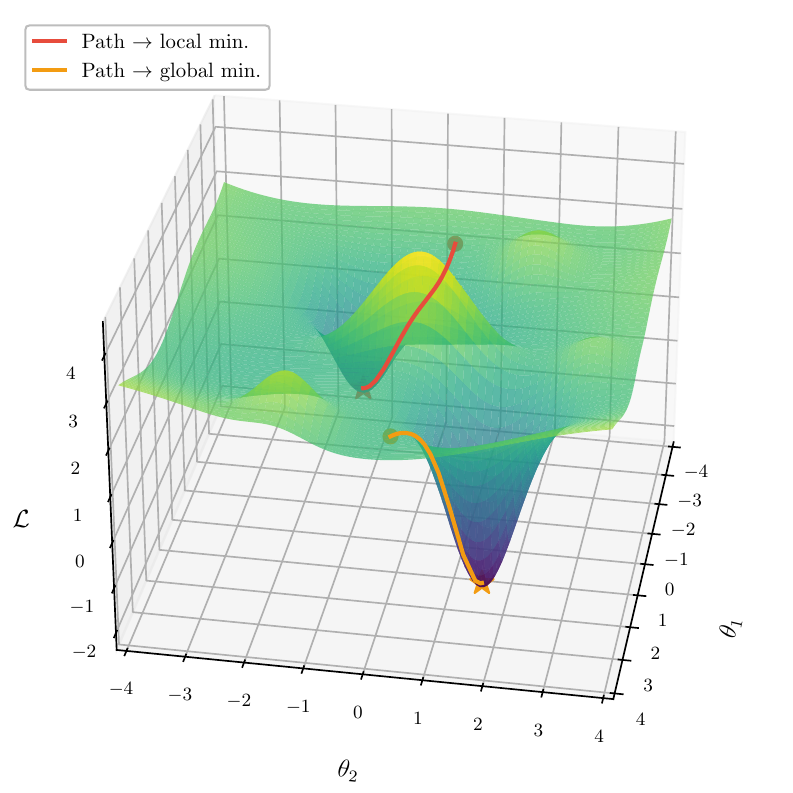}
    \caption{Geometric intuition for gradient descent on a two-dimensional loss surface. Starting from an initial parameter configuration, each update moves in the direction of steepest descent, converging toward a minimum. Gradient descent does not guarantee convergence to the true global minimum. The optimization can depend strongly on the random initialization of the network weights and biases as illustrated by the red and orange paths converging to different minima. Figure generated with Claude Opus 4.6.}
    \label{fig:gradient_descent}
\end{figure}

Once gradients of the loss with respect to all network parameters have been calculated, the parameters can be updated by gradient descent:
\begin{equation}
    \theta \leftarrow \theta - \eta \, \nabla_\theta \mathcal{L}(\theta),
    \label{eqn:gd}
\end{equation}
where $\eta > 0$ is the learning rate.
Intuitively, the negative gradient points in the direction of steepest descent on the loss surface, visualized in two dimensions in \Cref{fig:gradient_descent}.
Each update steps the parameters in the direction that minimizes the loss.
This simple procedure is effective but has limitations.
It has no guarantees about convergence to the global minimum.
Optimization can get stuck in local minima, and also depend strongly on the random initialization of the network weights and biases.
This effect is illustrated in \Cref{fig:gradient_descent}, where the red path converges to a local minimum while the orange path converges to the true global minimum.
The paths differ only by the initialization of the two parameters $\theta_1$ and $\theta_2$. 
Note however that these limitations can matter much less in practice as the fraction of critical points that are minima becomes vanishingly small as the dimensionality of the optimization increases.
This is one of the theoretical explanations for why large neural networks are easier to optimize than small ones.

In \Cref{eqn:gd}, the loss is computed over the full dataset before gradients are computed.
This requires running a forward and backward pass over the network for each data point in the training set before making a single parameter update, which is computationally inefficient and intractable for large datasets.
\textit{Stochastic gradient descent} (SGD) replaces the losses and gradient calculation over the full dataset with the same calculation over a randomly sampled \textit{mini-batch} of $B$ examples.
This procedure dramatically reduces the per-parameter-update computational cost, but introduces variance into the gradient estimates.
Empirically this variance often aids generalization of trained networks.
The learning rate $\eta$ (followed by the batch size $B$) is usually the most critical hyperparameter for any neural network training.
In modern practice, a \textit{learning rate scheduler} that adjusts $\eta$ over the course of training is typically applied.

In SGD, a single learning rate is used for each parameter in a network.
A more advanced alternative is \textit{adaptive optimizers}, which maintain per-parameter estimates of gradient statistics over the past few mini-batches, and adjust the learning rate per parameter based on the size of past gradients.
The commonly used Adam optimizer~\cite{kingma2014adam} computes exponential moving averages of the gradient $m_t$ and the squared gradient $v_t$:
\begin{align}
    m_t &= \beta_1 m_{t-1} + (1 - \beta_1) g_t, \label{eqn:adam_m} \\
    v_t &= \beta_2 v_{t-1} + (1 - \beta_2) g_t^2, \label{eqn:adam_v}
\end{align}
where $g_t = \nabla_\theta \mathcal{L}$ is the gradient at step $t$ and $\beta_1, \beta_2 \in (0,1)$ are decay rates.
Bias-corrected estimates $\hat{m}_t = m_t / (1 - \beta_1^t)$ and $\hat{v}_t = v_t / (1-\beta_2^t)$ are used to compute the parameter update $\theta \leftarrow \theta - \eta \, \hat{m}_t / (\sqrt{\hat{v}_t} + \epsilon)$.
The intuition is that parameters with consistently large (small) gradients receive smaller (larger) updates due to the factor of $\sqrt{v_t}$ in the denominator, allowing networks with a large range of gradient scales to be trained efficiently.

\subsubsection{Generalization and Regularization}

\begin{figure}[t]
    \centering
    \includegraphics[width=0.8\linewidth, alt={Three scatter plots with fitted curves illustrating the bias--variance tradeoff. The left plot shows an underfit model with a straight line that misses the curved trend in the data (high bias). The center plot shows a well-fit smooth curve that captures the underlying trend. The right plot shows an overfit model with a highly wiggly curve that passes through noise (high variance).}]{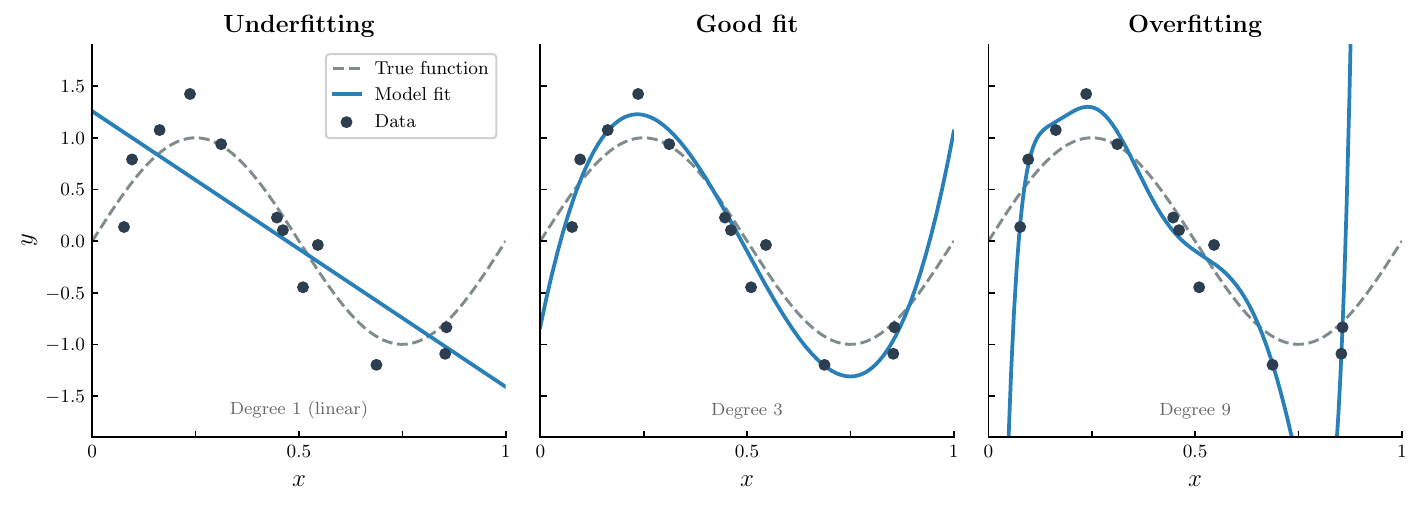}
    \caption{Illustration of the bias--variance tradeoff in the context of a 1D regression problem fitting a polynomial model. An underfit model with one polynomial degree (left) has high bias and fails to capture the true structure in the data. An overfit model with many polynomial degrees (right) has high variance and memorizes noise rather than the underlying function. The goal is to find the intermediate regime (center) that generalizes well. Figure generated by Claude Sonnet 4.5.}
    \label{fig:bias_variance}
\end{figure}

A trained model may not perform well on unseen data if it has ``memorized'' the specific data points rather than learning the underlying function.
This is called \textit{overfitting}, and is best understood as one extreme of the \textit{bias--variance tradeoff}.
A model with low capacity will be unable to fit the training data (high bias) and will simply predict a best guess for all data points (low variance).
A model with high capacity is able to memorize specific training examples (low bias), but can overfit to the data and yield very different results for only slightly different inputs (high variance).
\Cref{fig:bias_variance} illustrates overfitting and the bias--variance tradeoff in a 1D regression problem solved with a polynomial fit.
A model with too little capacity (e.g. too few polynomial degrees) underfits and has high bias, while a model with too much capacity (e.g. too many polynomial degrees) that is not properly regularized overfits and has high variance.

\begin{figure}[t]
    \centering
    \includegraphics[width=0.7\linewidth, alt={Line plot showing training loss and validation loss as a function of training epoch. The training loss decreases monotonically. The validation loss decreases initially then rises after reaching a minimum. A vertical dashed line marks the epoch of minimum validation loss, labeled as the early stopping point.}]{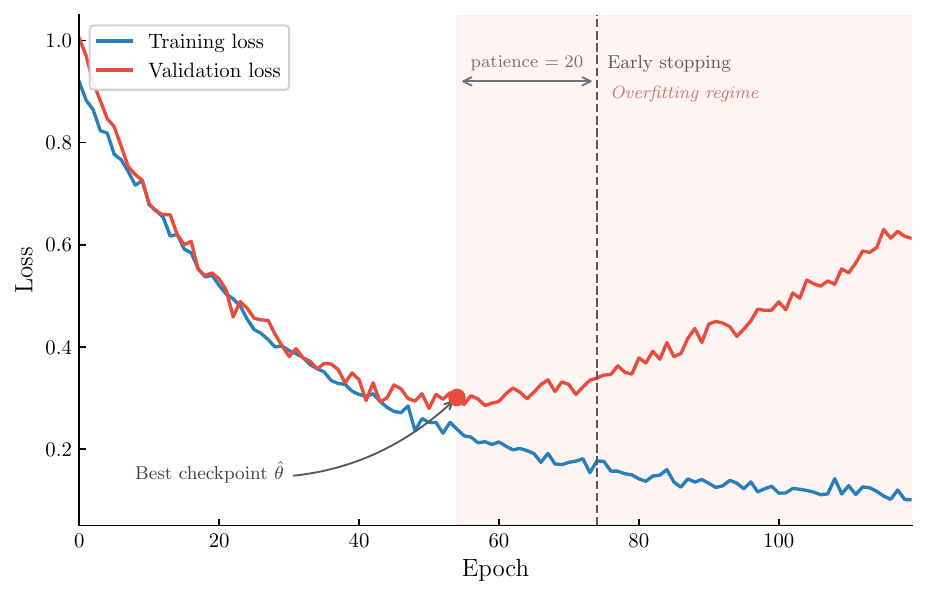}
    \caption{Training and validation loss curves illustrating early stopping. The training loss is plotted in blue and the validation loss in red. The configuration of model parameters that minimizes the validation loss is marked. Training past this point results in sub-optimal generalization. Early stopping with a patience of 20 epochs is marked by the vertical dashed line, indicating where model training would be stopped. Figure generated by Claude Sonnet 4.5.}
    \label{fig:early_stopping}
\end{figure}

\textit{Generalization} is a model's ability to produce accurate predictions on unseen data.
Standard practice to ensure generalization is to partition available data into three disjoint subsets: a \textit{training set} used to train the model, a \textit{validation set} used to monitor performance on held-out data during training, and a \textit{test set} used for a single final evaluation after training is complete.
Gradients are never computed using the validation or test sets.
A classic signal of overfitting is when the loss, calculated over the validation set, begins to decrease with further training, while the training loss always decreases since it is directly minimized.
\textit{Regularization} is a set of methods for preventing the model from overfitting.
A simple regularization strategy is early stopping, in which training is terminated when the validation loss stops decreasing, preventing the network from entering the high variance overfitting regime.
\Cref{fig:early_stopping} shows a typical training curve illustrating this procedure.
The training loss (blue curve) decreases monotonically except for small noise caused by the stochastic gradient estimates computed on mini-batches.
The validation loss (red curve) follows the training loss initially before flattening out and beginning to increase as the model overfits.
The configuration of model weights that offers the best generalization (lowest validation loss) is marked in the Figure.
In early stopping, training is stopped after the validation loss fails to decrease for some number of epochs, called the \textit{patience}, to account for the noise in the validation loss.\footnote{The relationship between model capacity, dataset size, and generalization in modern overparameterized networks with many more parameters than training examples is more subtle than the usual bias--variance tradeoff picture. Very large models that are trained well past the point of interpolating the training data sometimes exhibit \textit{double descent}: a second regime in which increasing capacity or training time actually improves generalization.}

Two other widely used regularization techniques are \textit{dropout} and \textit{weight decay}.
Dropout~\cite{hinton2012improving} randomly zeros a set of activation functions with probability $p$, called the \textit{dropout probability}, during training.
At inference time, all activations are retained.
Weight decay adds an L2 norm penalty on the parameters of the network $\|\theta\|^2$ to the loss.
This discourages large weights and limits the capacity of the network to overfit.
Weight decay can be understood as placing a zero-mean Gaussian prior on the network parameters.

\subsection{Convolutional Neural Networks}
\label{sec:cnn}

MLPs operate on flat vectors of inputs and have no \textit{implicit bias}, or architectural specialization to a given known structure in the input data.
Convolutional neural networks (CNNs) encode \textit{translational symmetry} into the network architecture, meaning that a feature at one spatial location should not affect the network output if shifted to another location.
This is achieved by \textit{weight sharing}, in which a set of weights and biases, often called a \textit{filter}, is scanned across the inputs rather than applying a separate set of weights per position.
CNNs were first developed for computer vision tasks using image data as input.
Specifically AlexNet was an early example of a CNN architecture.

Concretely, a CNN layer applies a set of $F$ learned filters $\{k_f\}_{f=1}^F$.
The set of learned weights that define each filter are often called a \textit{kernel}.
The kernel has a spatial extent $K$ (or $K \times K$ for 2D convolutions) which gives the number of input positions it considers at one time, and a depth which is equal to the number of \textit{channels} attached to each input position $C_{\text{in}}$.
The number of output channels is equal to the number of filters $F = C_{\text{out}}$.
For the specific case of a 2D convolution, the output at spatial position $(i, j)$ and output channel $f$ is
\begin{equation}
    h_{f,i,j} = \sigma\!\left( \sum_{c=1}^{C_{\text{in}}} \sum_{p=0}^{K-1} \sum_{q=0}^{K-1} k_{f,c,p,q} \cdot x_{c,\, i+p,\, j+q} + b_f \right),
    \label{eqn:conv2d}
\end{equation}
where $x_{c,i,j}$ is the input at channel $c$ and position $(i,j)$, $k_{f,c,p,q}$ are the kernel weights, and $b_f$ is the kernel bias.
The number of channels $C$ sets how many floating point numbers are used to represent each spatial location.
For the input layer in image data this is typically 3 (RGB color channels), but it grows to hundreds in the deeper layers of large networks.
This channel dimension is also directly analogous to the \textit{hidden dimension} of transformer models, which will be discussed in \Cref{sec:attention} and used extensively in this thesis.
The filters of a given convolutional layer are sometimes applied with a \textit{stride} $s > 1$, meaning they advance by $s$ positions at each step.
This makes the number of spatial sites in the output smaller than in the input, allowing the network to extract increasingly broad summary features.

\begin{figure}[htbp]
    \centering
    \includegraphics[width=0.85\linewidth, alt={Two-part diagram illustrating CNN operations. On the left, a 2D convolution is shown: a small learned filter slides over an input feature map and computes dot products at each position to produce a smaller output feature map. On the right, max pooling is shown: the feature map is divided into non-overlapping regions and the maximum value in each region is retained, reducing the spatial dimensions by a factor of two.}]{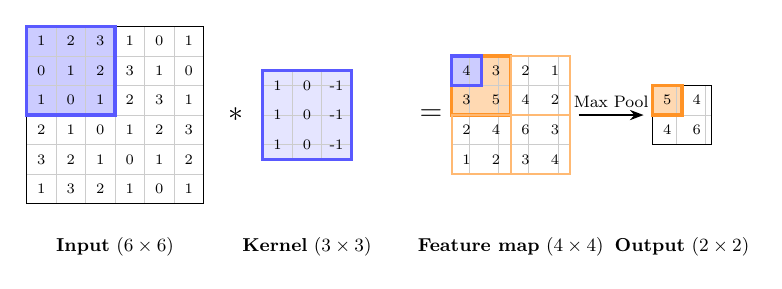}
    \caption{The two core spatial operations in a convolutional neural network. \textit{Left}: a learned $K \times K$ kernel is convolved with the input feature map to produce a new feature map, with each output value capturing local spatial structure. \textit{Right}: max pooling partitions the feature map into non-overlapping windows and retains only the maximum activation in each, reducing spatial resolution and providing local translation invariance. Figure generated by Claude Opus 4.6.}
    \label{fig:convolution}
\end{figure}

Another important operation in CNNs is \textit{pooling}, in which an aggregation operation like maximum (\textit{max pooling}) or average (\textit{average pooling}) is applied across a spatial region.
As with setting a filter's stride to be greater than one, pooling reduces the spatial dimensions of feature maps and encourages the network to extract broader summary features as depth increases.
Both operations are illustrated in \Cref{fig:convolution}.

\begin{figure}[ht]
    \centering
    \includegraphics[width=0.35\linewidth, alt={Diagram of a residual block. An input arrow enters from the top and splits into two paths: one passes through two convolutional layers with a ReLU activation in between, and the other bypasses these layers via a skip connection. The two paths are recombined by addition before a final ReLU activation produces the block output.}]{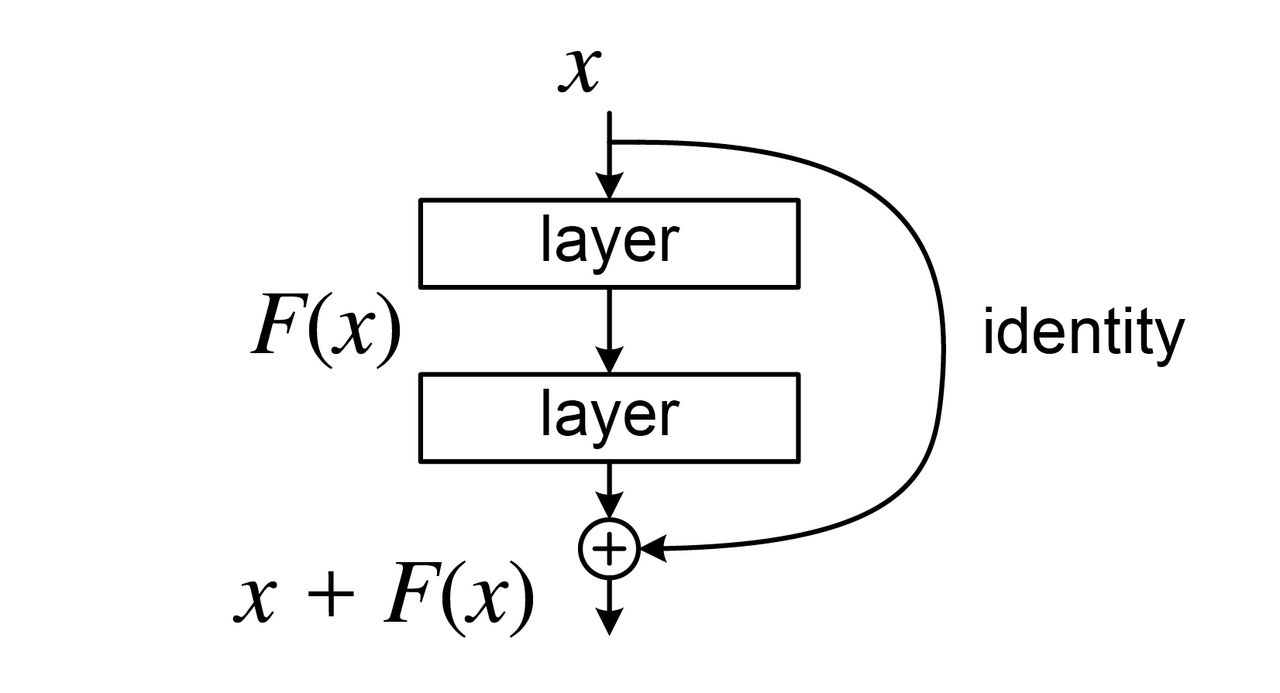}
    \caption{A single residual block. The block applies two layers of operations (convolution or otherwise) to its input and adds the result to the original input via a skip connection. This design allows gradient signals to propagate cleanly through very deep networks.}
    \label{fig:resblock}
\end{figure}

Just like with MLPs, very deep CNNs were found to suffer from vanishing gradients.
One method developed to address this was the \textit{residual} or \textit{skip} connection~\cite{He:2015resnet}, in which the output of a set of operations is added to its input:
\begin{equation}
    \mathbf{h}_{\text{out}} = \mathcal{F}(\mathbf{h}_{\text{in}}) + \mathbf{h}_{\text{in}},
    \label{eqn:residual}
\end{equation}
where $\mathcal{F}$ denotes one or more operations\footnote{Re-arranging this equation implies that the role of the operations is to learn \textit{residual} corrections on the data representation needed to minimize the loss. This idea connects to the idea of \textit{boosting} in tree models and to the diffusion models introduced in \Cref{sec:diffusion}.}.
The skip connection creates a direct gradient path from the output to the input of each block, making it possible to train networks with hundreds of layers.
The structure of a single residual block is shown in \Cref{fig:resblock}.

\begin{figure}[ht]
    \centering
    \includegraphics[width=0.9\linewidth, alt={Diagram of the full ResNet architecture. An input image enters a convolutional layer followed by batch normalization and ReLU, then passes through four groups of stacked residual blocks with increasing numbers of output channels and decreasing spatial resolution. The final feature map is globally average-pooled and passed to a fully connected layer that produces class scores.}]{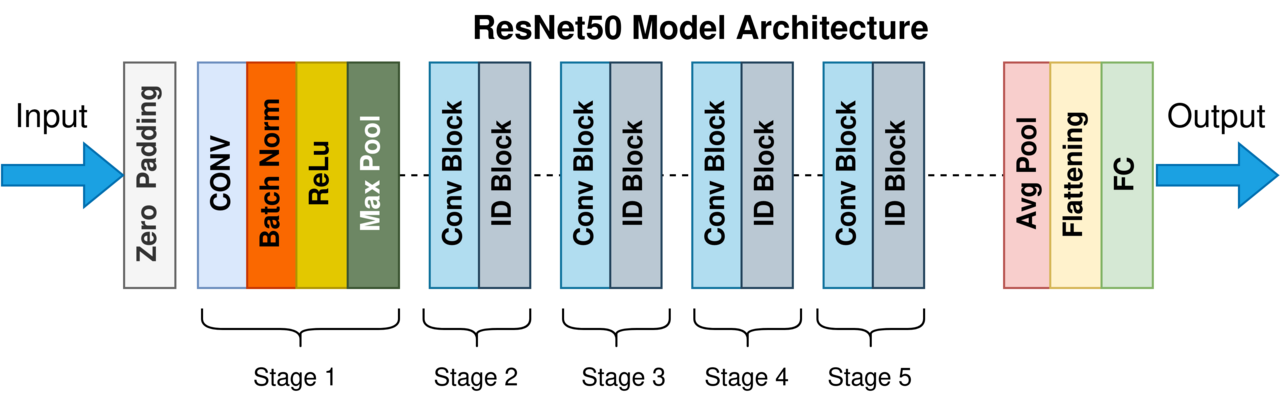}
    \caption{The ResNet architecture, composed of stacked residual blocks which alternate convolution operations and identity blocks (skip connections). The ResNets in the original paper had hundreds of layers and tens of millions of parameters, which were extraordinarily large in late 2015. Reproduced from \citet{gorlapraveen123_resnet50_2021} via Wikimedia Commons under a \href{https://creativecommons.org/licenses/by-sa/4.0/}{CC~BY~4.0} license.}
    \label{fig:resnet}
\end{figure}

The \textit{ResNet} architecture~\cite{He:2015resnet} stacks many residual blocks on top of each other as shown in \Cref{fig:resnet}.
The ``ID blocks'' shown in the Figure correspond to skip connections.
ResNet was the dominant image classification architecture for several years after it was introduced in 2015.
The first application of CNNs in HEP was for jet tagging tasks as discussed in \Cref{ch:tagging}, where the energy deposits in a calorimeter were interpreted as a 2D image and used as input to CNN models~\cite{Cogan:2014oua,deOliveira:2017pjk}.
ResNet-based models showed nearly state-of-the-art performance on jet classification tasks as recently as 2019~\cite{Kasieczka:2019dbj}.
They were ultimately superseded by graph neural networks and then transformers, but CNNs are an important part of the history of deep learning applications in HEP.

\subsection{Graph Neural Networks}
\label{sec:gnn}

After CNNs, the next major step in the history of deep learning in HEP was to interpret jets as \textit{point clouds} rather than images.
A point cloud is an unordered set of points in a high-dimensional space, for example the four-momenta of the particles in a jet.
Graph neural networks (GNNs)~\cite{Battaglia:2018mnb} interpret sets as graphs, in which each item in the set is a node connected to other nodes in the set by edges.

Most GNNs used in HEP are instances of the \textit{message-passing} framework, where the nodes of an input graph are updated by a series of operations which aggregate \textit{messages} passed along all of the node's edges.
The update at node $i$ after one message-passing step is
\begin{equation}
    \mathbf{h}_i' = \phi_{\text{node}}\!\left( \mathbf{h}_i,\; \bigoplus_{j \in \mathcal{N}(i)} \phi_{\text{edge}}(\mathbf{h}_i, \mathbf{h}_j) \right),
    \label{eqn:message_passing}
\end{equation}
where $\mathbf{h}_i$ is the current feature vector of node $i$, $\mathcal{N}(i)$ is the set of nodes connected to node $i$, $\phi_{\text{edge}}$ is a learned edge function that computes a message from each neighbor, $\bigoplus$ is a permutation-invariant aggregation (e.g.\ sum or max), and $\phi_{\text{node}}$ is a learned update function.
Both $\phi_{\text{edge}}$ and $\phi_{\text{node}}$ are typically implemented as MLPs.

The ParticleNet architecture~\cite{Qu:2019gqs}, which will be applied to a jet classification task in \Cref{ch:tagging}, uses a specific message-passing operation called \textit{edge convolution} (EdgeConv).
In EdgeConv, the message passed along the edge from node $j$ to node $i$ is computed from both the absolute feature of node $j$ and the \textit{relative} feature between $j$ and $i$:
\begin{equation}
    \mathbf{h}_i' = \frac{1}{|\mathcal{N}(i)|} \sum_{j \in \mathcal{N}(i)} \phi_{\text{edge}}\!\left( \mathbf{h}_i,\; \mathbf{h}_j - \mathbf{h}_i \right),
    \label{eqn:edgeconv}
\end{equation}
where $\phi_{\text{edge}}$ is a small MLP.
The relative features $\mathbf{h}_j - \mathbf{h}_i$ capture pairwise distances between inputs and are empirically found to be very useful in HEP, for example because the difference between two particles' four-momenta is physically meaningful.

Another important type of message passing GNN is graph attention networks (GATs), which generalize this pattern by replacing the uniform averaging in \Cref{eqn:edgeconv} with learned, input-dependent weights over the neighborhood.
This is identical to the attention mechanism covered in \Cref{sec:attention}, except with some entries of the attention matrix zeroed such that only inputs within a connected neighborhood of each other can attend.
The better expressivity and better scaling properties of full attention, in which the strength of the connection between two inputs is allowed to be a learned feature, has led attention-based networks to largely supplant GNNs in HEP applications.

\subsection{Attention and Transformers}
\label{sec:attention}

The attention mechanism, and the transformer architecture built on top of it, underlie most consumer AI products in use today.
They have seen equally broad adoption in HEP, so this Section will introduce them in some detail.

The origins of attention were in augmenting recurrent neural networks (RNNs) to perform better on machine translation tasks~\cite{bahdanau2014neural, luong2015effective, cheng2016long, lin2017structured}.
In early versions of attention the decoder of an RNN, whose job is to generate the next output token given the network's current internal representation, could selectively weight all positions in the input sequence according to their relevance to the current decoding step.
This was found to be more performant than compressing every token in the input into a fixed-size vector to provide context for the decoder.
The next conceptual step was to remove the recurrence entirely and simply compute attention between all items in the input sequence, with the relative positions of the items provided to the attention mechanism as positional encodings.
This was the contribution of the famous ``Attention Is All You Need'' paper~\cite{Vaswani:2017attention}\footnote{Arguably the citation count of this paper relative to the initial papers that actually introduced attention is evidence of the Matthew effect in science, in which papers with high citation counts attract even more citations as time goes on. The Vaswani paper became the point of reference for early transformer models like BERT and GPT. That made it the starting point for anyone learning about attention and ensured it would keep being cited in the future. Despite this success Vaswani et al. did not actually invent attention; that credit belongs to Bahdanau et al. and Luong et al.}.
This self-attention operation, without any additional recurrent operations computed on the input sequence, is the backbone of the transformer architecture applied at many points in this thesis and so is worth covering in detail.

The first step of self-attention is to compute three linear projections of the input set.
The input is represented as set of $N$ vectors, each with dimension $d$ and arranged as a matrix $X \in \mathbb{R}^{N \times d}$.
The linear projections are:
\begin{equation}
    Q = X W_Q, \quad K = X W_K, \quad V = X W_V,
    \label{eqn:qkv}
\end{equation}
where $W_Q, W_K, W_V \in \mathbb{R}^{d \times d_k}$ are learned weight matrices, producing three matrices known as the queries $Q$, keys $K$, and values $V$.
$d_k$ is the \textit{key dimension} and is typically set to a small value (e.g. 64) to reduce the computational cost of the attention operation.
Self-attention is then computed using these three matrices as
\begin{equation}
    \text{Attention}(Q, K, V) = \text{softmax}\!\left( \frac{Q K^\top}{\sqrt{d_k}} \right) V.
    \label{eqn:attention}
\end{equation}
The matrix $Q K^\top / \sqrt{d_k} \in \mathbb{R}^{N \times N}$ is the \textit{attention matrix}.
The $(i,j)$-th entry of this matrix can be interpreted as setting how much the input at position $i$ (query) should affect the item at position $j$ (key).
Applying a softmax row-wise ensures that these attention weights for a given query sum to one.
Multiplying by $V$ produces a weighted sum of value vectors for each position, which is the output of the attention operation.
Note that the vectors which determine the attention weights (keys) are not the same as the vectors which are being attended to (values).
This is a key difference between the modern, scaled dot-product attention operation and the original attention operation~\cite{bahdanau2014neural}.
The $\sqrt{d_k}$ scaling factor exists to prevent the variance of the entries of the attention matrix from growing with $d_k$.
Without this factor, the standard deviation of the entries of the pre-softmax attention matrix would grow with the square-root of $d_k$, meaning larger $d_k$ would imply different post-softmax statistics.

\begin{figure}[t]
    \centering
    \includegraphics[width=0.85\linewidth, alt={Diagram of scaled dot-product attention and multi-head attention. On the left, the single-head attention mechanism shows queries, keys, and values fed through a matrix multiply, scale, softmax, and final matrix multiply. On the right, the multi-head variant shows H parallel attention heads whose outputs are concatenated and linearly projected.}]{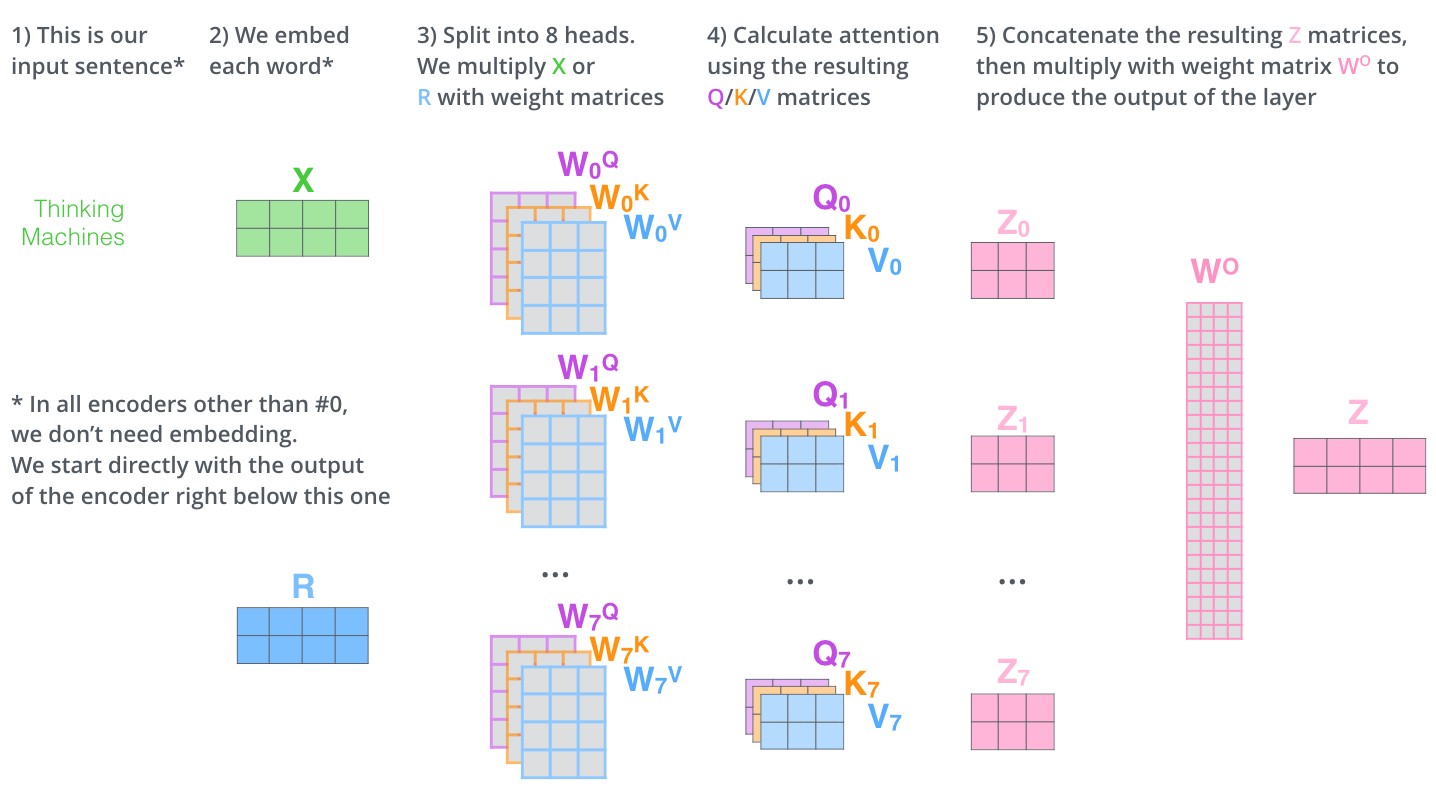}
    \caption{Scaled dot-product attention (left) and multi-head attention (right). Multi-head attention runs $H$ independent attention heads in parallel, concatenates their outputs, and applies a learned linear projection. Adapted from \citet{Alammar2018} under a \href{https://creativecommons.org/licenses/by-sa/4.0/}{CC~BY~4.0} license.}
    \label{fig:multihead_attention}
\end{figure}

In modern practice the self-attention operation is typically applied several times in parallel across $H$ independent \textit{attention heads}, each with its own projection matrices and a smaller key dimension $d_k = d / H$.
Outputs of the heads get concatenated together and finally projected back to dimension $d$:
\begin{equation}
    \text{MultiHead}(Q, K, V) = \text{Concat}(\text{head}_1, \ldots, \text{head}_H) \, W_O,
    \label{eqn:multihead}
\end{equation}
where $\text{head}_h = \text{Attention}(X W_{Q,h}, X W_{K,h}, X W_{V,h})$ and $W_O \in \mathbb{R}^{d \times d}$ is a learned output projection.
The benefit of multi-head attention is that the network can attend to inputs based on different aspects of the relationships between elements.
A diagram of the entire multi-head attention operation is shown in \Cref{fig:multihead_attention}.

\begin{figure}[htbp]
    \centering
    \includegraphics[width=0.3\linewidth, alt={Diagram of a single transformer block. Tokens enter from the bottom and pass through a multi-head self-attention sublayer followed by layer normalization with a residual connection, then through a feed-forward network sublayer followed by a second layer normalization with another residual connection, before exiting at the top.}]{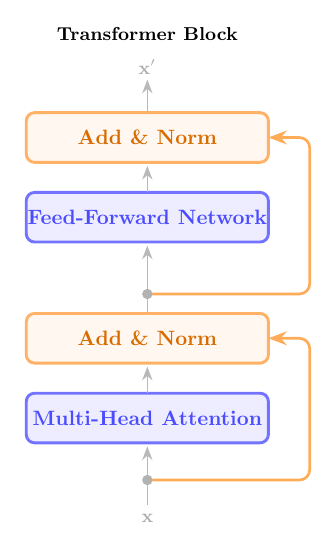}
    \caption{A single transformer block. Each block applies multi-head self-attention and a position-wise feed-forward network (FFN), each wrapped in a residual connection and layer normalization. A full transformer is formed by stacking $L$ such blocks. Generated by Claude Sonnet 4.6.}
    \label{fig:transformer_block}
\end{figure}

Self-attention is the most important operation in the \textit{transformer block}.
It captures relationships between inputs, but does not allow the element-wise processing of representations.
To address this, the \textit{transformer block} stacks an element-wise MLP on top of the attention mechanism, in addition to layer normalization and residual connection operations:
\begin{align}
    \mathbf{X}' &= \text{LayerNorm}\!\left( X + \text{MultiHead}(X) \right), \label{eqn:transformer_attn} \\
    \mathbf{X}'' &= \text{LayerNorm}\!\left( \mathbf{X}' + \text{FFN}(\mathbf{X}') \right). \label{eqn:transformer_ffn}
\end{align}
A transformer network is built by stacking $L$ such blocks, with the output of each block serving as the input to the next\footnote{More specifically the network described here is a transformer \textit{encoder}. The transformer architecture as originally published is meant specifically for machine translation and includes a decoder network for producing output sequences. However in current terminology, any network that utilizes the transformer block can be called a transformer.}.
The transformer block is illustrated in \Cref{fig:transformer_block}.

Importantly self-attention as defined in \Cref{eqn:attention} is \textit{permutation equivariant}, meaning permuting the rows of $X$ permutes the rows of the output without changing the actual vector representing the items in the set.
In natural language processing this is a problem because word order carries meaning.
This is why Ref.~\cite{Vaswani:2017attention} added \textit{positional encodings} to the input that break permutation symmetry.
However for HEP data permutation equivariance is often a very useful inductive bias~\cite{Komiske:2018cqr}.
Jets and events are composed of unordered sets of particles, where the order of the particles is typically not meaningful\footnote{An exception to this will be covered in the set-to-set unfolding work in \Cref{ch:unfolding}.}.
This means the transformer architecture can be applied to HEP data directly~\cite{Mikuni:2020wpr}, which is why transformers are so widely used in HEP whereas their direct predecessors, RNNs, are not.

%% file: chapter3_generative.tex
\section{Generative Models}
\label{sec:generative}

The previous section introduced several neural network architectures and used classification as a representative task.
However neural networks are universal function approximators and can be trained to perform many different tasks.
This section covers generative modeling, which falls under the umbrella of unsupervised learning.
The goal is to model the probability density $p_\text{data}(x)$ from which a training dataset is drawn.
Generative models can be used to draw new samples from the data generating distribution, estimate the density of a given point under the learned distribution, or learn a compressed latent representation of the data.
Additionally generative models can be \textit{conditional} on additional information, such as a set of labels or a text prompt.
These models seek to learn the joint distribution $p_\text{data}(x, y)$ of the data and the additional information.

Generative modeling has a long history predating the deep learning era.
In what follows I will cover three specific generative models of the post-AlexNet era which have seen significant application in HEP.

\subsection{Generative Adversarial Networks}
\label{sec:gan}

Generative Adversarial Networks (GANs)~\cite{Goodfellow:2014gan} were introduced in 2014.
GANs are unique since they do not try to explicitly model a likelihood of the training data.
Instead training a GAN involves setting up an adversarial game in which a \textit{generator} $G_\phi(z)$ that maps a noise vector $z \sim p(z)$ (typically a standard Gaussian) to a sample in data space, is trained to fool a \textit{discriminator} $D_\psi(x) \in (0,1)$ that acts as a judge determining whether a given data sample is a true draw from the training data.
The two networks are trained simultaneously according to the minimax objective
\begin{equation}
    \min_\phi \max_\psi \; \mathbb{E}_{x \sim p_\text{data}}\!\left[\log D_\psi(x)\right]
    + \mathbb{E}_{z \sim p(z)}\!\left[\log\!\left(1 - D_\psi(G_\phi(z))\right)\right].
    \label{eqn:gan}
\end{equation}
This loss function can be interpreted as training the discriminator to distinguish real from generated samples, and the generator to produce samples that fool the discriminator.
Assuming the game is ``well balanced'' and the generator has sufficient capacity, the generator distribution will eventually match $p_\text{data}$ and the discriminator can do no better than random guessing.
There is no explicit likelihood in \Cref{eqn:gan}.
The generator is never required to assign a density to the data, only to produce samples that are indistinguishable from it, so GANs are not capable of density estimation for example.
GANs are conceptually simple and capable of producing high-quality samples, but the training is notoriously difficult.
Common issues include mode collapse, where the generator learns to produce only a limited subset of the data distribution, and non-convergence, where the generator and discriminator cycle but never converge to a stable equilibrium.

GANs were among the first generative architectures applied to HEP data.
They were the first generative models to produce realistic jets and calorimeter showers~\cite{Paganini:2017hrr,Paganini:2017dwg,deOliveira:2017pjk,Butter:2020qhk,Buhmann:2023pmh} or perform unfolding~\cite{Bellagente:2019uyp}. 
GAN-based fast calorimeter simulation tools are still in active use by the ATLAS experiment~\cite{ATLAS:2021pzo}.

\subsection{Variational Autoencoders}
\label{sec:vae}

In contrast to GANs, Variational Autoencoders (VAEs)~\cite{Kingma:2013vae} attempt to explicitly model the likelihood of the training data.
Given this approach is important to the unfolding work in \Cref{ch:unfolding}, it is worth reviewing the basics of VAEs here.
VAEs are \textit{latent variable models}, in which generation of each data point $x$ is governed by a latent variable $z \sim p(z)$ from a prior (typically a Gaussian) and a conditional likelihood $p(x | z)$\footnote{The choice to introduce a latent vector $z$ relates to the manifold hypothesis discussed in \cref{sec:id}. This latent vector is typically much lower dimensional than the data representation, implying that the manifold the data cluster on can be modeled using these latent variables.}.
The complicated object is the conditional likelihood, which we attempt to model with a generative model, $p_{\theta}(x | z)$ with parameters $\theta$.
The marginal likelihood is
\begin{equation}
    p_\theta(x) = \int p_\theta(x | z)\, p(z)\, dz.
    \label{eqn:marginal}
\end{equation}
Direct maximum likelihood training would require evaluating this integral using an MCMC estimator for every mini-batch of data, which is intractable for high-dimensional $z$.
Attempting to model the posterior instead would produce the same issue, since by Bayes rule the posterior is inversely proportional to $p(x)$ which is the exact density we seek to model to begin with.
The VAE sidesteps these problems by using an \textit{encoder} network $q_\phi(z | x)$ to approximate the true posterior $p_\theta(z | x)$, and then optimizing a lower bound on the log-likelihood.
This lower bound is derived in the next section.

\subsubsection{Derivation of the ELBO}

The goal is to maximize the log-likelihood of a single data point $x$.
The first step is to introduce $q_\phi(z|x)$ by multiplying and dividing inside the expectation:
\begin{align}
    \log p_\theta(x)
    &= \log \int p_\theta(x|z)\,p(z)\,dz \notag \\
    &= \log \int q_\phi(z|x) \frac{p_\theta(x|z)\,p(z)}{q_\phi(z|x)}\,dz \notag \\
    &= \log \mathbb{E}_{z \sim q_\phi(z|x)}\!\left[\frac{p_\theta(x|z)\,p(z)}{q_\phi(z|x)}\right].
    \label{eqn:elbo_step1}
\end{align}
Applying Jensen's inequality ($\log \mathbb{E}[Y] \geq \mathbb{E}[\log Y]$) to move the log through the expectation gives
\begin{align}
    \log p_\theta(x)
    &\geq \mathbb{E}_{z \sim q_\phi(z|x)}\!\left[\log \frac{p_\theta(x|z)\,p(z)}{q_\phi(z|x)}\right] \notag \\
    &= \mathbb{E}_{z \sim q_\phi(z|x)}\!\left[\log p_\theta(x|z)\right]
       - \mathbb{E}_{z \sim q_\phi(z|x)}\!\left[\log \frac{q_\phi(z|x)}{p(z)}\right] \notag \\
    &= \underbrace{\mathbb{E}_{z \sim q_\phi(z|x)}\!\left[\log p_\theta(x|z)\right]}_{\text{reconstruction}}
       - \underbrace{D_\text{KL}\!\left(q_\phi(z|x) \,\|\, p(z)\right)}_{\text{regularization}}.
    \label{eqn:elbo}
\end{align}
The right side of the expression is a lower bound on the object we want to maximize, and is called the \textit{Evidence Lower Bound} (ELBO).
The ELBO can be maximized with respect to both $\theta$ and $\phi$, pushing the density provided by the decoder $p_{\theta}(x | z)$ toward the true log-likelihood.
If the approximate posterior $q_\phi(z|x)$ exactly equals the true posterior $p_\theta(z|x)$, then the expression reduces to an equality and the loss function equates to directly maximizing the log-likelihood of the data as desired.

Each term of the ELBO has a natural interpretation.
The \textit{reconstruction term} $\mathbb{E}_{z \sim q_\phi}[\log p_\theta(x|z)]$ ensures the decoder accurately reconstructs the data point $x$ given a sample from the latent distribution.
The \textit{regularization term} $D_\text{KL}(q_\phi(z|x) \| p(z))$ constrains the approximate posterior to be close to the prior.
This constraint term imparts some structure on the latent space and prevents the model from overfitting to the training data.
Without the regularization term the VAE reduces to a standard autoencoder.
Standard autoencoders can reconstruct the training data, but given their latent space is not constrained to follow any known distribution sampling is difficult, making them difficult to use as generative models.

In a standard VAE model, the prior $p(z) = \mathcal{N}(0, I)$ and the approximate posterior $q_\phi(z|x) = \mathcal{N}(\mu_\phi(x), \text{diag}(\sigma^2_\phi(x)))$ are taken to be Gaussian.
Note that the mean and variance of the approximate posterior are predicted by the encoder network with parameters $\phi$.
Under this choice the KL divergence has a closed form:
\begin{equation}
    D_\text{KL}\!\left(\mathcal{N}(\mu, \sigma^2 I) \,\|\, \mathcal{N}(0, I)\right)
    = \frac{1}{2} \sum_{j=1}^{d} \left(\mu_j^2 + \sigma_j^2 - \log \sigma_j^2 - 1\right),
    \label{eqn:kl_gaussian}
\end{equation}
where $d$ is the dimensionality of the latent space, and $\mu_j$ and $\sigma_j$ are the mean and standard deviation of the $j$th component predicted by the encoder.
This term can be computed analytically and differentiated without any Monte Carlo estimation.

\subsubsection{The Reparameterization Trick}

Taking the derivative of the reconstruction term in \Cref{eqn:elbo} requires back-propagating through the operations that sample from the approximate posterior $z \sim q_\phi(z|x)$.
This is not possible given the sampling procedure is stochastic.
The \textit{reparameterization trick} solves this issue by rewriting the sample as a deterministic function of the encoder output and some sampled noise:
\begin{equation}
    z = \mu_\phi(x) + \sigma_\phi(x) \odot \varepsilon, \quad \varepsilon \sim \mathcal{N}(0, I).
    \label{eqn:reparam}
\end{equation}
The random variable in this equation is $\varepsilon$, which does not depend on $\phi$, while the predictions of the encoder have completely deterministic effects on the sampled $z$.
This trick allows gradients to flow through $\mu_\phi$ and $\sigma_\phi$ to the encoder parameters without difficulty.

\subsubsection{Reconstruction Loss and Limitations}

\begin{figure}[ht]
    \centering
    \includegraphics[width=0.85\linewidth, alt={Schematic of a variational autoencoder. An input x is fed into an encoder network that outputs the mean and variance of a Gaussian distribution over a latent variable z. A latent sample z is drawn via the reparameterization trick and passed to a decoder network that reconstructs x. Arrows indicate the forward pass, and a dashed arrow from the prior distribution to the KL divergence term indicates the regularization loss.}]{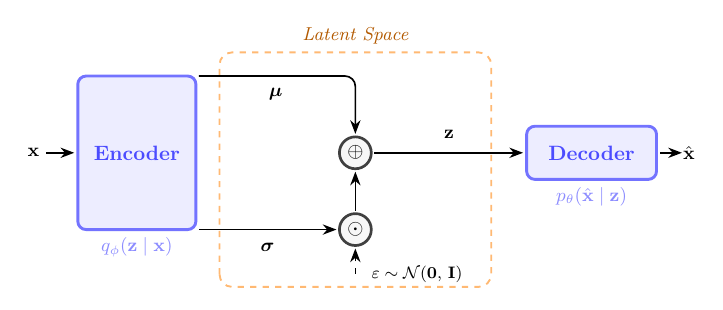}
    \caption{Schematic of the variational autoencoder. The encoder $q_\phi(z|x)$ maps an input $x$ to a distribution over latent codes $z$. A sample is drawn via the reparameterization trick and passed to the decoder $p_\theta(x|z)$, which reconstructs $x$. Generation of new samples is done by sampling from the latent space prior $p(z)$, and then passing the samples through the decoder network. Generated by Claude Sonnet 4.6.}
    \label{fig:vae}
\end{figure}

For continuous data, standard practice is to choose the decoder likelihood as a Gaussian $p_\theta(x|z) = \mathcal{N}(x; \mu_\theta(z), \sigma^2 I)$, where all neural network parameters serve only to calculate the mean $\mu_{\theta}(z)$.
This choice reduces the reconstruction term to a mean squared error (MSE) loss between the input and the decoder's predicted mean $\mu_\theta(z)$.
The full ELBO training objective is then
\begin{equation}
    \mathcal{L}_\text{VAE}(\theta, \phi; x) = \frac{1}{2\sigma^2}\left\|x - \mu_\theta(z)\right\|^2
    + D_\text{KL}\!\left(q_\phi(z|x)\,\|\,p(z)\right).
    \label{eqn:vae_loss}
\end{equation}
Just like with \Cref{eqn:bce,eqn:gan}, this loss function can be minimized or maximized by gradient descent over mini-batches of training examples.
A diagram of the overall encoder-decoder structure is illustrated in \Cref{fig:vae}.

One possible failure mode of the VAE stemming from the assumed Gaussian likelihood is worth noting explicitly.
Under the MSE loss, the encoder learns to predict the mean of the distribution $p_{\text{data}}(x|z)$.
If $p_\text{data}(x|z)$ is multimodal, the decoder will learn to predict an average across all modes, which could be a location with low likelihood.
In practice this can result in poor sample quality.
This limitation will be relevant for the unfolding application in \Cref{ch:unfolding}.

Despite this limitation, the easy training of the VAEs compared to GANs meant they were quickly adapted within HEP for generative tasks such as fast calorimeter simulation, unfolding, and density estimation~\cite{Howard:2021pos,Tsan:2021brw,Cheng:2020dal}.

\subsection{Normalizing Flows}
\label{sec:flows}

In contrast to VAEs, normalizing flows model the likelihood of the data directly, rather than using an auxiliary network to approximate the posterior.
They handle the intractable integral in \Cref{eqn:marginal} by insisting that the modeled likelihood is \textit{invertible}, making the integral over $z$ trivial to compute because only one $z$ maps to a particular $x$.
Normalizing flows define a \textit{bijective} mapping $f_\theta: \mathcal{X} \to \mathcal{Z}$ between the data space $\mathcal{X}$ and a latent space $\mathcal{Z}$ of equal dimension.
Given a prior $p(z)$, typically taken to be Gaussian, and this bijective mapping, an exact expression for the data likelihood can be written using the change of variables formula:
\begin{equation}
    \log p_\theta(x) = \log p(f_\theta(x)) + \log\left|\det J_{f_\theta}(x)\right|,
    \label{eqn:flow_likelihood}
\end{equation}
where $J_{f_\theta}(x) = \partial f_\theta / \partial x$ is the Jacobian of the transformation.
As with VAEs, the training objective is to maximize the log-likelihood of the model over the training data.
Unlike VAEs, there is no approximate posterior and the quantity of interest is optimized directly.
Given the exact loss function, once trained, flow models support sampling (by drawing $z \sim p(z)$ and computing $f_\theta^{-1}(z)$) and density estimation of any point $x$.
This ability to evaluate $p_\theta(x)$ without MCMC sampling or other tricks is a unique ability of flow models.

\begin{figure}[ht]
    \centering
    \includegraphics[width=0.75\linewidth, alt={Diagram of a normalizing flow composed of coupling layers. The input vector is split into two halves shown as separate channels. In each coupling layer, one half passes through unchanged while the other is transformed by a function conditioned on the first half. The roles of the two halves alternate between layers, ensuring all dimensions are eventually transformed. The final output is the mapped latent variable z.}]{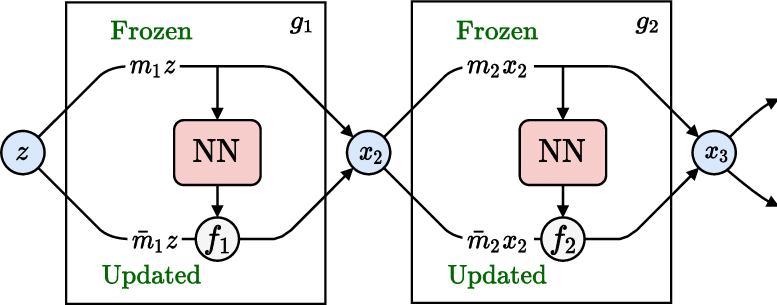}
    \caption{Illustration of two coupling layers in a normalizing flow. At each layer the input is split into two halves. The conditioning half determines the transformation applied to the other. Switching the roles of the halves across layers ensures all dimensions are transformed. Figure from~\cite{Albergo:2021bna} reproduced under a CC~BY~4.0 license.}
    \label{fig:coupling_flow}
\end{figure}

In order to evaluate \Cref{eqn:flow_likelihood}, a bijective-mapping $f_\theta$ whose Jacobian is practical to compute is needed.
For an MLP layer the Jacobian is a generic $n \times n$ whose determinant costs $\mathcal{O}(n^3)$ to compute, which is prohibitive for high-dimensional $x$.
Coupling layers~\cite{dinh2016density} avoid this issue by requiring a lower triangular Jacobian by construction, making the determinant the product of the diagonal entries of the Jacobian.
The lower triangular Jacobian is achieved by splitting the input $x$ into two halves, $x_{1:k}$ and $x_{k+1:n}$.
The first half passes through the coupling layer unchanged.
An elementwise transformation is applied to the second half whose parameters are determined by the first half.
The \textit{affine coupling layer} applies the transformation
\begin{align}
    z_{1:k} &= x_{1:k}, \label{eqn:coupling_id} \\
    z_{k+1:n} &= x_{k+1:n} \odot \exp\!\left(s_\theta(x_{1:k})\right) + t_\theta(x_{1:k}),
    \label{eqn:coupling_affine}
\end{align}
where $s_\theta$ and $t_\theta$ are arbitrary neural networks (typically MLPs) that act only on the unchanged half of the input.
The outputs of the coupling layer $z_{k+1:n}$ are elementwise functions of $x_{k+1:n}$, so the Jacobian of the transformation is lower triangular.
Its log-determinant is the sum along the diagonal $\sum_i s_\theta(x_{1:k})_i$.
This operation leaves the other half of the inputs unchanged, so the conditioning and transformed halves are swapped in the next coupling layer allowing the full input to be transformed after two layers.
This construction of two consecutive coupling layers is illustrated in \Cref{fig:coupling_flow}.

The affine transformation in \Cref{eqn:coupling_affine} is simple, computationally efficient, and elegant, but has limited expressivity compared to modern alternatives.
State-of-the-art flow models replace the affine map with a monotone \textit{rational-quadratic spline} (RQS)~\cite{Durkan:2019nsf}, where the neural networks applied to the conditioning half of the inputs parameterize a piecewise rational function that is applied to the other half.
RQS coupling layers are the current standard in most flow-based models being applied in HEP.

Despite their elegance, flow based models are something of a niche topic in the broader deep learning community.
For many applications in computer vision and natural language, the requirement that the latent and data spaces have the same dimensionality is highly restrictive.
Flow models do not provide the dimensionality reduction that makes VAEs and other latent variable models useful in such domains.
This architectural constraint also limits scalability to very high-dimensional data.
However, HEP data are relatively low-dimensional compared to other AI application domains.
Jet and event data typically involve tens to a few hundred dimensions rather than the millions of pixels in natural images, and flow models perform very well on these medium dimensional tasks.
Normalizing flows have been applied across a broad range of HEP tasks, including fast detector simulation~\cite{Krause:2021ilc}, unfolding~\cite{Bellagente:2020piv}, and anomaly detection~\cite{Hallin:2021wme}.
They have only recently been displaced as the default generative model for HEP applications by diffusion and flow-matching models, which I will turn to next.

%% file: chapter3_diffusion.tex
\section{Diffusion and Flow-Matching Models}
\label{sec:diffusion}

Diffusion and flow-matching models are the current state of the art in many generative tasks such as image generation.
Both classes of models decompose the complex single-step transformation between a latent-variable distribution and the data distribution modeled by the decoder network of a VAE into many small, tractable steps.
This is similar to flow based models, except the ``forward'' mapping between the latent-variable and data distributions is not required to be bijective.
Instead the mapping from the data distribution to the latent-variable distribution is fixed, not learned, and the model learns only the reverse mapping.

\subsection{Denoising Diffusion Probabilistic Models}
\label{sec:ddpm}

In diffusion models~\cite{sohl2015deep,Ho:2020ddpm}, the forward process is the progressive addition of Gaussian noise to a data sample.
More specifically this is a Markov chain over $T$ discrete time steps:
\begin{equation}
    q(x_t \mid x_{t-1}) = \mathcal{N}\!\left(x_t;\; \sqrt{1 - \beta_t}\, x_{t-1},\; \beta_t I\right),
    \label{eqn:forward_step}
\end{equation}
where $\{\beta_t\}_{t=1}^T$ is a fixed \textit{noise schedule} with $0 < \beta_t \ll 1$.
The joint forward distribution after $T$ steps is a product over \Cref{eqn:forward_step} for each time step:
\begin{equation}
    q(x_{1:T} \mid x_0) = \prod_{t=1}^{T} q(x_t \mid x_{t-1}).
    \label{eqn:forward_joint}
\end{equation}
Under any noise schedule $\beta_t$ that satisfies $\prod_{t=1}^T (1 - \beta_t) \to 0$, the joint forward distribution becomes a standard normal $q(x_T \mid x_0) \approx \mathcal{N}(0, I)$ for sufficiently large $T$.
In other words, adding increasingly high-variance noise to the data sample will eventually result in a pure-noise sample.
This pure noise distribution can be sampled from easily, analogous to the standard Gaussian latent variable distribution in a normalizing flow model.
The goal of a diffusion model is then to learn the reverse of this process, that evolves a pure-noise sample back to the data distribution.
In what follows the definitions $\alpha_t = 1 - \beta_t$ and $\bar\alpha_t = \prod_{s=1}^t \alpha_s$ will be used in place of $\beta_t$ to simplify the notation.

\subsubsection{The DDPM Objective as a Hierarchical ELBO}

The goal of the denoising diffusion probabilistic model (DDPM) objective is to learn the reverse process
\begin{equation}
    p_\theta(x_{t-1} \mid x_t) = \mathcal{N}\!\left(x_{t-1};\; \mu_\theta(x_t, t),\; \sigma_t^2 I\right),
    \label{eqn:reverse_step}
\end{equation}
which is parameterized by a neural network $\mu_\theta$ and a diagonal variance matrix which evolves with time $\sigma_t^2 I$.
Given a starting sample from $p(x_T) = \mathcal{N}(0, I)$, repeatedly sampling from the learned distribution in \Cref{eqn:reverse_step} for time steps $t = T, T-1, \ldots, 1$ eventually yields a sample from the learned data distribution.
This repeated sampling is a weakness of diffusion and flow-matching models: inference on a neural network must be run $T$ times to generate just one sample from the data distribution.

As was the case for VAEs, diffusion models are trained by maximizing a lower bound (ELBO) on $\log p_\theta(x_0)$~\cite{kingma2021variational}.
The structure is identical to the one derived in \Cref{sec:vae}, but instead of a single KL-term enforcing a prior constraint on the latent variable $z$, there are many KL-terms each enforcing a prior constraint on the many latent variables $x_1, \ldots, x_T$:
\begin{align}
    \log p_\theta(x_0)
    &\geq \mathbb{E}_{q}\!\left[\log p_\theta(x_0 \mid x_1)\right]
      - D_\text{KL}\!\left(q(x_T \mid x_0) \;\|\; p(x_T)\right) \notag \\
    &\quad - \sum_{t=2}^{T} \underbrace{D_\text{KL}\!\left(q(x_{t-1} \mid x_t, x_0) \;\|\; p_\theta(x_{t-1} \mid x_t)\right)}_{L_{t-1}}.
    \label{eqn:ddpm_elbo}
\end{align}
The first term in this Equation is identical to the reconstruction term in \Cref{eqn:elbo}.
The second term enforces a Gaussian prior (pure noise) on the distribution of the highest-noise latent variable $x_T$.
This is a constant term that does not depend on the model parameters $\theta$ given the noise process is not learned in the standard DDPM objective.
The important terms are contained in the sum over $t$.
Each $L_{t-1}$ enforces that the \textit{learned} posterior $p_\theta(x_{t-1} \mid x_t)$ matches the \textit{fixed} forward process posterior $q(x_{t-1} \mid x_t, x_0)$ that can be computed in closed form.
This is possible because the posterior of a product of Gaussian likelihoods is also a Gaussian:
\begin{equation}
    q(x_{t-1} \mid x_t, x_0)
    = \mathcal{N}\!\left(x_{t-1};\; \tilde\mu_t(x_t, x_0),\; \tilde\beta_t I\right),
    \label{eqn:forward_posterior}
\end{equation}
with
\begin{equation}
    \tilde\mu_t(x_t, x_0)
    = \frac{\sqrt{\bar\alpha_{t-1}}\,\beta_t}{1 - \bar\alpha_t}\,x_0
      + \frac{\sqrt{\alpha_t}(1 - \bar\alpha_{t-1})}{1 - \bar\alpha_t}\,x_t,
    \qquad
    \tilde\beta_t = \frac{1 - \bar\alpha_{t-1}}{1 - \bar\alpha_t}\,\beta_t.
    \label{eqn:posterior_mean}
\end{equation}
Under these equations the KL-terms in \Cref{eqn:ddpm_elbo} can be computed in closed form as the MSE loss between the means of the learned posterior and the fixed forward process posterior.

\subsubsection{The Noise Prediction Objective}

\begin{figure}[ht]
    \centering
    \includegraphics[width=\linewidth, alt={Diagram of the forward and reverse processes of a denoising diffusion probabilistic model. The top row shows the forward process progressively adding Gaussian noise to a data sample until it becomes pure noise. The bottom row shows the reverse process, in which a neural network iteratively denoises a pure-noise sample back into a data sample.}]{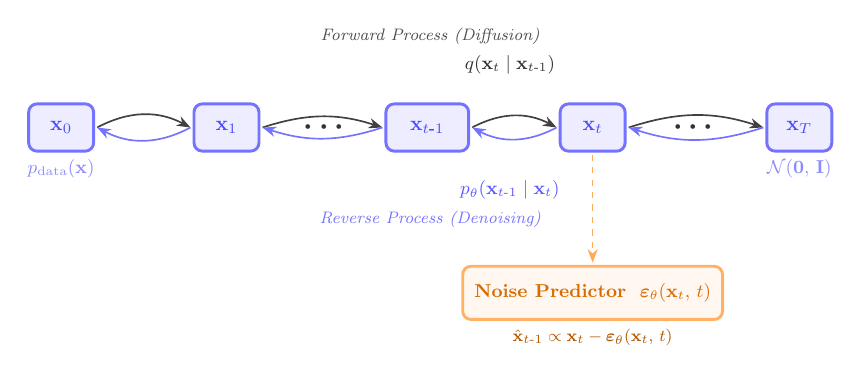}
    \caption{The forward (top) and reverse (bottom) processes of a denoising diffusion probabilistic model. The forward process $q$ progressively adds Gaussian noise to a data sample $x_0$ until the data sample is a draw from a Gaussian distribution. A neural network $p_\theta$ models the reverse process  $p_\theta$, iteratively denoising $x_T$ back to a data sample. Figure adapted from~\cite{Ho:2020ddpm} using Claude Sonnet 4.6.}
    \label{fig:ddpm}
\end{figure}

DDPM models typically do not attempt to predict the mean of $\tilde\mu_t(x_t, x_0)$ directly, but instead predict the noise that was added at time step $t$.
This is just a reparametrization, somewhat similar to the reparametrization trick used in VAEs but used here only because it empirically gives better sample quality.
Given the forward process is fixed in DDPM there is no need to backpropagate through the random sampling to train the posterior, as in a VAE.
The reparametrization is
\begin{equation}
    x_t = \sqrt{\bar\alpha_t}\,x_0 + \sqrt{1 - \bar\alpha_t}\,\varepsilon,
    \label{eqn:xt_reparam}
\end{equation}
where $\varepsilon \sim \mathcal{N}(0, I)$.
Substituting \Cref{eqn:xt_reparam} into the posterior mean \Cref{eqn:posterior_mean} and parameterizing the reverse process mean as
\begin{equation}
    \mu_\theta(x_t, t)
    = \frac{1}{\sqrt{\alpha_t}}\!\left(x_t - \frac{\beta_t}{\sqrt{1 - \bar\alpha_t}}\,\varepsilon_\theta(x_t, t)\right),
    \label{eqn:reverse_mean_reparam}
\end{equation}
where $\varepsilon_\theta$ is a neural network that predicts the noise $\varepsilon$, each KL term $L_{t-1}$ simplifies to
\begin{equation}
    L_{t-1} \propto \mathbb{E}_{x_0,\,\varepsilon}\!\left[\left\|\varepsilon - \varepsilon_\theta\!\left(\sqrt{\bar\alpha_t}\,x_0 + \sqrt{1-\bar\alpha_t}\,\varepsilon,\; t\right)\right\|^2\right].
    \label{eqn:ddpm_loss}
\end{equation}
The final training objective is then an MSE loss comparing the network's predicted noise to the true noise injected at time step $t$.
Typically uniform random sampling over the time step $t$ is used in training, rather than ensuring all terms are summed to calculate the loss before taking an optimizer step.
The training algorithm is:
\begin{enumerate}
    \item Sample $x_0 \sim p_\text{data}$, $t \sim \text{Uniform}(1, T)$, $\varepsilon \sim \mathcal{N}(0, I)$.
    \item Construct $x_t = \sqrt{\bar\alpha_t}\,x_0 + \sqrt{1 - \bar\alpha_t}\,\varepsilon$ via \Cref{eqn:xt_reparam}.
    \item Take a gradient step to minimize $\|\varepsilon - \varepsilon_\theta(x_t, t)\|^2$.
\end{enumerate}
A diagram giving an overview of this process is provided in \Cref{fig:ddpm}.

In practice DDPM models produce much higher sample quality than VAEs, despite the fact they are optimizing very similar objectives.
One explanation for this is that DDPM models predict only the mean of the forward-process posterior $q(x_{t-1} \mid x_t, x_0)$ at each time step, which is unimodal and Gaussian by construction.
This side-steps the inability of VAEs to model multimodal likelihoods as discussed in \Cref{sec:vae}.
The drawback is that model inference must be run many times to generate a new sample.

\subsubsection{DDPM Sampling}

To generate a sample from the data distribution with a trained DDPM model, a sample of noise is drawn from $x_T \sim \mathcal{N}(0, I)$ and is repeatedly passed through the neural network for time steps $t = T, \ldots, 1$.
Formally the forward process \Cref{eqn:forward_step} can be interpreted as the discretization of a \textit{stochastic differential equation} (SDE)~\cite{Song:2021score}.
The reverse of this process is also an SDE, which can be solved by numerical methods using the noise predictions from the model as input.
An example is the PNDM solver~\cite{liu2022pseudo} used in Ref.~\cite{Shmakov:2023kjj}, but many other solvers exist.
Ref.~\cite{Song:2021score} also shows that every forward SDE has a corresponding \textit{probability flow ODE} with identical marginal distributions $q(x_t)$ at every time $t$:
\begin{equation}
    dx = \left[f(x, t) - \tfrac{1}{2}g(t)^2\,\nabla_x \log q_t(x)\right] dt,
    \label{eqn:prob_flow_ode}
\end{equation}
where $f$ and $g$ are the drift and diffusion coefficients of the forward SDE and $\nabla_x \log q_t(x)$ is the \textit{score function} of the marginal distribution at time $t$.
Since $\varepsilon_\theta(x_t, t) \approx -\sqrt{1 - \bar\alpha_t}\,\nabla_{x_t} \log q_t(x_t)$, the same neural network can be used to solve either the stochastic SDE or deterministic ODE.
The ODE solution can be understood as the continuous-time generalization of normalizing flows~\cite{chen2018neural}.
This connection is made more apparent in the flow-matching models in the next section.

\subsection{Flow Matching}
\label{sec:flow_matching}

Flow matching~\cite{lipman2022flow} is a closely related class of models.
Instead of predicting the noise added to a data sample, flow matching models predict a \textit{vector field} that transports the noise distribution to the data distribution.
Given a time-dependent density $p_t(x)$ evolving under the vector field $v_t(x)$, the continuity equation supplies a constraint to keep the density normalized as with any solution to a PDE:
\begin{equation}
    \frac{\partial p_t}{\partial t} + \nabla \cdot (p_t\,v_t) = 0.
    \label{eqn:continuity}
\end{equation}
At $t = 0$, $p_0 = \mathcal{N}(0, I)$ and at $t=1$, $p_1 = p_\text{data}$.
The goal is then to train a neural network $u_\theta(x, t)$ that approximates the vector field $v_t$ which connects these two endpoints as the source distribution evolves under the continuity equation.

\begin{figure}[ht]
    \centering
    \includegraphics[width=0.8\linewidth, alt={Two side-by-side diagrams comparing sampling trajectories between a noise distribution and a data distribution. The diffusion model trajectory follows a curved path. The linear flow-matching trajectory follows a straight path.}]{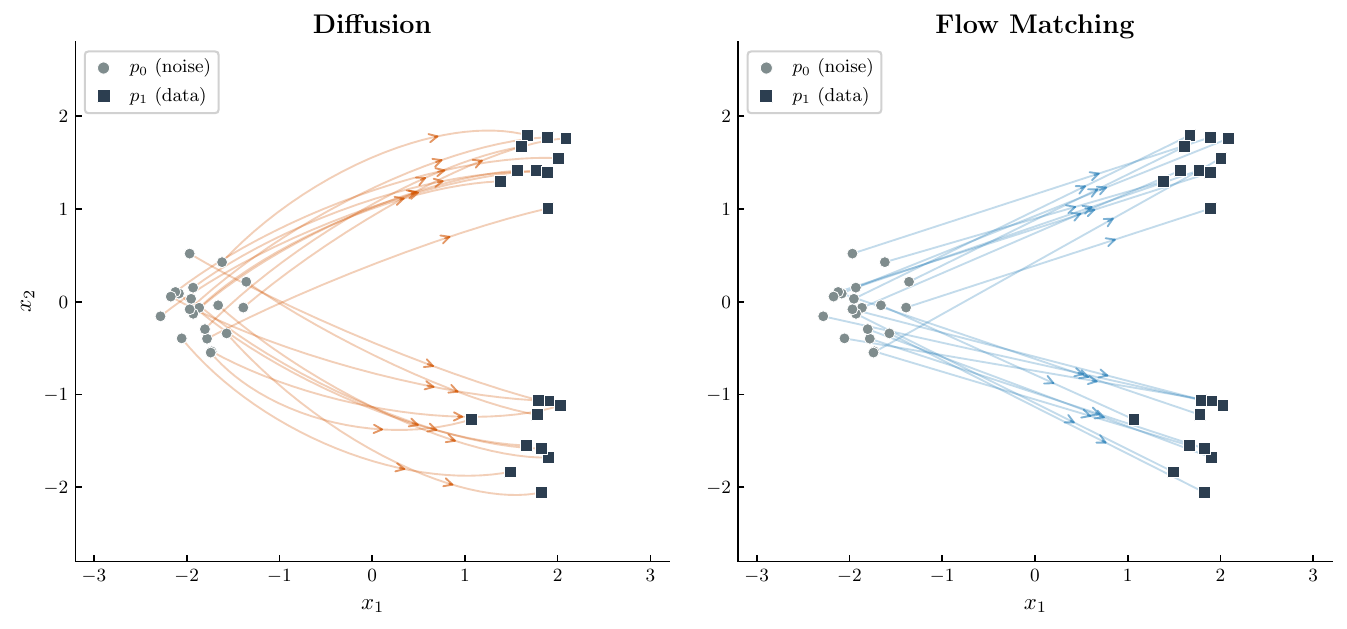}
    \caption{Comparison of probability flow trajectories for diffusion models (curved paths) and linear flow matching (straight paths). Straighter paths require fewer neural function evaluations to integrate accurately, making flow matching more computationally efficient at inference time. Figure generated by Claude Sonnet 4.6.}
    \label{fig:flow_matching}
\end{figure}

The marginal vector field $v_t(x)$ is intractable, but the \textit{conditional} vector field $u_t(x \mid x_1)$ for a single data point $x_1$ has the same expected gradient as the marginal vector field and yields an equivalent training objective~\cite{lipman2022flow}:
\begin{equation}
    \mathcal{L}_\text{CFM}(\theta)
    = \mathbb{E}_{t,\,x_1 \sim p_\text{data},\,x_t \sim p_t(\cdot \mid x_1)}
      \left\| u_\theta(x_t, t) - u_t(x_t \mid x_1) \right\|^2.
    \label{eqn:cfm_loss}
\end{equation}
Different choices of conditional probability paths imply different conditional vector fields and different training objectives.
Under Gaussian conditional paths $p_t(x \mid x_1) = \mathcal{N}(x;\, \mu_t(x_1),\, \sigma_t^2 I)$ the CFM objective becomes equivalent to the DDPM objective~\cite{lipman2022flow}.
Alternatively one could just interpolate between the noise and target data points,
\begin{equation}
    x_t = (1 - t)\,x_0 + t\,x_1, \qquad x_0 \sim \mathcal{N}(0, I),
    \label{eqn:linear_path}
\end{equation}
for which the conditional vector field takes the simple form $u_t(x \mid x_1) = x_1 - x_0$.
The straight paths in data space are much simpler to integrate than the curved trajectories that result from Gaussian conditional paths.
This means flow-matching models trained with these paths typically require fewer neural function evaluations to generate a sample accurately.
\Cref{fig:flow_matching} shows the difference between diffusion trajectories and linear flow-matching paths.

\subsection{Applications in HEP and Comparative Performance}
\label{sec:diffusion_hep}

\begin{figure}[t]
    \centering
    \includegraphics[width=1.0\linewidth, alt={Scatter plot of log-posterior score versus generation time per shower for CaloChallenge Dataset 2 submissions. Diffusion and flow-matching models achieve the highest log-posterior scores but have the longest generation times. Normalizing-flow-based models achieve a favorable balance of moderate score and low generation time.}]{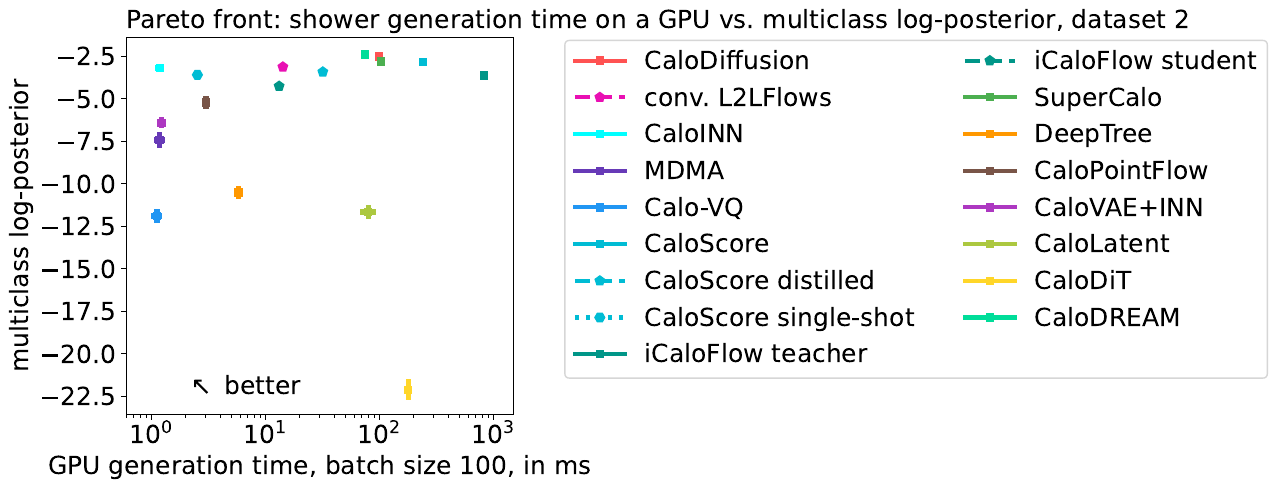}
    \caption{CaloChallenge Dataset 2: log-posterior score (higher is better) versus generation time per shower (lower is better) for all CaloChallenge submissions. The diffusion (CaloDiffusion) and flow-matching models (CaloDREAM) show the best sample quality, but have high sampling time. The normalizing-flow-based models (CaloINN) offer a good compromise. Reproduced from~\cite{Krause:2024avx}, \textit{Rep. Prog. Phys.}, \textbf{88}, 116201, 2025. \copyright~IOP Publishing Ltd. All rights reserved.}
    \label{fig:calochallenge_ds2}
\end{figure}

\begin{figure}[t]
    \centering
    \includegraphics[width=1.0\linewidth, alt={Scatter plot of log-posterior score versus generation time per shower for CaloChallenge Dataset 3 submissions. Diffusion and flow-matching models achieve the highest log-posterior scores but have the longest generation times. Normalizing-flow-based models achieve a favorable balance of moderate score and low generation time.}]{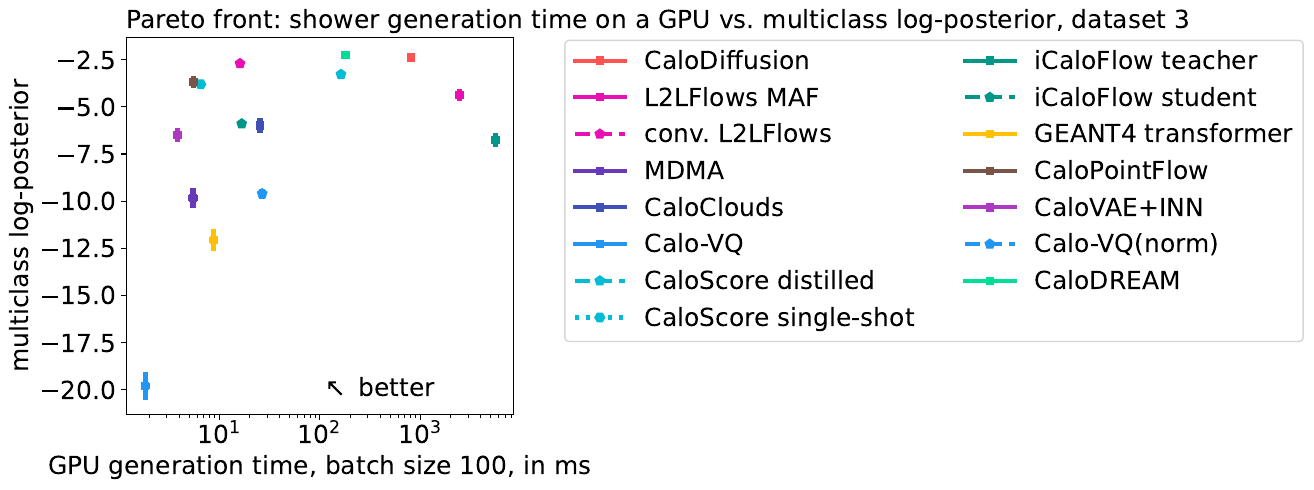}
    \caption{CaloChallenge Dataset 3: log-posterior score (higher is better) versus generation time per shower (lower is better) for all CaloChallenge submissions. The diffusion (CaloDiffusion) and flow-matching models (CaloDREAM) show the best sample quality, but have high sampling time. The normalizing-flow-based models (CaloPointFlow) offer a good compromise. Reproduced from~\cite{Krause:2024avx}, \textit{Rep. Prog. Phys.}, \textbf{88}, 116201, 2025. \copyright~IOP Publishing Ltd. All rights reserved.}
    \label{fig:calochallenge_ds3}
\end{figure}

Diffusion models were first applied in HEP after they demonstrated state-of-the-art performance on image generation tasks in 2021~\cite{Ho:2020ddpm,Song:2021score}.
The first applications targeted fast calorimeter shower simulation~\cite{Mikuni:2022xry,Amram:2023onf}, a task which aims to train a model to generate calorimeter showers based on a training set generated from the realistic but expensive \textsc{Geant4} simulation used by most particle physics experimental collaborations~\cite{GEANT4:2002zbu}.
Diffusion and flow-matching models have since been applied to anomaly detection and unfolding, the latter being a central topic of \Cref{ch:unfolding}.
Given calorimeter simulation is the most obviously useful application of generative models in HEP and provides a very useful benchmark, it is worth showing some results from a community challenge in this domain here.
The Calorimeter Simulation Challenge (CaloChallenge)~\cite{Krause:2024avx} required participants to train models that could generate calorimeter showers for three increasingly complicated and granular calorimeter geometries.
Some results from datasets 2 and 3 are shown in \Cref{fig:calochallenge_ds2,fig:calochallenge_ds3}.
The plots show the multi-class log posterior (a measure of sample quality, with higher being more similar to the ground truth \textsc{Geant4} simulation) versus generation time per shower (lower is better).
The takeaway from these plots is that diffusion (such as CaloDiffusion~\cite{Amram:2023onf}) and flow-matching (such as CaloDREAM~\cite{Favaro:2024rle}) models achieve the best sample quality, but at the cost of relatively slow sampling.
Of course they are still far faster than the \textsc{Geant4} simulation.
Interestingly normalizing-flow-based methods like CaloPointFlow~\cite{Schnake:2024mip} are competitive and substantially faster to sample from.
There is some possibility that diffusion and flow-matching models are actually generating samples with higher quality than is needed for downstream tasks, in which case using simpler models would be more efficient.
However in other generative tasks in HEP, such as anomaly detection or the unfolding task discussed in \Cref{ch:unfolding}, sample quality is essential and sampling speed is essentially irrelevant.
In these tasks, diffusion and flow-matching models are the current state of the art.

%% file: chapter3_dre.tex
\section{Density Ratio Estimation}
\label{sec:dre}

Generative models are able to sample from $p_\text{data}(x)$, typically by developing some explicit model for the data distribution given some latent variables.
This section introduces an elegant trick that allows for very simple estimation of the \textit{ratio} of two densities.
This trick allows a high-dimensional reweighting to be derived between two densities, which is very useful in HEP.
A central example is the Omnifold unfolding method discussed in \Cref{ch:unfolding}, which is used for the cross-section measurement presented in \Cref{ch:zjets}.

\subsection{The Likelihood Ratio Trick}
\label{sec:lrt}

The likelihood ratio between two densities $p(x)$ and $q(x)$ is simply
\begin{equation}
    r(x) = \frac{p(x)}{q(x)}.
    \label{eqn:likelihood_ratio}
\end{equation}
The likelihood ratio trick is a method for computing $r(x)$ that only requires training a neural network classifier with binary cross-entropy (BCE) loss\footnote{Other loss functions such as the MSE loss are also minimized by the likelihood ratio and so can be used in this way as well.}.
A proof that the likelihood ratio can be obtained from the function that minimizes the BCE loss is provided in what follows.

\subsubsection{Proof: the Optimal BCE Classifier Recovers the Likelihood Ratio}

Consider a classifier $f_\theta(x) \in (0,1)$ trained to minimize the BCE loss between samples from $p(x)$ (label $y=1$) and $q(x)$ (label $y=0$):
\begin{equation}
    \mathcal{L}(f) = -\mathbb{E}_{x \sim p}\!\left[\log f(x)\right]
                    -\mathbb{E}_{x \sim q}\!\left[\log(1 - f(x))\right].
    \label{eqn:bce_dre}
\end{equation}
An optimization algorithm like gradient descent hopes to find the function $f^*$ that minimizes the loss.
To find this function, first write the expectations as integrals:
\begin{equation}
    \mathcal{L}(f) = -\int \!\left[p(x)\log f(x) + q(x)\log(1-f(x))\right] dx.
    \label{eqn:bce_integral}
\end{equation}
To find the function that minimizes this loss function, we take the functional derivative of $\mathcal{L}$ with respect to $f$.
Perturbing $f \to f + \epsilon\,\eta$ for an arbitrary test function $\eta$, the two logarithms in the integrand expand to first order in $\epsilon$ as
\begin{align}
    \log\!\left(f(x) + \epsilon\,\eta(x)\right)
        &= \log f(x) + \frac{\epsilon\,\eta(x)}{f(x)} + \mathcal{O}(\epsilon^2), \label{eqn:log_expand1}\\
    \log\!\left(1 - f(x) - \epsilon\,\eta(x)\right)
        &= \log(1-f(x)) - \frac{\epsilon\,\eta(x)}{1-f(x)} + \mathcal{O}(\epsilon^2). \label{eqn:log_expand2}
\end{align}
Substituting into $\mathcal{L}(f+\epsilon\,\eta)$ and collecting terms by order in $\epsilon$,
\begin{align}
    \mathcal{L}(f + \epsilon\,\eta)
    &= -\int\!\Bigl[p(x)\Bigl(\log f(x) + \frac{\epsilon\,\eta(x)}{f(x)}\Bigr)
       \notag\\
    &\qquad + q(x)\Bigl(\log(1-f(x)) - \frac{\epsilon\,\eta(x)}{1-f(x)}\Bigr)\Bigr]\,dx
       + \mathcal{O}(\epsilon^2) \notag\\
    &= \mathcal{L}(f)
       - \epsilon\int\!\left[\frac{p(x)}{f(x)} - \frac{q(x)}{1-f(x)}\right]\eta(x)\,dx
       + \mathcal{O}(\epsilon^2).
    \label{eqn:functional_perturb}
\end{align}
The functional derivative is read off as the coefficient of $\eta(x)$ in the first-order term:
\begin{equation}
    \frac{\delta \mathcal{L}}{\delta f(x)} = -\frac{p(x)}{f(x)} + \frac{q(x)}{1-f(x)}.
    \label{eqn:functional_deriv}
\end{equation}
Setting $\delta\mathcal{L}/\delta f(x) = 0$ for all $x$ gives the stationary point condition for the functional $\mathcal{L}$:
\begin{equation}
    \frac{p(x)}{f^*(x)} = \frac{q(x)}{1-f^*(x)}
    \;\implies\;
    f^*(x) = \frac{p(x)}{p(x) + q(x)}.
    \label{eqn:optimal_classifier}
\end{equation}
This is a global minimum since the second functional derivative evaluated at $f^*$ is positive\footnote{In fact the second derivative is positive everywhere, making optimization of the BCE loss a ``convex'' problem. Working out the second functional derivative and showing this is left as an exercise for the reader. Note however that this does not mean that the loss with respect to the parameters of the classifier is convex!}.
The likelihood ratio is then recovered from the optimal classifier output by a simple algebraic rearrangement:
\begin{equation}
    r(x) = \frac{p(x)}{q(x)} = \frac{f^*(x)}{1 - f^*(x)}.
    \label{eqn:ratio_from_classifier}
\end{equation}
Any classifier trained to minimize BCE loss is implicitly learning to estimate $r(x)$, up to this monotonic rescaling.
Using a network trained with a BCE loss to estimate the likelihood ratio is sometimes called the \textit{likelihood ratio trick}~\cite{Cranmer:2015bka,Andreassen:2019nnm}.

\subsection{Reweighting and Applications in HEP}
\label{sec:dre_hep}

Practically, \Cref{eqn:ratio_from_classifier} implies that the output of a trained classifier can be used to reweight samples from $q(x)$ to instead be samples from $p(x)$, or vice versa since the inverse of the likelihood ratio is also a monotonic rescaling of $f^*(x)$.
Each sample $x_i \sim q$ is assigned a weight $w_i = r(x_i) = f^*(x_i) / (1 - f^*(x_i))$.
If the samples are, for example, binned with the contribution of each sample weighted appropriately, the resulting histogram will be an estimate of $p(x)$.

Reweighting samples from one probability distribution to samples from another is not a new idea in HEP.
A standard method is to derive correction factors as ratios of one-dimensional histograms, which is a simple form of density ratio estimation in a single variable.
Of course this does not generalize to more than a few dimensions given the curse of dimensionality.
The likelihood ratio trick bypasses the need to bin and so scales to very high dimensions.
The distributions can even be \textit{variable-dimensional} and the method is still applicable, since neural network architectures like graphs and transformers naturally handle variable-length sets.
In the context of this thesis, the most relevant application of the likelihood ratio trick is the Omnifold algorithm~\cite{Andreassen:2019cjw}, which is the method used for the cross-section measurement in \Cref{ch:zjets}.

%% file: chapter3_uq.tex
\section{Uncertainty Quantification}
\label{sec:uq}

\begin{figure}[t]
    \centering
    \includegraphics[width=\linewidth, alt={Three scatter plots with fitted curves illustrating aleatoric uncertainty, epistemic uncertainty, and domain shift on a toy regression problem. The left plot shows noisy data scattered around a true function, representing irreducible aleatoric noise. The center plot shows training data covering only part of the input range, with growing model uncertainty in the region with no training data, representing epistemic uncertainty. The right plot shows training data (blue) and evaluation data (red) drawn from different regions of the input space, representing domain shift.}]{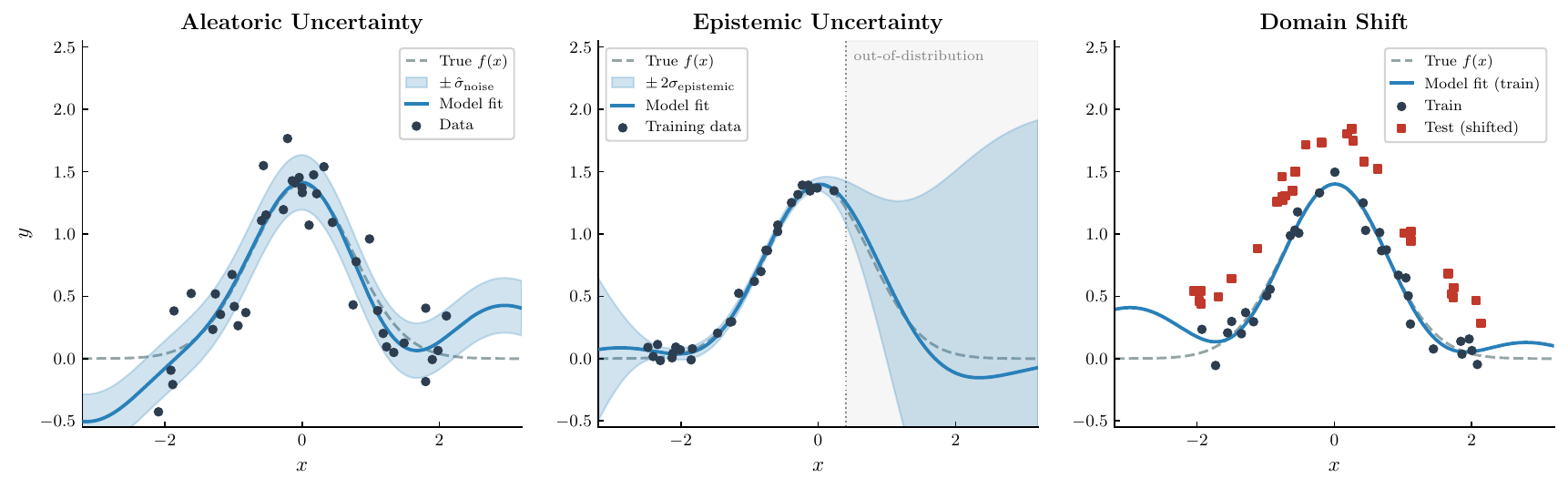}
    \caption{Illustration of the three categories of uncertainty on a toy regression problem: aleatoric uncertainty (irreducible scatter in the data), epistemic uncertainty (model uncertainty due to limited training data, for example in regions of phase space where no data is available), and domain shift (mismatch between training and evaluation distributions). Figure generated by Claude Sonnet 4.6.}
    \label{fig:uncertainty_types}
\end{figure}

Uncertainty quantification for HEP applications of deep learning is very problem-specific, so it is difficult to give a general overview of the methods.
This section covers the main types of uncertainty considered in the deep learning literature and how they map on to HEP problems.
The primary goal is to discuss the quantification methods, which are widely used in \Cref{ch:tagging,ch:zjets}.
\Cref{fig:uncertainty_types} provides visual intuition for each category of uncertainties on a toy problem.

\subsection{Aleatoric Uncertainty}
\label{sec:aleatoric}

Aleatoric uncertainty is the uncertainty inherent in the data-generating process.
This uncertainty is a property of the problem, not the data or methods.
The left hand panel of \cref{fig:uncertainty_types} shows a toy regression problem with the aleatoric uncertainty corresponding to the noise in the data points that make them deviate from the true function shown in gray.
This noise could have many sources in real world problems.
As a concrete example, in the jet classification task discussed in \Cref{ch:tagging} both top quark initiated and light quark or gluon initiated jets have a non-zero probability of producing a given jet configuration.
Some configurations are much more likely to be due to a top quark versus a light quark or gluon, but there is no way to decisively answer within the problem as posed.
The correct model output, which for binary classification is related to the likelihood as shown in \cref{sec:dre}, reflects this and assigns some non-zero probability to both classes.
This spread is an example of aleatoric uncertainty.
Reducing this uncertainty requires reframing the problem, for example by adding a new detector subsystem that provides additional discriminating information.
Given aleatoric uncertainty is a feature of the question rather than the method, quantification of it is usually not important in HEP, so these uncertainties will not be discussed much in this thesis.

\subsection{Epistemic Uncertainty}
\label{sec:epistemic}

Epistemic uncertainty arises from limitations of the machine learning model itself.
This can include failures due to finite training data, undertrained models, or optimization failures.
An example of epistemic uncertainty is the inability of a model to generalize outside of the support of its training data, as shown in the center panel of \cref{fig:uncertainty_types}.
Epistemic uncertainty is in principle reducible by collecting more data, using a more expressive architecture, or improving training.
In HEP, the relevance of epistemic uncertainty depends heavily on the task.
In jet tagging, epistemic uncertainties are present and result in sub-optimal classification performance, but the model is still useful.
For example using very simple cut-based jet tagging methods would have very large epistemic uncertainty since they are poor estimators of the likelihood ratio, but this does not prevent them from being used to define signal regions and perform measurements.
However in applications like unfolding, neural simulation-based inference (NSBI), and anomaly detection, the model outputs are used directly in the statistical analysis and epistemic uncertainty can have a significant impact on the result.
For example in NSBI, the model outputs are used directly as a test statistic in statistical tests to determine confidence intervals on the parameter of interest.
Failure of the model to fully capture the true test statistic (usually the likelihood ratio) will produce an unaccounted-for systematic bias in the measurement.
In this case assessing and eliminating epistemic uncertainty is crucial.
A few methods are used to estimate epistemic uncertainty in practice, but given there is limited theoretical understanding of gradient descent applied to neural networks there are no fully robust methods.
In HEP applications where epistemic uncertainty is important, auxiliary cross checks are typically needed to ensure the epistemic uncertainty is small or captured by appropriate systematics, as will be the case in \Cref{ch:zjets}.

\paragraph{Ensembling.}
The simplest method is to train many independent but identical models and treat the variance over their outputs as an estimate of the epistemic uncertainty.
The models are independent in that they have different weight initializations and different mini-batch sampling in the optimization process.
This approach has been previously applied in ATLAS measurements, for example in Refs.~\cite{ATLAS:2024xxl, ATLAS:2025clx}.
Ensembling is an easy and generally applicable method, but it does not capture the full epistemic uncertainty.
For example if the models lack sufficient capacity to capture the true optimum function, this will not be reflected by the ensemble.

\paragraph{Bayesian neural networks.}
Another approach is to assume a prior distribution over the model weights and frame the optimization problem as computing the posterior given the training data.
The posterior encodes a specific type of epistemic uncertainty due to limited training data.
Bayesian neural networks have seen limited use in HEP~\cite{Bollweg:2019skg} relative to ensembling, partly because the epistemic uncertainty due to limited training data is typically a small consideration in HEP tasks where simulated training data is plentiful.

\subsection{Domain Shift}
\label{sec:domain_shift}

Domain shift occurs when the probability distribution governing the data used in evaluation differs from the distribution governing the data used in training.
This type of uncertainty is visualized in the right hand panel of \cref{fig:uncertainty_types} where the testing data (red) is not drawn from the same distribution as the training data (blue).
This is the most common type of uncertainty in HEP.
In particular experimental systematic uncertainties typically fall in this category.
A particularly relevant example for \Cref{ch:tagging} is that jet tagging models are typically trained on simulation but applied on data.
Typically models show worse performance on data and this difference needs to be quantified by a \textit{scale factor} for the tagger to be used in measurements and searches.
Many other domain shift uncertainties will be handled in \Cref{ch:zjets}.
Several strategies for estimating or reducing domain shift exist.

\paragraph{Propagating uncertainties.}
The simplest strategy is to propagate possible domain shifts on the input $x$ through the trained model to obtain an uncertainty on the model output $f(x)$.
This accounts for the effect of the domain shift on the model output, but does not account for how a domain shift would adjust the training process, sometimes resulting in a sub-optimal model.
It also assumes that correct systematic uncertainties on the model inputs are available which is not always the case.
However in many applications this is easiest and most correct approach, and is used in both \Cref{ch:tagging,ch:zjets}.

\paragraph{Domain adaptation.}
An alternative is to train the model to be robust to known sources of domain shift.
This can be done by including data with known domain shifts applied in the training set.
An additional step that can be taken to reduce the domain shift uncertainty is to explicitly decorrelate the model output against these shifts.
This is often done in HEP so that systematic variations do not affect the model output~\cite{Kasieczka:2020yyl,ATLAS:2025dkv,Ghosh:2021hrh}.
Decorrelation results in smaller domain shift uncertainties but typically at the cost of performance.
This may still be a desirable tradeoff, for example jet taggers are typically decorrelated against the jet mass observable to avoid artificial sculpting of mass distributions.

\paragraph{Parametrized models.}
Typically the optimal approach is to incorporate the domain shift directly into the model by conditioning on the nuisance parameter or systematic variation as an additional input~\cite{Louppe:2016ylz,Baldi:2016fzo,Ghosh:2021roe}.
A model trained this way can learn the correct function across the full range of systematic variations, and the uncertainty on the output can be assessed by varying the conditioning parameter passed into the model.
This technique is powerful, but at a minimum requires detailed knowledge of the types of domain shifts that will be observed to be known beforehand, and may not be applicable even if this is the case.
For example there is no well-established method for parametrizing the iterative unfolding methods discussed in \Cref{ch:unfolding}, even if all systematic uncertainties are known and understood beforehand.

%% file: chapter4.tex
\chapter{Jet Classification at the LHC}
\label{ch:tagging}

Jet classification, commonly referred to as ``jet tagging'', was one of the first applications of deep learning in particle physics and remains arguably the most important.
The techniques covered in this chapter are used by both ATLAS and CMS and are driving improvements in the search for di-Higgs production, which is the main goal of the current LHC physics program.
This chapter will introduce jet tagging as a problem, then cover an ATLAS study on top tagging that is the first original research contribution in this thesis~\cite{ATLAS:2024rua}.
It closes with a discussion of how jet tagging has progressed since the publication of the ATLAS top tagging paper and prospects for the high-luminosity LHC.

\section{Jet Tagging}
\label{sec:tagging-intro}

Most searches and measurements at the LHC are concerned with the \textit{hard scatter}, the high energy scattering of quarks and gluons that results in final state quarks, gluons, and leptons which are possibly the decay products of other more massive SM or BSM particles.
Ideally one would directly measure the identity and kinematics of the individual quarks, gluons, and leptons produced in the hard scatter.
In practice, the parton shower and hadronization processes make this impossible for quarks and gluons.
Detectors can only observe their outcome, which are collimated sprays of hadrons known as jets.
See \Cref{ch:qcd} for a full discussion.
These jets are typically used as proxies for the initiating parton.
For example the kinematics of the initiating parton can be inferred from the kinematics of the jet, obtained by summing the four-momenta of the jet's constituent hadrons.
Jet tagging is the problem of inferring the type of parton that initiated the jet from measurements of the jet constituents.

\begin{figure}[tb]
    \centering
    \includegraphics[width=0.85\linewidth, alt={Schematic diagram of five jet types arranged clockwise from top left: light quark and gluon jets with no substructure, bottom and charm quark jets with a displaced secondary vertex, W and Z boson jets with two-pronged substructure, Higgs boson jets decaying to two b quarks with two-pronged substructure and displaced vertices, and top quark jets with three-pronged substructure and one displaced vertex.}]{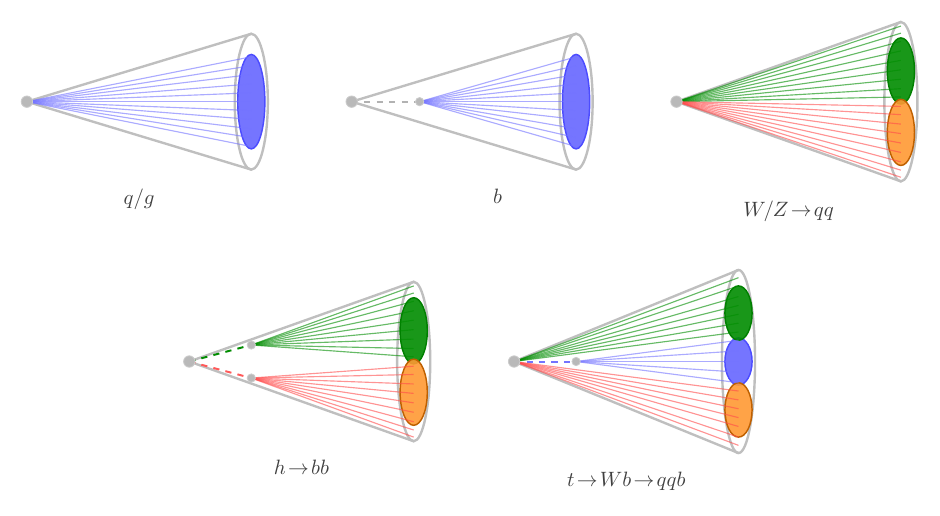}
    \caption{Schematic illustration of the two broad categories of jets produced in LHC collisions. From the top left working clockwise: light quark and gluon initiated jets are the most abundant jets with no substructure, bottom quark (and to some extent charm quark) initiated jets contain a displaced vertex due to the long-lived B and D hadrons, jets initiated by the hadronic decay of the W and Z bosons contain a two-pronged substructure, jets initiated by the decay of a Higgs boson to two b quarks results in a two-pronged substructure with displaced vertices, and the hadronic decay of a top quark results in a three-pronged substructure with one displaced vertex. Figure adapted from Ref.~\cite{Moreno:2019bmu} by Claude Sonnet 4.6.}
    \label{fig:jet-types}
\end{figure}

Historically, jet tagging at the LHC has been divided into two distinct problems that utilize different sets of information, though these problems have now largely been unified by the deep learning methods that are the topic of this chapter.
The first is \textit{boosted jet tagging}, which seeks to identify the hadronic decay modes of massive SM particles such as top quarks, $W$ and $Z$ bosons, or Higgs bosons.
When one of these massive particles is produced with a transverse momentum much larger than its mass ($p_T \gg m$), the jets initiated by each parton produced by the decay become collimated and merge into a single jet with distinct substructure.
$W/Z/H \to q\bar{q}$ decays produce a two-pronged substructure, and a $t \to Wbq\bar{q}$ decay produces a three-pronged substructure.
Both types of substructure are different from the single-pronged jets produced by light quarks and gluons.
Boosted jet tagging methods exploit these substructure differences to isolate the jets produced by massive SM particles against the very large background of light quark and gluon jets. 
Early methods for this classification task relied on multivariate discriminants such as BDTs or shallow neural networks that operated on a fixed-length set of hand-engineered substructure observables~\cite{Altheimer:2013yza, Thaler:2008ju}.
An example use case for boosted jet tagging is to search for BSM resonances that decay to boosted SM objects like top quarks or W/Z bosons~\cite{ATLAS:2022enb,CMS:2026ogg}.

The second problem is \textit{flavor tagging}.
Charm and bottom quarks are second and third generation particles which can be produced through QCD processes or by the decay of a massive SM particle such as the Higgs boson.
When charm and bottom quarks are produced they initiate a jet like any quark, but these jets typically contain at least one $D$ or $B$ hadron that carries the heavy-flavor quark.
These hadrons are long-lived relative to the much more common light hadrons such as pions, so they frequently travel a measurable distance before decaying, leaving a displaced \textit{secondary vertex} inside the jet.
Identifying jets that contain $B$ or $D$ hadrons, and so were likely initiated by $b$ or $c$ quarks, is called $b$- or $c$-tagging.
Standard flavor tagging methods center on reconstructing displaced vertices explicitly and using information about these vertices as input to a BDT or neural network~\cite{ATLAS:2016gsw}.
These approaches depended on hand-engineered features, just like with early boosted jet tagging methods.

The types of jets are illustrated in \Cref{fig:jet-types}.
Boosted jet tagging aims to identify top, W/Z, and Higgs jets against the background of light-quark and gluon jets, and flavor tagging aims to identify $b$ and $c$ jets\footnote{Note that $c$ jets are not shown in the figure but are a distinct category of jets from $b$ jets though their signatures are very similar.}.
Both are classification tasks, where one must distinguish a signal jet from a background jet using the kinematics of the jet constituents that are measured by the detector.
The distinction is what subset of the kinematics of the jet constituents are targeted by the hand-engineered features designed to help solve these tasks.
The features used in boosted tagging depend on the global spatial structure of constituent particles within a jet, while features used for flavor tagging depend on the longitudinal and transverse impact parameter information of the tracks within a jet that are used to reconstruct secondary vertices.
These sets of information are not independent, but do factorize somewhat into signals coming from the inner tracking and calorimeter sub-systems of the ATLAS and CMS detectors.

Both problems have a long history in collider experiments.
In runs one and two of the LHC, both were solved by combining hand-engineered expert features with a simple multivariate discriminant provided by a BDT or small neural network.
One story in the top tagging paper presented here is that larger neural networks that operate directly on the jet constituent kinematics rather than on hand-engineered features can achieve much better performance.
While this is illustrated in a boosted jet tagging task, the same is true for flavor tagging.

The paper featured in this chapter focuses on boosted top tagging (or just \textit{top tagging}), a particular boosted jet tagging task that classifies top quark initiated jets against the light quark and gluon initiated background.
These background jets are sometimes referred to as \textit{QCD jets}.
Top tagging is the prototypical boosted jet tagging problem since the three-prong substructure of boosted top jets is very distinct.
This is why it was the subject of many early deep learning applications in particle physics.
The study presented here was carried out in 2022--2023 and was the first original research contribution of my Ph.D.
Though the performance presented is somewhat below the current state-of-the-art, the lessons drawn remain relevant to large multi-class taggers of today, to which I will return in \Cref{sec:tagging-future}.

\section{Constituent Based Top Quark Tagging}
\label{sec:constituent-top-tagging}

\begin{figure}[ht]
    \centering
    \includegraphics[width=0.8\linewidth, alt={ROC curve plot comparing background rejection versus signal efficiency for a range of top tagging classifiers evaluated on a common benchmark dataset, showing constituent-based methods achieving higher background rejection than methods using hand-engineered features.}]{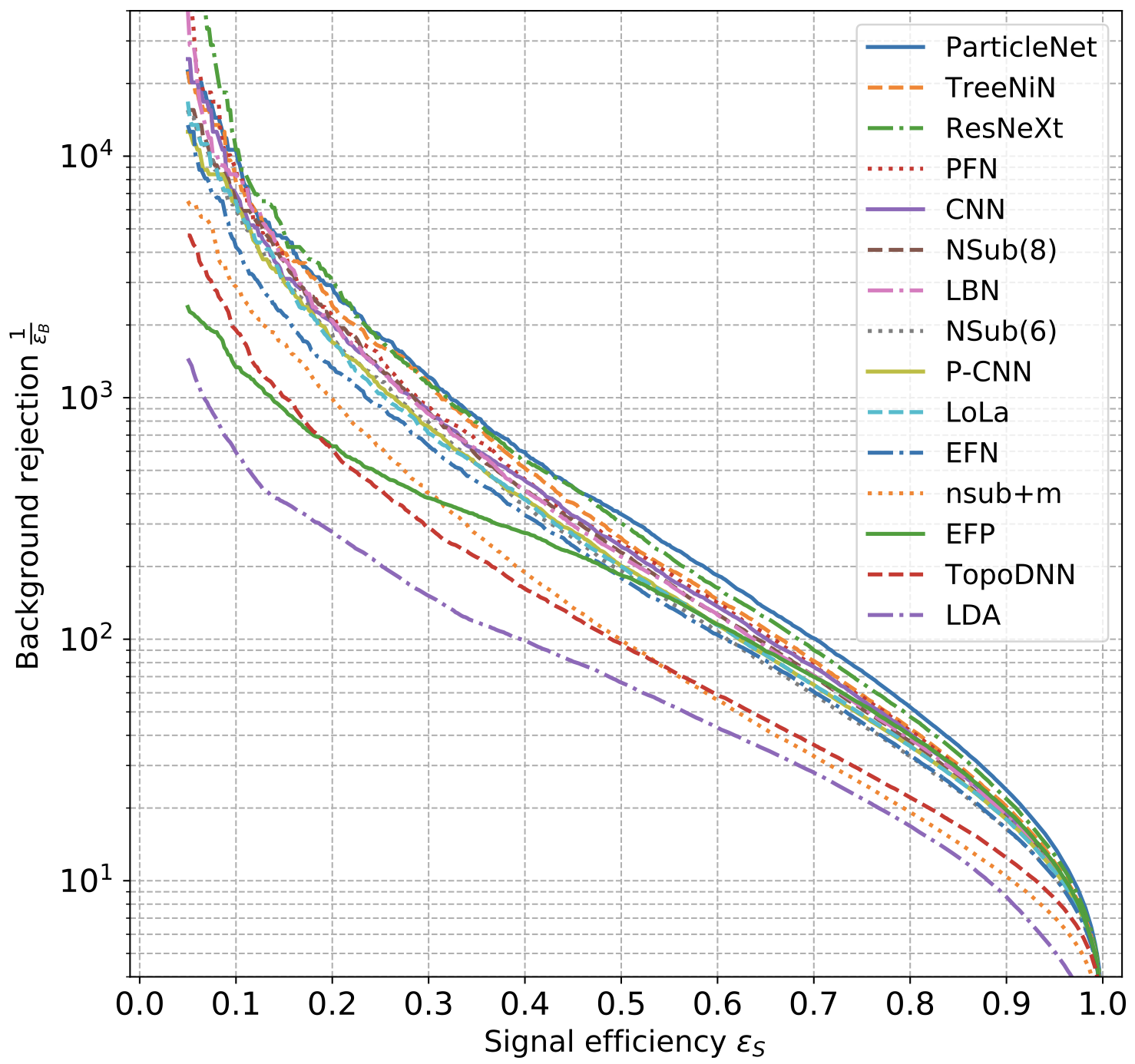}
    \caption{ROC curves for a range of top tagging classifiers evaluated on a common dataset, reproduced from Ref.~\cite{Kasieczka:2019dbj} under the CC~BY~4.0 license.}
    \label{fig:landscape-roc}
\end{figure}

Some context is needed to understand the research contribution of the paper.
A 2019 community comparison study~\cite{Kasieczka:2019dbj} benchmarked a wide variety of deep-learning-based top tagging methods on a single dataset to make a direct comparison of their performance. 
The result is summarized by the ROC (receiver operating characteristic) curve shown in \Cref{fig:landscape-roc}, which plots the signal efficiency (true positive rate) on the horizontal axis against background rejection (the inverse of the false positive rate) on the vertical axis.
As a cut on the tagger output is swept from its minimum to maximum value, the signal efficiency and background rejection will increase.
A tagger that does not separate signal and background at all will produce a flat line, since all thresholds yield the same signal efficiency and background rejection.
The curve traces by increasingly powerful taggers will move upward and to the right, as more background is removed at the same signal efficiency.
The area under the curve\footnote{The curve referred to in this performance metric is a different parametrization of the ROC curve than the one used in this thesis.} (AUC) summarizes overall classification power in a single number, with random guessing producing an AUC of 0.5 and a perfect classifier producing an AUC of 1.

There are two important takeways from this figure.
First, taggers that process the jet constituent kinematic information directly rather than relying on hand-engineered expert features produce the best performance.
This is the same lesson as the AlexNet moment in a jet tagging context.
It is more performant to dump all available information into a single neural network than to train a smaller network on a set of well understood features that have been developed by domain experts.
Second, the dataset was generated with the fast but approximate detector simulation software \textsc{Delphes}~\cite{deFavereau:2013fsa}.
This tool is commonly used when developing deep learning methods for HEP because it is easy to use and fast, in contrast to the much more realistic \textsc{Geant4}-based full simulation used by ATLAS and CMS.
However there is no guarantee methods that perform well on datasets generated with \textsc{Delphes} will perform well in full simulation or in experimental data.
Furthermore, the paper only compared raw classification power as measured by the ROC curve and AUC metric.
Actual deployment of a top tagging method in a physics analysis requires some consideration of systematic uncertainties.
This could not be done with \textsc{Delphes} simulation which does not maintain a set of systematic variations.
The primary objective of Refs.~\cite{ATLAS:2022qby, ATLAS:2024rua} is to address these gaps by training and evaluating constituent-based top taggers on ATLAS full simulation, assessing the systematic uncertainties on the tagger performance, and providing a high-quality public dataset to support future studies.

\FloatBarrier

\subsection{Dataset and Jet Selection}

The first step was to construct a dataset of labeled jets, taken from simulated events, that can be used to train and evaluate the top taggers.
Both ATLAS and CMS train taggers on simulated events and then apply them on data, because truth labeling schemes are straightforward to define for simulated events.
Setting truth labels on data is in principle possible but much more complicated.
In this study, top quark signal jets are obtained from simulated $Z' \rightarrow t\bar{t}$ events with $m_{Z'} = 2$~TeV, where the $Z'$ is a hypothetical massive resonance used solely as a convenient source of boosted top quarks.
These events are generated at leading-order (LO) with \textsc{PYTHIA8}~\cite{Sjostrand:2014zea} using the \textsc{NNPDF2.3LO}~\cite{Ball:2012cx} set of parton distribution functions (PDFs) and the A14~\cite{ATL-PHYS-PUB-2014-021} set of tuned parameters.
The production cross section is reweighted to produce an approximately flat jet $p_T$ distribution, which ensures that the tagger can learn to classify top jets across a wide range of jet \pt values.
Background jets are obtained from simulated QCD dijet events, which are simulated in slices of jet $p_T$ to avoid simulating only very low $p_T$ jets.
All simulated samples are passed through a \textsc{Geant4}~\cite{GEANT4:2002zbu}-based simulation of the ATLAS detector.
For more details on the ATLAS detector, see Ref.~\cite{ATLAS:2008xda}.

\input{tab_top_jet_selection.tex}

\subsection{Tagger Architectures}

Six classifiers are considered in the study.
The neural network architectures were introduced in \Cref{ch:ml}, so only the jet tagging specific details are described here\footnote{An obvious exclusion from this list is a transformer architecture. These models were just starting to be applied in HEP as this research was being done. Based on all evidence the transformer should slightly outperform ParticleNet.}.

\textbf{hlDNN.} A multi-layer perceptron that uses 15 hand-engineered substructure variables as inputs.
This serves as the baseline for this study, and a very similar neural network is still in use in various ATLAS analyses today.
For a list of the features see Table 3 in Ref.~\cite{ATLAS:2024rua}.

\textbf{DNN.} A multi-layer perceptron that uses the four-momenta of the jet constituents as inputs.
All expert-engineered features are calculable from this information.
However the information is variable dimensional because the number of constituents varies from jet to jet.
This is handled by setting an arbitrary number, where the inputs for jets with fewer constituents would be zero-padded and the inputs for jets with more constituents would be truncated.
In practice this number was taken to be 80 to balance avoiding truncation and preventing most inputs from being zero for most jets.
With the seven input features discussed below used per-constituent, this implies a 560 dimensional input vector.

\textbf{EFN and PFN.} The Energy Flow Network and Particle Flow Network are set processing neural networks made popular by their adoption for jet physics~\cite{Komiske:2018cqr}.
They are \textit{deep set} networks, which consist of two MLPs.
The first MLP acts on the four-vector of a single jet constituent, and outputs a vector of latent features.
The latent vectors for all constituents are then aggregated using a permutation invariant operation, usually a sum, to produce a single fixed-length representation of the jet.
This representation is then passed to the second MLP, which outputs the network output.

In a PFN, the inputs to the constituent-level MLPs are unconstrained so each of the seven input features described below is used.
In an EFN, only the angular information about a constituent, for example its $\eta$ and $\phi$ coordinates is used as inputs.
Further the sum over constituent representations is \textit{energy weighted} by the energy of each constituent, so higher energy constituents have much more impact on the output of the sum.
For hadron collider data, typically the $p_T$ is used instead of the constituent energy.
These two design constraints ensure that the output of the EFN is infrared and collinear (IRC) safe.
They limit the information that the EFN is able to use to distinguish signal from background jets, but, as will be discussed in the following section, give it greater robustness against systematic uncertainties.

\begin{figure}[!ht]
    \centering
    \begin{subfigure}[b]{0.48\textwidth}
        \centering
        \includegraphics[width=\linewidth, alt={Grayscale 2D histogram in the eta-phi plane showing the average pixel intensity of top quark jet images, with three distinct over-dense regions visible corresponding to the three-pronged substructure of the top quark decay.}]{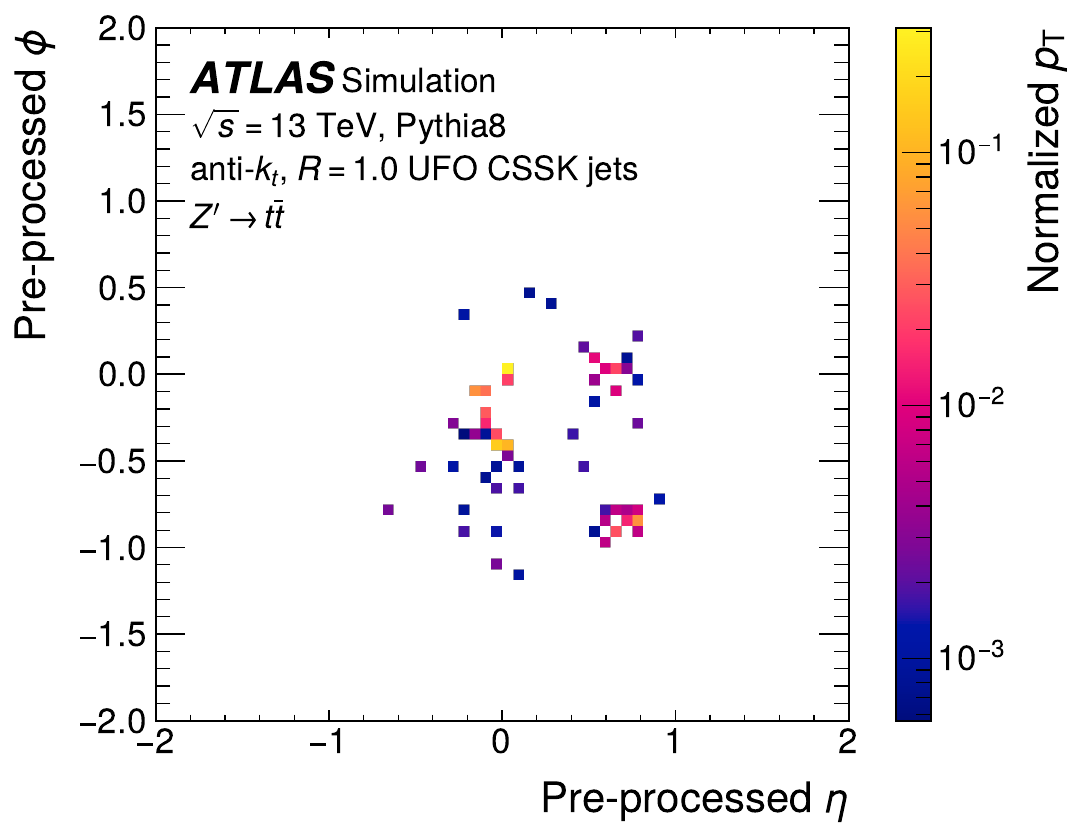}
        \caption{Average top jet image.}
        \label{fig:top-jet-image}
    \end{subfigure}
    \hfill
    \begin{subfigure}[b]{0.48\textwidth}
        \centering
        \includegraphics[width=\linewidth, alt={Grayscale 2D histogram in the eta-phi plane showing the average pixel intensity of QCD background jet images, concentrated near the jet axis with no distinct substructure.}]{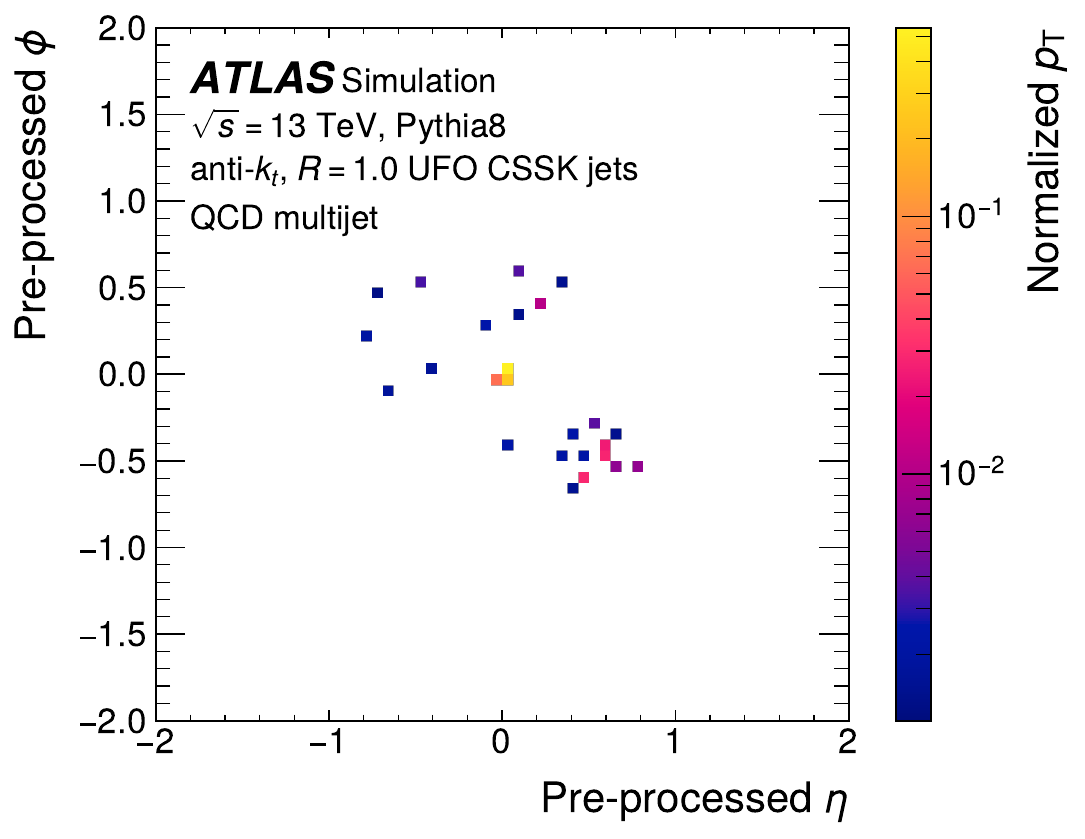}
        \caption{Average QCD jet image.}
        \label{fig:qcd-jet-image}
    \end{subfigure} \\
    \begin{subfigure}[b]{0.48\textwidth}
        \centering
        \includegraphics[width=\linewidth, alt={Grayscale 2D histogram showing the pixel-wise ratio of the average top jet image to the average QCD jet image, with excess intensity surrounding the jet axis indicating the additional substructure of top jets.}]{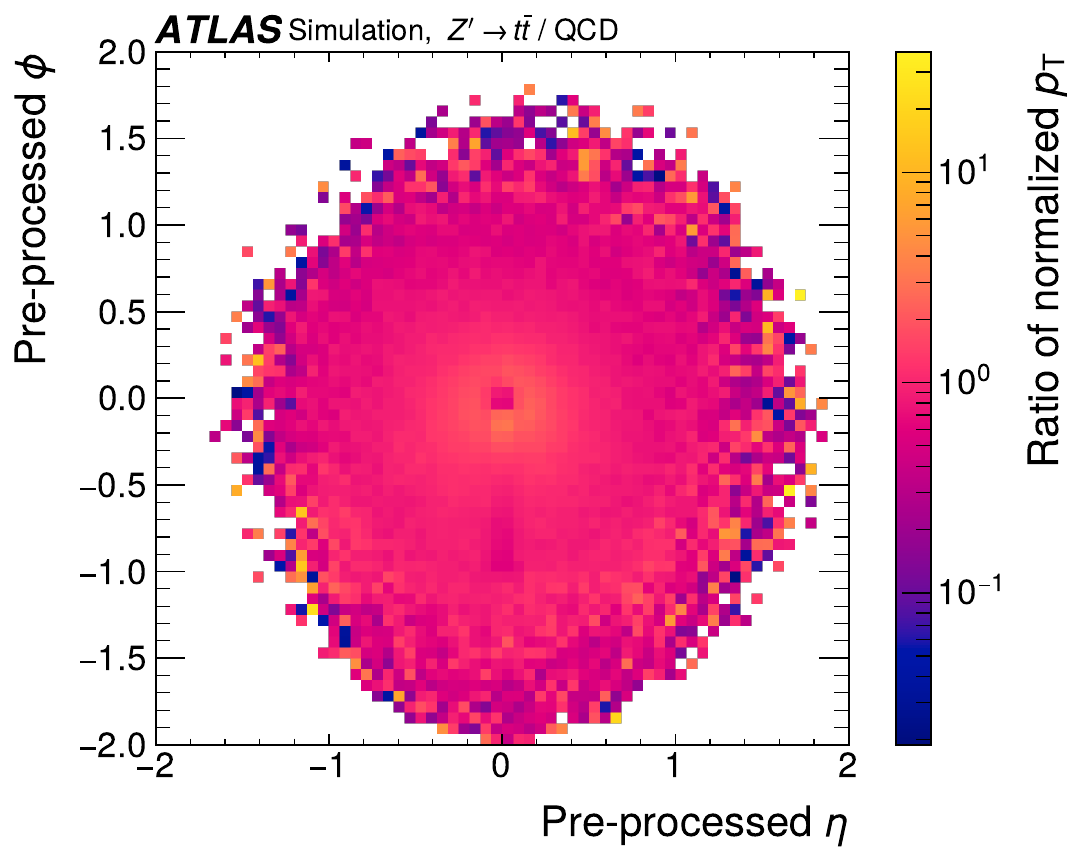}
        \caption{Ratio: top / QCD.}
        \label{fig:jet-image-ratio}
    \end{subfigure}
    \caption{Average jet images for top quark jets (top left) and QCD jets (top right), and their pixel-wise ratio (bottom). Pixel intensities are shown on a logarithmic scale. The additional substructure of the top quark decay is visible as the over-dense regions surrounding the jet center in the ratio image. Reproduced from Ref.~\cite{ATLAS:2024rua} under the CC~BY~4.0 license.}
    \label{fig:jet-images}
\end{figure}

\textbf{ResNet50.} A convolutional neural network that was designed for computer vision tasks~\cite{He:2015resnet}.
Its inputs are \textit{jet images}, which are 2D representations of jets constructed by binning the jet constituent information in a 2D grid in the $\eta$--$\phi$ plane.
In this study, a 64x64 bin histogram was used, and the contribution of a single constituent is weighted by the constituent $p_T$.
A logarithmic rescaling is then applied to the pixel intensities to enhance the contrast of the substructure.
Example jet images are shown in the top row of \Cref{fig:jet-images}, with the left example being a top jet and the right example being a QCD jet.
Note that jet images are very sparse since most jets have several tens of constituents that do not fill the plane.
The three-prong substructure of the top jet is visible as three over-dense regions in the top left panel.
The bottom row of \Cref{fig:jet-images} shows the pixel-wise ratio of the average top jet image to the average QCD jet image, with the average taken over the entire test set.
The additional substructure of the top jets is visible as the excess of normalized \pt surrounding the jet axis\footnote{The jet axis is defined using the ``winner take all'' scheme, where it is set to the $\eta$ and $\phi$ coordinates of the constituent with the highest \pt.}.
These jet images, a 64x64 grid of pixel intensities with one channel (i.e. the images are grayscale), are then passed to a ResNet50 classifier.

\textbf{ParticleNet.} A graph neural network~\cite{Qu:2019gqs} that applies the EdgeConv operation discussed in \Cref{sec:gnn}.
To interpret the jet constituent kinematic information as a graph, each constituent is represented as a node with the seven input features described below as node features.
In the first EdgeConv layer of the network, edges are assigned such that each node is connected to its $k=18$ nearest neighbors in the $\eta$--$\phi$ plane.
This means that spatially close constituents are allowed to influence each other's features through the message passing operations.
In the second and all subsequent EdgeConv layers, the edges are re-assigned to connect the $k=18$ nearest neighbors in the full feature space, rather than only the $\eta$--$\phi$ plane.
This means the graph edges change with each layer, so ParticleNet is a \textit{dynamic} graph network.
ParticleNet is an extremely popular deep learning architecture in HEP.
The CMS collaboration in particular has deployed it to perform both classification and regression tasks in a large range of measurements and searches.

\subsection{Input Features and Training}

Each constituent-based tagger, save ResNet50 and the EFN, uses a common set of per-constituent input features.
These are: the $\eta$ and $\phi$ coordinates relative to the jet axis, the logarithms of the constituent $p_T$ and energy $E$, the logarithms of the constituent $p_T$ and $E$ normalized to the total jet $p_T$ and $E$ respectively, and the angular distance $\Delta R$ between the constituent and the jet axis.
The EFN uses only the $\eta$, $\phi$, and logarithm of the $p_T$ as discussed above, and ResNet50 uses the jet images.

All taggers are trained using the Adam optimizer to minimize the binary cross-entropy loss.
The training continues until the validation loss fails to decrease for 20 consecutive epochs.
The training, validation, and test sets contain 9 million, 1 million, and 3.8 million jets respectively.

\subsection{Results}

\begin{figure}[ht]
    \centering
    \includegraphics[width=0.75\linewidth, alt={ROC curve plot showing inverse background rejection versus signal efficiency for six top quark taggers evaluated on the ATLAS full simulation test set, with ParticleNet achieving the highest background rejection, DNN and PFN also outperforming the hlDNN baseline, and ResNet50 performing worst.}]{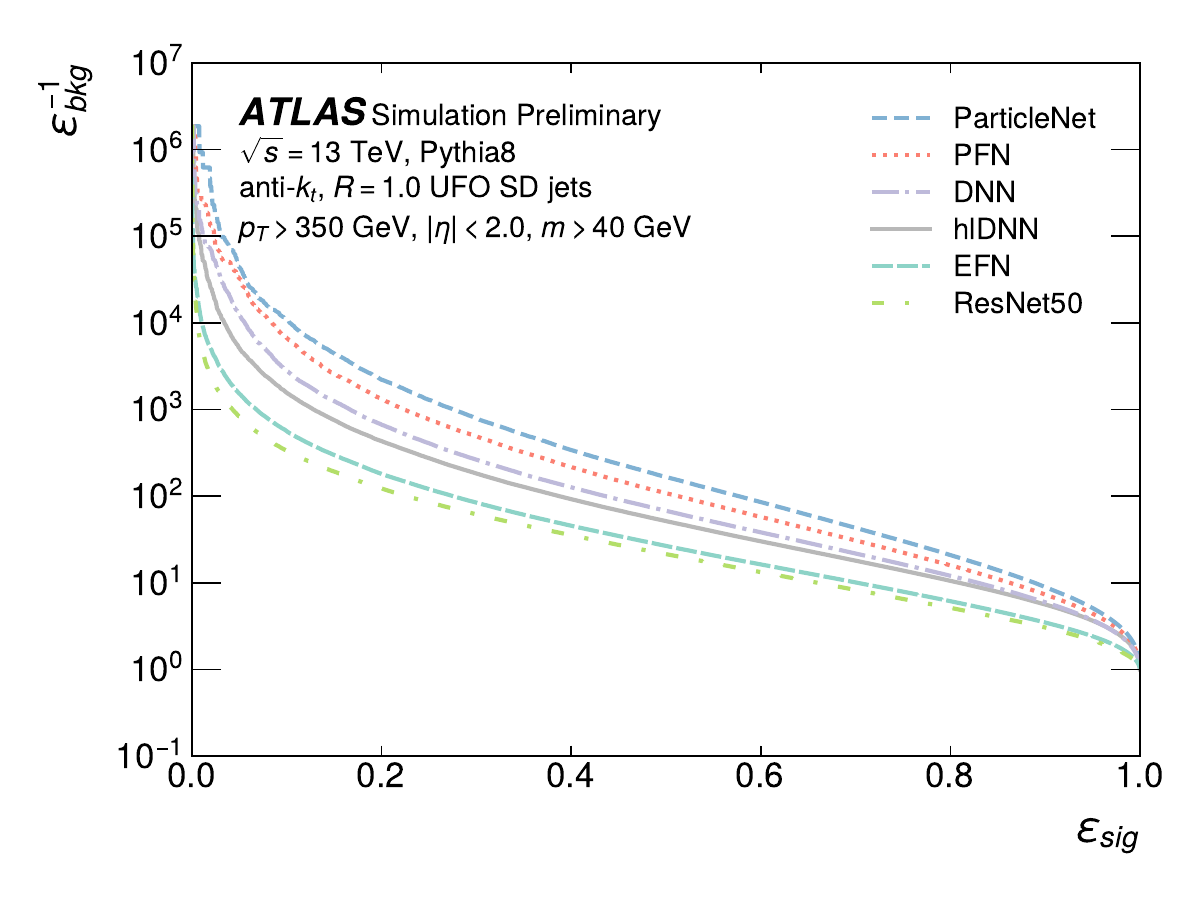}
    \caption{ROC curves for the six top quark taggers evaluated on the full simulation test set. The vertical axis shows the inverse background rejection $\varepsilon^{-1}_\text{bkg}$ and the horizontal axis shows the signal efficiency $\varepsilon_\text{sig}$. The DNN, PFN, and ParticleNet constituent-based taggers outperform the hlDNN expert-feature baseline, with ParticleNet achieving the highest background rejection across all working points. ResNet50 performs significantly worse than the other constituent-based taggers, likely due to the pixelization of the jet image. Reproduced from Ref.~\cite{ATLAS:2024rua} under the CC~BY~4.0 license.}
    \label{fig:roc}
\end{figure}

The classification performance of each tagger is summarized in \Cref{tab:performance_numbers} and \Cref{fig:roc}.
Four numeric metrics are used to characterize classification power in this study.
The AUC is discussed above.
The accuracy (ACC) is the fraction of correctly identified signal jets (true positive rate).
The inverse background rejection $\varepsilon^{-1}_\text{bkg}$ is the inverse of the false positive rate.
This quantity is plotted on the vertical axis in \Cref{fig:landscape-roc}.
However when a top tagger is used in a physics analysis, a threshold must be set above which jets are considered signal and below which jets are considered background.
This threshold is called a \textit{working point} and is typically set to fix a desired signal efficiency.
In this study working points of 50\% and 80\% signal efficiency are considered, and the background rejection at these working points are the last two metrics.

The background rejection metrics are the most important in this study, because fewer background jets in the signal region means more sensitivity to the signal for an arbitrary analysis.
The downside of the inverse background rejection as a metric is that it diverges when strong cuts are applied on the tagger output and almost no background jets survive the cut.
This will not occur for the top taggers discussed here, but can happen in flavor tagging applications as discussed in \Cref{sec:tagging-future}.

\input{tab_top_metrics.tex}

ParticleNet produces the best classification performance across all metrics.
It outperforms the hlDNN baseline by a factor of 2--3 in total background rejection.
This confirms the results of the community comparison study (\Cref{fig:landscape-roc}): constituent-based top taggers outperform those that rely on expert features, even in the highly realistic ATLAS full simulation.

A major surprise of the study is that ResNet50 significantly underperforms the hlDNN baseline, whereas it was among the top performers in the Kasieczka et al. study (\Cref{fig:landscape-roc}).
Interpretation of this result is difficult, but a possible explanation is that building jet images imposes a fixed spatial grid on top of the non-uniform geometry of the ATLAS calorimeter.
This is a secondary discretization on top of the one required by the calorimeter cells, which could produce significant distortions in the jet image.
Unlike the ATLAS full simulation, \textsc{Delphes} does not model individual calorimeter cells and only applies smearing to particle energy measurements in line with the resolution of real-life calorimeters.
Therefore this discretization would have less effect on jets in a dataset generated with \textsc{Delphes}.

The other constituent-based tagger that underperforms the hlDNN baseline is the EFN.
This can be interpreted as a direct result of the IRC safety constraint.
By weighting constituent features by their $p_T$ fraction before aggregation, the EFN is prevented from utilizing the low \pt details of the jet substructure.
Apparently these details carry significant discriminating power for top versus QCD classification.
This is an important lesson from the application of deep learning to jet tagging that took some time to be absorbed by the community.
Event generators simulate observables that are not IRC safe and cannot be computed from first principles given our current understanding of QCD, but are still very useful for jet tagging.
Both ATLAS and CMS now use IRC unsafe information in their jet tagging algorithms, despite the lack of rigorous theoretical understanding for why this information is useful.
This leads to impressive performance but also creates significant systematic uncertainties arising from mis-modeling of this IRC unsafe information in Monte Carlo simulations.
This issue is the main topic of the next section.

\begin{figure}[ht]
    \centering
    \begin{subfigure}[b]{0.48\textwidth}
        \centering
        \includegraphics[width=\linewidth, alt={Line plot of background rejection versus jet transverse momentum at the 50 percent signal efficiency working point for six top taggers, showing the hlDNN baseline decreasing with jet pT while constituent-based taggers peak at intermediate pT.}]{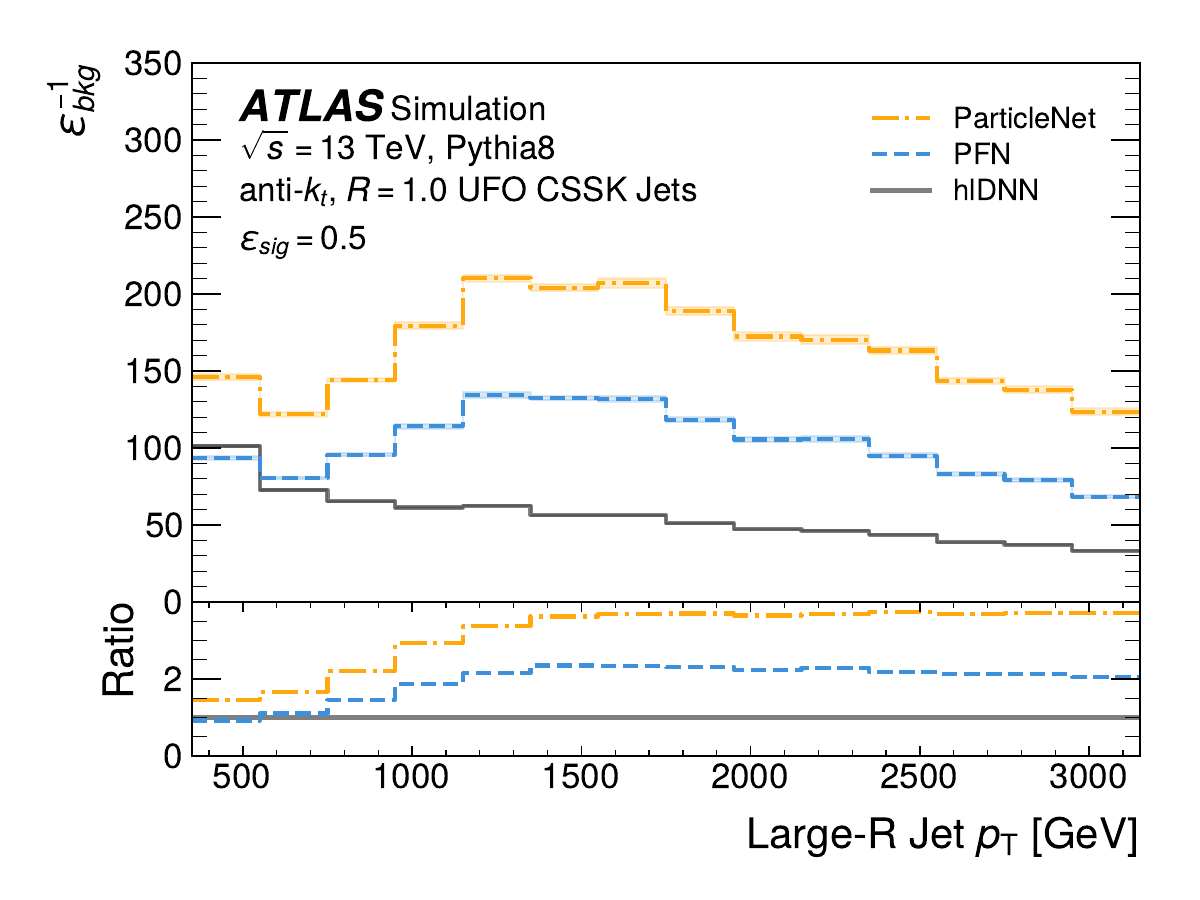}
        \caption{Background rejection at $\varepsilon_\text{sig} = 50\%$.}
        \label{fig:br-50}
    \end{subfigure}
    \hfill
    \begin{subfigure}[b]{0.48\textwidth}
        \centering
        \includegraphics[width=\linewidth, alt={Line plot of background rejection versus jet transverse momentum at the 80 percent signal efficiency working point for six top taggers, showing the hlDNN baseline decreasing with jet pT while constituent-based taggers peak at intermediate pT.}]{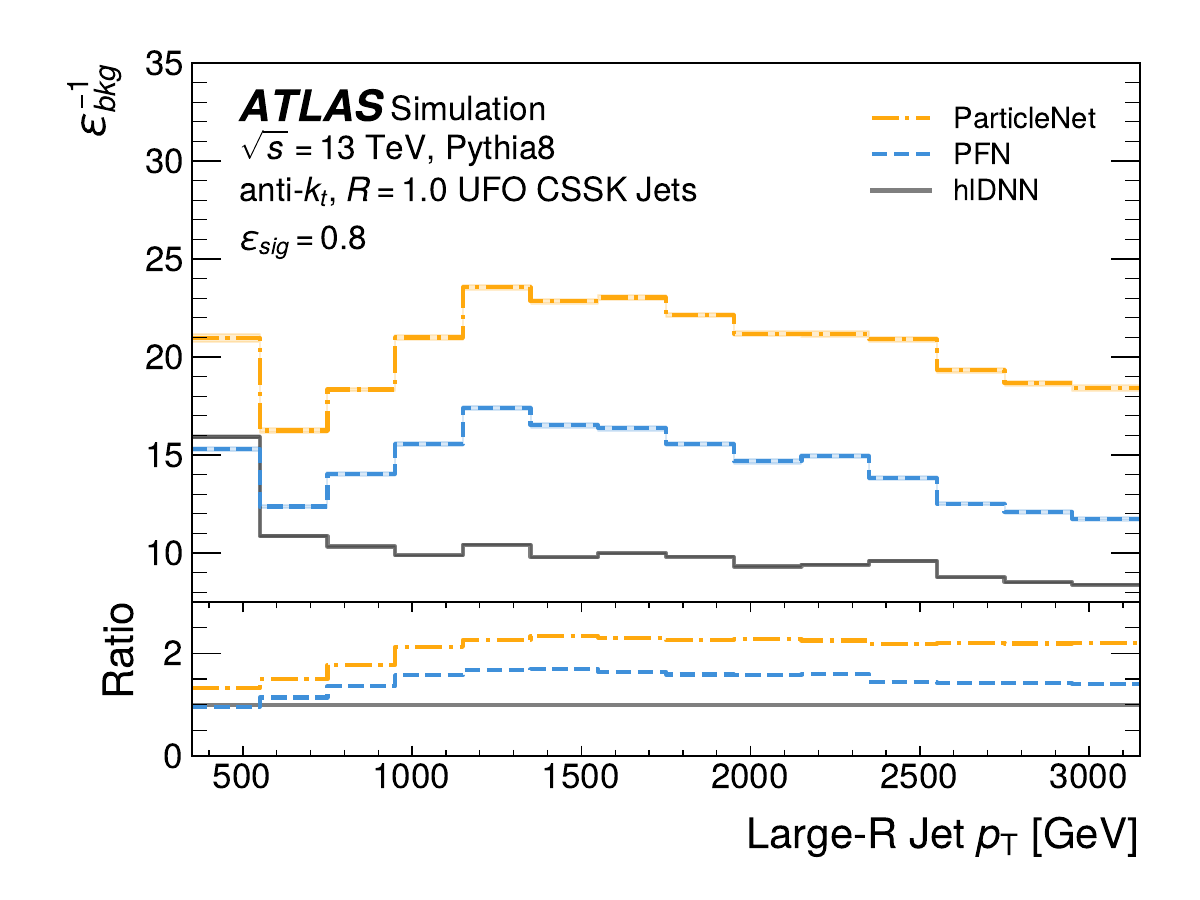}
        \caption{Background rejection at $\varepsilon_\text{sig} = 80\%$.}
        \label{fig:br-80}
    \end{subfigure}
    \caption{Background rejection as a function of jet $p_T$ at the fixed signal efficiency working points of 50\% (left) and 80\% (right). The hlDNN background rejection decreases with increasing jet $p_T$, consistent with the expectation that the expert features lose information as the jet becomes more boosted and the constituents are more tightly collimated. Constituent-based taggers demonstrate a different evolution with jet $p_T$, with peak performance in the intermediate $p_T$ range. This trend is not yet fully understood. Reproduced from Ref.~\cite{ATLAS:2024rua} under the CC~BY~4.0 license.}
    \label{fig:br-pt}
\end{figure}

Finally, \Cref{fig:br-pt} shows the background rejection at the fixed signal efficiency working points in bins of jet $p_T$.
The hlDNN background rejection falls monotonically with increasing jet $p_T$.
This is because the three-prong substructure of the top decay becomes increasingly difficult to resolve as the quarks from the three-body decay become more collimated.
In the limit of very large $p_T$, all decay products would be contained within a single calorimeter cell, and top and QCD jets would become indistinguishable\footnote{This is a real concern for tagging extremely boosted top quarks, for example at a future proton-proton collider at $\sqrt{s} = 100$~TeV. Top quarks with $p_T > 5$~TeV would be contained within a single cell of the calorimeters currently in use at the LHC, making them indistinguishable from QCD jets. Developing high granularity calorimeters is necessary for jet tagging at very high $p_T$.}.
The background rejection of the constituent-based taggers peaks in the intermediate \pt range of 1--2 TeV.
The degradation of the performance at very high \pt is expected, but the better performance at intermediate \pt is not fully understood.
One plausible contributing factor is that the training dataset contains the most jets in the intermediate \pt range so the networks learn to classify better in this region.

\section{Systematic Uncertainty Quantification}
\label{sec:systematic-uq}

The results of \Cref{sec:constituent-top-tagging} demonstrate that constituent-based top taggers substantially outperform top taggers based on expert features, even when using the realistic ATLAS full simulation.
The next objective of the ATLAS top tagging study is to quantify the systematic uncertainties on tagger performance.
ATLAS and CMS both train their taggers on simulation and apply them in data.
However the simulation is not a perfect description of the data.
The extent to which the imperfections in the simulation affect the tagger performance is an important consideration in designing and deploying jet taggers in collider experiments.

\subsection{Scale Factor Calibration}

The standard approach to quantify differences between simulation and data is to measure a \textit{scale factor}, which is a generic term used to describe any ratio between two quantities calculated in data and simulation.
In jet tagging, scale factors are typically calculated for the signal and background efficiency of a classifier at a given working point (fixed signal efficiency in simulation).
These scale factors can be applied as a multiplicative weight to event yields measured in simulation so that they are in agreement with the same event yields for data.

To measure a scale factor on the signal efficiency, a sample of relatively pure signal jets must be isolated in data by independent means rather than using the tagger itself.
The same must be done for the background efficiency, but given background jets are extremely common this is typically straightforward.
A pure sample of boosted top (signal) jets can be obtained using semi-leptonic $t\bar{t}$ events, where one top quark decays leptonically and produces an easy-to-identify lepton along with a $b$ jet and a neutrino, while the other top quark decays hadronically and produces a boosted top jet.
Some straightforward event selections on the lepton can isolate a sample of recoiling jets that can be considered to be due to top quarks to a good approximation.
The tagger efficiency can then be measured by performing a likelihood fit on the ratio of jet yields in data and simulation.
This is typically done in bins of jet \pt.
Extracting a scale factor amounts to measuring a set of parameters via simulation-based inference, but the parameters are features of the tagger rather than some parameter of the Standard Model for example.
As such, systematic uncertainties must be propagated through the measurement to produce a scale factor with associated uncertainties.

Scale factors are the gold standard method for quantifying systematic uncertainties in jet tagging.
Virtually every ATLAS and CMS search or measurement that relies on jet tagging applies scale factor corrections, and the corresponding uncertainties can often be the dominant sources of uncertainty in many analyses.
Scale factors do not eliminate the need for uncertainty quantification.
They just shift the need for uncertainty quantification to the scale factor as opposed to the tagger output.

\subsection{The Bottom-Up Approach}

An alternative to scale factors is to propagate systematic uncertainties through the tagger by applying variations to the tagger inputs and recording the variation in the tagger output.
This is the approach used in Refs.~\cite{ATLAS:2022qby,ATLAS:2024rua}, which is sometimes referred to as the \textit{bottom-up approach}.

There are a few benefits to the bottom-up approach compared to scale factors.
First the variations on the tagger inputs need only be defined once, and then the uncertainty on an arbitrary tagger can be determined by passing the variations through the network.
Setting uncertainties is only a matter of running inference, rather than the more complex likelihood fit required for scale factors\footnote{Note that the effect of systematic uncertainties does not need to be propagated through the neural network training because the trained neural network is arbitrary from the perspective of the event selection and yields. In more advanced applications where the network outputs have a direct statistical interpretation the training procedure must also be considered, as will be the case in \Cref{ch:zjets}.}.
Second, the bottom-up method does not require isolating a pure sample of signal or background jets in data.
For top taggers there are known methods for obtaining these samples as described above, but for taggers designed to identify jets from BSM or exotic processes it may be difficult or impossible.
This is an issue for anomaly detection searches, where the signal topology is by definition unknown~\cite{ATLAS:2025obc, CMS:2025sch} so data-driven background estimation techniques must be used.
See Ref.~\cite{CMS:2025eyd} for an alternative, but similar, approach to uncertainty quantification in anomalous jet tagging.

The downside of the bottom-up approach is that a reliable set of systematic variations must be available on the tagger inputs.
This is particularly difficult for experimental systematic uncertainties, which often are based on auxiliary measurements of the detector response as opposed to simply varying features of the simulation.
In particular ATLAS does not maintain a set of validated experimental uncertainties for the jet constituent objects used in this study.
One of the novel aspects of this study is to develop such a set of variations.
However it should be emphasized that these variations are approximate, intended only to estimate the magnitude of experimental effects on tagger performance rather than to serve as the basis for a precision measurement.
The theoretical uncertainties on tagger performance will be sufficiently dominant that the approximations made are acceptable.

\subsection{Experimental Uncertainties}

\begin{figure}[tb]
    \centering
    \includegraphics[width=0.85\linewidth, alt={Flow diagram showing the construction of Unified Flow Objects from inner detector tracks and particle flow objects, illustrating the cluster-splitting step that produces charged, neutral, and merged UFO types.}]{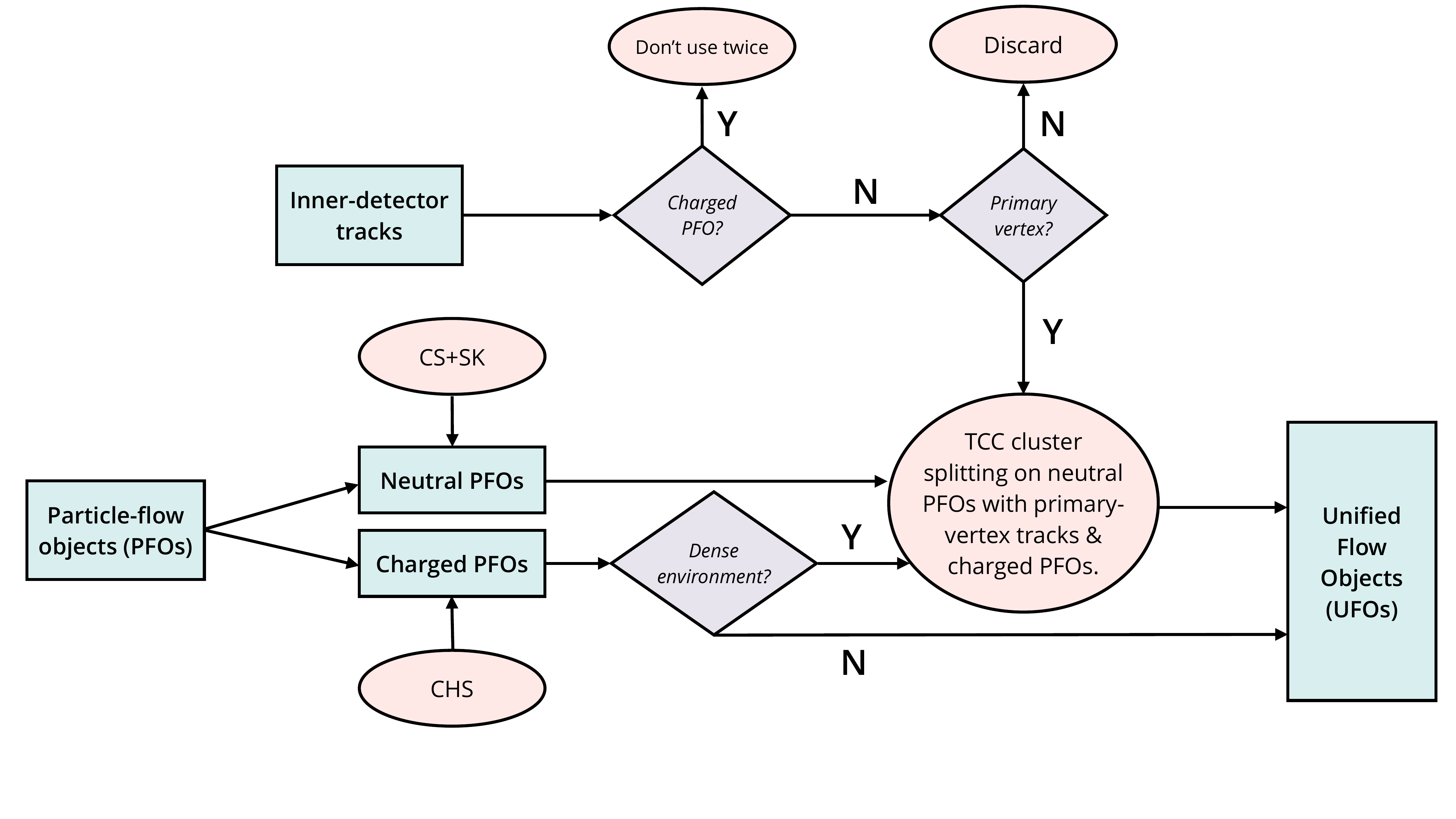}
    \caption{Flow diagram illustrating the construction of Unified Flow Objects from inner detector tracks and particle flow objects, reproduced from Ref.~\cite{ATLAS:2020gwe} under the CC~BY~4.0 license. The splitting step applied to clusters in dense regions produces three distinct UFO types sensitive to different systematic uncertainties.}
    \label{fig:ufo-diagram}
\end{figure}

The procedure for building UFOs is illustrated in \Cref{fig:ufo-diagram}, reproduced from Ref.~\cite{ATLAS:2020gwe}.
UFOs are built from particle flow objects (PFOs), which are a non-trivial combination of information from inner detector (ID) tracks and calorimeter clusters~\cite{ATLAS:2017ghe}.
This is further complicated by a splitting step in which charged PFOs in dense environments are used to split apart calorimeter showers that were likely formed by overlapping showers.
The distinction between charged and neutral PFOs, in combination with the splitting step, produces three distinct types of UFOs, each sensitive to different systematic uncertainties.

Charged UFOs are produced by charged PFOs that are not in dense environments where multiple tracks are associated with the same calorimeter cluster.
Given the properties of these objects are mostly determined by the underlying ID tracks, these constituents are assigned the set of tracking systematic uncertainties.
ID track uncertainties are well understood in ATLAS and the bottom-up approach is reliable for this class of constituent.
The bottom-up methodology with tracking uncertainties will be used to produce a cross-section measurement in \Cref{ch:zjets}.

Neutral UFOs arise from neutral PFOs, which have no associated track and are formed from calorimeter clusters alone.
However this picture is complicated by the cluster-splitting step in which ID tracks can indirectly influence neutral UFOs even though no track is directly associated.
In this study that indirect effect is neglected, and neutral UFOs are assigned a set of calorimeter cluster systematic uncertainties.
These uncertainties are less rigorously understood than the ID track uncertainties, but they have been used to perform some measurements, for example Ref.~\cite{ATLAS:2017zda}.

Merged UFOs arise from the charged PFOs in dense regions that are used as inputs to the cluster splitting step.
The final kinematics of these objects are set by a complicated interplay between inner detector and calorimeter information.
Here the \pt measurement of merged UFOs is treated as a calorimeter-dominated quantity and assigned the same calorimeter cluster uncertainties as neutral UFOs.
The angular measurements ($\eta$ and $\phi$) are treated as track-dominated, and so have negligible uncertainties.

\subsection{Theoretical Uncertainties}

Theoretical uncertainties exist due to approximations in the calculations that underlie event generators (see \Cref{ch:qcd}).
For boosted jet tagging that depends largely on the jet substructure, the most important uncertainties relate to the parton shower and hadronization processes.

In this study these uncertainties are handled separately for the signal and background jets.
For the signal jets, parton shower and hadronization uncertainties are assessed using two samples of SM $t\bar{t}$ events.
Both use \textsc{Powheg Box v2}~\cite{Frixione:2007vw, Nason:2004rx, Alioli:2010xd} to calculate matrix elements at NLO and the \textsc{NNPDF3.0nlo}~\cite{Ball:2014uwa} PDF set.
Decays of bottom and charm hadrons are modeled by \textsc{EvtGen 1.6.0}~\cite{Lange:2001uf}.
The two samples differ in the model used for parton shower and hadronization.
One uses \textsc{Pythia 8}~\cite{Sjostrand:2014zea} and the other uses \textsc{Herwig 7}~\cite{Bellm:2015jjp, Bahr:2008pv}.
\textsc{Pythia} and \textsc{Herwig} use different parton showers and hadronization models, so changing one to the other involves varying both elements of the simulation simultaneously.
The full difference in tagger performance between the two samples is taken as the signal modeling uncertainty.

For the background jets, the parton shower and hadronization uncertainties are assessed independently.
All samples are generated with matrix elements at LO using the \textsc{NNPDF3.0lo}~\cite{Ball:2014uwa} PDF set.
The parton shower uncertainty is assessed by comparing the tagger performance between two samples generated with \textsc{Herwig}, one which uses the default angular-ordered parton shower model and the other which uses an alternative dipole parton shower model.
Both \textsc{Herwig} samples use the default cluster-based hadronization model~\cite{Winter:2003tt}.
The hadronization uncertainty is assessed by comparing the tagger performance between two samples generated with \textsc{Sherpa 2.2}~\cite{Sherpa:2019gpd}, which share the same $p_T$-ordered parton shower.
One uses the default cluster-based hadronization model, and the other uses the \textsc{Sherpa} interface to the Lund string fragmentation model of \textsc{Pythia 6}~\cite{Sjostrand:2006za}.
Two separate background modeling uncertainties are assessed by varying the parton shower and hadronization models independently across these four samples.
These are the dominant theoretical uncertainties for the tagger performance.
Other sub-leading uncertainties, such as from the choice of renormalization and factorization scales in the parton shower, are assessed by on-the-fly event reweighting.

\subsection{Validation and Results}

Before applying the bottom-up uncertainties to the constituent-based taggers, they are validated by comparing their predictions to the well-established scale factor approach for the hlDNN tagger.
Scale factors for the hlDNN were measured using semi-leptonic $t\bar{t}$ events~\cite{ATLAS:2018wis}.
The total bottom-up uncertainty on the hlDNN is an upper bound on the possible performance difference between data and simulation, so the scale factor is expected to be of a similar magnitude but not any larger than the total bottom-up uncertainty. 

\begin{figure}[tb]
    \centering
    \includegraphics[width=0.75\linewidth, alt={Plot comparing the hlDNN scale factor measured in data, shown as points with error bars, against the total bottom-up systematic uncertainty band, shown as a shaded region, as a function of jet transverse momentum, with the scale factor consistently below 1.0 and contained within the uncertainty band.}]{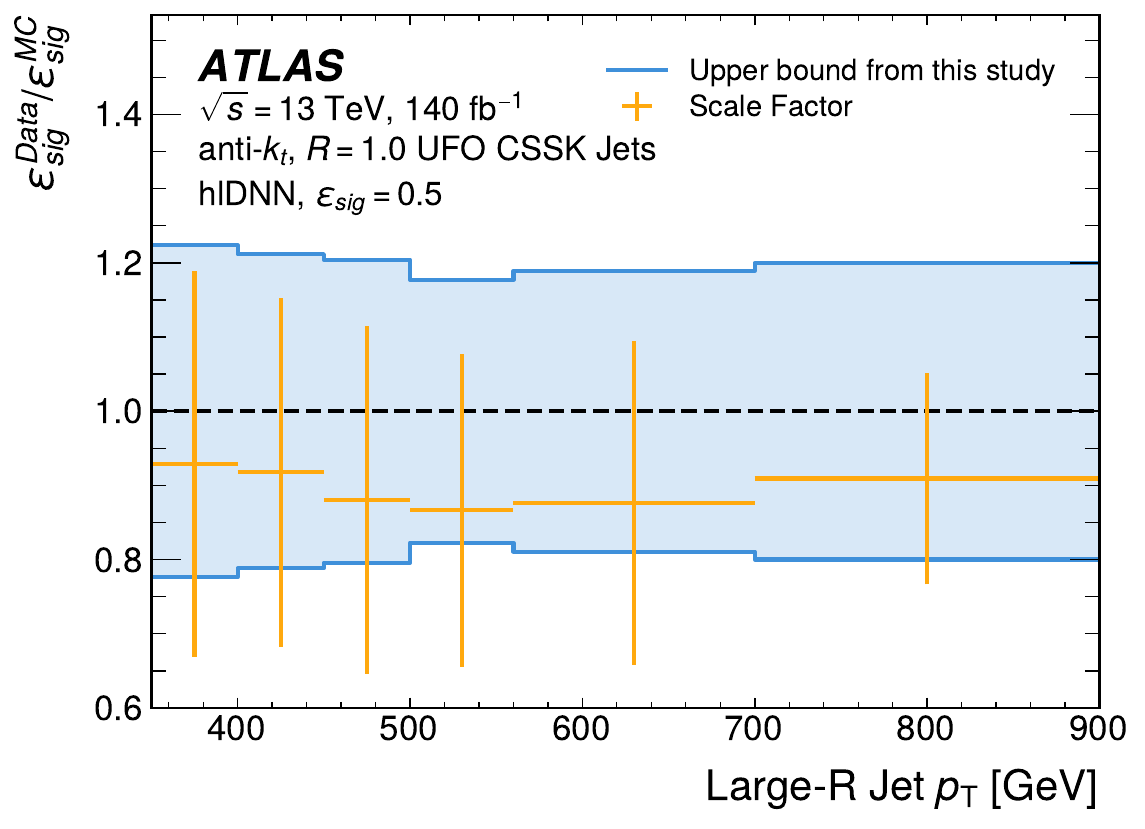}
    \caption{Comparison of the hlDNN scale factor measured in data (points with error bars) against the total bottom-up uncertainty (shaded region), as a function of jet $p_T$. The bottom-up uncertainty covers the measured scale factor in all $p_T$ bins, validating the method. The scale factor is consistently below 1.0, indicating that tagger efficiency is degraded in data relative to simulation. Reproduced from Ref.~\cite{ATLAS:2024rua} under the CC~BY~4.0 license.}
    \label{fig:sf-comparison}
\end{figure}

\Cref{fig:sf-comparison} plots the total bottom-up uncertainty and scale factor in bins of jet $p_T$.
The bottom-up uncertainty band is greater than the measured scale factor in all bins of jet $p_T$, confirming that the bottom-up uncertainties provide a conservative but reasonable bound on the possible performance differences between data and simulation.
The scale factor is consistently below 1.0 across the full $p_T$ range, showing that the performance in data is worse than in simulation.
This is expected since the tagger is trained on the simulation.
The bottom-up uncertainties are symmetrized and therefore predict improvements in data as well as degradations, but improvements are not expected.

\begin{figure}[tb]
    \centering
    \begin{subfigure}[b]{0.48\textwidth}
        \centering
        \includegraphics[width=\linewidth, alt={Plot of systematic uncertainty magnitude versus jet transverse momentum for the EFN tagger, showing signal modeling uncertainty as the dominant contribution among several colored uncertainty groups.}]{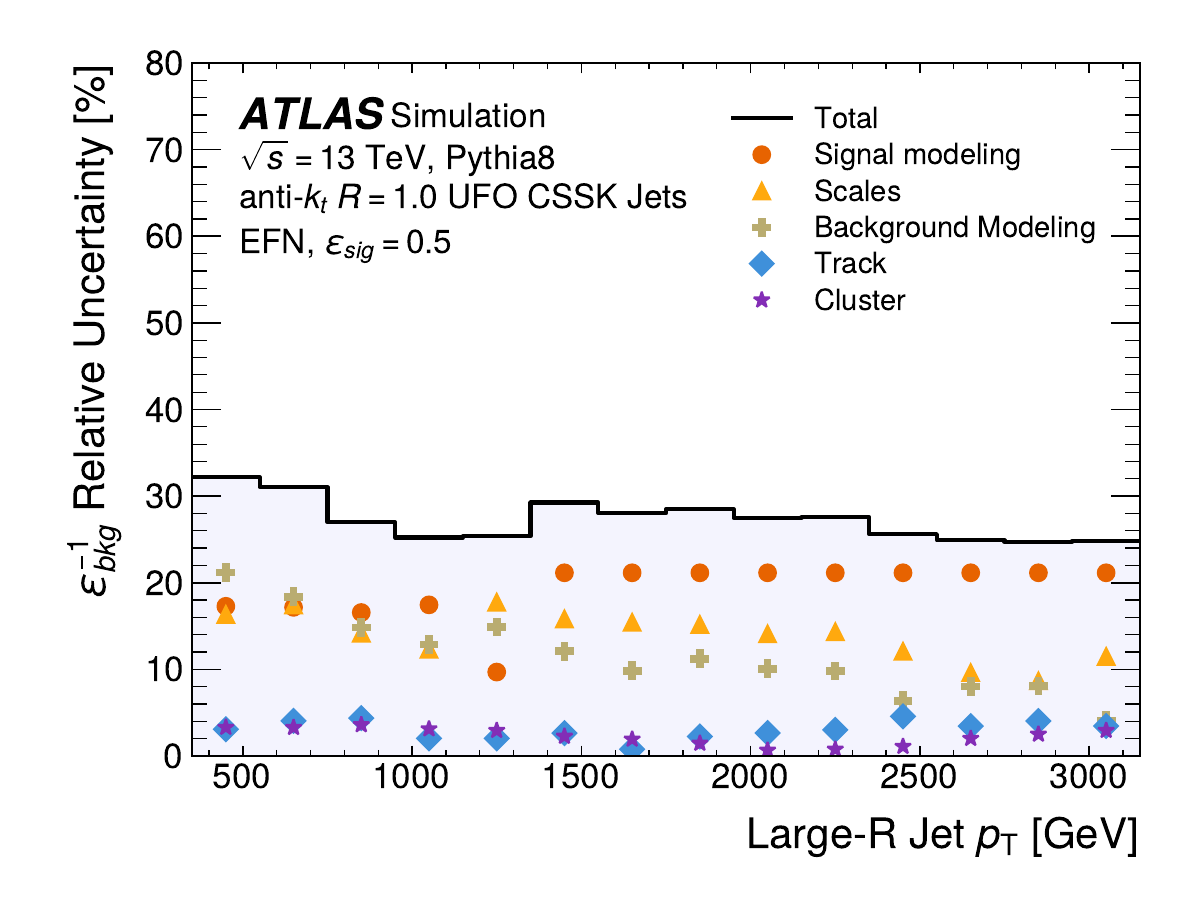}
        \caption{EFN uncertainty breakdown.}
        \label{fig:efn-uncert}
    \end{subfigure}
    \hfill
    \begin{subfigure}[b]{0.48\textwidth}
        \centering
        \includegraphics[width=\linewidth, alt={Plot of systematic uncertainty magnitude versus jet transverse momentum for the ParticleNet tagger, showing signal modeling uncertainty as the dominant contribution and larger overall uncertainties than the EFN.}]{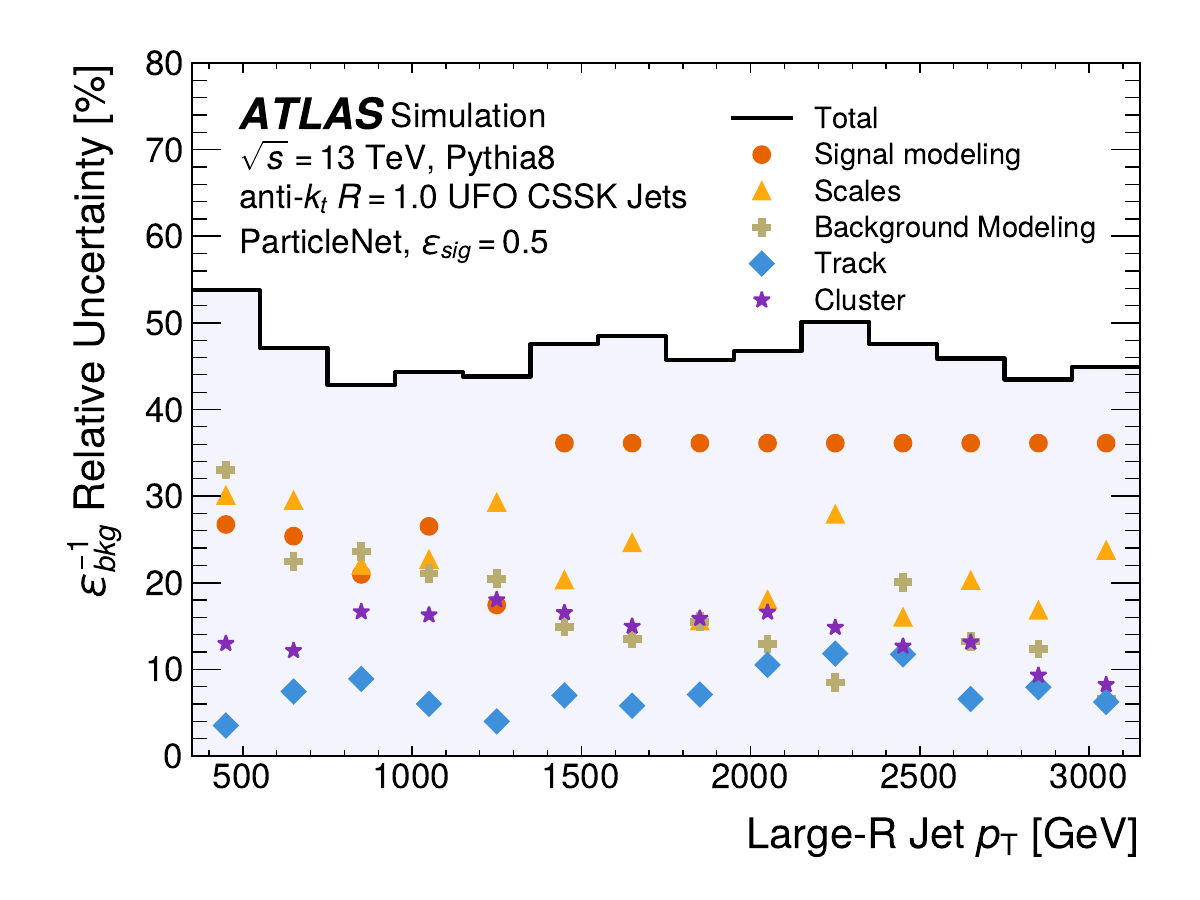}
        \caption{ParticleNet uncertainty breakdown.}
        \label{fig:particlenet-uncert}
    \end{subfigure}
    \caption{Systematic uncertainty budgets for the EFN (left) and ParticleNet (right) taggers as a function of jet $p_T$, reproduced from Ref.~\cite{ATLAS:2024rua} under the CC~BY~4.0 license. The magnitudes of various uncertainty groups are shown as colored points, with the total uncertainty given by their quadrature sum. The signal modeling uncertainty is dominant for both taggers, and all uncertainties are smaller for the EFN than for ParticleNet. Reproduced from Ref.~\cite{ATLAS:2024rua} under the CC~BY~4.0 license.}
    \label{fig:uncert-budgets}
\end{figure}

Following validation the bottom-up uncertainties were used to place bounds on the impact of systematic uncertainties for all taggers.
\Cref{fig:efn-uncert,fig:particlenet-uncert} show the total uncertainty budget for the EFN and ParticleNet respectively.

The dominant source of uncertainty for both taggers is the signal modeling uncertainty, or the difference in tagger performance between the \textsc{Pythia 8} and \textsc{Herwig 7} generated SM $t\bar{t}$ samples.
This is a perfect example of a domain shift uncertainty.
The taggers learn features of the particular parton shower and hadronization model used to produce their training samples, and those features do not generalize perfectly to alternative models or to data.
Importantly, all uncertainties are smaller for the EFN than for ParticleNet.
An explanation for this behavior in the signal modeling uncertainty in particular is that the IRC safety constraint prevents the EFN from exploiting the soft and collinear radiation patterns that are most sensitive to parton shower modeling.
The experimental uncertainties are sub-leading for both taggers, but they are also generally larger for ParticleNet.
This is more evidence that ParticleNet is sensitive to fine details of the jet substructure.

\begin{figure}[tb]
    \centering
    \includegraphics[width=0.7\linewidth, alt={Scatter plot of total systematic uncertainty versus background rejection at the 50 percent signal efficiency working point for six top taggers, with ParticleNet in the upper right showing highest performance and uncertainty, the EFN in the lower left showing lowest performance and uncertainty, and ResNet50 as an outlier with poor performance and large uncertainty.}]{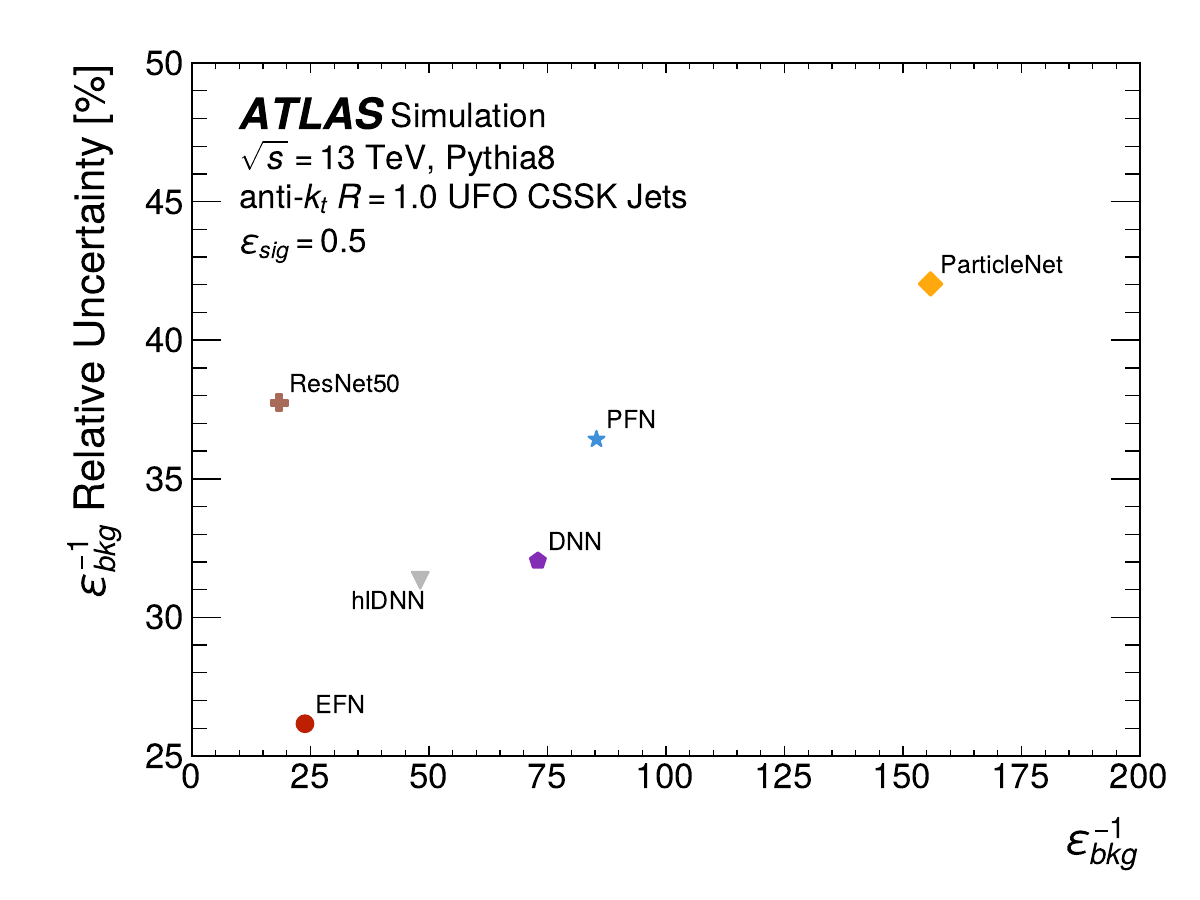}
    \caption{Total systematic uncertainty on the inverse background rejection at the 50\% signal efficiency working point plotted against the background rejection itself. Each point corresponds to one of the taggers. ParticleNet (upper right) has the highest classification performance but also the largest uncertainty. The EFN (lower left) has the smallest uncertainty but at the cost of lower performance. The ResNet is an outlier with both poor performance and large uncertainty. The other taggers trace a roughly monotonic performance--robustness frontier. Reproduced from Ref.~\cite{ATLAS:2024rua} under the CC~BY~4.0 license.}
    \label{fig:uncert-vs-perf}
\end{figure}

Calculating the total uncertainty for each tagger over the full range of jet $p_T$ provided a single numerical metric to quantify sensitivity to systematic uncertainties.
\Cref{fig:uncert-vs-perf} plots this quantity against background rejection for each tagger at the 50\% signal efficiency working point.
In this figure, a new axis beyond raw classification power is introduced to compare top taggers.
The ideal tagger would sit in the lower right hand corner, with high background rejection but low uncertainties.
ParticleNet occupies the upper right corner with the highest background rejection and greatest performance, but also the highest uncertainty.
The EFN occupies the lower left with the lowest uncertainty, but also low background rejection.
The remaining taggers fall roughly on a monotonic performance--robustness frontier between these two endpoints.
ResNet50 is the exception, which achieves neither good performance nor low uncertainty.
Recent results have produced further evidence for this frontier for many different taggers~\cite{Gambhir:2025xim}.

The correct performance versus robustness tradeoff when using a jet tagger for a search or measurement will be very specific to the particular use case.
As discussed above large uncertainties on the tagger performance will manifest practically in larger scale factor uncertainties.
The size of these uncertainties relative to the other systematic uncertainties and the gains in sensitivity due to higher background rejection will depend on the specifics of a given analysis.
A search with few events in the signal region may benefit most from the tagger with the highest background rejection regardless of uncertainty, since statistical uncertainties will likely be dominant anyway.
A measurement with many events in the signal region may have small statistical uncertainties, making a more robust tagger produce more accurate results.
In general the jet tagging and broader LHC community has been interested in pursuing higher background rejection at the expense of larger scale factor uncertainties.
A concrete example and some explanation of this will be given in the next section.
However it is possible that some precise measurements of SM processes would forgo a very powerful tagger in favor of one with small scale factor uncertainties.

A last contribution of the top tagging study is to publish all of the datasets used in the construction of these results~\cite{top_tagging_data}.
Most other existing benchmark datasets for deep learning methods development in HEP have no treatment of systematic uncertainties.
This dataset is somewhat unique in that it lets jet tagging methods be developed with an eye toward systematic uncertainties.
Additionally this dataset consists of roughly 200 million jets all together, which is a large number of jets generated with ATLAS full simulation.
This has made the data additionally useful for the development of foundation models, for example Ref.~\cite{Bhimji:2025isp}.

\section{Jet Tagging Prospects for HL-LHC}
\label{sec:tagging-future}

The top tagging study described in this chapter was completed in 2022--2023.
There has been a large amount of progress in the field since then.
Two broad research directions are worth highlighting heading into the high-luminosity (HL) runs of the LHC.

\subsection{Symmetry-Constrained Architectures}

The first direction seeks to encode known symmetries of particle physics datasets into neural networks, either through direct constraints or loss-term penalties.
The physics of LHC collisions are invariant under Lorentz boosts along the beam axis and rotations around it.
Symmetry-constrained (sometimes called \textit{equivariant}) architectures respect these symmetries exactly rather than learning to approximate them from data.
A number of equivariant architectures have been proposed~\cite{Gong:2022lye, Bogatskiy:2022czk, Brehmer:2024yqw}.
Apart from pure performance, another practical motivation is that enforcing symmetries can reduce the number of trainable parameters required to reach a certain performance, which can allow for fast inference times or deployment on FPGAs as is needed for trigger level jet tagging~\cite{Petitjean:2025zjf}.
Equivariant networks are a model-level development produced by the HEP community rather than an import from the broader ML research community.
Almost no other types of data have such well-known symmetries, making this research something only the HEP community would be motivated to pursue.
A notable open question is how robustly equivariant taggers behave under the systematic variations discussed in \Cref{sec:systematic-uq}.
This has not yet been studied in any depth, though there is some evidence that they are very sensitive to \textsc{Delphes} versus \textsc{Geant4}-based detector simulation as will be discussed in \Cref{ch:zjets}.

\subsection{Scaling and Foundation Models}

The second direction follows the AI community's focus on scaling, where larger and larger models are trained on larger and larger datasets.
An early effort to scale HEP specific neural networks introduced the Particle Transformer (ParT) architecture, which consisted of a few million parameter model trained on 100 million jets~\cite{Qu:2022mxj}.
Apart from introducing some architectural tricks that have become very common in HEP applications, this paper was also notable because it demonstrated state-of-the-art performance on the community top tagging dataset from Ref.~\cite{Kasieczka:2019dbj} could be achieved by pretraining ParT on a larger auxiliary dataset and then fine-tuning on the much smaller benchmark dataset.
At the time of the publication this was referred to as \textit{transfer learning}, but is now generally called building a \textit{foundation model}, which is a large model that is pretrained on large datasets and fine-tuned on more specific tasks.
Recently, many foundation models for jet physics or LHC physics more broadly have been proposed.
A few of these used only supervised learning techniques~\cite{Mikuni:2024qsr,Hsu:2026sww}, but many others used self-supervised objectives like masked particle modeling, next-token predictions, or contrastive learning~\cite{Mikuni:2024qsr, Birk:2024knn, Golling:2024abg, Katel:2024ygn, Li:2024htp, Leigh:2024ked, Birk:2025fbs}.
These models represent interesting methodological direction, but as of writing none have been deployed by ATLAS or CMS on actual collision data, in part because their performance on jet classification tasks is not dramatically larger than scratch trained models.
See \Cref{ch:zjets} for an application where pretraining does offer significant performance gains.

\begin{figure}[tb]
    \centering
    \includegraphics[width=0.78\linewidth, alt={Scatter plot of background rejection on the community top tagging benchmark dataset versus publication year for many deep learning models, showing pretrained models achieving the highest background rejection and the equivariant L-GATr model leading among non-pretrained models.}]{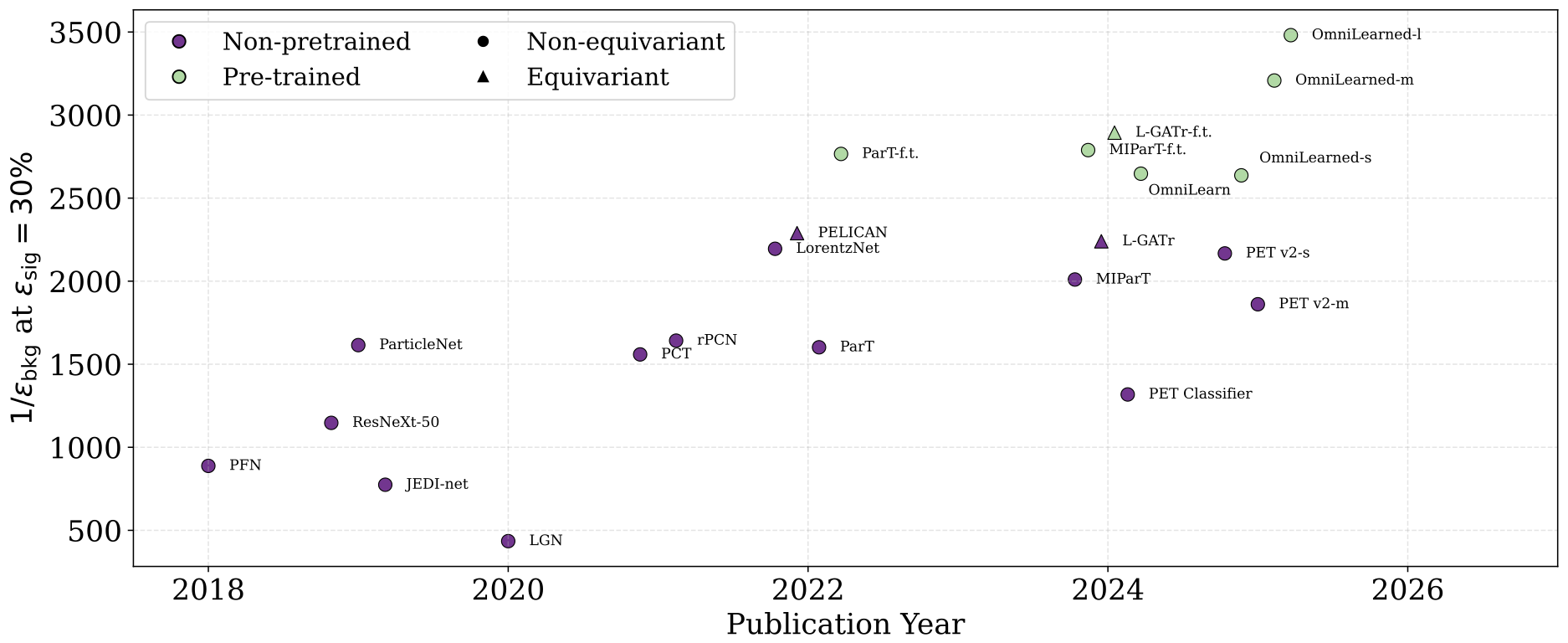}
    \caption{Background rejection on the community top tagging dataset from Ref.~\cite{Kasieczka:2019dbj} versus the year of publication for many different top tagging neural networks. Pretrained models achieve the highest background rejection because they exploit auxiliary training data instead of only the benchmark dataset. The equivariant L-GATr model is among the best performers of the non-pretrained networks. Reproduced from Ref.~\cite{Bhimji:2025isp} under the CC~BY~4.0 license.}
    \label{fig:br-vs-year}
\end{figure}

\Cref{fig:br-vs-year} shows the background rejection on the community top tagging benchmark for a variety of deep learning based models as function of their publication date\footnote{Note that background rejection performance metric on this dataset has become somewhat unreliable as the models have become powerful enough to nearly saturate performance. The performance differences between the top-performing models likely come down to how they sort a small subset of the most difficult to classify jets.}.
Pretrained models are consistently the best performers, since they leverage training data beyond what is included in the benchmark.
Equivariant models such as L-GATr exhibit the best performance among non-pretrained models.
This plot illustrates that equivariant priors can be useful, but are not as useful as larger scale training data.
This echoes the ``bitter lesson'' of AI: that methods which leverage larger scale always outperform methods which leverage human knowledge in the long run.
My expectation is that this pattern will continue and the primary direction of improvement in jet tagging will come from larger scale rather than more clever methods for adding domain knowledge to neural networks by hand.
Whether equivariant inductive biases will be useful for tasks other than classification will be discussed briefly in \Cref{ch:zjets}.

\subsection{Flavor Tagging at Scale}

A concrete example of the pursuit of larger scale within ATLAS and CMS comes from comparing recent state-of-the-art flavor tagging methods.
Both ATLAS and CMS now use transformer neural networks trained on low level ID track and calorimeter information.
These neural networks are trained with $\mathcal{O}(10^8)$ jets and have parameter counts in the low millions~\cite{ATLAS:2025dkv, Sarkar:2024vjz}.
These models are especially important because the dominant Higgs boson decay mode is $H \to b\bar{b}$, making $b$-jet identification central to many measurements and searches involving the Higgs, and essentially all searches for di-Higgs production.

\begin{figure}[tb]
    \centering
    \begin{subfigure}[b]{0.48\textwidth}
        \centering
        \includegraphics[width=\linewidth, alt={Bar or line plot showing the improvement in light-jet rejection of successive ATLAS flavor tagging algorithms relative to earlier baselines, including light-jet rejection measured in data, with roughly an order of magnitude improvement from BDT and early deep learning methods to the GN2 transformer.}]{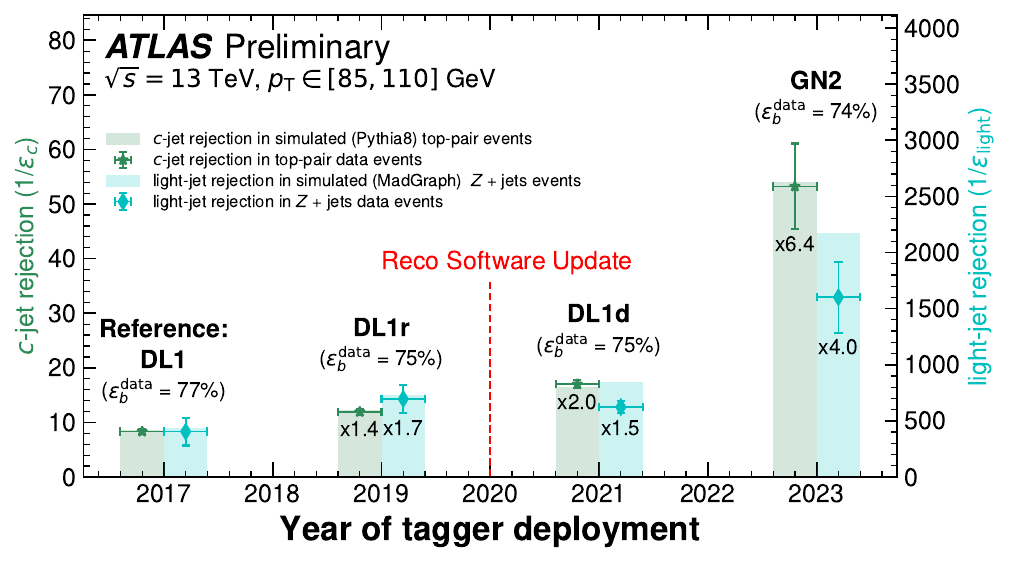}
        \caption{ATLAS flavor tagger performance evolution.}
        \label{fig:ftag-atlas}
    \end{subfigure}
    \hfill
    \begin{subfigure}[b]{0.48\textwidth}
        \centering
        \includegraphics[width=\linewidth, alt={Bar or line plot showing the improvement in light-jet rejection of successive CMS flavor tagging algorithms relative to a Run 1 BDT-based baseline, with roughly two orders of magnitude improvement for current transformer-based taggers.}]{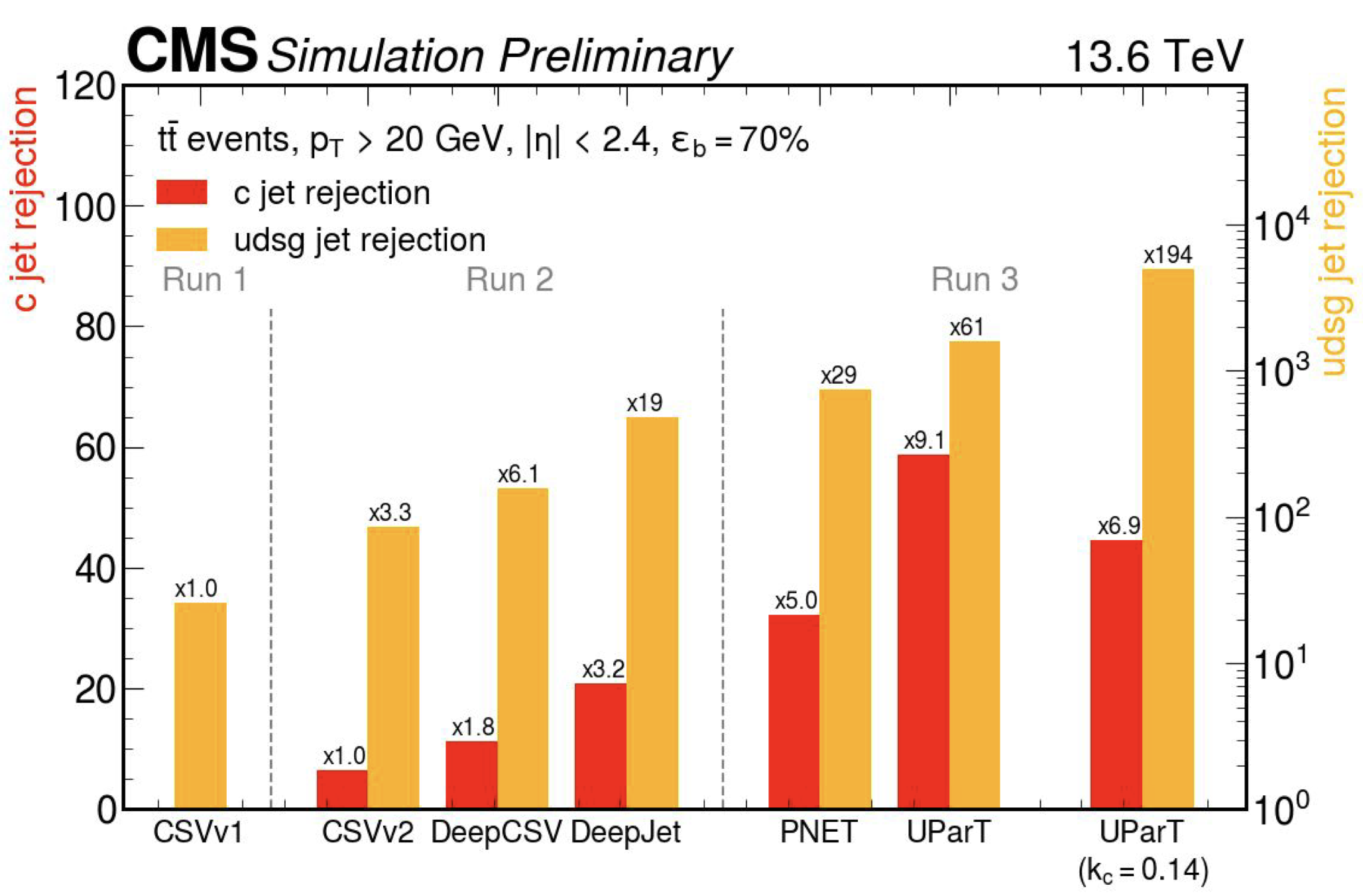}
        \caption{CMS flavor tagger performance evolution.}
        \label{fig:ftag-cms}
    \end{subfigure}
    \caption{Background rejection of ATLAS (left) and CMS (right) flavor tagging algorithms relative to earlier baselines. The CMS figure uses an early Run 1 flavor tagging algorithm as its reference and so shows larger multipliers for the current taggers. Both experiments have seen roughly an order of magnitude improvement in light jet rejection by transitioning from BDT-based and early deep learning methods to transformer networks. The ATLAS figure additionally shows the light jet rejection as measured in data, showing that performance gains have also produced larger differences in performance between data and simulation, consistent with the results of \Cref{sec:systematic-uq}. Reproduced from Refs.~\cite{ATLAS:2025dkv} and~\cite{Sarkar:2024vjz} under the CC~BY~4.0 license.}
    \label{fig:ftag-comparison}
\end{figure}

\Cref{fig:ftag-atlas,fig:ftag-cms} show the light jet rejection of the latest ATLAS and CMS flavor taggers at fixed signal efficiencies relative to earlier baselines.
In the ATLAS figure, transitioning from the DL1r architecture (an MLP acting on expert features plus a recurrent mechanism operating on displaced vertices) to the GN2 large-scale transformer trained on low-level inputs yields a roughly 5x improvement in raw light-jet rejection.
The CMS figure uses BDT-based discriminants trained in run 1 as a baseline, and shows roughly two orders of magnitude improvement in raw light-jet rejection.
The ATLAS figure also shows the light jet rejection as measured in data for each tagger.
The differences in performance between data and simulation and the size of the scale-factor uncertainties are larger for the transformer-based methods being used today than for their predecessors, in agreement with the results of \Cref{sec:systematic-uq}.
Larger, more expressive networks exploit more of the available information in simulation, including information that is not perfectly modeled, and the performance gap between simulation and data grows accordingly.

At the moment both experiments are focused solely on performance in this performance--bias tradeoff.
The reason is that a current focus of the LHC physics program is the search for di-Higgs production.
Di-Higgs production has a very small cross section, so searches for it are limited by statistics rather than systematic uncertainties such as scale factor uncertainties.
Maximizing performance is therefore the correct strategy to improve sensitivity to di-Higgs production.
Robustness considerations can be important to measurements of SM processes, but these use-cases do not set the priorities of the combined performance groups that train and calibrate the flavor taggers in ATLAS and CMS.

\subsection{Prospects for the High-Luminosity LHC}

Further scaling of jet classification neural networks is likely to continue, enabling even greater performance when analyzing the datasets produced by the HL-LHC.
Very recently, ATLAS released a note detailing GN3, a larger model than GN2 trained on additional information and a larger dataset~\cite{ATLAS:2026vyw}.
Recently dedicated studies of scaling laws for jet classification tasks have appeared~\cite{ATLAS:2026xxx, Vigl:2026ppx}, including a study of how pretraining data composition affects jet classification scaling laws that I published with collaborators~\cite{Uslu:2026ywh}.
The takeaway from these studies is that there are a lot of performance gains available from scaling jet flavor tagging models and training them on billions or even trillions of jets.
However it is not clear that these performance gains in simulation will translate to performance gains in data.
The left-hand panel of \Cref{fig:ftag-comparison} suggests that rejection in data does not scale as favorably as rejection in simulation.
At some point this gap might limit the usefulness of simply scaling supervised training on the cross entropy loss, and additional training objectives like self-supervised or data-driven approaches may become relevant.

These improvements are critical for the success of the HL-LHC.
The projected sensitivity for observing di-Higgs production at the HL-LHC has improved substantially over the past few years.
The 2019 CERN Yellow Report~\cite{Cepeda:2019klc} predicted that ATLAS and CMS would produce a \textit{combined} significance of roughly $4\sigma$ with the $3$ $\mathrm{ab}^{-1}$ of luminosity expected from runs 4 and 5.
In the most recent ATLAS and CMS projections~\cite{ATLAS:2025eii}, the same figure is greater than $7\sigma$.
Deep learning based $b$-tagging has been one of the stronger drivers of this improvement.
This is especially true in channels that depend heavily on b-tagging like $HH \to b\bar{b}b\bar{b}$.
The ATLAS GN2 tagger alone has improved projected HH sensitivity by up to 30\% in very $b$-tagging-sensitive channels relative to its MLP predecessor~\cite{ATLAS:2025dkv}.
It now seems likely that the unambiguous observation of di-Higgs production with the HL-LHC datasets will be the first large physics payoff from the adoption of deep learning methods.
I will return to these prospects in \Cref{sec:conclusion-dihiggs}.

%% file: tab_top_jet_selection.tex
\begin{table}[]
    \centering
    \begin{tabular}{l|l}
        \hline
        \hline
        Jet Requirements & Top quark jet requirements \\ \hline 
        Jet  $|\eta| < 2.0$ & $dR(\text{jet, truth jet}) < 0.75$ \\
        Jet  $p_{\text{T},\text{truth}} > 350$ GeV & $dR(\text{truth jet, top parton}) < 0.75$ \\
        Number of constituents $\geq 3$ & Ungroomed truth jet mass $> 140$ GeV \\
        Jet mass $> 40$ GeV & Number ghost associated $b$-hadrons $\geq 1$ \\
         & Truth jet $\sqrt{d_{23}} > \exp (3.3 - 6.98 \times 10^{-4} \times \text{truth jet} \: p_T)$ \\ \hline \hline
    \end{tabular}
    \caption{A summary of the requirements applied on all of the jets in the simulation samples to produce the training and testing sets. The additional top quark jet requirements constitute the truth labeling strategy, and are only applied to jets in the sample of simulated top quarks.}
    \label{tab:cuts}
\end{table}

%% file: tab_top_metrics.tex
\begin{table}[]
    \centering
    \caption{The performance of each top quark tagger is measured with several metrics evaluated on the testing set. AUC is the area under the receiving-operator-characteristic curve, ACC is the accuracy, and $\varepsilon^{-1}_{bkg}$ is the inverse background efficiency (or background rejection) evaluated at working points which yield a given signal efficiency ($\varepsilon_{sig}$) across the entire testing set. For all metrics, a higher value means better performance, and the table is sorted by increasing AUC. The uncertainty reported on the metrics is the quadrature sum of the uncertainty from the finite statistics of the testing set and the error from the random initialization of network weights and the stochastic nature of network training.}
    \begin{tabular}{c|c|c|c|c}
        \hline \hline
        Tagger & AUC & ACC & $\varepsilon^{-1}_{bkg}$ @ $\varepsilon_{sig} = 0.5$ & $\varepsilon^{-1}_{bkg}$ @ $\varepsilon_{sig} = 0.8$ \\ \hline
        ResNet 50 & $0.872 \pm 0.006$ & $0.787 \pm 0.006$ & $18.4 \pm 1.1$ & $4.63 \pm 0.2$ \\ 
        EFN & $0.894 \pm 0.001$ & $0.810 \pm 0.001$ & $23.8 \pm 0.5$ & $5.74 \pm 0.07$ \\  
        hlDNN & $0.9374 \pm 0.0001$ & $0.8628 \pm 0.0002$ & $47.2 \pm 0.4$ & $10.36 \pm 0.03$ \\ 
        DNN & $0.9447 \pm 0.0004$ & $0.8715 \pm 0.0008$ & $73.0 \pm 1.3$ & $12.5 \pm 0.1$ \\
        PFN & $0.9502 \pm 0.0004$ & $0.878 \pm 0.001$ & $92.7 \pm 1.8$ & $14.6 \pm 0.2$ \\ 
        ParticleNet & $0.9614 \pm 0.0005$ & $0.895 \pm 0.001$ & $155.8 \pm 3.8$ & $20.6 \pm 0.4$ \\ \hline \hline
    \end{tabular}
    \label{tab:performance_numbers}
\end{table}

%% file: chapter5.tex
\chapter{Unfolding}
\label{ch:unfolding}

This chapter is a detailed treatment of \textit{unfolding}, a problem in particle physics data analysis that has been a primary focus of my Ph.D. research.
It will introduce the problem and why it is important in LHC physics, then illustrate conventional solutions that have been in use for many years.
Then it will turn to AI/ML based unfolding methods, highlighting a few methods that I played a role in developing.
These methods enable very high dimensional cross section measurements with many downstream physics applications.
The results have been published in Refs.~\cite{Shmakov:2023kjj,Shmakov:2024gkd,Huetsch:2024quz,Petitjean:2025tgk}.
The Chapter closes with an overview of the \Omnifold method.
This method is applied to experimental data to produce such a high dimensional cross section measurement in \Cref{ch:zjets}.

\section{Introduction to Unfolding}
\label{sec:unfolding-intro}

The detectors used in particle physics experiments are not perfect.
Every measurement they make has finite resolution and acceptance.
For example the ATLAS calorimeters are divided into calorimeter cells whose size places a firm lower bound on the angular resolution, and ATLAS as a whole has limited acceptance in the forward region because detectors cannot be placed inside the beampipe.
These imperfect measurements introduce \textit{detector effects} which must be accounted for in order to compare cross section measurements to theoretical predictions\footnote{Note that it is possible to neglect these effects if the measurement is performed in sufficiently coarse bins or in regions of phase space where acceptance effects are negligible. In this case the response matrix (see \Cref{sec:binned-unfolding-methods}) is completely diagonal and the detector-level and truth-level histograms are identical. These cases are rare because measurements are typically designed to make full use of the detector response, and so use finer binning and more inclusive phase space selections at the expense of introducing significant detector effects.}.
There are two general approaches to accounting for detector effects.
The first is \textit{forward folding} in which theoretical predictions, specifically the output of an MC event generator (see \Cref{ch:qcd}) that specifies the final state four-momenta of all long-lived particles produced in a collision, are passed through a simulation of the detector response and compared to the raw experimental data.
In this case the data and theoretical predictions are said to exist at \textit{detector level}, since the detector distortions are present in both.
The second strategy is \textit{unfolding}.
Instead of passing the theoretical predictions through a detector simulation, statistical methods are used to construct an approximate inversion of the detector response, which is then applied to the data~\cite{Cowan:2002in}.
This removes detector effects from the measurement so that it can be directly compared to theoretical predictions without passing the predictions through the detector simulation.
This process of removing distortions is called \textit{deconvolution} in other fields, since one can think of detector effects as the convolution of a noise function with the spectrum of interest.
In the case that the unfolding adjusts only for detector distortions, the cross section measurement is said to exist at \textit{particle level}, since it makes a claim about the configuration of final state particles before interaction with the detector.
Some unfolding applications additionally account for the distortions produced by the parton shower and hadronization processes (see \Cref{ch:qcd}).
In this case the cross section measurement is said to exist at \textit{parton level}, since it makes a claim about the configuration of partons produced in the underlying hard scatter.
Both of these are unfolding and the methods discussed in this chapter apply to both.

In practice detector distortions are not the only effects that must be considered when unfolding.
A complete list is:
\begin{enumerate}[label={(\arabic*)}]
    \item Acceptance and efficiency: particles produced may not be measured, or migrate out of the fiducial volume targeted by the measurement between truth level and detector level.
    \item Fakes: particles measured may not be from real particles, or migrate in to the fiducial volume targeted by the measurement between truth level and detector level.
    \item Background processes: if one wants to measure the differential cross section of a particular process, then one may want to subtract the contributions from background processes.
    \item Combinatorics: if there are $n$ particles and one wants to measure the properties of a particular order (e.g. leading $p_T$), the detector effects can change the order.
    \item Detector distortions: the detector response introduces bias and resolution effects.
\end{enumerate}
In a well-structured measurement, the unfolding will primarily need to address effect (5).
For the measurement presented in \Cref{ch:zjets}, item (3) is dealt with but is a very sub-leading effect, and items (1,2,4) are irrelevant.
This is not generally the case for all particle physics measurements, but for the purposes of this thesis those items will be ignored.

Forward folding and unfolding are both actively used in cross section measurements and have distinct advantages and disadvantages.
Forward folding is strictly more precise than unfolding because unfolding methods will always incur additional systematic uncertainties.
These may be very sub-leading in which case the precision of both approaches is equivalent, but the unfolding approach is almost never more accurate than forward folding\footnote{Unfolded measurements can be more precise than forward folding if unfolding removes the need for some systematic uncertainties. This is sometimes the case for theoretical modeling uncertainties, where the difference between two MC generators is understood as an uncertainty for detector level spectra but only an arbitrary choice of prior for unfolded spectra. This ambiguity stems from the fact that modeling uncertainties are not true systematic uncertainties with a nuisance parameter that can be varied. Whether or not they should be considered in a measurement is often a somewhat philosophical question.}.
Despite this limitation, unfolding is often worth the added uncertainty for a few reasons which are worth spelling out in detail:

\begin{itemize}
    \item Unfolded measurements can be compared directly to theoretical predictions furnished by MC event generators without the need to pass the simulated events through the detector simulation. This is an enormous practical benefit. The general hope with any cross section measurement is that it will be used to improve our knowledge of particle physics, for example by constraining a free parameter of the Standard Model or tuning a hadronization model used to provide theoretical predictions. All of these downstream tasks are greatly simplified when detector simulation is not needed. For LHC datasets, the detector simulation is mostly proprietary, computationally expensive, and requires substantial expertise to run, making the forward folding approach inconvenient\footnote{Note that ``inconvenient'' can mean anything from hard to do within the personnel and funding constraints of a research group, to computationally infeasible with existing computational resources. Some downstream applications, for example fitting many Wilson coefficients of an effective field theory to measured spectra, require large amounts of simulation that may be impossible to produce at detector level.}. For legacy datasets, such as those produced by experiments at LEP or SLC, the detector simulation is no longer available, making the forward folding approach impossible. Unfolding can then be understood as a way of preserving cross section measurements for future unforeseen use cases, which might be understood only many years after the initial data taking of an experiment is complete.
    \item Unfolded measurements can be compared to non-fully-inclusive theory predictions. In particular they can be compared to resummed calculations of jet substructure observables, which are defined at particle level but are represented as smooth functions rather than lists of final state particle four-momenta. This is an underappreciated motivation for unfolding. It is \textbf{required} to compare data to the rigorously understood theoretical predictions provided by state-of-the-art QCD calculations.
    \item Finally unfolded measurements are much easier to combine with other measurements, perhaps from different experiments, to exploit additional statistical power. Combining forward folded measurements is possible, but requires access to the detector simulation for proper correlation of uncertainties. This becomes especially difficult if results from different experiments are combined. Combining unfolded spectra is much easier. See Ref.~\cite{Heimel:2024drk} for an excellent example.
\end{itemize}

\begin{figure}[tb]
    \centering
    \includegraphics[width=0.66\linewidth, alt={Overlaid histograms of citation counts received by ATLAS and CMS physics results, comparing searches, measurements that forward fold, and measurements that unfold.}]{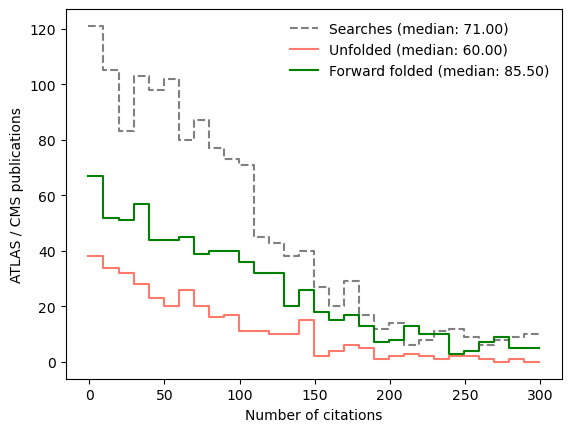}
    \caption{Histograms of all ATLAS and CMS physics results in the number of citations received. The tail of the distribution at very high citation counts is not shown. Histograms are shown for searches, measurements which forward fold, and measurements which unfold. Data scraped from the InspireHEP archive, and sorted based on keywords in abstracts and whether they cite key papers. Note these data are very approximate given unfolding is not always labeled explicitly.}
    \label{fig:paper-citations}
\end{figure}

Given that these motivations may or may not apply depending on the goals of a given measurement, the decision to use forward folding or unfolding is highly context specific.
It is worth mentioning a few contexts from LHC physics before moving on.
Searches for beyond-Standard-Model (BSM) particles almost never unfold.
Such searches look for coherent shapes like resonances in low-statistics regions of phase space, and unfolding would substantially reduce sensitivity.
It is true that a search using unfolded spectra would be vastly easier to re-interpret, but there is little practical motivation for doing this since well-established procedures exist within the experimental collaborations for re-interpreting searches via forward folding.
Among measurements of Standard Model processes, unfolding is used in roughly one third of published ATLAS and CMS measurements (excluding for example detector performance papers and searches), as illustrated in \Cref{fig:paper-citations}.
Extractions of parameters such as particle masses or Wilson coefficients typically do not unfold unless the result is intended to be used in combinations, while differential cross section measurements typically do.
This picture is complicated by the fact that some unfolded measurements are not described as such explicitly.
The Simplified Template Cross Section (STXS) framework used in ATLAS and CMS Higgs measurements~\cite{Berger:2019wnu}, for example, is an instance of profile likelihood unfolding (see \Cref{sec:binned-unfolding-methods}), but is rarely labeled as such in publications.
In summary, unfolding occupies a niche but important role in the LHC physics program, and is essential for data preservation and comparisons to resummed theory calculations.

\section{Binned Unfolding Methods}
\label{sec:binned-unfolding-methods}

For the past century of physics research, cross section measurements have been reported as bin counts in histograms.
Given the goal is to report a histogram, the unfolding task reduces to inferring a truth-level histogram given two inputs: the detector-level histogram produced by binning the data, and a \textit{response matrix} that encodes how events migrate between the defined bins at truth-level and detector-level.
This response matrix is constructed by binning the truth-level and detector-level configurations of a sample of simulated events provided by an event generator.

The unfolding problem can be formalized as a system of linear equations
\begin{align}
\label{eq:foldingequation}
\left(\textbf{R}\cdot (\textbf{t}\odot \textbf{c})\right)\odot 1/\textbf{f}+\textbf{b}=\textbf{d}\,,
\end{align}
where bold letters denote vectors or matrices, $\cdot$ is the matrix product, $\odot$ is the component-wise product, and division is defined component-wise.\footnote{An equivalent way to write \Cref{eq:foldingequation} is $\left(\sum_{i}R_{ij}\,t_i\,c_i\right)/f_j+b_j=d_j$, where $i$ and $j$ are indices for the truth-level and detector-level bins, respectively.}
Here $\textbf{t}$ is the particle-level distribution, $\textbf{d}$ is the detector-level distribution, $\textbf{b}$ is the background contribution at detector level, and $\textbf{R}$ is the response matrix.
The correction factors $\textbf{c}$ account for events whose particle-level configuration passes the particle-level selection but whose detector-level configuration fails the detector-level selection (efficiency and acceptance losses corresponding to item (1) in \Cref{sec:unfolding-intro}).
The fake factors $\textbf{f}$ account for the inverse: events whose detector-level configuration passes the detector-level selection but whose particle-level configuration does not, corresponding to item (2).

The maximum likelihood solution to \Cref{eq:foldingequation} is
\begin{align}
\label{eq:unfolding}
\textbf{t}_\text{measured} = \textbf{R}^{-1}\left( (\textbf{d}-\textbf{b})\odot \textbf{f}\right)\odot 1/\textbf{c}.
\end{align}
In practice this maximum likelihood estimator is often not useful, because the response matrix $\textbf{R}$ is often \textit{ill-conditioned}, meaning that its smallest singular values are close to zero.
The maximum likelihood solution amplifies statistical fluctuations in the detector-level histogram by factors proportional to the condition number of the response matrix.
If the matrix is ill-conditioned and the condition number is large, the maximum likelihood estimator for the truth histogram has large variance.
This means that the naive solution to \Cref{eq:foldingequation} is highly unstable, in that small changes to the detector level histogram produce large and unphysical fluctuations in the predicted truth level histogram.
The most commonly used unfolding methods reduce this large variance by injecting a small bias into the estimator, a procedure known as \textit{regularization} with conceptual ties to the bias--variance tradeoff discussed in \Cref{sec:ml-overview}.
In HEP applications this bias is typically estimated and propagated as a systematic uncertainty on the final result.
Reviews of binned unfolding methods in the context of high-energy physics can be found in Refs.~\cite{Cowan:2002in, Blobel:2011fih}.
A common feature of all methods is that they do not scale beyond unfolding a few simultaneous dimensions.
This is due to the curse of dimensionality, where the number of bins in a histogram grows exponentially with the number of dimensions making populating the bins of a high-dimensional histogram intractable.
The inability of binned unfolding methods to scale to high dimensions is one motivation for unbinned unfolding, which will be discussed in the next Section.

\subsection{Profile Likelihood Unfolding}
\label{sec:plu}

Profile likelihood unfolding and its variants replace the naive matrix inversion solution with a likelihood model.
First write the expected detector-level bin counts as a function of the truth-level bin counts,
\begin{equation}
    \lambda_j = \sum_i^{N_t} R_{ij} t_i ,
    \label{eq:unfolding_expected_yield}
\end{equation}
with $R_{ij}$ the response matrix element for detector-level bin $j$ given truth-level bin $i$ and $N_t$ the number of truth-level bins.
Then assuming Poisson distributed bin counts, the log-likelihood is
\begin{equation}
    \mathcal{L}(\textbf{t} | \textbf{d}) = \log \prod_{j=1}^{N_d} \frac{e^{-\lambda_j}\lambda_j^{d_j}}{d_j!} = \sum_{j=1}^{N_d} \left( d_j \log \lambda_j - \lambda_j - \log d_j! \right) .
    \label{eq:unfolding_poisson_likelihood}
\end{equation}
where $N_d$ is the number of detector-level bins.
This replaces the linear system in \Cref{eq:unfolding} with a non-linear system that can be solved numerically.
Note that although the profile likelihood approach does not require explicit matrix inversion, it does not circumvent the ill-conditioning problem.
If $\mathbf{R}$ is ill-conditioned, there exist directions in $\mathbf{t}$ along which the likelihood is essentially flat, and the same numerical instabilities can result.
This is why additional regularization of the likelihood is often required.
See for example Ref.~\cite{Tikhonov:1963}.

The main advantage of profile likelihood methods is that nuisance parameters describing systematic uncertainties can be constrained simultaneously with the bin counts of the truth level histogram.
This avoids the need to propagate systematic uncertainties by re-running the unfolding as is done in measurements performed with the IBU method in the next section.
The normalization of irreducible backgrounds can also be treated as a nuisance parameter and included in the unfolding.
The main drawback is that the regularization, though principled, can be delicate to tune in practice.
A common fix to these issues is to simply use coarser bins, so the response matrix is better conditioned but the measurement loses resolution.

Profile likelihood unfolding is becoming an increasingly popular tool in LHC physics.
Part of the reason for this is that profile likelihood fits are already a common tool for parameter estimation in HEP, so using the same method for unfolding is intuitive for the community.
See for example Ref.~\cite{CMS:2024mke} or any Higgs boson measurements within the STXS framework, which is a prominent example of profile likelihood unfolding as noted in \Cref{sec:unfolding-intro}.
See Ref.~\cite{Collaboration:2944725} for an example of relatively high-dimensional profile likelihood unfolding with an explicit regularization term added to the likelihood.

\subsection{Iterative Bayesian Unfolding}
\label{sec:ibu}

The standard binned unfolding method in HEP is \textit{Iterative Bayesian Unfolding} (IBU)~\cite{DAgostini:1994fjx}, also known as Lucy-Richardson deconvolution in other fields~\cite{Lucy:1974yx, Richardson:1972}.
IBU is an example of an expectation-maximization (EM) algorithm, where a likelihood over the bin counts for both the detector-level and truth-level histograms is approximated with an expectation value (expectation step) and then maximized (maximization step).
Note that the expectation step is required because we do not know the truth-level bin counts by construction.
Instead we have a prior furnished by MC simulation that is iteratively updated, which is why the method is referred to as Bayesian even though it is fundamentally attempting to maximize a likelihood.
For simplicity the equations below set $\textbf{b}=\textbf{0}$ and $\textbf{f}=\textbf{c}=\textbf{1}$, so \Cref{eq:foldingequation} reduces to $\textbf{R}\cdot\textbf{t} = \textbf{d}$.

Here the solution to one iteration of the expectation-maximization process will just be quoted without derivation.
The IBU estimate at iteration $n$ is
\begin{align}
t_i^{(n)}&=\sum_j \Pr(t_i|d_j)\, d_j \label{eq:ibu1}\\
&=\sum_j \frac{\Pr(d_j|t_i)\,\Pr^{(n)}(t_i)}{\sum_{i'} \Pr(d_j|t_{i'})\,\Pr^{(n)}(t_{i'})}\, d_j \label{eq:ibu2}\\
&=\sum_j \frac{R_{ji}\,\Pr^{(n)}(t_i)}{\sum_{i'} R_{ji'}\,\Pr^{(n)}(t_{i'})}\, d_j, \label{eq:ibu3}
\end{align}
where $\Pr^{(n)}(t_i) \propto t_i^{(n-1)}$ is the prior on the truth-level bin $i$ at iteration $n$.
This is updated at each step using the prior from the previous iteration as a starting point.
The prior for the first iteration is taken from simulation, $\Pr^{(0)}(t_i) \propto t_i^{(0)}$.
It can be shown that as $n \to \infty$ the truth-level histogram estimates produced by IBU converge to the maximum likelihood estimator~\cite{Shepp:1982}.

The EM algorithm structure can be read off from these equations.
First, in \Cref{eq:ibu2} the current prior $\Pr^{(n)}(t_i)$ is held fixed and the posterior probability $\Pr(t_i | d_j)$ that a detector-level event in bin $j$ originated from truth-level bin $i$ is calculated using Bayes' theorem.
The response matrix elements $R_{ji}$ are interpreted as the likelihood $\Pr(d_j|t_i)$.
This posterior probability is then multiplied by the observed data $d_j$, producing an estimate of the joint probability density $\Pr(d_j, t_i)$.
This is the expectation step, since the method is calculating the expectation value of latent observables (how the events in bin $j$ of the detector-level histogram split across the truth-level bin $i$, or the joint density) given the current parameters $t_i^{(0)}$ and observations $d_j$.
The maximization step is then simply the sum over $j$, which aggregates the contribution to the truth-level bin $i$ over all detector-level bins $j$.
The estimate produced by the M-step then serves as a new prior and the process repeats.
The number of iterations controls how far the estimate departs from the prior taken from simulation.
Few iterations keep the result close to the prior corresponding to high regularization (high bias, low variance), while many iterations approach the maximum likelihood solution and zero regularization (low bias, high variance).
With infinite iterations, the procedure converges to the maximum likelihood estimate and the solution is prior independent, but not useful due to the large variance.

IBU is commonly used in HEP because it is flexible and easily interpretable.
Best practices for choosing when to stop the iterations have been documented and implemented in HEP specific software packages~\cite{Brenner:2019lmf}.
IBU is often the fastest and easiest way to get from detector-level to truth-level spectra.

\section{Unbinned and High-Dimensional Unfolding}
\label{sec:ai-unfolding-motivation}

The binned methods of \Cref{sec:binned-unfolding-methods} have been used in particle physics for decades but have limitations that bottleneck the flexibility of cross section measurements.
This section motivates adopting unbinned and high-dimensional unfolding techniques and surveys existing experimental applications.

\subsection{Motivation}
\label{sec:unbinned-motivation}

There are three key challenges of the binned unfolding methods, all of which can be addressed by unbinned and high-dimensional unfolding.
First, with binned methods the binning must be fixed before the measurement is complete and cannot be changed without rerunning the analysis code which requires access to the data.
This is limiting because cross section measurements can be used for different purposes, and a binning optimized for one application will often be inappropriate for another.
This is particularly relevant if one wishes to compute statistical moments of measured distributions, since binning introduces artifacts that will bias the result.
Unbinned unfolding avoids all of this by allowing the binning to be adjusted post-hoc or avoided entirely in the case of computing statistical moments.
This trick will be used extensively in \Cref{ch:zjets}.
Note that unbinned but low-dimensional unfolding methods that do not used deep learning have been proposed~\cite{Lindemann:1995ut, Aslan:2003vu, Glazov:2017vni}, as well as a deep learning approach for unfolding statistical moments directly~\cite{Desai:2024yft}.

The second challenge is that binned unfolding methods do not scale to high dimensions.
This means that all cross section measurements performed with them are differential in typically no more than two observables.
As discussed in \Cref{sec:binned-unfolding-methods}, extending binned unfolding to more dimensions is not possible due to the curse of dimensionality.
High-dimensional unfolding makes measurements of very highly differential cross sections possible.
In practice this requires the unfolding be unbinned, and rely on machine learning methods since they are a natural tool for processing high-dimensional data.

The third challenge is that low-dimensional unfoldings do not take into account all possible auxiliary features that control the detector response.
This means that there can be \textit{hidden variables}, which the unfolding does not constrain and must marginalize over, which can limit the accuracy of the unfolding and increase unfolding uncertainties.
This is a subtle point, but the takeaway is that high-dimensional unfolding has the capacity to be more accurate than lower dimensional unfolding that can't utilize all available information.

A simple application of high-dimensional unfolding is to extend the length of the list of observables of interest to be anywhere from a handful to several tens.
This would produce a fixed but high dimensional cross section measurement that can be projected to any input observable to produce a measurement.
Additionally the correlations between observables would be preserved, so projecting to functions of the input observables is also supported.
Measurements of many observables can then be accomplished with a single analysis.
However the most ambitious application of high-dimensional unfolding is to extend the measurement to the \textit{full phase space}, unfolding the kinematics of every particle measured by the detector.
This approach is maximally information preserving, in that the primary measurements actually made by the detector are corrected for detector effects rather than individual summary observables.
Note that because the number of particles produced in collisions is variable, this application defines a cross section measurement on a variable-dimensional space.
Such a measurement implicitly contains every observable computable from the final state, making the potential for downstream applications immense.
Future observables of interest can be quickly measured and future theoretical calculations can be quickly compared to data, without the need to conduct a new analysis, which can either be time and resource intensive or impossible if the detector simulation is no longer available.
Realizing this greatly increased flexibility and physics potential requires two major technical advances: deep learning methods powerful enough to unfold variable-length sets of particle four-momenta, and systematic uncertainties defined at the level of individual particle measurements.
The deep learning methods described below have already been shown to be capable of unfolding in high and variable dimensions.
The requirement for systematic uncertainties on individual particles is already satisfied for charged particles but difficult for neutral particles, which is why the measurement in \Cref{ch:zjets} uses charged particles only.
This also links directly to the bottom-up approach to systematic uncertainties in jet tagging discussed in \Cref{sec:constituent-top-tagging}.

\subsection{Experimental Results}
\label{sec:unbinned-experimental-results}

\input{tab_unbinned_unfolding_measurements.tex}

A growing number of experimental measurements, summarized in \Cref{tab:unbinned-unfolding-measurements}, have been performed with deep learning based unfolding methods.
All of them use the \Omnifold algorithm, which is described in detail in \Cref{sec:omnifold}.
The measurements span a range of collision systems, including $pp$, $ep$, $e^+e^-$, and neutrino scattering, illustrating that these methods are general purpose and not only for use in LHC experiments.
The world record for the highest-dimensional cross section measurement is held by a preliminary H1 result~\cite{H1:prelim}, which targets the full phase space of high-$Q^2$ DIS events that reaches a maximum of a few hundred dimensions.
This preliminary result was a proof of concept for full-phase-space measurements.
The rest of the measurements in \Cref{tab:unbinned-unfolding-measurements} target fixed-length cross sections ranging from 4 to 24 simultaneous dimensions.
Only a few of these measurements are from LHC collaborations, and most have been carried out in the past two to three years.
Remaining barriers to broader adoption of these methods by the community will be discussed in \Cref{ch:conclusion}.

\section{Review of Existing Methods}
\label{sec:unfolding-methods}

For unbinned methods it is useful to recast the unfolding problem in continuous notation.
The forward folding equation of \Cref{eq:foldingequation} becomes
\begin{align}
\label{eq:folding-continuous}
p_{\mathrm{data}}(x) = \int p(x \mid z)\, p_{\mathrm{unfold}}(z)\, dz,
\end{align}
where $x$ denotes the detector-level event representation, $z$ the truth-level event representation, $p(x \mid z)$ is the conditional probability which can be sampled from using the detector simulation, and $p_{\mathrm{data}}(x)$ the measured data distribution.
In the continuous setting, the goal of unfolding is to infer the distribution $p_{\mathrm{unfold}}(z)$ that when convolved with the detector simulation produces the data distribution.
The analog for the ill-conditioned response matrix is that there are many distributions which satisfy this condition, including many with large and unphysical fluctuations, and convolution operation is information destroying in that there is no way to guarantee that a given solution is the true data generating distribution $p_{\mathrm{true}}(z)$.

The binned IBU estimates the true distribution through iterative expectation and maximization steps.
The continuous analogs of \Cref{eq:ibu2,eq:ibu3} are
\begin{align}
\label{eq:ibu-continuous-estep}
p^{(n+1)}(z, \, x) = p^{(n)}(z \mid x)\, p_{\mathrm{data}}(x) = \frac{p(x \mid z)\, p^{(n)}(z)}{\int p(x \mid z')\, p^{(n)}(z')\, dz'}\, p_{\mathrm{data}}(x), \quad \text{(E-step)}
\end{align}
\begin{align}
\label{eq:ibu-continuous-mstep}
p^{(n+1)}(z) = \int p^{(n+1)}(z, \, x) \, dx. \quad \text{(M-step)}
\end{align}
The E-step involves using Bayes' theorem to compute the posterior $p^{(n)}(z \mid x)$ using the current prior $p^{(n)}(z)$ and the detector response (likelihood) $p(x \mid z)$.
This posterior is then used to calculate an estimate of the joint density $p^{(n+1)}(z, \, x)$.
The M-step updates the truth-level distribution by marginalizing over $x$ in the estimate of the joint density.
As in the binned case, the prior is initialized using the (now unbinned) output of an event generator and the regularization is provided by truncating the procedure at finite iterations.

The ML-based unfolding methods discussed either solve the integral equation in \Cref{eq:folding-continuous} directly, or use the EM procedure described above.
The first class of methods are in some sense unbinned generalizations of profile likelihood unfolding, and the second class are unbinned generalizations of IBU.
This is true in a precise sense for \Omnifold as will be shown in \Cref{sec:omnifold}.
This Section should illustrate that there are many unbinned and high dimensional unfolding methods available, all of which work~\cite{Huetsch:2024quz}.
What is currently needed is much more application to experimental data.

\subsection{Direct Methods}
\label{sec:direct-unfolding-methods}

\Cref{eq:folding-continuous} can be solved directly either with generative models that model the relevant distributions or with likelihood ratio estimation via classifiers.
The generative approaches model the unfolded distribution as $q_\phi(z)$ and optimize $\phi$ so that samples from $q_\phi(z)$ match the data.
This requires a differentiable surrogate for the detector response $p(x \mid z)$, typically another generative model trained to approximate $q_\theta(x \mid z)$.
Since neural networks are differentiable by construction, gradients can flow through $q_\theta(x \mid z)$ allowing for the direct optimization of $q_\phi(z)$ using a log-likelihood loss as in \Cref{sec:generative}.
The models $q_\theta$ and $q_\phi$ can be optimized independently or jointly end-to-end~\cite{Vandegar:2020yvw, Butter:2025via}.
These methods are attractive for their direct approach and the ability to include nuisance parameters in the unfolding problem as with profile likelihood unfolding.
However the optimization can be difficult, especially in high- and variable-dimensional settings where point-cloud generative models must be trained.

The classifier-based approach, which has only been recently explored, replaces the generative models with classifiers.
Dividing both sides of \Cref{eq:folding-continuous} by a reference density provided by an alternative simulation $p_{\mathrm{sim}}(x)$ yields a ratio equation, where the ratios can be learned with classifiers via the likelihood ratio trick discussed in \Cref{ch:ml}.
A novel loss function enforces that the learned ratios solve the integral equation.
This method is purely classifier-based and solves the integral equation directly without EM iteration~\cite{Ore:2026qgp}.
This new method carries substantial promise because it does not require training generative models and the lack of iterations makes the method substantially less computationally expensive than \Omnifold, which is the only other purely-classifier unfolding method.
See in addition Ref.~\cite{Qureshi:2026oiv} for a recent method that derives a similar reweighting using an adversarial loss rather than classification.

\subsection{Generative Unfolding Methods}
\label{sec:generative-unfolding-methods}

A large class of ML unfolding methods performs EM, as in IBU, but use a generative model to learn the posterior $p(z \mid x)$ in the left hand side of \Cref{eq:ibu-continuous-estep}.
This replaces the binned response matrix in IBU.
Several different generative architectures have been applied for this purpose, including GANs~\cite{Datta:2018mwd,Bellagente:2019uyp}, VAEs~\cite{Howard:2021pos}, normalizing flows~\cite{Bellagente:2020piv,Diefenbacher:2023wec, Butter:2024vbx}, and diffusion and flow matching models~\cite{Butter:2025mek}.
These methods use various different generative models, but the underlying principle of using a generative model to solve the E-step is common to all of them\footnote{Many of these papers do not demonstrate using the EM procedure explicitly, but it is a known solution that can be assumed once the more difficult task of modeling the posterior is solved.}.
The diffusion and flow matching based methods I have contributed to are discussed in \Cref{sec:diffusion-unfolding}.

Two different methods for performing the M-step with the modeled posterior have been proposed.
Both start by using the measured data to condition generation, obtaining samples from the learned posterior under the data distribution.
The first approach then reweights an alternative MC simulation sample to match the posterior samples by training a classifier~\cite{Backes:2022sph}.
This is very similar to the second step of \Omnifold discussed below.
The second trains another generative model to reproduce samples from the learned posterior under the data distribution.
This defines a truth-level distribution $p^{(n+1)}(z)$ that is used as the prior in the next EM iteration.
This approach requires passing samples from the updated truth-level distribution through the detector simulation, possibly via a surrogate model~\cite{Butter:2025mek}.

These methods are quite mature, but as of this writing none have been applied to experimental data.
This is partially because they require additional machine learning complexity (the training of generative models) but reduce to the same statistical procedure as the classifier-based \Omnifold.
These methods do have lighter assumptions than \Omnifold on the support of the MC simulations used in the training, but it is unclear whether these assumptions are ever actually limiting for \Omnifold and how much better the generative-model-based approaches would perform if such a case were found.
This is doubly true because the same MC simulation support assumptions enter if the classifier-based approach is used in the M-step.
The purely generative-model-based approach of Ref.~\cite{Butter:2025mek} is explicitly free of them, but still implicitly carries assumptions given generative models have difficulty generalizing out of distribution.
For another alternative, see Ref.~\cite{Acosta:2025shf} which forgoes the EM procedure entirely, and treats unfolding as a Bayesian problem with a fixed prior.
Despite these issues applications of methods that replace parts of the EM procedure with generative models should be pursued, since the pros and cons of different methods often only appear in realistic circumstances.

\subsection{Classifier-Based EM}
\label{sec:omnifold-preview}

The final class of methods uses the EM procedure but uses classifiers to estimate density ratios rather than generative models to estimate densities.
\Omnifold~\cite{Andreassen:2019cjw} and its variants~\cite{Chan:2023tbf, Zhu:2024drd} are the prime example of these techniques.
Given \Omnifold is central to the $Z$+jets measurement detailed in \Cref{ch:zjets}, a full discussion is provided in \Cref{sec:omnifold}.

\section{Unfolding with Diffusion and Flow Matching}
\label{sec:diffusion-unfolding}

\subsection{Variational Latent Diffusion Models}
\label{sec:vld-unfolding}

Ref.~\cite{Shmakov:2023kjj}, the second major research contribution of this thesis, was the first paper to apply diffusion models to unfolding.
It considers only the E-step of the EM algorithm, and uses a diffusion model framework to model the posterior on the left-hand-side of \Cref{eq:ibu-continuous-estep}.
Before this study, the normalizing-flow-based approach of Ref.~\cite{Bellagente:2020piv} was the state-of-the-art result for this task.
Rather than use a standard DDPM as discussed in \Cref{sec:ddpm} to model the posterior, we chose to use a novel framework termed \textit{variational latent diffusion} (VLD), which to our knowledge had never appeared in the generative modeling literature before this paper.

The VLD model stems from latent diffusion models (LDMs), which are a popular class of models that are roughly speaking the current state of the art for image generation tasks.
Unlike the standard DDPM, the latent diffusion model proposed in Ref.~\cite{rombach2022high} conducts the diffusion process in the latent space of a pre-trained VAE.
This reduced the computational cost of sampling, give the VAE latent space was lower dimensional than the data space, while maintaining or improving sample quality.
Subsequent work demonstrated the LDMs can be used to condition image generation on different data modalities, for example text.
This is done using a contrastive learning objective like CLIP~\cite{radford2021learning} which embeds samples from two different data modalities in the same representation.
These multi-modal embeddings were ultimately not used in modeling the posterior for the unfolding task.
However they were an early motivation to explore latent diffusion, because the posterior must condition generation of one modality of data (the truth-level event configuration) on another modality of data (the detector-level event configuration).

However the LDMs in Ref.~\cite{rombach2022high} train the VAE and diffusion models independently.
This approach is simple and easy, but it intuitively lowers the capacity of the end-to-end model given the VAE latent space is not optimized to be a useful representation in which to train a diffusion model.
A possible improvement is to optimize all aspects of the end-to-end model (the VAE and the diffusion model) simultaneously.
This requires constructing a single unified training objective for both models, which is the topic of the next section.

\subsubsection{The VLD Loss Function}

\begin{figure}[t]
    \centering
    \includegraphics[width=0.88\linewidth, alt={Block diagram of the variational latent diffusion model showing the conditional detector encoder, parton VAE encoder, diffusion denoising network, and parton VAE decoder connected in a unified end-to-end architecture conditioned on detector-level information}]{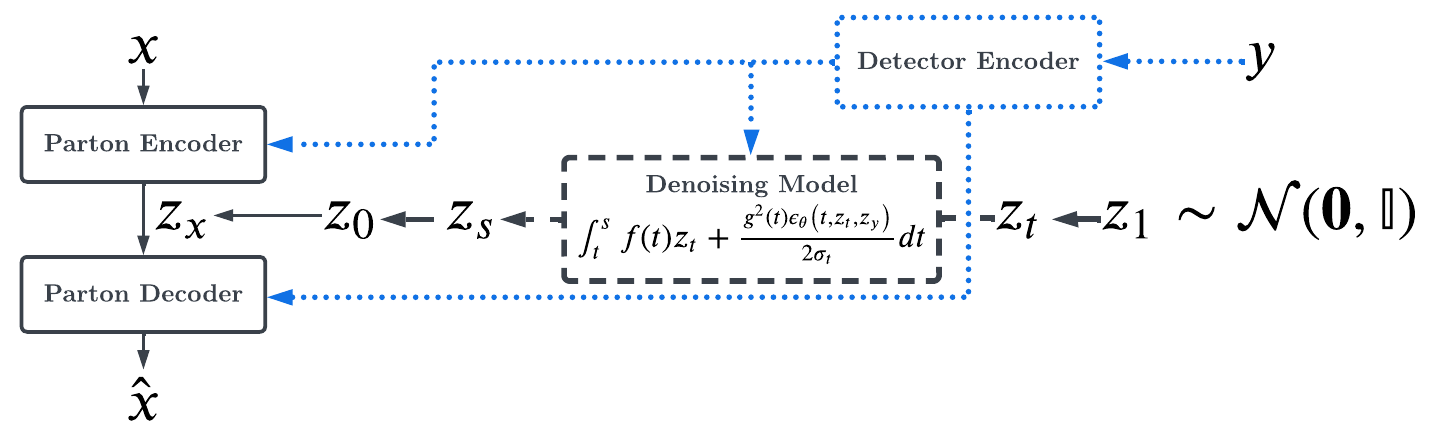}
    \caption{Block diagram of the variational latent diffusion (VLD) model. The detector encoder (top right) embeds the detector-level event $y$ into a fixed-size latent vector that conditions the rest of the network. During training, the VAE encoder maps the truth-level event $x$ to a latent sample $z_0$. The forward diffusion process adds noise to this sample to produce $z_t \sim \mathcal{N}(0, I)$. The denoising network predicts the added noise conditioned on $z_t$, the current time $t$, and the detector embedding. Inference is performed by sampling a new latent variable $z_t$ and using it to generate $z_0$ via the trained diffusion model. The VAE decoder then maps $z_0$ to the generated truth-level event configuration. Figure reproduced from Ref.~\cite{Shmakov:2023kjj}.}
    \label{fig:vld-diagram}
\end{figure}

A schematic of the VLD model is provided in \Cref{fig:vld-diagram}.
A few notational differences between this section and the rest of the thesis should be stated explicitly.
The variables $x$ and $y$ denote the truth-level and detector-level event configurations respectively.
Like all diffusion models, the VLD framework introduces a set of latent variables $z_t$, parametrized by a continuous diffusion time $t \in [0, 1]$.
$t = 0$ corresponds to zero added noise, and $z_0$ is a draw from the VAE encoder posterior $q(z_0 \mid x, y)$.
As $t$ increases, more Gaussian noise is added to the sample so that $z_1$ becomes pure noise and can be considered to be a sample from $p(z_1) = \mathcal{N}(0, I)$.
If a trained noise-prediction model is in hand, samples from the posterior we seek to model can be obtained by sampling a new configuration of the latent variable $z_1$, then using either an SDE or ODE solver as outlined in \Cref{sec:generative-unfolding-methods} to reverse the diffusion process and yield a new sample from the VAE latent space.
Finally the sample is passed through the VAE decoder which models $p(x \mid z_0, y)$.
The latent variable $z_0$ can be marginalized over by repeating this process many times, producing a set of samples from the desired posterior.

Producing this model requires training three independent neural networks: the VAE encoder and decoder, and the noise-prediction model.
The main contribution of Ref.~\cite{Shmakov:2023kjj} to the generative modeling literature is to propose that this can all be done simultaneously via the loss function:
\begin{align}
    \mathcal{L}_{\text{VLD}}
    &= D_\text{KL}\!\left(q(z_1 \mid x, y) \;\|\; p(z_1)\right)
     + \mathbb{E}_{q(z_0 \mid x, y)}\!\left[-\log p(x \mid z_0, y)\right] \notag \\
    &\quad + D_\text{KL}\!\left(q(z_0 \mid x, y) \;\|\; p(z_0 \mid z_1)\right)
     + \mathbb{E}_{\substack{\varepsilon \sim \mathcal{N}(0, I) \\ t \sim \mathcal{U}(0,1)}}
       \!\left[\gamma_\phi'(t)\, \norm{\varepsilon - \hat{\varepsilon}_\theta(z_t, t, y)}^2\right].
    \label{eq:loss_lvd}
\end{align}
A term-by-term interpretation of this loss is as follows.
The first term is the prior loss, and serves the same purpose as the second term in \Cref{eqn:ddpm_elbo}.
It ensures that the end-point of the diffusion process is close to Gaussian noise, analogous to the Gaussian KL penalty term in the VAE objective.
The second term is the reconstruction loss.
This ensures that the VAE decoder recovers samples from the distribution we wish to model.
As with the first term, similar terms appear in the VAE and DDPM training objectives.
The third term is the novel contribution that couples the latent space of the VAE to the start-point of the diffusion process.
This term ensures that the latent space of the VAE and the data distribution that is targeted by the diffusion model are identical.
The fourth term is the standard noise prediction objective.
Translating to the notation of \Cref{sec:ddpm}, $\hat{\varepsilon}_\theta$ is the noise prediction network that predicts the noise $\varepsilon$ added to a latent sample $z_0$ at diffusion time $t$, and $\gamma_\phi'(t)$ is the derivative of the noise schedule.

\paragraph{Noise schedule.}
When training the VLD model, we follow Ref.~\cite{kingma2021variational} and parametrize the noise schedule as 
\begin{equation}
    \sigma_t^2 = \textrm{sigmoid}(\gamma_\eta(t)),
    \label{eq:noise-schedule}
\end{equation}
where $\gamma_\eta(t)$ is a monotonic neural network with learned parameters $\eta$.
We use the variance-preserving diffusion process, so the parametrization is mapped to the DDPM notation of \Cref{sec:ddpm} via the relations
\begin{equation}
    \alpha_t = \sqrt{1 - \sigma_t^2}
    \qquad \beta_t = 1 - \alpha_t .
    \label{eq:vld-schedule}
\end{equation}
The monotonic neural network is trained to minimize the variance of the diffusion loss, the last term in \Cref{eq:loss_lvd}.
See the Appendix of Ref.~\cite{kingma2021variational} for more details.
At large $t$, $\gamma_\eta(t)$ is constrained to be large so that $\beta_t \approx 1$ and $z_1 \approx \mathcal{N}(0, I)$.
The derivative $\gamma_\phi'(t)$ appearing in the final term of \Cref{eq:loss_lvd} provides a weighting of the diffusion loss term that up-weights the contributions to the loss when the noise is changing rapidly.

\paragraph{Physics-informed consistency loss.}
An important practical challenge for unfolding with generative models is that HEP datasets often contain sharply peaked or \textit{resonant} distributions.
For example in the parton-level $t\bar{t}$ unfolding task described below, the mass distributions of the hadronic $W$ boson, leptonic $W$ boson, hadronic top, leptonic top, and the full $t\bar{t}$ system all have resonant features that correspond to known particle masses.
Generative models are known to struggle with reproducing very sharp features in distributions.
In Ref.~\cite{Shmakov:2023kjj}, we employed a physics-informed consistency loss to help alleviate this failure mode.
In particular when predicting the truth-level four-vector for a massive particle, we use a MAE loss that enforces the correct relativistic relation between the particle's mass, energy, and total momentum. 
As shown below, the VLD model is trained to furnish predictions for each of these quantities, denoted $\hat{M}$, $\hat{E}$, and $\hat{\mathbf{p}}$.
The consistency loss is
\begin{equation}
    \mathcal{L}_C = \lambda_C\,\abs{\hat{M}^2 - \bigl(\hat{E}^2 - \norm{\hat{\mathbf{p}}}^2\bigr)},
    \label{eq:vld-consistency}
\end{equation}
with hyperparameter $\lambda_C$ controlling the weight of this loss relative to the main VLD objective.
This constraint can be thought of as a second term in the reconstruction loss term in \Cref{eq:loss_lvd}.
This constraint was found to be very useful for accurately modeling the very sharp peaks in the parton-level mass distributions of heavy resonances.

\subsubsection{Model Architecture}

The model architectures needed for the four neural networks in \Cref{fig:vld-diagram} are relatively simple, given the VLD model is only designed to target fixed-length unfolding tasks where each sample from the posterior has the same dimensionality.
The number of leptons and jets in the conditioning detector-level events is not fixed, so the detector encoder architecture borrows the gated transformer design of SPANet~\cite{Shmakov:2021qdz}.
The remaining neural networks are all gated inverted-bottleneck feedforward networks (similar to MLPs), since more advanced set processing architectures are not required.

\subsubsection{Unfolding Semi-leptonic $t\bar{t}$ Events}

\begin{figure}[t]
    \centering
    \tikzset{
        photon/.style={decorate, decoration={snake}, draw=red},
        electron/.style={draw=blue, postaction={decorate},
            decoration={markings, mark=at position .55 with {\arrow[draw=blue]{>}}}},
        gluon/.style={decorate, draw=magenta,
            decoration={coil, amplitude=4pt, segment length=5pt}}
    }
    \input{ch5_semi_leptonic_ttbar}
    \caption{Feynman diagram for semi-leptonic $t\bar{t}$ production. Each top quark decays to a $W$ boson and a $b$ quark. Here the hadronic $W$ decays to $u\bar{d}$, and the leptonic $W$ decays to $e^- \bar{\nu}$. Decays of the leptonic $W$ to muons are also considered. Figure reproduced from Ref.~\cite{Shmakov:2023kjj}.}
    \label{fig:ttbar-feynman}
\end{figure}
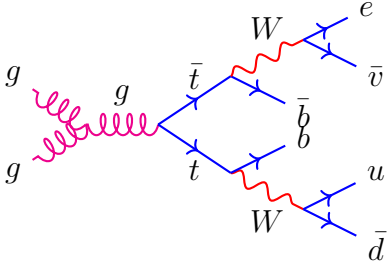

For a benchmark unfolding task, we chose the parton-level unfolding of semi-leptonic top quark pair ($t\bar{t}$) production.
The majority of top quarks are produced in pairs at the LHC, with both a top and anti-top quark, and each of these decays to a $W$ boson and a $b$ quark.
In the semi-leptonic channel, one of the produced $W$ bosons decays to a quark and anti-quark (hadronic decay), while the other decays to a charged lepton and a neutrino (leptonic decay).
A Feynman diagram of this process is shown in \Cref{fig:ttbar-feynman}.

A full description of a semi-leptonic $t\bar{t}$ event at parton-level consists of the four-momenta of a six final-state partons (two $b$ quarks, two light quarks, one charged lepton, one neutrino).
We additionally include the four-momenta of the four intermediate resonances ($W_\text{had}$, $W_\text{lep}$, $t$, $\bar{t}$) and the $t\bar{t}$ system in the parton-level event description.
These quantities are calculable from the four-momenta of the final-state partons, but are included so that they are accurately predicted.
See \Cref{sec:vl-vld} for a discussion of how problems in which the intermediate resonances are not known explicitly can be solved.
Finally each of these 11 four-vectors is extended to be represented by five quantities $(M, \log E, p_x, p_y, p_z)$, notably including the mass in addition to the standard four-momenta.
This ensures that the mass is accurately predicted as well as the other quantities.
The physics informed consistency loss ensures the proper relativistic relationship between the mass and the other quantities is maintained.
The end result is 55 observables that must be predicted by the VLD model.

\paragraph{Dataset and training.}
All data used in this study are generated with Monte Carlo simulation at $\sqrt{s} = 13$~TeV.
Matrix elements are evaluated using \textsc{MadGraph\_aMC@NLO}~\cite{Alwall:2014hca} with a top mass of $m_t = 173$~GeV.
Parton showering and hadronization are performed with \textsc{Pythia8}~\cite{Sjostrand:2014zea} and detector response is approximated using \textsc{Delphes}~\cite{deFavereau:2013fsa} with the CMS detector card.
Jets are clustered with the anti-$k_t$ algorithm~\cite{Cacciari:2008gp} with radius parameter $R = 0.5$.
Events are required to contain exactly one lepton and at least four jets with $p_T > 25$~GeV and $|\eta| < 2.5$, of which at least two must be $b$-tagged.
Approximately 9.9 million events are used for training and 1.3 million for testing.
Given we only demonstrate the modeling of the posterior in the EM algorithm, only one Monte Carlo simulation is needed.

\paragraph{Baseline models.}
Six baseline models are trained for comparison.
The CVAE and CINN are variational autoencoder and normalizing flow models that represent the prior state of the art in modeling in the posterior in the EM algorithm~\cite{Bellagente:2020piv}.
Two ablation studies determine whether the VLD framework is more expressive than a standard diffusion model (VDM) and a latent diffusion model with a pre-trained VAE (LDM).
Two additional ablations test whether providing the detector-level embedding to the VAE encoder and decoder is useful (C-VLD and UC-VLD).
Both of these last two ablations can be understood as variations of the VLD framework.

\subsubsection{Results}

\Cref{tab:vld-distances} presents distribution-free distances between the models' unfolded samples and the truth parton-level distributions, summed across all 55 observables included in the parton-level event representation.
Each of these six distances is computed differently, but all attempt to quantify the distance between distributions using only samples.
For all, lower means that the distributions are more similar (lower is better).
The VLD variants achieve the best performance across all metrics, motivating their use for modeling the posterior when unfolding.
Interestingly the VAE outperforms both the CINN and the standard diffusion model on this task.

\begin{table}[t]
    \caption{Total distribution-free distance metrics, summed across all 55 observables in the parton-level event representation, for each generative model. Lower values indicate better agreement with the truth distribution. Bold entries mark the best performance in each column.}
    \label{tab:vld-distances}
    \centering
    \input{tab_vld_distance_metrics.tex}
\end{table}

\begin{figure}[t]
    \centering
    \includegraphics[width=\linewidth, alt={Comparison of unfolded kinematic distributions for selected parton-level components including top quark masses energies and W boson masses comparing VLD UC-VLD LDM CINN and VDM models against truth distributions with ratio panels}]{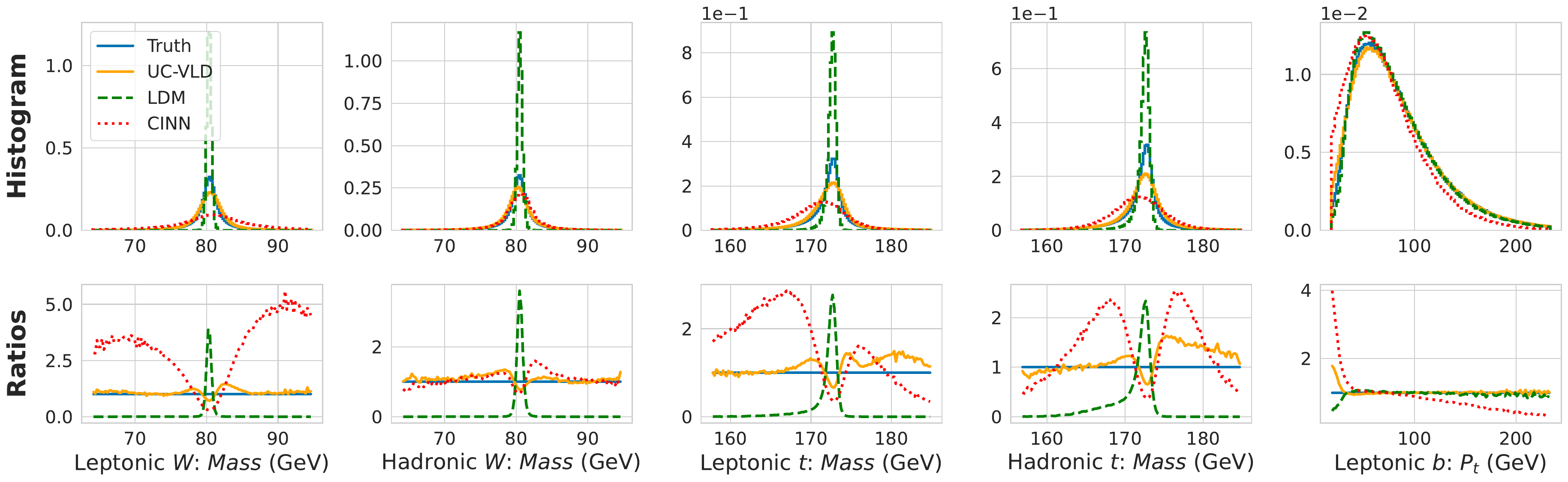}
    \caption{Highlighted unfolded distributions for five of the 55 unfolded observables in the parton-level event configuration. Upper panels show the predicted and truth-level distributions, and lower panels show the ratio of predicted to truth. Truth distributions are shown in blue, the results of the proposed VLD model are shown in orange, and the results of the LDM and CINN models are shown in green and red respectively. Figure reproduced from Ref.~\cite{Shmakov:2023kjj}.}
    \label{fig:particle-highlights}
\end{figure}

Distributions for a selection of parton-level observables are shown in \Cref{fig:particle-highlights}.
The VLD model is able to reproduce the truth distributions for all observables.
This includes distributions with difficult-to-model peaks such as the masses of the $W$ bosons and top quarks.
The agreement with the truth is not perfect but it is superior to all baselines.
Modeling mass peaks at parton level is very difficult since they are sharp, whereas the same observables are smeared out by parton shower and hadronization effects at particle level.
In practice few analyses attempt to perform this type of parton-level unfolding.
The performance shown in this Figure was state-of-the-art at the time of publication, but is probably not good enough to support a realistic measurement using LHC data.
It would be interesting to return to this problem with some of the current state-of-the-art generative unfolding methods described below.

\subsubsection{Per-Event Posterior Distributions}

\begin{figure}[t]
    \centering
    \includegraphics[width=\linewidth, alt={Per-event posterior distributions for selected parton-level components from several example events showing the VLD model posterior compared against an empirical brute-force estimate with the neutrino pseudorapidity showing a bimodal structure}]{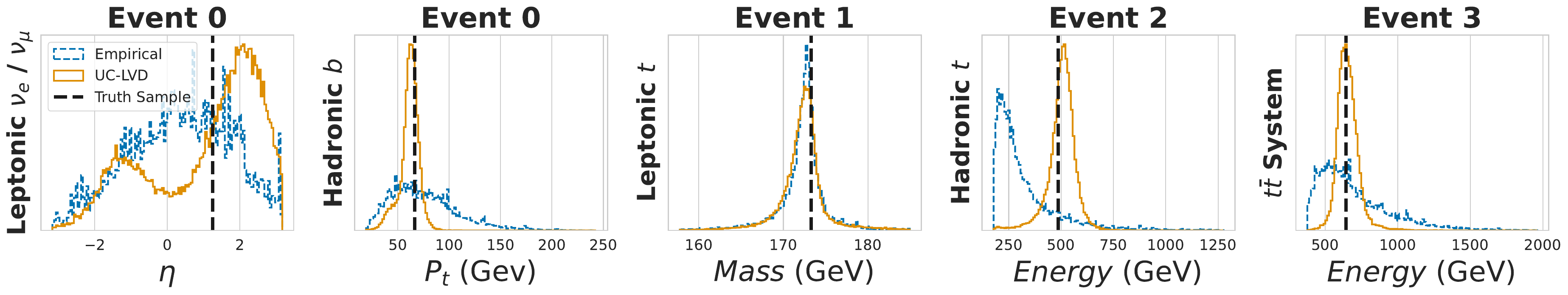}
    \caption{Highlighted per-event posterior distributions for several example events and parton-level components. The VLD model posteriors (orange) are compared against an empirical brute-force posterior estimate (blue) derived from the training data via exponential distance weighting. The neutrino pseudorapidity $\eta_\nu$ displays a clear bimodal structure in the VLD samples, reflecting the two solutions to the on-shell $W$ boson constraint on the neutrino longitudinal momentum. Figure reproduced from Ref.~\cite{Shmakov:2023kjj}.}
    \label{fig:vld-posteriors}
\end{figure}

The trained generative models produce samples from the posterior $p(x \mid y)$.
These posteriors are marginalized over $x$ to produce the distributions in \Cref{fig:particle-highlights}, as would be done in the EM algorithm, but it is also interesting to examine the posterior distributions themselves.
This can be done by repeatedly sampling from the generative model with a fixed detector-level event configuration $y$, and is a unique ability of unfolding methods which have an explicit model for the posterior.
This is also sometimes called \textit{oversampling}, because multiple truth-level configurations are sampled for a single detector-level event and it can amplify the statistics of small data samples~\cite{Butter:2020qhk}.
Example per-event posteriors for several different parton-level observables are shown in \Cref{fig:vld-posteriors} using samples from the trained VLD model.
The most interesting posterior is that of the neutrino pseudorapidity, for which the VLD model predicts a bimodal shape.
This shape is not produced by an empirical estimate of the posterior based on an exponential distance weighting derived at detector level.
Neutrinos escape LHC detector without interacting, so their four-momenta can only be constrained by examining the missing transverse momentum in an event.
Typically there is no constraint on the longitudinal momentum, which is related to the pseudorapidity, however in semi-leptonic $t\bar{t}$ events there is a constraint provided by requiring the leptonic $W$ boson to satisfy its relativistic mass--energy relation.
This constraint typically admits two solutions, so two neutrino configurations are often simultaneously consistent with energy and momentum conservation for the same detector-level semi-leptonic $t\bar{t}$ event.
The VLD model appears to have correctly learned such constraints from data and produced a reasonable bimodal posterior.
Note that a method that constrains the posterior to be Gaussian (for example a naive MSE regression) would be unable to capture this behavior.

\subsubsection{Summary and Key Contributions}

Ref.~\cite{Shmakov:2023kjj} was the first work to apply diffusion and flow matching models to the modeling of the posterior in an unfolding problem.
This class of models remains state-of-the-art for this task, though the exact approach has been improved upon.
The unified end-to-end VLD training objective was additionally a novel contribution to the machine learning literature, and could be useful in tasks outside of physics.
A latent diffusion model trained with this objective achieved greater accuracy than all considered baselines, and achieved roughly a factor of 3 improvement in distribution-free metrics relative to a latent diffusion model with VAE and diffusion model optimized independently.
This work also demonstrated that generative AI could accurately model posteriors in several tens of dimensions.
The parton-level phase space is described by 55 dimensions, which was the highest-dimensional posterior-modeling task ever considered at the time of publication.
With this posterior model, extension of these methods to a proper unfolding method using either of the approaches described in \Cref{sec:generative-unfolding-methods} should be possible.
Despite this progress, the VLD model is still not capable of modeling a variable-dimensional posterior, in which the number of components of the feature vector $x$ is not fixed.
Extension of the VLD framework to this variable-dimensional setting is the subject of the next section.

\subsection{Variable Length VLD}
\label{sec:vl-vld}

The VLD model described in the last section is capable of accurately estimating a 55-dimensional posterior, but still has a few important limitations.
Most importantly the model has no way to model a variable-dimensional truth-level phase space configuration.
The generation of variable-length outputs is a prerequisite for the full-phase-space unfolding application described in \Cref{sec:unbinned-motivation}.
The parton-level task in the last section does estimate the posterior for essentially all of the parton-level phase space.
This is possible because parton-level event configurations are typically fixed length: every event is assummed to be a semi-leptonic $t\bar{t}$ event and have exactly the set of partons described in the previous section.
However the vast majority of LHC measurements target the particle-level phase space because unfolding to the parton-level phase space incurs significant modeling uncertainties due to the assummed parton shower and hadronization models.
The full particle-level phase space is inherently variable dimensional, so an unfolding must either target a fixed subset of observables or accommodate variable dimensions.
All of this strongly motivates developing methods for estimating variable-dimensional posteriors.
A final limitation of the previous section is that only the E-step of the EM procedure was considered.
The full EM procedure can be understood as removing the \textit{prior dependence} from the modeled posterior.
A demonstration that this prior dependence is not too large would back up the claim that the EM procedure can be applied.
The section that follows summarizes Ref.~\cite{Shmakov:2024gkd} which addresses all of these limitations and is the third original research contribution in this thesis.

\subsubsection{Model Architecture}

\begin{figure}[tb]
  \centering
  \includegraphics[width=\linewidth, alt={Flow diagram of the VL-VLD model components: particle encoder and decoder (VAE), detector encoder, multiplicity predictor, and the latent diffusion process, along with arrows indicating the data flow during training and inference.}]{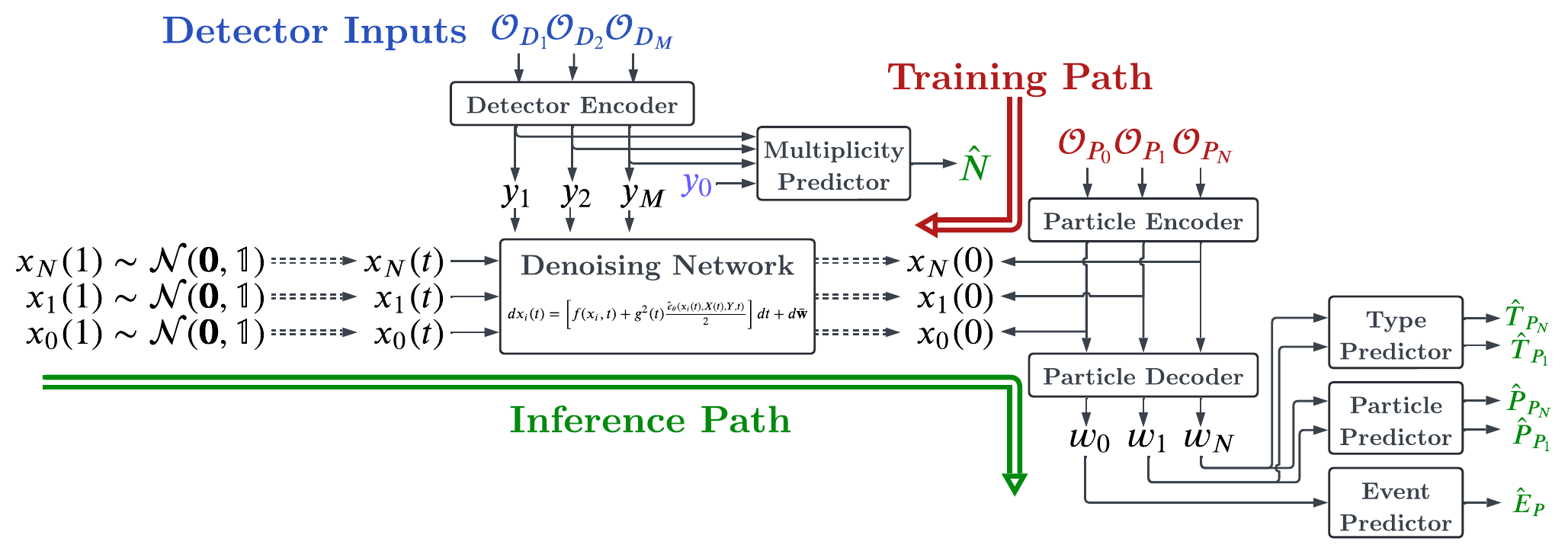}
  \caption{A Flow diagram of the VL-VLD model.
  Pairs of particle level (${\mathcal{O}_P}$) and detector level (${\mathcal{O}_D}$) events are used to train the model using the loss functions introduced in Ref.~\cite{Shmakov:2023kjj}.
  At inference time, a detector level event is used to produce a multiplicity prediction and mapped to a latent embedding through the detector encoder.
  The multiplicity prediction and the latent representation of the detector level event are then used to condition the diffusion process, resulting in a sample from the latent space of the particle VAE.
  A particle decoder is then applied to produce a sample from the learned conditional distribution $P(X|Y)$.}
  \label{fig:vl_vld_diagram}
\end{figure}

When modeling a variable-dimensional posterior $p(x \mid y)$ for unfolding with the EM algorithm, both the detector-level condition $y$ and the truth-level sample $x$ are variable-dimensional vectors which are not necessarily of the same length.
In the remainder of this section, the truth-level phase space will be referred to as the particle-level phase space, since the VL-VLD model is specifically designed to model the posterior in particle level unfolding tasks.
Modeling this posterior can be described as a set-conditional set-generation task.
Specifically given a set of $M$ detector-level physics analysis objects $\mathcal{O}_D = \{y_1, y_2, \ldots, y_M\}$, the model must generate a set of $N$ truth-level objects $\mathcal{O}_P = \{x_1, x_2, \ldots, x_N\}$.
In general $N \neq M$, so the model must have some method for predicting $N$ given the set $\mathcal{O}_D$.
A diagram of the variable-length variational latent diffusion (VL-VLD) model which is capable of modeling this variable dimensional posterior is shown in \Cref{fig:vl_vld_diagram}.

Given all sets involved are variable-dimensional, all of the major neural networks in the model need to be set-processing.
Transformer neural networks are used for the VAE encoder and decoder, the detector encoder, and the denoising network.
The main addition of this model compared to the previous section is the multiplicity predictor network, which is also a transformer.
Brief overviews of the purpose and design of each of these networks is detailed in the following.

\paragraph{Particle VAE.}
The particle VAE consists of an encoder and decoder.
The encoder is used only during training, and maps the true particle-level set of objects into the latent space.
Each object in the true particle-level set is described by a vector of its kinematic properties $P_{P_i}$, concatenated with a vector of one-hot encodings $T_{P_i}$ that identify the type of object.
Event-level observables are projected into the model dimension and appended to the set of objects as $\mathcal{O}_{P_0}$, making the full input set $\mathcal{O}_P = \{\mathcal{O}_{P_0}, \mathcal{O}_{P_1}, \ldots, \mathcal{O}_{P_N}\}$.
This set is processed by a transformer encoder to produce a sample from the VAE latent space $X = \{x_0, x_1, \ldots, x_N\}$ with $N+1$ $D_\text{Latent}$-dimensional vectors for an event with $N$ input objects.
The decoder network follows a similar structure, using a transformer to process the latent vectors the passing each object in the set to dedicated MLPs which reconstruct the kinematic features, type labels, and event-level observables.
Unlike the encoder the decoder is used during inference to furnish the particle-level event configuration that is interpreted as a sample from the posterior.

\paragraph{Detector Encoder.}
An identical transformer architecture is used to process the detector-level event that conditions generation.
The detector-level event is represented by the set of objects $\{\mathcal{O}_{D_1}, \ldots, \mathcal{O}_{D_M}\}$ with each vector consisting of the same kinematic features and one-hot encodings as used at particle level.
A transformer processes this set into a latent set $Y = \{y_1, \ldots, y_M\}$ which provides context for both the multiplicity predictor and diffusion models.

\paragraph{Multiplicity Predictor.}
There are two standard methods for generating variable length objects with neural networks.
The autogressive method, used by LLMs for example, has the network recursively generate objects until it outputs a special \textit{stop token}.
See Ref.~\cite{Butter:2023fov} for an application of this approach in HEP.
The more common approach, which is taken here, is to use a multiplicity predictor network that predicts the number of objects in the sample and conditions generation with the appropriate number of latent variables.
In the posterior modeling task, the multiplicity predictor network uses the latent set $Y$ as a condition.
This set is augmented by a learnable vector $y_0$ (analogous to a class token for example as used in Ref.~\cite{Qu:2022mxj}), and is then processed by a small transformer.
After the transformer, the object corresponding to the learnable vector is passed into an MLP that predicts the shape and scale parameters of a Gamma distribution.
The prediction for $N$ is then sampled from this distribution:
\begin{equation}
  \hat{N} \sim \mathrm{Gamma}\!\left(MLP_k(z),\, MLP_\theta(z)\right).
  \label{eq:vl_vld_multiplicity}
\end{equation}

\paragraph{Latent Diffusion.}
As in the previous section, the diffusion model is trained in the latent space of the particle VAE.
However the set the diffusion model must generate now has no inherent ordering, adding some ambiguity to the denoising loss term.
Specifically at low signal-to-noise ratios where the samples are primarily noise, each of the latent vectors will look nearly identical and there is no way to assign a object-to-object mapping between the true noise added to the true sample from the VAE latent space furnished by the encoder, and the predicted noise.
This difficulty with generating unordered sets was addressed by introducing a \pt-ordering for the set during the diffusion process only.
This breaks the permutation equivariance of the diffusion processes because the order of the elements is critical: high \pt objects, which are the most important for a given event, are listed first in the sequence.
This \pt ordering is used only for the noise prediction network, leaving all other transformer networks in the VL-VLD model fully permutation equivariant.

Once the ordering of the objects in the generated set is fixed, each object $i$ in the sequence can also be assigned its own learned noise schedule $\gamma_\phi(i, t)$.
These schedules are implemented with the same method as in \Cref{sec:vld-unfolding}, but each object is assigned an independent noise schedule by the monotonic neural network.
In practice this is achieved by including the index $i$ as an additional input via a learned positional embedding.
All predicted schedules share the same global endpoints $\gamma_\text{min}$ and $\gamma_\text{max}$, but the paths taken between these endpoints are different depending on the position in the sequence.
The diffusion flow for each object is then
\begin{equation}
  x_i(t) \sim \mathcal{N}\!\left(\alpha_i(t)\, x_i,\; \sigma_i^2(t)\, \mathbf{I}\right),
\end{equation}
where $\alpha_i(t) = \sqrt{\mathrm{sigmoid}(\gamma_\phi(i,t))}$ and $\sigma_i(t) = \sqrt{\mathrm{sigmoid}(-\gamma_\phi(i,t))}$.

\begin{figure}[tb]
  \centering
  \includegraphics[width=0.6\linewidth, alt={Block diagram of the VL-VLD denoising network. Noisy particle latent vectors x_i(t) and detector conditioning vectors y_i are both fed into a shared transformer encoder. Detector outputs are dropped; particle outputs give the noise predictions.}]{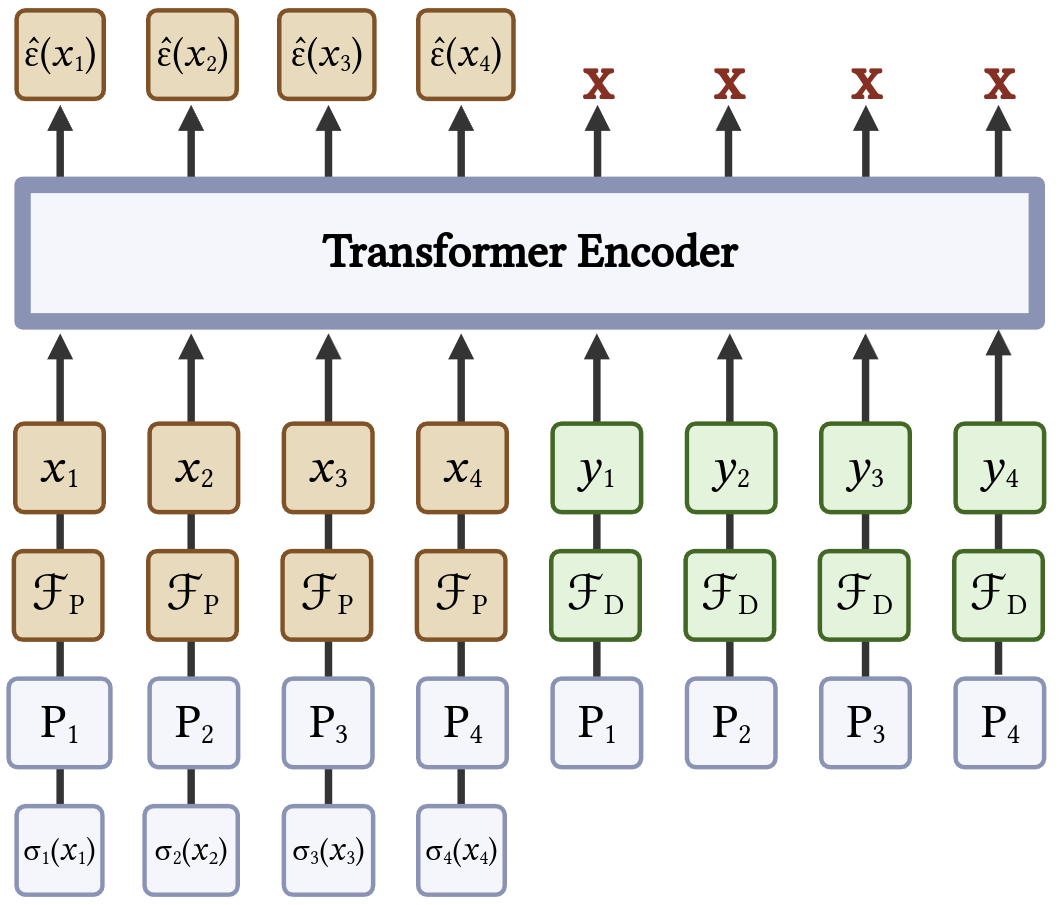}
  \caption{A block diagram of the denoising network used in the VL-VLD model.
  Both the noisy particle-level latent vectors $X(t)$ and the detector-level condition $Y$ are provided as inputs to the transformer encoder.
  The outputs corresponding to the condition are simply discarded, and the outputs corresponding to the noisy latent vectors are taken to be the noise prediction $\hat{\epsilon}_\theta$.
  The condition and latent vector inputs are distinguished by fourier positional features for the input ordering in \pt, the learned noise scales $\sigma_i(t)$, and a dedicated one-hot encoding.}
  \label{fig:vl_vld_block_diagram}
\end{figure}

The denoising network $\hat{\epsilon}_\theta(X(t), Y, t)$ is implemented as a transformer that uses all available information as input: the noisy particle-level latent vectors $X(t)$, the detector-level condition $Y$, and the timestep $t$.
This lets the network condition the noise predicted for a given $x_i(t)$ on the rest of the noisy particle-level objects and the detector-level condition, allowing it to exploit correlations between objects and use context from the rest of the event.
A block diagram of the denoising network and its inputs is shown in \Cref{fig:vl_vld_block_diagram}.

\subsubsection{Training Objective and Inference}

Similar to the fixed-length VLD model in the previous section, all networks are optimized simultaneously with an extension of the evidence lower bound:

\begin{align}
  \mathcal{L} &= \sum_{i=0}^{N} D_\text{KL}\!\left(q(x_i(1)\,|\,\mathcal{O}_P, \mathcal{O}_D) \;\|\; p(x_i(1))\right) & &\textsc{Prior} \notag \\
  &+ \sum_{i=0}^{N} \mathbb{E}_{q(x_i(0)\,|\,\mathcal{O}_P)}\!\left[-\log p(\hat{\mathcal{O}}_P \,|\, x_i(0))\right] & &\textsc{Reconstruction} \notag \\
  &+ \sum_{i=0}^{N} \mathbb{E}_{\varepsilon,\, t}\!\left[\gamma_\phi'(i,t)\left\|\varepsilon - \hat{\epsilon}_\theta(x_i(t), X(t), Y, t)\right\|^2\right] & &\textsc{Denoising} \notag \\
  &- \log p(\hat{N} = N \,|\, \mathcal{O}_D). & &\textsc{Multiplicity}
  \label{eq:vl_vld_loss}
\end{align}

The statistical interpretation of the prior, reconstruction, and denoising loss terms are identical to the previous section, but they are written in slightly different notation.
A Gaussian prior is assumed for $p(x_i(1))$ in keeping with standard diffusion models.
Unlike the previous section, the predicted particle-level event configuration contains both continuous kinematic features and discrete categorical features.
The reconstruction term expands into the sum of Gaussian losses for the continuous features and cross-entropy losses for the discrete features.
The denoising term is essentially identical to the previous section, save for the weighting by the derivative of the per-position log signal-to-noise ratio.
The multiplicity term is new to the variable length model, and is simply the negative log-likelihood of the true multiplicity $N$ under the Gamma distribution predicted by the network.
Crucially, the \pt ordering breaks the permutation invariance of the denoising loss term, but the remaining loss terms remain fully position equivariant.
Finally it should be noted that the third term in \Cref{eq:loss_lvd} is absent here.
This term was reduced to essentially zero during network training, and ablation studies showed no difference in performance when training with or without this term.
Therefore for simplicity the term was dropped, but the VAE and diffusion models were always optimizeed jointly.

Once the VL-VLD model is trained, sampling from the posterior using only a detector-level event $\mathcal{O}_D$ as a condition proceeds as follows:
\begin{enumerate}
  \item Encode the detector event: $Y = \textsc{Transformer}_\textsc{Det}(\mathcal{O}_D)$.
  \item Extract the multiplicity latent: $z = \textsc{Transformer}_\textsc{Mult}(y_0 \,\|\, Y)_0$.
  \item Sample the multiplicity: $\hat{N} \sim \mathrm{Gamma}(MLP_k(z), MLP_\theta(z))$, rounded to the nearest integer $N$.
  \item Sample $N$ standard normal vectors from the prior: $x_i(1) \sim \mathcal{N}(0, \mathbf{I})$ for $i \in \{0, 1, \ldots, N\}$.
  \item Perform the reverse diffusion process with an ODE solver to obtain $x_i(0)$.
  \item Decode $\{x_i(0)\}$ with the particle decoder to obtain the particle-level observables $\hat{\mathcal{O}}_P$.
\end{enumerate}
The particle-level observables are samples from the posterior and can be used in the EM algorithm to perform a full unfolding.

\subsubsection{Application to Particle-Level \texorpdfstring{$t\bar{t}$}{ttbar} Unfolding}

The performance of the VL-VLD model is evaluated using the same semi-leptonic $t\bar{t}$ example as the previous section, but now targeting the particle level rather than the parton level.
The most advanced use-case is to model the posterior of the kinematics of all final state particles.
This requires predicting a set of features with cardinality in the several hundreds.
Early experiments attempting to model this very high dimensional posterior with VL-VLD were not successful, so all stable non-leptonic particles are first clustered into jets using the anti-$k_T$ algorithm~\cite{Cacciari:2008gp} with radius $R = 0.5$.
The particle-level event configuration to be predicted by VL-VLD is then a set of jets and leptons.
Jet clustering reduces the dimensionality of the required posterior, but this simplified representation is still variable dimensional.
Events contain between four and ten jets at particle level, corresponding to a posterior with several tens of dimensions on average.

As in the previous section, simulated $t\bar{t}$ events from proton--proton collisions are generated with the Standard Model (SM) at a center-of-mass energy of $\sqrt{s}=13$~TeV using \textsc{MadGraph\_aMC@NLO}~\cite{Alwall:2014hca} for the matrix element calculation and \textsc{Pythia8}~\cite{Sjostrand:2014zea} for the parton showering and hadronization.
Interaction with the experimental apparatus is simulated with \textsc{Delphes}~\cite{deFavereau:2013fsa} using the default CMS detector card.
Events are required to have one electron or muon and at least four jets, of which at least two are $b$-tagged using the standard \textsc{Delphes} b-tagging approximation.
Electrons, muons, and jets are required to have a transverse momentum \pt $> 25$~GeV and absolute pseudo-rapidity $|\eta| < 2.5$.
All events must pass these selections at both detector level and particle level.
In total 14 million events are used for training and 1 million events are used for testing.

Different particle-level and detector-level event representations are used, given these representations have different roles in the model.
Particle-level objects have five kinematic features $(P_x, P_y, P_z, \log(E+1), \log(M+1))$ and four one-hot encodings which tell whether the object is a light-quark jet, b-quark jet, electron, or muon.
Event-level observables are taken to be the magnitude and azimuthal angle of the missing transverse momentum, and the kinematics of the neutrino $(P_x^\nu, P_y^\nu, P_z^\nu, E^\nu)$.
At detector level, two representations of each of the objects kinematics are provided to the detector encoder.
These are the cartesian representation used for the particle-level objects, and a polar coordinate representation.
The same four one-hot encoded features are also included at detector level.
The event-level observables are only the magnitude and azimuthal angle of the missing transverse momentum ($E_T^\text{miss}$ and $\phi^\text{miss}$), since the neutrino kinematics are not constrained at detector level.

\subsubsection{Results}

\paragraph{Learned Noise Schedules.}

\begin{figure}[tb]
  \centering
  \begin{subfigure}{0.47\linewidth}
    \centering
    \includegraphics[width=\linewidth, alt={Learned log signal-to-noise ratio as a function of diffusion time t, shown separately for each particle-level object ordered by pT. The two highest-pT objects have the fastest-rising SNR curves.}]{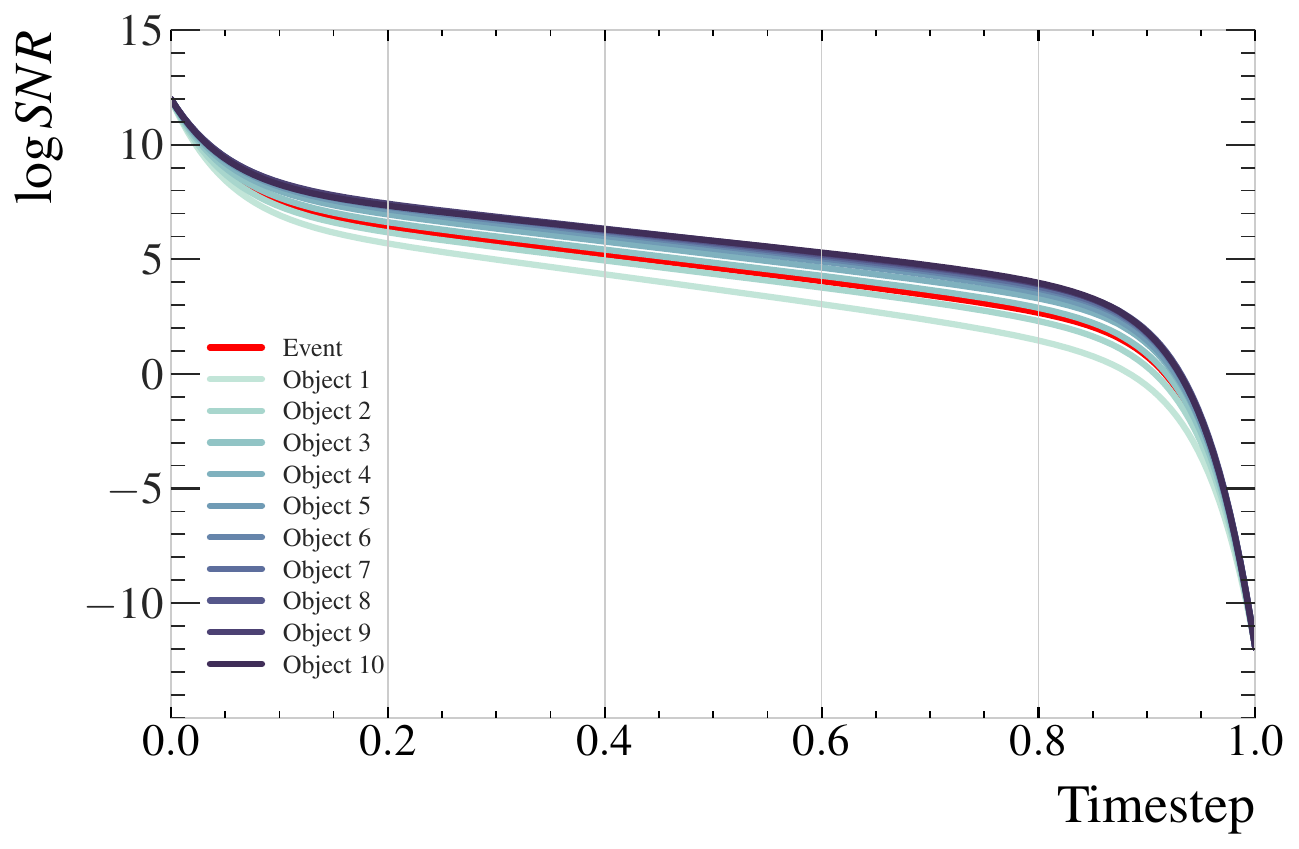}
    \caption{}
    \label{subfig:vl_vld_snr}
  \end{subfigure}\hfill
  \begin{subfigure}{0.47\linewidth}
    \centering
    \includegraphics[width=\linewidth, alt={Beta schedule as a function of diffusion time step for each particle-level object, showing the per-object noise rates during inference in the DDPM framework.}]{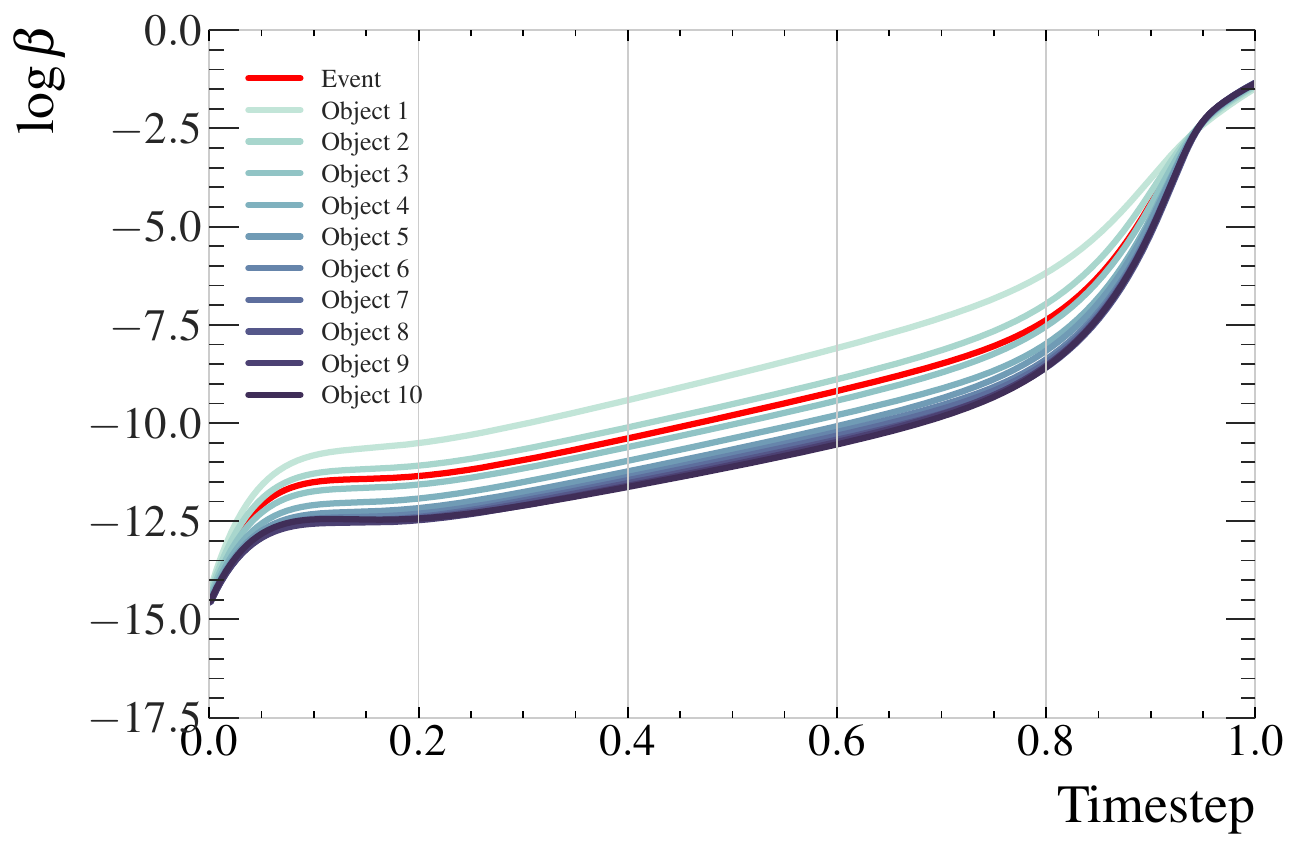}
    \caption{}
    \label{subfig:vl_vld_beta}
  \end{subfigure}
  \caption{Learned noise schedules for each object in the VL-VLD model, with the index in the legend reverse-ordered by \pt.
  The ``Event'' curve corresponds to the event-level observable element in the sequence.
  (\subref{subfig:vl_vld_snr}): The learned log signal-to-noise ratios (SNR) as a function of diffusion time $t$.
  (\subref{subfig:vl_vld_beta}) The corresponding $\beta$ schedule used during inference in the DDPM framework~\cite{Ho:2020ddpm}.}
  \label{fig:vl_vld_noise_schedule}
\end{figure}

The learned log signal-to-noise (SNR) ratios $\gamma_\phi(t)$ and the corresponding $\beta$ schedules for each \pt-ordered position $i$ are shown in \Cref{fig:vl_vld_noise_schedule}.
Objects are ordered by decreasing \pt so object 1 is the highest \pt object in the event, with the 0th-element of the generated sequence that encodes event-level observables labeled as ``Event''.
As $t$ decreases away from 1 early in the reverse diffusion process, the model increases the log SNR fastest for the lowest \pt objects in the event.
This can be understood as the model adjusting the amount of added noise to be correctly proportional to the size of the energy and momentum kinematic quantities for each object.
When $t$ is in the range of 0.8 to 0.2, the log SNR increases roughly equivalently for all objects, before increasing rapidly for the highest \pt objects as $t$ approaches 0.
In other words the model chooses to fix the highest \pt objects last when building the particle-level event representation.

\paragraph{Jet, Lepton, and Event Observables.}

\begin{figure}[phtb]
  \centering
  \begin{subfigure}{0.3\linewidth}
    \centering
    \includegraphics[width=\linewidth, alt={Inclusive jet transverse momentum distribution comparing truth, unfolded, and detector levels.}]{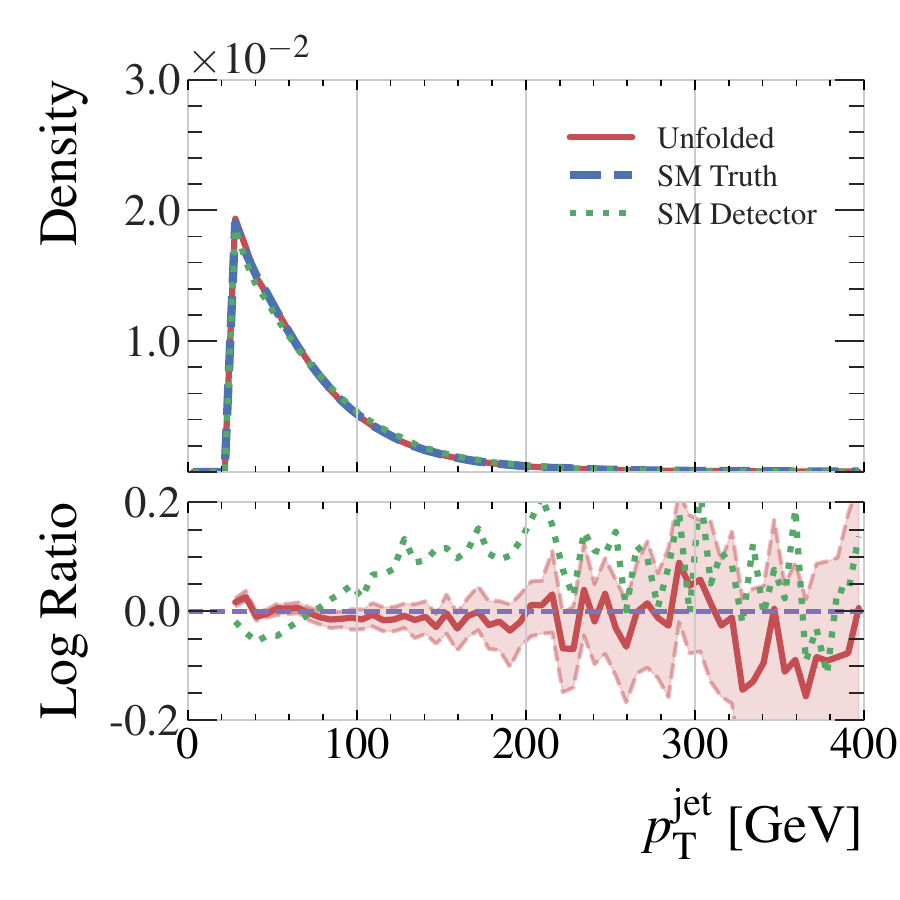}
    \caption{}
  \end{subfigure}\hfill
  \begin{subfigure}{0.3\linewidth}
    \centering
    \includegraphics[width=\linewidth, alt={Inclusive jet pseudorapidity distribution comparing truth, unfolded, and detector levels.}]{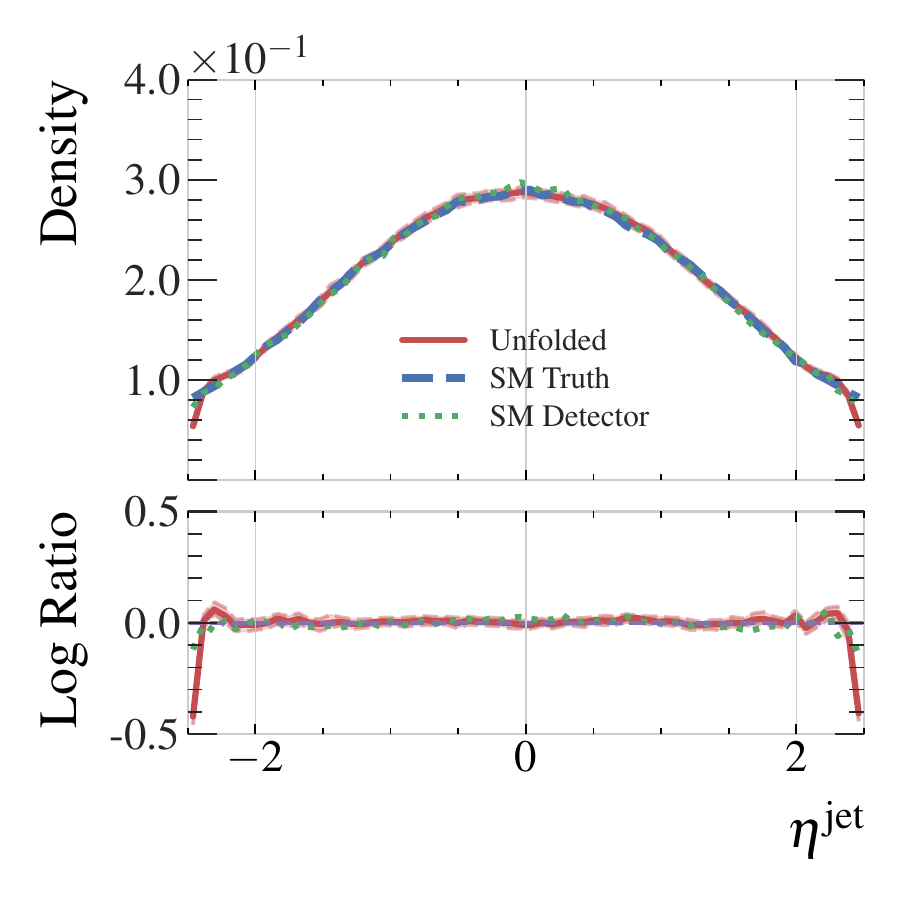}
    \caption{}
  \end{subfigure}\hfill
  \begin{subfigure}{0.3\linewidth}
    \centering
    \includegraphics[width=\linewidth, alt={Inclusive jet azimuthal angle distribution comparing truth, unfolded, and detector levels.}]{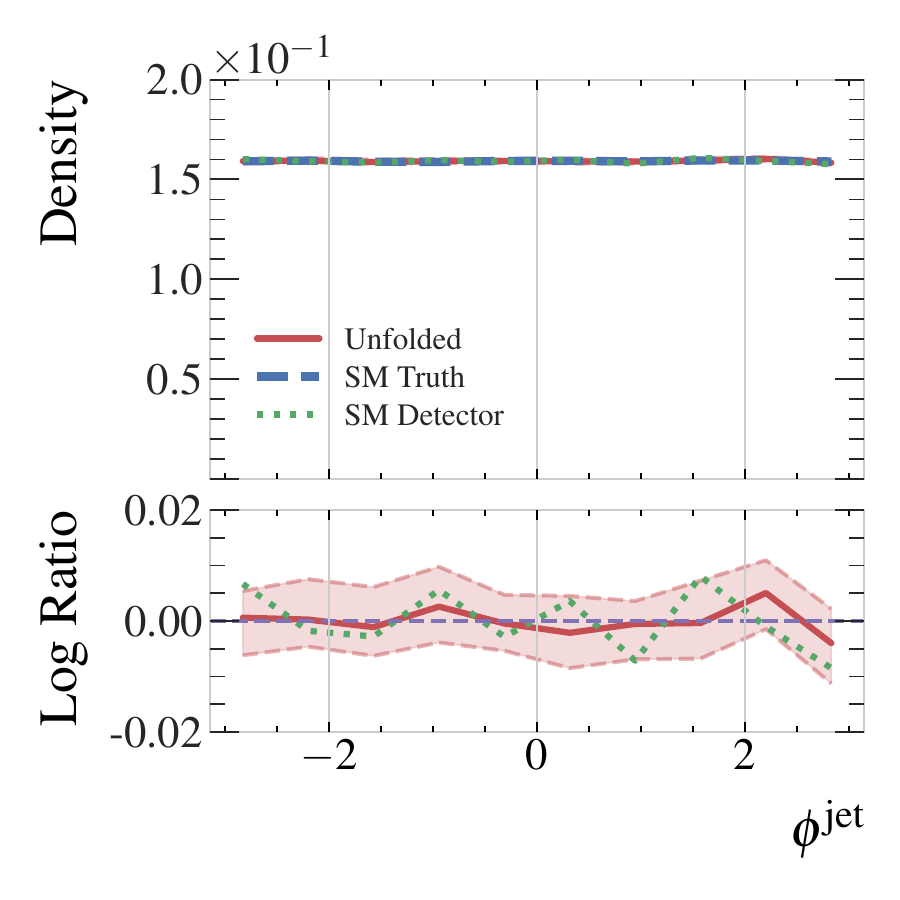}
    \caption{}
    \label{subfig:vl_vld_kinematics_jets_eta}
  \end{subfigure}
  \par\vspace{8pt}
  \begin{subfigure}{0.3\linewidth}
    \centering
    \includegraphics[width=\linewidth, alt={Inclusive jet mass distribution comparing truth, unfolded, and detector levels.}]{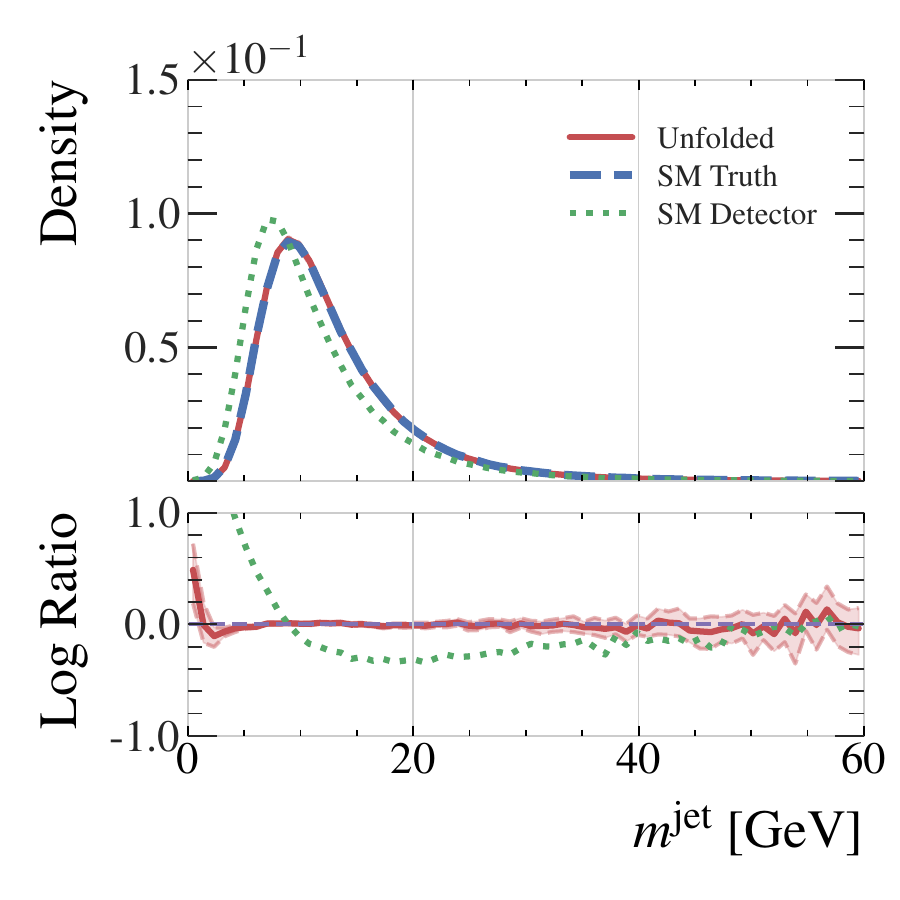}
    \caption{}
    \label{subfig:vl_vld_kinematics_jets_mass}
  \end{subfigure}
  \begin{subfigure}{0.3\linewidth}
    \centering
    \includegraphics[width=\linewidth, alt={Inclusive jet energy distribution comparing truth, unfolded, and detector levels.}]{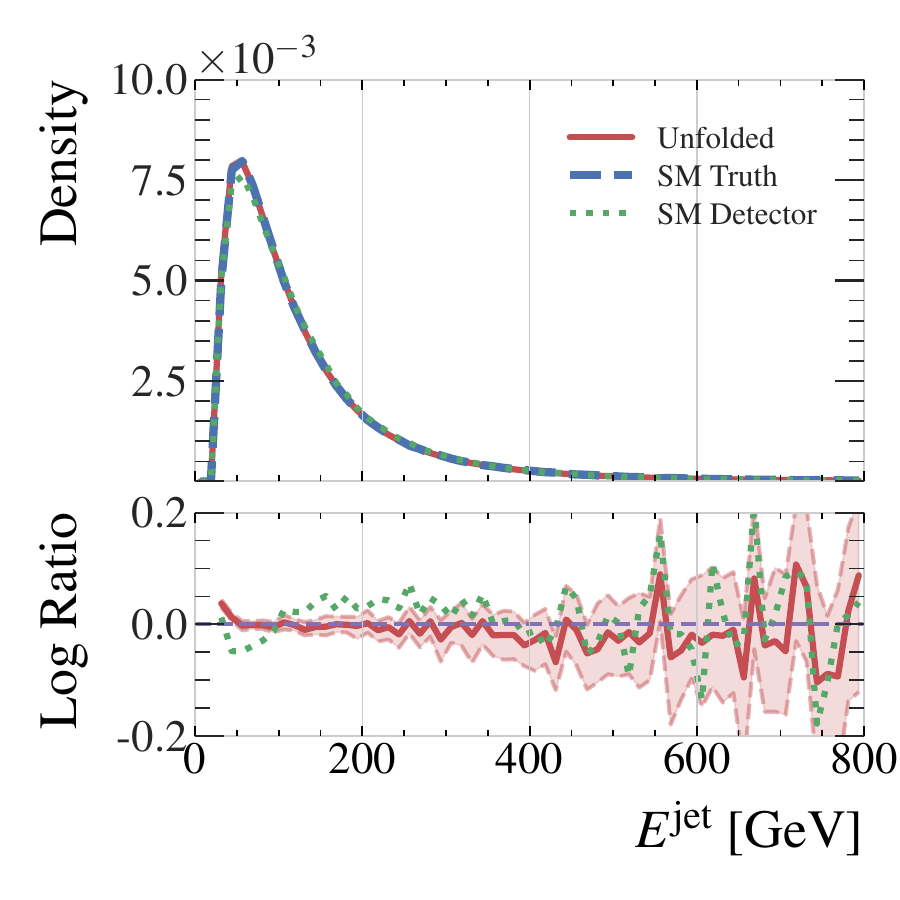}
    \caption{}
  \end{subfigure}
  \caption{Inclusive kinematic distributions for jets in the test dataset.
  Shown are the true particle-level distribution (dashed blue), the distribution obtained by sampling the learned posterior (solid red), and the detector-level distribution (dotted green).
  Uncertainty bands on the unfolded distributions are estimated by sampling each event 128 times.}
  \label{fig:vl_vld_kinematics_jets}
\end{figure}

\begin{figure}[phtb]
  \centering
  \begin{subfigure}{0.3\linewidth}
    \centering
    \includegraphics[width=\linewidth, alt={Inclusive lepton transverse momentum distribution comparing truth, unfolded, and detector levels.}]{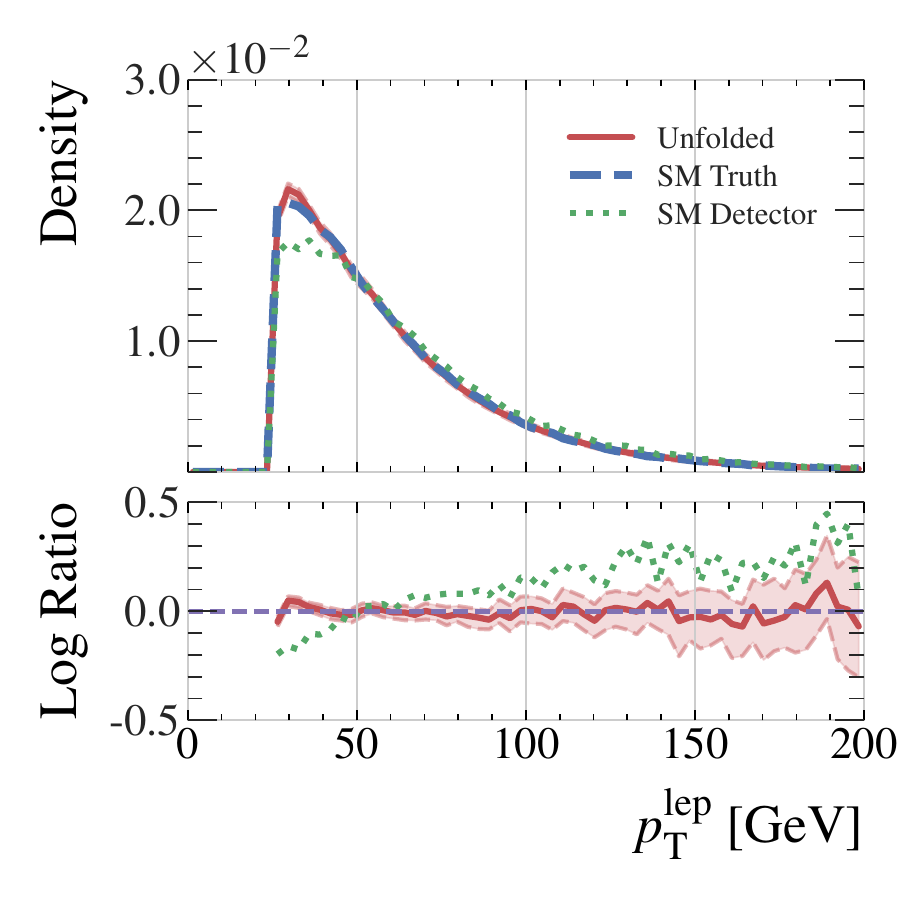}
    \caption{}
  \end{subfigure}\hfill
  \begin{subfigure}{0.3\linewidth}
    \centering
    \includegraphics[width=\linewidth, alt={Inclusive lepton pseudorapidity distribution comparing truth, unfolded, and detector levels.}]{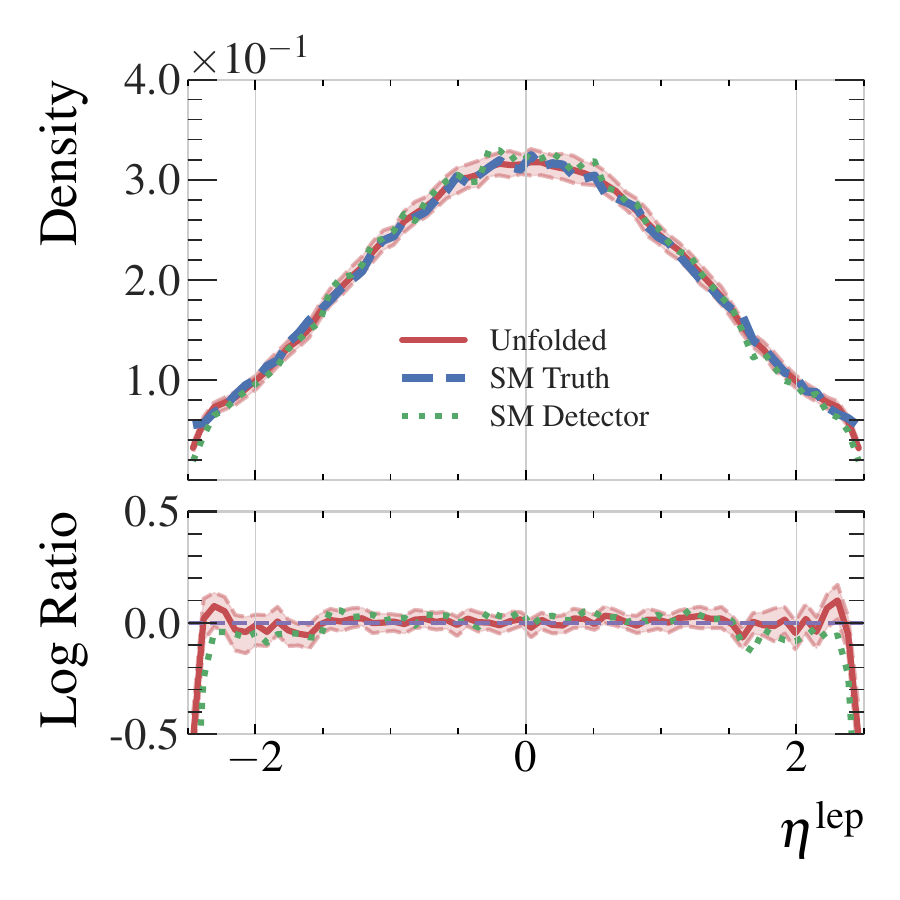}
    \caption{}
    \label{subfig:vl_vld_kinematics_leptons_eta}
  \end{subfigure}\hfill
  \begin{subfigure}{0.3\linewidth}
    \centering
    \includegraphics[width=\linewidth, alt={Inclusive lepton azimuthal angle distribution comparing truth, unfolded, and detector levels.}]{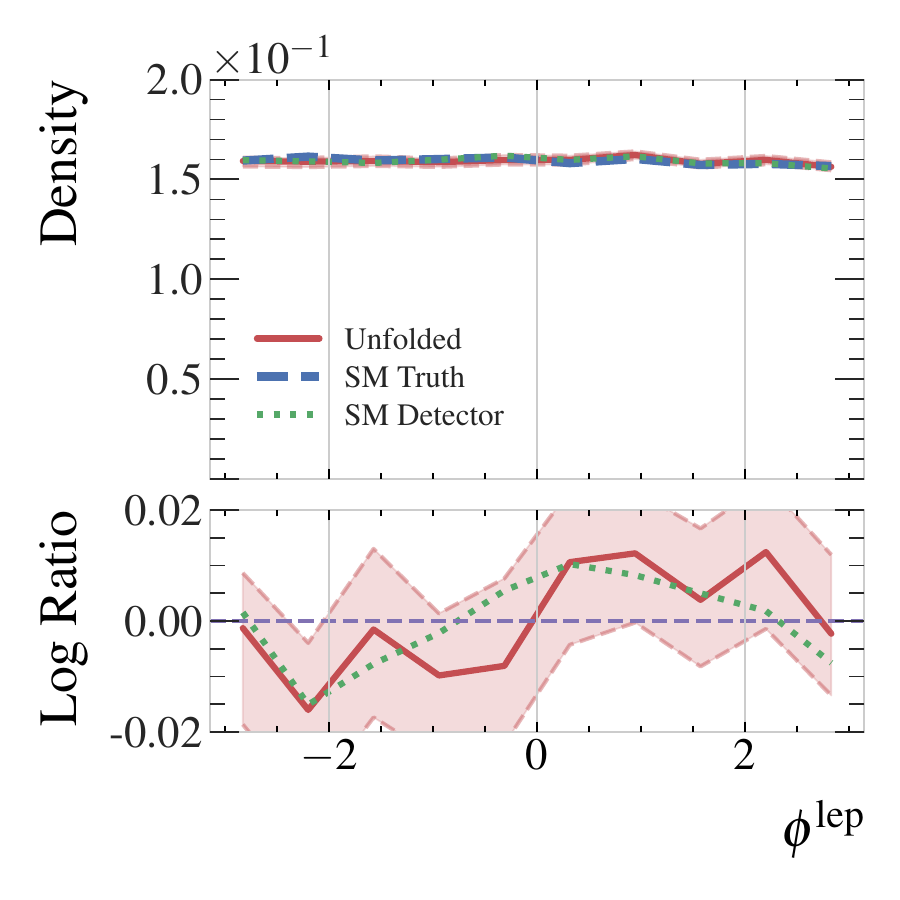}
    \caption{}
  \end{subfigure}
  \caption{Inclusive kinematic distributions for leptons in the test dataset.
  Shown are the true particle-level distribution (dashed blue), the distribution obtained by sampling the learned posterior (solid red), and the detector-level distribution (dotted green).
  Uncertainty bands on the unfolded distributions are estimated by sampling each event 128 times.}
  \label{fig:vl_vld_kinematics_leptons}
\end{figure}

\Cref{fig:vl_vld_kinematics_jets,fig:vl_vld_kinematics_leptons} show the inclusive kinematic distributions for the particle-level jets and leptons respectively.
Distributions are shown for the true particle-level objects (dashed blue), the detector-level objects (dotted green), and the particle-level objects sampled from the learned posterior (solid red).
Shaded uncertainty bands on the posterior samples are estimated by sampling each detector-level event 128 times and re-binning.
Ideally, the solid red distribution would match the dashed blue exactly\footnote{The term ``Unfolded'' used to describe the samples from the learned posterior is an unfortunate misnomer, since modeling the posterior used in the E-step of the EM algorithm is only one part of unfolding. This term, used in the figures and tables that follow, should always be understood to refer to the learned posterior.}.

\begin{figure}[phtb]
  \centering
  \begin{subfigure}{0.3\linewidth}
    \centering
    \includegraphics[width=\linewidth, alt={Missing transverse energy distribution comparing truth, unfolded, and detector levels.}]{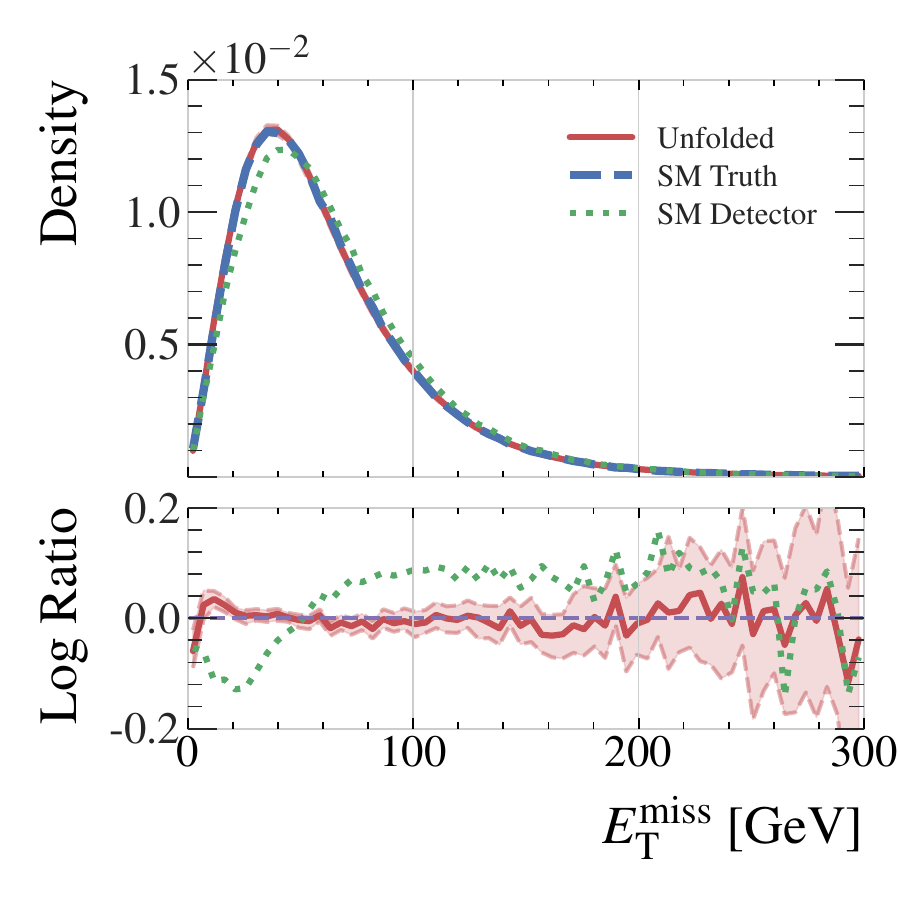}
    \caption{}
    \label{subfig:vl_vld_event_obs_met}
  \end{subfigure}\hfill
  \begin{subfigure}{0.3\linewidth}
    \centering
    \includegraphics[width=\linewidth, alt={Missing transverse energy azimuthal angle comparing truth, unfolded, and detector levels.}]{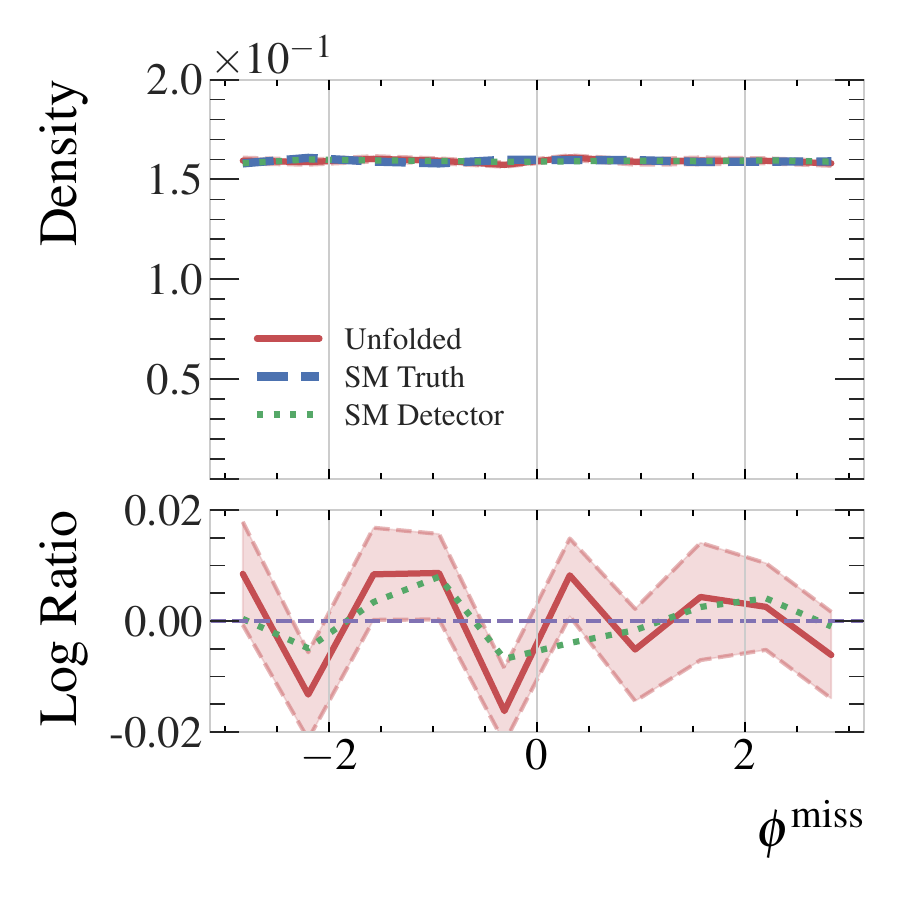}
    \caption{}
    \label{subfig:vl_vld_event_obs_metphi}
  \end{subfigure}\hfill
  \begin{subfigure}{0.3\linewidth}
    \centering
    \includegraphics[width=\linewidth, alt={Neutrino pseudorapidity distribution comparing truth, unfolded, and detector levels, showing a slight excess near zero in the unfolded distribution.}]{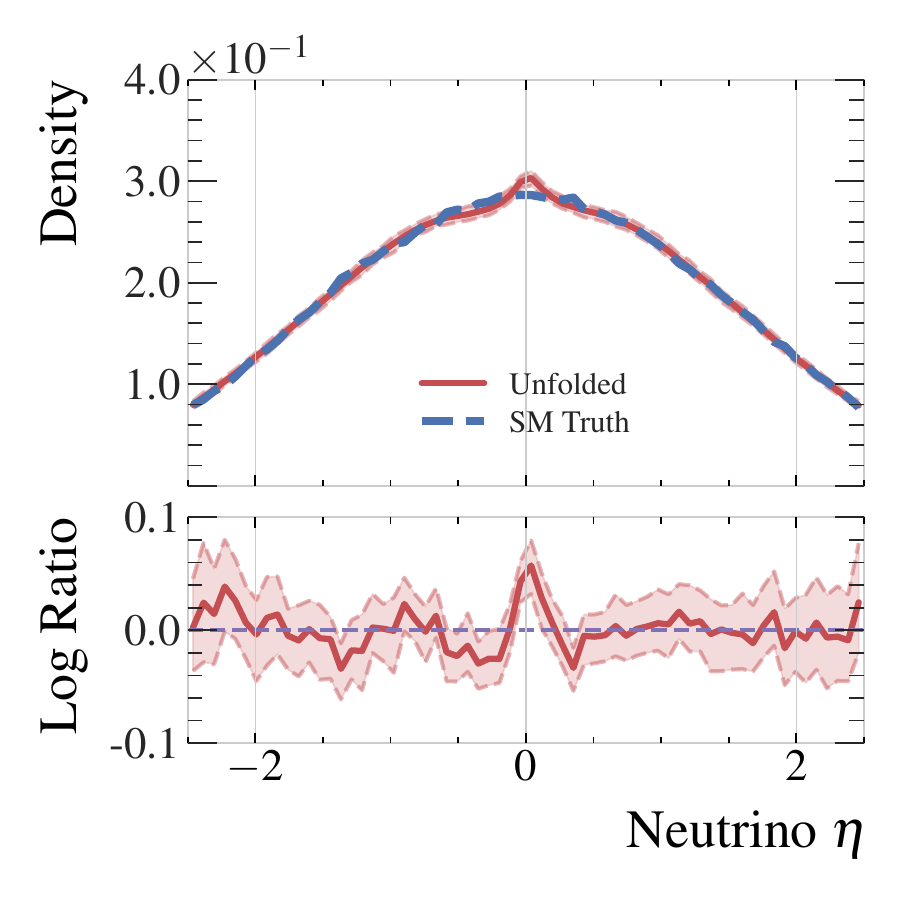}
    \caption{}
    \label{subfig:vl_vld_event_obs_etanu}
  \end{subfigure}
  \caption{Distributions for event-level observables in the test dataset.
  Shown are the true particle-level distribution (dashed blue), the distribution obtained by sampling the learned posterior (solid red), and the detector-level distribution (dotted green).
  Uncertainty bands on the unfolded distributions are estimated by sampling each event 128 times.}
  \label{fig:vl_vld_event_obs}
\end{figure}

Generally excellent agreement between the distributions built from the learned posterior and the truth particle-level events is observed.
What disagreements do exist are typically observed at the edges of kinematic distributions.
For example there is mis-modeling at low jet mass in \Cref{subfig:vl_vld_kinematics_jets_mass} and at extreme $\eta$ in \Cref{subfig:vl_vld_kinematics_jets_mass,subfig:vl_vld_kinematics_leptons_eta}.
This can be understood as the result of a lack of training examples, for example at low jet mass, or events migrating across the event selection boundaries between particle and detector level, for example at extreme $\eta$.
These issues could be fixed by larger-scale training and careful consideration of the event selections applied to the training samples in a realistic analysis.
\Cref{fig:vl_vld_event_obs} shows the sample distributions for the event-level observables $E_T^\text{miss}$, $\phi^\text{miss}$, and the neutrino pseudorapidity $\eta_\nu$.
Faithful reproduction of the truth particle-level distributions is observed for the first two, while a slight excess of density around $\eta_\nu = 0$ is observed in \Cref{subfig:vl_vld_event_obs_etanu}.
As discussed in the previous section, $\eta_\nu$ is not strongly constrained at detector level.
This excess density can be interpreted as events where the detector-level constraint is very weak so the model simply returns the mean of the target distribution.

\begin{figure}[tb]
  \centering
  \begin{subfigure}{0.45\linewidth}
    \centering
    \includegraphics[width=\linewidth, alt={Jet multiplicity distribution comparing truth, unfolded, and detector levels. Good agreement is shown between truth and unfolded.}]{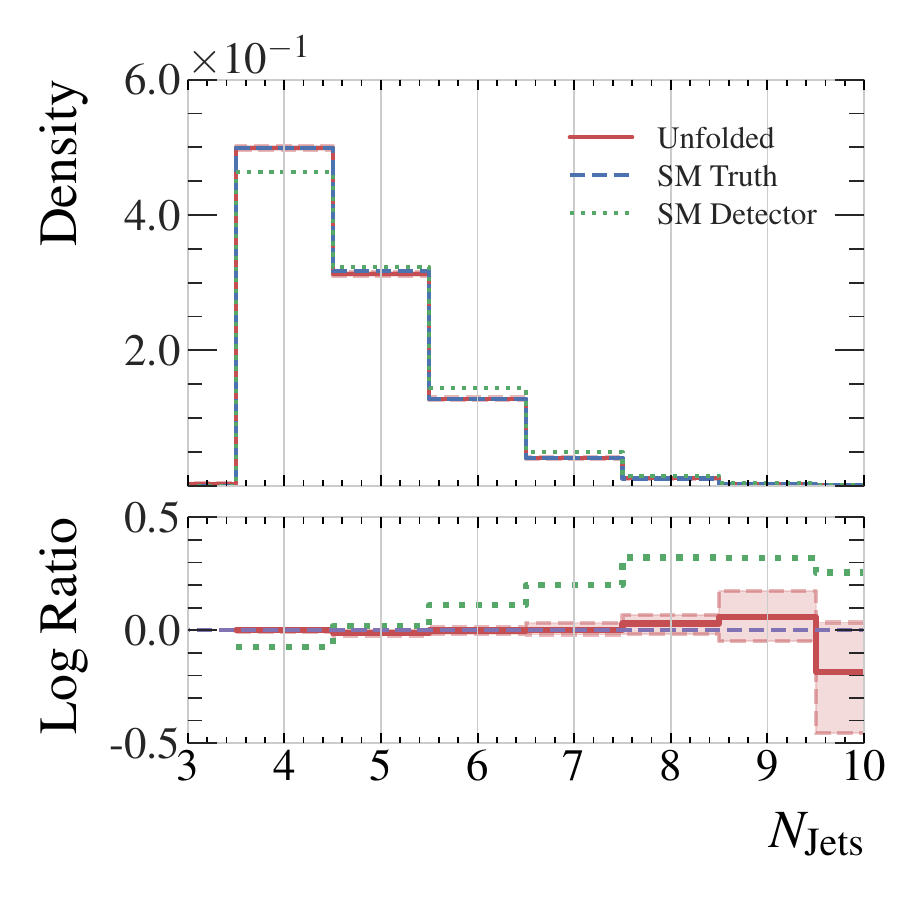}
    \caption{}
  \end{subfigure}\hfill
  \begin{subfigure}{0.45\linewidth}
    \centering
    \includegraphics[width=\linewidth, alt={H_T (scalar pT sum) distribution comparing truth, unfolded, and detector levels, with a slight disagreement at low values.}]{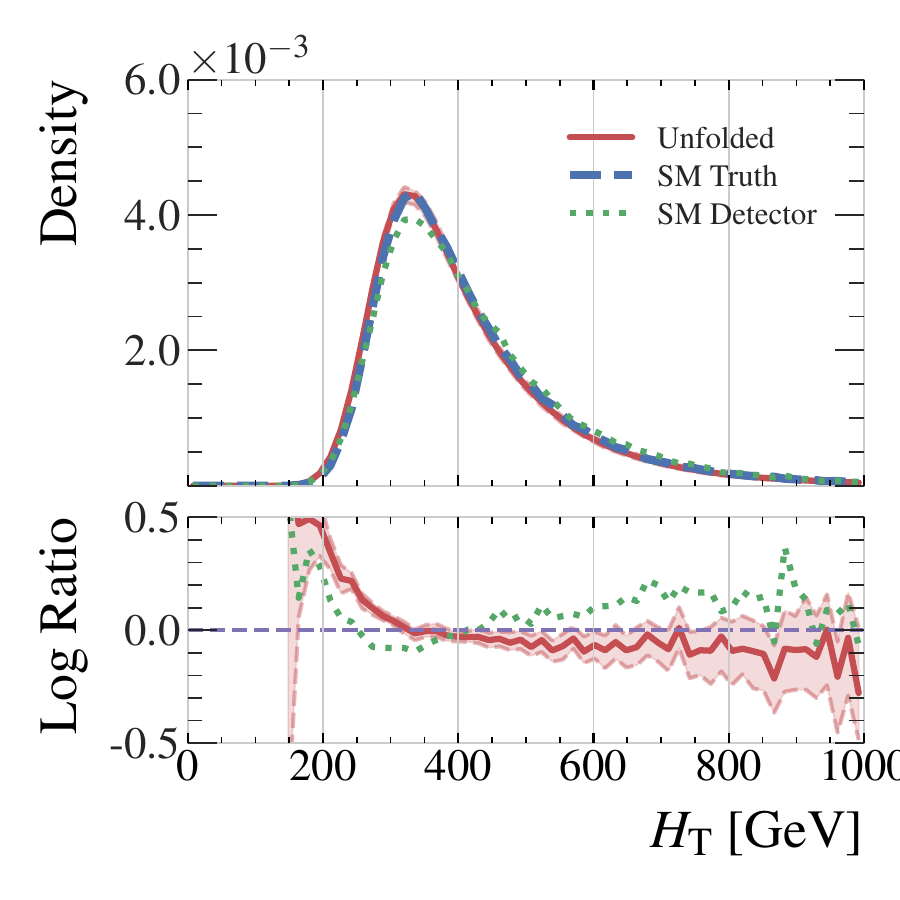}
    \caption{}
    \label{subfig:vl_vld_multiplicity_ht}
  \end{subfigure}
  \caption{Distributions of the jet multiplicity and the \HT observables in the test dataset.
  Shown are the true particle-level distribution (dashed blue), the distribution obtained by sampling the learned posterior (solid red), and the detector-level distribution (dotted green).
  Uncertainty bands on the unfolded distributions are estimated by sampling each event 128 times.}
  \label{fig:vl_vld_multiplicity}
\end{figure}

Distributions of the jet multiplicity and the scalar sum of the \pt of all objects in an event (\HT) are shown in \Cref{fig:vl_vld_multiplicity}.
The particle-level jet multiplicity distribution is reproduced with high accuracy, showing that the VL-VLD model is capable of accommodating a variable-dimensional generative task.
As with the jet mass distributions, the \HT distributions in \Cref{subfig:vl_vld_multiplicity_ht} shows some mismodeling in the low statistics tails, reflecting the lack of training examples in this region of phase space.

\Cref{tab:vl_vld_kinematics_distance} shows the Wasserstein, energy, and Kullback--Leibler distances between two pairs of distributions: the truth particle-level distribution and the distribution predicted by the learned posterior (Unfolded), and the truth particle-level distribution and the detector-level distribution (Detector).
In almost all cases the predicted distributions are closer to the truth than the detector-level distributions which are used to condition the generation.
This is further evidence that the VL-VLD model is correctly modeling the posterior.
Given this is the first method that has been used to solve a variable-dimensional posterior-modeling task, no comparison to existing baselines is possible.

\begin{table}[tb]
  \caption{Wasserstein, energy, and Kullback--Leibler distance metrics between two pairs of distributions: the particle-level truth and the learned posterior (Unfolded) and the particle-level truth and the detector-level (Detector). Metrics are calculated for observables grouped into three categories: jet kinematics (top), lepton kinematics (middle), and event-level observables (bottom).
  Uncertainties on the unfolded distances are estimated by resampling the generative model and recalculating the distances 128 times.}
  \label{tab:vl_vld_kinematics_distance}
  \centering
  \input{tab_vl_vld_jet_lepton_kinematics}
\end{table}

\paragraph{Derived Observables: Top Quark Kinematics.}

\begin{figure}[htb]
  \centering
  \begin{subfigure}{0.3\linewidth}
    \centering
    \includegraphics[width=\linewidth, alt={Hadronic top quark transverse momentum comparing truth, unfolded, and detector levels.}]{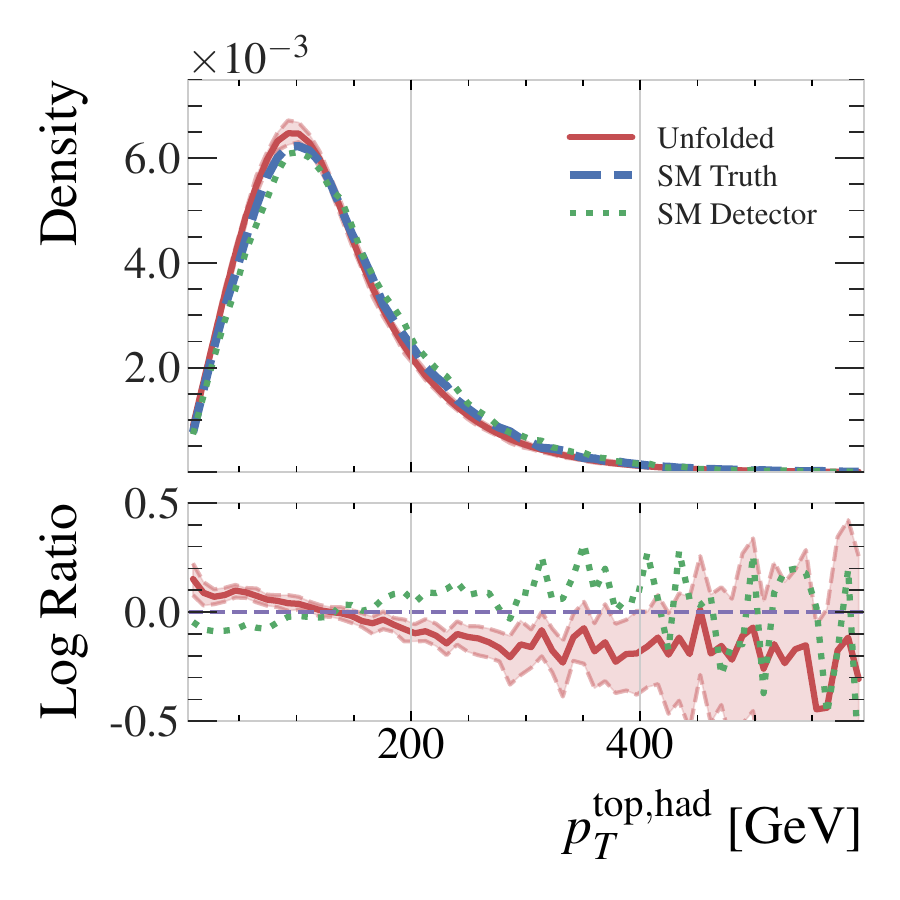}
    \caption{}
  \end{subfigure}\hfill
  \begin{subfigure}{0.3\linewidth}
    \centering
    \includegraphics[width=\linewidth, alt={Hadronic top quark pseudorapidity comparing truth, unfolded, and detector levels.}]{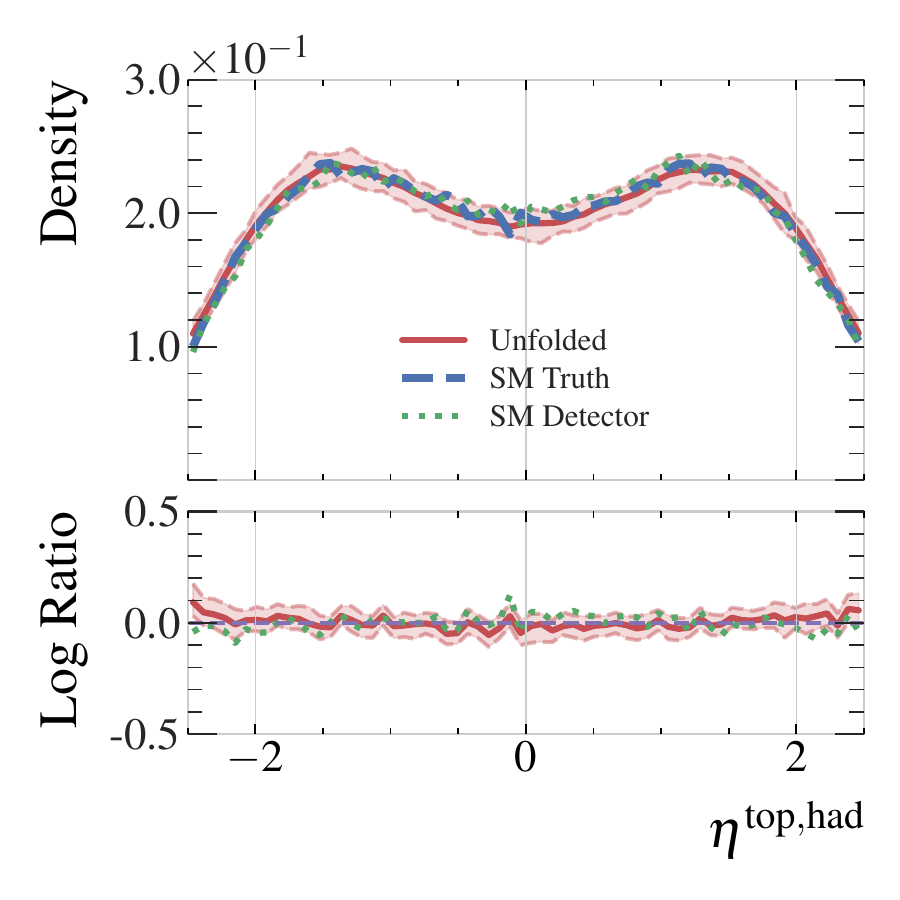}
    \caption{}
  \end{subfigure}\hfill
  \begin{subfigure}{0.3\linewidth}
    \centering
    \includegraphics[width=\linewidth, alt={Hadronic top quark mass distribution comparing truth, unfolded, and detector levels. The sharp truth-level peak near the top mass is not well reproduced.}]{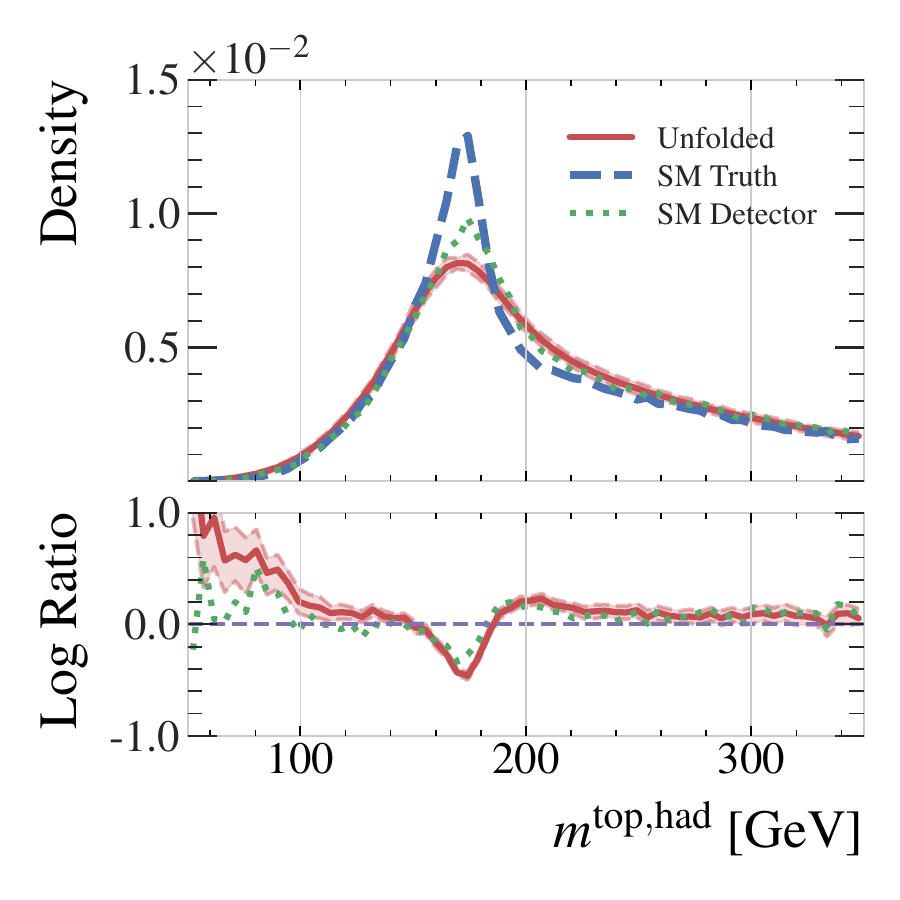}
    \caption{}
  \end{subfigure}
  \par\vspace{8pt}
  \begin{subfigure}{0.3\linewidth}
    \centering
    \includegraphics[width=\linewidth, alt={Leptonic top quark transverse momentum comparing truth, unfolded, and detector levels.}]{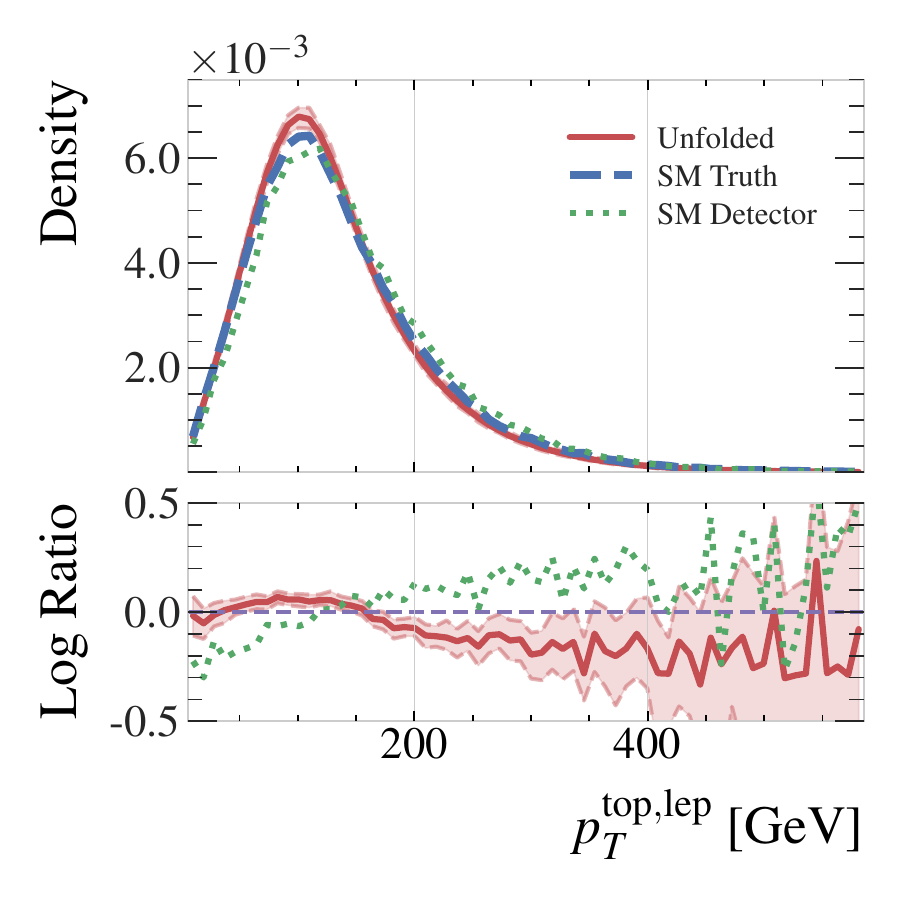}
    \caption{}
  \end{subfigure}\hfill
  \begin{subfigure}{0.3\linewidth}
    \centering
    \includegraphics[width=\linewidth, alt={Leptonic top quark pseudorapidity comparing truth, unfolded, and detector levels.}]{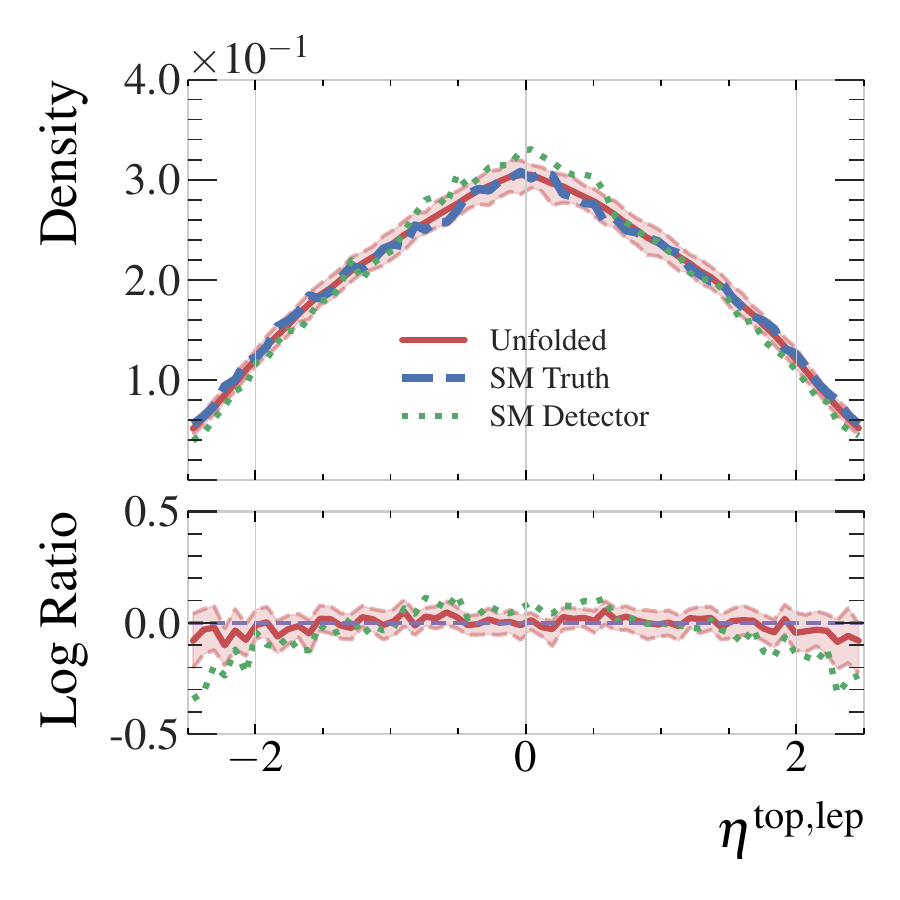}
    \caption{}
  \end{subfigure}\hfill
  \begin{subfigure}{0.3\linewidth}
    \centering
    \includegraphics[width=\linewidth, alt={Leptonic top quark mass distribution comparing truth, unfolded, and detector levels. The sharp truth-level peak is not well reproduced.}]{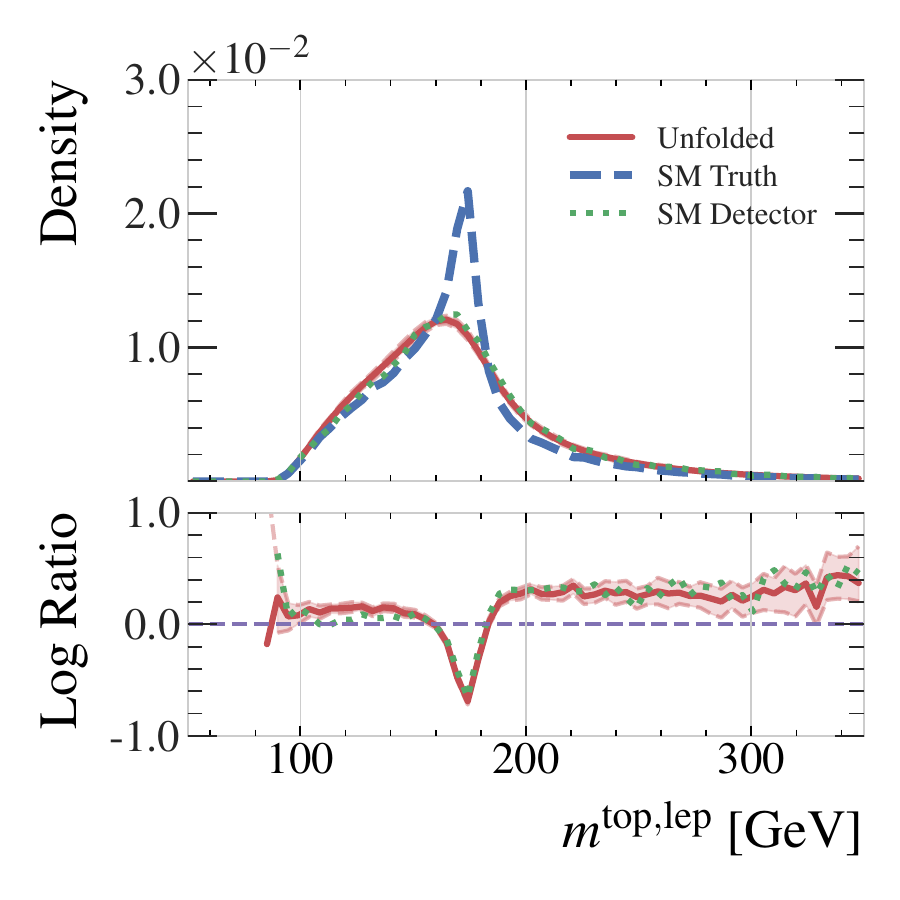}
    \caption{}
  \end{subfigure}
  \par\vspace{8pt}
  \begin{subfigure}{0.3\linewidth}
    \centering
    \includegraphics[width=\linewidth, alt={ttbar system transverse momentum comparing truth, unfolded, and detector levels.}]{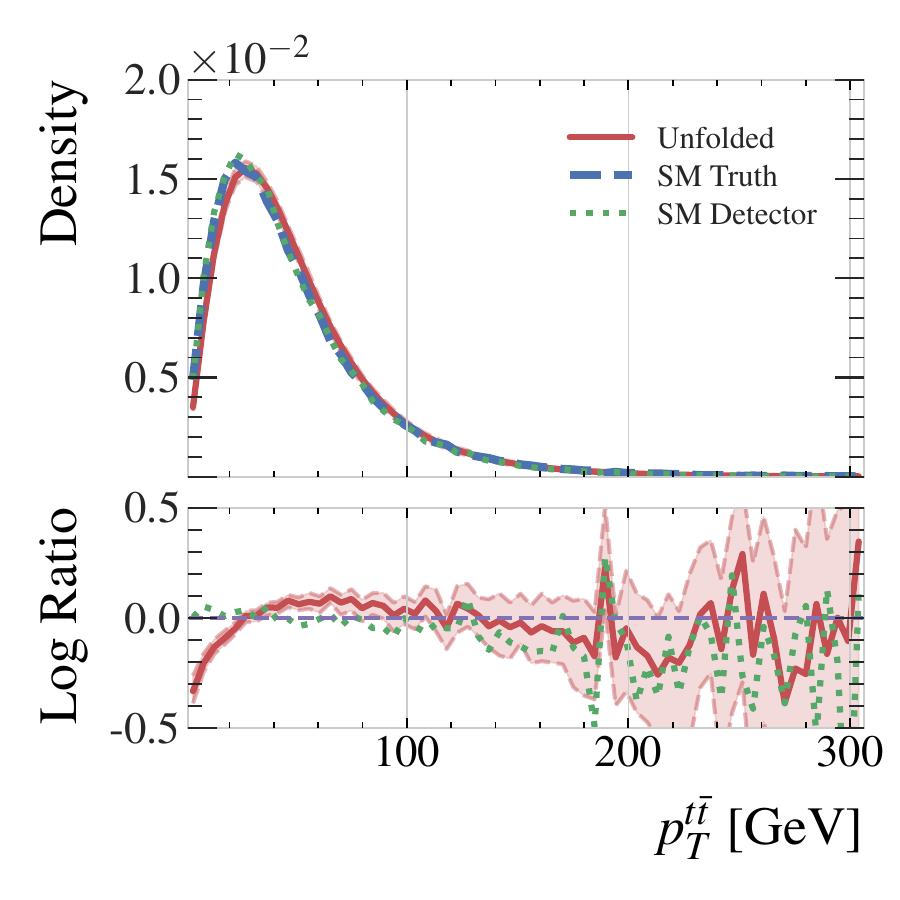}
    \caption{}
  \end{subfigure}\hfill
  \begin{subfigure}{0.3\linewidth}
    \centering
    \includegraphics[width=\linewidth, alt={ttbar system pseudorapidity comparing truth, unfolded, and detector levels.}]{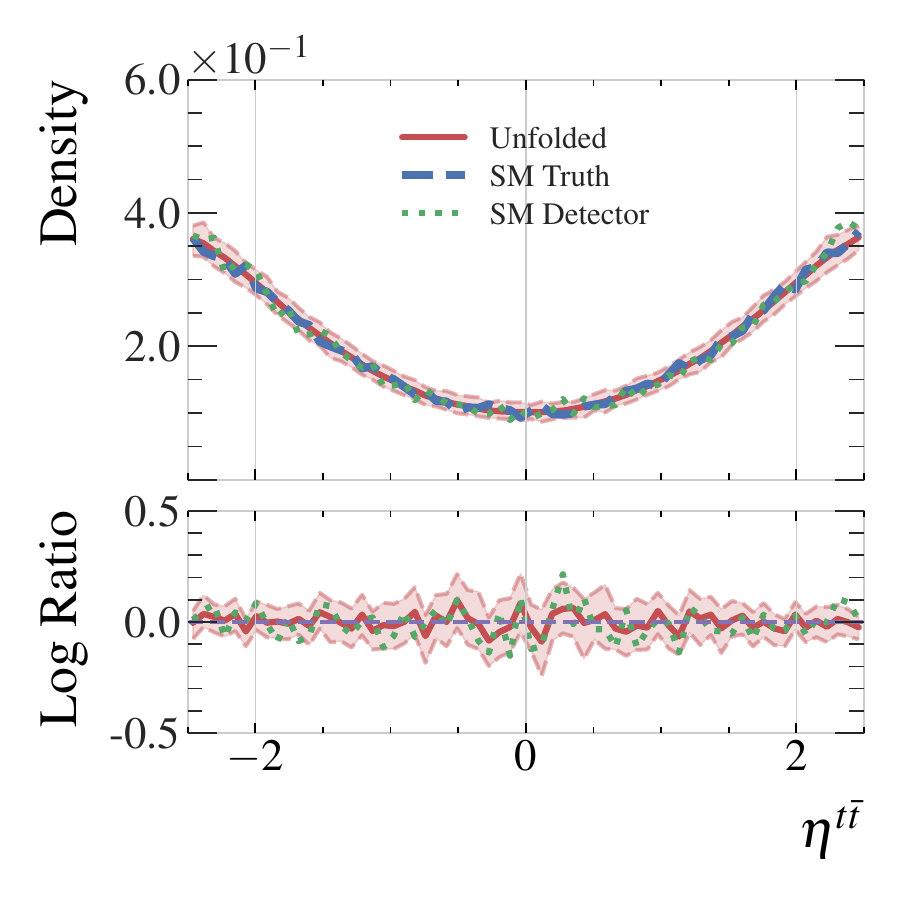}
    \caption{}
  \end{subfigure}\hfill
  \begin{subfigure}{0.3\linewidth}
    \centering
    \includegraphics[width=\linewidth, alt={ttbar invariant mass comparing truth, unfolded, and detector levels.}]{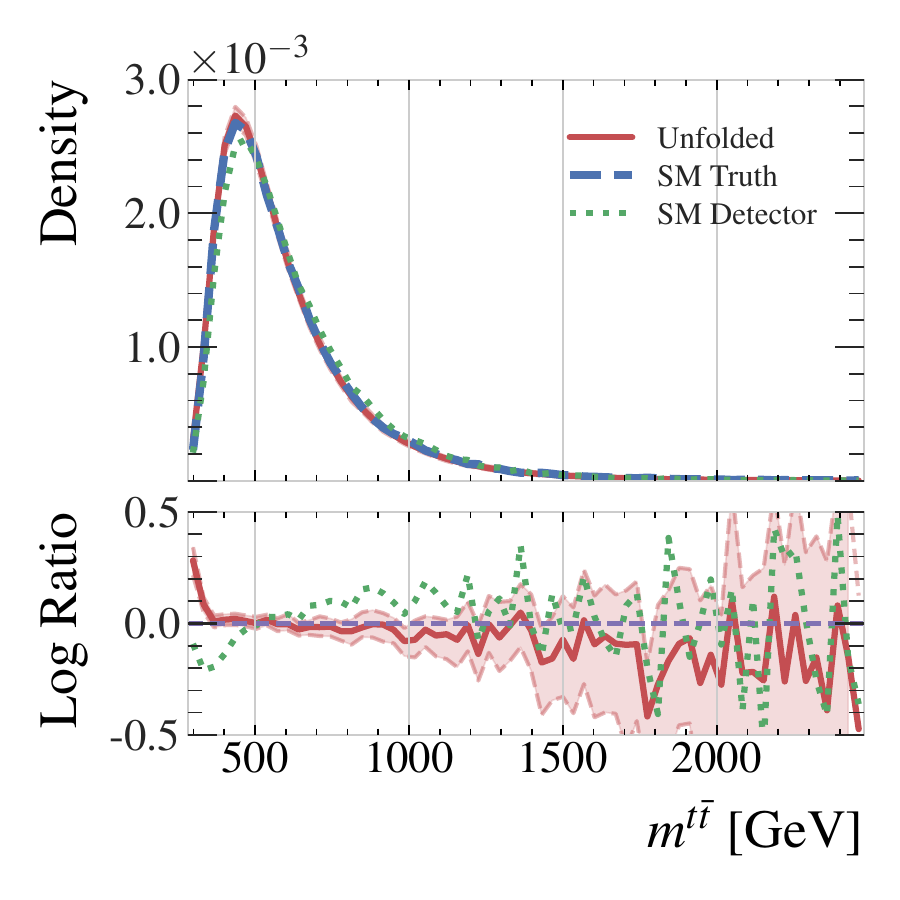}
    \caption{}
  \end{subfigure}
  \caption{Distributions of the kinematics of top quark candidates, assuming the pseudo-top jet parton assignment algorithm.
  All results are built using the test dataset.
  Shown are the true particle-level distribution (dashed blue), the distribution obtained by sampling the learned posterior (solid red), and the detector-level distribution (dotted green).
  Uncertainty bands on the unfolded distributions are estimated by sampling each event 128 times.}
  \label{fig:vl_vld_top_kinematics}
\end{figure}

The motivation for developing a variable-dimensional estimate of the posterior was to allow full-event unfolding with generative models.
Full event unfolding requires that all \textit{derived observables}, or observables that are functions of the object kinematics directly predicted by the neural network, be accurately modeled in the posterior.
In the case of particle-level unfolding of semi-leptonic $t\bar{t}$ production, key derived observables are the kinematics of the top quarks and $t\bar{t}$ system.
These observables were directly optimized in the previous section's parton-level task, but here they are functions of the object kinematics which are not directly optimized.
\Cref{fig:vl_vld_top_kinematics} shows kinematic distributions for the hadronically and leptonically decaying top quark candidates and the $t\bar{t}$ system.
Note that assigning jets and leptons to the hadronic top and leptonic top is not necessarily trivial.
In these results the pseudo-top algorithm is used to assign jets and leptons to these partons, but a key feature of the full-event posterior is that choosing a different algorithm would not require retraining the model.
The agreement between the truth particle-level distributions and the model predictions is considerably worse in these observables than for the directly-optimized observables discussed above.
In particular the mass peaks of the hadronic and leptonic top quarks are not well modeled.
The posterior samples follow essentially the same distribution as the detector-level samples, indicating that the model has learned essentially the identity function in this observable.
As mentioned above, the modeling of sharply peaked distributions is a known difficulty for generative models.
Here the difficulty is exacerbated by the fact that these distributions can not be optimized directly, or constrained through use of a physics-informed loss function as was done in the previous section.
Other derived observables, such as the \pt of the top quarks and the mass of the $t\bar{t}$ system, are better modeled but still have significant disagreements with the particle-level truth.
These issues with derived observables is a significant limitation of the current approach, and motivates the additional methodological developments in the next section.

\paragraph{Two-Dimensional Correlations.}

\begin{figure}[tb]
  \centering
  \begin{subfigure}{0.5\linewidth}
    \centering
    \includegraphics[width=\linewidth, alt={Corner plot of hadronic top quark kinematic variables from the truth-level test set, showing the two-dimensional distributions and marginals for pt, eta, and mass.}]{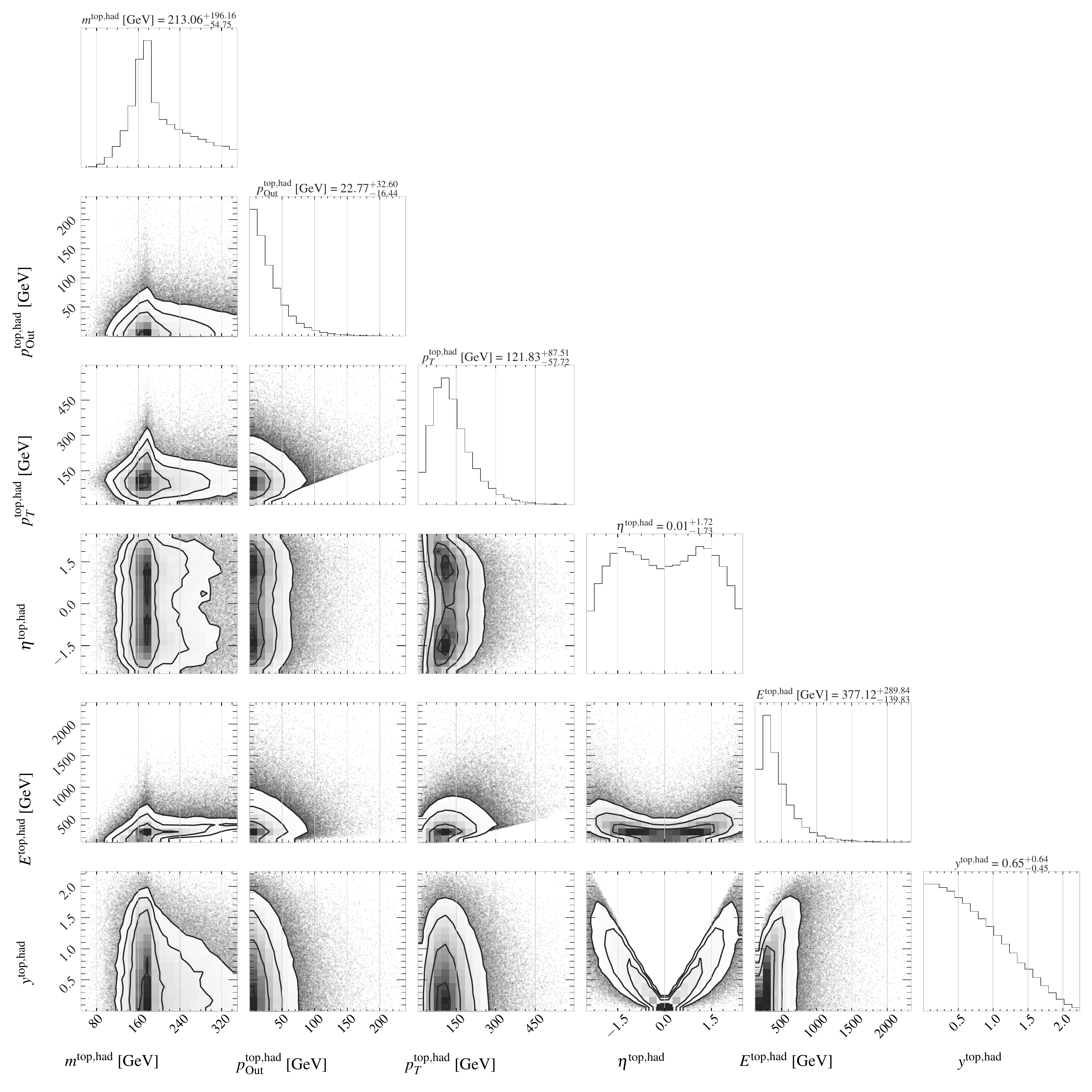}
    \caption{Truth}
  \end{subfigure} \\
  \begin{subfigure}{0.5\linewidth}
    \centering
    \includegraphics[width=\linewidth, alt={Corner plot of hadronic top quark kinematic variables from the VL-VLD model predictions, showing the two-dimensional distributions and marginals for pt, eta, and mass.}]{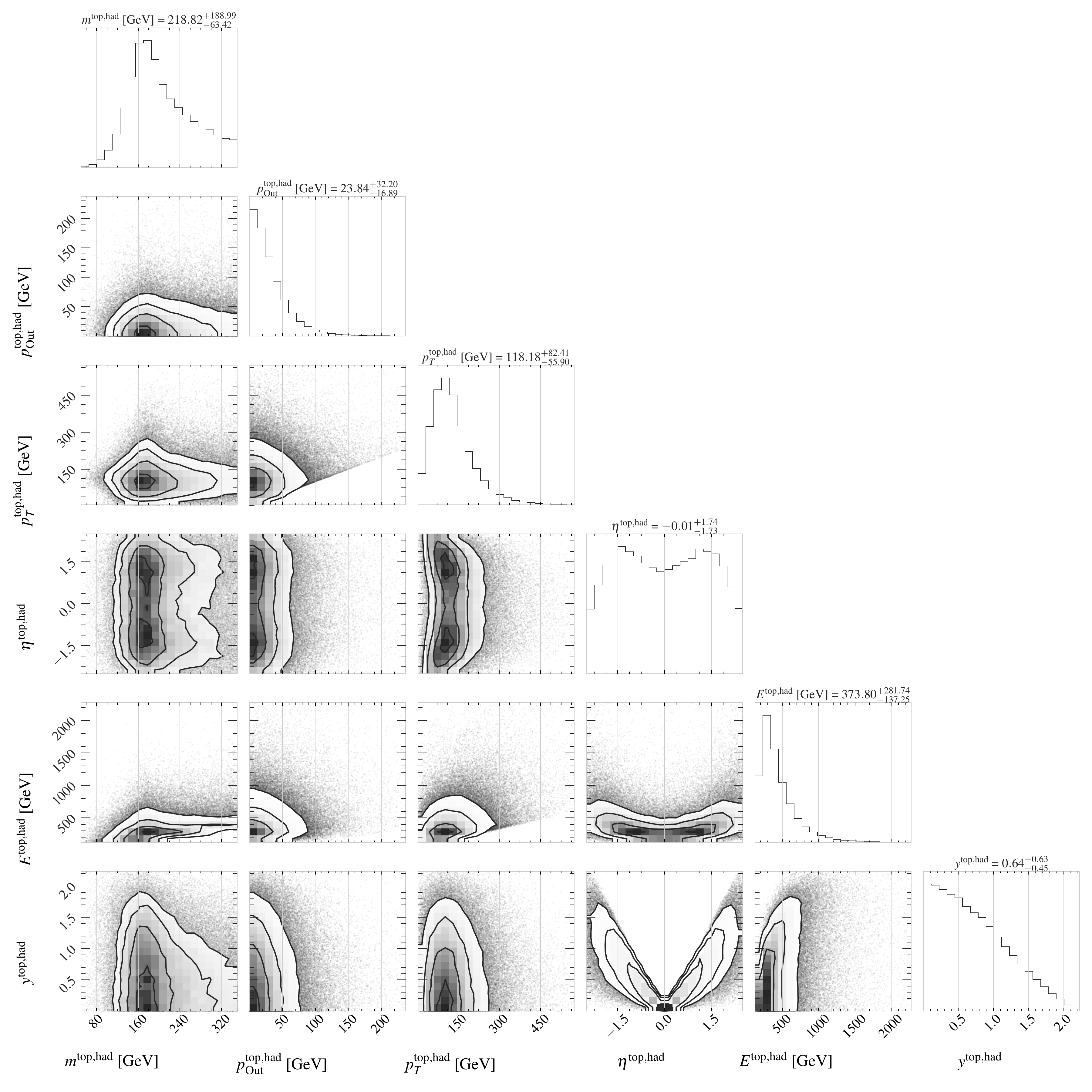}
    \caption{Unfolded}
  \end{subfigure}
  \caption{Corner plots of the hadronic top quark kinematic observables, shown for the truth particle-level distribution (top) and the distribution obtained by sampling from the learned posterior (bottom).
  The two-dimensional joint distributions between each kinematic observable are visualized with contour plots.}
  \label{fig:vl_vld_corner}
\end{figure}

In addition to inspecting the one-dimensional marginal distributions of the top quark kinematics, it is also useful to inspect two-dimensional distributions.
This ensures that the correlations between different observables are properly captured in the modeled posterior.
\Cref{fig:vl_vld_corner} shows corner plots of the hadronic top quark kinematic observables for both the truth particle-level distributions and for the VL-VLD model predictions.
The correlations are visibly very similar, indicating that the VL-VLD model is correctly capturing the correlations between these observables and not only reproducing the correct marginal distributions.
The VL-VLD model is able to capture the overall structure of the ground truth posterior, but the precision of the learned model is lacking.
Greater precision often boils down to finding the correct parametrization for the generative task, as will be shown in the next section.

\paragraph{Performance on a BSM Prior.}

Modeling of the posterior is only the first step in the EM algorithm needed to produce a proper unfolding.
The M-step and subsequent iterations exist to remove the prior dependence from the learned posterior and arrive at the maximum likelihood solution.
While we did not perform the iterative procedure explicitly in Ref.~\cite{Shmakov:2024gkd}, we did explore this direction by examining the prior dependence of the learned posterior.
Specifically the VL-VLD network trained on Standard Model $t\bar{t}$ events is evaluated on an alternative test sample containing BSM physics injected through a non-zero EFT coefficient in the top-gluon vertex ($c_{tg} = 25$).
This is a large coefficient and is rather unphysical, but it produces a sizeable shift in the kinematic distributions of the top quarks and $t\bar{t}$ system that is useful for assessing the prior dependence.

\begin{figure}[htb]
  \centering
  \begin{subfigure}{0.3\linewidth}
    \centering
    \includegraphics[width=\linewidth, alt={Hadronic top quark transverse momentum in the EFT test dataset, comparing EFT truth and unfolded with SM truth and unfolded for reference.}]{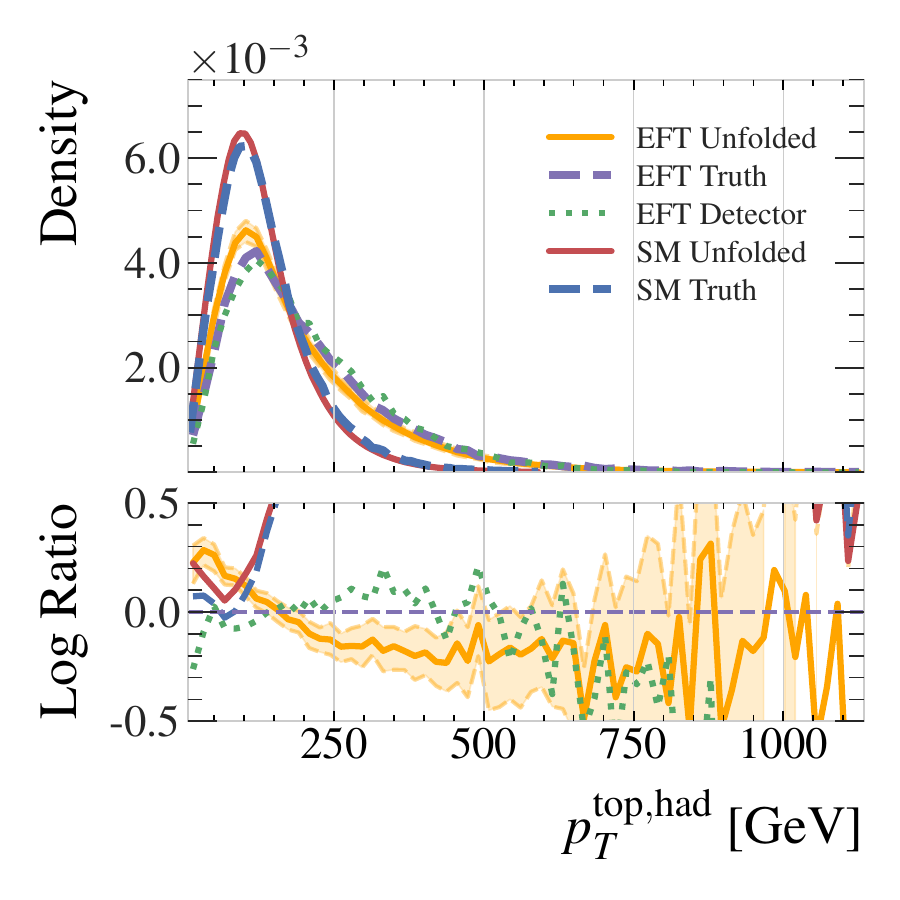}
    \caption{}
  \end{subfigure}\hfill
  \begin{subfigure}{0.3\linewidth}
    \centering
    \includegraphics[width=\linewidth, alt={Hadronic top quark pseudorapidity in the EFT dataset.}]{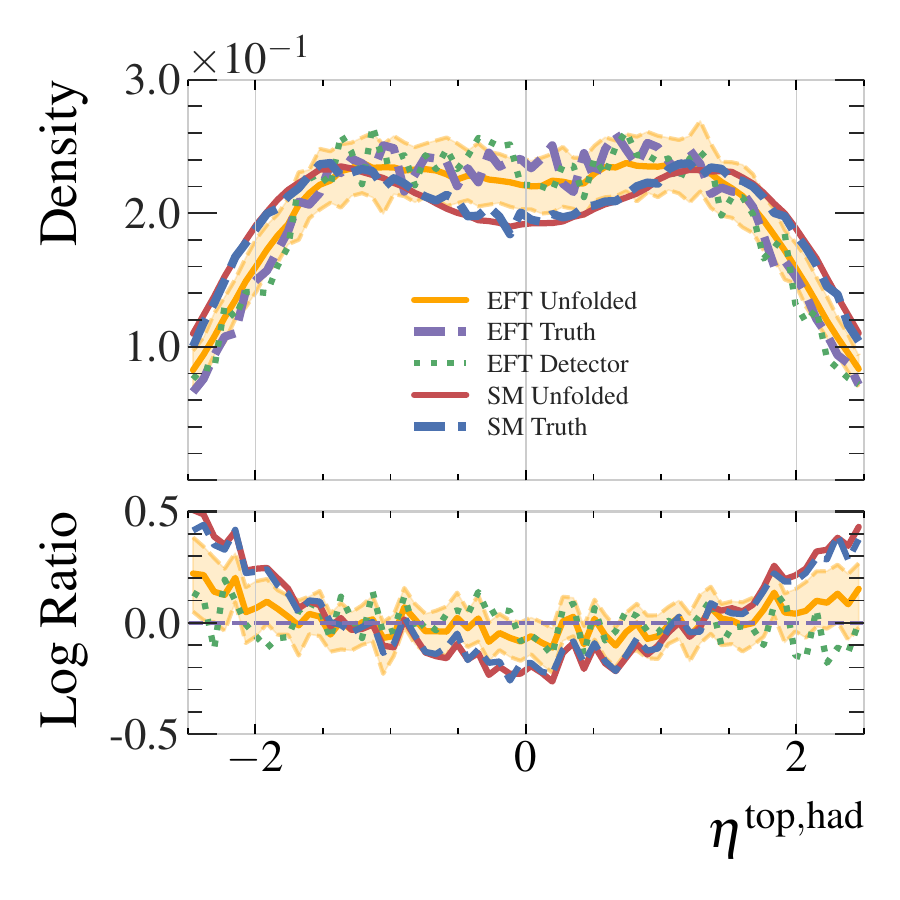}
    \caption{}
  \end{subfigure}\hfill
  \begin{subfigure}{0.3\linewidth}
    \centering
    \includegraphics[width=\linewidth, alt={Hadronic top quark mass in the EFT dataset.}]{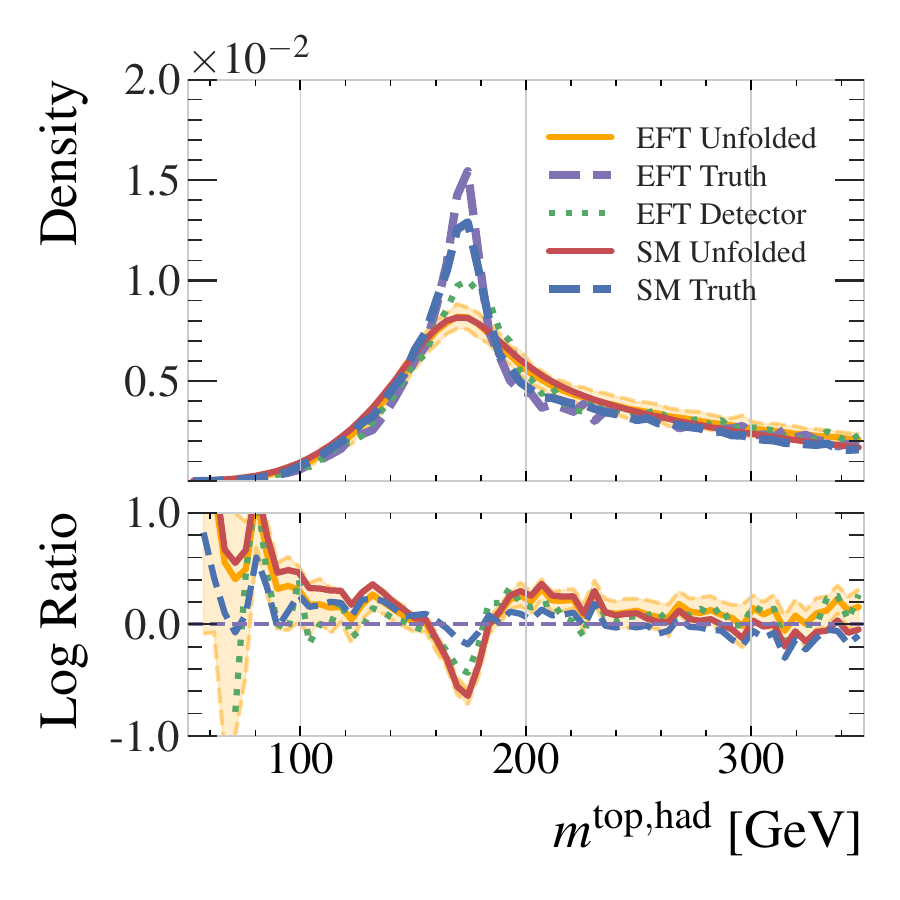}
    \caption{}
  \end{subfigure}
  \par\vspace{8pt}
  \begin{subfigure}{0.3\linewidth}
    \centering
    \includegraphics[width=\linewidth, alt={Leptonic top quark transverse momentum in the EFT dataset.}]{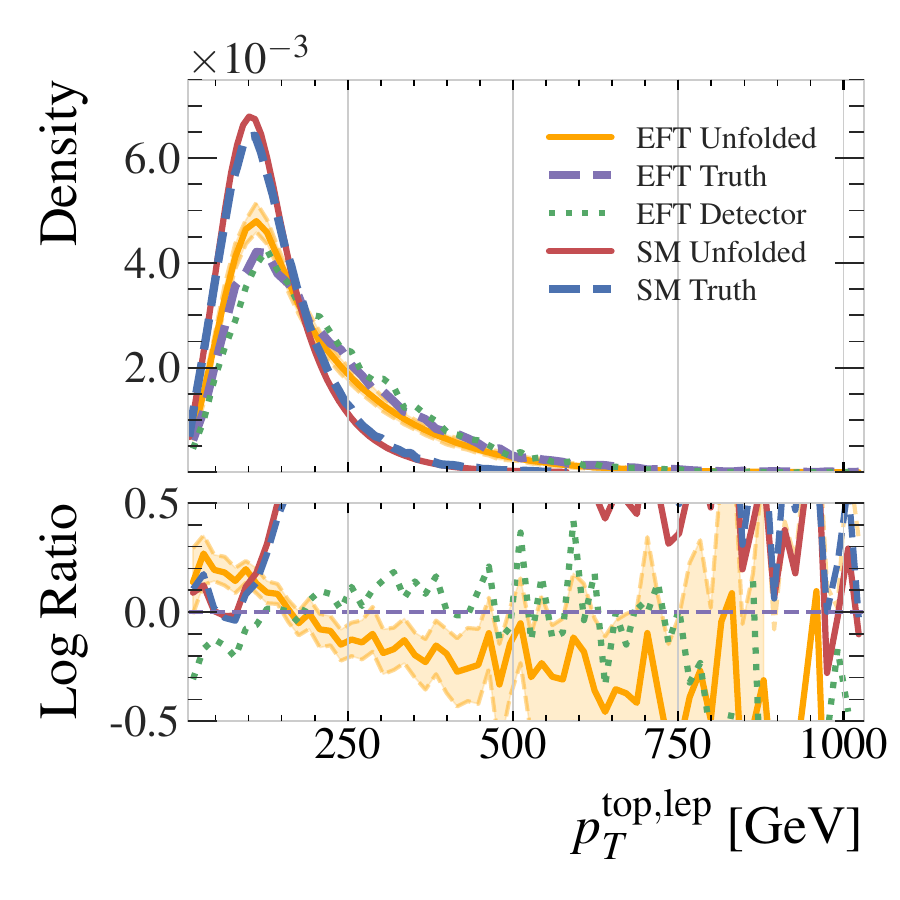}
    \caption{}
  \end{subfigure}\hfill
  \begin{subfigure}{0.3\linewidth}
    \centering
    \includegraphics[width=\linewidth, alt={Leptonic top quark pseudorapidity in the EFT dataset.}]{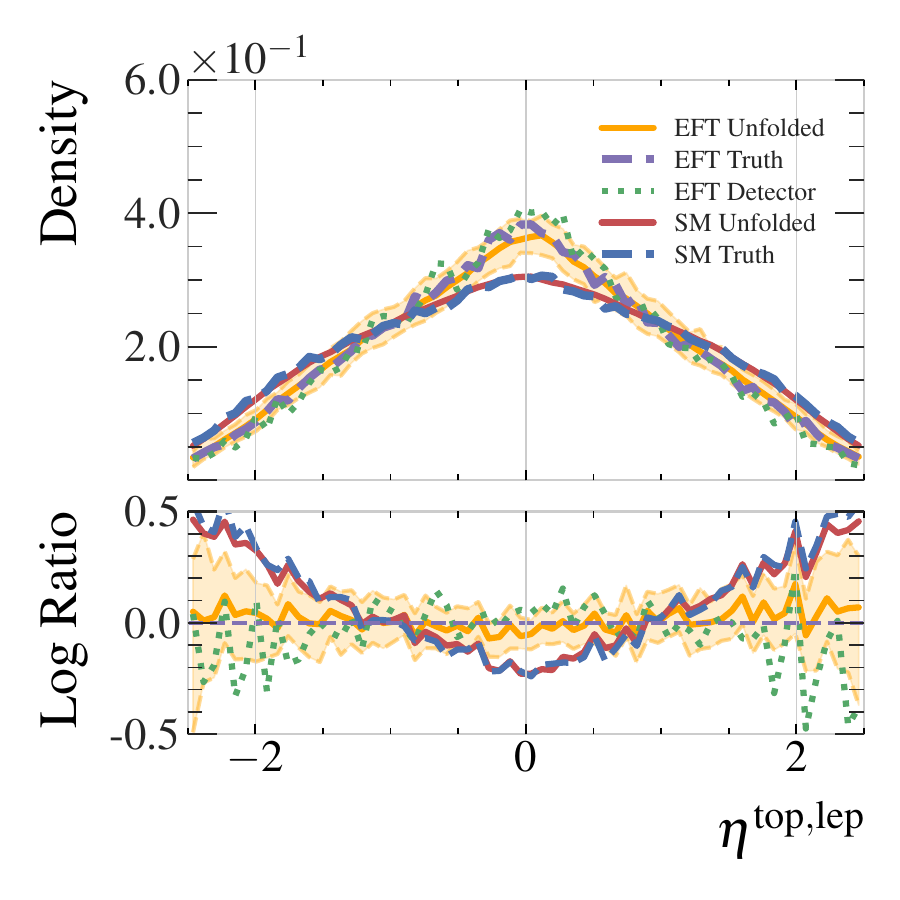}
    \caption{}
  \end{subfigure}\hfill
  \begin{subfigure}{0.3\linewidth}
    \centering
    \includegraphics[width=\linewidth, alt={Leptonic top quark mass in the EFT dataset.}]{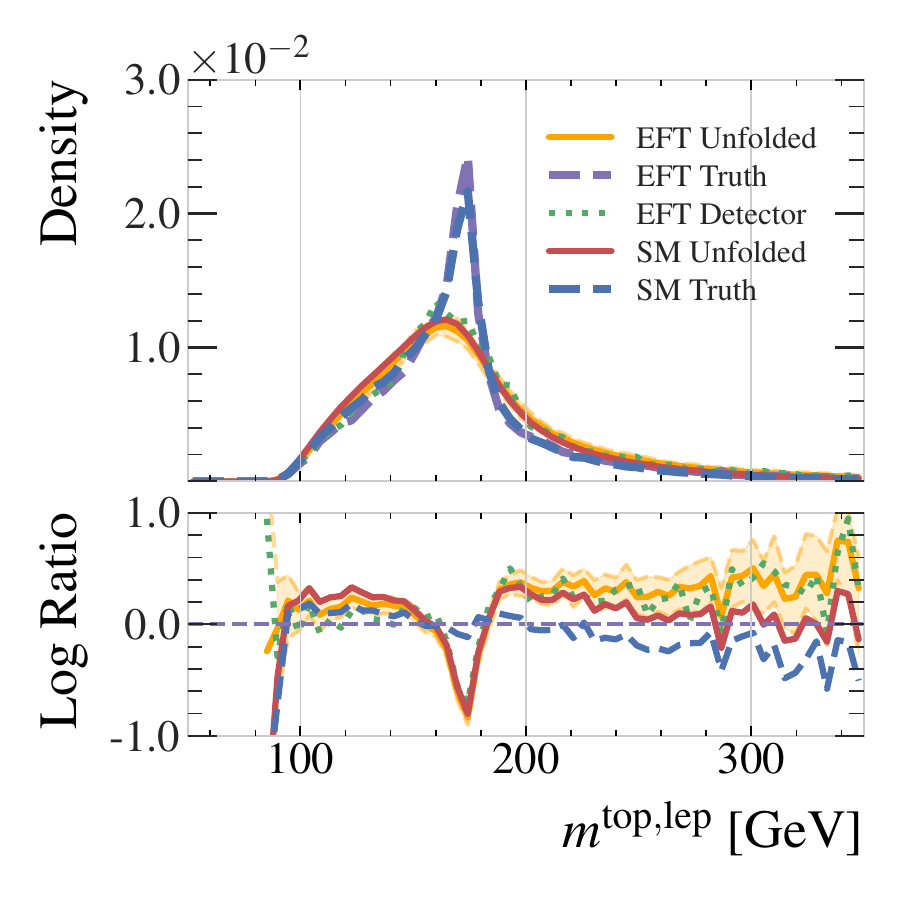}
    \caption{}
  \end{subfigure}
  \par\vspace{8pt}
  \begin{subfigure}{0.3\linewidth}
    \centering
    \includegraphics[width=\linewidth, alt={ttbar system transverse momentum in the EFT dataset.}]{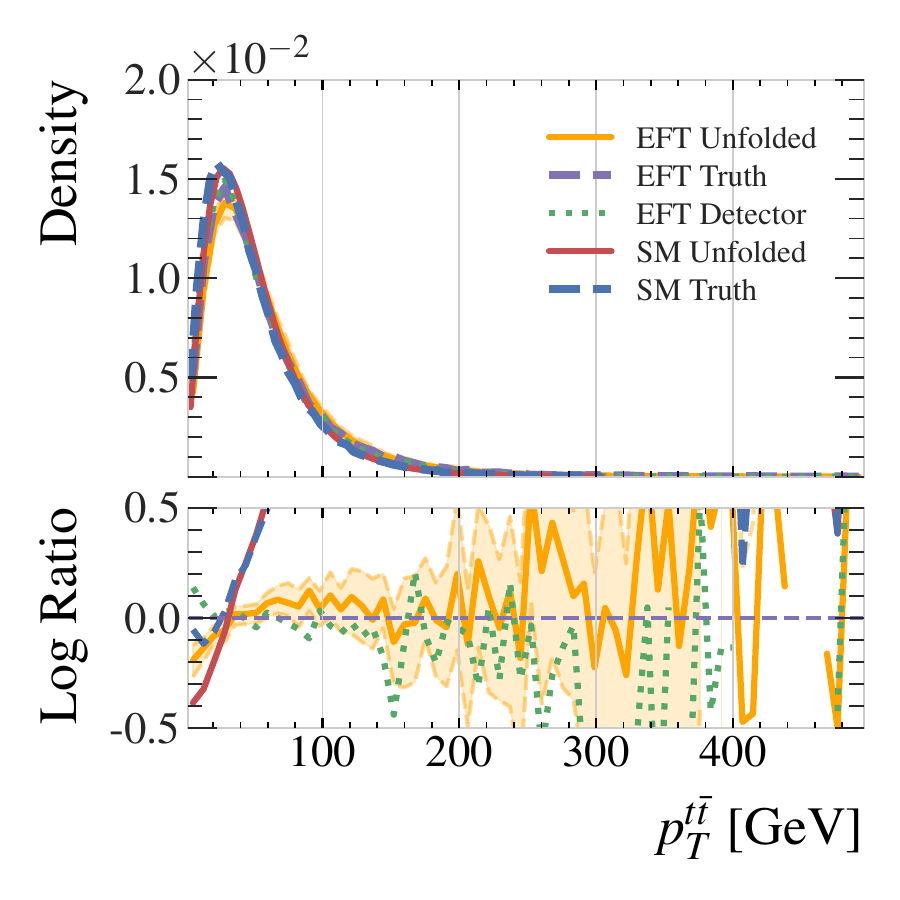}
    \caption{}
  \end{subfigure}\hfill
  \begin{subfigure}{0.3\linewidth}
    \centering
    \includegraphics[width=\linewidth, alt={ttbar pseudorapidity in the EFT dataset.}]{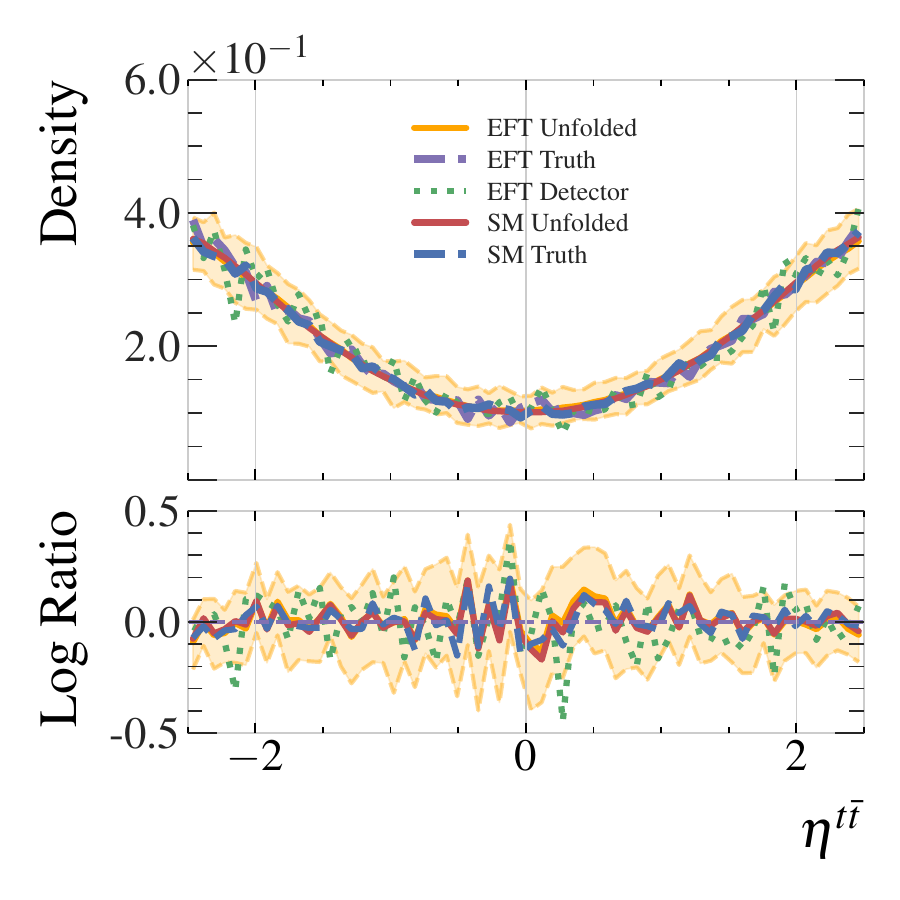}
    \caption{}
  \end{subfigure}\hfill
  \begin{subfigure}{0.3\linewidth}
    \centering
    \includegraphics[width=\linewidth, alt={ttbar invariant mass in the EFT dataset.}]{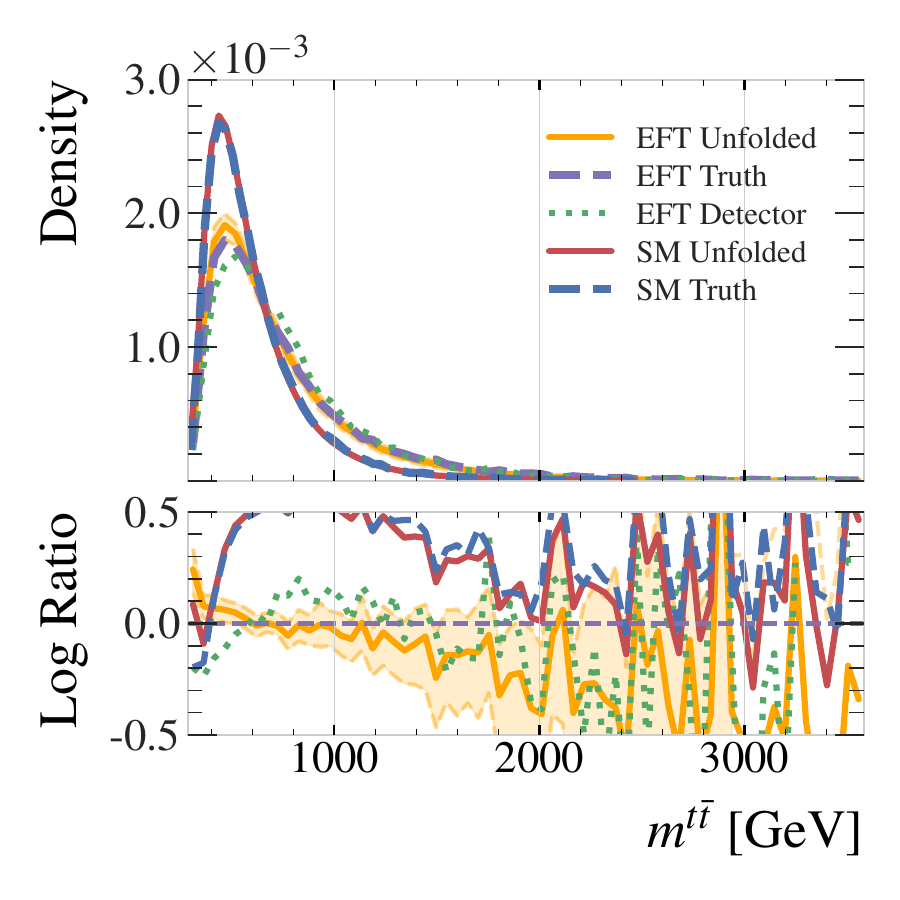}
    \caption{}
  \end{subfigure}
  \caption{Distributions of selected kinematic observables relating to the top quark candidates and $t\bar{t}$ system, analogous to those shown in \Cref{fig:vl_vld_top_kinematics}. Shown are the EFT particle-level truth distribution (dashed purple), the distribution obtained by sampling the learned posterior with the EFT test set (solid orange), and the EFT detector-level distribution (dotted green). The SM particle-level truth distribution (dashed blue) and the SM modeled distribution (solid red) are also provided for visual comparison.
  Uncertainty bands on the modeled distributions are estimated by sampling each event 128 times.}
  \label{fig:vl_vld_eft}
\end{figure}

\Cref{fig:vl_vld_eft} shows the distributions of various top quark and $t\bar{t}$ system kinematic observables in the EFT test dataset, in addition to the SM distributions previously presented in \Cref{fig:vl_vld_top_kinematics}.
While there are visible disagreements between the EFT particle-level truth and the learned posterior sampled using the EFT test set, these disagreements are in general no worse than the disagreements between the truth and modeled distributions in the SM test set.
Generally there is a very low level of prior dependence visible in these observables, despite the large prior-shift produced by the non-zero EFT operator which can be visualized as the difference between the EFT and SM truth distributions.
This is strong evidence that the EM algorithm should be capable of removing any residual prior dependence and producing a proper unfolding.
Applying the EM algorithm would still be necessary in an application of this method to data, but this preliminary check shows that the prior dependence is at least under control.

\subsubsection{Summary}

Ref.~\cite{Shmakov:2024gkd} presents the first method for modeling a variable-dimensional posterior in the context of unfolding.
The VLD model of Ref.~\cite{Shmakov:2023kjj} is extended with transformer-based neural networks to handle position-equivariant set data types, per-position noise schedules, and a multiplicity predictor network to ensure variable-length generation.
The VL-VLD model is applied to model the posterior for particle-level event configurations conditioned on the detector-level data for semi-leptonic $t\bar{t}$ events.
Unlike the posterior modeling task in the previous section, this task is inherently variable dimensional and the VL-VLD model is able to accurately reproduce observables relating to this variable dimensionality such as the jet multiplicity.
Other kinematic observables that are directly optimized show excellent agreement with the particle-level truth, while derived observables such as those relating to the kinematics of candidate top quarks are not captured as accurately.
This shortcoming motivates the work in the next section which investigates how clever parametrizations of the generative task can improve accuracy.
Finally the prior dependence of the learned posterior is assessed by sampling the posterior trained on an SM sample using a sample that has been adjusted by a large non-zero EFT operator.
Despite the sizeable shift in the truth-level distributions, the VL-VLD model is able to adjust its predictions based on the condition provided by the detector-level data.
This is strong evidence that the EM algorithm could be applied to build a complete unfolding method using the VL-VLD model.

\FloatBarrier

\subsection{Unfolding Jet Substructure with Flow Matching}
\label{sec:cfm-jets}

The results discussed in the previous two sections demonstrated that generative models can be used to model rather complicated posterior distributions in up to several tens of dimensions.
However accuracy gaps remained, most notably in the derived observables calculated using the variable dimensional particle-level posterior in the previous section.
Ref.~\cite{Petitjean:2025tgk} addresses these gaps within the particular task of modeling the particle-level posterior describing the substructure of a single large-radius jet, and is the fourth original research contribution of this thesis.
While the results in the previous section clustered hadronic particles into jets and modeled the posterior as a function of these jets, the present results instead model the posterior as a function of all particles within a single jet.
As in \Cref{ch:tagging}, jets at the LHC contain several tens of constituent particles each described by a four-vector.
Therefore the target particle-level phase space has $\mathcal{O}(100)$ dimensions.
Extension of the methods presented here from single jets to entire event should be possible, but is left to future work.

\subsubsection{Three-Stage Unfolding Framework}

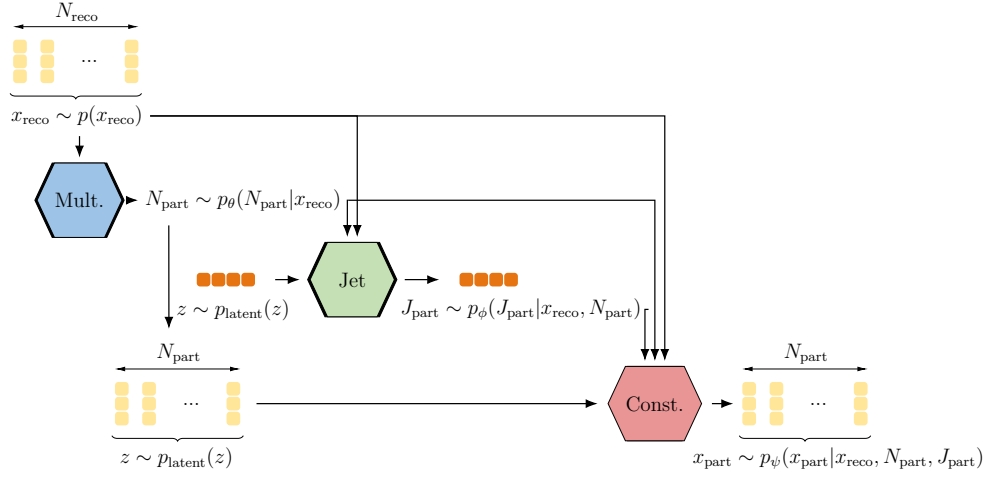
\begin{figure}[tb]
  \centering
  \begingroup
  \def\xp{x_\text{part}}%
  \def\xr{x_\text{reco}}%
  \def\Npart{N_\text{part}}%
  \def\Ndet{N_\text{reco}}%
  \def\Jpart{J_\text{part}}%
  \def\Jdet{J_\text{reco}}%
  \scalebox{0.82}{\input{ch5_cfm_jets_model.tex}}
  \endgroup
  \caption{A diagram of the three-stage posterior decomposition developed in Ref.~\cite{Petitjean:2025tgk}.
  The posterior is conditional in only the detector-level jet represented by $x_\text{reco}$.
  This is used to condition the multiplicity predictor (Mult) which provides the particle-level constituent count $N_\text{part}$.
  All available information then conditions the jet predictor (Jet) which produces the particle-level jet kinematics $J_\text{part}$.
  Finally the constituent predictor (Const) uses all available information to predict the particle-level jet constituents, in the parametrization provided by $J_\text{part}$.
  The orange squares represent the jet four-momentum components, and the yellow squares represent the constituent three-momenta $(\delta_{\log p_T}, \delta_\phi, \delta_\eta)$.}
  \label{fig:cfm_model}
\end{figure}

The goal of this section is to model the posterior of the kinematics of each constituent particle of a particle-level jet, conditioned on the same quantities at detector level.
As in the previous section this is a set-conditional set generation task, and the conditioning and target sets need not have the same cardinality.
The previous section developed an effective machine learning model that could learn variable-dimensional posteriors, but used a somewhat basic representation of the data at detector and particle-level.
In Ref.~\cite{Petitjean:2025tgk}, we sought to improve on the accuracy by developing a physics-inspired autoregressive factorization of the desired posterior:
\begin{align}
  p_{\theta,\phi,\psi}(x_\text{part}, J_\text{part}, N_\text{part} \,|\, x_\text{reco}) &=
  p_\theta(N_\text{part} \,|\, x_\text{reco}) \nonumber \\
  &\times p_\phi(J_\text{part} \,|\, x_\text{reco}, N_\text{part}) \nonumber \\
  &\times p_\psi(x_\text{part} \,|\, x_\text{reco}, N_\text{part}, J_\text{part}),
  \label{eq:cfm_factorization}
\end{align}
where $N_\text{part}$ is the particle-level constituent multiplicity and $J_\text{part}$ the particle-level jet kinematics.
Each of the factors is modeled by an independent neural network with trainable parameters $\theta$, $\phi$, and $\psi$ respectively.
The network that models $p_\theta(N_\text{part} \,|\, x_\text{reco})$ serves the same role as the multiplicity predictor network in the previous section.
The novel contribution of this section is therefore the factorization of the second two terms: the posterior of the jet kinematics and the posterior of the constituent kinematics conditional on the jet kinematics.
The overall framework is illustrated in \Cref{fig:cfm_model} and each neural network is implemented as follows:
\begin{enumerate}
  \item \textbf{Multiplicity predictor}: Similar to the previous section, the multiplicity predictor is implemented as a transformer encoder that processes the detector-level condition.
  However instead of parametrizing a Gamma distribution, the outputs of the network parametrize a Gaussian mixture model, composed of five Gaussians with learnable means, variances, and weights.
  The multiplicity prediction is obtained by sampling from the distribution and rounding to the nearest integer.
  \item \textbf{Jet predictor}: A CFM model is used to model the posterior of the jet kinematics, conditioned on the detector-level jet kinematics and $N_\text{part}$.
  The jet four-momentum are represented as $J_\text{part} = (\log p_{T,J},\, \phi_J,\, \eta_J,\, \log m_J^2)$.
  Note that given we wish to unfold the jet substructure, $J_\text{part}$ is not the final product of the model.
  Instead it is used autoregressively to define a physically motivated parametrization of the constituent posterior.
  \item \textbf{Constituent predictor}: A second, larger CFM network models the posterior of the jet constituents, conditional on the detector-level jet kinematics, $N_\text{part}$, and $J_\text{part}$.
  The particle-level jet kinematics are additionally used to define relative coordinates in which all jet kinematics are represented\footnote{The detector-level jet kinematics are represented in the equivalent coordinates, but using the detector-level jet kinematics.}:
  \begin{equation}
    (\delta_{\log p_T},\, \delta_\phi,\, \delta_\eta) = (\log p_T - \log p_{T,J},\, \phi - \phi_J,\, \eta - \eta_J).
    \label{eq:cfm_relative}
  \end{equation}
  There are two benefits to this coordinate system.
  First, the distributions of the jet kinematics are close to normally distributed, in that they have close to zero mean and unit standard deviation, which is known to be useful for generative modeling.
  Second, the invariance of jet configurations to boosts and rotations of the collision system is used to factorize the generation process so the constituent predictor must only learn jet substructure and not more global distribution such as the jet rapidity.
  The constituents predicted by this network are the final output of the model.
\end{enumerate}

\subsubsection{Application to \texorpdfstring{$Z$}{Z}+Jets}

The posterior modeling framework is applied to two different example datasets.
The first is the well known unfolding benchmark dataset of Ref.~\cite{Andreassen:2019cjw}.
It consists of light-flavor jets produced in association with a $Z$ boson in proton--proton collisions at $\sqrt{s} = 14$~TeV.
Strong $Z$ boson production via $pp \to Z + \text{jets}$ is simulated using \textsc{Pythia}~8.243~\cite{Sjostrand:2014zea}, with detector response modeled by \textsc{Delphes}~3.4.2~\cite{deFavereau:2013fsa} using the default CMS card.
The $Z$ boson is required to satisfy $p_T^Z > 150$~GeV to ensure a sufficient hadronic recoil.
Detector-level and particle-level jets are clustered with the anti-$k_T$ algorithm~\cite{Cacciari:2008gp} with $R = 0.4$.
In total 1.6 million events are saved, with 80\% used for training and 20\% used for evaluation in what follows.
CFM models are trained using both a standard position-equivariant transformer and a Lorentz-equivariant L-GATr as the underlying neural networks.

\begin{figure}[phtb]
  \centering
  \includegraphics[width=0.325\linewidth, alt={Unfolded jet transverse momentum for Z+jets, showing truth and unfolded distributions with a ratio panel. Percent-level agreement is seen in the bulk.}, page=6]{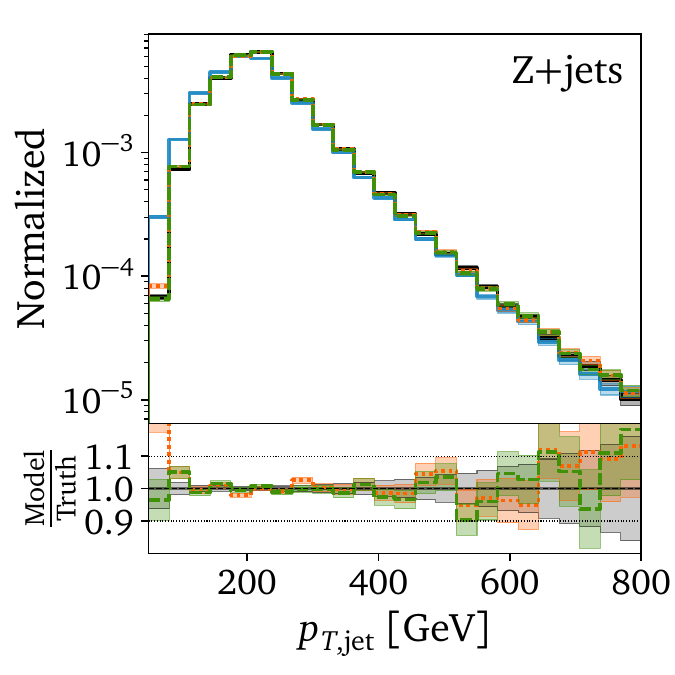}
  \includegraphics[width=0.325\linewidth, alt={Unfolded jet pseudorapidity for Z+jets.}, page=8]{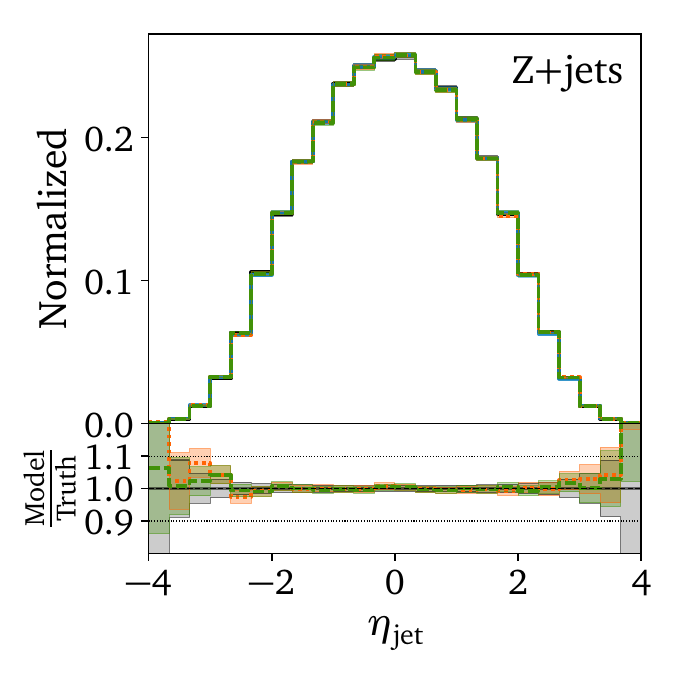}
  \includegraphics[width=0.325\linewidth, alt={Unfolded jet mass for Z+jets.}, page=9]{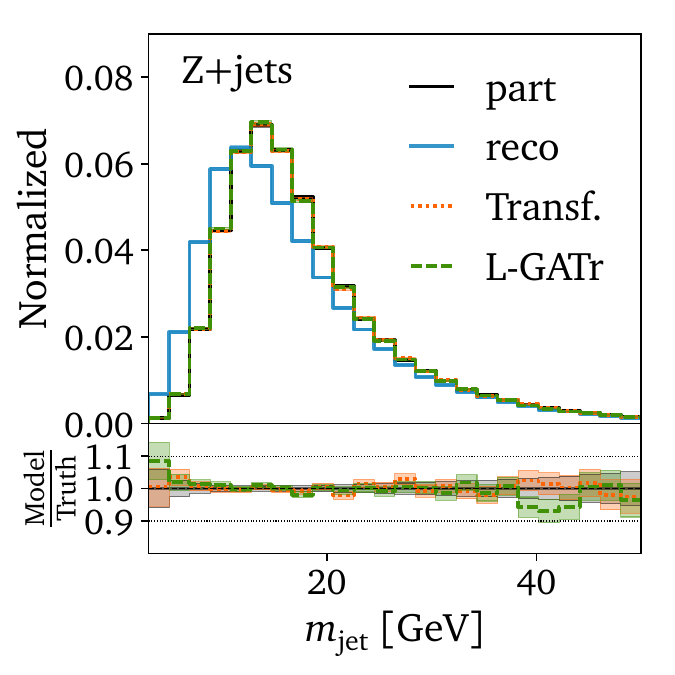}\\[6pt]
  \includegraphics[width=0.325\linewidth, alt={Groomed momentum fraction z_g for Z+jets, showing percent-level agreement between truth and unfolded.}, page=30]{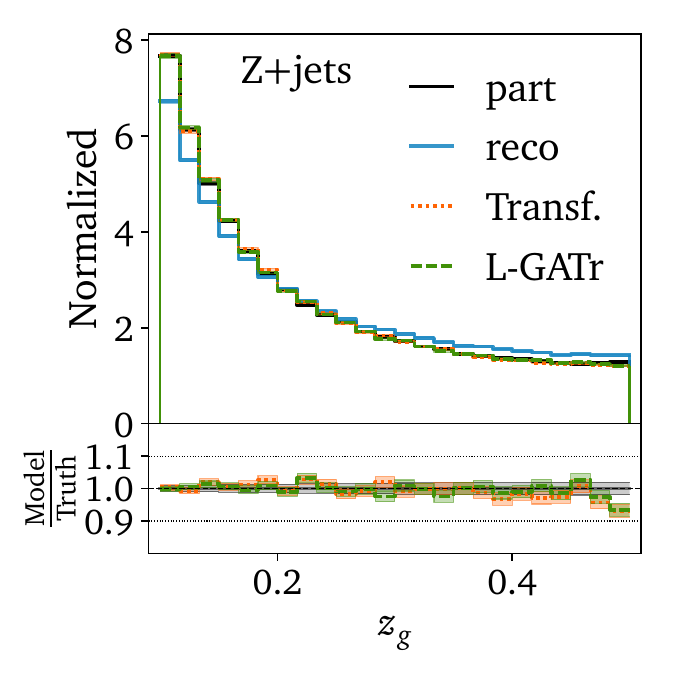}
  \includegraphics[width=0.325\linewidth, alt={Soft drop mass log rho for Z+jets.}, page=29]{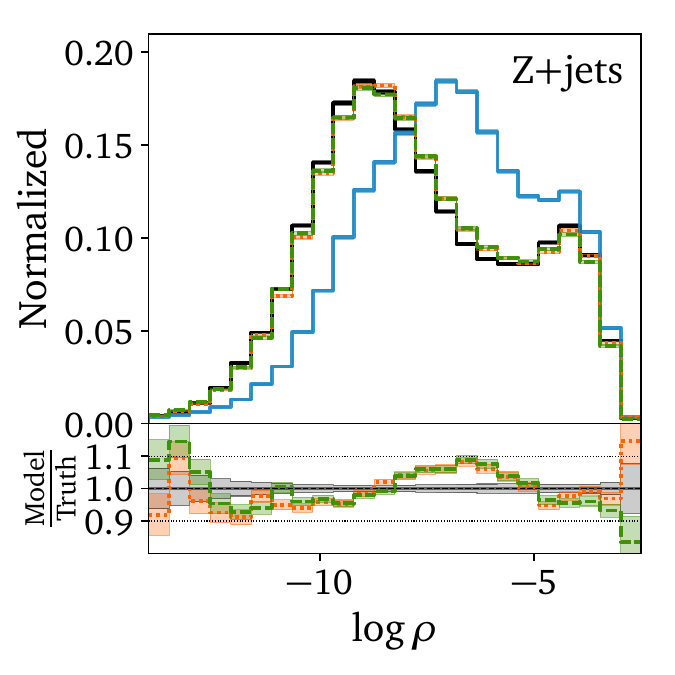}
  \includegraphics[width=0.325\linewidth, alt={Two-point energy-energy correlator EEC for Z+jets.}, page=46]{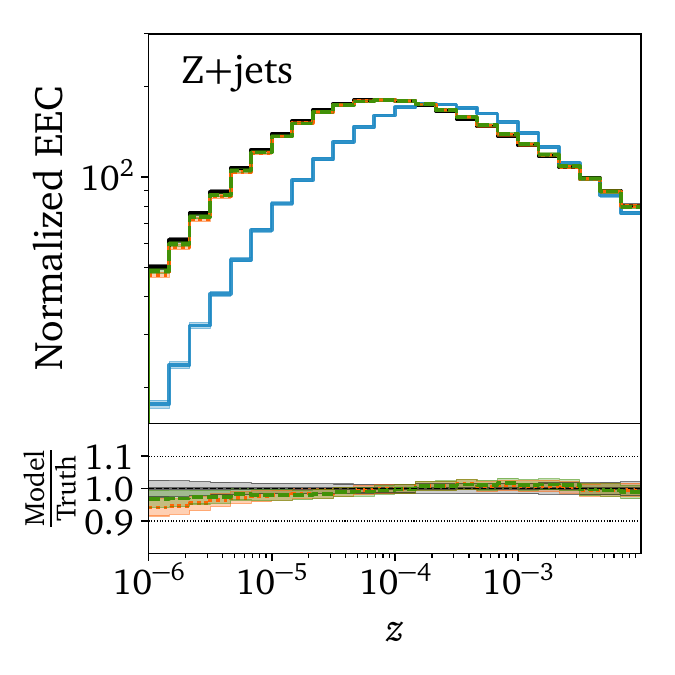}
  \caption{Comparisons between the particle-level distribution (black) and the distribution obtained by sampling the modeled prior for a set of six jet-level observables.
  Predictions are shown for the transformer-based (dotted orange) and L-GATr-based (dashed green) models.
  The detector-level distributions (blue) are also provided for visual comparison.
  All observables are calculated from the jet constituents, rather than predicted by the jet predictor network.
  The top row shows jet kinematics observables: the \pt, $\eta$, and mass.
  The bottom row shows jet substructure observables: groomed momentum fraction $z_g$, soft drop mass $\log\rho$, and two-point energy correlator.
  See \Cref{sec:eec} for a discussion of the EEC.}
  \label{fig:cfm_z_constituents1}
\end{figure}

Both the multiplicity predictor and jet predictor networks, once trained, are able to reproduce the particle-level distributions at percent-level accuracy.
This is expected since both of these modeling tasks are low dimensional and the CFM models are more than capable of high precision in this setting.
The difficult task is training the constituent predictor network, and in particular producing accurate distributions for derived observables, which were not accurately modeled in the previous section.
The top row of \cref{fig:cfm_z_constituents1} compares the particle-level truth and modeled distributions for the jet \pt, $\eta$, and mass observables computed from the constituent kinematics rather than as-modeled by the jet predictor network.
This is a non-trivial check because the constituent predictor network must ensure that the constituent kinematics produce a jet that has the correct properties when they are clustered together.
The particle-level truth is reproduced with percent-level accuracy for each of these observables by both the transformer-based and L-GATr-based models.
The bottom row of \cref{fig:cfm_z_constituents1} shows equivalent plots for three other derived observables relating to the jet substructure.
These are the groomed momentum fraction $z_g$~\cite{Larkoski:2014wba}, the logarithm of the soft drop mass $\log\rho$, and the two-point energy correlator (EEC)~\cite{Moult:2025nhu}.
The EEC in particular is an observable of current interest to the QCD community.
See \Cref{sec:eec}.
The $z_g$ and EEC observables are predicted with percent accuracy, while the $\log\rho$ observable shows deviations up to 10\%.
These deviations are common to the transformer-based and L-GATr-based models.

\begin{figure}[tb]
  \centering
  \includegraphics[width=0.325\linewidth, alt={Jet width (tau_1) for Z+jets.}, page=22]{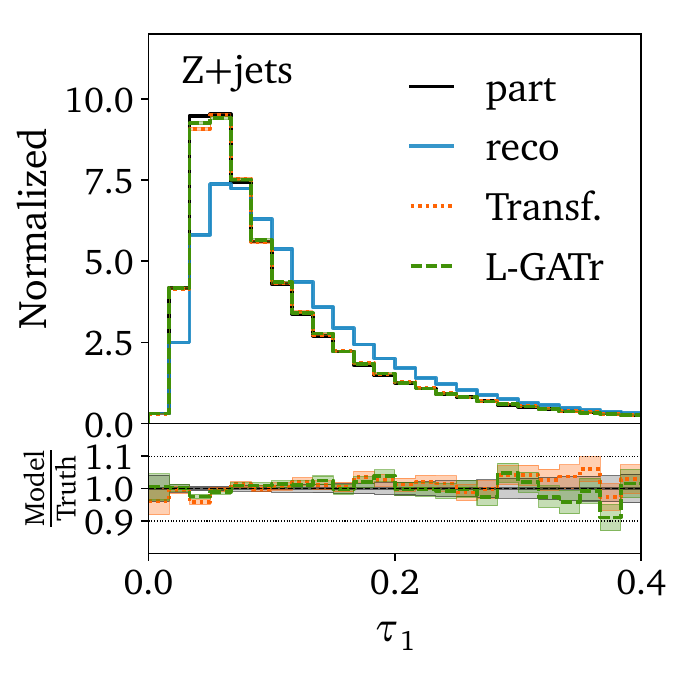}
  \includegraphics[width=0.325\linewidth, alt={2-subjettiness (tau_2) for Z+jets.}, page=23]{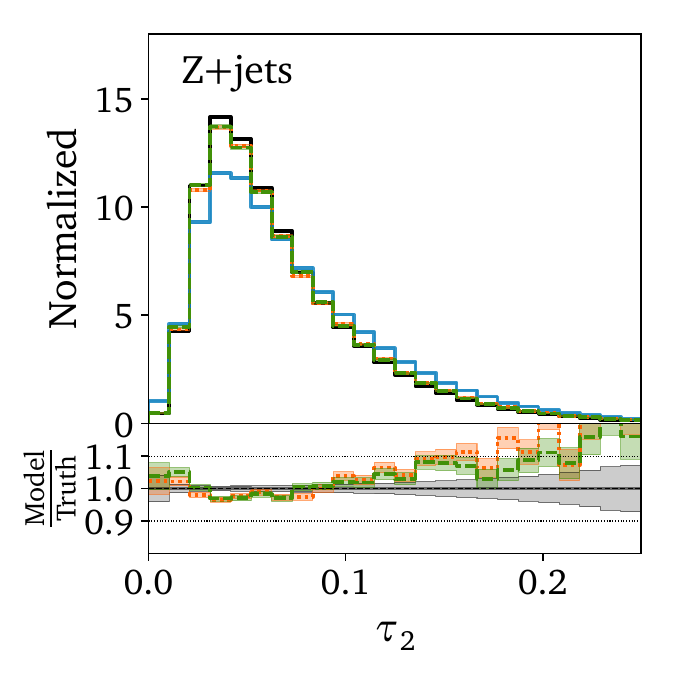}
  \includegraphics[width=0.325\linewidth, alt={21-subjettiness ratio (tau_21) for Z+jets, showing the most challenging observable with 5-10 percent deviations.}, page=26]{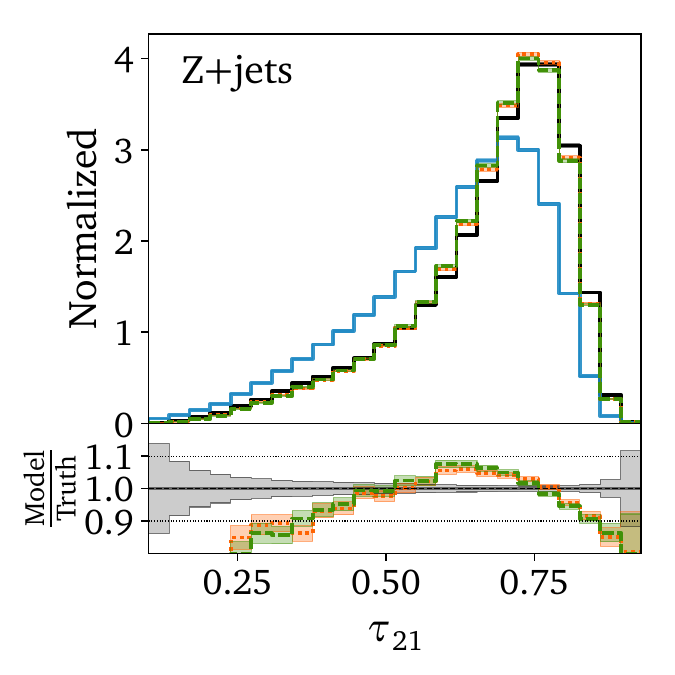}
  \caption{Comparisons between the particle-level distribution (black) and the distribution obtained by sampling the modeled posteriors for three $n$-subjettiness observables: $1$-subjettiness ($\tau_1$), $2$-subjettiness ($\tau_2$), and the ratio of the $2$-subjettiness to the $1$-subjettiness ($\tau_{21}$).
  Predictions are shown for the transformer-based (dotted orange) and L-GATr-based (dashed green) models.
  The detector-level distributions (blue) are also provided for visual comparison.}
  \label{fig:cfm_z_constituents2}
\end{figure}

Plots illustrating the agreement of the sampled posterior with the particle-level truth in a few $n$-subjettiness observables~\cite{Thaler:2010tr} are provided in \Cref{fig:cfm_z_constituents2}.
The $1$-subjettiness is modeled with percent-level precision, while the $2$-subjettiness shows a deviation between the modeled and truth distributions in the tail.
$\tau_{21}$, or the ratio of the $2$-subjettiness to the $1$-subjettiness, shows deviations between the modeled distributions and the ground truth upwards of 10\%.
As of this writing, this performance is state-of-the-art for high- and variable-dimensional posterior modeling in the context of unfolding, but there is still room for improvement in these methods.
As with previous observables, no differences between the transformer based and L-GATr based models is visible, indicating that the equivariant prior is not useful but also not harmful for this particular task.

\subsubsection{Application to Boosted Top Jets}

As a second test of the method, we also consider modeling the posterior of all particle-level jet constituents in top quark initiated jets~\cite{Favaro:2025psi}.
The process $pp \to t\bar{t} \to (bq\bar{q}') (\bar{b}\ell\bar{\nu}) + \text{c.c.}$ is simulated.
Unlike in the previous section, the goal is to gather a sample of boosted top jets (see \Cref{ch:tagging}) produced by the hadronic decay of a high \pt top quark.
Matrix elements are calculated with \textsc{MadGraph}~5~\cite{Alwall:2011uj} assuming $m_t = 173$~GeV, parton showering and hadronization are modeled with \textsc{Pythia}~8.313~\cite{Sjostrand:2014zea}, and the detector response is simulated with \textsc{Delphes}~3.5.0~\cite{deFavereau:2013fsa} using the default CMS card.
Detector and particle level jets are clustered with the anti-$k_T$ algorithm~\cite{Cacciari:2008gp} with a large radius $R = 1.2$ so that all products of the hadronic top decay are included in a single jet.
Events are required to have exactly one lepton with $p_T > 60$~GeV and $|\eta| < 2.4$.
The jet with the largest angular distance to this lepton is taken to be the boosted top jet, and this jet is required to have $p_T > 400$~GeV and $|\eta| < 2.4$ to enter the dataset.
In total 6 million simulated top jets are collected, with 90\% used for training and the rest for evaluation in what follows.

Modeling the posterior of top jets is more challenging than the light quark and gluon jets for three reasons.
First, there are generally more jet constituents in the sample of top jets because the large \pt of the initiating top quark implies a longer evolution of the parton shower and thus more final-state hadrons.
Second, the three-body decay of the initiating top quark produces the distinct three-prong substructure that must be reproduced by the model.
Third, the jet mass of top jets is peaked near the mass of the top quark, in contrast to light quark and gluon jets which follow a smoothly falling mass distribution.
Given all of this, modeling the posterior of top jets is a higher dimensional task with more complex distributions that must be reproduced at particle level.

\begin{figure}[p]
  \centering
  \includegraphics[width=0.28\linewidth, alt={Unfolded jet transverse momentum for top jets.}, page=11]{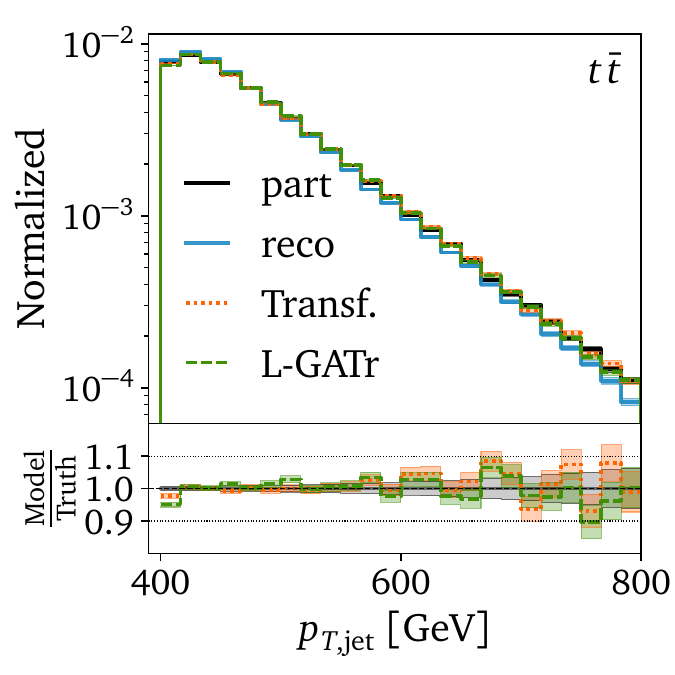}
  \includegraphics[width=0.28\linewidth, alt={Unfolded jet mass for top jets, showing sharply peaked truth distribution at the top mass.}, page=13]{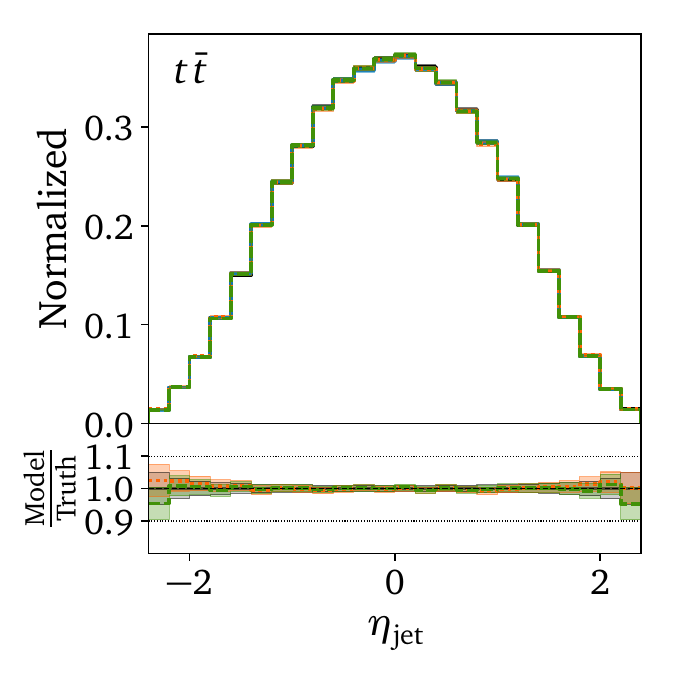}
  \includegraphics[width=0.28\linewidth, alt={Unfolded jet pseudorapidity for top jets.}, page=14]{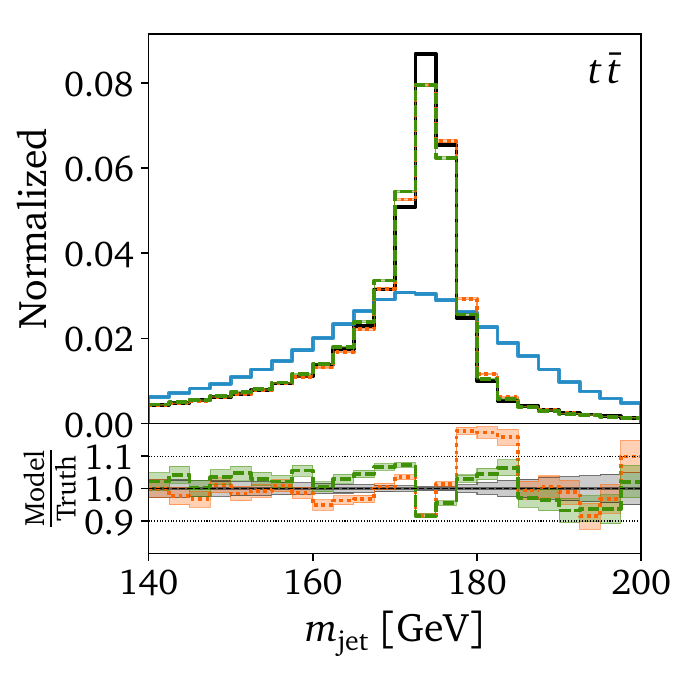}\\[3pt]
  \includegraphics[width=0.28\linewidth, alt={Groomed momentum fraction z_g for top jets.}, page=35]{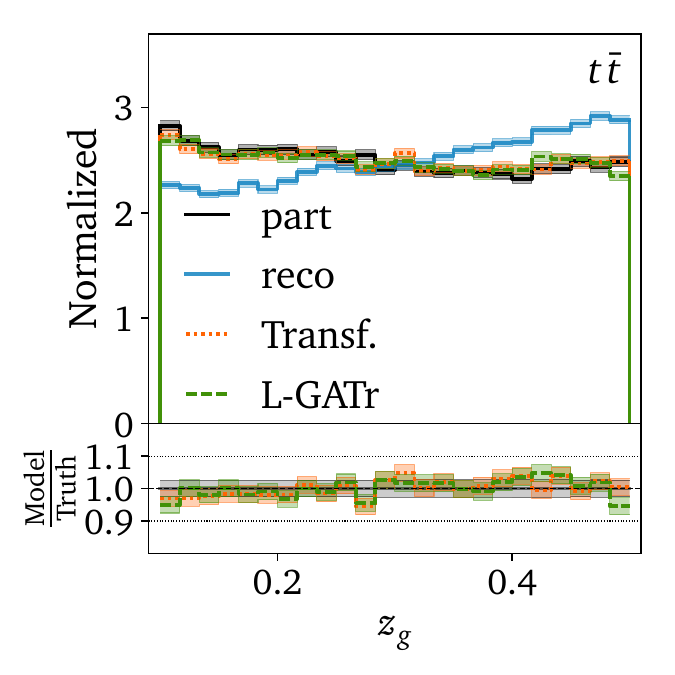}
  \includegraphics[width=0.28\linewidth, alt={Soft drop mass for top jets.}, page=34]{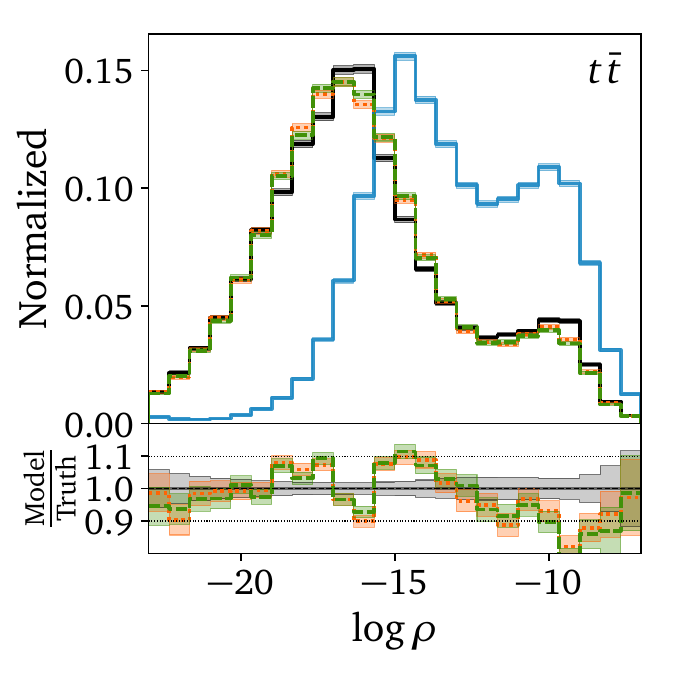}
  \includegraphics[width=0.28\linewidth, alt={Energy-energy correlator EEC for top jets.}, page=51]{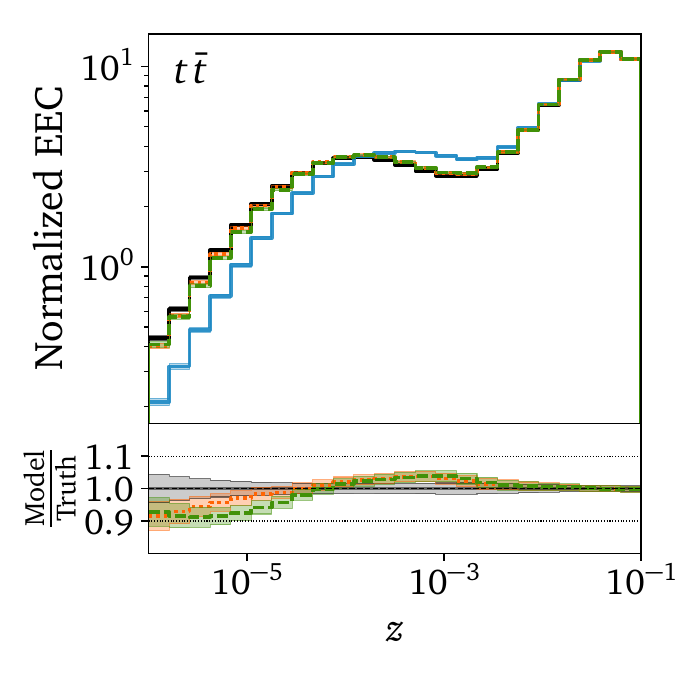}\\[3pt]
  \includegraphics[width=0.28\linewidth, alt={Jet width (tau_1) for top jets.}, page=28]{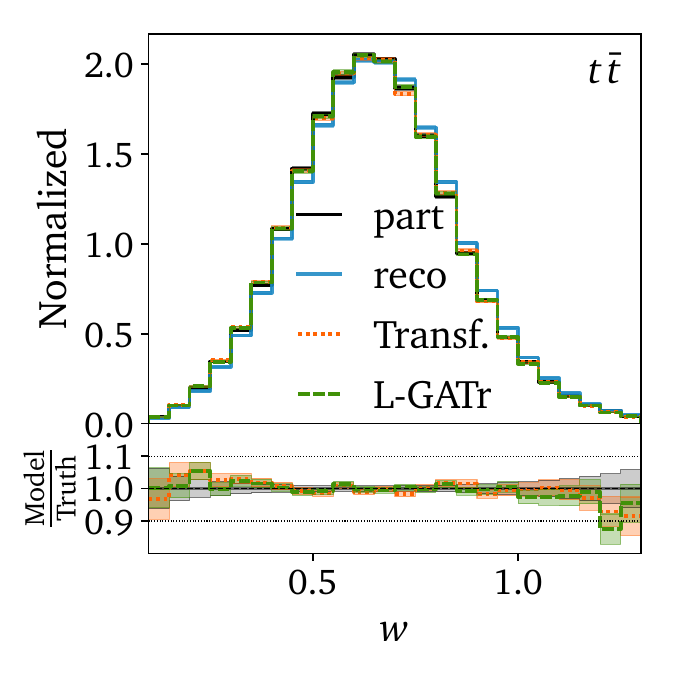}
  \includegraphics[width=0.28\linewidth, alt={2-subjettiness (tau_2) for top jets.}, page=29]{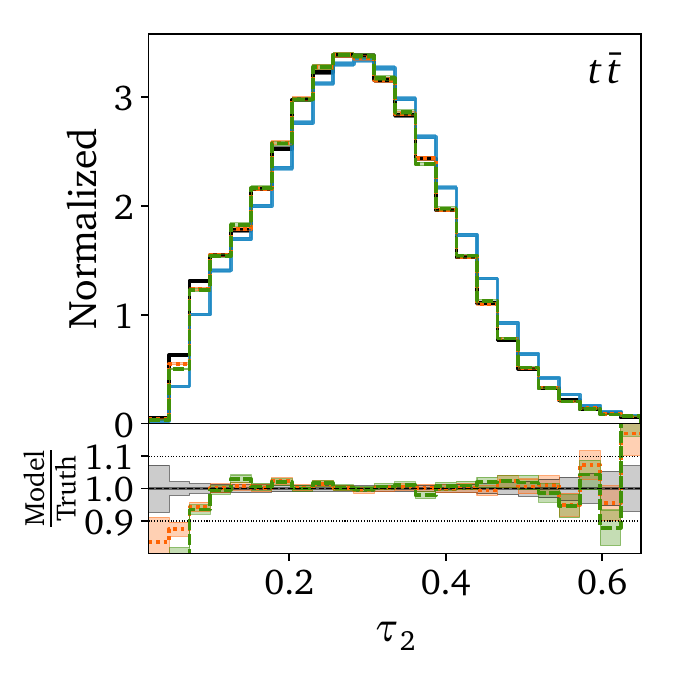}
  \includegraphics[width=0.28\linewidth, alt={3-subjettiness (tau_3) for top jets.}, page=31]{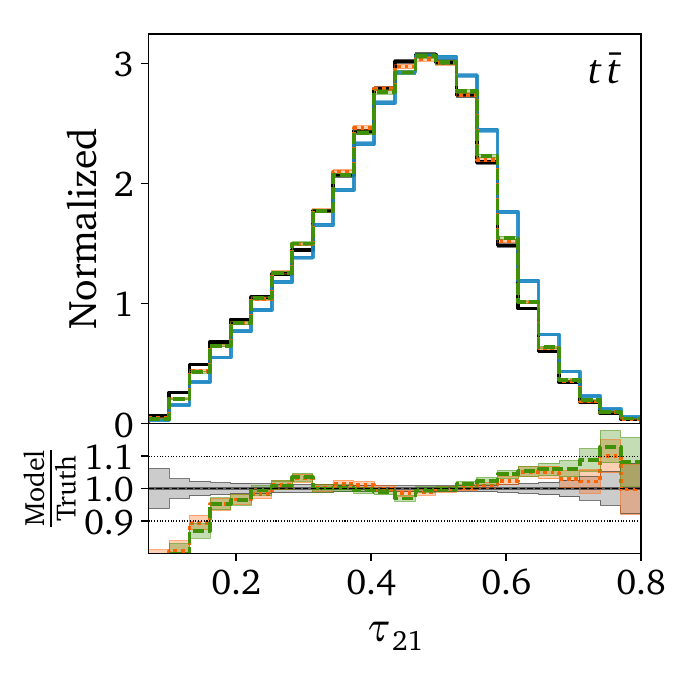}\\[3pt]
  \includegraphics[width=0.28\linewidth, alt={21-subjettiness ratio (tau_21) for top jets.}, page=30]{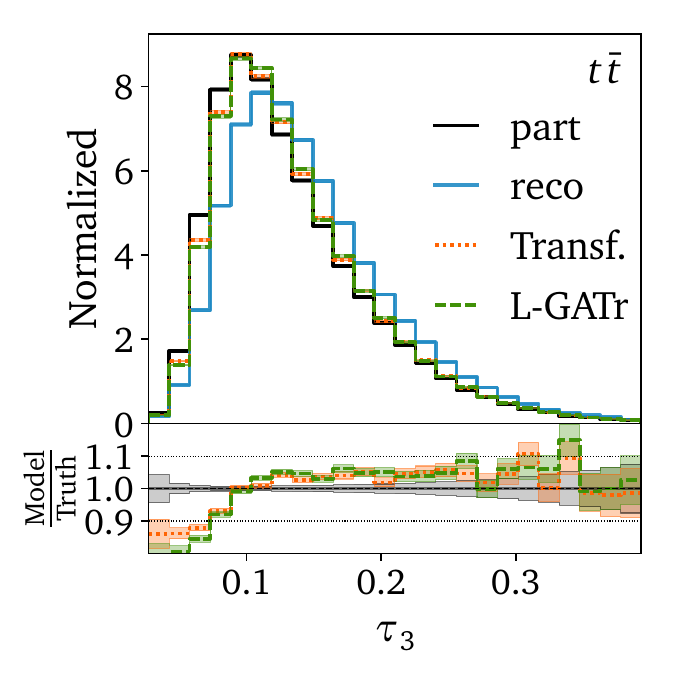}
  \includegraphics[width=0.28\linewidth, alt={32-subjettiness ratio (tau_32) for top jets.}, page=32]{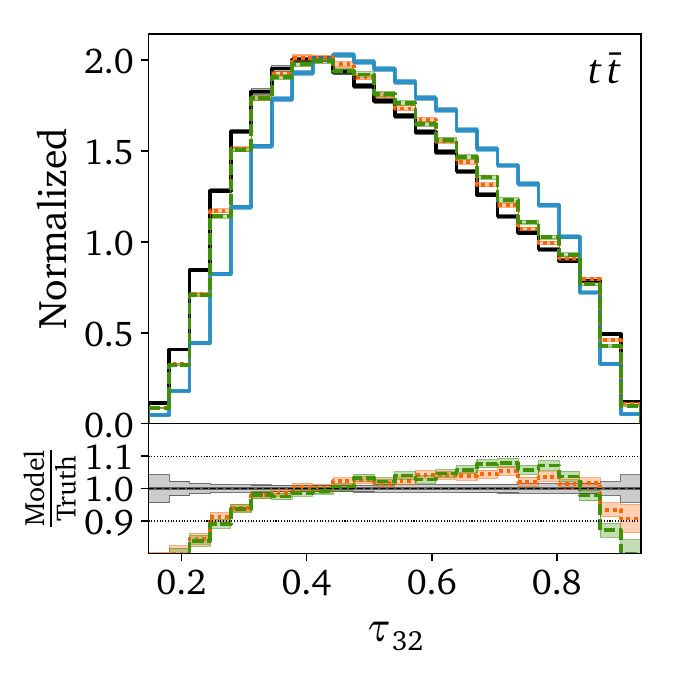}
  \includegraphics[width=0.28\linewidth, alt={43-subjettiness ratio (tau_43) for top jets.}, page=33]{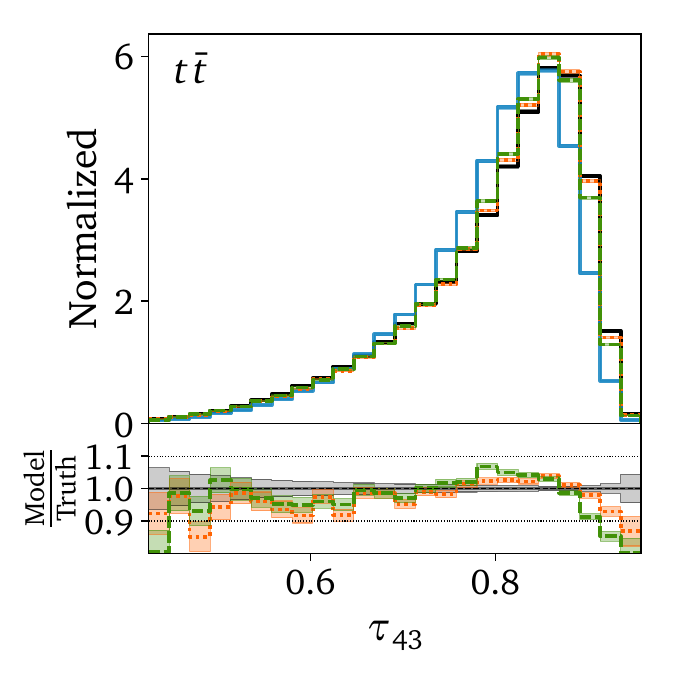}
  \caption{Comparisons between the particle-level distribution (black) and the distribution obtained by sampling the modeled posteriors for many different jet-level observables.
  All observables are calculated from the jet constituents, rather than predicted by the jet predictor network.
  Predictions are shown for the transformer-based (dotted orange) and L-GATr-based (dashed green) models.
  The detector-level distributions (blue) are also provided for visual comparison.}
  \label{fig:cfm_t_constituents}
\end{figure}

Plots analogous to those shown for the light quark and gluon task above are shown in \cref{fig:cfm_t_constituents}.
As in the previous application, the constituent multiplicity and jet kinematics are unfolded very accurately.
The most important observable shown in the Figure is the jet mass (top right), whose particle-level truth distribution is strongly peaked similar to the top quark mass distributions in the previous section.
The CFM model using the factorized posterior method is able to achieve much better precision.
The transformer-based model shows deviations of up to 20\%, while the L-GATr-based model keeps all deviations from the truth below 10\%.
This is impressive performance considering that reproducing this distribution requires accurately modeling the substructure of complicated top quark initiated jets with many tens of constituents.
These results also suggest that the equivariant prior of the L-GATr network may be useful for modeling the overall kinematics of jet constituent point clouds.

Other observables, such as the groomed momentum fraction $z_g$ and the jet width $w$ are reproducedd at percent-level precision.
Substantially better performance for the $n$-subjettiness class observables ($\tau_1$, $\tau_2$, $\tau_3$ and the ratios $\tau_{21}$, $\tau_{32}$, $\tau_{43}$) is obtained compared to the previous application.
Generally deviations in these observables are below 10\% except in the low statistics tails.
The EEC observable is captured slightly worse than in the previous section, possibly simply due to the high constituent multiplicity.
In all of these other observables, the accuracy of the transformer-based and L-GATr-based models is comparable.

\subsubsection{Summary}

Ref.~\cite{Petitjean:2025tgk} builds on the results of the previous sections and demonstrates that high- and variable-dimensional posteriors can be accurately modeled with machine learning methods.
This ability opens the possibility to perform a full unfolding, using the EM algorithm, with generative methods rather than the classifier-based \Omnifold method.
Motivation for using generative methods is discussed in \Cref{sec:generative-unfolding-methods}.
The key innovation to increase the accuracy of the modeled posterior was an auto-regressive factorization into three pieces which simplifies the point-cloud generative task.
Most jet kinematic and jet substructure observables are modeled to within a few percent precision in both posterior modeling tasks considered.
The most challenging observables relate to $n$-subjettiness observables, in particular the ratio $\tau_{21}$ in the light quark and gluon jets application.
The top mass peak in the boosted top jet task was recreated to within 10\%, which is approaching the precision needed to perform a realistic measurement at the LHC.
As of this writing, the methods presented in this section are currently state-of-the-art for modeling posteriors of jet data with generative models.

\FloatBarrier

\section{Omnifold}
\label{sec:omnifold}

\Omnifold~\cite{Andreassen:2019cjw} is a classifier-based implementation of the EM algorithm that relies on reweighting distributions rather than generative models.
Some discussion of \Omnifold and its place in the taxonomy of machine-learning-based unfolding algorithms is provided in \Cref{sec:unfolding-methods}.
This section will provided statistical details, including motivating \Omnifold as the zero-bin-width limit of the popular IBU binned unfolding algorithm.

\subsection{EM Structure}
\label{sec:omnifold-em}

Instead of modeling the posterior $p(t \mid m)$ directly as in \Cref{sec:diffusion-unfolding}\footnote{In this section, the truth-level event configuration is denoted $t$ and the detector-level event configuration is denoted $m$.}, \Omnifold relies on modeling two density ratios to build the required pieces for the EM algorithm.
In what follows these density ratios will be called $w_{n}(m)$ for the detector-level ratio and $\nu_{n}(t)$ for the truth-level ratio.
For iteration $n+1$ which uses the result of iteration $n$ as a starting point, performing the E-step requires the likelihood ratio
\begin{align}
\label{eq:omnifold-estep}
w_{n+1}(m) = \frac{p_{\mathrm{data}}(m)}{\displaystyle\int p(m \mid t)\, \nu^{\mathrm{of}}_{n}(t)\, p_{\mathrm{gen}}(t)\, dt},
\end{align}
where $p_{\mathrm{gen}}(t)$ is the truth-level distribution of the MC generator, $p_{\mathrm{data}}(m)$ is the data distribution, and $\nu^{\mathrm{of}}_{n}(t)$ is the \textit{unfolding function} that gives the current best estimate truth-level distribution at iteration $n$.
The likelihood ratio is then between the detector-level data and the current best-estimate truth-level distribution pushed to detector level via the detector simulation, and can be intuitively understood as a reweighting function that shifts the detector-level MC simulation to match the data.
In practice $w_{n+1}(m)$ is estimated by training a classifier to distinguish the detector-level data from the detector-level simulation.
The ratio estimate can then be used to compute the E-step in \Cref{eq:ibu-continuous-estep} by noting that it is exactly the denominator of the fraction multiplied by the data density.
To estimate the joint density, all that is left is to multiply this ratio by the numerator of the fraction, which is proportional to the posterior $p(t \mid m)$ via Bayes' theorem.
This is done by pulling events at detector-level back through the detector simulation to produce truth-level events that have been reweighted by the density ratio $w_{n+1}(m)$.
Note the posterior is no longer explicitly modeled.
Instead the one-to-one mapping between truth-level and detector-level events is interpreted as the posterior.
The end result is an estimate of the joint density $p_{(n+1)}(t, \, m)$ provided by a sample of Monte Carlo events that has been reweighted at detector level to match the data.

Performing the M-step of iteration $n+1$ requires the likelihood ratio
\begin{align}
\label{eq:omnifold-mstep}
\nu_{n+1}(t) = \frac{\displaystyle\int w_n(m)\, p(t \mid m)\, dm}{\nu^{\mathrm{of}}_{n}(t)\, p_{\mathrm{gen}}(t)}.
\end{align}
The numerator is the result of the E-step integrated over the detector-level events $m$ as in \Cref{eq:ibu-continuous-mstep}.
The denominator is the current best-estimate truth-level distribution.
The likelihood ratio is then between the particle-level MC sample, reweighted by the ratio in \Cref{eq:omnifold-estep}, and the particle-level MC sample reweighted by the current unfolding function $\nu^{\mathrm{of}}_{n}(t)$.
This can be interpreted as a correction to the reweighting function that matches it to the new estimate of the joint distribution provided by the E-step.
In practice $\nu_n(t)$ is estimated by training a second classifier to distinguish the truth-level MC events reweighted by the pulled $w_{n+1}(m)$ weights from the truth-level MC events reweighted by the unfolding function.
The unfolding function is then updated via
\begin{align}
\label{eq:omnifold-prior-update}
\nu^{\mathrm{of}}_{n+1}(t) = \nu_{n+1}(t)\, \nu^{\mathrm{of}}_{n}(t).
\end{align}
After the desired number of iterations $N$ has been run, the final output of \Omnifold is the reweighting function $\nu^{\mathrm{of}}_N(t)$.
This procedure has been shown to converge to the maximum likelihood solution of the folding equation in the Appendix of Ref.~\cite{Andreassen:2019cjw}. 
Alternatively, this can also be understood as the zero-bin-width limit of the analogous arguments for IBU~\cite{Shepp:1982}.

\subsection{Practical Procedure}
\label{sec:omnifold-procedure}

\begin{figure}[tb]
    \centering
    \includegraphics[width=1.0\linewidth, alt={Schematic flowchart of the Omnifold procedure, showing a detector-level classifier that reweights simulation to data in step 1 and a truth-level classifier that reweights the simulation to itself in step 2, with the updated weights carried into the next iteration.}]{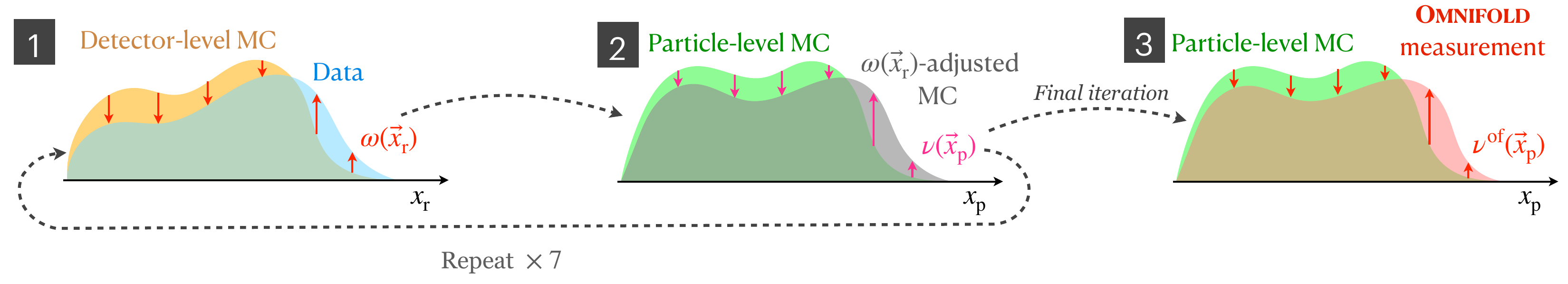}
    \caption{Schematic illustration of the \Omnifold procedure.
    Two classifiers are trained per iteration.
    Step~1 trains a classifier at detector level to reweight MC simulation to data.
    Step~2 trains a classifier at truth level to reweight the MC simulation to itself with the weights derived in step 1 applied.
    The MC simulation with weights calculated in step 2 then serve as the prior for the next iteration.
    The choice of seven iterations is to align with the implementation of \Omnifold in \Cref{ch:zjets}.}
    \label{fig:omnifold-schematic}
\end{figure}

The practical steps needed for the EM algorithm described above are illustrated in \Cref{fig:omnifold-schematic} and described in what follows.

\textbf{Step 1.}
Train a classifier to distinguish the data from a detector-level MC sample.
If this is the first iteration the MC sample carries the raw event weights produced by the event generator.
If it is not the first iteration, the MC sample carries the event weights formed by multiplying the event generator weights with the current particle-level unfolding function pushed to detector-level.
The output of the classifier is interpreted as a reweighting function $w_n(m)$ that adjusts the detector-level MC distribution to match the data.

\textbf{Pull to truth level.}
Pull the detector-level weights for the MC sample assigned by $w_n(m)$ to truth-level using the one-to-one mapping between detector-level and truth-level events\footnote{Events that pass the truth-level event selection but fail the detector-level selection can be assigned the weight of the prior, which is provided by the event generator.}.

\textbf{Step 2.}
Train a classifier to distinguish the truth-level MC sample reweighted by the pulled weights from the truth-level MC sample reweighted by the current unfolding function.
In the first iteration the unfolding function is simply the raw event weights produced by the event generator.
In the second and subsequent iterations the unfolding function is the composition of all previous step 2 reweighting functions.
Importantly in step 2 the classifier is trained to distinguish the truth-level MC sample from itself, with the only difference between the classes being the event weights.
The output of the classifier is interpreted as a reweighting function $\nu_n(t)$ that adjusts the unfolded distribution to match the pulled-weights distribution.
The weights provided by this function are multiplied by the weights provided by all previous step 2 reweightings to produce the updated \Omnifold weights $\nu^{\mathrm{of}}_n(t)$.

\textbf{Push to detector level.}
If the desired number of iterations has been run the procedure terminates and the \Omnifold weights are the result.
Otherwise the \Omnifold weights are pushed to detector level, again using the one-to-one mapping between detector-level and truth-level events\footnote{Events that pass the detector-level event selection but fail the truth-level selection can be assigned the weight of the prior at this step as well. Note that every event will receive an adjustment by the step 1 or step 2 trainings, but only the events that pass the truth-level selection will be present in the final result.}.
These pushed weights are used as the starting point for the next iteration.

\subsection{Properties}
\label{sec:omnifold-properties}

A few features of \Omnifold should be noted explicitly before moving on.
\begin{itemize}
  \item \textbf{Physical consistency.}
    The output of the procedure is the reweighting function $\nu^{\mathrm{of}}_n(t)$ which is applied to a set of truth-level simulated events.
    These events can be binned to construct a cross section measurement as discussed in \Cref{sec:zjets-method-norm}.
    Given the events are produced by a theory-motivated event generator, they are constrained to satisfy conservation laws such as energy and momentum by construction.
    This is not necessarily true for some generative approaches to unfolding, where samples from a generative model are not constrained to lie exactly on the manifold of physical event configurations.

  \item \textbf{Variable-dimensional distributions.}
    Extension of \Omnifold to variable-dimensional settings requires only replacing list-processing classifiers with set-processing classifiers, for example the neural networks used for jet classification in \Cref{ch:tagging}.
    These classifiers are substantially easier to design and train than the set conditional and set generative models developed in \Cref{sec:diffusion-unfolding}.

  \item \textbf{Flexibility in classifier choice.}
    Neural networks are used to estimate the likelihood ratios in \Cref{ch:zjets}, but in many applications, especially low- and fixed-dimensional measurements, they are not required and may even be sub-optimal.
    Simpler and computationally cheaper methods such as boosted decision trees can also be used to estimate likelihood ratios in these cases.

  \item \textbf{Terminology.}
    The terminology surrounding \Omnifold can be confusing.
    In the original paper~\cite{Andreassen:2019cjw}, utilizing the procedure above to unfold one observable in an unbinned fashion was called \textit{Unifold}, utilizing the procedure to unfold a fixed-length list of several observables was called \textit{Multifold}, and utilizing the procedure to produce a full-phase-space unfolding was called \textit{Omnifold}.
    However over time the name \Omnifold has become an umbrella term that refers to the procedure itself, and covers each of the uses above.
    In this thesis, the small-caps \Omnifold refers to the procedure while Multifold refers to the specific application of the procedure to unfold a fixed-length list of observables.
    All existing experimental measurements in \Cref{tab:unbinned-unfolding-measurements}, with the exception of the H1 preliminary result~\cite{H1:prelim}, utilize Multifold.
\end{itemize}

The application of \Omnifold to produce a full-phase-space measurement of $Z$+jets production is the subject of the next chapter.

%% file: tab_unbinned_unfolding_measurements.tex
%
%

\begin{table}
  \caption{%
    Overview of experimental measurements using unbinned unfolding methods,
    adapted from Ref.~\cite{Canelli:2025ybb} with the addition of the ALEPH thrust
    measurement~\cite{Electron-PositronAlliance:2025nbl}.
    For most results the unfolded dimensionality is the same at reconstruction
    and truth level; $*$ indicates analyses where the reconstruction-level
    dimensionality spanned the full phase space but the truth-level result is
    8-dimensional.
    Phase space definitions, including $\eta$ requirements, are given in
    the individual papers.
    Adapted under CC~BY~4.0.%
  }
  \label{tab:unbinned-unfolding-measurements}
  \centering
  \resizebox{\linewidth}{!}{%
  \begin{tabular}{@{} l l c l l @{}}
    \toprule
    Experiment & arXiv & Dim.\ & Final state & Kinematic selection \\
    \midrule
    ATLAS~\cite{ATLAS:2024xxl}
      & \href{https://arxiv.org/abs/2405.20041}{2405.20041}
      & 24
      & $Z$+jets
      & $p_{T}^{\ell\ell} > 200\,\text{GeV}$ \\
    ATLAS~\cite{ATLAS:2025qtv}
      & \href{https://arxiv.org/abs/2502.02062}{2502.02062}
      & 6
      & Dijets
      & $p_{T}^{j_{1}} > 240\,\text{GeV}$, $p_{T}^{j_{1}} < 1.5\,p_{T}^{j_{2}}$ \\
    CMS~\cite{CMS:2025sws}
      & \href{https://arxiv.org/abs/2505.17850}{2505.17850}
      & 8
      & Minimum bias
      & ${>}2$ charged particles, \pt $> 0.5\,\text{GeV}$ \\
    H1~\cite{H1:2021wkz}
      & \href{https://arxiv.org/abs/2108.12376}{2108.12376}
      & $8^{*}$
      & High-$Q^{2}$ DIS
      & $Q^{2} > 150\,\text{GeV}^{2}$ \\
    H1~\cite{H1:2023fzk}
      & \href{https://arxiv.org/abs/2303.13620}{2303.13620}
      & 10
      & High-$Q^{2}$ DIS
      & $Q^{2} > 150\,\text{GeV}^{2}$ \\
    H1~\cite{H1:2024mox}
      & \href{https://arxiv.org/abs/2412.14092}{2412.14092}
      & $8^{*}$
      & High-$Q^{2}$ DIS
      & $Q^{2} > 150\,\text{GeV}^{2}$ \\
    H1~\cite{H1:prelim}
      & --
      & var.
      & High-$Q^{2}$ DIS
      & $Q^{2} > 150\,\text{GeV}^{2}$ \\
    LHCb~\cite{LHCb:2022rky}
      & \href{https://arxiv.org/abs/2208.11691}{2208.11691}
      & 4
      & $Z$+hadrons in jets
      & $20 < p_{T}^{j} < 100\,\text{GeV}$, $p_{T}^{h} > 0.25\,\text{GeV}$ \\
    STAR~\cite{Song:2023sxb}
      & \href{https://arxiv.org/abs/2307.07718}{2307.07718}
      & 6
      & Jets
      & $20 < p_{T}^{j} < 50\,\text{GeV}$ \\
    STAR~\cite{Pani:2024mgy}
      & \href{https://arxiv.org/abs/2403.13921}{2403.13921}
      & 7
      & Jets (heavy ions)
      & $20 < p_{T}^{j} < 45\,\text{GeV}$ \\
    T2K~\cite{Huang:2025ziq}
      & \href{https://arxiv.org/abs/2504.06857}{2504.06857}
      & 6
      & Muon $+$ proton
      & $p_{p} > 450\,\text{MeV}$ (single transverse variables) \\
    \midrule
    ALEPH~\cite{Electron-PositronAlliance:2025nbl}
      & \href{https://arxiv.org/abs/2507.14349}{2507.14349}
      & 1
      & Hadronic $Z$ decays
      & $E_{ch} > 15$~GeV  \\
    \bottomrule
  \end{tabular}%
  }
\end{table}

%% file: ch5_semi_leptonic_ttbar.tex
\begin{tikzpicture}[
        thick,
        scale=0.80,
        level/.style={level distance=1.2cm},
        level 2/.style={sibling distance=1.6cm},
        level 3/.style={sibling distance=1.2cm}
    ]
    \coordinate
        child[grow=left]{
            child {
                node {$g$}
                edge from parent [gluon]
            }
            child {
                node {$g$}
                edge from parent [gluon]
            }
            edge from parent [gluon] node [above=3pt] {$g$}
        }
        child[grow=right, level distance=0pt] {
        child  {
            child {
                child {
                    node {$\bar{d}$}
                    edge from parent [electron]
                }
                child {
                    node {$u$}
                    edge from parent [electron]
                }
                edge from parent [photon] node [below=3pt] {$W$}
            }
            child {
                node {$b$}
                edge from parent [electron]
            }
            edge from parent [electron]
            node [below] {$t$}
        }
        child {
            child {
                node {$\bar{b}$}
                edge from parent [electron]
            }
            child {
                child {
                    node {$\bar{v}$}
                    edge from parent [electron]
                }
                child {
                    node {$e^{-}$}
                    edge from parent [electron]
                }
                edge from parent [photon] node [above=3pt] {$W$}
            }
            edge from parent [electron]
            node [above] {$\bar{t}$}
        }
    };
\end{tikzpicture}

%% file: tab_vld_distance_metrics.tex
\begin{tabular}{lrrrrrr}
\toprule
 & Wasserstein & Energy & K-S & $KL_{64}$ & $KL_{128}$ & $KL_{256}$ \\
\midrule
\textbf{VLD} & 108.76 & 7.59 & 4.08 & \textbf{3.47} & \textbf{3.74} & \textbf{4.53} \\
\textbf{UC-VLD} & \textbf{73.56} & \textbf{6.35} & \textbf{3.41} & 5.77 & 7.10 & 8.48 \\
\textbf{C-VLD} & 389.62 & 25.39 & 4.65 & 9.54 & 10.09 & 10.79 \\
LDM & 402.32 & 24.09 & 5.91 & 14.71 & 16.34 & 17.92 \\
VDM & 2478.35 & 181.35 & 17.14 & 29.28 & 32.29 & 35.60 \\
CVAE & 484.56 & 32.29 & 6.37 & 7.79 & 9.17 & 10.60 \\
CINN & 3009.08 & 185.13 & 15.74 & 28.55 & 30.19 & 32.37 \\
\bottomrule
\end{tabular}

%% file: tab_vl_vld_jet_lepton_kinematics.tex
\begin{tabular}{cl|ccc}
  \toprule
  Observable & Level & Wasserstein & Energy & KL \\
  \midrule
  $p_\mathrm{T}^\mathrm{Jet}$ & \makecell[l]{Unfolded \\ Detector} & \makecell[l]{$0.36 \pm 0.08$ \\ $2.30$} & \makecell[l]{$0.04 \pm 0.01$ \\ $0.27$} & \makecell[l]{$0.02 \pm 0.01$ \\ $0.26$}\\
  \midrule
  $\eta^\mathrm{Jet}$ & \makecell[l]{Unfolded \\ Detector} & \makecell[l]{$0.01 \pm 0.00$ \\ $0.01$} & \makecell[l]{$0.00 \pm 0.00$ \\ $0.01$} & \makecell[l]{$13.03 \pm 1.18$ \\ $3.44$}\\
  \midrule
  $\phi^\mathrm{Jet}$ & \makecell[l]{Unfolded \\ Detector} & \makecell[l]{$0.00 \pm 0.00$ \\ $0.00$} & \makecell[l]{$0.00 \pm 0.00$ \\ $0.00$} & \makecell[l]{$0.01 \pm 0.01$ \\ $0.02$}\\
  \midrule
  $m^\mathrm{Jet}$ & \makecell[l]{Unfolded \\ Detector} & \makecell[l]{$0.03 \pm 0.01$ \\ $1.52$} & \makecell[l]{$0.01 \pm 0.00$ \\ $0.52$} & \makecell[l]{$0.17 \pm 0.05$ \\ $61.66$}\\
  \midrule
  $E^\mathrm{Jet}$ & \makecell[l]{Unfolded \\ Detector} & \makecell[l]{$0.66 \pm 0.26$ \\ $2.05$} & \makecell[l]{$0.05 \pm 0.02$ \\ $0.20$} & \makecell[l]{$0.01 \pm 0.00$ \\ $0.06$}\\
  \midrule\midrule
  $p_\mathrm{T}^\mathrm{Lepton}$ & \makecell[l]{Unfolded \\ Detector} & \makecell[l]{$0.27 \pm 0.10$ \\ $3.89$} & \makecell[l]{$0.05 \pm 0.02$ \\ $0.53$} & \makecell[l]{$0.17 \pm 0.05$ \\ $2.64$}\\
  \midrule
  $\eta^\mathrm{Lepton}$ & \makecell[l]{Unfolded \\ Detector} & \makecell[l]{$0.01 \pm 0.00$ \\ $0.03$} & \makecell[l]{$0.01 \pm 0.00$ \\ $0.02$} & \makecell[l]{$16.72 \pm 3.27$ \\ $56.01$}\\
  \midrule
  $\phi^\mathrm{Lepton}$ & \makecell[l]{Unfolded \\ Detector} & \makecell[l]{$0.01 \pm 0.01$ \\ $0.01$} & \makecell[l]{$0.01 \pm 0.00$ \\ $0.01$} & \makecell[l]{$0.08 \pm 0.05$ \\ $0.02$}\\
  \midrule\midrule
  $E_\mathrm{T}^\mathrm{miss}$ & \makecell[l]{Unfolded \\ Detector} & \makecell[l]{$0.31 \pm 0.06$ \\ $3.03$} & \makecell[l]{$0.04 \pm 0.01$ \\ $0.41$} & \makecell[l]{$0.04 \pm 0.01$ \\ $0.95$}\\
  \midrule
  $H_\mathrm{T}$ & \makecell[l]{Unfolded \\ Detector} & \makecell[l]{$7.45 \pm 0.54$ \\ $8.54$} & \makecell[l]{$0.51 \pm 0.03$ \\ $0.59$} & \makecell[l]{$0.20 \pm 0.02$ \\ $0.20$}\\
  \bottomrule
\end{tabular}

%% file: ch5_cfm_jets_model.tex
\usetikzlibrary{arrows, shapes.geometric, arrows.meta, shapes, decorations.pathreplacing, fit, patterns, patterns.meta,positioning,calc}

\definecolor{Rcolor}{HTML}{E99595}
\definecolor{Gcolor}{HTML}{C5E0B4}
\definecolor{Gcolor_light}{HTML}{F1F8ED}
\definecolor{Gcolor_dark}{HTML}{9CB391}
\definecolor{Bcolor}{HTML}{9DC3E6}
\definecolor{Ycolor}{HTML}{FFE699}
\definecolor{Ycolor_light}{HTML}{FFF7DE}
\definecolor{Ocolor}{HTML}{E67410}

\tikzstyle{expr} = [rectangle, rounded corners=0.3ex, minimum width=1.5cm, minimum height=1cm, text centered, align=center, inner sep=0, fill=Ycolor, font=\large, draw]
\tikzstyle{small_cinn} = [double arrow, double arrow head extend=0cm, double arrow tip angle=130, inner sep=0, align=center, minimum width=1.5cm, minimum height=1.7cm, fill=Rcolor, draw]
\tikzstyle{small_cinn_black} = [small_cinn, minimum height=1.8cm, fill=black]
\tikzstyle{transformer} = [rectangle, rounded corners, minimum width=6cm, minimum height=2.4cm, font=\large, fill=Gcolor_light, draw]
\tikzstyle{attention} = [rectangle, rounded corners=0.3ex, minimum width=5.5cm, minimum height=1.2cm, align=center, fill=Gcolor, draw, font=\large]
\tikzstyle{transformer_huge} = [rectangle, rounded corners, minimum width=1cm, minimum height=1cm, fill=Gcolor]
\tikzstyle{input} = [rectangle, rounded corners, minimum width=1cm, minimum height=1cm, draw=black]
\tikzstyle{particle_component} = [rectangle, minimum width = 0.25cm, minimum height=0.25cm, font = \large, fill=Ycolor, rounded corners=0.3ex]
\tikzstyle{attention_huge} = [rectangle, rounded corners=0.3ex, minimum width=8cm, minimum height=1.2cm, align=center, fill=Gcolor, draw, font=\large]
\tikzstyle{txt_huge} = [align=center, font=\Huge, scale=2]
\tikzstyle{txt} = [align=center, minimum height=1cm]
\tikzstyle{arrow} = [thick,-{Latex[scale=1.0]}, line width=0.2mm, color=black]
\tikzstyle{line} = [thick, line width=0.2mm, color=black]

\begin{tikzpicture}
[node distance=0.3cm, scale=0.8, every node/.style={transform shape}]
\node (particle_1_1) [particle_component]{};
\node (particle_1_2) [particle_component, below of = particle_1_1]{};
\node (particle_1_3) [particle_component, below of = particle_1_2]{};
\node (particle_2_1) [particle_component, right of = particle_1_1, xshift=0.25cm]{};
\node (particle_2_2) [particle_component, below of = particle_2_1]{};
\node (particle_2_3) [particle_component, below of = particle_2_2]{};
\node (and) [txt, right of = particle_2_2, xshift = 0.55cm]{...};
\node (particle_N_2) [particle_component, right of = and, xshift = 0.55cm]{};
\node (particle_N_1) [particle_component, above of = particle_N_2]{};
\node (particle_N_3) [particle_component, below of = particle_N_2]{};
\draw[{Latex[width=0.75mm]}-{Latex[width=0.75mm]}] ([yshift=0.4cm]particle_1_1.west) -- ([yshift=0.4cm]particle_N_1.east);
\draw [decorate,decoration={brace, mirror}]
  ([yshift=-0.2cm, xshift=-0.2cm]particle_1_3.south) -- ([yshift=-0.2cm, xshift=0.2cm]particle_N_3.south){};
\node (N_reco) [txt, above of = and, yshift=0.7cm, xshift=-0.25cm]{$\Ndet$};
\node (xreco)[txt, below of = N_reco, yshift=-1.8cm]{$\xr \sim p(\xr)$};

\node (Multi_b)[small_cinn_black, below of = and, xshift=-0.2cm, yshift=-2.5cm]{};
\node (Multi)[small_cinn, above of=Multi_b, yshift=-0.3cm, fill=Bcolor]{Mult.};
\node (xmult) [txt, right of=Multi, xshift=3.0cm]{$\Npart \sim p_\theta (\Npart|\xr)$};

\node (CFM_b_jet) [small_cinn_black, right of = xmult, xshift=1.9cm, yshift=-1.6cm]{};
\node (CFM_jet) at (CFM_b_jet) [small_cinn, fill=Gcolor]{Jet};

\node (base_jet_1) [particle_component,left of = CFM_jet, xshift=-1.8cm, fill=Ocolor]{};
\node (base_jet_2) [particle_component,left of = base_jet_1, fill=Ocolor]{};
\node (base_jet_3) [particle_component,left of = base_jet_2, fill=Ocolor]{};
\node (base_jet_4) [particle_component,left of = base_jet_3, fill=Ocolor]{};
\node (xbasejet)[txt, below of=base_jet_1, yshift=-0.3cm, xshift=-0.3cm]{$z \sim p_{\text{latent}}(z)$};

\node (final_jet_1) [particle_component,right of = CFM_jet, xshift=2.0cm, fill=Ocolor]{};
\node (final_jet_2) [particle_component,right of = final_jet_1, fill=Ocolor]{};
\node (final_jet_3) [particle_component,right of = final_jet_2, fill=Ocolor]{};
\node (final_jet_4) [particle_component,right of = final_jet_3, fill=Ocolor]{};
\node (xjet)[txt, below of = final_jet_1, xshift=1.1cm, yshift=-0.3cm]{$\Jpart \sim p_\phi(\Jpart|\xr, \Npart)$};

\node (CFM_b) [small_cinn_black, right of = CFM_jet, xshift=5.8cm, yshift=-2.5cm]{};
\node (CFM) at (CFM_b) [small_cinn]{Const.};

\node (base_node) [txt, left of=CFM, xshift=-9.05cm]{...};
\node (base_2_2) [particle_component,left of = base_node, xshift=-0.55cm, fill=Ycolor]{};
\node (base_2_1) [particle_component,above of = base_2_2, fill=Ycolor]{};
\node (base_2_3) [particle_component,below of = base_2_2, fill=Ycolor]{};
\node (base_1_2) [particle_component,left of = base_2_2, xshift=-0.25cm, fill=Ycolor]{};
\node (base_1_1) [particle_component, above of = base_1_2 , fill=Ycolor]{};
\node (base_1_3) [particle_component, below of = base_1_2, fill=Ycolor]{};
\node (base_N_2) [particle_component,right of = base_node,xshift = 0.55cm, fill=Ycolor]{};
\node (base_N_1) [particle_component,above of = base_N_2, fill=Ycolor]{};
\node (base_N_3) [particle_component,below of = base_N_2, fill=Ycolor]{};
\node (pgauss) [txt, below of = base_node, yshift=-0.8cm, xshift=-0.3cm]{$z\sim p_\text{latent}(z)$};
\node (N_reco) [txt, above of = base_node, yshift=0.7cm, xshift=-0.25cm]{$\Npart$};
\draw [decorate,decoration={brace, mirror}]
  ([yshift=-0.2cm, xshift=-0.2cm]base_1_3.south) -- ([yshift=-0.2cm, xshift=0.2cm]base_N_3.south){};
\draw[{Latex[width=0.75mm]}-{Latex[width=0.75mm]}] ([yshift=0.4cm]base_1_1.west) -- ([yshift=0.4cm]base_N_1.east);

\node (final_node) [txt, right of=CFM, xshift=3cm]{...};
\node (final_2_2) [particle_component,left of = final_node, xshift=-0.55cm, fill=Ycolor]{};
\node (final_2_1) [particle_component,above of = final_2_2, fill=Ycolor]{};
\node (final_2_3) [particle_component,below of = final_2_2, fill=Ycolor]{};
\node (final_1_2) [particle_component,left of = final_2_2, xshift=-0.25cm, fill=Ycolor]{};
\node (final_1_1) [particle_component, above of = final_1_2 , fill=Ycolor]{};
\node (final_1_3) [particle_component, below of = final_1_2, fill=Ycolor]{};
\node (final_N_2) [particle_component,right of = final_node,xshift = 0.55cm, fill=Ycolor]{};
\node (final_N_1) [particle_component,above of = final_N_2, fill=Ycolor]{};
\node (final_N_3) [particle_component,below of = final_N_2, fill=Ycolor]{};
\node (ppart) [txt, below of = final_node, yshift=-0.8cm, xshift=0.4cm]{$\xp \sim p_\psi(\xp|\xr, \Npart, \Jpart)$};
\node (N_reco2) [txt, above of = final_node, yshift=0.7cm, xshift=-0.25cm]{$\Npart$};
\draw [decorate,decoration={brace, mirror}]
  ([yshift=-0.2cm, xshift=-0.2cm]final_1_3.south) -- ([yshift=-0.2cm, xshift=0.2cm]final_N_3.south){};
\draw[{Latex[width=0.75mm]}-{Latex[width=0.75mm]}] ([yshift=0.4cm]final_1_1.west) -- ([yshift=0.4cm]final_N_1.east);

\draw [arrow, color=black] ([yshift=0.1cm,xshift=0.05cm]xreco.south) -- ([yshift=0.1cm]Multi.north);
\draw [arrow, color=black] ([xshift=0.2cm]Multi.east) -- (xmult.west);

\draw [arrow, color=black] (xreco.east) -- ([xshift=0.1cm]CFM_jet |- xreco.east) -- ([yshift=0.1cm, xshift=0.1cm]CFM_jet.north);
\draw [arrow, color=black] (xmult.east) -- ([xshift=-0.1cm]CFM_jet |- xmult.east) -- ([yshift=0.1cm, xshift=-0.1cm]CFM_jet.north);
\draw [arrow, color=black] ([xshift=0.4cm]base_jet_1.east) -- ([xshift=-0.2cm]CFM_jet.west);
\draw [arrow, color=black] ([xshift=0.2cm]CFM_jet.east) -- ([xshift=-0.4cm]final_jet_1.west);

\draw [arrow, color=black] (xreco.east) -- ([xshift=0.2cm]CFM |- xreco.east) -- ([yshift=0.1cm, xshift=0.2cm]CFM.north);
\draw [arrow, color=black] (xmult.east) -- (CFM |- xmult.east) -- ([yshift=0.1cm]CFM.north);
\draw [arrow, color=black] (xjet.east) -- ([xshift=-0.2cm]CFM |- xjet.east) -- ([xshift=-0.2cm,yshift=0.1cm]CFM.north);

\draw [arrow, color=black] ([xshift=1.0cm]base_node.east) -- ([xshift=-0.2cm]CFM.west);
\draw [arrow, color=black] ([xshift=0.2cm]CFM.east) -- ([xshift=-1.4cm]final_node.west);
\draw [arrow, color=black] ([xshift=-1.5cm]xmult.south) -- ([xshift=-0.2cm]N_reco.north);

\end{tikzpicture}

%% file: chapter6.tex
\chapter{Z+Jets Cross Section Measurement}
\label{ch:zjets}

This chapter presents the application of the \Omnifold unfolding method to ATLAS data and the fifth original research contribution of this thesis.
The analysis is currently in ATLAS review and the paper is expected to be published in September of 2026.
A conference note summarizing the results of the analysis was recently released.
Where possible, approved plots taken from this conference note are used.
The rest of the plots should be considered a work in progress.

\section{Measurement Overview}
\label{sec:zjets-overview}

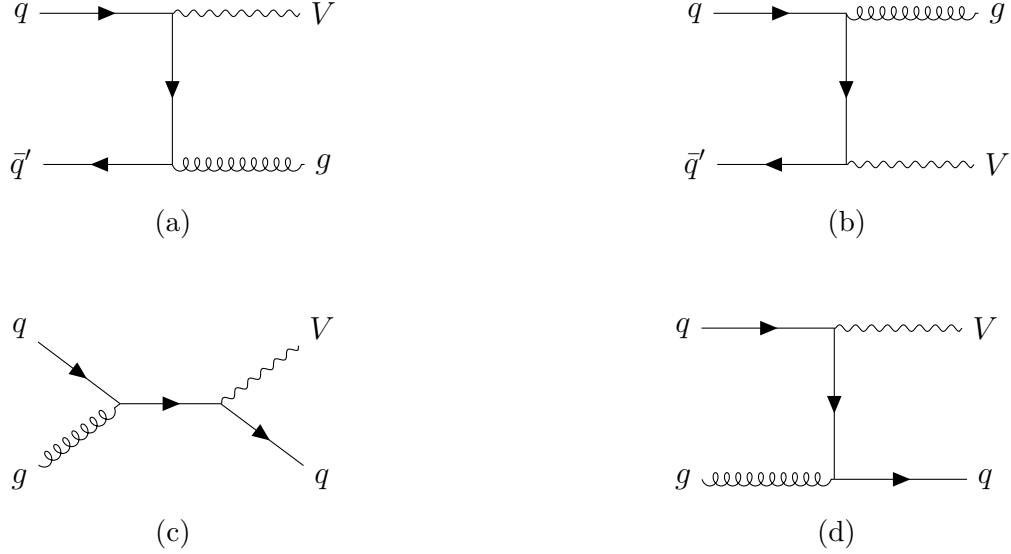
\begin{figure}[htb]
    \centering
    \input{ch6_lo_feynman_diagrams}
    \caption{Leading order Feynman diagrams for $V$+jets production ($V = Z$ or $W$) in proton--proton collisions.
    (a,\,b) The quark--antiquark annihilation channel $q\bar{q}' \to V g$, with the vector boson emitted from the quark leg (a) or the antiquark leg (b).
    (c,d) The quark--gluon Compton channel $qg \to Vq$, with the vector boson emitted after (c) or before (d) the gluon is absorbed.
    For $Z$ production $q$ and $\bar{q}'$ are the same flavour and for $W$ production they are different flavours.
    Together these two partonic channels, each with two diagrams, constitute the complete set of $\mathcal{O}(\alpha_s)$ contributions to the inclusive $V$+1-jet cross section.}
    \label{fig:zjets-lo-diagrams}
\end{figure}

\begin{figure}[htb]
    \centering
    \includegraphics[width=0.85\linewidth, alt={Diagram of Z+jets production in proton--proton collisions, illustrating the hard scatter that produces the Z boson and associated partons, the parton shower, and the hadronization into final-state charged and neutral hadrons that are observed in the detector.}]{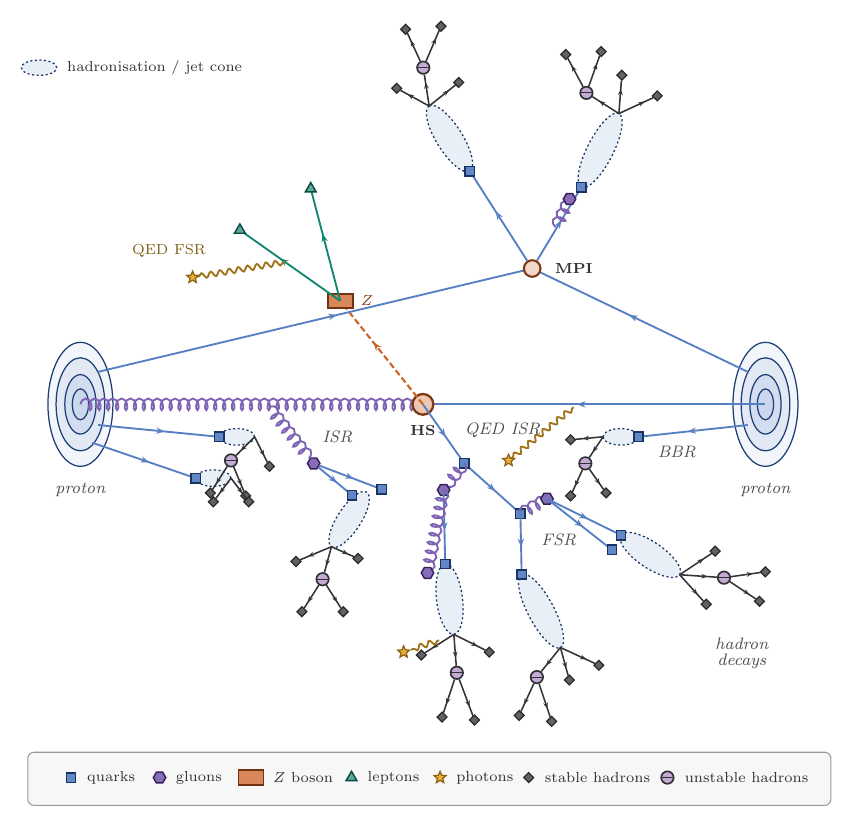}
    \caption{Schematic illustration of $Z$+jets production and other possible processes in a single proton--proton collision. Two partons, a gluon from the left proton and a quark from the right, participate in a hard scatter producing a $Z$ boson recoiling against a quark. The $Z$ boson decays to two muons. The quark initiates the parton shower and hadronization processes, resulting in a collimated spray of hadrons. Additional processes produce radiation that did not result from the hard scatter. Beam--beam remnants (BBR) are hadrons resulting from the parton shower and hadronization of partons that did not participate in the hard scatter. Initial and final-state radiation (ISR and FSR) result in radiation from the initial and final states of the hard scatter process. Finally multiple parton interactions (MPI) can occur between partons in the colliding protons that do not participate in the hard scatter. Figure generated with Claude Opus 4.7.}
    \label{fig:zjets-production}
\end{figure}

This chapter presents a measurement of $Z(\to\mu\mu)$+jets production in proton--proton collisions at $\sqrt{s} = 13$~TeV, using the full Run~2 dataset collected by the ATLAS experiment in 2015--2018 and corresponding to an integrated luminosity of $140.1~\mathrm{fb}^{-1}$.
The measurement is differential in the full phase space of all final-state charged particles.
This is the first measurement of its kind to be performed at the LHC.
The target phase space has a minimum of six, a maximum of 843, and a mean of 150 dimensions.

The primary physics goal is to measure many different jet observables in an unbiased sample of high-\pt jets obtained by placing selections on the dimuon system.
High \pt muons offer a very clean experimental signature that can be used for triggering and event selection.
This allows $Z$+jets events to be selected with high efficiency and almost no background, and the hadronic recoil in the event can then be studied without consideration of jet trigger thresholds or other complications.
Several measurements of the hadronic recoil in $Z$+jets events have been performed at the LHC~\cite{ATLAS:2022nrp,CMS:2022ilp}.
Most importantly, a previous ATLAS measurement targeting an identical phase space as the measurement presented here, used the Multifold variant of \Omnifold to simultaneously unfold 24 observables~\cite{ATLAS:2024xxl}.
The measurement presented here is a direct follow-up to this work.
The previous measurement will be referred to as the ``Multifold measurement'', and the current measurement will be referred to as the ``Omnifold measurement'' in what follows.

The Multifold measurement was the first to apply an unbinned and high-dimensional unfolding method to LHC data, and illustrated that these techniques can be used in a realistic context.
The follow-up measurement discussed here addresses two shortcomings.
First, there are many jet substructure observables, and it is impossible to enumerate all of them and construct a measurement differential in each.
As discussed in \Cref{sec:unbinned-motivation}, full-phase-space measurements are maximally information preserving and allow for unforeseen observables of interest to be measured without requiring a new analysis.
Second, some classes of observables cannot be trivially unfolded with a fixed-length set of observables.
Two such observables are measured in this chapter: the two-point energy correlator and nearest neighbor intrinsic dimension observables introduced in \Cref{sec:eec,sec:id} respectively.

A full-phase-space measurement of $Z$+jets production is substantially more complicated than a fixed-length measurement.
Measuring the full phase space requires capturing the cross section including all of the complicated processes that occur within proton--proton collisions at the LHC.
The leading order Feynman diagrams for $Z$+jets production at a hadron collider are illustrated in \Cref{fig:zjets-lo-diagrams}.
The top (bottom) row diagrams result in events where the $Z$ boson recoils against a gluon (quark).
Given the large gluon PDF at LHC energies the bottom $qg$ diagrams tend to dominate, so the hadronic recoil in $Z$+jets is primarily quark jets.
However higher order diagrams, electroweak $Z+jj$ production, or diboson $VZ$ production all can produce events with two or more jets.
The Multifold measurement only considered observables relating to the leading and sub-leading jets, but when measuring the full phase space all jets are implicitly measured.
Beyond this additional sensitivity to the hard scatter, the full phase space is also sensitive to other processes in LHC collisions that can produce additional hadronic activity.
A diagrammatic representation of some of these processes is provided in \Cref{fig:zjets-production}.
After the parton shower and hadronization processes, a $Z$+jets event can contain several tens of stable charged particles that are measured by the detector.
These could be the muons from the $Z$ boson decay, or hadrons from the hard scatter recoil, beam--beam remnants, initial or final state radiation, or multiple parton interactions.
In principle a full-phase-space measurement is sensitive to all of these processes\footnote{In practice measuring the kinematics of beam--beam remnants is typically not possible given the limited coverage of ATLAS in the forward region.}, making it a powerful tool for improving the theoretical modeling of proton--proton collisions.

The remainder of this chapter is organized as follows.
\Cref{sec:zjets-eventsel} describes the event selection and Monte Carlo samples used for the measurement.
\Cref{sec:zjets-omnifold} details the \Omnifold implementation, including the neural network classifiers, the pretraining procedure, and the unbinned background subtraction.
\Cref{sec:zjets-uncert} describes the uncertainty propagation, and \Cref{sec:zjets-validation} presents the statistical validation including comparisons to alternative measurements, in particular IBU based measurements and the previous round Multifold result.
\Cref{sec:zjets-results} then presents measurements of four classes of observables obtained from the full-phase-space measurement.
In order, these are simple event-level observables that are poorly modeled by MC generators, jet-shape observables, the energy-energy correlator (EEC), and the nearest-neighbor intrinsic dimension (NNID) of jets.
\Cref{sec:zjets-discussion} draws conclusions and offers some lessons learned that should be applied to future measurements which use AI/ML based unfolding techniques.

\FloatBarrier

\section{Event Selection and Simulation Samples}
\label{sec:zjets-eventsel}

The data sample consists of the full Run~2 $pp$ collision dataset collected by the ATLAS detector between 2015 and 2018, corresponding to an integrated luminosity of $140.1~\mathrm{fb}^{-1}$ after application of the standard ATLAS good runs lists.
\Cref{tab:eventsel} describes the event selections applied to all data and Monte Carlo simulation.
Events are selected with a single-muon trigger~\cite{TRIG-2018-01} and given the large dimuon \pt requirement described below all triggers are fully efficient in the target phase space.
Standard event cleaning requirements reject events with detector errors or data-taking faults.
To select $Z$+jets events, two opposite-sign muons are required which satisfy $81~\mathrm{GeV} < m_{\mu\mu} < 101~\mathrm{GeV}$ and $p_T^{\mu\mu} > 190~\mathrm{GeV}$.
The first requirement ensures that the dimuon system falls within the $Z$ boson mass peak window and the second requirement ensures the dimuon system is produced with large \pt.
\Cref{tab:ObjCuts} describes object-level requirements applied to all muons.
The muon requirements are additional event selections given each event is required to have two opposite-sign muons.
The track requirements simply determine which tracks are included in the detector-level event description.
The most important requirement is that all tracks must satisfy $p_T > 0.5$~GeV, the minimum \pt supported by ATLAS tracking measurements, and do not originate from muons.
After all selections, approximately 247 thousand events enter the signal region.

\begin{table}[htb]
    \centering
    \begin{tabular}{l|l}
         \hline
    \textbf{Event Selection} & \textbf{Description} \\ \hline
    Good Runs List & Event must be part of GRL \\ \hline
    Event Cleaning & No LAr, tile calorimeter, or tracker errors. Event is complete. \\ \hline
    Trigger & Single muon trigger: \\
    & \texttt{HLT\_mu20\_iloose\_L1MU15\_OR\_HLT\_mu50} (2015) \\
    & \texttt{HLT\_mu26\_ivarmedium\_OR\_HLT\_mu50} (2016-18) \\ \hline
    Primary Vertex & $N_{\text{PV}}\geq1$ \\ \hline
    Muons & $N_{\text{muons}}\geq2$ \\
          & Opposite charges \\
          & Pass TTVA recommendations for muons \\
          & $81\text{ GeV}\leq m_{\mu\mu}\leq 101\text{ GeV}$ \\
          & $p_T^{\mu\mu}\geq 190$ GeV \\ \hline
    \end{tabular}
    \caption{Overview of the event selection applied to both data and MC simulation.}
    \label{tab:eventsel}
\end{table}

\begin{table}[htb]
    \centering
    \begin{tabular}{l|l}
    \hline
     \textbf{Object} & \textbf{Additional selection criteria} \\ \hline
     Muons & $p_T>25~\text{GeV}$, $|\eta|<2.4$ \\
           & Medium quality, pass isolation \texttt{PflowLoose\_VarRad} \\ \hline
     Tracks & $p_{\text{T}} > 500$ MeV \\
            & Loose Quality \\
            & Tight TTVA \\
            & Do not originate from a muon \\ \hline
    \end{tabular}
    \caption{Requirements placed on individual analysis objects for them to enter the event record.}
    \label{tab:ObjCuts}
\end{table}

The fiducial volume targeted by the measurement is defined at particle level and is differential in the kinematics of all stable ($c\tau > 10~\mathrm{mm}$) charged particles.
In general the particle-level requirements mirror the detector-level requirements with the muon requirements applied to the muons produced by the $Z$ boson decay, and the track requirements applied to the remaining stable charged particles.
The only difference is that the dimuon \pt must be greater than 200~GeV.
As described below the \Omnifold procedure is run with all events that satisfy $p_T^{\mu\mu} > 190$~GeV and then the tighter selection is applied post-unfolding to reduce acceptance effects.
An important feature of this measurement is that no requirements are placed on jets or on the number or configuration of charged particles (excepting muons).
Event selections that rely on jets often introduce substantial acceptance effects in which events migrate across the event selection between detector level and particle level.
These complications are sufficiently small to be ignored in this analysis due to the simple event selection.
Additionally, avoiding event selections on jets or charged particles yields an unbiased sample of jets that can be used to constrain jet observables.

Monte Carlo simulation plays a significant role in this analysis.
All unfolding techniques require at least one Monte Carlo simulation which has both particle-level and detector-level event configurations.
A second Monte Carlo simulation is additionally used in this analysis for construction of the simulation based closure check detailed in \Cref{sec:zjets-validation} and to set an unfolding uncertainty.
Finally Monte Carlo simulation is used to estimate the contribution of top processes to the signal region which is statistically subtracted prior to the unfolding.

The primary contribution to the $Z$+jets signal comes from the production of a $Z$ boson via the strong interactions (the leading order Feynman diagrams for this process are shown in \Cref{fig:zjets-lo-diagrams}).
The nominal strong $Z$+jets sample is produced with \texttt{MG5\_aMC@NLO}~v2.6.5 at NLO accuracy~\cite{Alwall:2014hca,Frederix:2012ps} with the NNPDF3.0 NNLO PDF set.
Parton shower and hadronization is modeled with \textsc{Pythia}~8.240 using the A14 tune.
This sample is referred to as \mgpy throughout.
An alternative strong $Z$+jets sample is provided by \textsc{Sherpa}~2.2.11~\cite{Bothmann:2019yzt} with NLO matrix elements using the NNPDF3.0 NNLO PDF set.
The parton shower and hadronization is modeled using the default \textsc{Sherpa} settings.
This sample is referred to as \sherpa throughout.

Sub-leading contributions to the $Z$+jets signal come from electroweak $Z(\to\mu\mu)jj$ production and diboson processes.
Both processes are modeled with \textsc{Sherpa}~2.2.1 or 2.2.2, and these samples are appended to both strong $Z$+jets samples described above to produce the nominal or alternative MC samples used for unfolding.
In what follows, any reference to the \mgpy or \sherpa MC samples should be understood to include these simulations of non-strong processes.
Background contributions from top processes ($t\bar{t}$, $tW$, and single top in $s$ and $t$ channels) are modeled nominally with \textsc{Powheg}~v2~\cite{Alioli:2010xd} interfaced to \textsc{Pythia}~8.230, and an alternative $t\bar{t}$ sample is produced with \textsc{Sherpa}~2.2.12 for evaluation of the background-modeling uncertainty.

All simulated events are passed through a \textsc{Geant4} simulation of the ATLAS detector~\cite{ATLAS:2010arf}, overlaid with simulated minimum-bias pileup, and reconstructed with the same algorithms as data.
After selection the signal sample is 95.3\% strong $Z$+jets production, 2.9\% diboson (mostly $ZV\to\mu\mu jj$), and 1.6\% electroweak $Zjj$.
The analysis measures the inclusive sum of these contributions without attempting to separate them.

\begin{figure}[tb]
    \centering
    \includegraphics[page=1, width=0.48\linewidth, alt={Comparison of charged track multiplicity between data, MG5 FxFx, and Sherpa 2.2.11 showing approximately 10 percent disagreement in either direction.}]{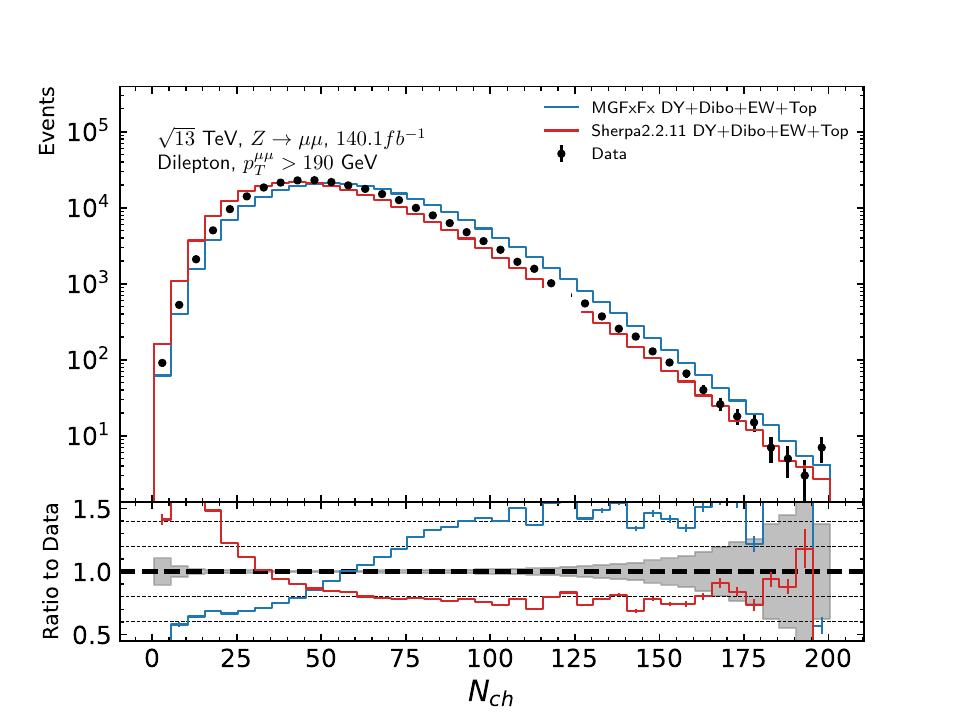}
    \includegraphics[page=2, width=0.48\linewidth, alt={Comparison of the scalar sum of charged track transverse momenta H_T between data, MG5 FxFx, and Sherpa 2.2.11.}]{ch6_6_MGFxFx_Sherpa2211_and_Others_Vs_Data.pdf}
    \caption{Comparison of the per-event charged-track multiplicity \nch (left) and the scalar sum of charged-track transverse momenta \HT (right) between data and the Monte Carlo samples used in this analysis.}
    \label{fig:zjets-trackmcdata}
\end{figure}

Histograms comparing the data and MC samples in two observables, the track multiplicity \nch and the scalar sum of transverse momomentum \HT, are provided in \Cref{fig:zjets-trackmcdata}.
The modeling of these observables is quite poor for both generators.
\mgpy overpredicts the \nch observable by almost 50\% in the tails, and \sherpa underpredicts by a similar margin.
The modeling for \HT is better, but there are still deviations of up to 25\%.
These observables are known to be particularly difficult to model accurately.
In addition to making them good observables to measure, the poor modeling also makes them valuable targets with which to assess the performance of \Omnifold because the method must produce a reweighting that can cover such large differences.

\begin{figure}[htb]
    \centering
    \includegraphics[page=1, width=0.48\linewidth, alt={Reconstructed-level event display showing two muons, reconstructed tracks, and track jets for a simulated Z plus jets event.}]{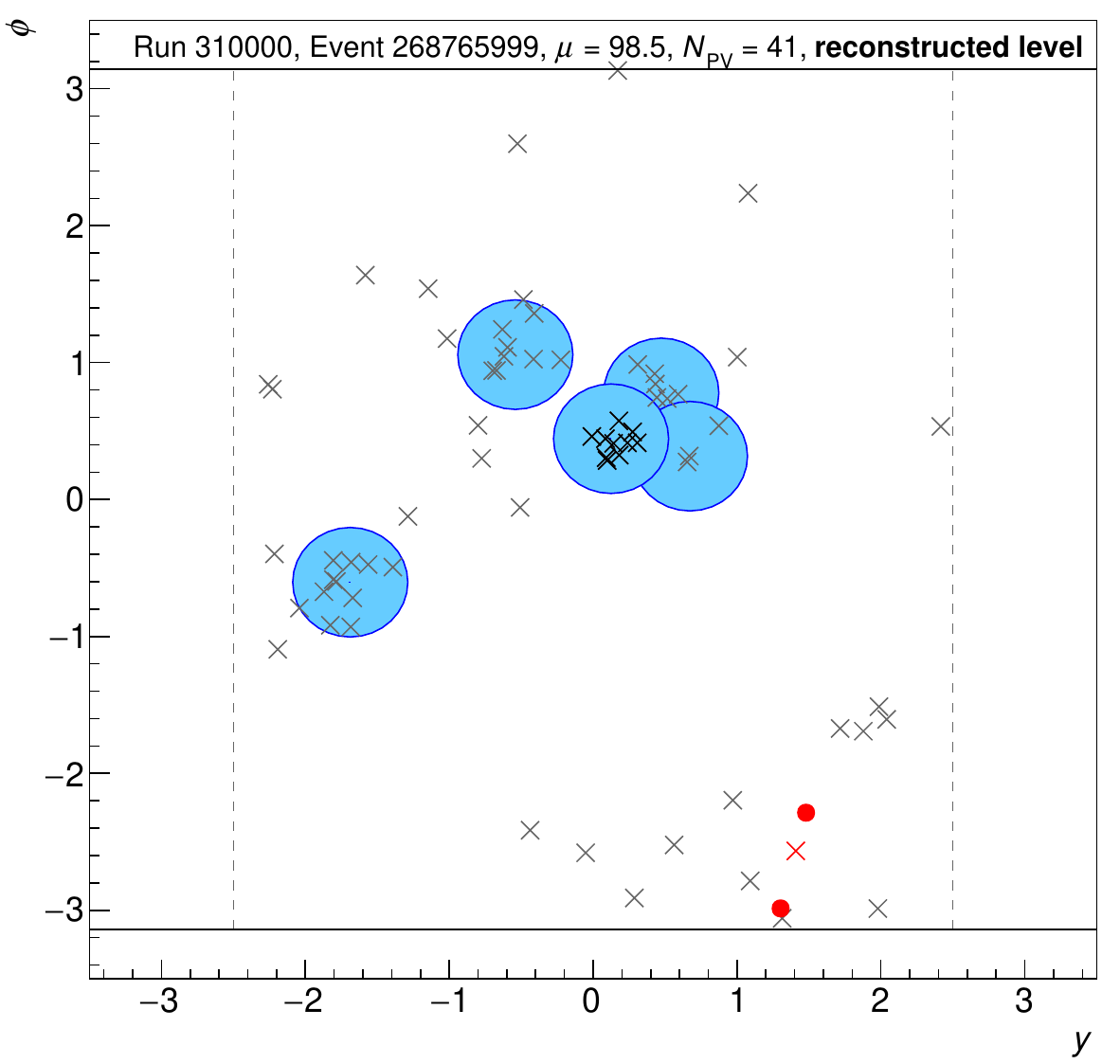}
    \includegraphics[page=2, width=0.48\linewidth, alt={Particle-level event display of the same Z plus jets event showing the dimuon pair and all charged hadrons.}]{ch6_EventDisplays.pdf}\\
    \includegraphics[page=3, width=0.48\linewidth, alt={Zoom-in of the leading jet for the simulated event showing the correspondence between reconstructed tracks and truth charged hadrons, including examples of missed and fake tracks.}]{ch6_EventDisplays.pdf}
    \caption{Event displays for a simulated event passing the detector-level and particle-level requirements. Top left: detector-level view showing muons (red dots), tracks (crosses), the dimuon system (red cross), and the leading five \ak track jets (blue circles). Top right: the same event but at particle-level. Bottom: zoom-in of the leading track jet showing the correspondence between reconstructed tracks (crosses) and truth hadrons (circles).}
    \label{fig:zjets-eventdisplay}
\end{figure}

\Cref{fig:zjets-eventdisplay} shows event displays for a single simulated event that passes the requirements at both detector and particle level.
Most tracks have a clear truth counterpart, but examples of both missed charged hadrons (inefficiencies) and tracks without a truth counterpart (fakes) are visible.
The separation between the dimuon system (and some possible final state radiation) and the hadronic recoil is also visible as the two large clusters of particles in both configurations.

\subsection{Pseudodata Construction}
\label{sec:zjets-pseudodata}

Given the novelty of the unfolding method used in this measurement, it is important to validate the unfolding using a simulated data sample with a known truth target.
This is done by constructing \textit{pseudodata} from MC simulation, which is then unfolded using the same MC samples and unfolding procedure as in the data measurement.
This auxiliary simulation-based measurement is called the \textit{pseudomeasurement}.
To construct the pseudodata, events were sampled from the \sherpa strong $Z$+jets, electroweak, and diboson samples along with the \textsc{Powheg}+\textsc{Pythia} $t\bar{t}$ sample.
This produced a Monte Carlo estimate of the data sample, which was then reweighted at particle level to match the data at detector level.
This reweighting was obtained with a series of one-dimensional Gaussian kernel reweighting functions, iteratively applied to many different observables.
Following this step the pseudodata was further divided into two sets, one to serve as the detector level pseudodata treated as ``data'' for the unfolding, and the other serves as particle level truth pseudodata which sets the target for the pseudomeasurement.
The pseudomeasurement is run using an identical implementation of \Omnifold as the data measurement.
Unlike the data measurement, the pseudomeasurement has a well-defined and fixed target, which can be used to validate the unfolding method.
See \Cref{sec:zjets-validation} for more details on the validation of the unfolding method using the pseudomeasurement.
The \Omnifold implementation described in the next section was designed using the pseudodata.
This is because the \textit{method bias}, or the difference between the unfolded distribution and the target distribution, can be explicitly checked when unfolding pseudodata and used to make design decisions.

\FloatBarrier

\section{OmniFold Implementation}
\label{sec:zjets-omnifold}

The \Omnifold algorithm is introduced in \Cref{sec:omnifold}.
Running the algorithm to produce a cross section measurement requires a large number of practical choices that are the subject of this section.
\Cref{sec:zjets-classifier} describes the machine learning details, specifically which neural network architecture is used for the classification tasks within \Omnifold.
\Cref{sec:zjets-pretraining} describes the pretraining used to stabilize the results.
The pretraining is a novel aspect of this analysis compared to Ref.~\cite{ATLAS:2024xxl} which has substantial implications for future AI-based unfolding applications.
\Cref{sec:zjets-bkg-subtraction} describes the unbinned background subtraction procedure, and \Cref{sec:zjets-method-norm} describes how the weights produced by \Omnifold are normalized to produce a cross section measurement.

\subsection{Classifier Architecture and Training}
\label{sec:zjets-classifier}

\begin{figure}[htb]
    \centering
    \subfloat{
        \includegraphics[width=0.45\textwidth, alt={Pseudodata unfolding result for the charged-track multiplicity using the ParT classifier, comparing unfolded density to target with a ratio pad and an uncertainty pad.}]{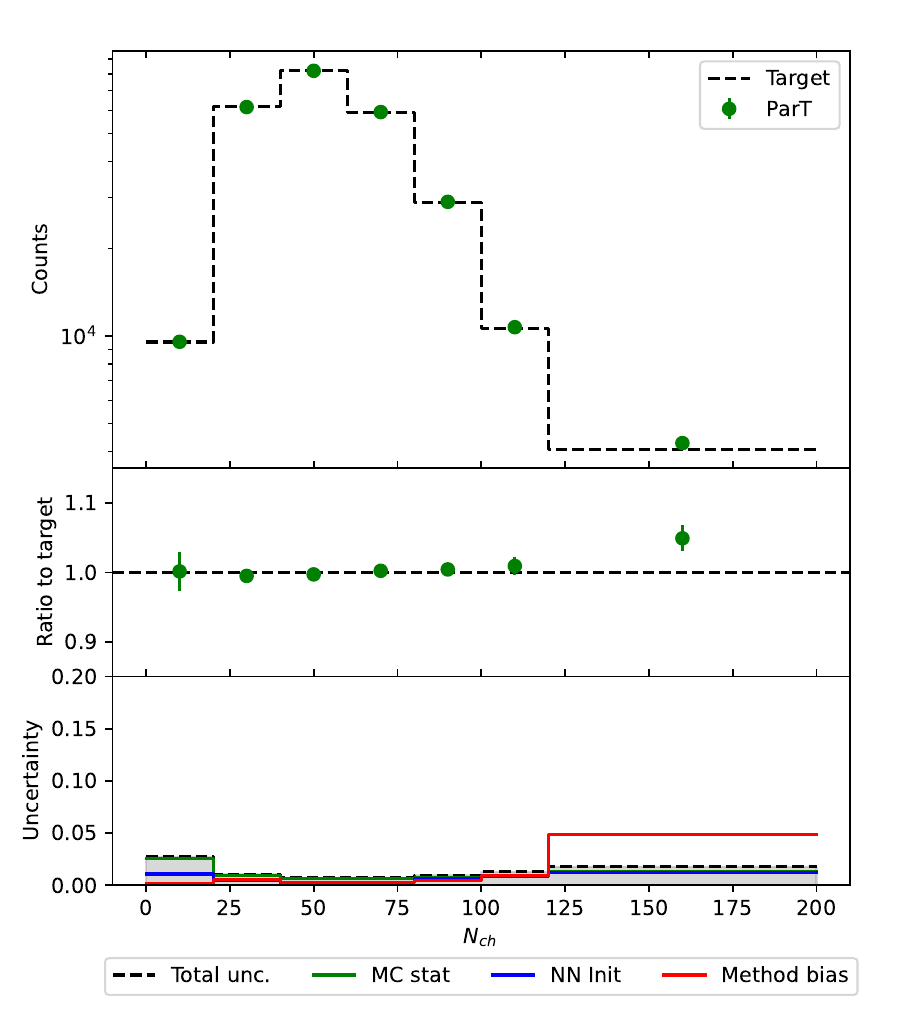}
    }
    \hfill
    \subfloat{
        \includegraphics[width=0.45\textwidth, alt={Pseudodata unfolding result for the charged-track multiplicity using the PET classifier, comparing unfolded density to target with a ratio pad and an uncertainty pad.}]{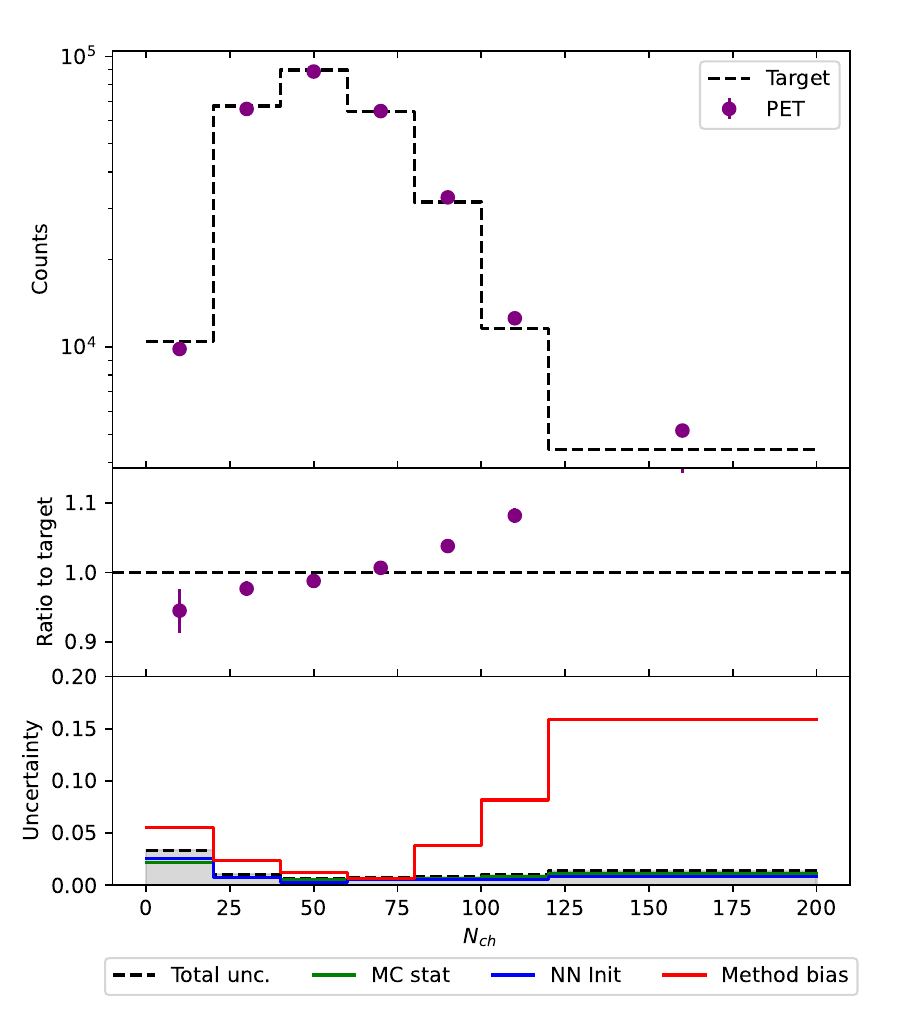}
    } \\
    \subfloat{
        \includegraphics[width=0.45\textwidth, alt={Pseudodata unfolding result for the scalar sum of track transverse momenta using the ParT classifier, comparing unfolded density to target with a ratio pad and an uncertainty pad.}]{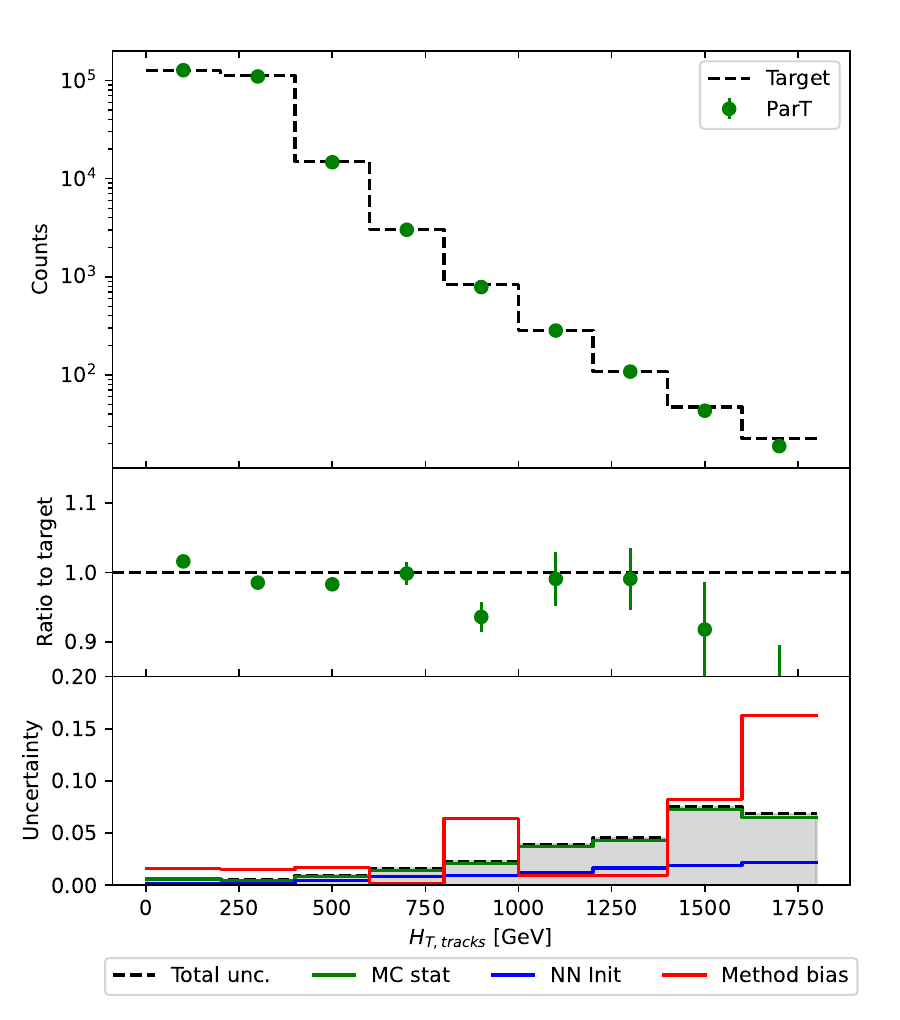}
    }
    \hfill
    \subfloat{
        \includegraphics[width=0.45\textwidth, alt={Pseudodata unfolding result for the scalar sum of track transverse momenta using the PET classifier, comparing unfolded density to target with a ratio pad and an uncertainty pad.}]{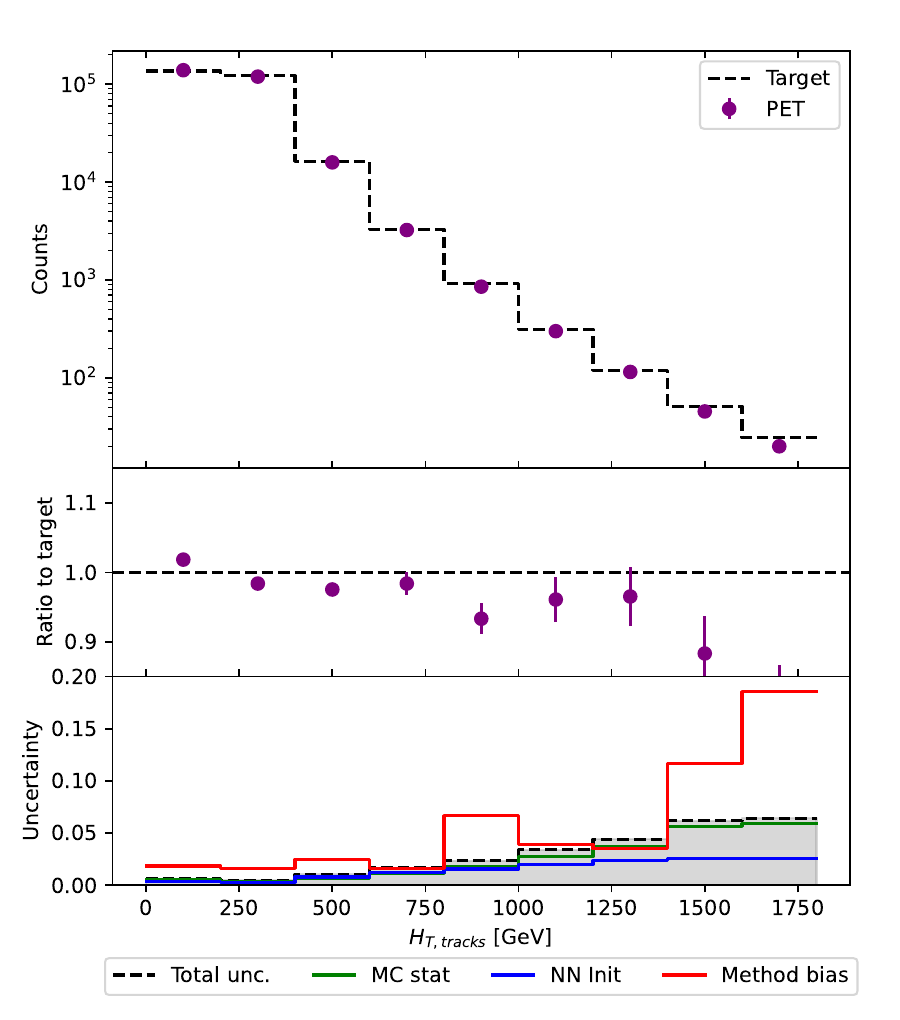}
    } \\
    \caption{A comparison of \Omnifold results in the pseudodata measurement using the ParT (left) and PET (right) classifier architectures for the \nch (top) and \HT (bottom) observables. The top pad in the plots compares the unfolded density to the target, the ratio pad shows the ratio of the unfolded density to the target, and the bottom pad shows the NN initialization and MC stat uncertainties, in addition to the method bias. The ensembles used to create these plots consist of 10 runs of Omnifold.}
    \label{fig:pet_unfolding1}
\end{figure}

\begin{figure}[htb]
    \centering
    \subfloat{
        \includegraphics[width=0.45\textwidth, alt={Pseudodata unfolding result for the sub-leading track-jet mass using the ParT classifier, comparing unfolded density to target with a ratio pad and an uncertainty pad.}]{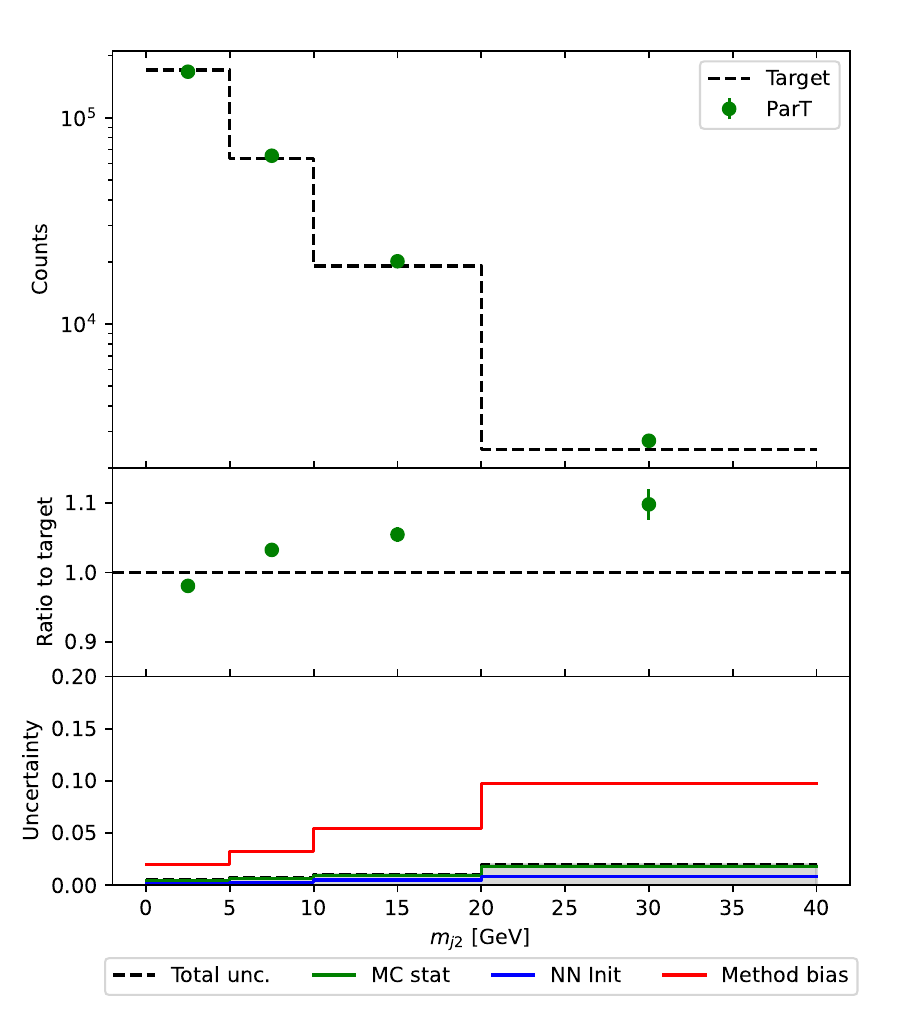}
    }
    \hfill
    \subfloat{
        \includegraphics[width=0.45\textwidth, alt={Pseudodata unfolding result for the sub-leading track-jet mass using the PET classifier, comparing unfolded density to target with a ratio pad and an uncertainty pad.}]{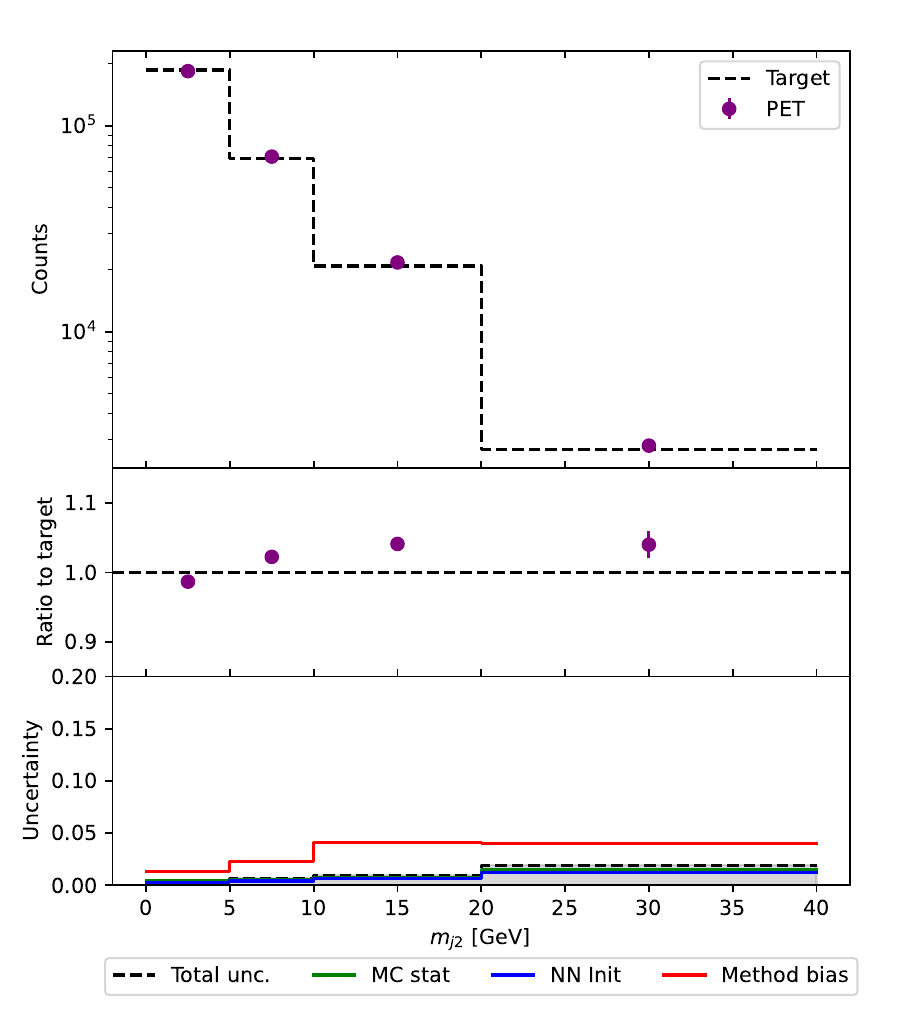}
    } \\
    \subfloat{
        \includegraphics[width=0.45\textwidth, alt={Pseudodata unfolding result for the leading jet pT clustered with the anti-kt fat-jet algorithm using the ParT classifier, comparing unfolded density to target with a ratio pad and an uncertainty pad.}]{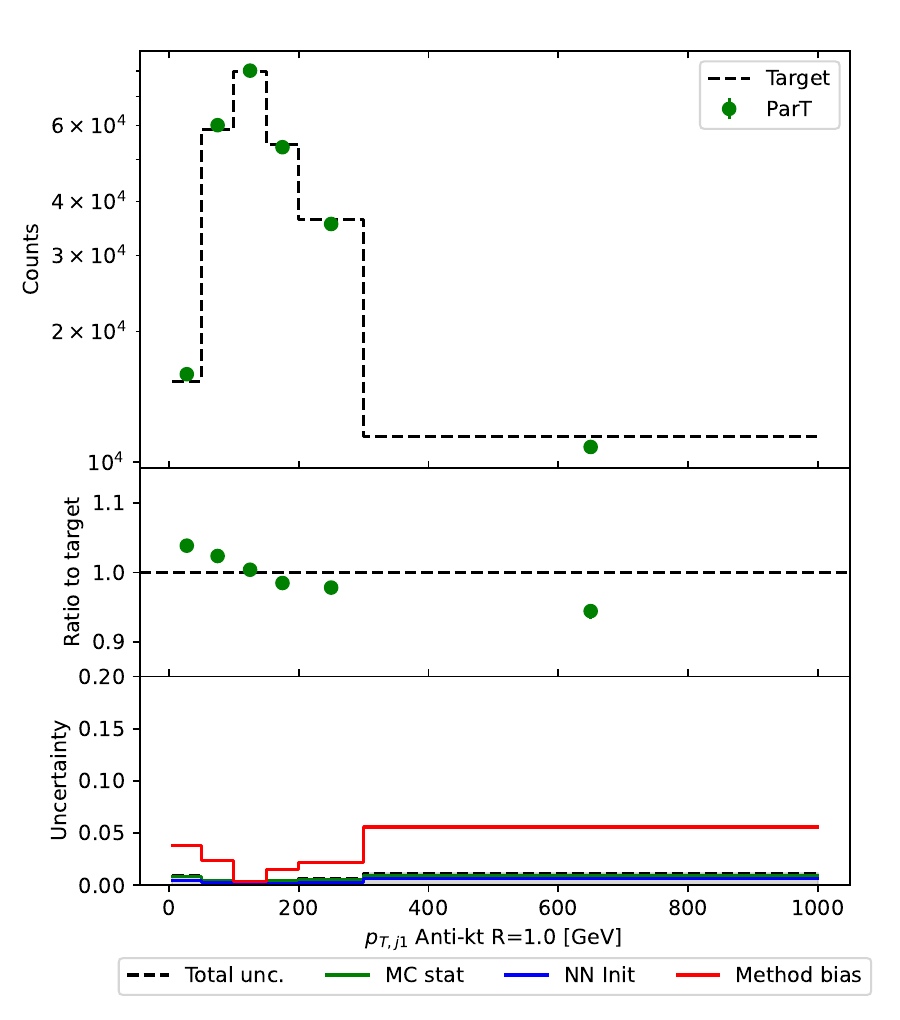}
    }
    \hfill
    \subfloat{
        \includegraphics[width=0.45\textwidth, alt={Pseudodata unfolding result for the leading jet pT clustered with the anti-kt fat-jet algorithm using the PET classifier, comparing unfolded density to target with a ratio pad and an uncertainty pad.}]{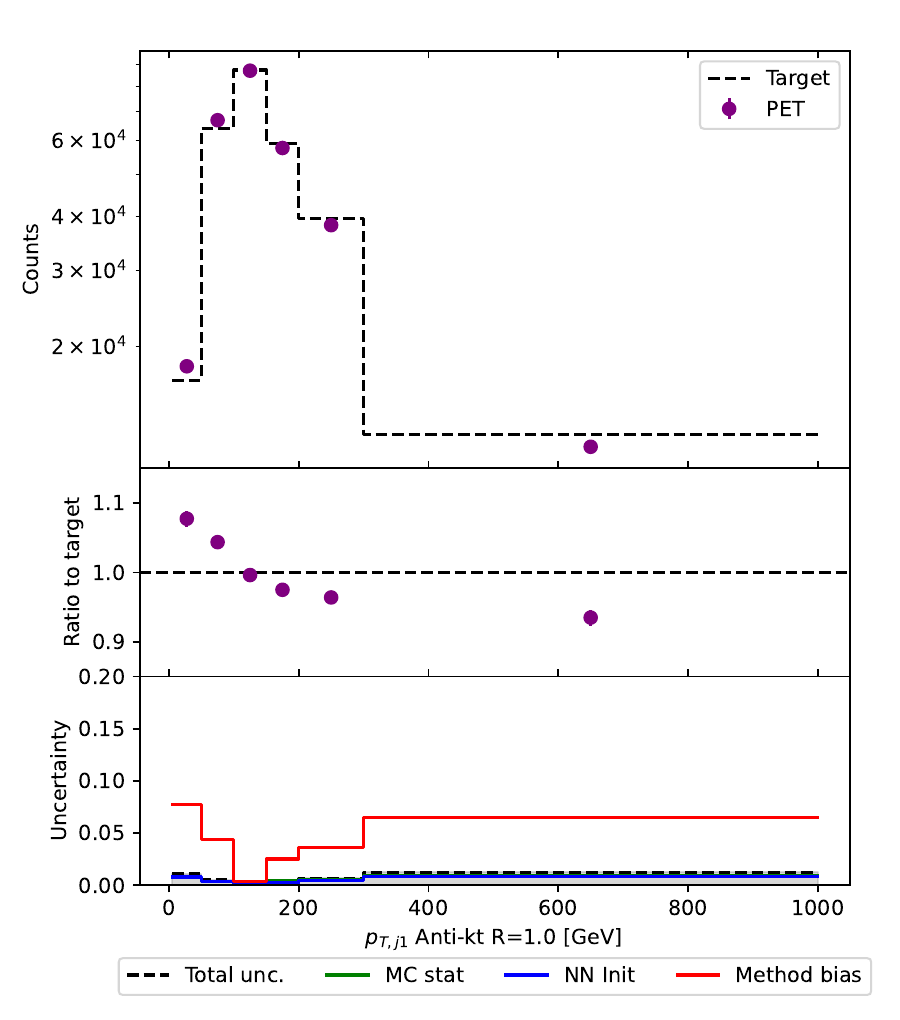}
    } \\
    \caption{A comparison of \Omnifold results in the pseudodata measurement using the ParT (left) and PET (right) classifier architectures. Results are shown for (top) the sub-leading track jet mass ($m_{j2}$) and (bottom) the \pt of the leading track jet clusted with the \akfat algorithm. The top pad in the plots compares the unfolded density to the target, the ratio pad shows the ratio of the unfolded density to the target, and the bottom pad shows the NN initialization and MC stat uncertainties, in addition to the method bias. The ensembles used to create these plots consist of 10 runs of Omnifold.}
    \label{fig:pet_unfolding2}
\end{figure}

Given both the detector-level and particle-level phase spaces in this measurement are variable-dimensional, set processing neural networks are needed.
This motivates using either graph or transformer networks.
Rather than design a network architecture from scratch, some of the most successful architectures on HEP tasks like jet tagging are considered.
These are the Particle Transformer (ParT)~\cite{Qu:2022mxj}, the Point-Edge Transformer (PET)~\cite{Mikuni:2024qsr}, the Lorentz-equivariant geometric transformer L-GATr~\cite{Brehmer:2024yqw}, and LundNet~\cite{Dreyer:2020brq}.
ParT, PET, and L-GATr are all transformer architectures, while LundNet is a graph network.
These models are chosen for the following motivations.
ParT is a very popular transformer architecture for HEP tasks.
It shows strong performance on jet tagging tasks and has been studied and used by both ATLAS and CMS.
PET is a similar model to ParT, with the primary distinction being that PET uses local graph attention embeddings in place of ParT's pairwise features.
Both of these networks have little inductive bias, or specialization for particle physics tasks.

L-GATr is a Lorentz equivariant network whose output is constrained to be invariant under known symmetries of collider datasets.
Prior to this work, no uses of equivariant networks for likelihood ratio estimation tasks were documented, so L-GATr is tested here to see if equivariance is helpful in this setting.
Finally, LundNet is another successful jet tagging network that includes a different physics prior.
It builds a graph using the Cambridge-Aachen jet clustering algorithm where the nodes are splittings rather than individual particles.
The node features are kinematic properties of these splittings.
This specifically incorporates features known to be very useful for understanding jet formation, which is a central concern of this measurement.
Though all of these models are originally designed to process individual jets rather than entire events as is needed here, the extension to full events is trivial for the transformer networks.
For LundNet, all tracks in the event are reclustered with the Cambridge-Aachen jet clustering algorithm and then the input graph is built with the clustering tree.
Muon kinematics are included as event-level features in the final MLP that processes the aggregated node features.
These processing steps add significant compute overhead to using LundNet.

ParT is chosen as the classifier architecture for all neural network trainings in the \Omnifold procedure given the following considerations.
An explicit comparison of the method bias produced by unfolding with the ParT and PET networks is shown in \Cref{fig:pet_unfolding1} and \Cref{fig:pet_unfolding2}.
The unfolding is shown in the observables \nch, \HT, $m_{j2}$, and the \pt of the leading jet clustered with the \akfat algorithm.
In general ParT and PET produce similar unfolding results, but ParT has overall better performance especially in the \nch observable.
Notably PET is more accurate in the $m_{j2}$ observable.
ParT is additionally slightly faster to train than PET, so is taken to be the best general purpose transformer architecture.

\FloatBarrier

\begin{figure}[htb]
    \centering
    \subfloat{
        \includegraphics[width=0.45\textwidth, alt={Validation loss curves for L-GATr and ParT as a function of training epoch.}]{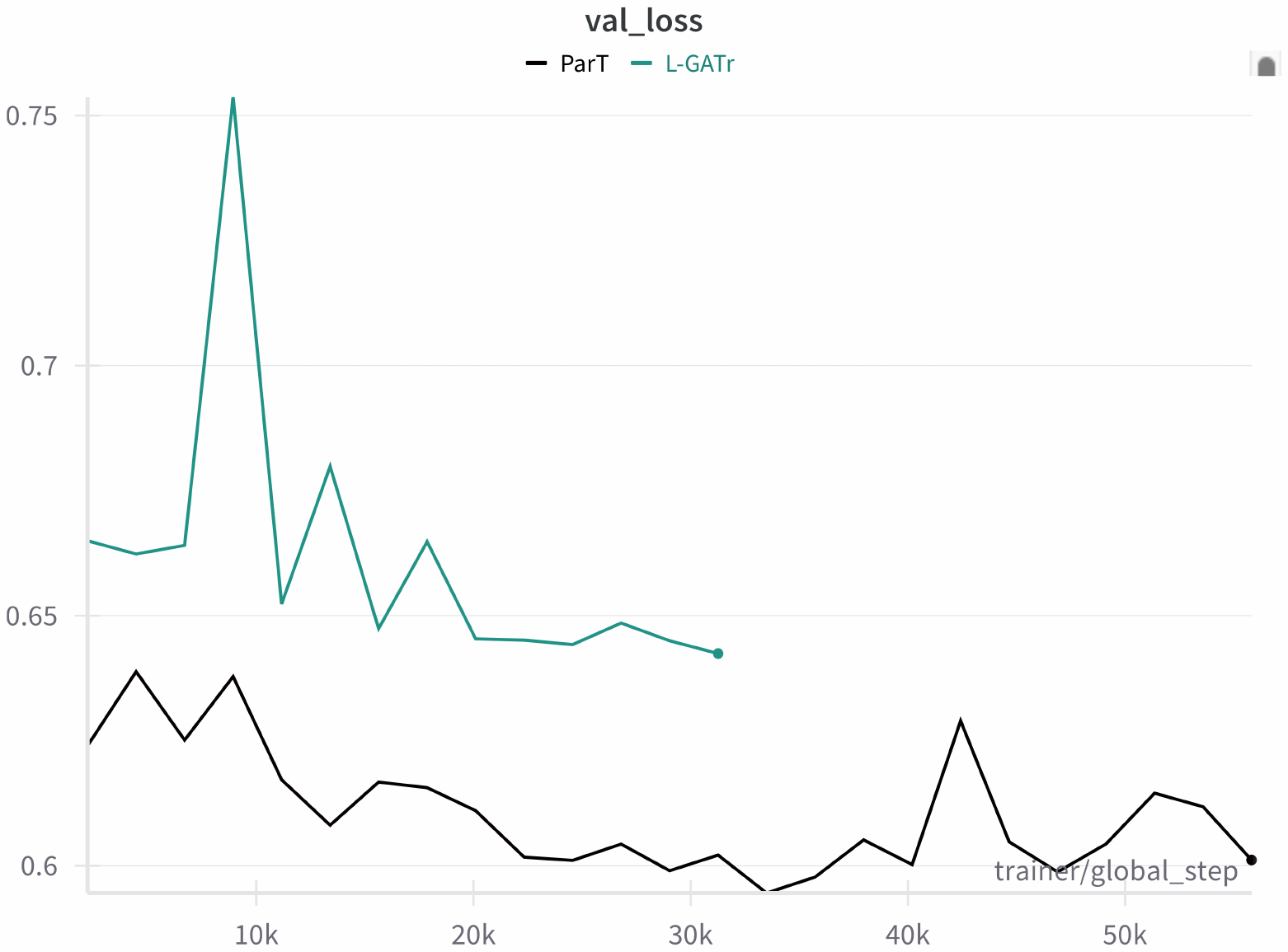}
        \label{fig:lgatr_loss}
    }
    \hfill
    \subfloat{
        \includegraphics[width=0.45\textwidth, alt={Epoch number versus wall-clock training time for L-GATr and ParT on identical GPU hardware, showing L-GATr trains more slowly.}]{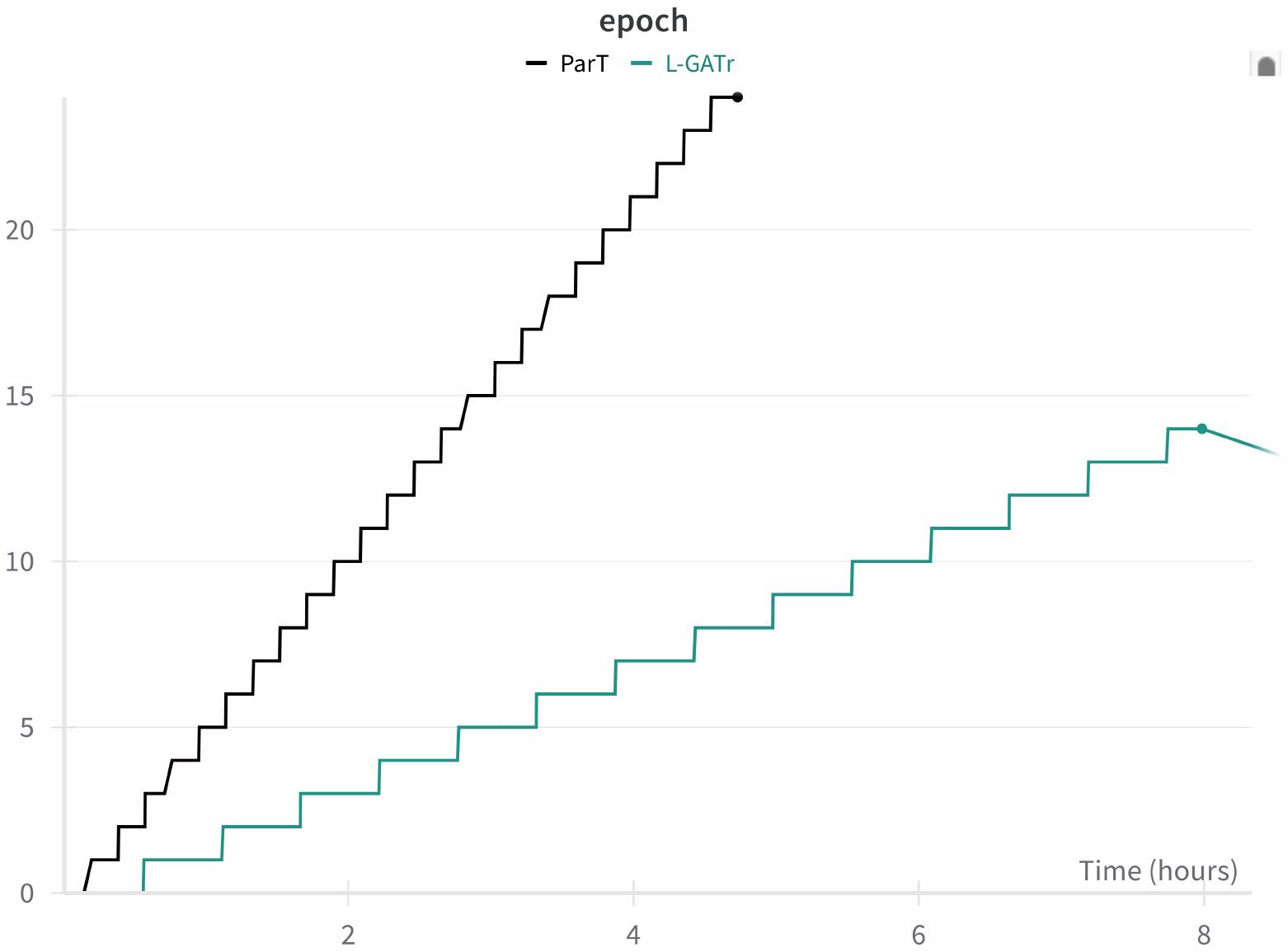}
        \label{fig:lgatr_time}
    }
    \caption{(Left) Validation loss curves for L-GATr (green) and ParT (black) as a function of epoch. (Right) The epoch number versus wall time for L-GATr (green) and ParT (black).}
    \label{fig:lgatr_train}
\end{figure}

\begin{figure}[htb]
    \centering
    \subfloat{
        \includegraphics[width=0.48\textwidth, alt={Distribution of reweighting weights produced by the ParT classifier in the pretraining task between MG5 FxFx and Sherpa detector-level samples.}]{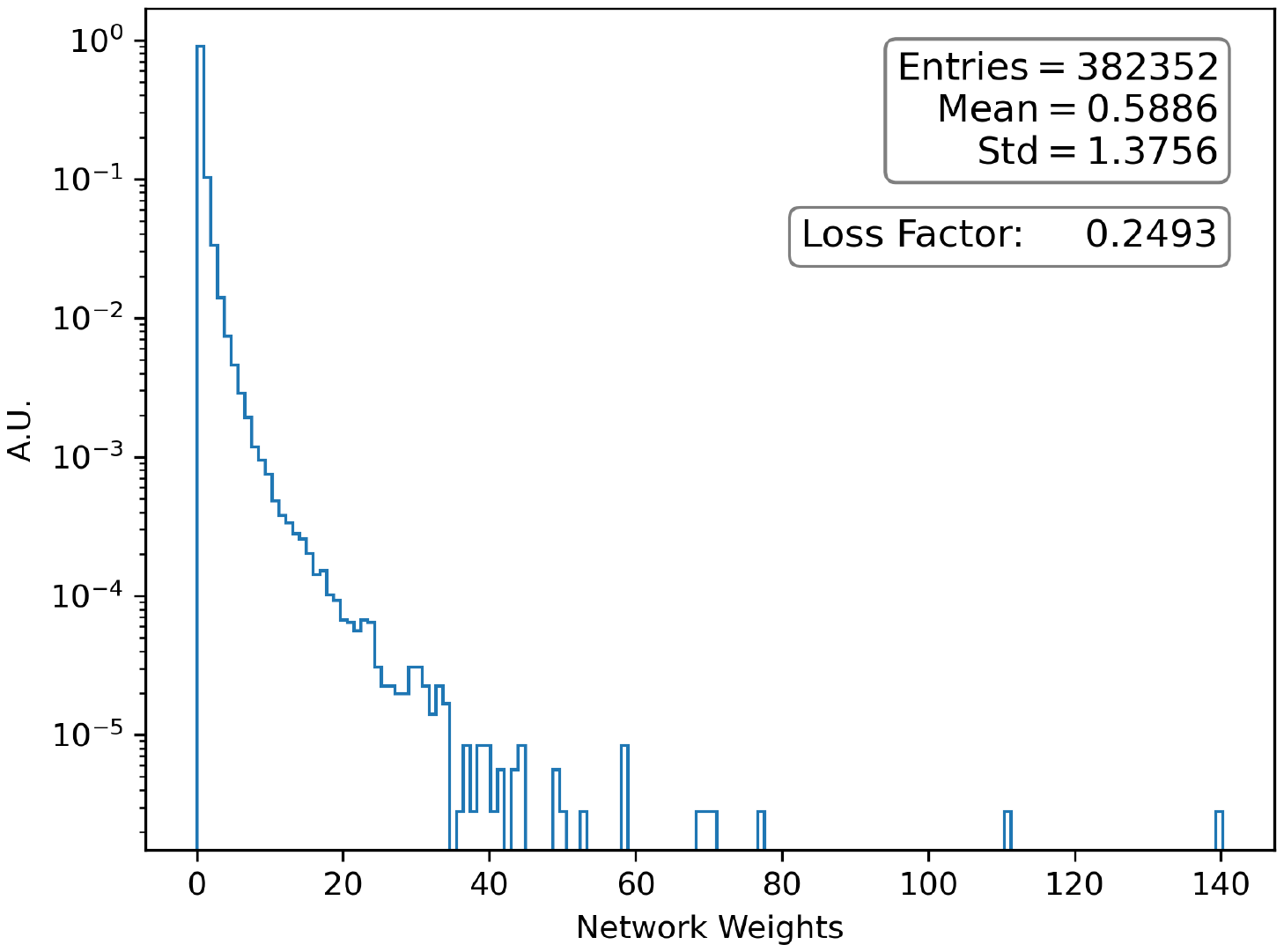}
        \label{fig:part_weights_v2}
    }
    \subfloat{
        \includegraphics[width=0.48\textwidth, alt={Distribution of reweighting weights produced by the L-GATr classifier in the pretraining task between MG5 FxFx and Sherpa detector-level samples.}]{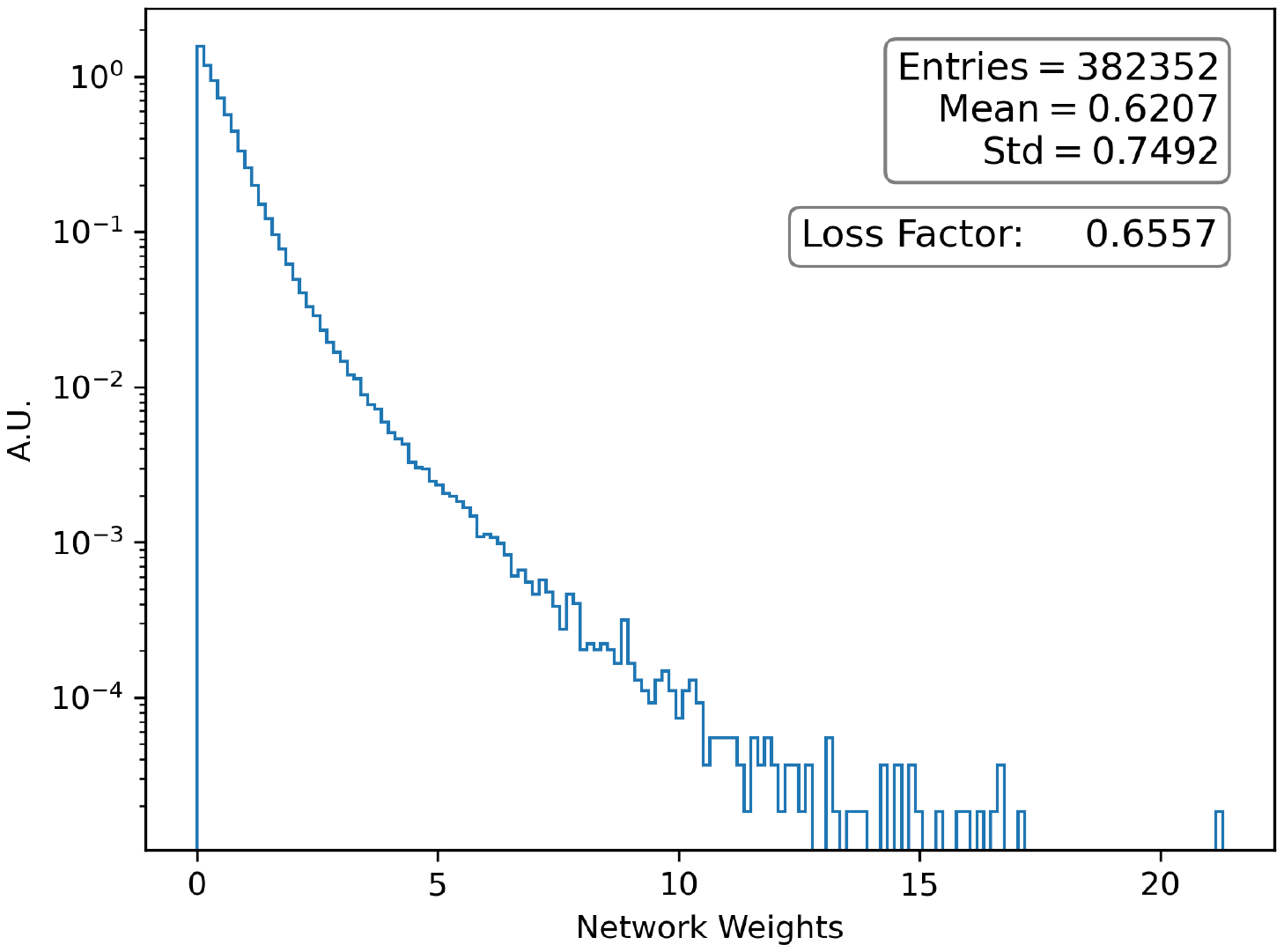}
        \label{fig:lgatr_weights}
    } \\
    \subfloat{
        \includegraphics[width=0.48\textwidth, alt={Reweighting produced by the ParT classifier in the sub-leading track-jet mass dimension, comparing MG5 FxFx and Sherpa detector-level distributions.}]{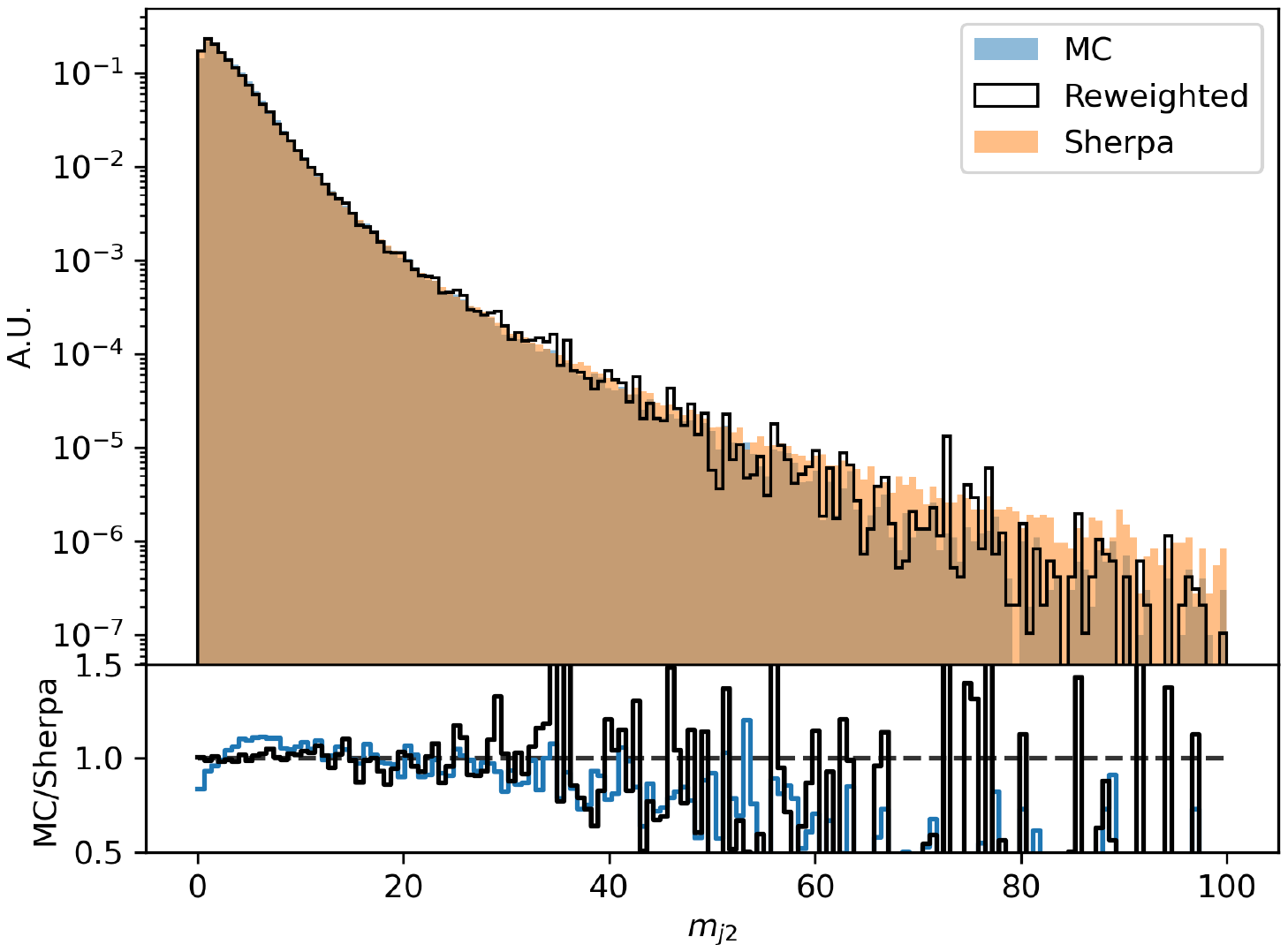}
        \label{fig:part_mj2_v2}
    }
    \subfloat{
        \includegraphics[width=0.48\textwidth, alt={Reweighting produced by the L-GATr classifier in the sub-leading track-jet mass dimension, comparing MG5 FxFx and Sherpa detector-level distributions, showing reduced accuracy relative to ParT.}]{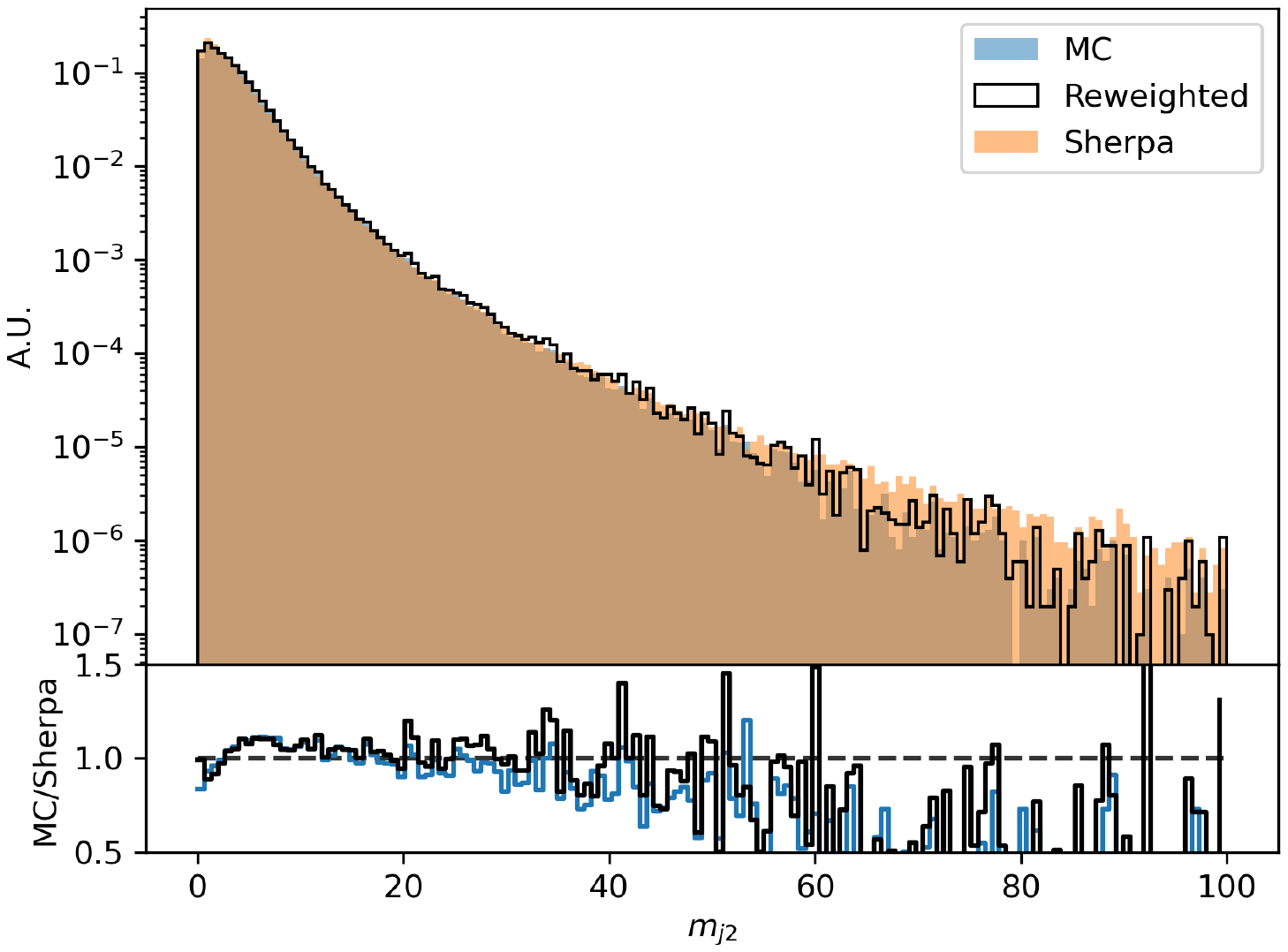}
        \label{fig:lgatr_mj2}
    }
    \caption{Plots of the reweighting between the detector-level \mgpy and \sherpa samples used in the pretraining task for example ParT (left) and L-GATr (right) architectures. The top row shows the weight distributions produced by each network, and the bottom row shows the reweighting in the $m_{j2}$ dimension.}
    \label{fig:lgatr_perf}
\end{figure}

Given this it remains to compare the performance of ParT to the more physics motivated architectures, LundNet and L-GATr.
Unlike ParT and PET, LundNet and L-GATr are only tested on a single classification task (iteration 1 step 1 for LundNet and the pretraining task described in \Cref{sec:zjets-pretraining} for L-GATr) of the \Omnifold procedure rather than the full procedure.
The reason is that both networks have substantial issues that make further studies not worth the effort.
LundNet shows worse performance in numerical classification metrics than ParT even after some tuning.
For example, the test AUC of LundNet is 0.734 versus 0.778 for ParT.
Examining the reweightings produced by the classifiers shows that LundNet mostly captures the likelihood ratio, but is significantly less accurate than ParT in a few observables, namely the jet mass observables.
This combined with the slow training of LundNet due to the need to recluster the tracks for each event make ParT preferable.

L-GATr has shown strong performance in jet classification tasks and has explicit physics constraints that should help it learn even with the somewhat limited amount of training data.
However much like LundNet, L-GATr is observed to train slower than ParT and produces only similar or slightly worse performance. 
\Cref{fig:lgatr_train} shows validation loss curves for L-GATr and ParT as a function of optimizer step, as well as the epoch number versus wall time for each network with identical GPU hardware.
\Cref{fig:lgatr_perf} shows the weight distributions and reweightings of the $m_{j2}$ observable for ParT and L-GATr.
When judging the quality of a reweighting, the primary concern is the reweighting accuracy.
The bottom row plots in \Cref{fig:lgatr_perf} shows that L-GATr is not as accurate as ParT when reweighting the $m_{j2}$ observable.
This is a surprising result since equivariant networks have been shown to be effective and highly data efficient in jet classification tasks conducted using \textsc{Delphes}-based detector simulation (see \Cref{sec:tagging-future}).
An additional consideration is the weight distribution produced by the neural network.
Significant departure of the derived weights from 1.0 degrades the statistical power of the dataset by reducing the number of effective events.
Within \Omnifold, this degradation is unavoidable but ideally should be controlled by avoiding arbitrarily large weights.
See \Cref{sec:zjets-method-norm} for some discussion of how this is avoided in practice.
In the context of comparing the ParT and L-GATr networks, L-GATr produces a weight distribution with a much smaller standard deviation compared to ParT, resulting in a much smaller degradation of the statistical power of the data capture by the \textit{loss factor}, or ratio of the number of effective events in the data post- and pre-reweighting.
This provides evidence that the underperformance of L-GATr may be because the network is constrained to be Lorentz equivariant, preventing it from exploiting non-Lorentz equivariant features of the data that turn out to be useful for modeling the likelihood ratio.
This results in both underperformant reweightings in some observables such as $m_{j2}$ and a weight distribution with a lower standard deviation than the weight distribution produced by the non-equivariant ParT.
Ultimately the better reweighting performance of ParT was determined to be more valuable than the lower statistical degradation of L-GATr.

This comparison of likelihood ratio estimation between ParT and L-GATr motivated me and my collaborators to publish Ref.~\cite{Breso-Pla:2026tlz}, that benchmarked L-GATr and the pretrained transformer model \textsc{OmniLearn} on a set of likelihood ratio estimation tasks.
The study sought to understand whether it is more performant to explicitly encode known physics constraints, for example Lorentz equivariance, into the network architecture as with L-GATr or to implicitly encode the same information through large-scale pretraining as with the \textsc{OmniLearn} model and the pretraining procedure discussed in \Cref{sec:zjets-pretraining}.
The result is that the performance is surprisingly similar on the tasks considered, but the pretraining based approach generally seems to have an edge.
It is also notable that the biggest performance difference between \textsc{OmniLearn} and L-GATr in this study is seen in the likelihood ratio estimation task on Monte Carlo generated with a realistic simulation of the H1 detector rather than \textsc{Delphes}.
This is more evidence in support of the hypothesis that equivariant architectures do not generalize well to realistic detector simulation.

\begin{figure}[htb]
    \centering
    \includegraphics[width=0.85\linewidth, alt={Schematic of the Particle Transformer architecture showing the particle embedding MLP, stack of particle attention blocks with pairwise physics-motivated features added to the attention logits, class attention block with learned class token, and final MLP producing a single output logit.}]{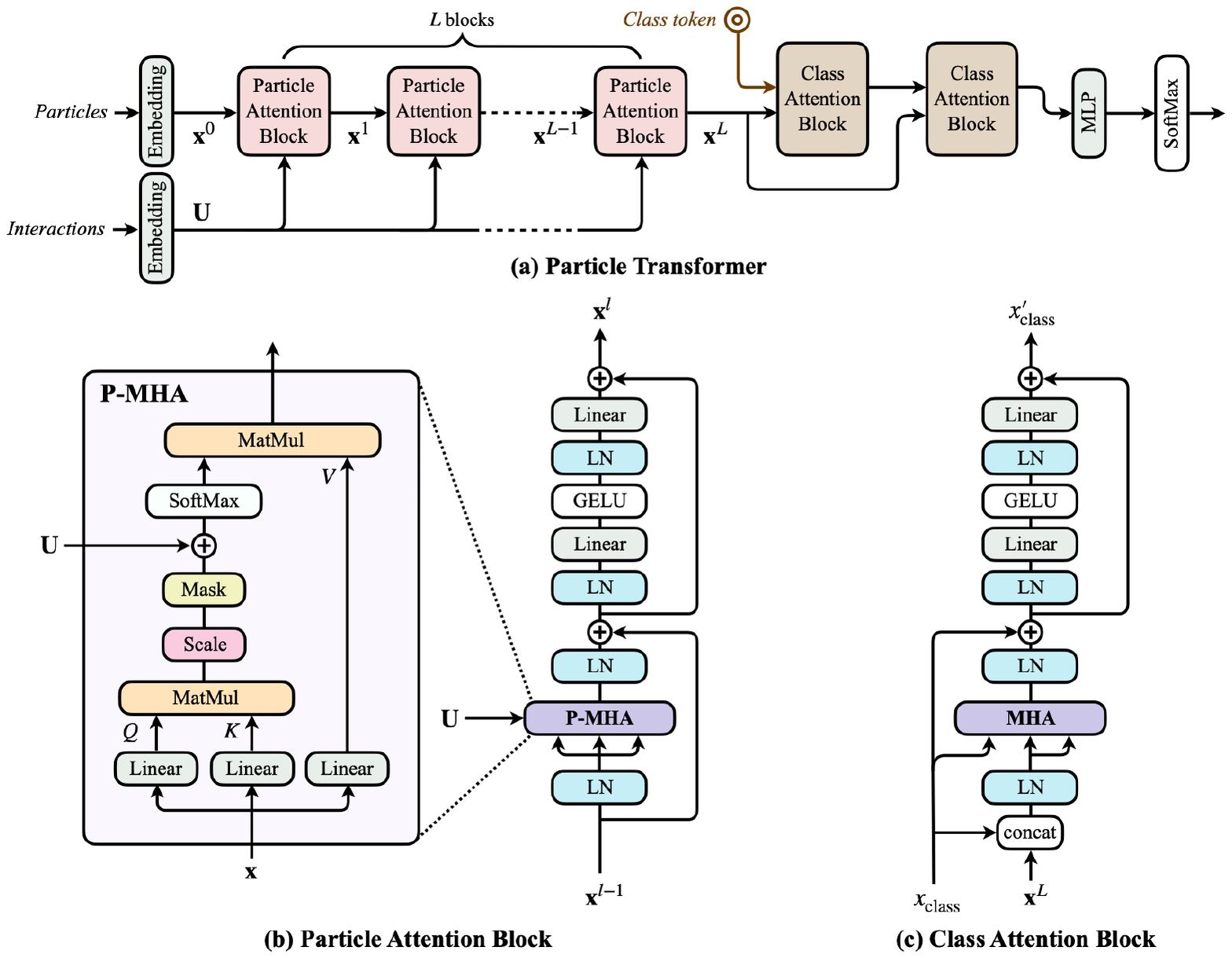}
    \caption{Schematic of the Particle Transformer architecture. The input list of particle features is embedded into a fixed-length feature vector per particle, passed through a stack of particle attention blocks in which physics-motivated pairwise features ($\log k_T$, $\log z$, $\log \Delta R$, $\log m^2$) are added to the attention logits before the softmax, and aggregated into a single class-token output by a class attention block. A final MLP produces the output logit whose transformation via the sigmoid is a monotonic function of the likelihood ratio. Reproduced from Ref.~\cite{Qu:2022mxj} under a CC~BY~4.0 license.}
    \label{fig:zjets-part-arch}
\end{figure}

\begin{figure}[htb]
    \centering
    \includegraphics[width=0.65\linewidth, alt={Plot of the learning rate schedules used in all trainings, with the pretraining schedule shown in red, step one in blue, and step two in gray, with fading colors indicating increasing iteration number of the Omnifold procedure.}]{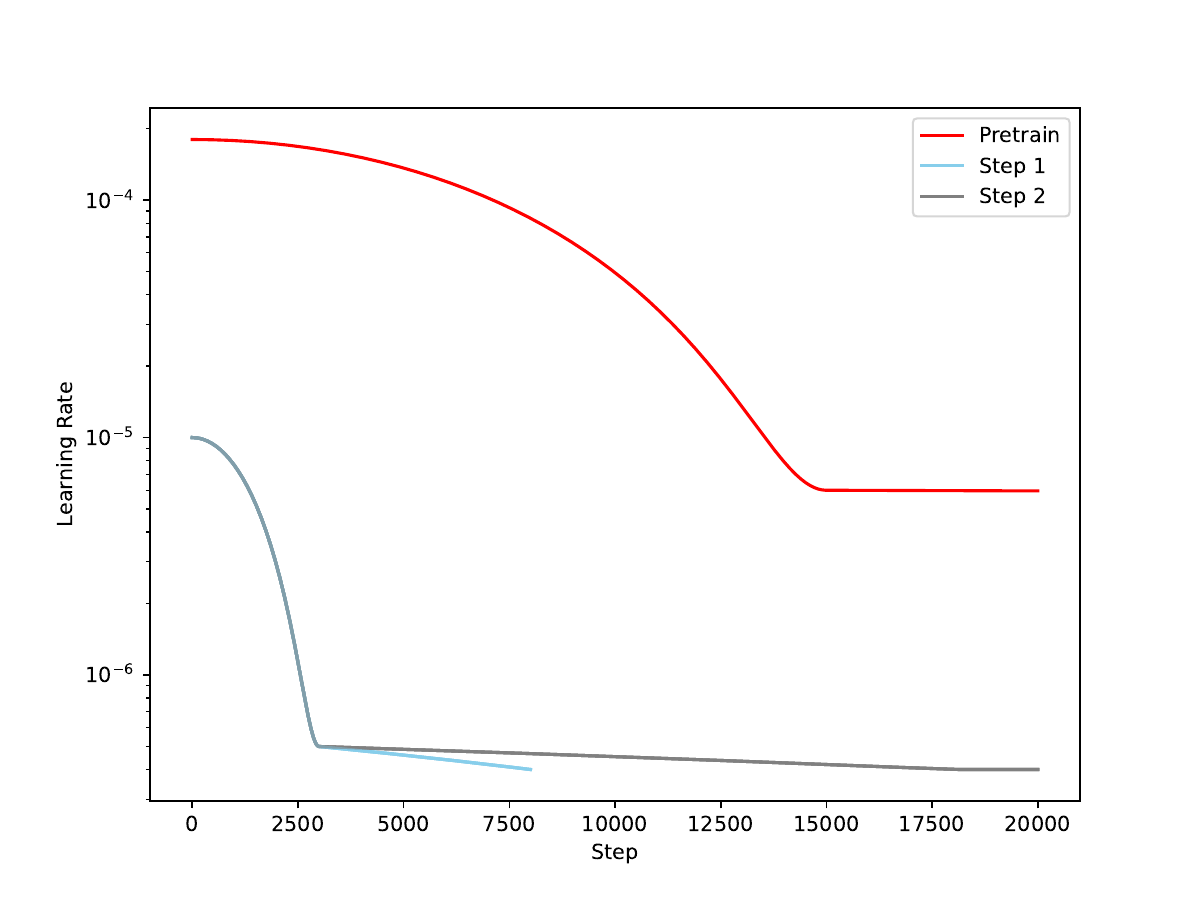}
    \caption{Learning rate schedules used in the pretraining task described below (red), as well as the step-one (blue) and step-two (gray) network trainings. Each schedule consists of a cosine annealing period followed by a linear decay to the minimum learning rate.}
    \label{fig:zjets-lr-schedule}
\end{figure}

Following these optimizations, the ParT network is selected as the classifier for the \Omnifold procedure.
\Cref{fig:zjets-part-arch} shows the ParT architecture used throughout the analysis.
The public PyTorch implementation of ParT is used\footnote{\url{https://github.com/jet-universe/particle_transformer}.}.
The input to the network is a set of particles, the two muons and the collection of charged hadrons, each described by a feature vector of ten entries.
The first four features are the particle kinematics $(\log p_T, \eta, \phi, m)$.
The \pt, $\eta$, and $\phi$ observables are directly measured by ATLAS at detector level, however ATLAS is insensitive to the masses of individual particles.
Given the charged pion is the most abundant charged particle produced in LHC collisions, the mass of every particle (besides the muons) at detector level is taken to be the pion mass.
This is a significant approximation that will have consequences for what follows.
At particle level the masses of the individual particles are available from the generator.
The remaining six features are one-hot encodings indicating whether the particle belongs to the leading, sub-leading, third, fourth, or fifth \ak track jet in the event, or to none of these jets.
No explicit flag is used to distinguish muons from hadrons because this is already accomplished by the particle mass.
Including these track jet indices as inputs is found to substantially improve the quality of the reweightings produced by the ParT network.
It was also checked that the kinematic features of jets clustered with alternative algorithms (different $R$ values or the Cambridge-Aachen algorithm) are unfolded with equal accuracy despite inputs specific to the \ak algorithm being used.

The ten per-particle features are passed through an MLP embedding which encodes the features into a latent vector.
These latent vectors are then stacked into a sequence which is passed through six \textit{particle attention} blocks.
Particle attention is distinguished from regular attention by the addition of physics-motivated pairwise features $(\log k_T, \log z, \log \Delta R, \log m^2)$, computed between every particle in the sequence, to the attention matrices.
This is accomplished by passing these pairwise features through an MLP to furnish a latent representation, and then adding the features to the attention logits before the softmax.
After the six particle attention blocks, a single class attention block then aggregates information from the entire sequence into a single latent vector.
This latent vector is passed through a final MLP to produce a single output which is interpreted as a logit classifying signal against background.
With the hyperparameters tuned via the procedure described below, the ParT networks used in \Omnifold have roughly 1.6 million trainable parameters.

All classifiers are trained with the binary cross-entropy loss of \Cref{eqn:bce} and the Lion optimizer~\cite{chen2023symbolic}.
Learning rate schedules consist of a cosine annealing ramp followed by a linear decay phase, with the schedule period tuned separately for the three types of trainings involved in the analysis: pretraining, and steps one and two of the \Omnifold procedure.
The resulting schedules are shown in \Cref{fig:zjets-lr-schedule}.
The learning rate schedule was the subject of a large amount of hyperparameter optimization, since it was found to have a significant impact on the method bias produced by \Omnifold.
The need to explicitly tune these schedules is one of the most difficult parts of implementing \Omnifold in practice.
In particular the reweighting quality produced by the step one classifiers is very sensitive to the maximum and minimum learning rates, as well as the cosine annealing period.

\begin{figure}[htb]
    \centering
    \includegraphics[width=0.48\linewidth, alt={Reweighting of the charged hadron H_T distribution derived from a checkpoint that minimizes the validation loss, showing visible over-correction in the tail.}]{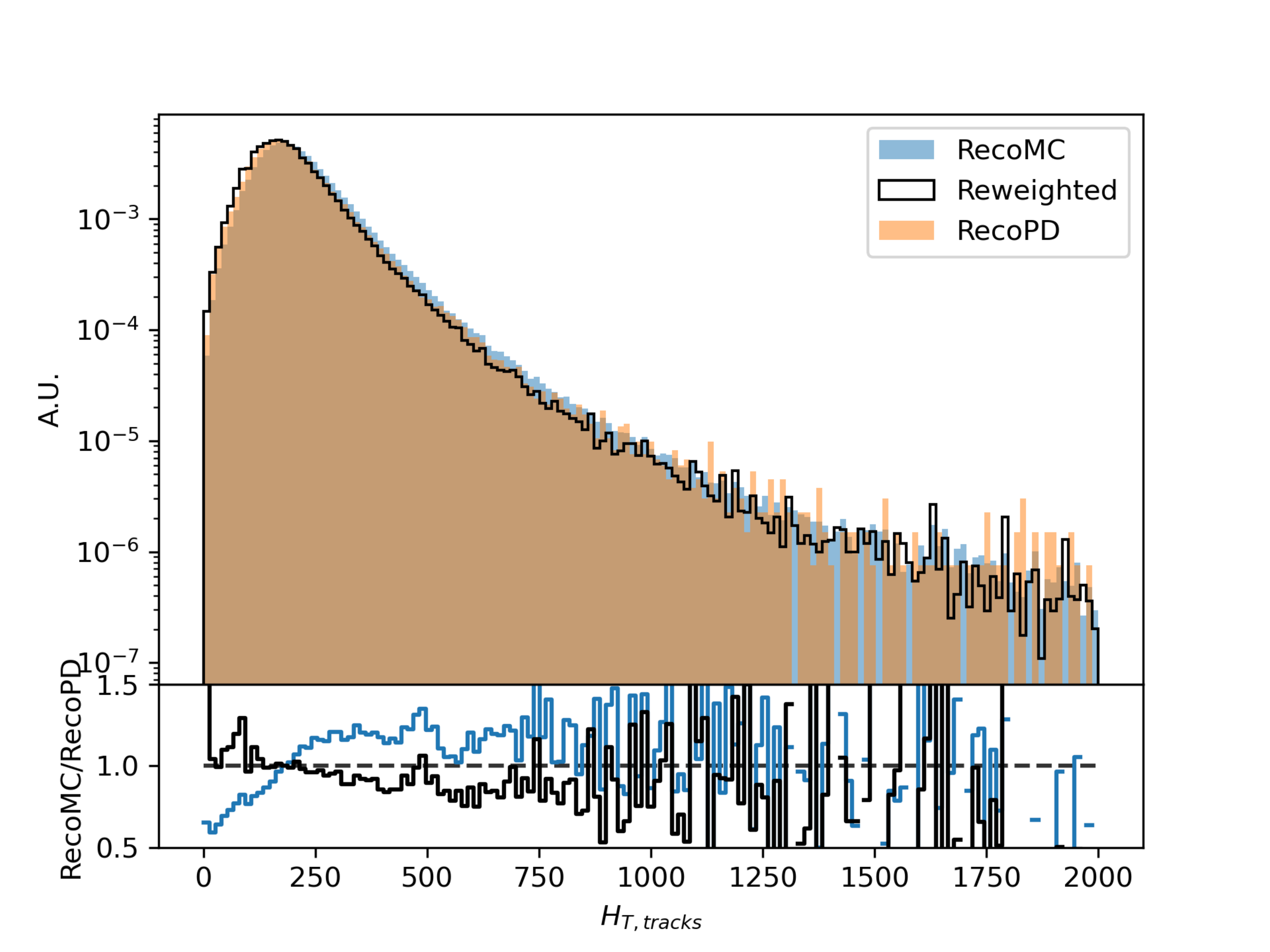}
    \includegraphics[width=0.48\linewidth, alt={Reweighting of the charged hadron H_T distribution derived from a checkpoint that minimizes the marginal Wasserstein distance, showing accurate closure without over-correction.}]{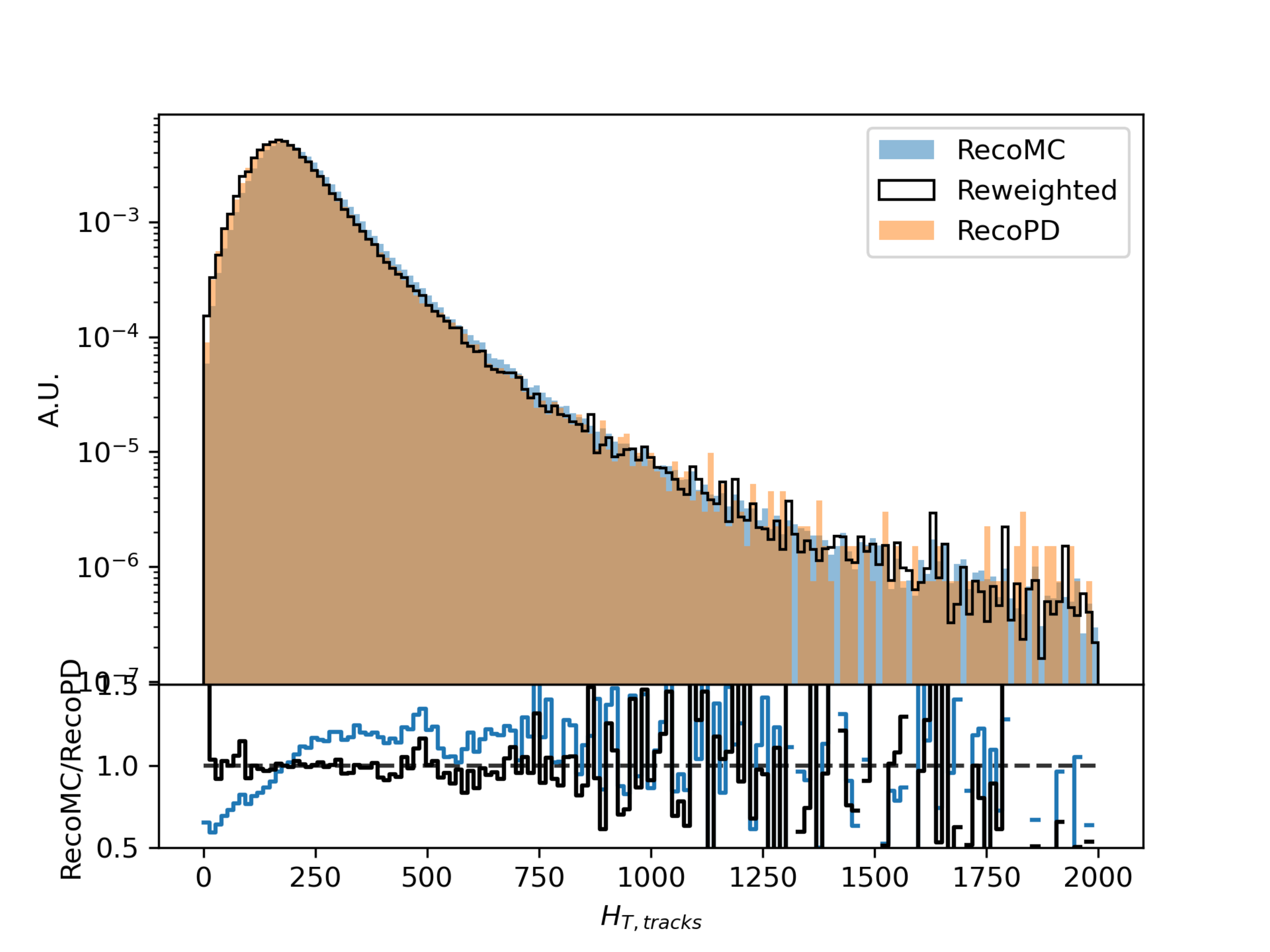}
    \caption{Reweightings of the charged \HT observable produced by two checkpoints from the same training. (Left) The checkpoint that minimizes the validation binary cross-entropy loss over-corrects the distribution in the tail. (Right) The checkpoint that minimizes the summed marginal 1D Wasserstein distance over 26 observables produces an accurate reweighting. The Wasserstein-based stopping criterion is used throughout this analysis.}
    \label{fig:zjets-overcorrection}
\end{figure}

A standard practice in neural network training is to select the final network weights by taking the checkpoint that minimizes the validation loss over the course of training.
This practice was found to be sub-optimal for training networks within \Omnifold, because checkpoints with the lowest validation loss frequently produced reweightings that \textit{over-corrected} in the tails of distributions where training data is sparse.
An example is shown in the left panel of \Cref{fig:zjets-overcorrection}, where the reweighting in the \HT observable produced by the checkpoint with the lowest validation loss down-weights the tail of the distribution beyond what is needed to match the pseudodata.
To select better checkpoints a different metric based on Wasserstein distances~\cite{panaretos2019statistical} is used.
The Wasserstein distance between two 1D distributions with cumulants $F$ and $G$ is
\begin{equation}
    W(p, q) = \int \abs{F(x) - G(x)} \, dx,
\end{equation}
and has the intuitive interpretation of the minimum cost to move the probability mass of one distribution onto the other.
High-dimensional generalizations of this one-dimensional expression exist but are computationally expensive.
Instead the 1D Wasserstein distance is computed for each of 26 marginal observables (the 24 used in the previous round Multifold measurement, plus \nch and \HT) and summed.
The network checkpoint minimizing this summed marginal Wasserstein distance is selected for the reweighting produced by each training within \Omnifold.
The right panel of \Cref{fig:zjets-overcorrection} shows the \HT reweighting produced by the checkpoint that minimizes this Wasserstein metric.
The marginal Wasserstein distance was also used as the figure of merit in hyperparameter optimization.
It was independently verified that checkpoints selected by this criterion produce accurate reweightings even for observables outside of the 26 used in the calculation.

Hyper-parameters were tuned through a random search guided by the Wasserstein metric above and by some visual inspection of the reweightings produced by the networks.
Importantly it is computationally infeasible to optimize all hyperparameters of \Omnifold at once using the method bias or the final uncertainty budget.
Instead three sub-tasks within \Omnifold were optimized: the pretraining step described below, and the step one and step two trainings of the first iteration.
The hyperparameters for future iterations were simply taken to be the optimal hyperparameters identified for the first iteration.
Architectural hyperparameters (number of attention blocks, number of class-attention blocks, dropout rates) were fixed by the pretraining optimization because the same architecture must be reused in all subsequent trainings.
Training hyperparameters (learning rate, weight decay, cosine annealing period, choice of optimizer) were tuned separately in the step-one and step-two optimizations.

\FloatBarrier

\subsection{Pretraining}
\label{sec:zjets-pretraining}

The 247~thousand data events passing the event selection constitute an order of magnitude less training data than is typically advised for training a 1.6 million parameter model from scratch.
The standard solution to this problem is to \textit{pretrain} the network on a large auxiliary dataset before fine-tuning it on the target task.
This is also known as transfer learning, where a related task with better data statistics is leveraged to find better performance on the target task.
Pretraining has become standard practice in many AI applications, especially language models which are pretrained on large corpora of text before being fine-tuned for specific applications.
The underlying principle is that the representations learned by the hidden layers of the network when solving the auxiliary task transfer to the target task.

The auxiliary task used here is MC-versus-MC classification, as opposed to the data-versus-MC classification task in the step one trainings of \Omnifold.
These step one trainings are the important ones since they are limited by the data statistics whereas the step two trainings are not.
In the pretraining task, ParT networks are trained from random initialization to discriminate between events drawn from the \mgpy and \sherpa strong $Z$+jets samples.
Dedicated partitions of both samples are used to avoid any concerns about overfitting in the unfolding procedure.
Roughly 14 million events are available between the two classes.
The pretraining is performed using both the detector-level and particle-level event configurations, and the resulting models are used as starting points for the step one and step two trainings respectively.
To avoid biasing the measurement by relying on the representations learned by a single pretrained starting point for each step of \Omnifold, ten independent ParT models are pretrained at both detector level and particle level using different random seeds.
Each \Omnifold step-one training begins from one of the ten detector-level pretrained models, and each step-two training from one of the ten particle-level pretrained models.

\begin{figure}[tb]
    \centering
    \includegraphics[width=0.48\linewidth, alt={Ten trajectories of marginal Wasserstein distance between reweighted truth-level MC and truth pseudodata as a function of Omnifold iteration for networks initialized without pretraining, showing high variance across runs and large final Wasserstein distance.}]{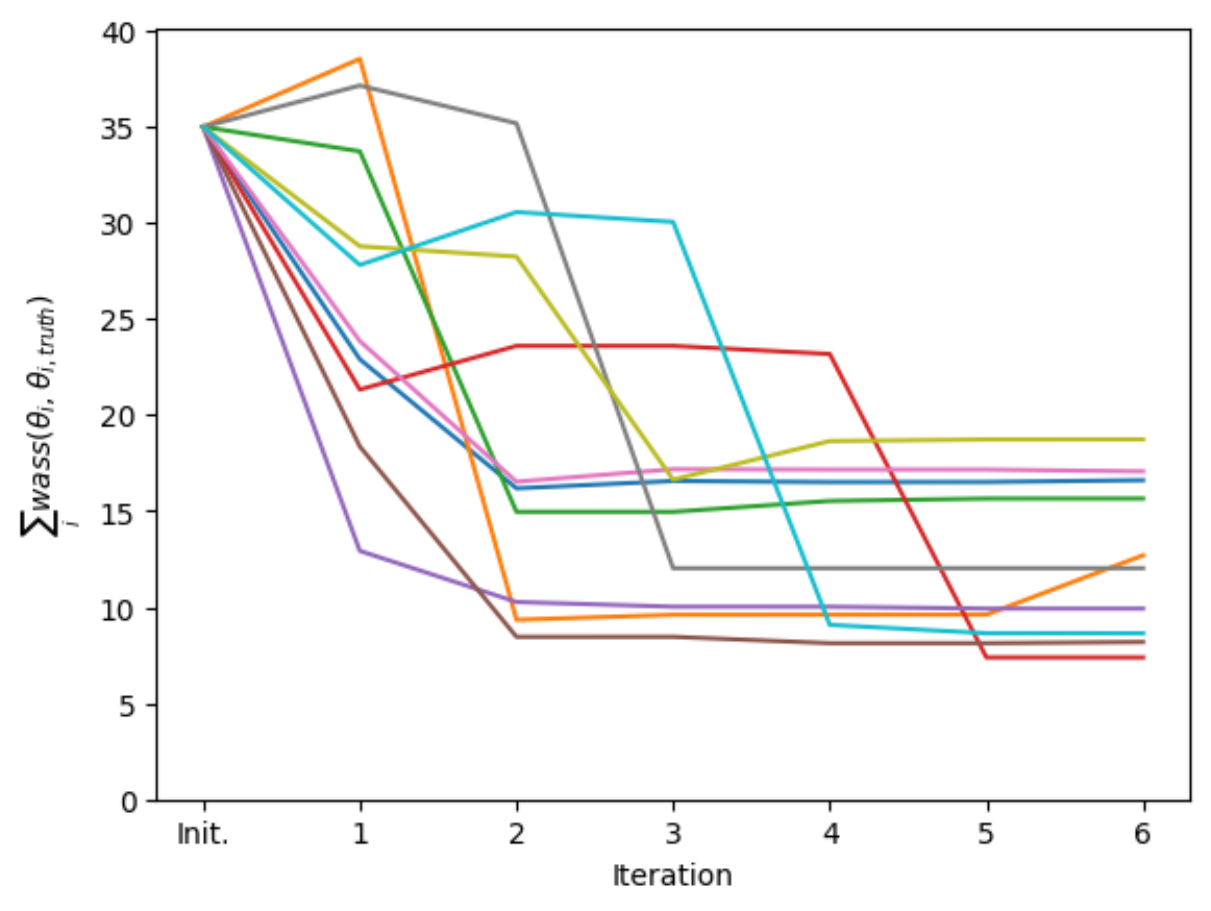}
    \includegraphics[width=0.48\linewidth, alt={Same ten trajectories for networks initialized from pretrained checkpoints, showing much lower final Wasserstein distance and significantly reduced variance across independent runs.}]{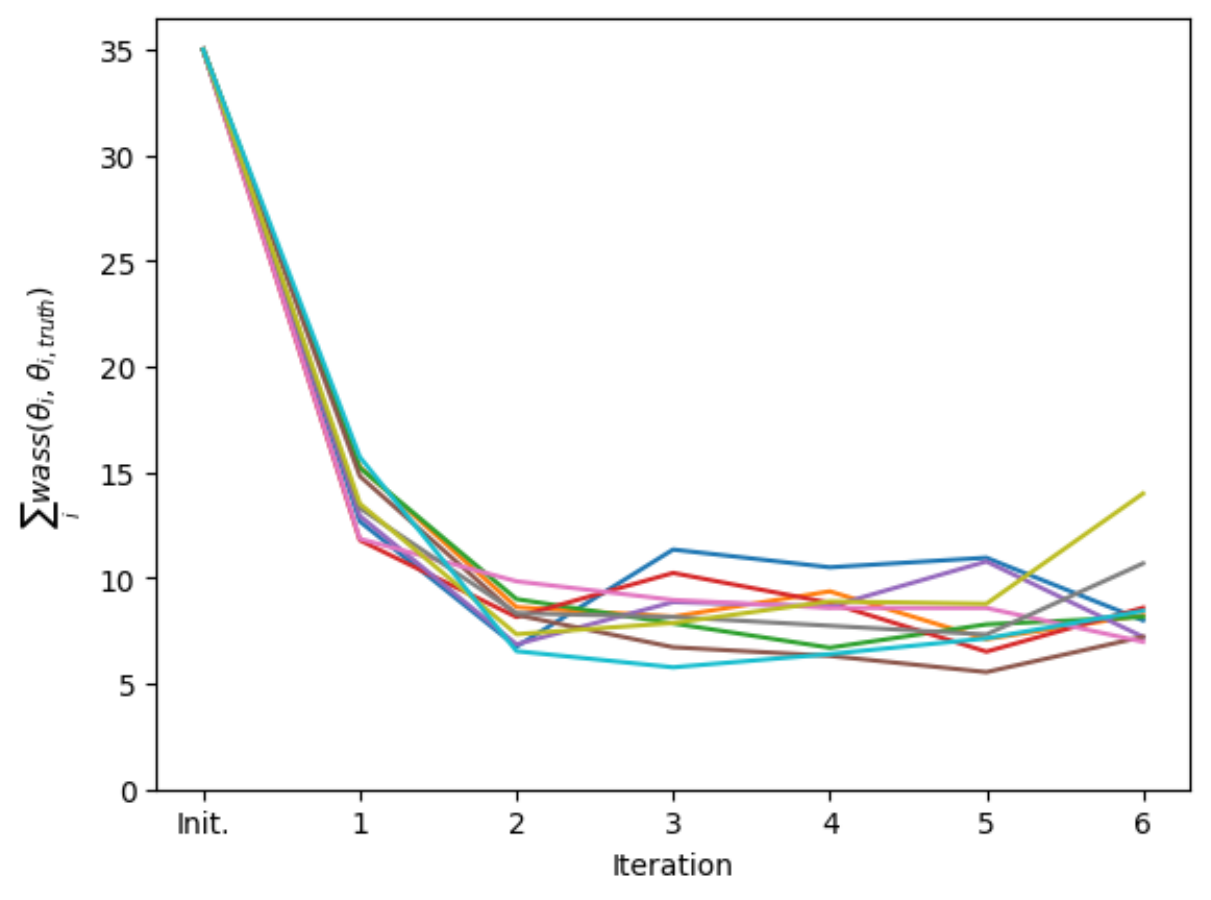}
    \caption{Summed marginal Wasserstein distance between the reweighted truth-level MC and the truth pseudodata as a function of the \Omnifold iteration number for ten independent runs. (Left) Runs initialized with random network weights. (Right) Runs initialized from pretrained checkpoints. Pretraining produces smaller final Wasserstein distances and markedly reduced run-to-run variance.}
    \label{fig:zjets-pretrain-wass}
\end{figure}

\begin{figure}[tb]
    \centering
    \includegraphics[width=0.48\linewidth, alt={Ratio plot of the reweighted truth-level MC to truth pseudodata for the charged hadron multiplicity without pretraining, showing larger method bias and larger network initialization uncertainty band.}]{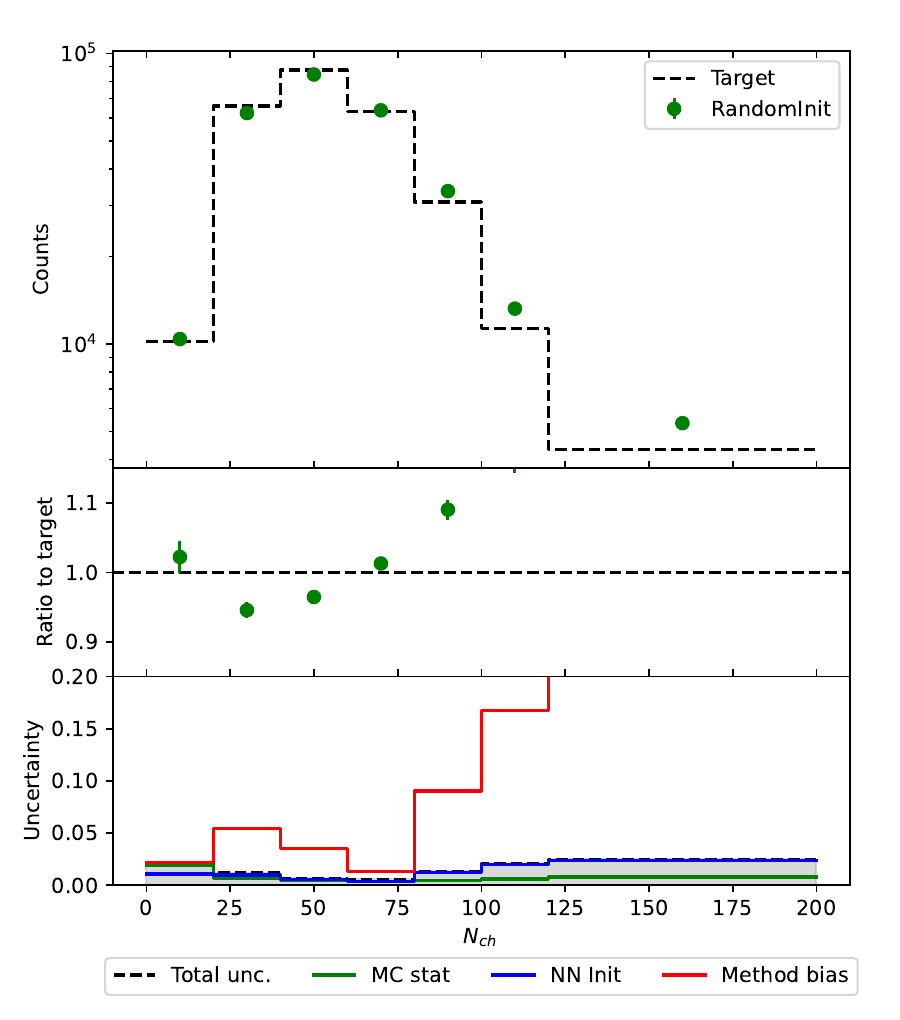}
    \includegraphics[width=0.48\linewidth, alt={Same ratio plot for the charged hadron multiplicity with pretraining, showing smaller method bias and a tighter initialization uncertainty band.}]{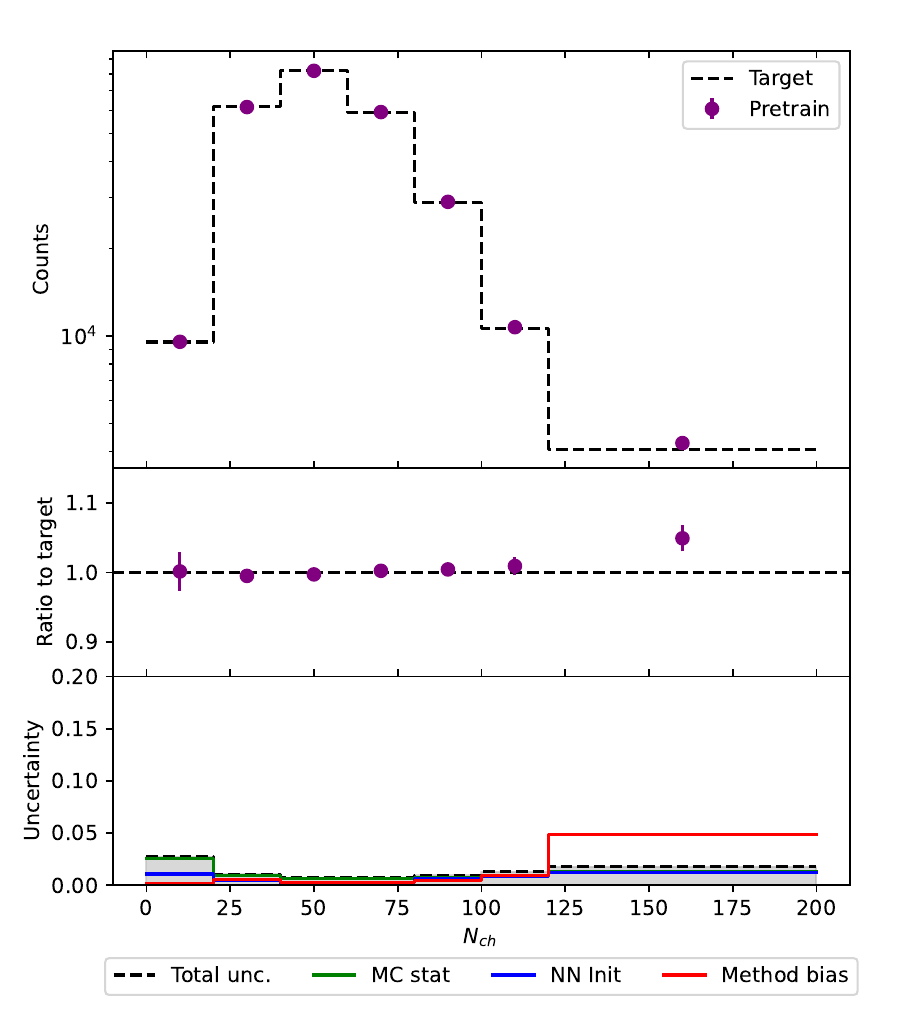}\\
    \includegraphics[width=0.48\linewidth, alt={Same ratio plot for the sub-leading track-jet mass without pretraining.}]{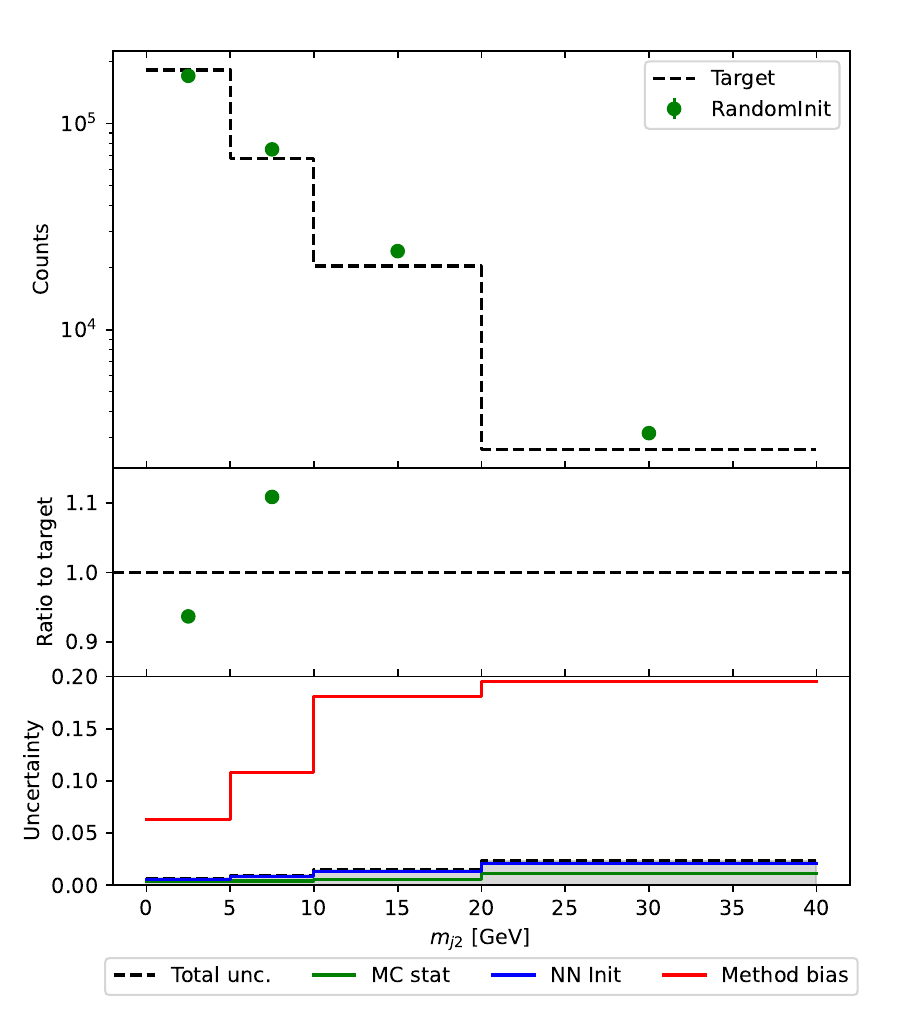}
    \includegraphics[width=0.48\linewidth, alt={Same ratio plot for the sub-leading track-jet mass with pretraining showing improved method bias and initialization uncertainty.}]{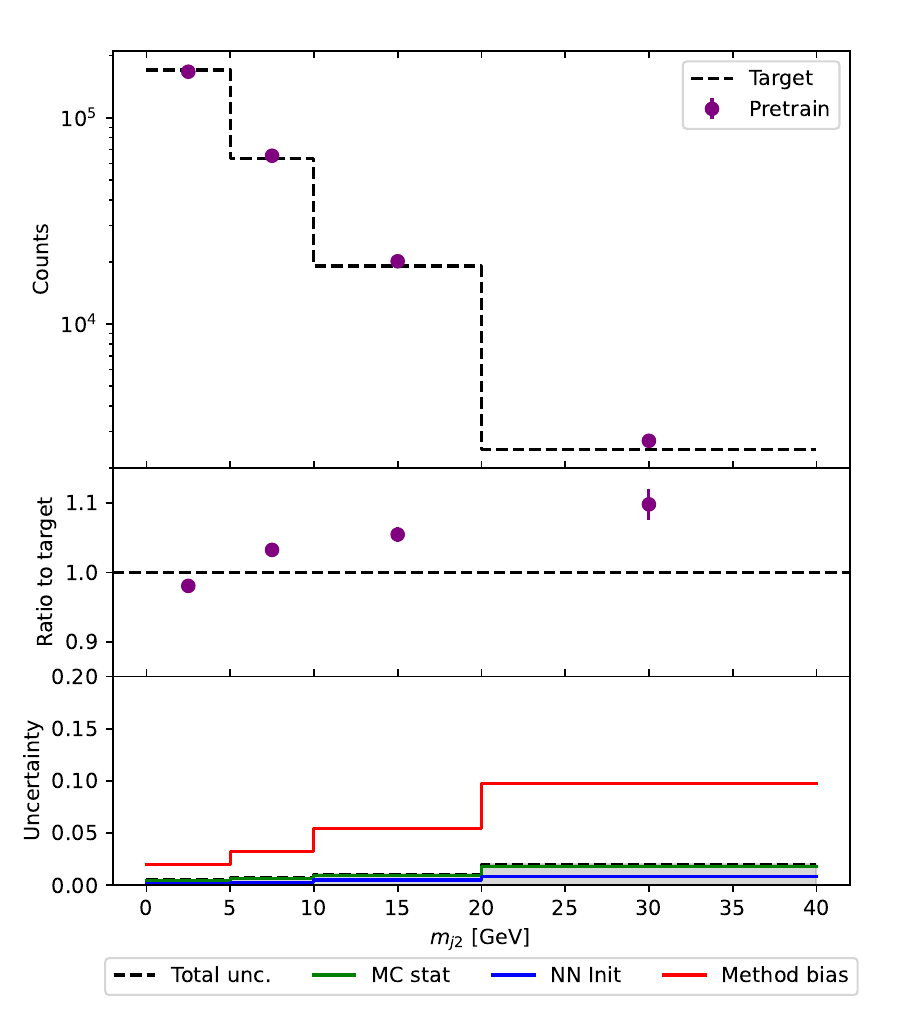}
    \caption{Preliminary results of the pseudodata measurement presented in two observables, \nch (top) and the sub-leading jet mass (bottom), presented both without (left) and with (right) the pretraining step. The figures compare the densities of the reweighted truth-level MC and the truth pseudodata in the top two pads, and plot the MC statistical uncertainty (green), neural network initialization uncertainty (blue), and the method bias (red) in the bottom pad.}
    \label{fig:zjets-pretrain-uncert}
\end{figure}

A first method of determining the effectiveness of the pretraining is to use the Wasserstein distance metric introduced in \Cref{sec:zjets-classifier}.
\Cref{fig:zjets-pretrain-wass} shows this distance, evaluated between the reweighted truth-level MC and the truth pseudodata (see \Cref{sec:zjets-validation}) as a function of \Omnifold iteration for ten independent runs of the procedure without (left) and with (right) pretraining.
Pretraining reduces the final Wasserstein distance and substantially reduces the variance across the ten runs, but in a metric which is known to be very sensitive to outlier events which might have spuriously large weights.
A more physical way to visualize the benefit of pretraining is to bin the reweighted truth-level MC and the truth pseudodata in some observables and plot the method bias and the variance.
The variance is defined as the spread in the reweighted truth-level MC when running \Omnifold with identical data inputs but different random seeds.
\Cref{fig:zjets-pretrain-uncert} plots preliminary results of the pseudodata measurement, performed without (left) and with (right) the pretraining step.
The pretraining step dramatically reduces the method bias, especially the tails of the distribution where the statistics are most limited.
It also produces a modest reduction in the neural network initialization uncertainty, which is directly related to the variance.
The method bias produced by \Omnifold without the pretraining is likely too large to make a publishable measurement in these particular observables, so the inclusion of the pretraining is a significant advance.
It was only after including the pretraining step that the method bias produced by the full-phase-space unfolding was close to the method bias produced by Multifold and IBU.
See \Cref{sec:zjets-validation}.
The substantial performance gains offered by pretraining also motivate data-limited likelihood ratio estimation tasks as an important application of HEP specific foundation models, despite the current literature's focus on jet classification as the primary application.
See \Cref{sec:tagging-future} for a discussion.

\FloatBarrier

\subsection{Unbinned Background Subtraction}
\label{sec:zjets-bkg-subtraction}

\begin{figure}[tb]
    \centering
    \includegraphics[width=\linewidth, alt={Flow diagram showing the background rejection and unfolding procedure in sequence, beginning with reconstructed data and simulated backgrounds, applying an unbinned background rejection step, and then passing the background-subtracted data through the Omnifold unfolding procedure to obtain particle-level results.}]{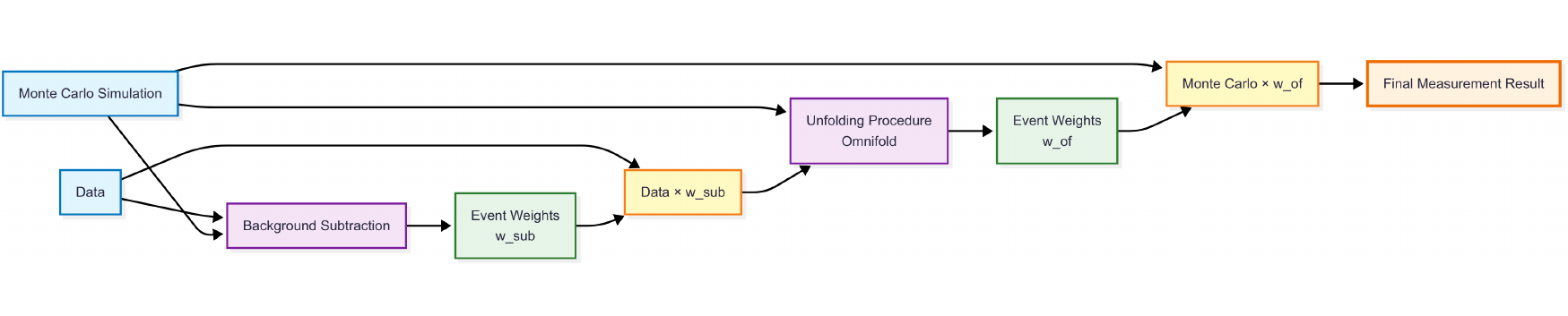}
    \caption{Schematic illustration of the sequential background rejection and unfolding procedure. The reconstructed data are first reweighted to reject the top-quark background contribution, and the resulting background-subtracted data are then used as the detector-level target for the \Omnifold unfolding.}
    \label{fig:zjets-br-and-unfolding-flow}
\end{figure}

The $p_T^{\mu\mu} > 200$~GeV fiducial volume contains almost entirely events from $Z$+jets processes.
Top-quark processes contribute about $0.2\%$ of the observed events, with the remainder split among strong $Z$+jets, diboson, and electroweak $Zjj$ production (all taken as signal).
In the previous round Multifold analysis, the top contribution was taken as an uncertainty and not subtracted explicitly.
This uncertainty was sub-leading in all observables and all fiducial volumes considered in the validation of the Multifold analysis.
The Omnifold measurement adds an unbinned background subtraction.
This addition is likely invisible in the final results because the top background is so small, but was added because explicit background rejections are expected in ATLAS cross section measurements, and because it serves as an illustration of how backgrounds can be handled in unbinned and high-dimensional measurements.
This is the first unbinned and high-dimensional cross section measurement to explicitly subtract a background.
Existing measurements (summarized in \Cref{tab:unbinned-unfolding-measurements}) have all been performed in phase spaces with negligible backgrounds.
The procedure for performing the unbinned subtraction and unfolding in sequence is illustrated in \Cref{fig:zjets-br-and-unfolding-flow}.
First data and detector-level MC simulation are used as inputs to the background rejection procedure, which outputs event weights $w_{\textrm{sub}}(\vec{x}_r)$ which remove the background when applied to the data.
The data with the event weights are then used as input to the unfolding procedure, along with the detector-level and particle-level MC simulation, which outputs event weights $w_{\textrm{of}}(y)$ that are the final result.
The data entering the unfolding with non-trivial event weights is analogous to the data entering the unfolding with non-integer bin counts in standard binned analyses.

The reweighting for the data is obtained by training a classifier to perform likelihood ratio estimation in the detector-level phase space.
The classifier is trained to separate the detector-level data (or pseudodata) from itself but with the detector-level MC estimate for the top background appended with weights turned negative.
In other words the classifier estimates the likelihood ratio $p_{\mathrm{D-B}}(x) / p_{\mathrm{D}}(x)$, with D being the data (or pseudodata) distribution that is composed of both signal and background events, and B being the detector-level MC estimate for the top background.
Since B is very subtle, the differences between the densities are very small.
Specifically B is known to be about 0.2\% of D, so the likelihood ratio should be roughly 0.998 integrated across the whole phase space.

Initially it was attempted to train a ParT classifier with the same architecture and inputs as described above to estimate the likelihood ratio directly.
However the differences between the classes were too subtle and the network failed to provide a meaningful background subtraction.
It was checked that if the normalization of the top background was artificially increased by a factor of 10 the network was able to pick up on the difference between the signal and background and provide a very accurate subtraction.
This is an interesting property of unbinned background subtraction techniques compared to standard binned approaches.
In binned approaches the subtraction accuracy is typically not dependent on the background normalization.
However unbinned subtraction with likelihood ratio estimation requires training a neural network classifier, and if the background is subtle the signal and background classes can be very similar making the classification task difficult.

\begin{figure}[tb]
    \centering
    \includegraphics[width=0.48\linewidth, alt={Histogram of the ParT top-versus-Zjets classifier logit on pseudodata constructed with the Sherpa 2.2.12 ttbar sample, showing similar separation between ttbar and Zjets events.}]{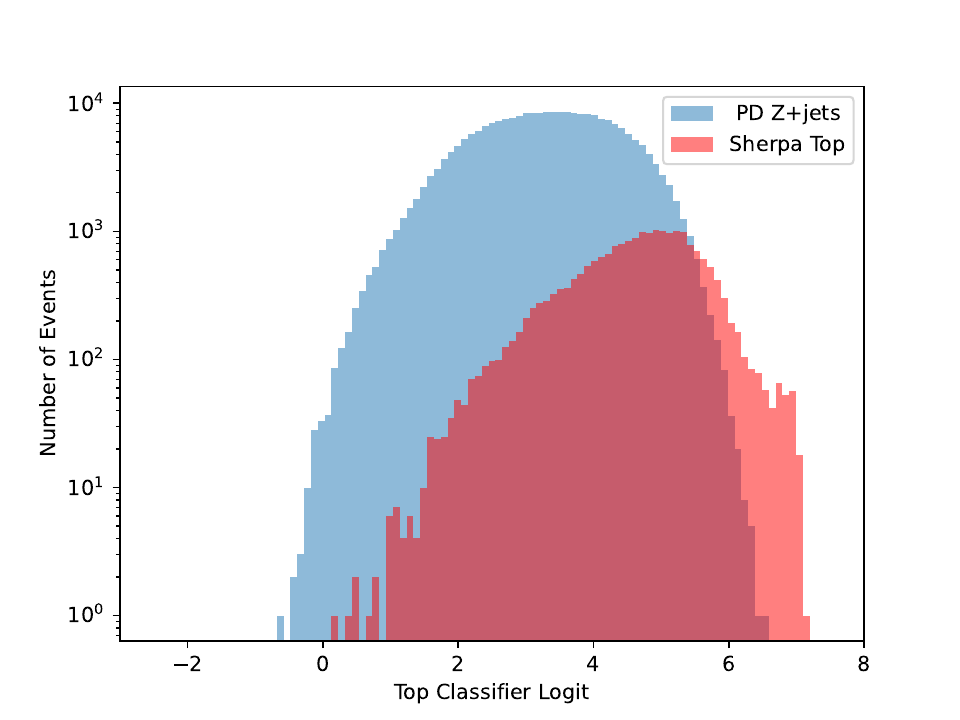}
    \caption{Output logit of the ParT classifier trained to discriminate the top background from the $Z$+jets signal, evaluated on the pseudodata with the top contribution removed (blue), and the \textsc{Sherpa}~2.2.12 top sample (red). The histograms indicate that the classes are not very easily separated, but the top events cluster toward higher logit values. This logit is used as an observable in the subsequent 5D unbinned subtraction.}
    \label{fig:zjets-toplogit}
\end{figure}

To overcome these issues, a two-step procedure is devised.
First, a ParT classifier with same architecture and inputs as described above is trained to discriminate top from $Z$+jets events in the detector level simulation.
The goal of this training is to provide a maximally expressive observable to separate the top events from the $Z$+jets events, not to provide any reweighting.
The training set uses all 21~thousand unweighted top events from the \textsc{Sherpa}~2.2.12 sample as the signal class and the pseudodata sample with the top events removed as the background class.
Histograms of the network output, referred to as the top classifier logit, are provided in \Cref{fig:zjets-toplogit}.
These histograms show that many of the top events populate the same phase space as the $Z$+jets signal, and so can only be removed statistically.
The high-logit tail of the distribution is populated almost entirely by top events.

The top classifier logit is then used as a particularly expressive observable in which to perform the top background subtraction, alongside four event-level observables, \nch, \HT, the leading track-jet \pt, and the leading track-jet mass.
The motivation for using these five observables is that the top contribution to the density for each observable is concentrated in the tails rather than being uniformly distributed in phase space.
This makes the top contribution easier to detect for a neural network classifier than it would be in the full phase space in which the unfolding is derived.
Neural network classifiers are then trained to estimate the desired likelihood ratio using these five observables as inputs.
Given this task is fixed-dimensional, two-hidden-layer MLPs with 512 nodes per layer and GELU activations are used rather than the transformer networks.
An additional benefit of performing the background subtraction in the five-dimensional space is that the simple MLP networks are very fast to train.
This allows ensembles of 100 networks to be trained, then their outputs are averaged to produce a more accurate likelihood ratio estimate.
Additionally the weights produced by each network are clipped to a maximum of 1.0, since the subtraction should only remove density and never add it.

\begin{figure}[tb]
    \centering
    \includegraphics[width=0.48\linewidth, alt={Pre-unfolding detector-level distribution of the leading track-jet pT on pseudodata before top subtraction, compared between pseudodata built from Powheg plus Pythia ttbar and Sherpa 2.2.12 ttbar.}]{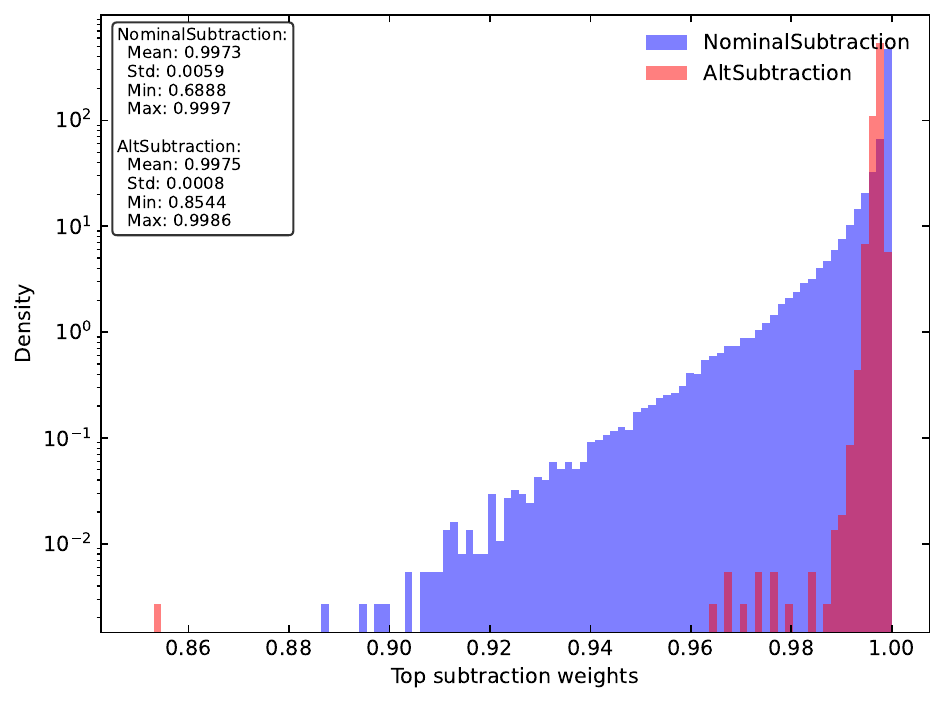}
    \includegraphics[width=0.48\linewidth, alt={Same distribution on real data before and after the unbinned top subtraction is applied, showing minor shape changes in the high track-jet pT tail.}]{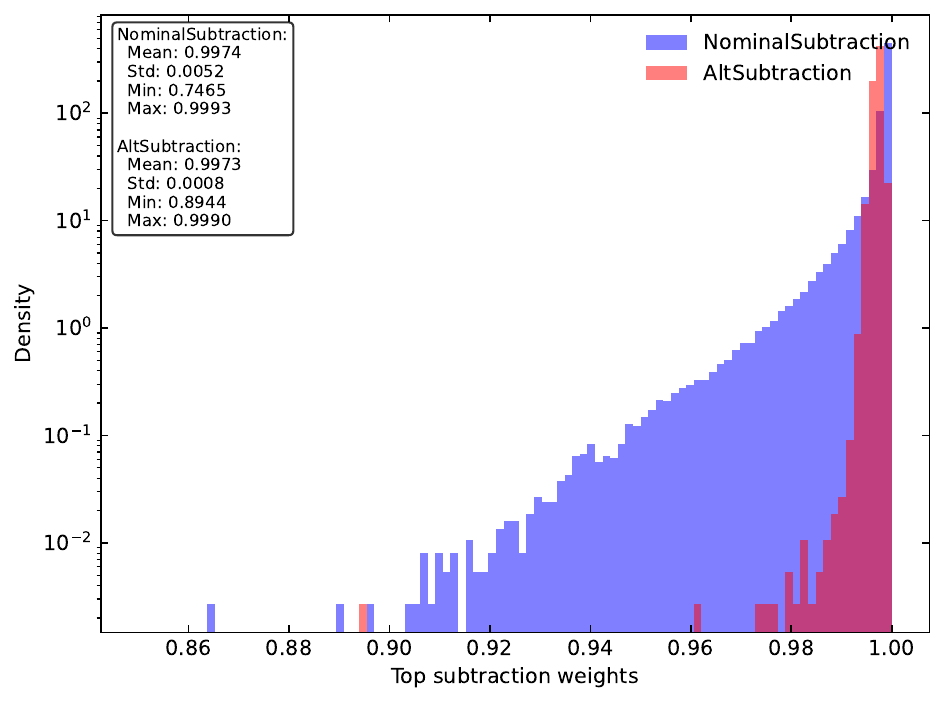}
    \caption{The weight distributions produced by the unbinned background subtraction procedure for the pseudodata (left) and data (right) measurements. Histograms are shown for subtractions performed with the nominal \textsc{Powheg}+\textsc{Pythia} top sample (blue) as well as the alternative \textsc{Sherpa}~2.2.12 top sample (red). The weights are generally very close to 1.0 given the subtle background, and all subtractions return weight distributions with a mean of about 0.998, reflecting an overall subtraction of 0.2\% as expected.}
    \label{fig:zjets-topsubtraction}
\end{figure}

\begin{figure}[tb]
    \centering
    \includegraphics[page=3, width=0.48\linewidth, alt={Pseudodata top-subtraction closure in a first representative observable.}]{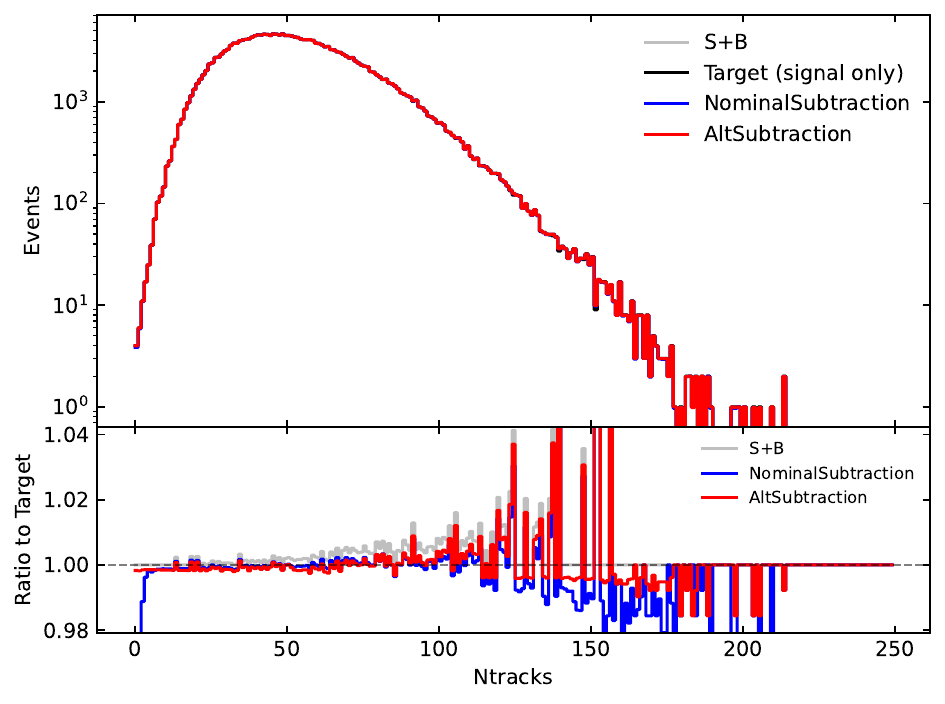}
    \includegraphics[page=4, width=0.48\linewidth, alt={Pseudodata top-subtraction closure in a second representative observable.}]{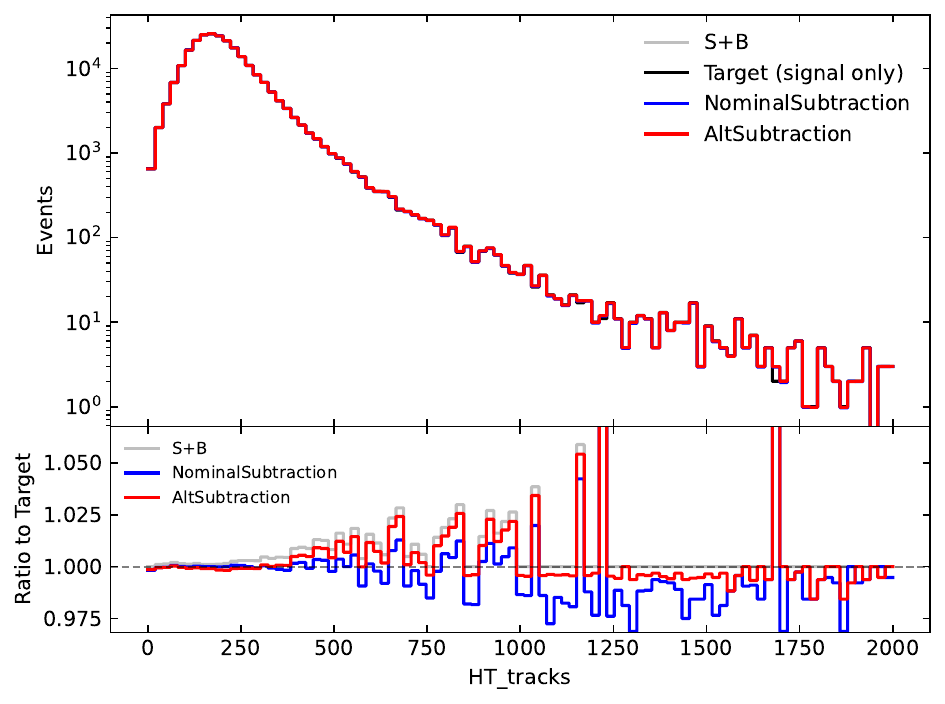}\\
    \includegraphics[page=5, width=0.48\linewidth, alt={Pseudodata top-subtraction closure in a third representative observable.}]{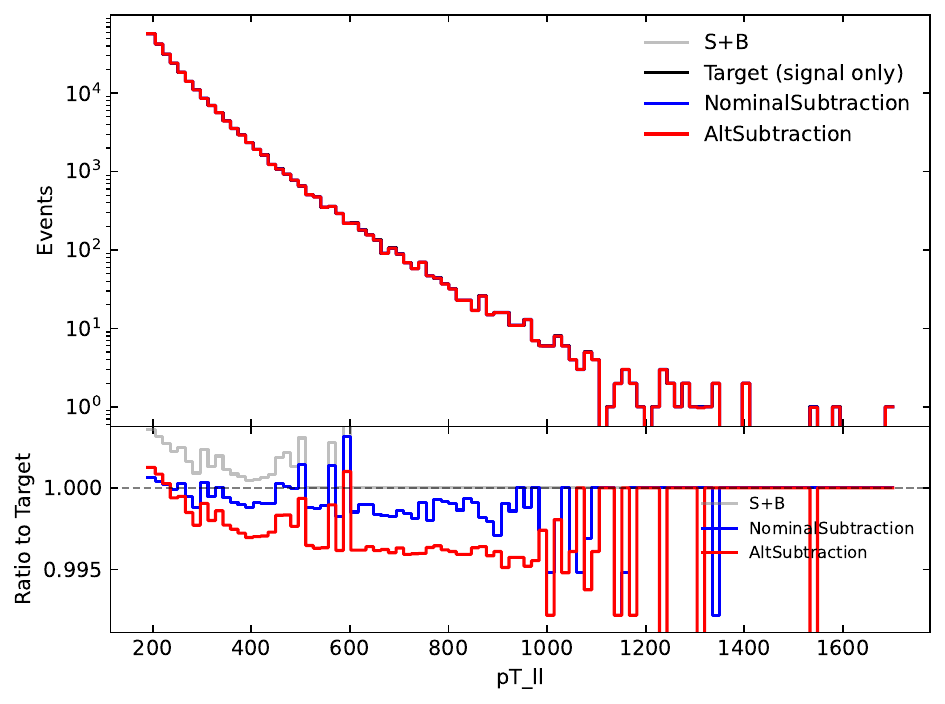}
    \includegraphics[page=6, width=0.48\linewidth, alt={Pseudodata top-subtraction closure in a fourth representative observable.}]{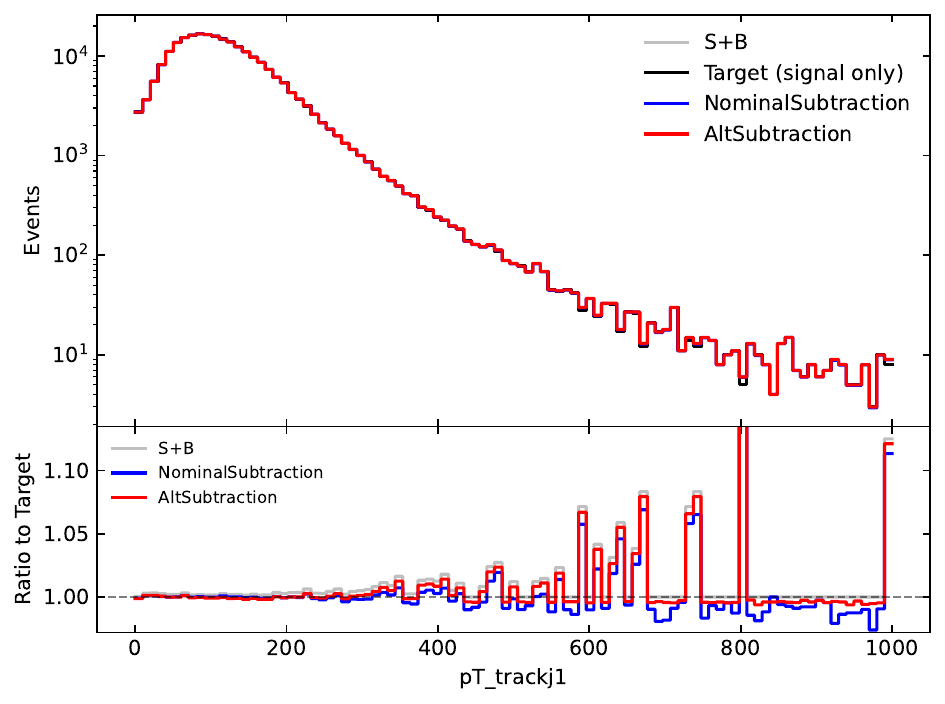}\\
    \includegraphics[page=7, width=0.48\linewidth, alt={Pseudodata top-subtraction closure in a fifth representative observable.}]{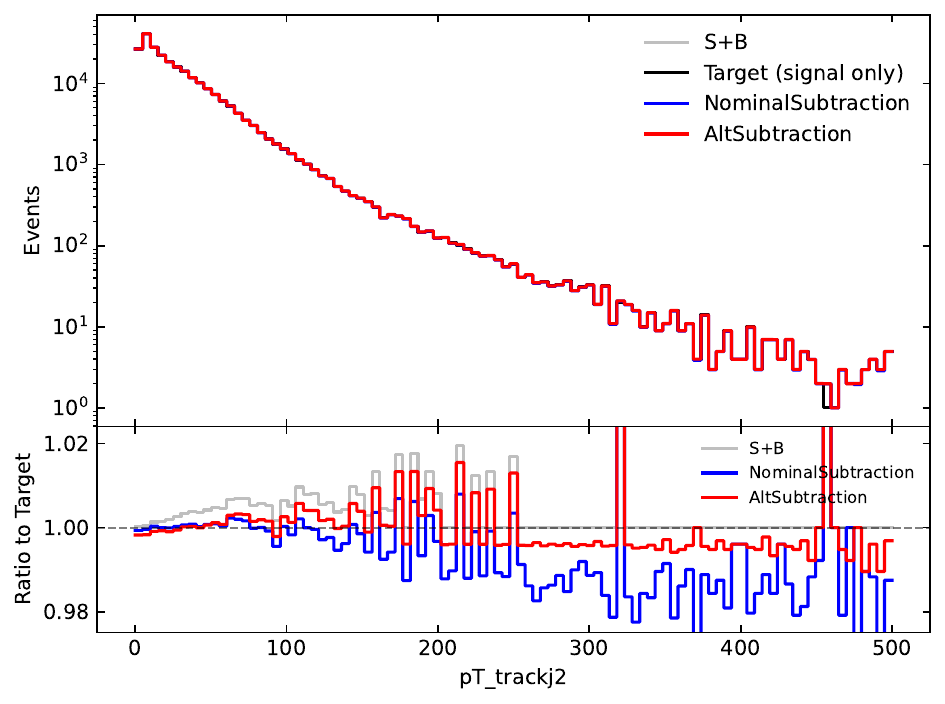}
    \includegraphics[page=8, width=0.48\linewidth, alt={Pseudodata top-subtraction closure in a sixth representative observable.}]{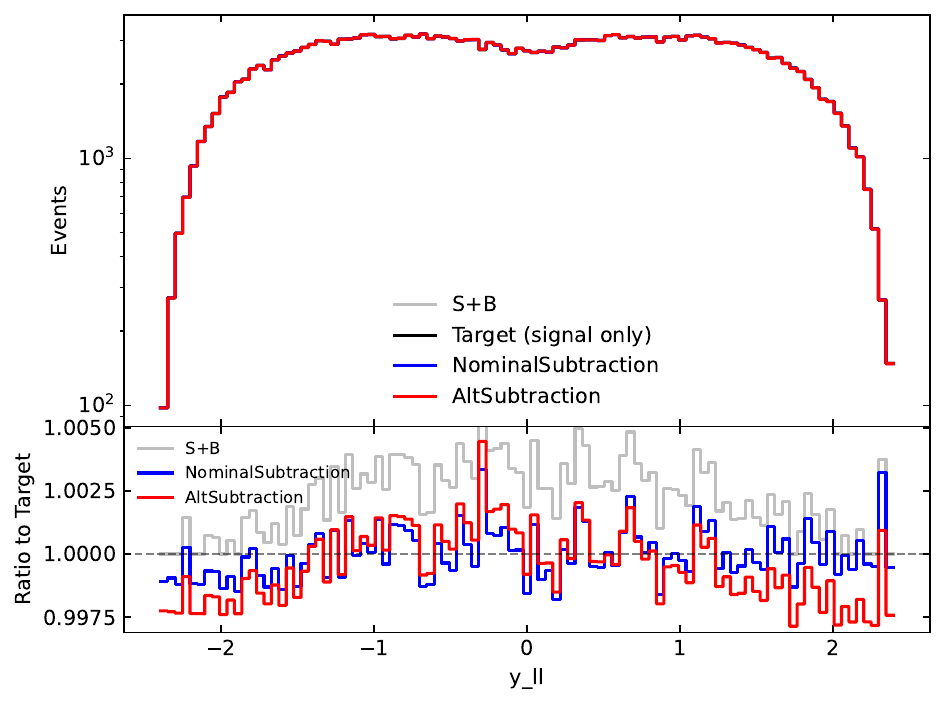}
    \caption{Results of the top subtraction in pseudodata shown in histograms of several physical observables. Histograms are shown for the detector-level pseudodata (gray), nominal and alternative top subtracted pseudodata (blue and red), and the signal only pseudodata (black). The ratio pads plot the ratio to the signal only pseudodata.}
    \label{fig:top-subtraction-obs1}
\end{figure}

\begin{figure}[tb]
    \centering
    \includegraphics[page=3, width=0.48\linewidth, alt={Data distribution in a first representative observable before and after top subtraction.}]{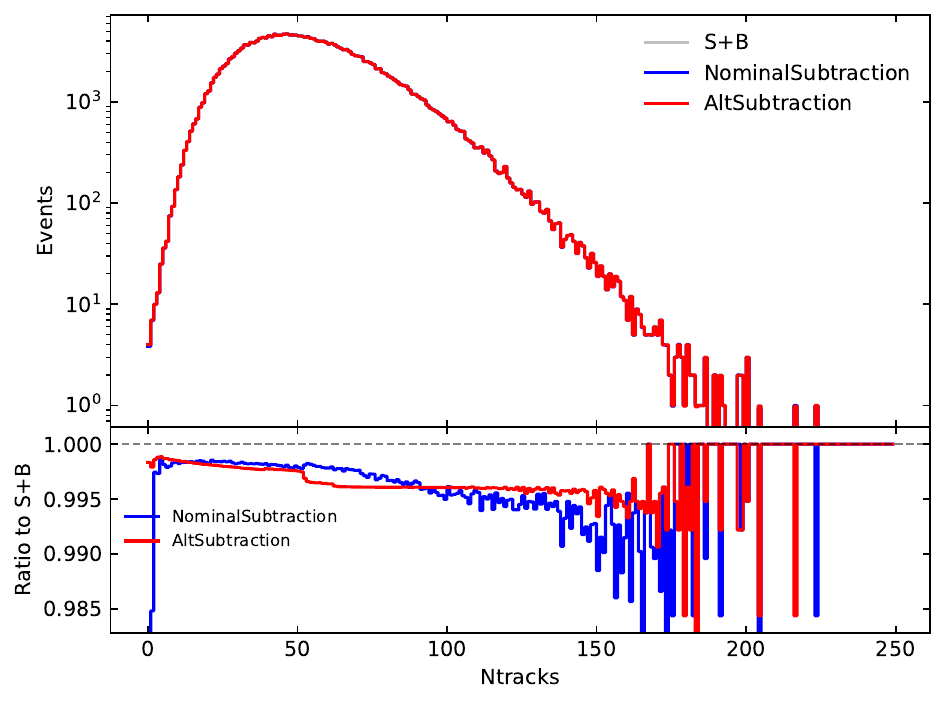}
    \includegraphics[page=4, width=0.48\linewidth, alt={Data distribution in a second representative observable before and after top subtraction.}]{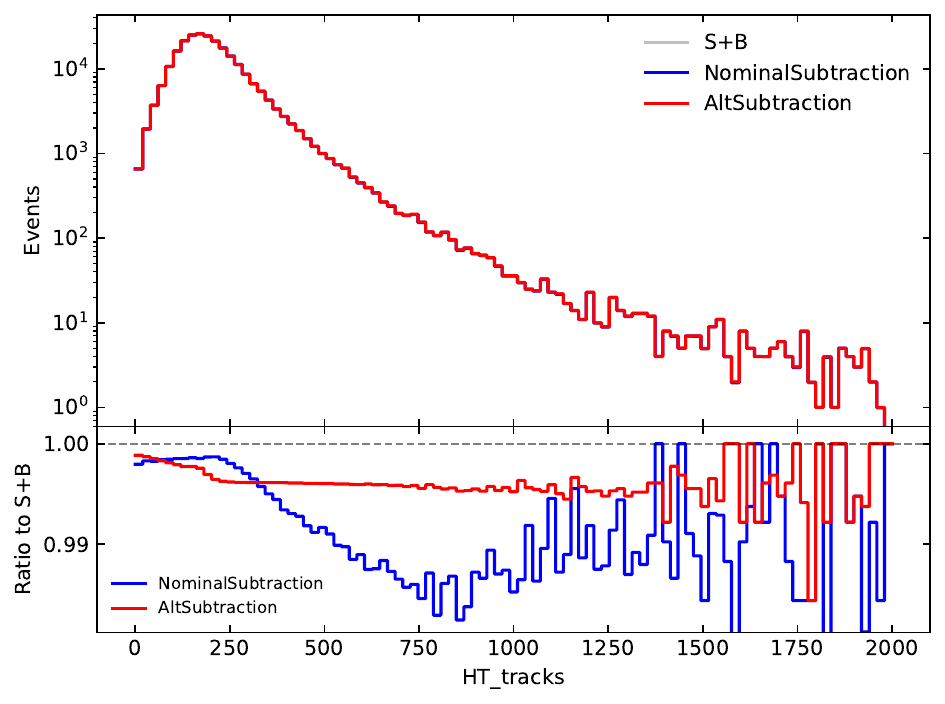}\\
    \includegraphics[page=5, width=0.48\linewidth, alt={Data distribution in a third representative observable before and after top subtraction.}]{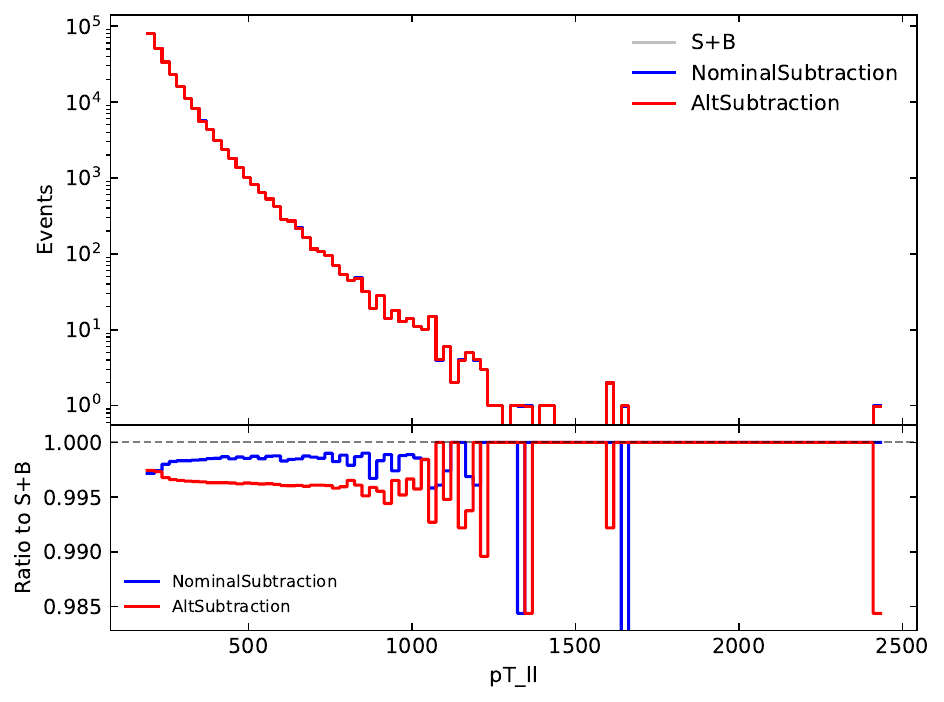}
    \includegraphics[page=6, width=0.48\linewidth, alt={Data distribution in a fourth representative observable before and after top subtraction.}]{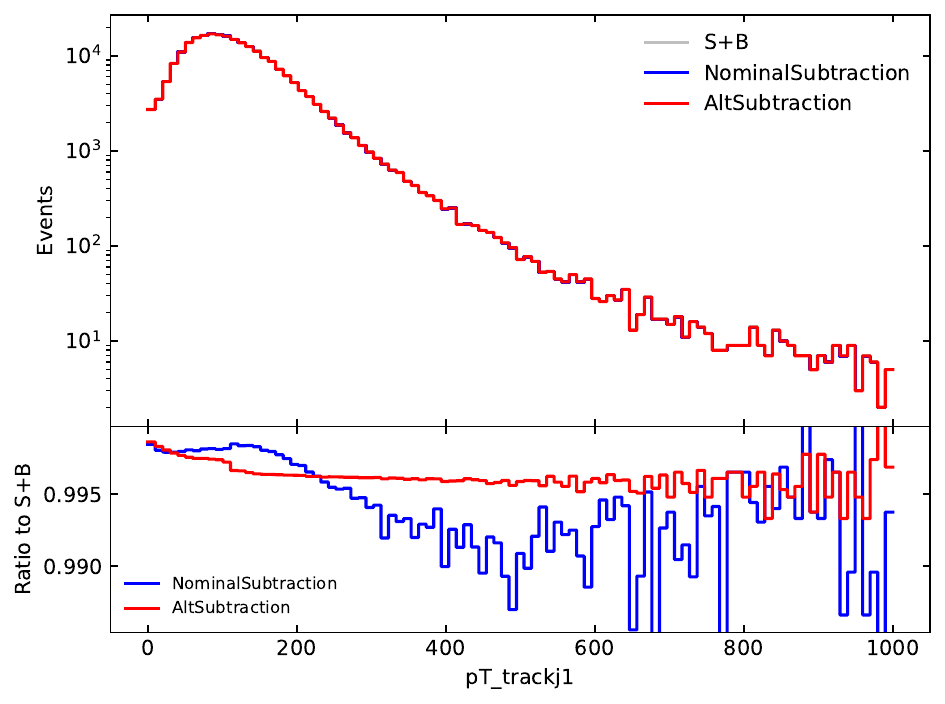}\\
    \includegraphics[page=7, width=0.48\linewidth, alt={Data distribution in a fifth representative observable before and after top subtraction.}]{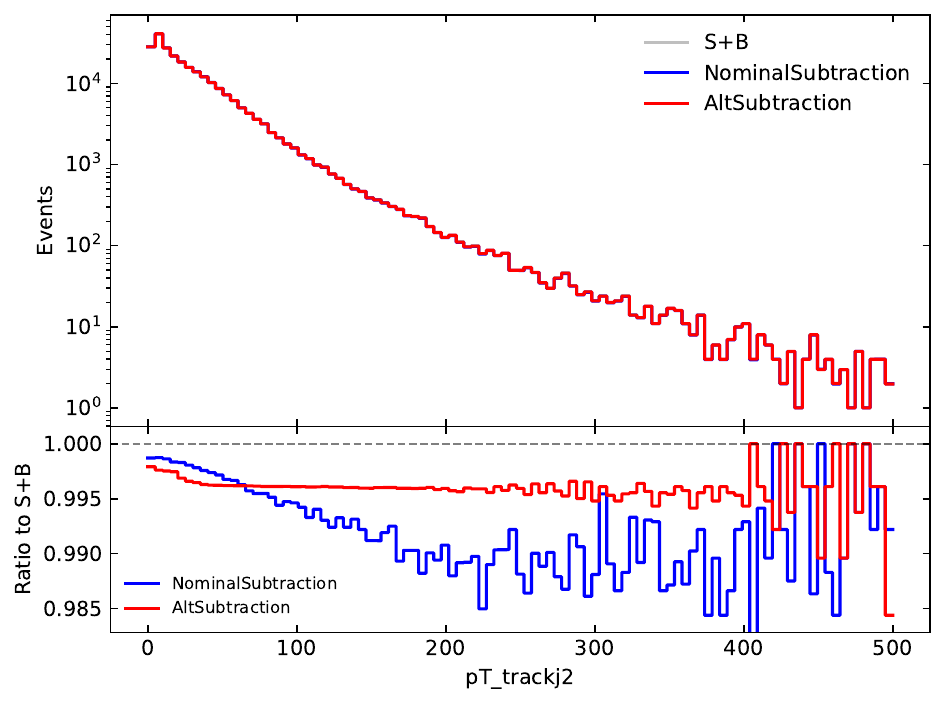}
    \includegraphics[page=8, width=0.48\linewidth, alt={Data distribution in a sixth representative observable before and after top subtraction.}]{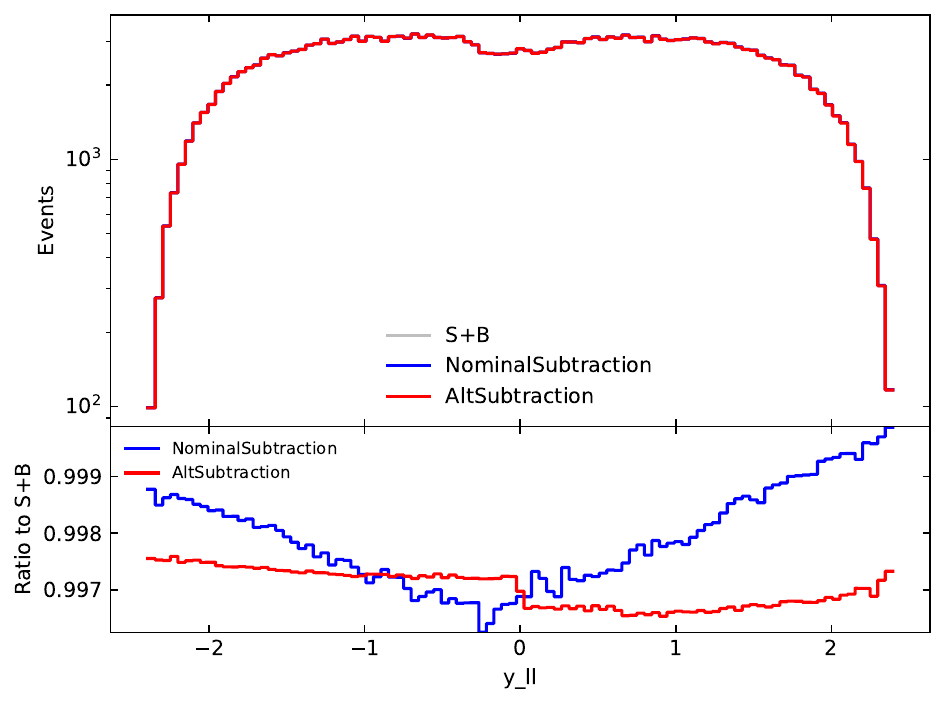}
    \caption{Results of the top subtraction in data shown in histograms of several physical observables. Histograms are shown for the detector-level data (gray), and the nominal and alternative top subtracted data (blue and red). The ratio pads plot the ratio to the detector-level data.}
    \label{fig:top-subtraction-obs1-data}
\end{figure}

\Cref{fig:zjets-topsubtraction} shows histograms of the weights produced by the background subtraction procedure for the pseudodata and data measurement.
Note that the subtraction step of the procedure is applied independently for both pseudodata and data, but the top classifier is trained using the pseudodata since the MC generator channel is needed for class labels.
The Figure shows the subtraction weights derived with the nominal \textsc{Powheg}+\textsc{Pythia} top sample, as well as the alternative \textsc{Sherpa}~2.2.12 sample used to set the top modeling uncertainty (see \Cref{sec:zjets-uncert-top}).
For all subtractions the mean of the subtraction weights is approximately 0.998, showing that the subtraction procedure succeeds in removing the correct amount of density integrated over phase space.
The larger variance of the nominal sample's weights indicates that the top contribution is more distinctive, and can be isolated to a smaller set of events, in the nominal sample compared to the alternative sample.
The results of the nominal and alternative top background subtractions in several physical observables are shown in \Cref{fig:top-subtraction-obs1,fig:top-subtraction-obs1-data} for pseudodata and data respectively.
Given the MC generator channel is available in pseudodata the subtractions are compared to the pseudodata with the top contribution removed (signal only).
In data the subtractions are simply compared to the detector-level data (S+B).
In general the subtraction is impressively accurate given how subtle the background is, but issues do appear.
For example subtraction with the nominal top sample produces a significant artifact at very low \nch as can be seen in the upper left panel of both Figures.
However this artifact does not appear in the alternative subtraction.
Given the full difference between the nominal and alternative subtraction is taken as the top modeling uncertainty, this artifact is covered by an appropriate uncertainty.
Generally the difference between the nominal and alternative subtractions is smaller than the difference between the pseudodata and the signal only pseudodata (gray and black) in \Cref{fig:top-subtraction-obs1}, indicating that the top subtraction procedure results in smaller uncertainties than simply taking the overall top background as an uncertainty as is done in the Multifold analysis.
The benefit would increase if the background were to become more significant.

\FloatBarrier

\subsection{Normalization to a Cross Section}
\label{sec:zjets-method-norm}

The final per-event weights produced by \Omnifold, called $w_{\textrm{of}}(y)$ above but referred to as $\nu(\vec{x})$ in what follows, encode the shape of the unfolded cross section but not its normalization.
To recover the normalization, these weights are re-scaled following the same procedure used in the previous round Multifold analysis~\cite{ATLAS:2024xxl}.
Each MC event $i$ is assigned a weight $w_i^\mathrm{y}$ by the MC generator.
It is further normalized to the cross section times the integrated luminosity (provided by an auxiliary measurement) to arrive at
\begin{equation}
    \label{eq:mc-norm}
    w^\mathrm{y}_i = w^\mathrm{MC}_i \frac{\mathcal{L}\,k\,\sigma_\mathrm{MC}\,\varepsilon}{\sum_i w^\mathrm{MC}_i},
  \end{equation}
where $\mathcal{L}$ is the integrated luminosity of the data, $k$ is the fiducial factor predicted by the simulation, $\sigma_\mathrm{MC}$ is the total cross section predicted by the MC event generator, and $\varepsilon$ is the filter efficiency.
The sum in the denominator runs over all generated MC events.
At detector level, the event weight is further multiplied by several corrections:
\begin{equation}
  \label{eq:mc-norm-reco}
  w^\mathrm{reco}_i = w^\mathrm{y}_i \prod_c\,w^\mathrm{corr}_{ci},
\end{equation}
where the product runs over five different corrections with index $c$: the pileup reweighting, the trigger efficiency scale factor, and the three muon efficiency scale factors.

The event weights $w^\mathrm{y}_i$ and $w^\mathrm{reco}_i$ produce the MC generator predictions for the particle-level and detector-level events respectively.
In what follows each detector-level event configuration is represented by $(w^\mathrm{reco},\vec{x}_\mathrm{reco})$ and each particle-level event configuration is represented by $(w^\mathrm{y},\vec{x})$.
The unfolding procedure corrects the truth-level event weights (keeping the truth-level features $\vec{x}$ the same): $(\nu(\vec{x})\,w^\mathrm{y},\vec{x})$.
In order to regain the overall normalization which might be altered by the weights produced by \Omnifold, the event weights are scaled by a factor equal to
\begin{align}
  \nu(\vec{x}_i)\,w^\mathrm{y}_{i,\text{ norm}}&=\frac{\nu(\vec{x}_i)\,w^\mathrm{y}_{i}}{\sum_i \nu(\vec{x}_i)\,w^\mathrm{y}_{i}}\times\frac{\sum_j w^y_j}{\sum_k w^\text{reco}_k}\times N_\text{data}\,,
\end{align}
where the $j$ and $k$ sums run over the truth and reconstructed events in the MC sample, respectively, and $N_\text{data}$ is the number of data events.
This restores the event yield predicted by the MC based on the input data.

The weights are then finally normalized by:
\begin{equation}
  \label{eq:xsec-norm}
  w_{i} = \frac{\nu(\vec{x}_i)\,w^\mathrm{y}_{i,\text{ norm}}}{\mathcal{L}},
\end{equation}
where $\mathcal{L}$ is the integrated luminosity.
The weights $w_i$ defined according to equation~\ref{eq:xsec-norm} have units of cross section and the sum of the weights over all events that fall in any fiducial region will produce a measurement of the fiducial cross section.
This can be written as:
\begin{equation}
  \label{eq:fid-xsec}
  \hat{\sigma}_{k} = \sum_{i}^{N_\mathrm{events}}w_{i}\,\mathcal{I}[b(\vec{x}_{i})=k] = \sum_{i\in B_k} w_i,
\end{equation}
where $b(\vec{x}_{i})$ returns an index $k\in\{0,1,2,...,n_\text{bins}\}$ corresponding to the fiducial region event $i$ falls in (based on its features $\vec{x}_i$), and $\mathcal{I}[\cdot]$ is the indicator function that is 1 when $\cdot$ is true and zero otherwise.
The last equality writes the same thing in different notation: the sum runs over all events $i$ that fall in bin $k$, where $B_k$ denote the set of events that fall in the bin (based on their $\vec{x}$).

\FloatBarrier

\section{Uncertainties}
\label{sec:zjets-uncert}

Systematic uncertainties are assessed on the particle-level cross section measurement by \textit{propagating} them through the unfolding procedure.
For each source $u$, an appropriate perturbation is introduced to either the simulation or the data and the full \Omnifold procedure is re-run from scratch on the perturbed input.
This defines an alternative set of normalized event weights $\nu_u(\vec{x})$.
The uncertainty on any observable $\mathcal{O}$ in any bin $k$ is the signed difference
\begin{equation}
    \Delta_{ku} \;=\; \hat{\sigma}_{ku} - \hat{\sigma}_k,
\end{equation}
and the total uncertainty on $\hat{\sigma}_k$ is the quadrature sum over the uncorrelated sources $u$.

In this thesis the experimental and theoretical systematic uncertainties will only be mentioned briefly.
Experimental uncertainties cover possible differences between data and simulation introduced by the imperfect modeling of the detector response.
This includes effects that impact the muon and track kinematics, as well as the overall normalization of the distributions.
Experimental uncertainties do not shift the particle-level quantities, so they can be thought of as uncertainties on the response matrix (or the unbinned equivalent of the response matrix).
The leading experimental uncertainty in this measurement is the tracking efficiency, where the efficiency with which the ATLAS tracking detector identifies tracks is slightly shifted.
In total there are 12 experimental uncertainties propagated to the final result.
Theoretical uncertainties cover possible differences between data and simulation introduced by imperfect modeling of physical processes in the theoretical calculations underlying event generators.
Unlike experimental uncertainties, theoretical uncertainties shift both the particle-level and detector-level distributions as opposed to only the detector-level.
This can be thought of as a shift in the particle-level prior provided to the unfolding rather than a shift in the response matrix.
The prior shifts can be large, for example altering the QCD factorization scales can result in shifts of up to 10\% in the $p_T^{\mu\mu}$ observable, but unfolding is broadly insensitive to choice of prior so the theory uncertainties are all sub-leading\footnote{The practice of interpreting theory uncertainties as a shift in the pre-unfolding prior rather than a full uncertainty to be applied to the detector-level distributions is the reason some unfolded measurements can actually be more accurate than forward folded measurements. See \Cref{sec:unfolding-intro} for a discussion.}.
There are 9 theoretical uncertainties propagated to the final result.

Statistical uncertainties are evaluated separately through a procedure known as \textit{bootstrapping}, where the data or MC are fluctuated within their statistical uncertainty and then used to repeat the unfolding.
To bootstrap an unbinned data sample, the event weights are multiplied by a factor sampled from a Poisson probability distribution with parameter $\lambda = 1$.
This is equivalent to resampling the data with replacement, so that each given event can appear in the bootstrap zero, one, or multiple times.
This procedure is repeated 100 times for the data statistical uncertainty (both for the pseudodata and data measurements), and 25 times for the MC training set statistical uncertainty.
The final uncertainty is then the variance in the result produced by this ensemble of bootstraps.

The remaining uncertainty categories are related to the background subtraction and unfolding methods and are the focus of the rest of this section.
The unfolding uncertainties in particular make ample use of an alternative reweighting method known as \textit{Omnisequential reweighting}, which is introduced briefly here.
OmniSequential is a Gaussian Kernel-based reweighting method which iteratively reweights a preset list of observables in one dimension without the use of machine learning.
First, a sensible binning is determined for each of the input variables. 
The bin width is chosen based on Scott's rule~\cite{scott1979optimal}, and if the variable has a skewness ($\mu_3=E[(x-\mu)^3]$) that is larger than 2 then log binning is used.

In order to reweight some source distribution to match a target, one iteration of the algorithm then proceeds as follows: 
\begin{itemize}
  \item The variable with the worst agreement between the source and target is determined as the one with the worst (largest) `chi-squared pull', using binned spectra ($n_\text{dof}=n_\text{bins}$):
  \begin{align}
  z_{\chi^2} &= \frac{\chi^2-E[\chi^2]}{\sigma[\chi^2]} = \frac{\chi^2-n_\text{dof}}{\sqrt{2n_\text{dof}}}.
  \end{align}
  \item A Gaussian Kernel fit is then performed for the identified variable, resulting in a smooth function that quantifies the target to source ratio
  \item The source events are then reweighted based on the result of the fit. Each source event weight is multiplied by the target to source ratio given by the Gaussian Kernel fit evaluated at the value of the variable in question
\end{itemize}
The iterations continue until closure is reached, meaning all 24 variables have $z_{\chi^2}<2$, or until a maximum number of iterations is run.
OmniSequential is an auxiliary reweighting method that does not require estimating a likelihood ratio through machine learned classifiers.
However like the machine learning approach, the end result is a set of event weights that shift a distribution to match a desired target.
This procedure was devised and used to avoid over-reliance on ML based likelihood ratio estimation techniques, though the accuracy of its reweightings is typically worse than the more advanced ML approach.

\subsection{Unfolding Uncertainties}
\label{sec:zjets-uncert-unfold}

The unfolding uncertainty prescription follows the standard ATLAS unfolding recommendations~\cite{Armbruster:1694351} but adapted to the unbinned setting.
There are two sources of uncertainty injected by the unfolding, each of which is covered by one or more dedicated uncertainties taken on the final result.
The first is the residual bias intrinsic to any unfolding procedure, necessary due to the bias--variance tradeoff discussed in \Cref{sec:binned-unfolding-methods}, which is covered by the data-driven unfolding uncertainty.
The second is possible hidden-variables, which are quantities that are not constrained by the unfolding procedure but can bias the detector response.
Concrete examples of such quantities will be given below.
Inaccuracies in the unfolding due to these hidden variables are covered by the quadrature sum of the hadron composition and hidden variable unfolding uncertainties.

The data-driven unfolding uncertainty is assessed by first running OmniSequential to reweight the particle-level \mgpy sample such that its detector-level distributions agree with data.
This reweighting is unique among all of the reweightings in this analysis in that it is defined at particle-level but optimized at detector-level.
The result of this reweighting is then applied to the detector-level \mgpy sample, and this data is used as pseudodata in the unfolding procedure.
The difference between the truth distribution predicted by the unfolding and the input truth is taken as the data-driven unfolding uncertainty.
Note this procedure is very similar to the pseudodata measurement itself.
For this reason the method bias in the pseudodata measurement and data-driven unfolding uncertainty are highly correlated.
Further the data-driven unfolding uncertainty is identical for both the pseudodata and data measurements, since for both measurements the data inputs are the reweighted \mgpy sample and the nominal MC.

The hidden variable unfolding uncertainty is also assessed by running OmniSequential, but this time to reweight the particle-level \sherpa samples to match the particle-level \mgpy sample.
This reweighting is performed in a set of observables that are all functions of the full phase space explicitly included in the \Omnifold unfolding procedure.
The result is an alternative MC sample that looks similar to the nominal MC sample, but has possibly different hidden variables.
Then this alternative MC sample is used as the MC sample in the unfolding routine, and the difference between the result produced with the alternative and nominal MC samples is taken as the uncertainty.
An assumption of this uncertainty is that all possible hidden variables that might bias the detector response are different between the nominal and alternative MC samples.
In practice this assumption is likely valid because changing the generator produces a rather large shift in the particle-level spectra and there is no reason to suppose this pattern would not hold for hidden variables.

\begin{figure}[htb]
    \centering
    \subfloat{
        \includegraphics[page=5, width=0.48\linewidth, alt={Hadron-composition comparison from page 5 showing truth-jet pT fractions for charged-hadron species in the signal samples.}]{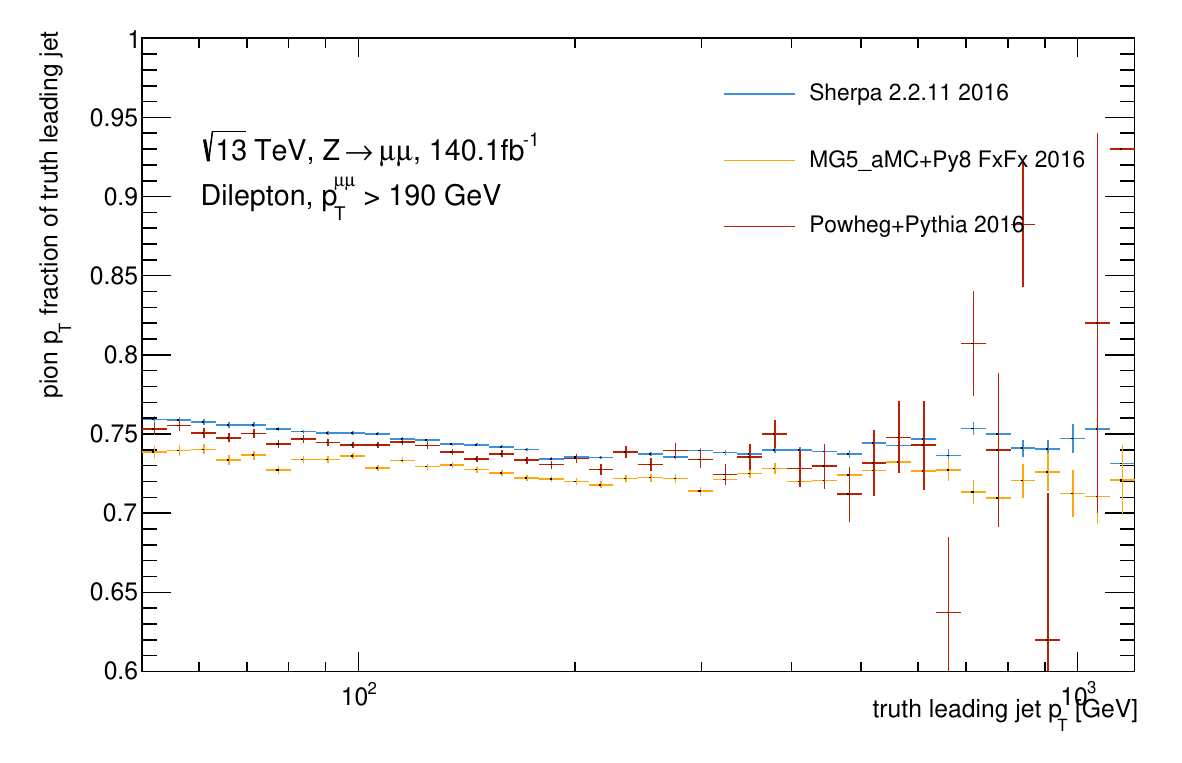}
    }
    \subfloat{
        \includegraphics[page=6, width=0.48\linewidth, alt={Hadron-composition comparison from page 6 showing truth-jet pT fractions for charged-hadron species in the signal samples.}]{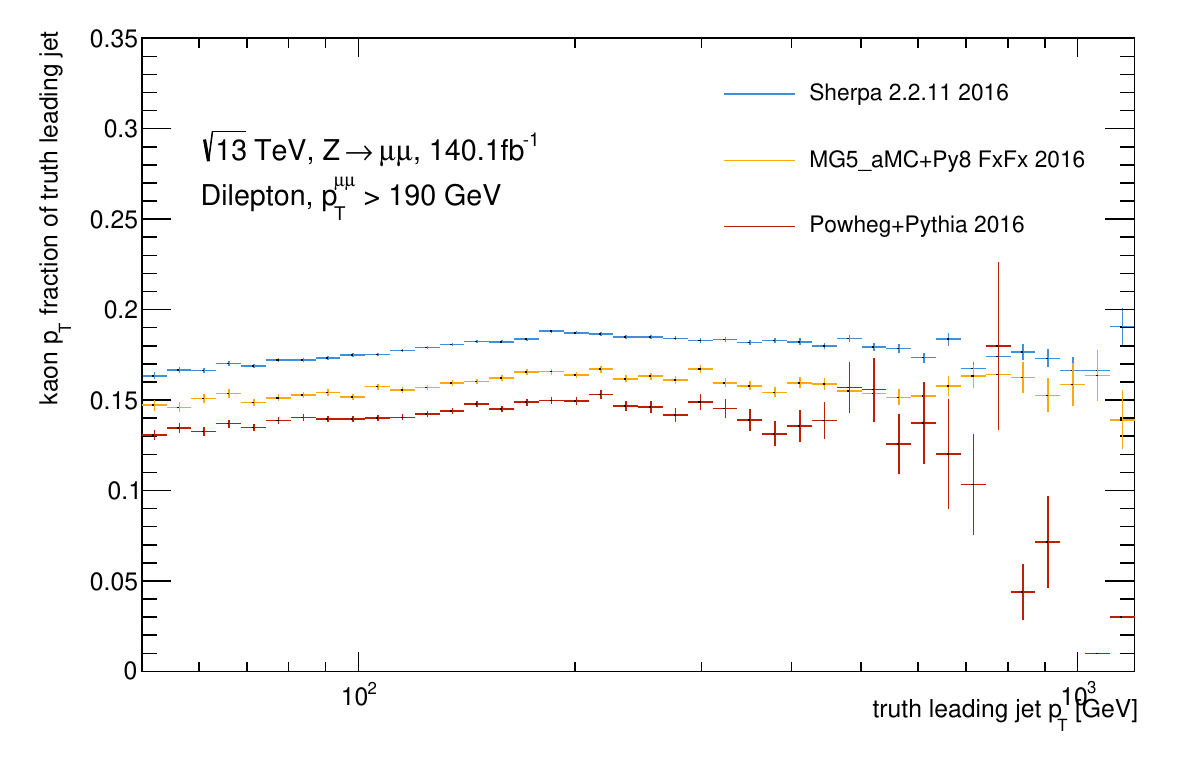}
    } \\
    \subfloat{
        \includegraphics[page=7, width=0.48\linewidth, alt={Hadron-composition comparison from page 7 showing truth-jet pT fractions for charged-hadron species in the signal samples.}]{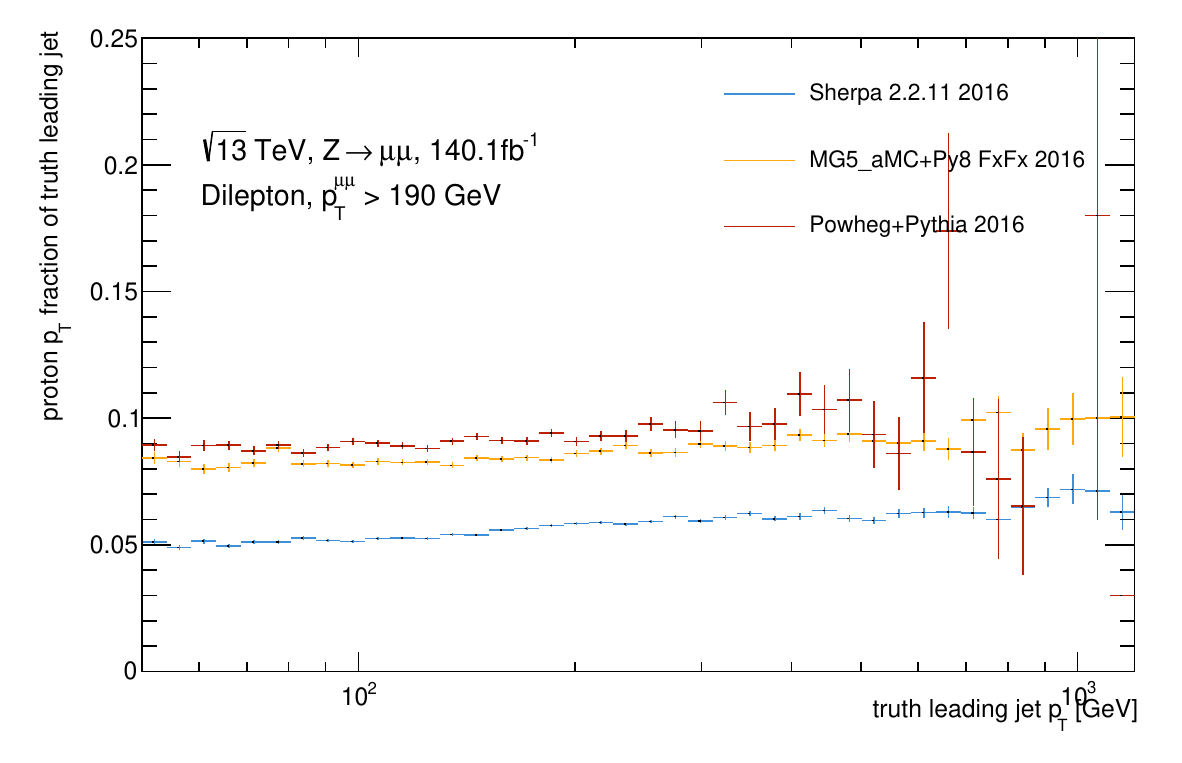}
    }
    \subfloat{
        \includegraphics[page=8, width=0.48\linewidth, alt={Hadron-composition comparison from page 8 showing truth-jet pT fractions for charged-hadron species in the signal samples.}]{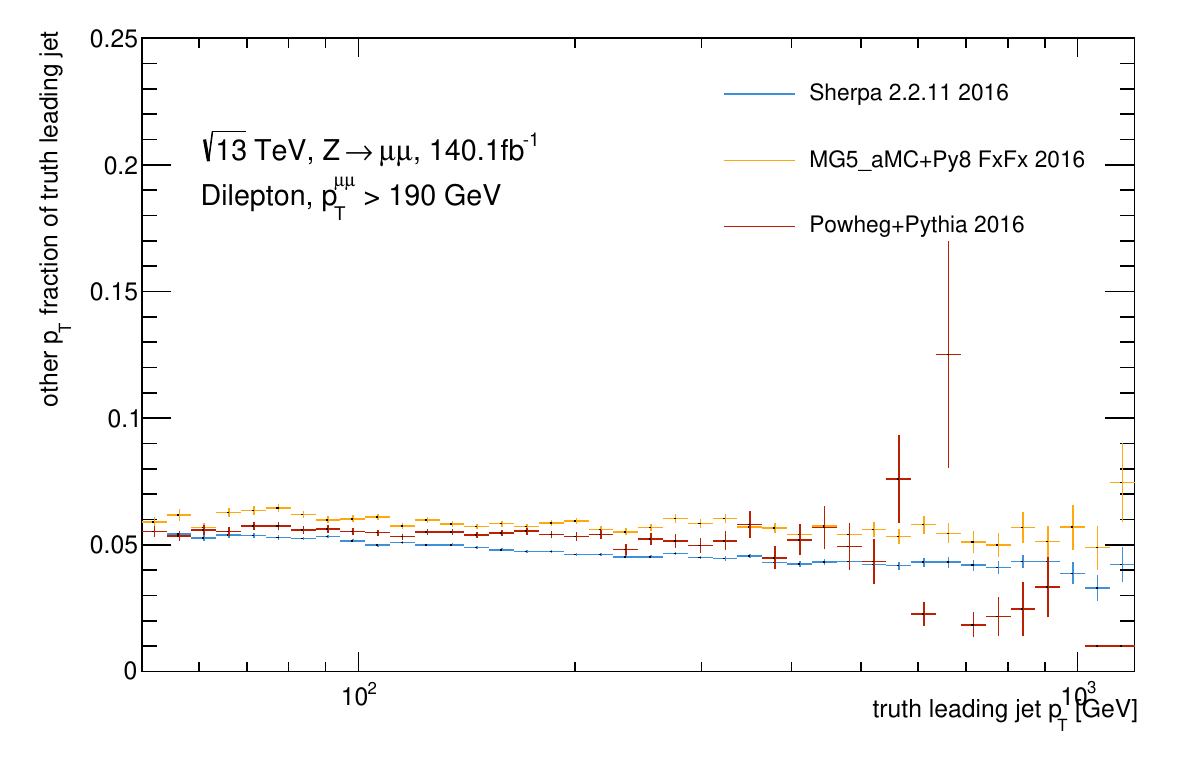}
    }
    \caption{Fraction of truth leading-jet \pt carried by pions, kaons, protons, and ``other'' charged hadrons as a function of truth leading-jet \pt, for the \mgpy and \sherpa strong $Z$+jets samples as well as a \textsc{Powheg}+\textsc{Pythia} sample included here for illustration. The panels show the fraction of the \pt within the leading particle-level jet carried by charged pions (top left), charged kaons (top right), protons (bottom left), and other charged hadrons (bottom right).}
    \label{fig:zjets-hadron-composition}
\end{figure}

\begin{figure}[tb]
    \centering
    \includegraphics[width=0.95\linewidth, alt={Four panels showing the event-level pT fraction carried by pions, kaons, protons, and other hadrons as a function of H_T in a pseudodata closure test; the unfolded distribution tracks the MG5 FxFx prior almost exactly and does not approach the Sherpa 2.2.11 truth.}]{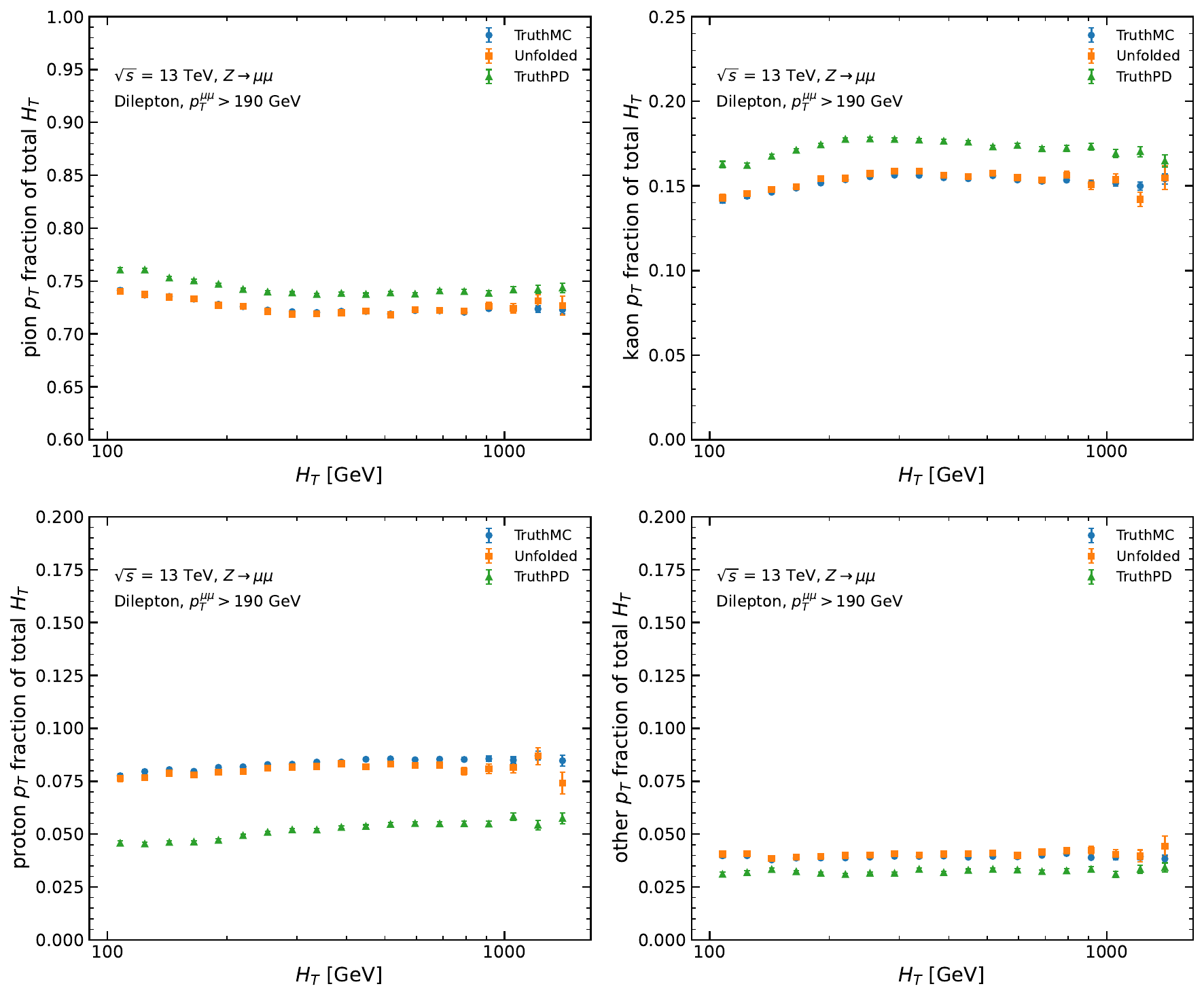}
    \caption{Event-level \pt fraction carried by pions, kaons, protons, and ``other'' charged hadrons as a function of the event \HT in the pseudodata measurement. Blue points indicate the unfolding prior, which is provided by the \mgpy sample, green points represent the post-unfolding result with event weights applied, and orange points represent the truth pseudodata provided by the \sherpa sample. The unfolding is clearly unable to distinguish the hadron fractions, as expected since this information is not provided at detector level.}
    \label{fig:zjets-unfold-hadron-fractions}
\end{figure}

Initial validation of the measurement following the procedures of \Cref{sec:zjets-validation-chi2} showed that the hidden variable uncertainty alone was insufficient to show good statistical agreement with the target in the pseudodata measurement for the leading and sub-leading jet mass observables $m_{j1}$ and $m_{j2}$.
The reason is a particular class of hidden variables that strongly bias the measurement of the jet masses: the truth-hadron fractions.
This hidden variable is the fraction of truth hadrons at particle-level that are charged pions, protons, kaons, or other charged hadrons.
Each generator produces slightly different truth hadron fractions.
\Cref{fig:zjets-hadron-composition} shows, for three strong $Z$+jets samples including the \mgpy and \sherpa samples, the fraction of truth-jet transverse momentum carried by each hadron species as a function of truth-jet \pt.
At particle-level, charged pions account for about 75\% of the jet \pt for all three generators, but differences emerge in the charged kaon and proton fractions.
The \mgpy sample predicts roughly 5\% more (fewer) protons (charged kaons) than the \sherpa sample.
The fraction of \pt carried by the other category, which is mainly composed of the strange baryons $\Omega$, $\psi$, and $\Sigma$, is roughly consistent between the generators.
Since the fraction of charged kaons versus protons are not constrained at detector level given the pion mass assumption, the unfolding returns the unfolding prior and does not alter the truth hadron fractions at all.
This is illustrated in \Cref{fig:zjets-unfold-hadron-fractions}, which plots the hadron fractions before and after unfolding, and compares to the truth pseudodata target.
The unfolding does not shift the hadron fractions at all, leaving this observable completely prior dependent.
However changing this hidden variable also adjusts the jet mass distributions, meaning an unfolding built with the \mgpy sample versus the \sherpa sample will yield slightly different results.
This effect was important enough that a dedicated uncertainty, the hadron composition uncertainty, was designed to cover it.
In practice this uncertainty is observed to be small except in the jet mass observables.

Before concluding this section, it is worth noting that a possible advantage of a high-dimensional unfolding is that it is less sensitive to hidden variables since it is differential in more observables.
For example if observable B biases the response of observable A, then performing a simultaneous unfolding in both observables should eliminate observable B as a hidden variable.
However this does not account for hidden variables that are not constrained by the detector altogether, for example the hadron fractions discussed above.
These are hidden variables even for a full-phase-space unfolding.
These considerations will be important for comparing the relative accuracy of the IBU and \Omnifold based unfoldings in \Cref{sec:zjets-validation-ibu}.

\FloatBarrier

\subsection{Network Initialization Uncertainty}
\label{sec:zjets-uncert-nninit}

Each neural network training within the \Omnifold procedure will yield slightly different results depending on the random seed used to initialize the network weights and order the training examples.
The spread across the \Omnifold results under variation of these seeds, or equivalently repeated runs with different random seeds, constitutes an uncertainty on the final result.
In this analysis, this uncertainty is estimated by performing ten independent runs of \Omnifold for the nominal result and each systematic uncertainty.
The unfolded weights are calculated as the event-by-event mean over these ten runs.
To calculate the network initialization uncertainty, each of the ten runs is binned and the uncertainty is the standard deviation in the bin counts produced across the ten runs.
Note binning each of the ten members of the ensemble is only required for the nominal results.
For the systematic uncertainties only the mean weights are used.
The previous round Multifold analysis used one hundred run ensembles rather than ten.
This reduction was possible because the variance was reduced by the pretraining procedure detailed in \Cref{sec:zjets-pretraining}, also producing a large savings in the computational cost of the measurement.

\subsection{Top Background Subtraction Uncertainty}
\label{sec:zjets-uncert-top}

The top background subtraction described in \Cref{sec:zjets-bkg-subtraction} relies on the choice of $t\bar{t}$ process MC sample to be subtracted from the data before unfolding.
As shown in \Cref{sec:zjets-bkg-subtraction}, the top subtraction is performed independently with the \textsc{Sherpa}~2.2.12 $t\bar{t}$ sample in place of the nominal \textsc{Powheg}+\textsc{Pythia} sample.
Only the five-dimensional MLP ensemble of \Cref{sec:zjets-bkg-subtraction} is re-trained with the alternative sample.
The ParT top-versus-$Z$+jets classifier is not retrained, because it serves only to define a highly expressive observable along which the subsequent subtraction is performed.
The \Omnifold procedure is then re-run with the alternative set of top subtraction weights applied to the data.
The difference between this result and the nominal result is taken as the uncertainty.
Given the top background is small the subtraction uncertainty is also small, but illustrates the procedure for propagating modeling uncertainties through an unbinned background subtraction.

\section{Unfolding Validation}
\label{sec:zjets-validation}

Given the novel nature of this cross section measurement extensive validation of the results is required.
Three different validation methods are used.
First, a set of binned $\chi^2$ goodness of fit tests between the pseudodata measurement and the truth pseudodata target are performed in the twenty four observables measured in the Multifold analysis plus the \nch and \HT observables.
This provides a first round of validation that crucially can be run alongside future applications of the measurement which are not covered here or in the eventual publication.
See \Cref{sec:zjets-discussion} for more discussion of this point.
Second, the measurement is compared to an Iterative Bayesian Unfolding (IBU) based measurement of these 26 observables which use the same data inputs but are only differential in one observable at a time.
This provides comparison to a well-used baseline unfolding method with much more limited scope than \Omnifold.
Third, the measurement is compared directly to the previous round Multifold result on the 24 observables common to both.
This provides a cross-check of the results against an existing ATLAS measurement performed in the same fiducial volume.

\subsection{Statistical Closure Tests}
\label{sec:zjets-validation-chi2}

\begin{figure}[tb]
    \centering
    \includegraphics[page=15, width=0.48\linewidth, alt={Signed uncertainty contributions used as inputs to the chi-squared closure test for the observable shown on page 15.}]{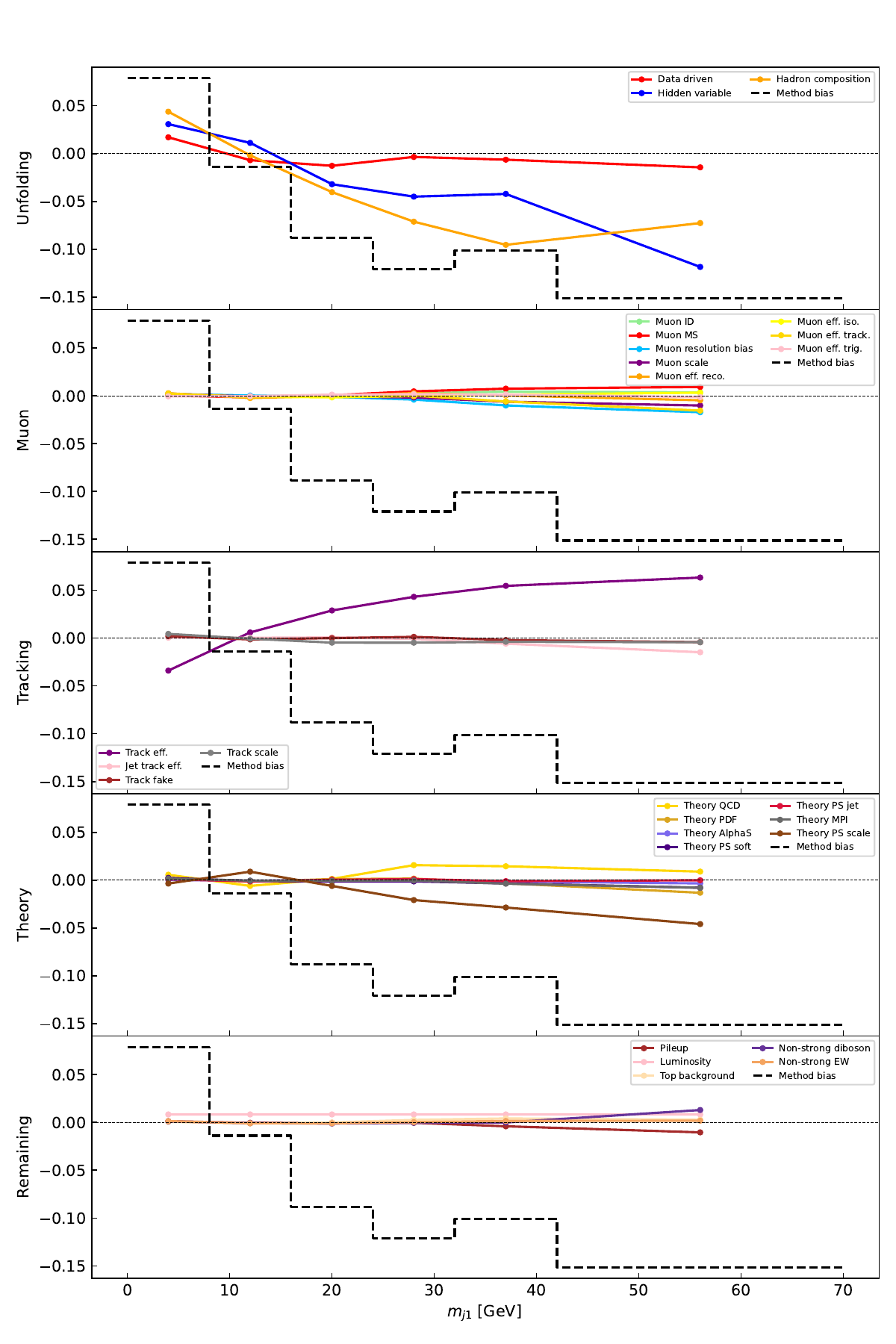}
    \includegraphics[page=15, width=0.48\linewidth, alt={Correlation matrix used as input to the chi-squared closure test for the observable shown on page 15.}]{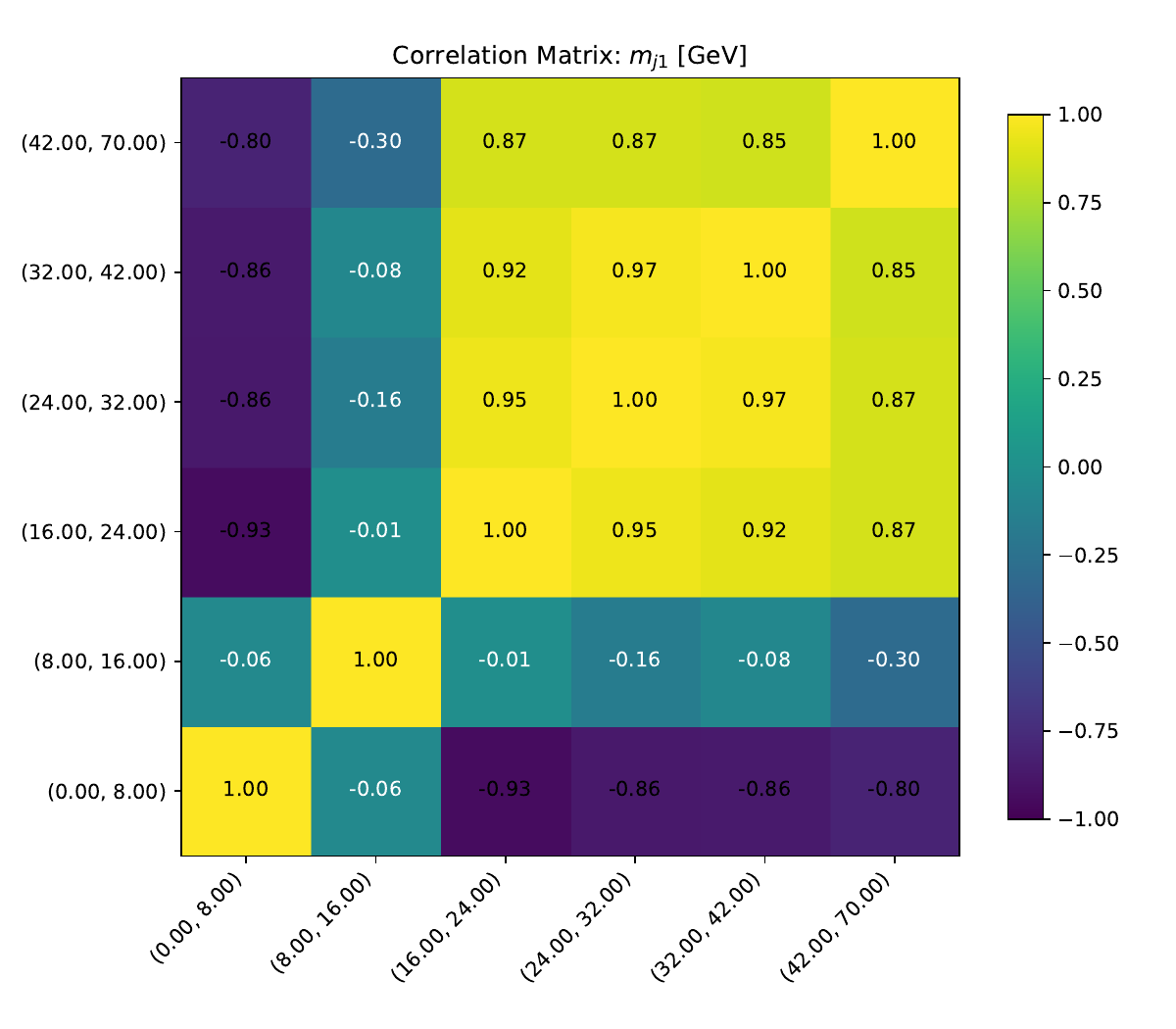}
    \caption{Signed uncertainty contributions (left) and covariance-matrix correlations (right) used as inputs to the binned $\chi^2$ closure test, shown for the $m_{j1}$ observable.}
    \label{fig:zjets-closure-uncert-page15}
\end{figure}

\begin{figure}[tb]
    \centering
    \includegraphics[page=25, width=0.48\linewidth, alt={Signed uncertainty contributions used as inputs to the chi-squared closure test for the observable shown on page 25.}]{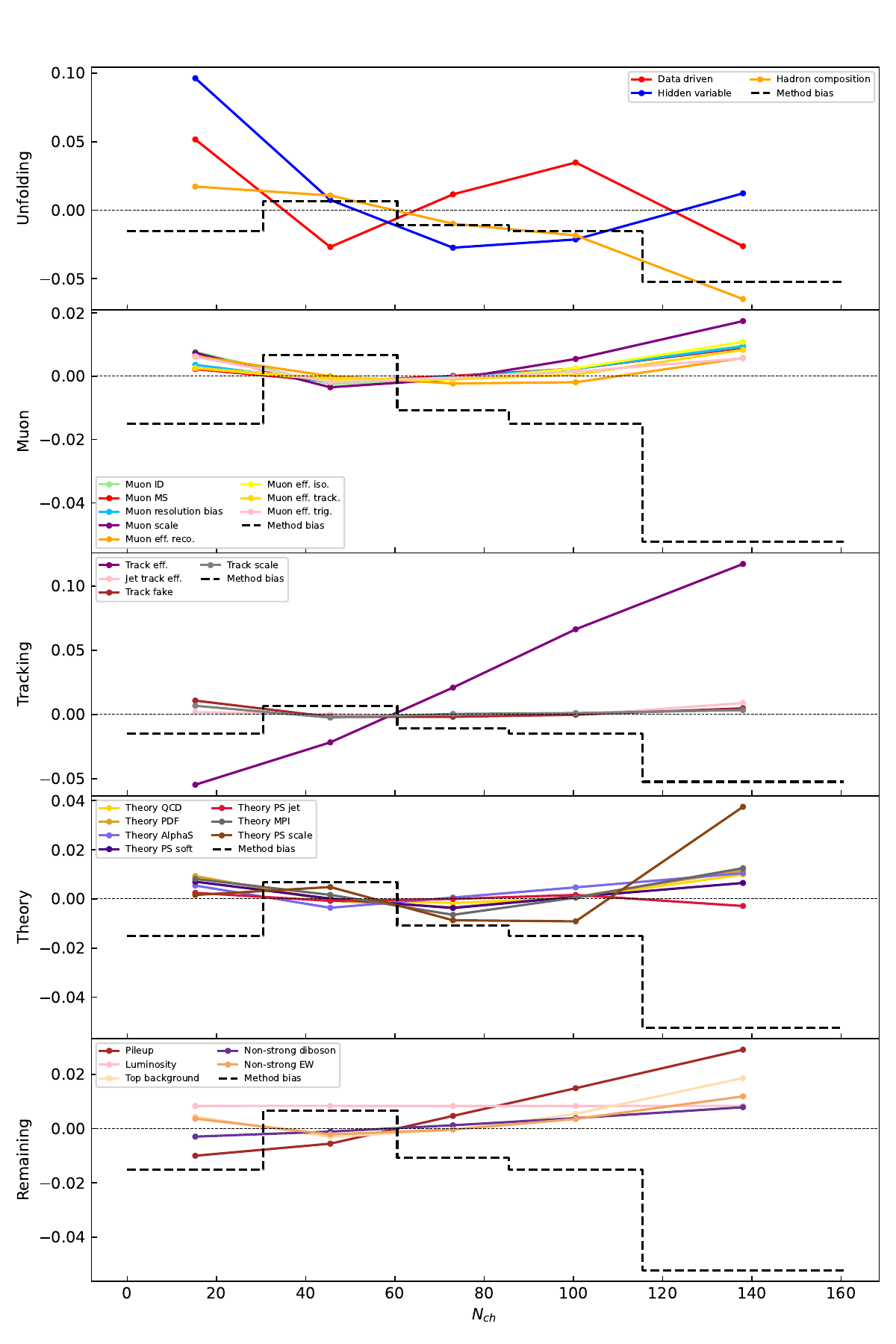}
    \includegraphics[page=25, width=0.48\linewidth, alt={Correlation matrix used as input to the chi-squared closure test for the observable shown on page 25.}]{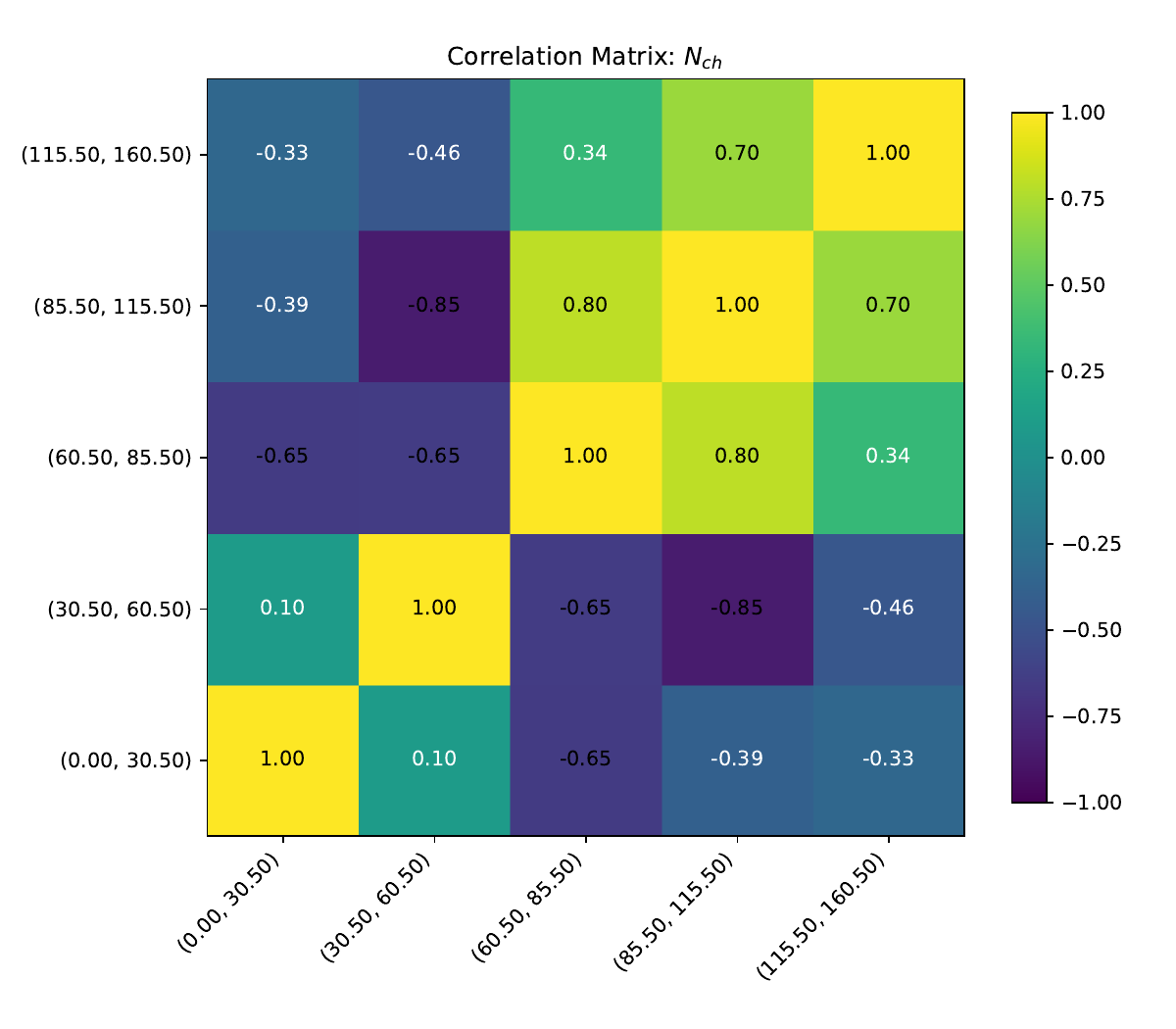}
    \caption{Signed uncertainty contributions (left) and covariance-matrix correlations (right) used as inputs to the binned $\chi^2$ closure test, shown for the \nch observable.}
    \label{fig:zjets-closure-uncert-page25}
\end{figure}

In the pseudodata measurement the desired truth distribution is known.
This allows the method bias and uncertainty model to be jointly assessed through a binned $\chi^2$ goodness of fit test between the measurement and the truth pseudodata.
This test can be run in many different observables to ensure the measurement is accurate in the full phase space as desired.
The $\chi^2$ test statistic is
\begin{equation}
    \chi^2 \;=\; (\hat{\vec{\sigma}} - \vec{T})^{\top} \Sigma^{-1} (\hat{\vec{\sigma}} - \vec{T}),
\end{equation}
where $\hat{\vec{\sigma}}$ is the vector of unfolded cross sections, $\vec{T}$ is the vector of pseudodata truth values, and $\Sigma$ is the covariance matrix of the unfolded cross section built from the uncertainties of \Cref{sec:zjets-uncert} excepting the experimental uncertainties since the nuisance parameters are fixed to their nominal values for the nominal MC and pseudodata.
The hidden variable uncertainty was observed to show some mild statistical fluctuations after propagation through the unfolding because it is built with an entirely independent sample of particle-level events.
To eliminate these a Gaussian kernel smoothing is applied to the hidden variable uncertainty before it is used to fill the covariance matrix, except for in the jet mass observables because the uncertainty has a sharp shape that smoothing was found to distort.
The signed uncertainties and covariance matrix used as input to the $\chi^2$ tests are shown for the \nch and $m_{j1}$ observables in \cref{fig:zjets-closure-uncert-page15,fig:zjets-closure-uncert-page25} respectively.
For most observables, as with the observables shown here, the unfolding uncertainties are leading, followed by the tracking uncertainties and then the remaining uncertainties are generally negligible.

Under the null hypothesis that the reweighted particle-level MC is drawn from the same distribution as the target truth pseudodata, this test statistic should follow a $\chi^2$ distribution with degrees of freedom equal to the number of bins used minus one.
Since running sufficient bootstraps to observe this is computationally infeasible, instead a p-value is calculated for each test.
The p-value quantifies the probability of observing a test statistic larger than the one actually observed under the null hypothesis.
A large p-value indicates that the probability of the observed test statistic is large under the null hypothesis and the null hypothesis can be accepted.
A small p-value indicates the opposite and that the null hypothesis should be rejected.
A typical threshold for acceptable p-values is $0.05$, indicating that there is a 5\% chance of the observed statistic or greater under the null hypothesis.
\Cref{tab:zjets-closure-pvalues} shows the resulting p-values for the full-phase-space \Omnifold measurement, together with the corresponding values from the previous round Multifold measurement on the 24 common observables.
For every observable, the \Omnifold p-value exceeds the $0.05$ threshold, indicating no significant tension between the unfolded distribution and the pseudodata truth.
The smallest \Omnifold p-value is $0.056$ for the sub-leading track-jet mass, and the smallest Multifold $p$-value is $0.17$ for the leading track-jet mass.

\begin{table}[tb]
    \centering
    \caption{Binned $\chi^2$ and $p$-value from the pseudodata closure test, for every observable measured in the previous round plus \nch and \HT. Results are shown for the full-phase-space \Omnifold measurement and for the previous round Multifold measurement on the 24 common observables. Empty cells correspond to observables not measured by a particular method. The closure test is performed in the fiducial volume restricted to events with sub-leading track-jet $p_T > 5$~GeV, with Gaussian-kernel smoothing applied to the hidden-variable unfolding uncertainty for every observable except the jet-mass observables.}
    \label{tab:zjets-closure-pvalues}
    \input{tab_ch6_closure_pvalues.tex}
\end{table}

\begin{figure}[tb]
    \centering
    \includegraphics[page=1, width=0.48\linewidth, alt={Pseudodata measurement of the charged particle multiplicity N_ch, showing the unfolded cross section compared with the pseudodata truth.}]{ch6_Ntracks_pseudodata.pdf}
    \includegraphics[page=1, width=0.48\linewidth, alt={Pseudodata measurement of the scalar sum H_T of charged track transverse momenta, showing the unfolded cross section compared with the pseudodata truth.}]{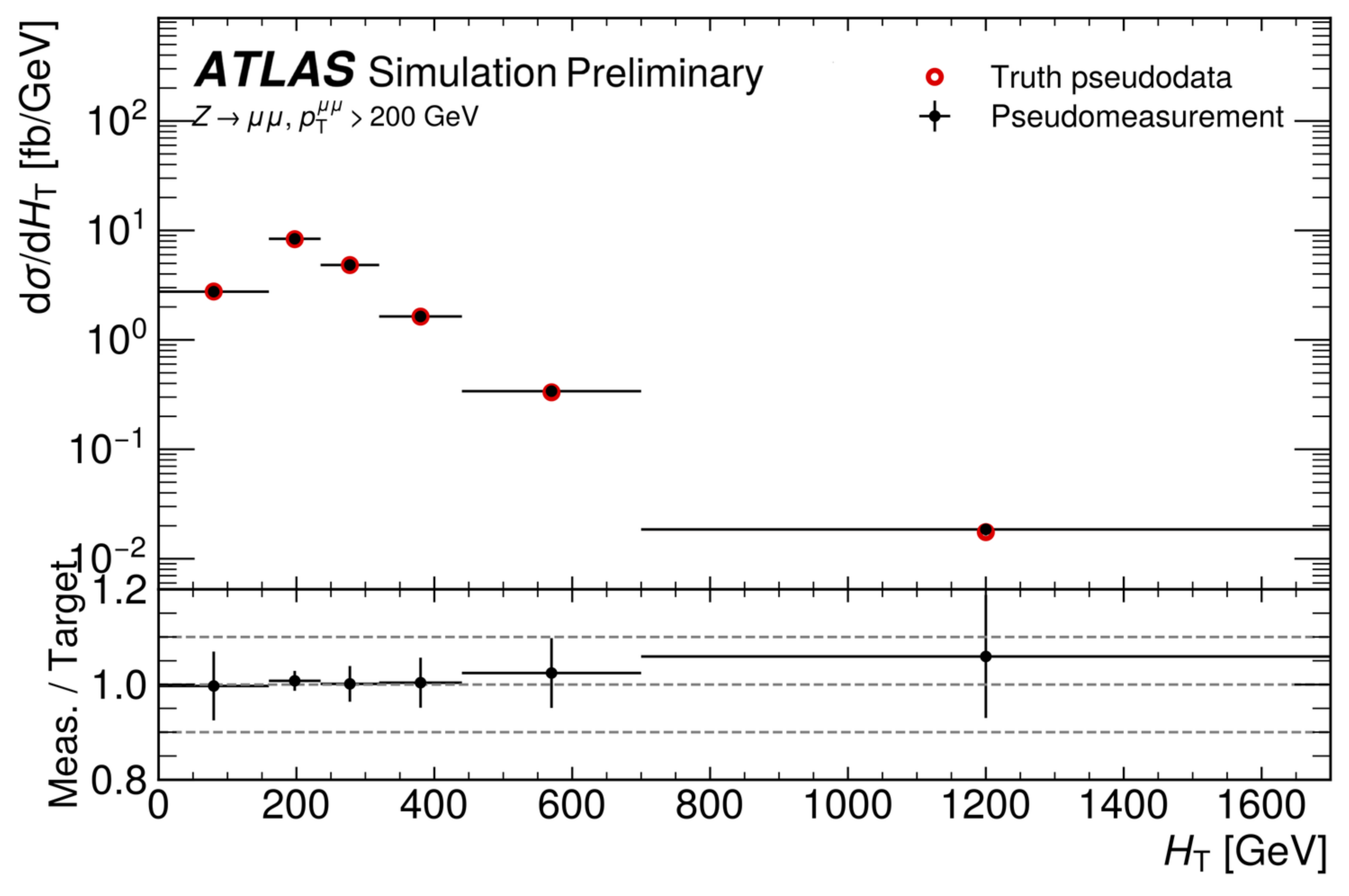} \\
    \includegraphics[page=2, width=0.48\linewidth, alt={Uncertainty budget for the pseudodata measurement of the charged particle multiplicity N_ch, showing dominant tracking and unfolding uncertainties and full coverage of the method bias.}]{ch6_Ntracks_pseudodata.pdf}
    \includegraphics[page=2, width=0.48\linewidth, alt={Uncertainty budget for the pseudodata measurement of the scalar sum H_T of charged track transverse momenta, showing dominant tracking and unfolding uncertainties.}]{ch6_HT_pseudodata.pdf}
    \caption{Pseudodata measurement of the event-level charged-particle multiplicity \nch (left) and scalar sum of charged-track transverse momenta \HT (right). The top row shows the unfolded cross sections compared with the pseudodata truth, and the bottom row shows the corresponding uncertainty budgets. Tracking and unfolding uncertainties dominate the budget, and the method bias is covered by the total uncertainty in every bin.}
    \label{fig:zjets-pd-ue}
\end{figure}

\begin{figure}[tb]
    \centering
    \includegraphics[page=1, width=0.48\linewidth, alt={Pseudodata measurement of the leading track-jet mass, showing the unfolded cross section compared with the pseudodata truth.}]{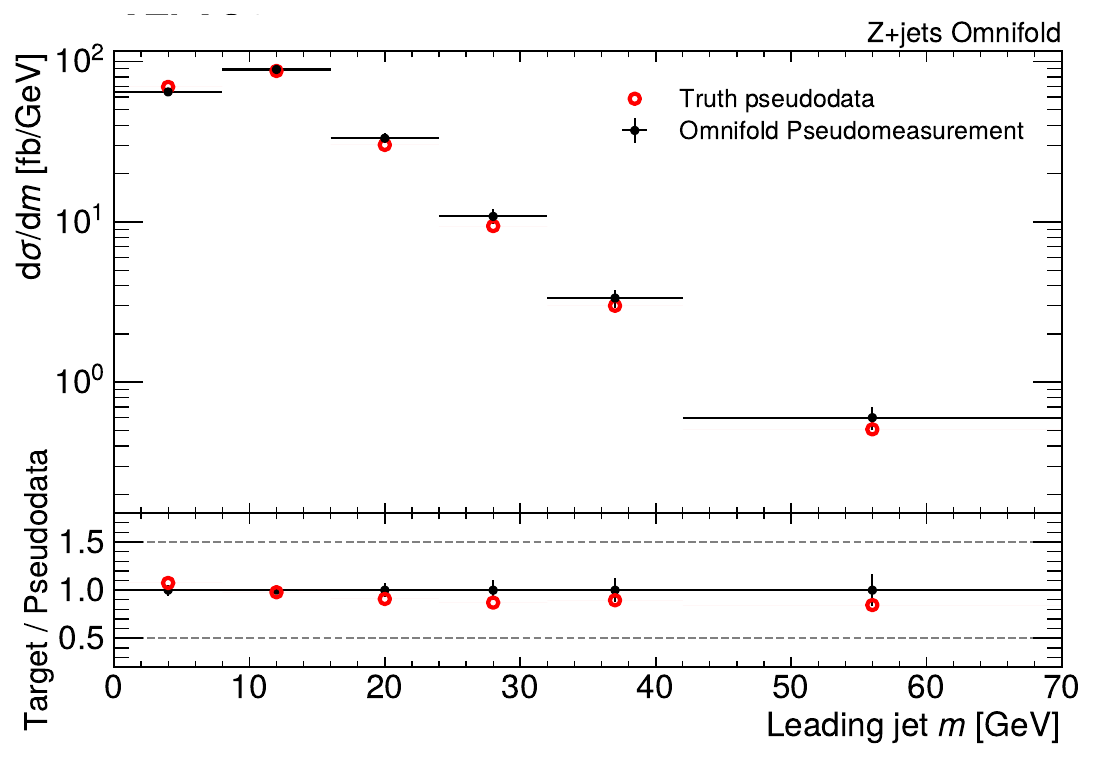}
    \includegraphics[page=1, width=0.48\linewidth, alt={Pseudodata measurement of the leading track-jet transverse momentum, showing the unfolded cross section compared with the pseudodata truth.}]{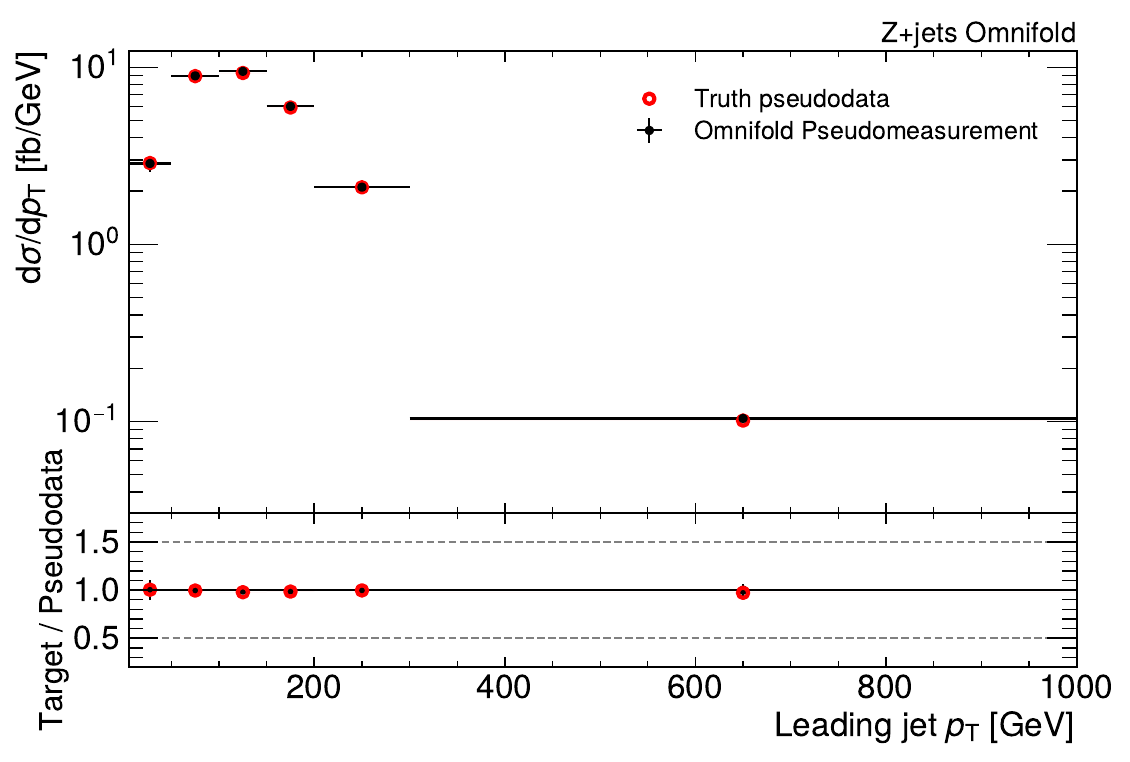} \\
    \includegraphics[page=2, width=0.48\linewidth, alt={Uncertainty budget for the pseudodata measurement of the leading track-jet mass, showing that the unfolding uncertainty carries a sharp shape to cover the method bias.}]{ch6_pd_disp_m_trackj1.pdf}
    \includegraphics[page=2, width=0.48\linewidth, alt={Uncertainty budget for the pseudodata measurement of the leading track-jet transverse momentum.}]{ch6_pd_disp_pT_trackj1.pdf}
    \caption{Pseudodata measurement of the leading track-jet mass $m_{j1}$ (left) and leading track-jet transverse momentum $p_T^{j1}$ (right). The top row shows the unfolded cross sections compared with the pseudodata truth, and the bottom row shows the corresponding uncertainty budgets. The hidden-variable unfolding uncertainty carries a sharp shape in the leading track-jet mass to cover the method bias at low mass, where hadron-composition effects in the jet interior are most pronounced.}
    \label{fig:zjets-pd-trackjet}
\end{figure}

These results are highly non-trivial.
Large differences between the nominal MC and truth pseudodata are present before unfolding, but these closure tests demonstrate that these are recovered by the unfolding such that the pseudodata measurement and the truth pseudodata agree within the quoted uncertainties.
\Cref{fig:zjets-pd-ue} shows the unfolded \nch and \HT spectra on pseudodata together with the uncertainty budgets.
The unfolding uncertainties are dominant for \nch and the tracking uncertainties are dominant for \HT.
Equivalent plots can be drawn for any other observable that is a function of the full phase space.
\Cref{fig:zjets-pd-trackjet} shows the results for two representative track-jet observables, the leading track-jet mass $m_{j1}$ and the leading track-jet transverse momentum $p_T^{j1}$.
Notably the method bias is larger than the total uncertainty in a few of the bins of the $m_{j1}$ observable, but the uncertainty budget produces an acceptable p-value of about 0.26 in the $\chi^2$ test.
The unfolding uncertainties, and in particular the hadron composition uncertainty, is dominant in the $m_{j1}$ observable while the tracking uncertainties are dominant in $p_{T,j1}$.

Additional closure tests also verified good agreement between the pseudodata measurement and the truth pseudodata target in restricted phase space regions, where cuts are placed on the unbinned spectra to further reduce the fiducial volume of the measurement.
Closure tests were also performed for the applications of the measurement presented in \Cref{sec:zjets-results}.
The usage recommendations for the public spectra (see \Cref{sec:zjets-discussion}) additionally require that these tests be performed to validate re-interpretations of the measurement on new observables.

\FloatBarrier

\subsection{Comparison to Iterative Bayesian Unfolding}
\label{sec:zjets-validation-ibu}

\begin{figure}[p]
    \centering
    \begin{tabular}{@{}ccc@{}}
        \textbf{\Omnifold{} vs. IBU} & \textbf{\Omnifold{} uncertainty} & \textbf{IBU uncertainty} \\
        \includegraphics[page=9, width=0.31\linewidth, alt={Data measurement comparison between Omnifold and IBU for the dimuon transverse momentum observable.}]{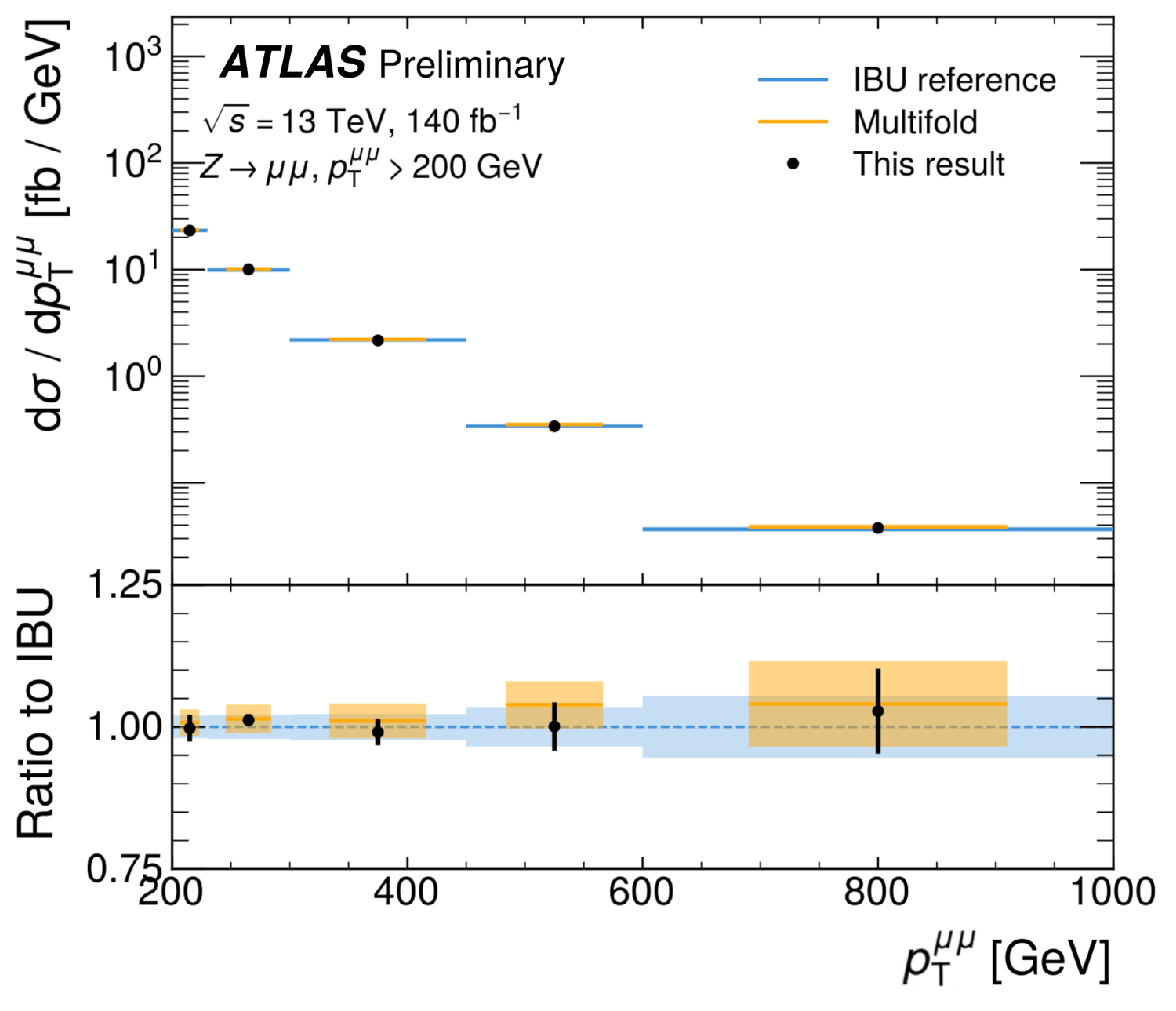} &
        \includegraphics[page=2, width=0.31\linewidth, alt={Uncertainty budget for the Omnifold data measurement of the dimuon transverse momentum observable.}]{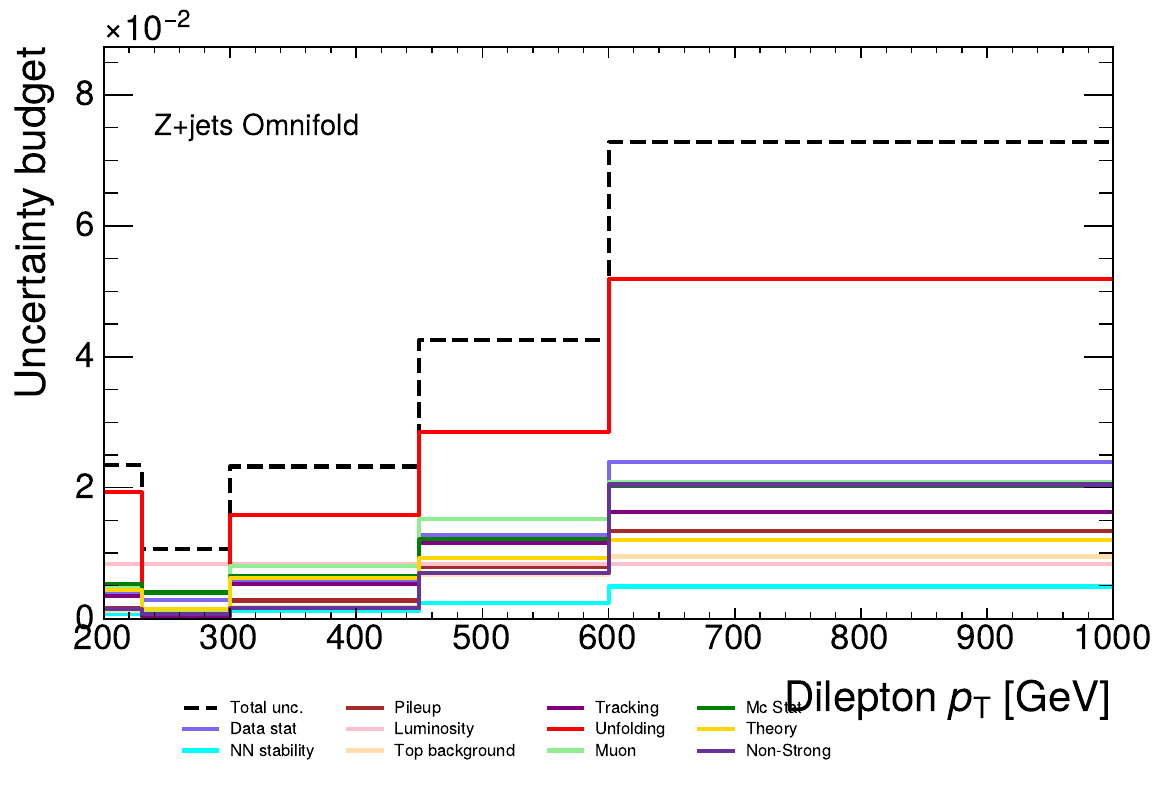} &
        \includegraphics[page=28, width=0.31\linewidth, alt={Uncertainty budget for the IBU data measurement of the dimuon transverse momentum observable.}]{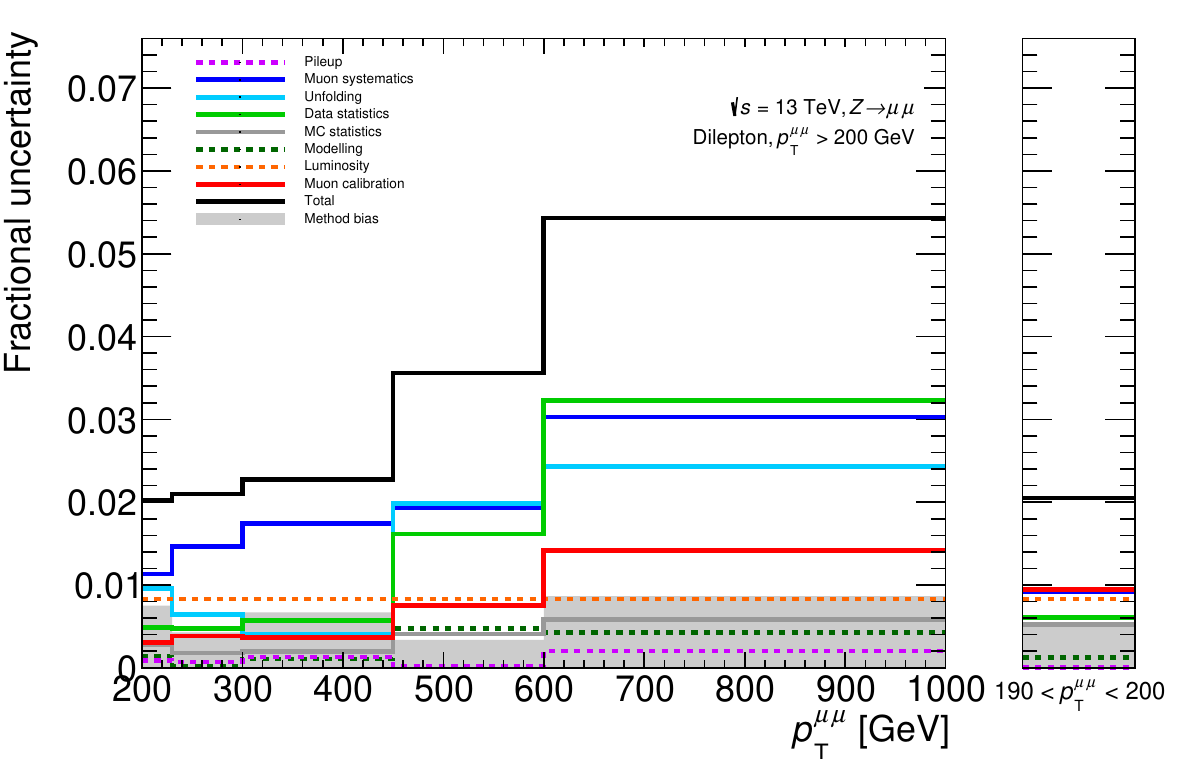} \\
        \includegraphics[page=5, width=0.31\linewidth, alt={Data measurement comparison between Omnifold and IBU for the leading track-jet mass observable.}]{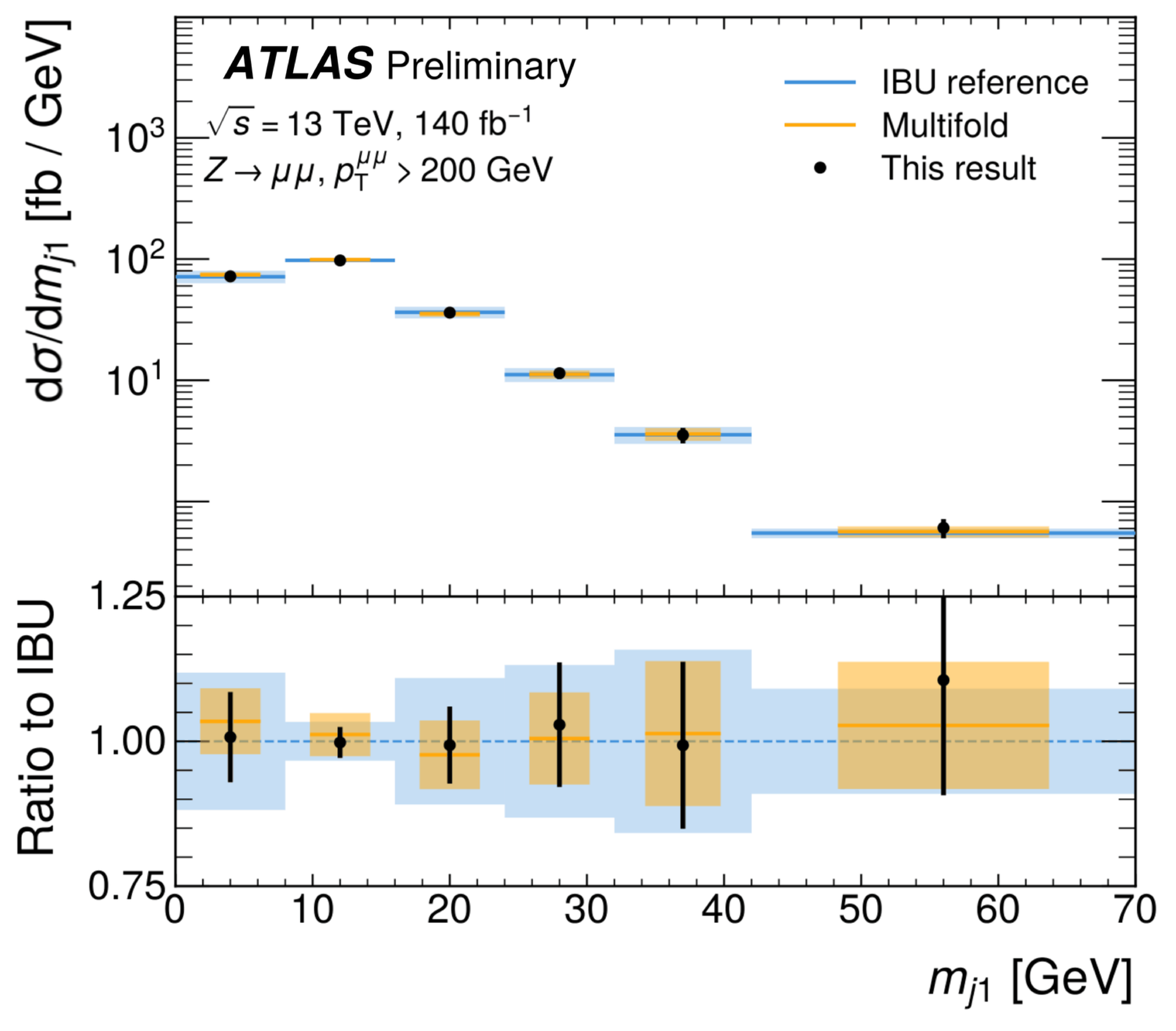} &
        \includegraphics[page=2, width=0.31\linewidth, alt={Uncertainty budget for the Omnifold data measurement of the leading track-jet mass observable.}]{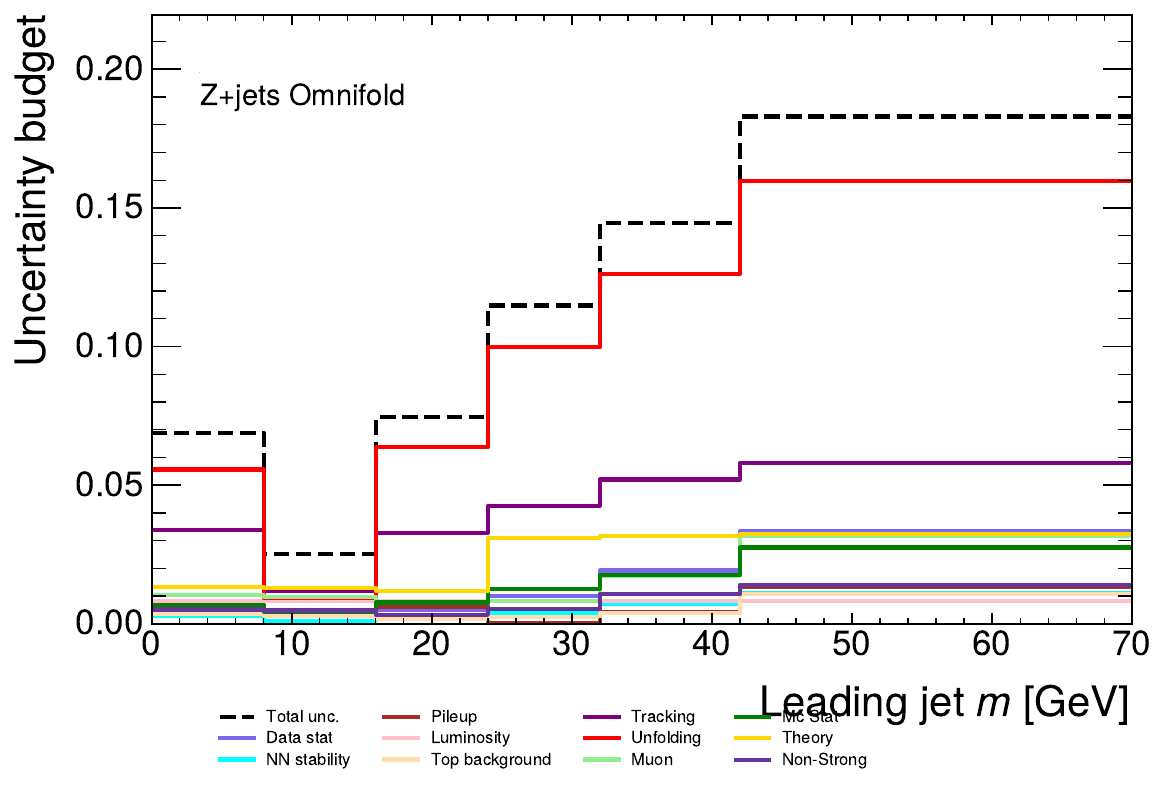} &
        \includegraphics[page=343, width=0.31\linewidth, alt={Uncertainty budget for the IBU data measurement of the leading track-jet mass observable.}]{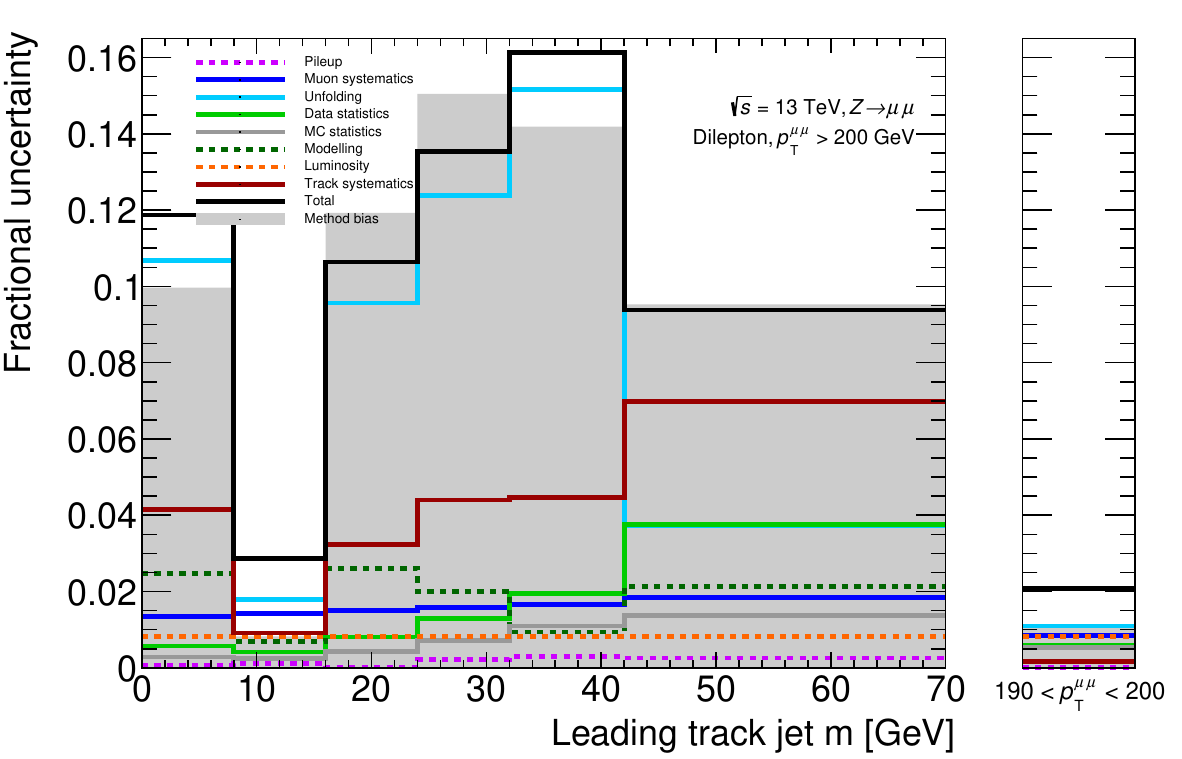} \\
        \includegraphics[page=24, width=0.31\linewidth, alt={Data measurement comparison between Omnifold and IBU for the sub-leading track-jet rapidity observable.}]{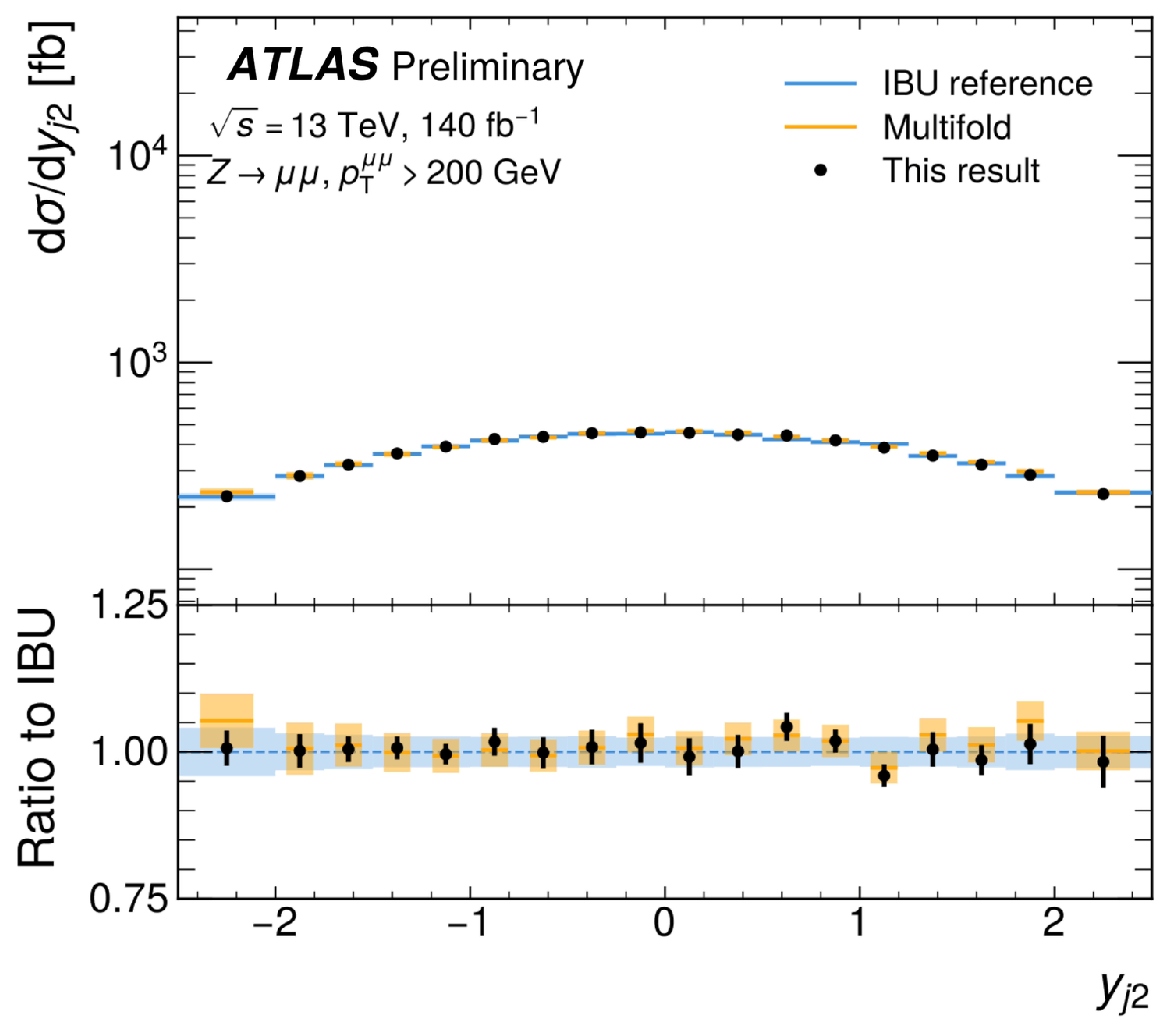} &
        \includegraphics[page=2, width=0.31\linewidth, alt={Uncertainty budget for the Omnifold data measurement of the sub-leading track-jet rapidity observable.}]{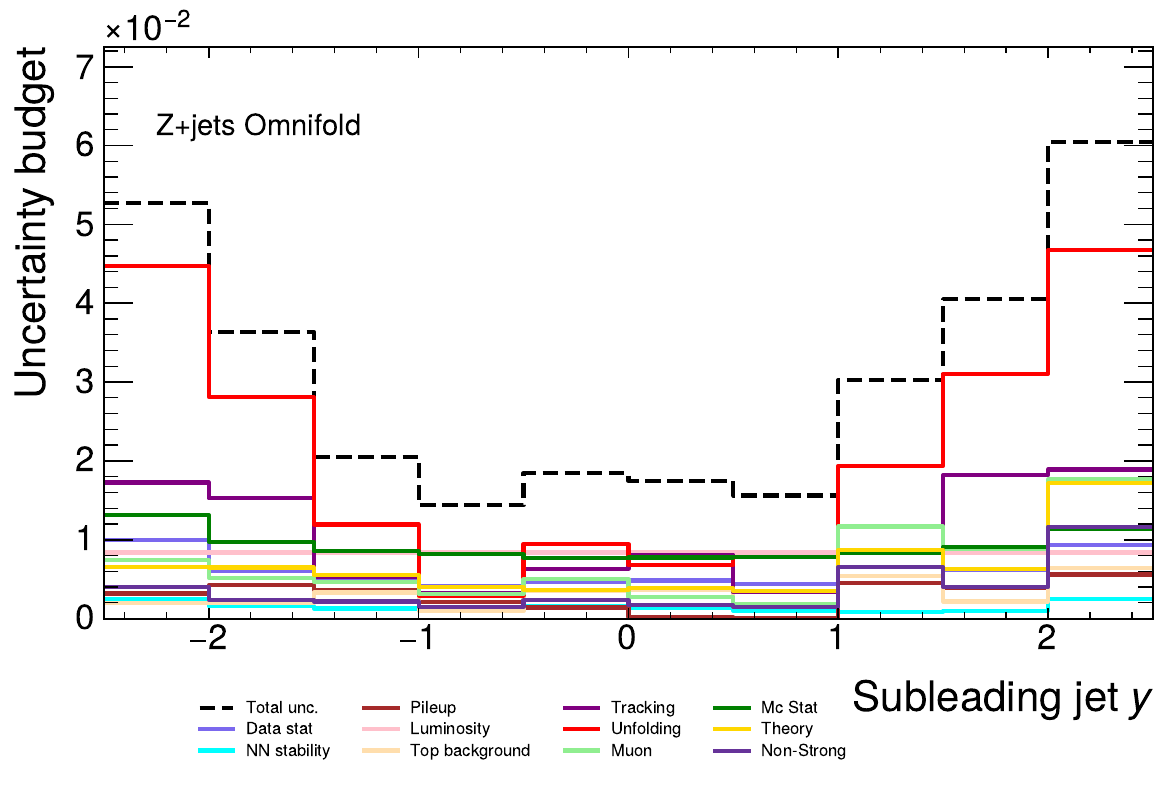} &
        \includegraphics[page=488, width=0.31\linewidth, alt={Uncertainty budget for the IBU data measurement of the sub-leading track-jet rapidity observable.}]{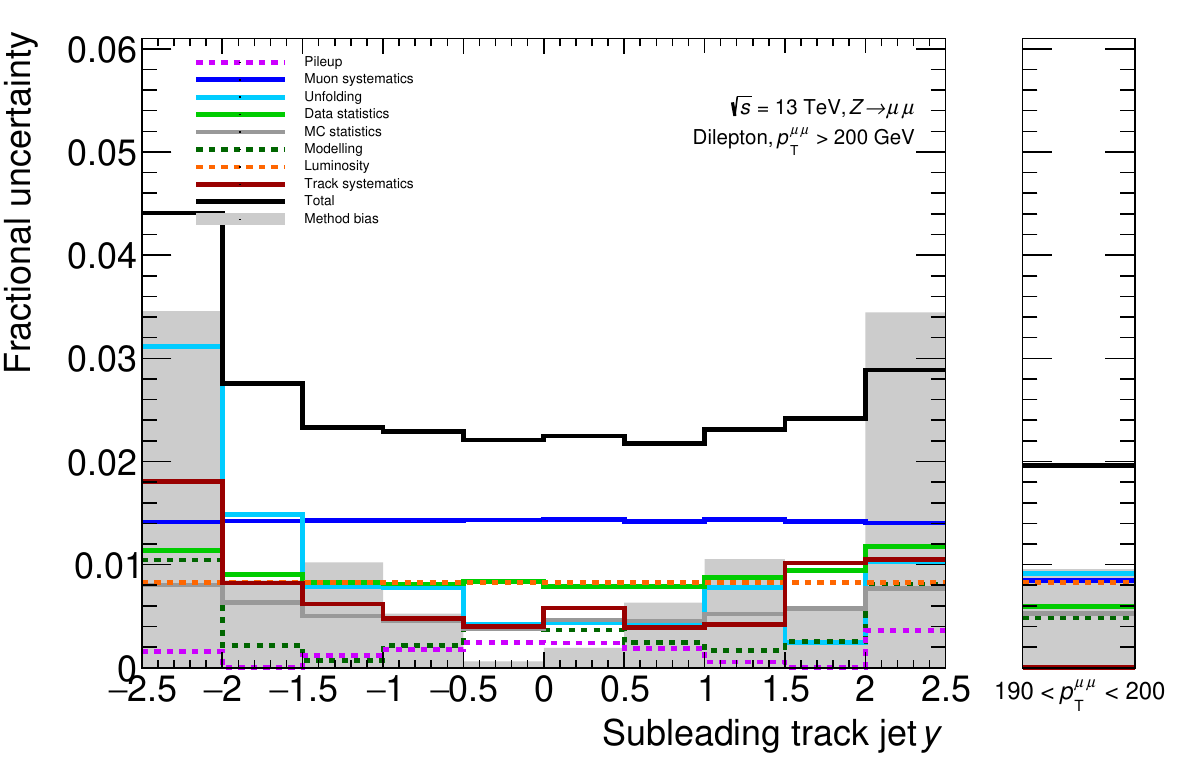} \\
        \includegraphics[page=21, width=0.31\linewidth, alt={Data measurement comparison between Omnifold and IBU for the sub-leading track-jet 3-subjettiness observable.}]{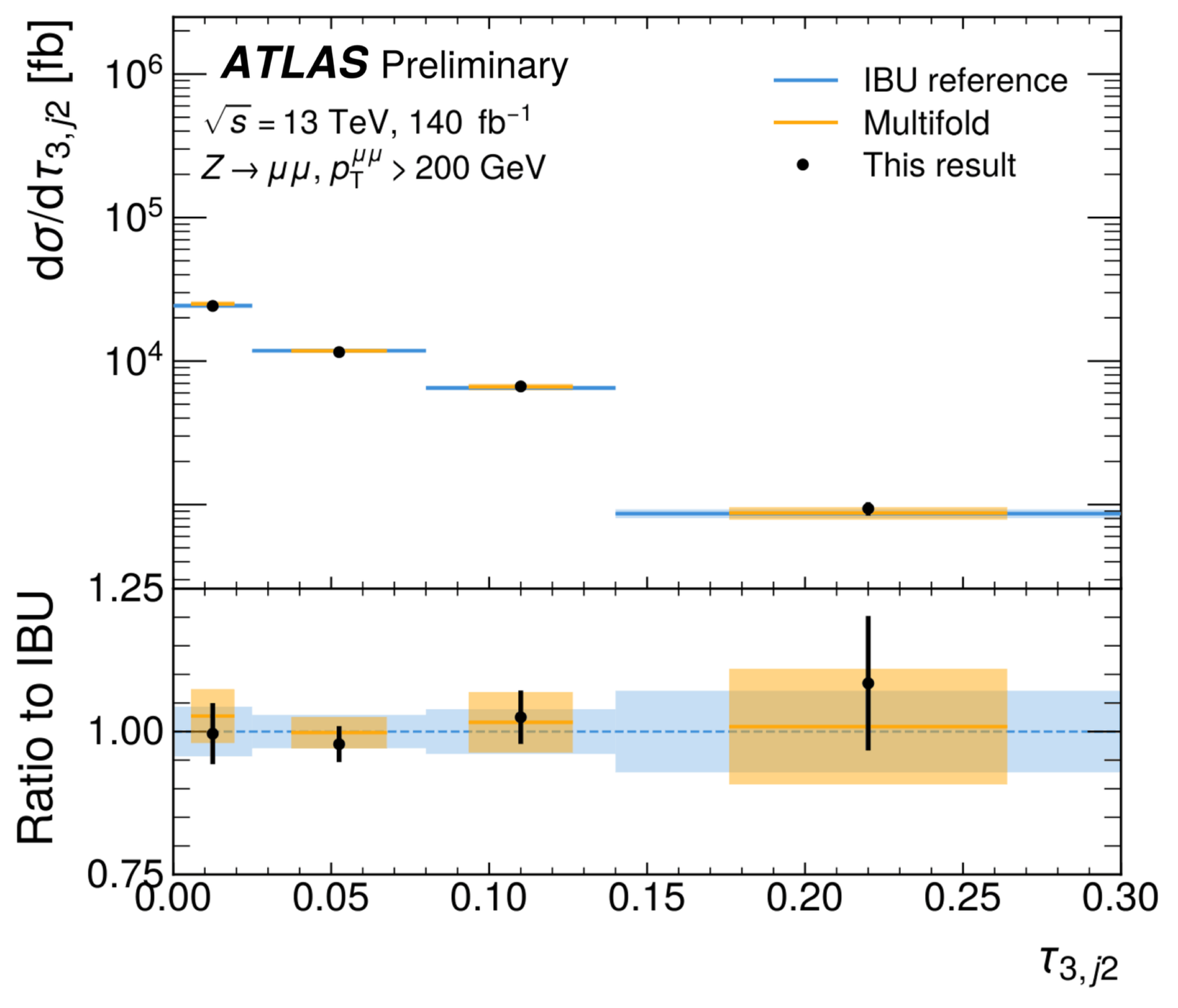} &
        \includegraphics[page=2, width=0.31\linewidth, alt={Uncertainty budget for the Omnifold data measurement of the sub-leading track-jet 3-subjettiness observable.}]{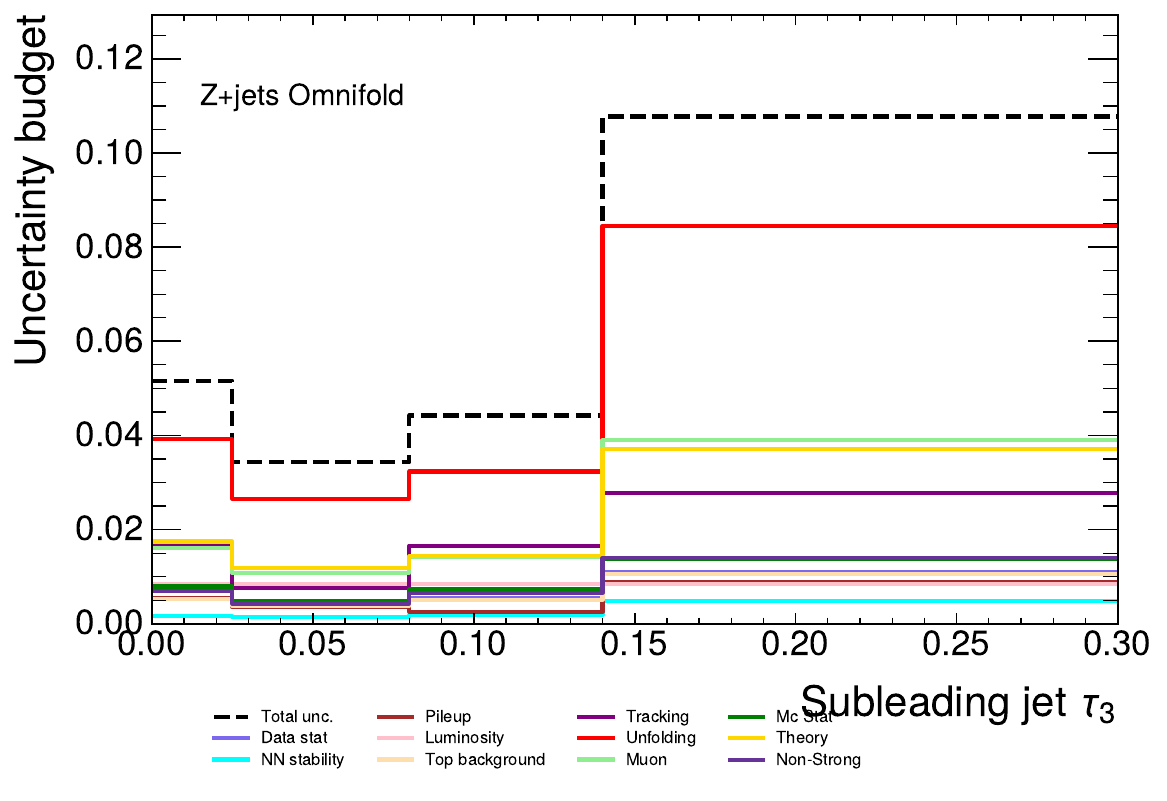} &
        \includegraphics[page=604, width=0.31\linewidth, alt={Uncertainty budget for the IBU data measurement of the sub-leading track-jet 3-subjettiness observable.}]{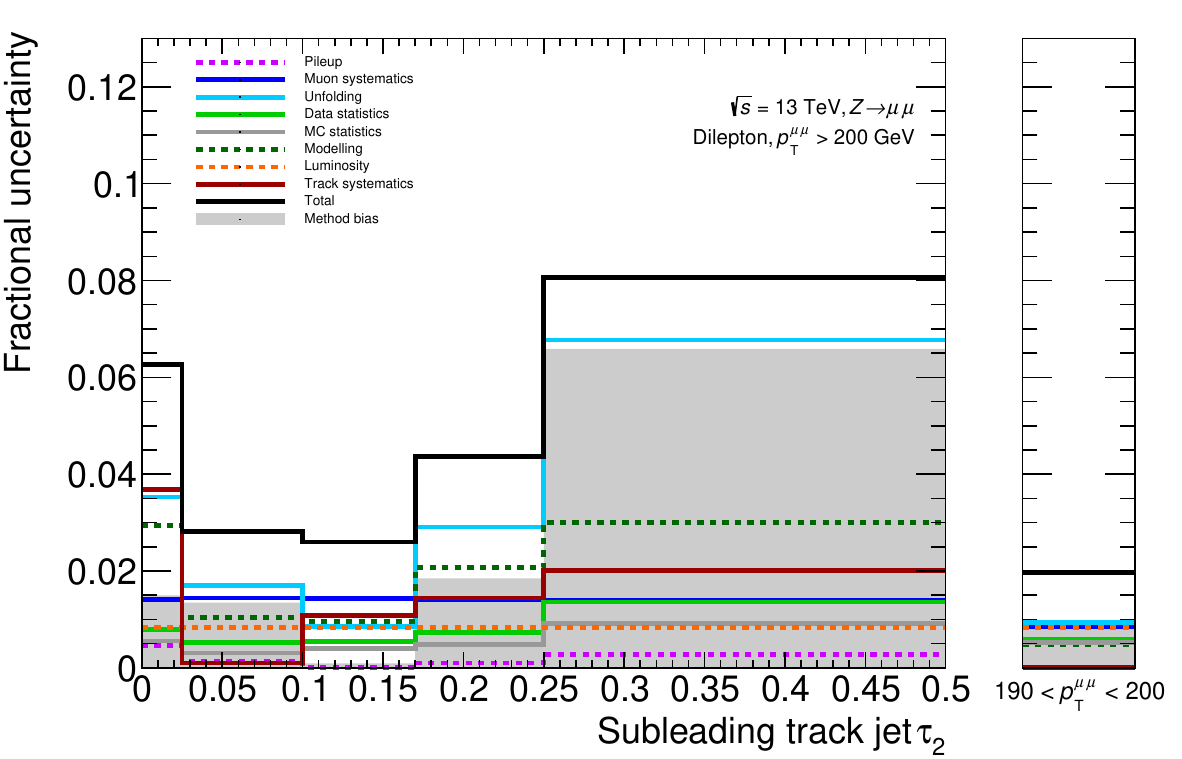}
    \end{tabular}
    \caption{Comparison of the \Omnifold data measurement with the corresponding IBU measurements for four representative observables. The first column compares the unfolded spectra (note the Multifold predictions are also included), the second column shows the \Omnifold uncertainty budget, and the third column shows the corresponding IBU uncertainty budget. Rows show $p_T^{\mu\mu}$, $m_{j1}$, $y_{j2}$, and $\tau_3^{j2}$. In the left column, the black (green) markers indicate the \Omnifold (IBU) measurement, and the blue and pink markers indicate the \mgpy and \sherpa generator predictions respectively.}
    \label{fig:zjets-ibu-data-comparison}
\end{figure}

A second method of validation is to compare the results to independent measurements of the 24 observables measured by the Multifold analysis, plus \nch and \HT, performed with IBU~\cite{DAgostini:1994fjx}.
See \Cref{sec:binned-unfolding-methods} for a review of IBU.
This comparison was done for both the pseudodata and data measurements, but only the results of the data measurement will be shown here since the pseudodata measurement can be independently validated using the $\chi^2$ tests in \Cref{sec:zjets-validation-chi2}.
It is important to note that the measurement produced by \Omnifold contains vastly more physical information than any of the one-dimensional IBU measurements presented in this section, and further enables entirely new types of observables to be measured as shown in \Cref{sec:zjets-results}.
Therefore direct comparison between IBU and \Omnifold based measurements can be misleading.
Even if \Omnifold results in drastically larger uncertainties, the measurement may still be worth performing given the huge potential for downstream applications.

Comparisons of the \Omnifold and IBU based measurements of four observables are shown  in \Cref{fig:zjets-ibu-data-comparison}.
In general excellent agreement between the \Omnifold and IBU measurements is observed, with differences in the central values always less than a few percent.
The largest difference occurs in the last bin of the $m_{j1}$ observable which has large uncertainties for both measurements.
The total uncertainties produced by both measurements are generally similar but have noticeable differences in a few of the bins shown.
For example the total uncertainty for \Omnifold is smaller than IBU in the bulk of the $m_{j1}$ and $\tau_{3}^{j2}$ distributions but larger in the tail.
All differences between the uncertainty budgets are due to the unfolding uncertainties.
Note that it is a non-trivial cross check that the \Omnifold measurement is able to nearly exactly reproduce the experimental, theoretical, and statistical uncertainties of IBU.
A possible explanation for the larger unfolding uncertainties of \Omnifold in the tails of distributions is that there are limited data statistics in this region.
IBU avoids this issue by binning over low statistics regions, but the binning in the \Omnifold results is done only after the unfolding is performed.
The unfolding itself is unbinned and so suffers from the lack of data statistics.
Practically this corresponds to a limited amount of data events to use in the step one neural network trainings.

\begin{figure}[p]
    \centering
    \begin{subfigure}[t]{0.8\linewidth}
        \centering
        \includegraphics[width=\linewidth, alt={Hidden-variable unfolding uncertainty comparison between IBU and Omnifold projected onto the observable families for data.}]{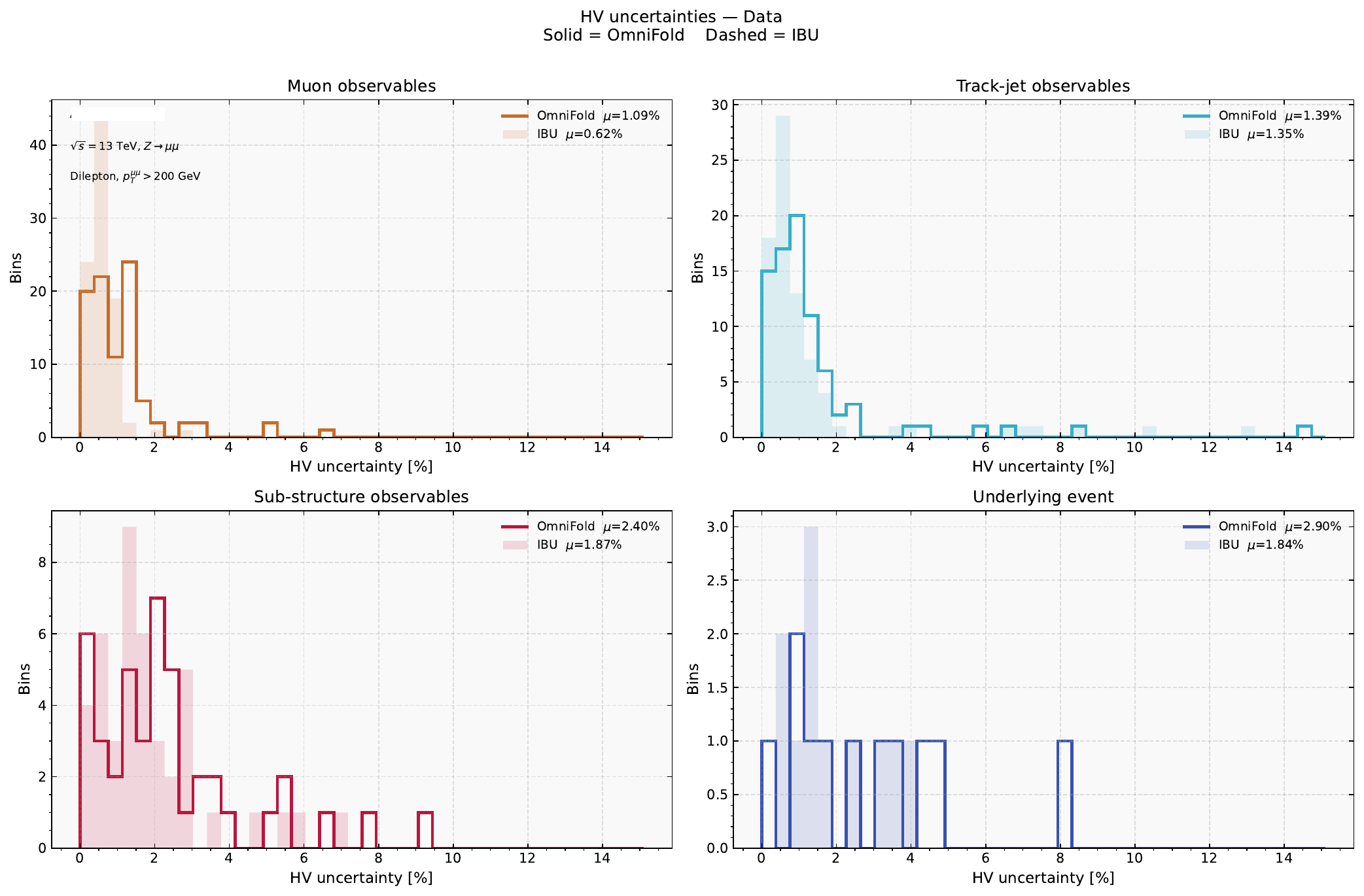}
        \caption{Hidden-variable unfolding}
    \end{subfigure}\\
    \begin{subfigure}[t]{0.8\linewidth}
        \centering
        \includegraphics[width=\linewidth, alt={Hadron-composition hidden-variable unfolding uncertainty comparison between IBU and Omnifold projected onto the observable families for data.}]{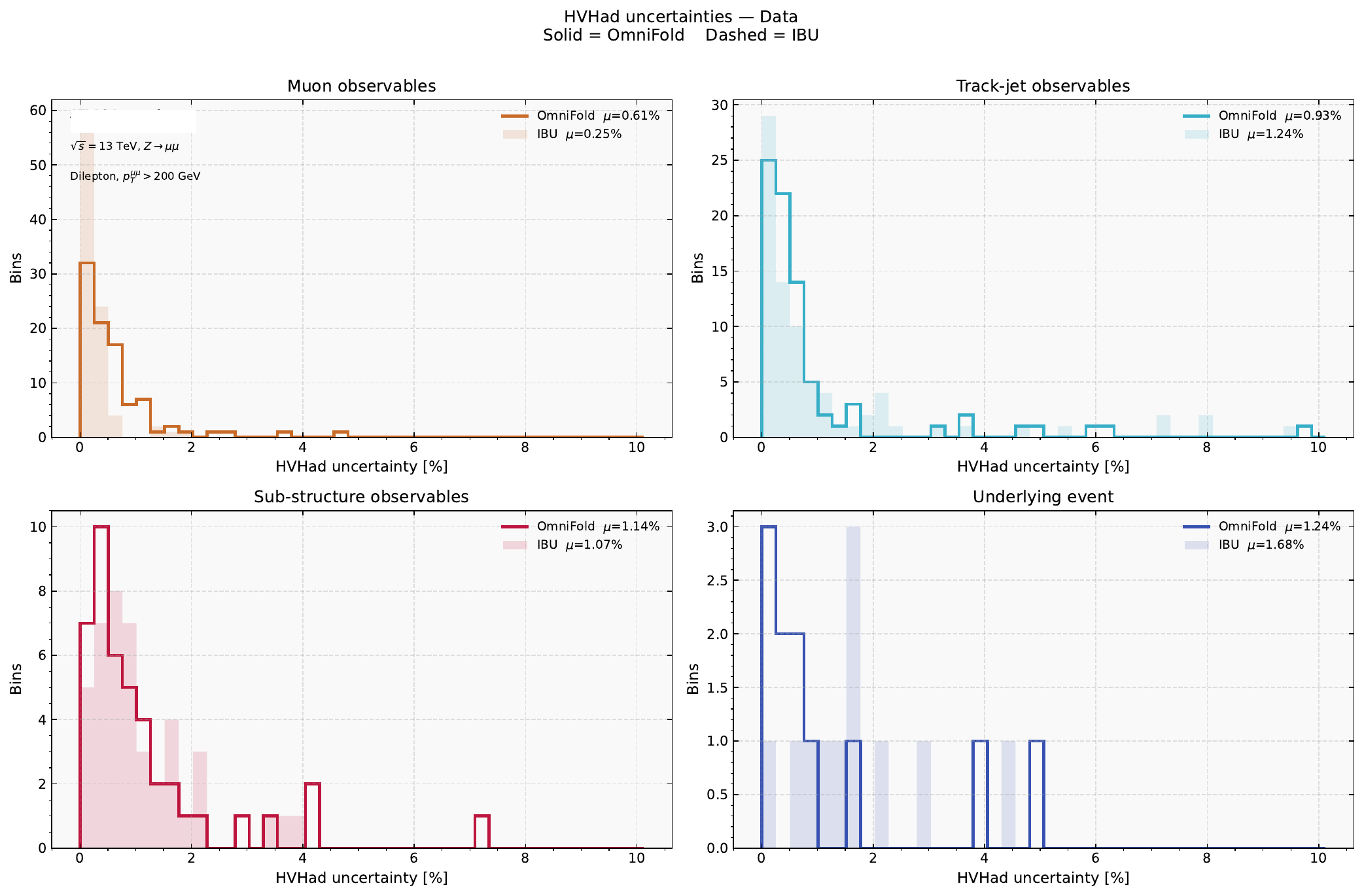}
        \caption{Hadron-composition hidden variable}
    \end{subfigure}
    \caption{Histograms of the hidden variable (top) and hadron composition (bottom) unfolding uncertainties, where each entry in the histogram corresponds to a bin in the \Omnifold or IBU based measurements of the 24 observables from the Multifold analysis plus \nch and \HT. In each panel, these observables are broken into muon (top left), track-jet (top right), substructure (bottom left), and underlying event (bottom right) groups. The shaded histograms represent the IBU uncertainty and the lines represent the \Omnifold uncertainty.}
    \label{fig:zjets-ibu-global-hidden-data}
\end{figure}

\begin{figure}[p]
    \centering
    \begin{subfigure}[t]{0.8\linewidth}
        \centering
        \includegraphics[width=\linewidth, alt={Data-driven unfolding uncertainty comparison between IBU and Omnifold projected onto the observable families for data.}]{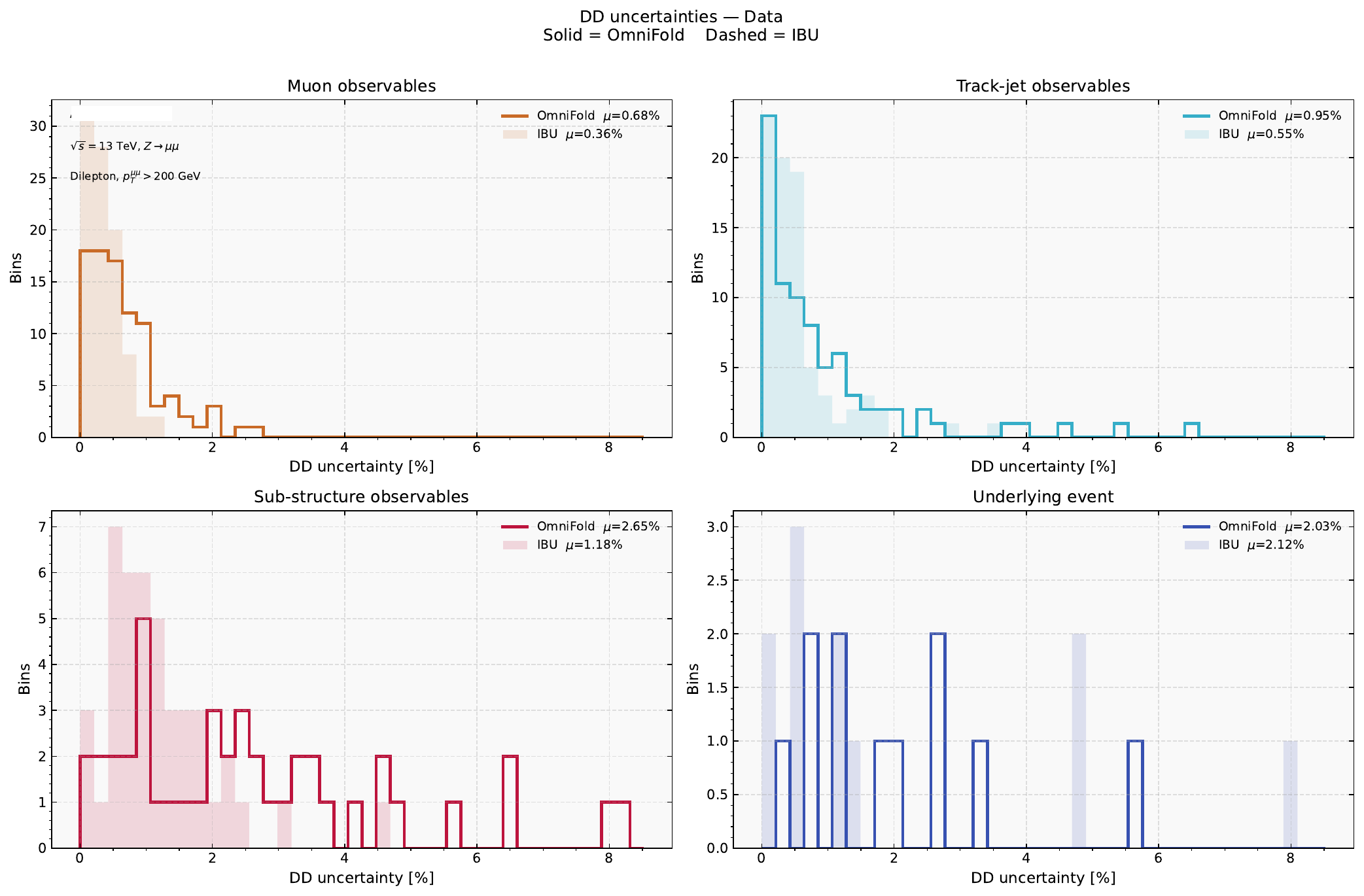}
        \caption{Data-driven unfolding}
    \end{subfigure}\\
    \begin{subfigure}[t]{0.8\linewidth}
        \centering
        \includegraphics[width=\linewidth, alt={Total unfolding uncertainty comparison between IBU and Omnifold projected onto the observable families for data.}]{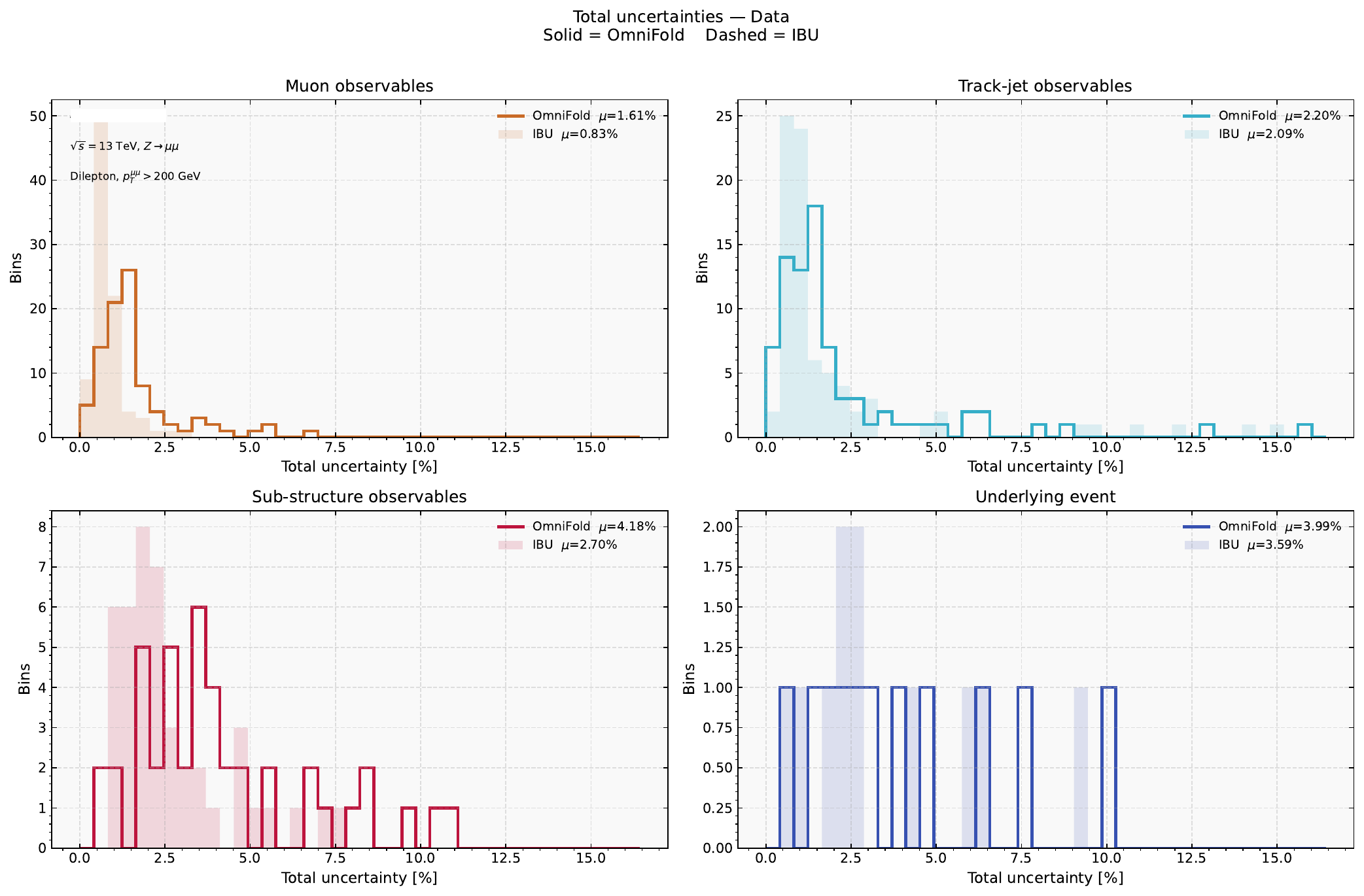}
        \caption{Total unfolding}
    \end{subfigure}
    \caption{Histograms of the data-driven (top) and total (bottom) unfolding uncertainties, where each entry in the histogram corresponds to a bin in the \Omnifold or IBU based measurements of the 24 observables from the Multifold analysis plus \nch and \HT. In each panel, these observables are broken into muon (top left), track-jet (top right), substructure (bottom left), and underlying event (bottom right) groups. The shaded histograms represent the IBU uncertainty and the lines represent the \Omnifold uncertainty.}
    \label{fig:zjets-ibu-global-unfold-data}
\end{figure}

Given the unfolding uncertainties are the only noticeable differences between the \Omnifold and IBU based measurements, it is also worth performing a global comparison of these uncertainties across measurements of many different observables.
This is done by binning the \Omnifold measurement in the same bins measured by IBU in the 24 + 2 observables, calculating the unfolding uncertainties, and using the magnitude of the unfolding uncertainties to fill a histogram.
Importantly the bin entries all correspond to the same high-dimensional measurement for \Omnifold, but to 26 independent single-dimensional measurements for IBU.
The results for the data measurement are shown in \Cref{fig:zjets-ibu-global-hidden-data,fig:zjets-ibu-global-unfold-data}.

Overall the magnitude of the uncertainties is very similar between the IBU and \Omnifold measurements, with \Omnifold showing slightly larger uncertainties globally.
This is particularly true in the hidden variable uncertainty for the substructure observables, which correspond to the $\tau_1$, $\tau_2$, and $\tau_3$ observables for the leading and sub-leading track jets.
From the discussion in \Cref{sec:zjets-uncert-unfold}, naively an \Omnifold based measurement should be \textit{less} sensitive to hidden variables and have smaller hidden variable uncertainties, but this expectation is not borne out in this measurement.
This unexpected behavior is understood as the combination of two factors.
The first is that \Omnifold is data limited in the tails of distributions given the need to train neural networks.
This explanation is backed up by the observation that the data-driven unfolding uncertainty, which roughly corresponds to the method bias produced by each unfolding method, is also slightly larger for \Omnifold in the substructure observables as seen in \Cref{fig:zjets-ibu-global-unfold-data}.
\Omnifold simply provides a less accurate unfolding in the tails of these distributions due to the limited data and this results in larger unfolding uncertainties.
The second factor is that \Omnifold only has an advantage with respect to hidden variables if the dominant hidden variables are constrained at detector-level by some observable used in the unfolding.
As noted in \Cref{sec:zjets-uncert-unfold}, the most dominant hidden variables in this measurement are the truth hadron fractions, which are not constrained at detector level and so remain hidden variables even for \Omnifold.
The first factor is a limitation of \Omnifold that is partially addressed by the pretraining procedure.
The second factor is a feature of the target phase space rather than \Omnifold itself, and could be more or less relevant in other measurements.

Importantly the total unfolding uncertainties shown in the bottom panel of \Cref{fig:zjets-ibu-global-unfold-data} are largely very similar between the two methods.
As mentioned above, most of the applications of the measurement shown in \Cref{sec:zjets-results} would be either difficult or impossible perform with IBU unfolding.
The \Omnifold measurement encodes vastly more information and unlocks entirely new types of measurements.
In this context, the slightly degraded accuracy of the measurement is a very small price to pay.

\FloatBarrier

\subsection{Comparison to the Multifold Measurement}
\label{sec:zjets-validation-mf}

\begin{figure}[p]
    \centering
    \begin{tabular}{@{}ccc@{}}
        \textbf{\Omnifold{} vs. Multifold} & \textbf{\Omnifold{} uncertainty} & \textbf{Multifold uncertainty} \\
        \includegraphics[page=9, width=0.31\linewidth, alt={Data measurement comparison between Omnifold and the previous round Multifold measurement for the dimuon transverse momentum observable.}]{ch6_cross_section_comparison.pdf} &
        \includegraphics[page=2, width=0.31\linewidth, alt={Uncertainty budget for the Omnifold data measurement of the dimuon transverse momentum observable.}]{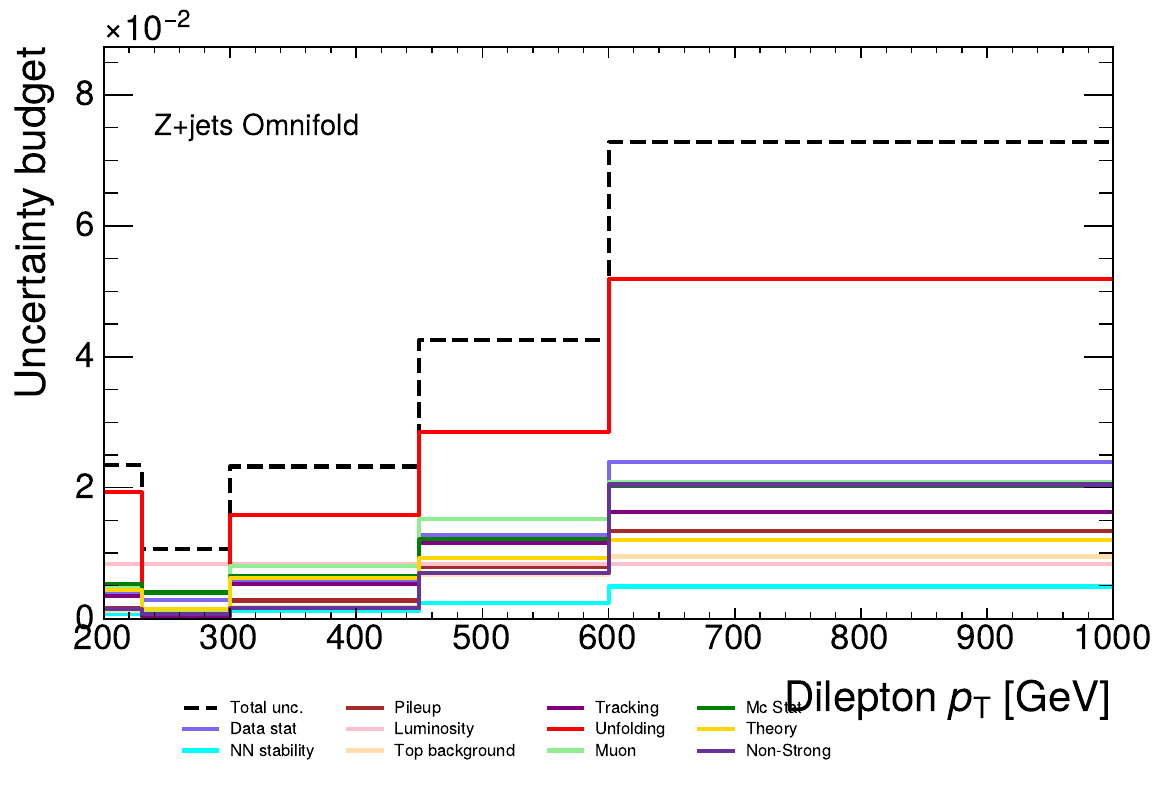} &
        \includegraphics[width=0.31\linewidth, alt={Uncertainty budget for the previous round Multifold data measurement of the dimuon transverse momentum observable.}]{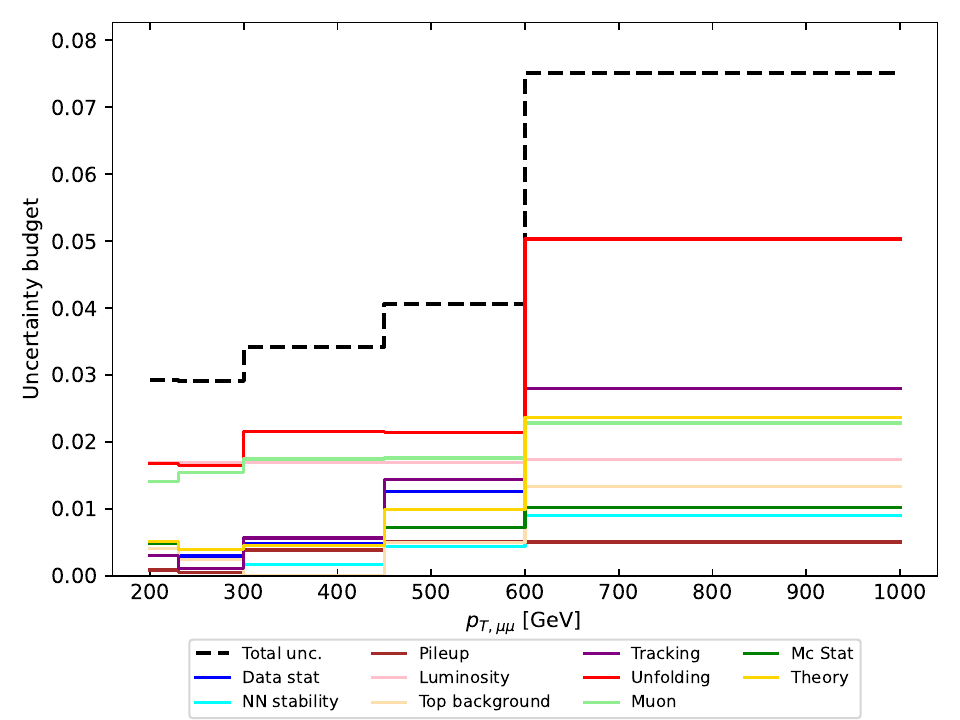} \\
        \includegraphics[page=5, width=0.31\linewidth, alt={Data measurement comparison between Omnifold and the previous round Multifold measurement for the leading track-jet mass observable.}]{ch6_cross_section_comparison.pdf} &
        \includegraphics[page=2, width=0.31\linewidth, alt={Uncertainty budget for the Omnifold data measurement of the leading track-jet mass observable.}]{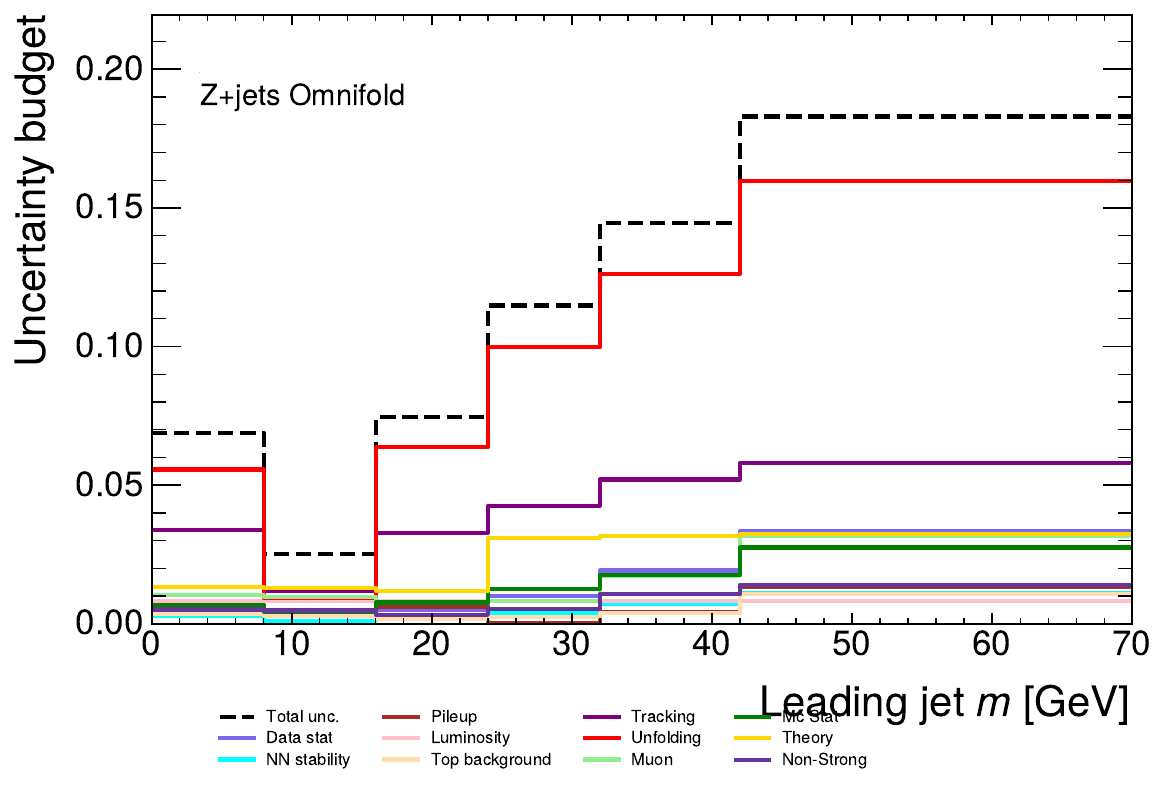} &
        \includegraphics[width=0.31\linewidth, alt={Uncertainty budget for the previous round Multifold data measurement of the leading track-jet mass observable.}]{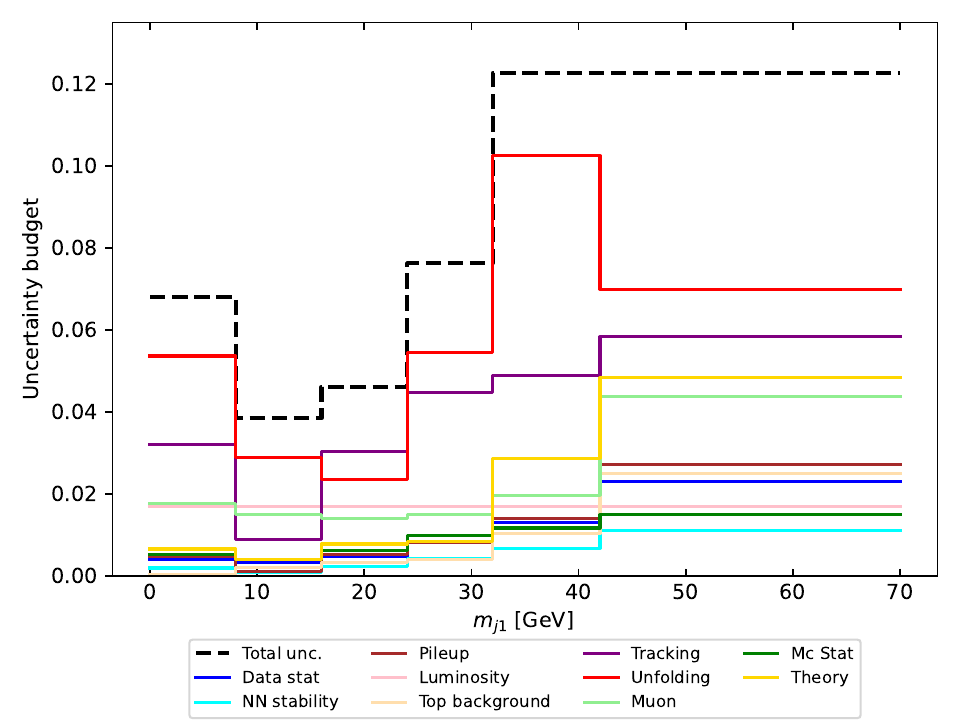} \\
        \includegraphics[page=24, width=0.31\linewidth, alt={Data measurement comparison between Omnifold and the previous round Multifold measurement for the sub-leading track-jet rapidity observable.}]{ch6_cross_section_comparison.pdf} &
        \includegraphics[page=2, width=0.31\linewidth, alt={Uncertainty budget for the Omnifold data measurement of the sub-leading track-jet rapidity observable.}]{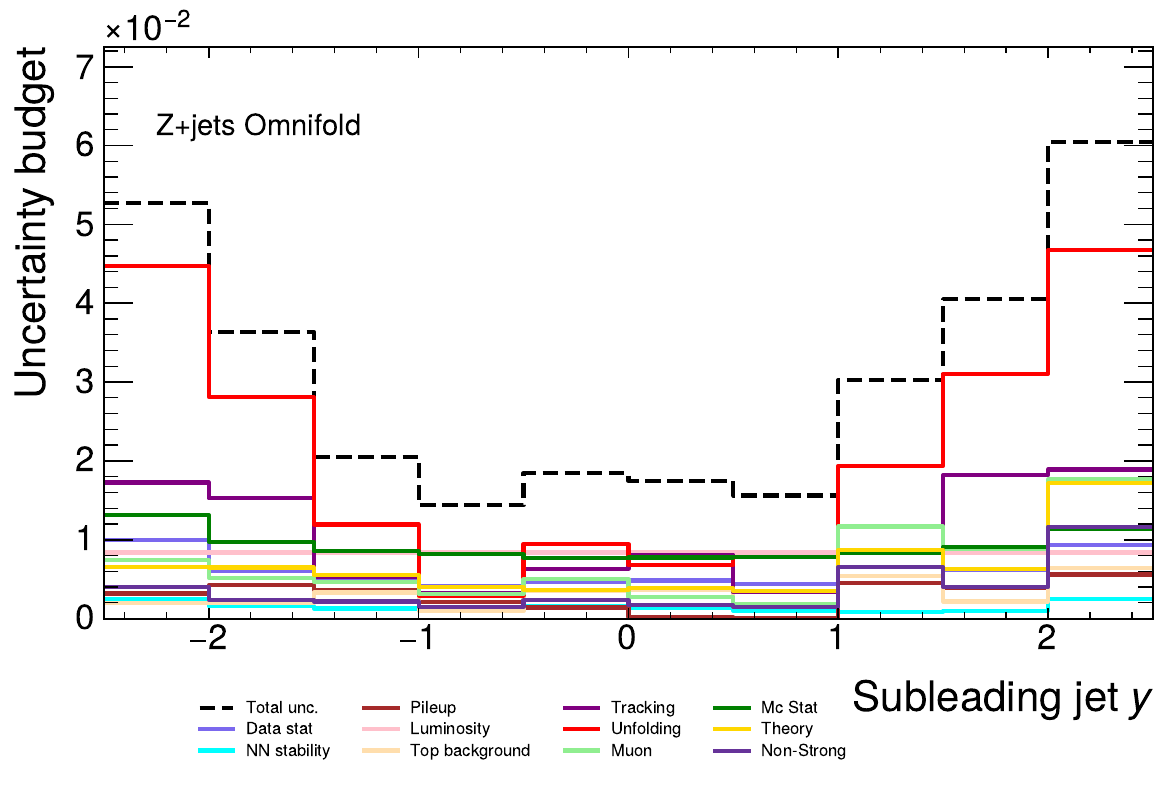} &
        \includegraphics[width=0.31\linewidth, alt={Uncertainty budget for the previous round Multifold data measurement of the sub-leading track-jet rapidity observable.}]{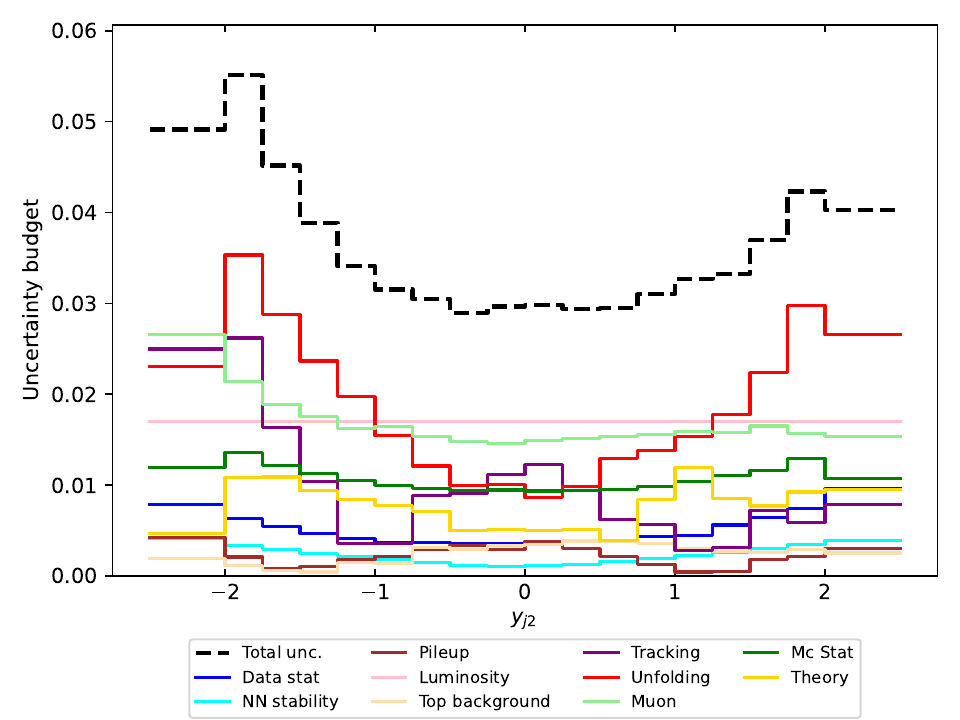} \\
        \includegraphics[page=21, width=0.31\linewidth, alt={Data measurement comparison between Omnifold and the previous round Multifold measurement for the sub-leading track-jet N-subjettiness observable.}]{ch6_cross_section_comparison.pdf} &
        \includegraphics[page=2, width=0.31\linewidth, alt={Uncertainty budget for the Omnifold data measurement of the sub-leading track-jet N-subjettiness observable.}]{ch6_data_disp_tau3_trackj2.pdf} &
        \includegraphics[width=0.31\linewidth, alt={Uncertainty budget for the previous round Multifold data measurement of the sub-leading track-jet N-subjettiness observable.}]{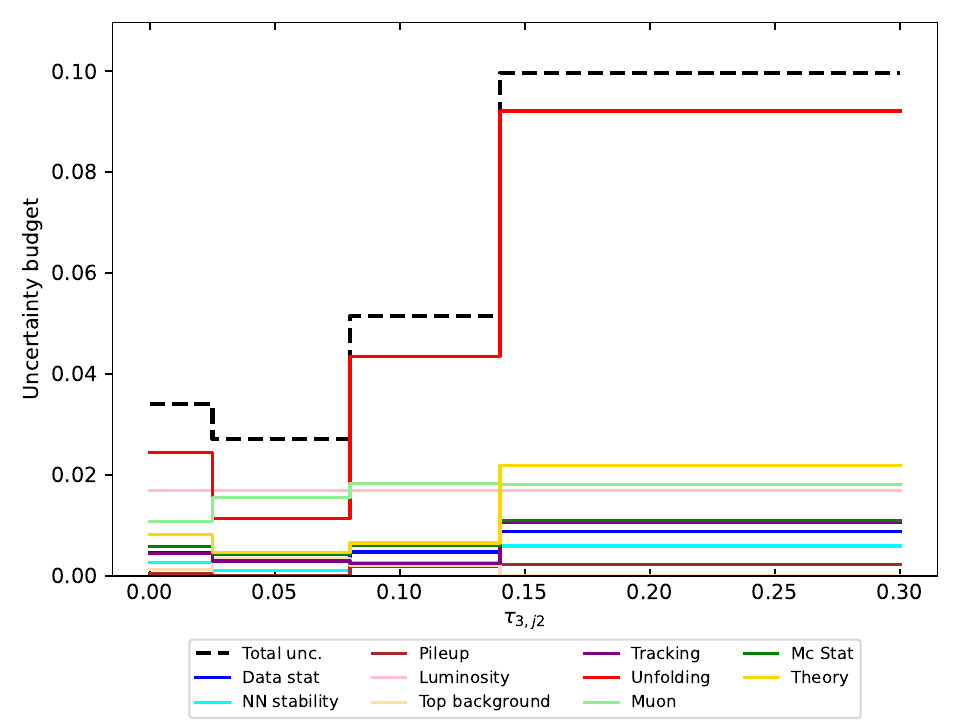}
    \end{tabular}
    \caption{Comparison of the present \Omnifold data measurement with the previous round Multifold measurement for four representative observables. The first column compares the unfolded spectra (note the IBU predictions are also shown), the second column shows the \Omnifold uncertainty budget, and the third column shows the Multifold uncertainty budget. Rows show $p_T^{\mu\mu}$, $m_{j1}$, $y_{j2}$, and $\tau_3^{j2}$.}
    \label{fig:zjets-mf-data}
\end{figure}

The previous round Multifold measurement~\cite{ATLAS:2024xxl} measured the same 24 observables in the same fiducial volume using the same input dataset and MC samples.
The unbinned spectra from the Multifold measurement are publicly available, allowing the results from the full-phase-space measurement to be compared to its smaller-scale counterpart.
The principal methodological differences between the measurements are the classifier (MLP operating on a fixed-length input versus ParT operating on variable-length inputs) and the absence of pretraining in the Multifold measurement.

\Cref{fig:zjets-mf-data} shows this comparison for the same four representative observables used in the IBU comparison.
Unlike the IBU comparison the central values predicted by the full-phase-space measurement are shifted uniformly up from the Multifold measurement.
This is because the full run 2 luminosity measurement was revised upward between the Multifold and Omnifold measurements~\cite{ATLAS:2022hro}.
Beyond this the measurements are very similar in these four observables and the other 20 observables which are not explicitly shown here.
This is expected given the data inputs and unfolding methodology are very similar.
These results show that with the methodological improvements discussed in \Cref{sec:zjets-omnifold}, unfolding the full phase space comes with almost no reduction in precision compared with unfolding the more limited 24-dimensional phase space.

\FloatBarrier

\section{Results: Four Physics Use Cases}
\label{sec:zjets-results}

Unlike standard binned cross section measurements, the unbinned full-phase-space measurement is best understood as a dataset that can be processed to produce measurements of arbitrary observables.
This section presents applications of this dataset to produce measurements of four observables.
The broad goal of each application is to provide measurements that can be used to improve the theoretical modeling of QCD processes, addressing one of the open problems discussed in \Cref{ch:intro}.
In some cases the measurements are immediately applicable to this goal since large deviations between the data and current theoretical models are observed.
In other cases the observables are of interest because they are highly novel, and measurement of them will hopefully spur more development of theoretical tools.
In each of these applications the \Omnifold spectrum is compared to the truth-level predictions of the \mgpy and \sherpa generators used throughout this analysis.
The first application, measurements of the underlying event observables \nch and \HT, could have equally well been performed using standard binned unfolding methods like IBU, but it is important to note the results presented here are simultaenous and can be re-binned on the fly.
The other three applications would either require significant approximations, or be simply impossible to unfold with binned methods.

\subsection{Underlying Event Observables}
\label{sec:zjets-results-ue}

\begin{figure}[tb]
    \centering
    \includegraphics[width=0.48\linewidth, alt={Unfolded differential cross section of the charged-particle multiplicity N_ch compared to the IBU measurement and to the MG5 FxFx and Sherpa 2.2.11 truth predictions.}]{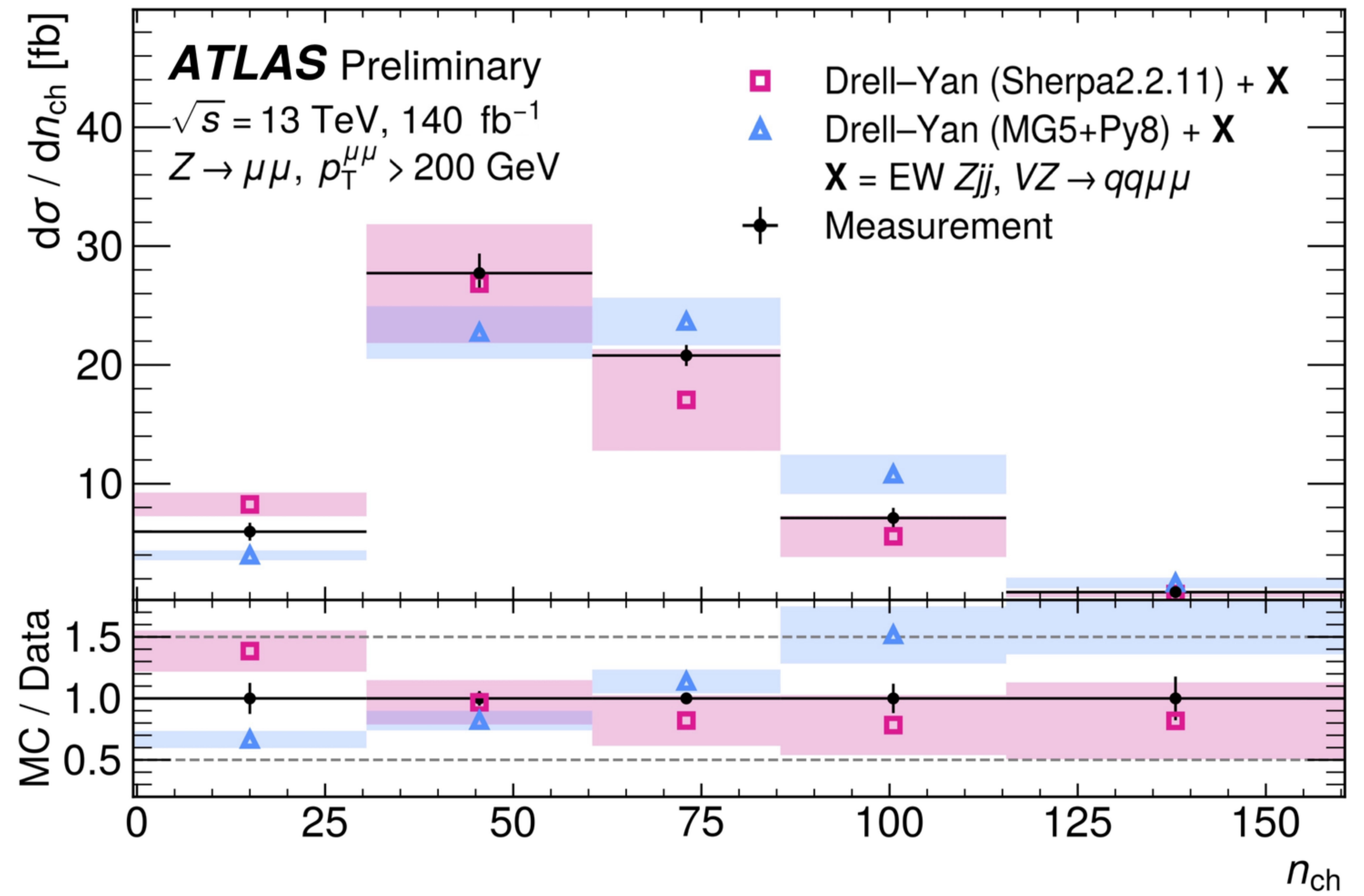}
    \includegraphics[width=0.48\linewidth, alt={Unfolded differential cross section of the scalar sum H_T of charged-particle transverse momenta compared to the IBU measurement and to the MG5 FxFx and Sherpa 2.2.11 truth predictions.}]{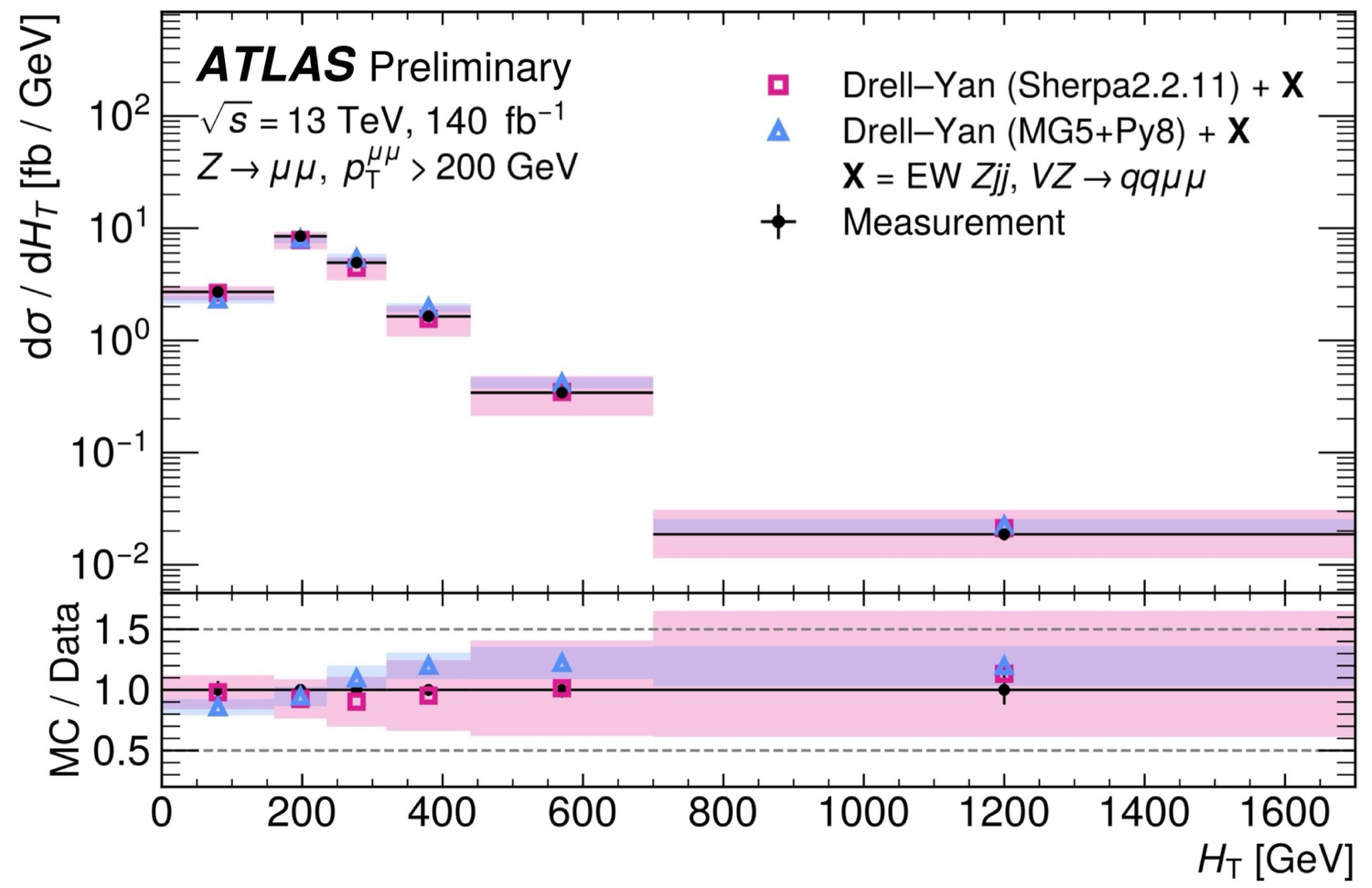}
    \caption{Differential cross section measurements of the \nch (left) and \HT (right) observables. The \Omnifold measurement (black points) is shown together with the particle-level predictions of the \mgpy (blue) and \sherpa (pink) samples. Error bars on the data show the total uncertainty and shaded boxes on the generator predictions show the theory uncertainty.}
    \label{fig:zjets-results-ue}
\end{figure}

Measurements of simple event-level observables can be produced en-masse with the full-phase-space measurement by simply calculating the observable and binning the data.
The resulting plots are very similar to IBU based measurements of the same observables as can be seen in \Cref{sec:zjets-validation-ibu}, but the underlying unbinned measurement is far more flexible.
For example, correlations between event-level observables are preserved and can be used in down-stream applications, the binning applied to each observable can be modified without the need to re-run the unfolding, and the observables can be measured in sub-regions of the full fiducial volume.
Two measurements of standard observables that show particularly poor modeling by the MC generators, \nch and \HT, are shown in this section.
\Cref{fig:zjets-results-ue} shows the measurement of both observables together with the truth-level predictions of the \mgpy and \sherpa samples.
As already shown in \Cref{fig:zjets-trackmcdata}, the generator predictions are very different for the \nch observable.
This is understood to be primarily due to the different hadronization models between the \mgpy and \sherpa samples.
The data points sit roughly between the predictions of the two generators, with both generators showing deviations from the data of 30\% or more in either direction.
In the \HT observable, substantial mismodeling of up to 20\% is observed for the \mgpy sample, but the \sherpa sample is able to reproduce the data to within 10\%.

Explicitly tuning the hadronization models to these measurements could substantially improve the theoretical modeling of these observables.
Since the measurement is simultaneous in both observables this can be done while taking correlations into account, which would not be possible with the independent IBU measurements in \Cref{sec:zjets-validation-ibu}.
The public release of the unbinned spectra (see \Cref{sec:zjets-discussion}) will also allow theorists from outside of the collaboration to adjust the binning of these observables as they see fit.
Further because this measurement itself is unbinned, it is possible to use it to tune entirely new types of hadronization models, specifically those built using deep learning techniques.
There is a growing body of work on this topic, many of which assume that unbinned measurements such as this one exist and are able to furnish training data~\cite{Assi:2025gog,Bierlich:2023fmh,Bierlich:2023zzd,Bierlich:2024xzg,Chan:2023ume,Ghosh:2022zdz,Ilten:2022jfm}.
These methods are beyond the scope of this thesis, but have the potential to produce much more accurate hadronization models which could broadly improve the accuracy of measurements and sensitivity of searches using the HL-LHC datasets.
As more unbinned measurements are performed in the future, more data will become available for training these models and hopefully these more accurate theoretical models will produce a broad impact on LHC physics.

\FloatBarrier

\subsection{Jet Shape Observables}
\label{sec:zjets-results-shape}

\begin{figure}[tb]
    \centering
    \includegraphics[page=1, width=0.48\linewidth, alt={Measurement of the charged-particle pT density profile rho(r) about the jet axis using the anti-kt R=0.4 algorithm, compared to the MG5 FxFx and Sherpa 2.2.11 truth predictions.}]{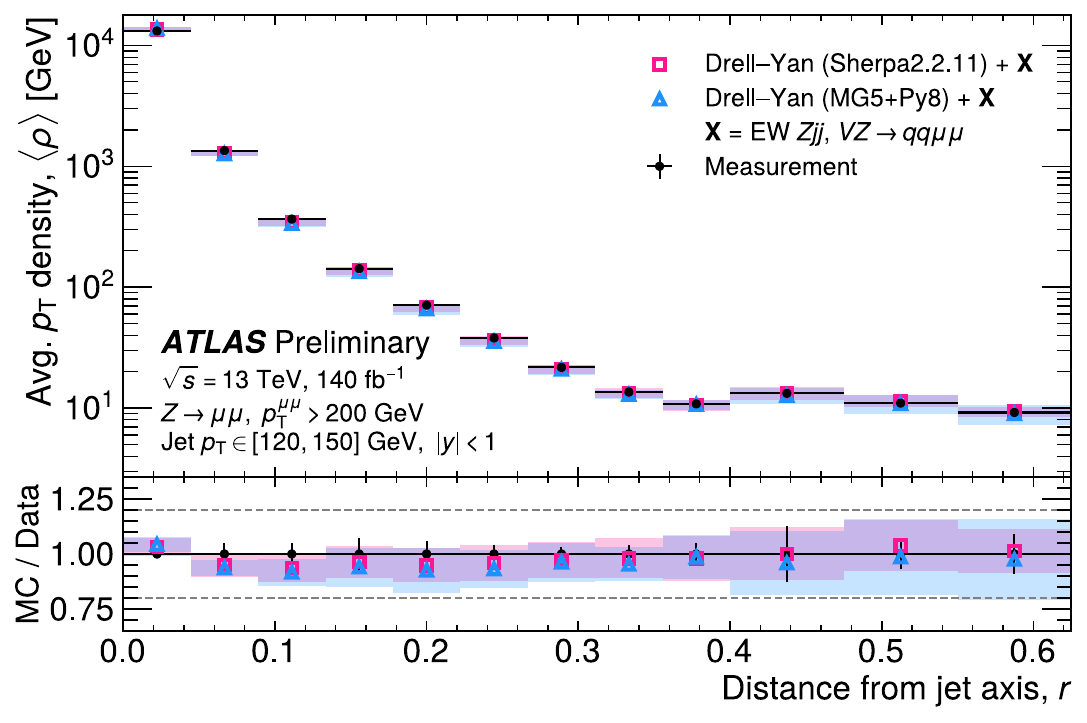}
    \includegraphics[width=0.48\linewidth, alt={Uncertainty budget for the charged-particle pT density profile measurement with the anti-kt R=0.4 algorithm.}]{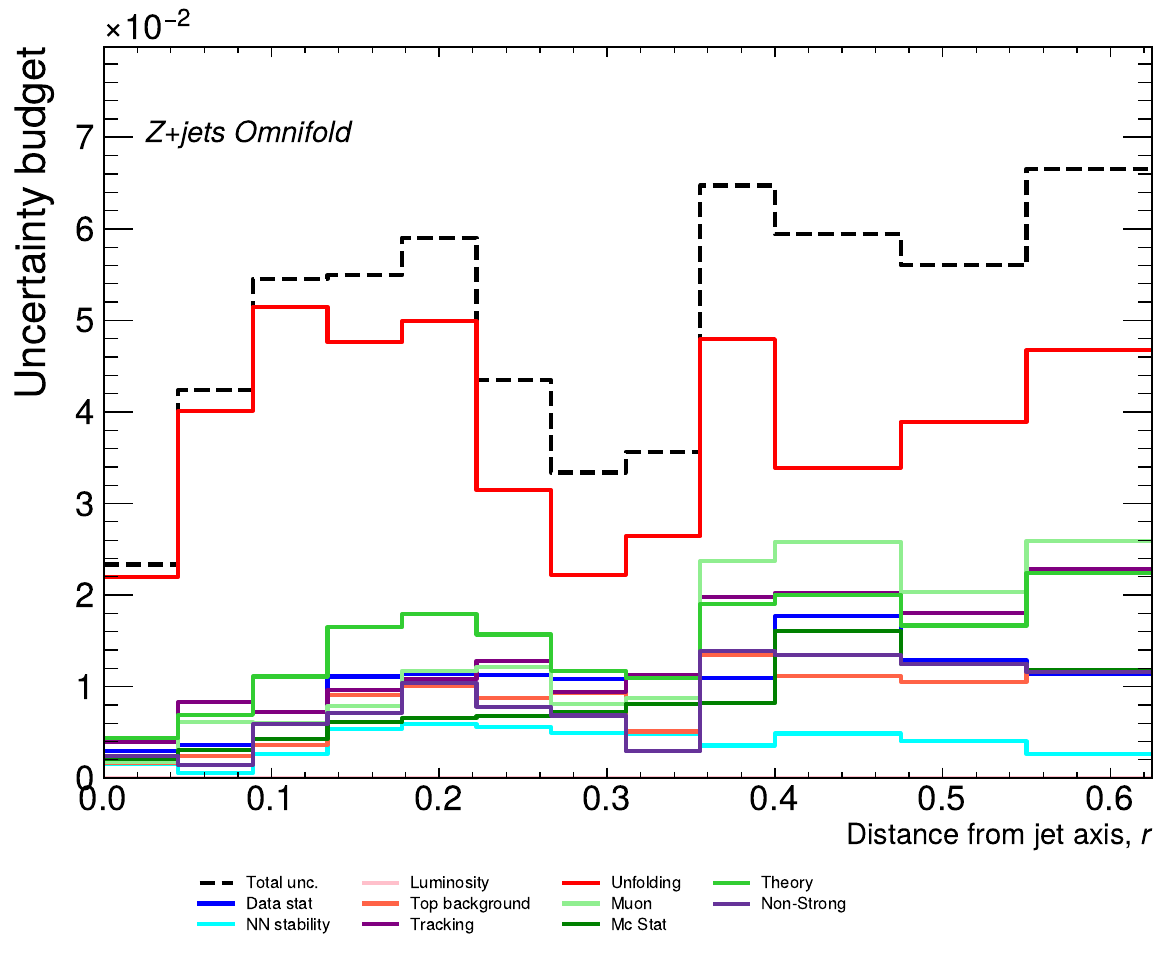}
    \caption{Unfolded charged-particle \pt density profile $\langle \rho(r) \rangle$ about the jet axis, for jets clustered with the anti-$k_t$ algorithm at $R = 0.4$ and satisfying $p_T^{\mathrm{jet}} \in [120, 150]$~GeV and $|y^{\mathrm{jet}}| < 1$. Left: the \Omnifold measurement (black points) compared to the truth-level predictions of the \mgpy (blue) and \sherpa (pink) samples. Right: the uncertainty budget, with the total uncertainty (black dashed line) and the individual uncertainty groups (colored bands). All charged particles in the event are included in the sum over the annulus, not only those assigned to the jet.}
    \label{fig:zjets-results-shape-akt}
\end{figure}

\begin{figure}[tb]
    \centering
    \includegraphics[page=2, width=0.75\linewidth, alt={Comparison of the unfolded charged-particle pT density profile rho(r) computed with the anti-kt, kt, and Cambridge-Aachen clustering algorithms all at R=0.4, showing the characteristic central dip is specific to anti-kt clustering.}]{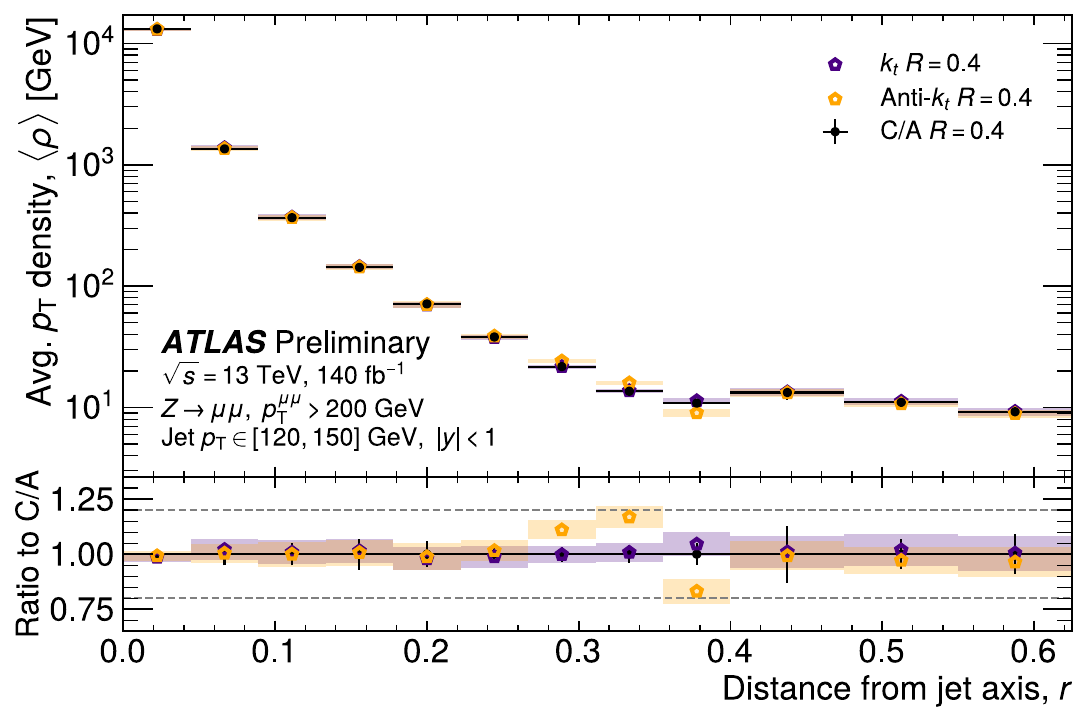}
    \caption{Unfolded charged-particle \pt density profile $\langle \rho(r) \rangle$ for three jet clustering algorithms, all at radius parameter $R = 0.4$: anti-$k_t$, $k_t$, and Cambridge--Aachen. The dip in $\langle \rho(r) \rangle$ at $r \approx R$ is pronounced for anti-$k_t$, absent for Cambridge--Aachen, and intermediate for $k_t$, demonstrating that the feature is driven by the choice of clustering algorithm rather than by the underlying radiation pattern. All three measurements are produced from the same unbinned \Omnifold output.}
    \label{fig:zjets-results-shape-algos}
\end{figure}

Because the unbinned measurement is differential in the kinematics of individual charged particles, jets can be clustered on the fly post-unfolding using any jet clustering algorithm and any radius parameter without the need to repeat the measurement with a different jet definition.
In a typical jet measurement, each combination of algorithm and radius requires its own dedicated jet energy scale and resolution calibration.
These calibrations are performed by dedicated performance groups working within the ATLAS and CMS calibrations, and are quite elaborate and time consuming~\cite{ATLAS:2023tyv}.
The limited time that human researchers have to produce these calibrations implies that each experiment's jet performance group can only support calibrations for a few types of jets.
Measurements utilizing jet definitions beyond these few supported types of jets are difficult, and few such measurements have been performed~\cite{ATLAS:2013uet}.
The result is that questions of how different jet definitions affect jet substructure are largely unexplored.

However in the full-phase-space measurement, the jet itself is not a calibrated input to the measurement.
It is a function constructed on the unfolded charged particles, just like any other event-level observable.
These charged particles can be clustered using an arbitrary jet definitions, and then these jets can be used to measure arbitrary substructure observables.
This allows questions about the interplay of jet substructure and jet definition, previously difficult and time consuming to pursue, to be answered very quickly, even without the need to conduct an additional measurement.

As an illustration of this ability, this section presents an unfolded measurement of the charged-particle \pt density profile using three different jet clustering algorithms.
The \pt density profile is given by,
\begin{equation}
    \langle \rho(r) \rangle \;=\; \frac{1}{N_{\mathrm{jets}}} \sum_{\mathrm{jets}} \frac{p_T(r - \Delta r / 2,\, r + \Delta r / 2)}{A_{\mathrm{annulus}}(r, \Delta r)},
    \label{eqn:zjets-rho-r}
\end{equation}
where $r$ is the angular distance from the jet axis in $(y, \phi)$ space, $\Delta r$ is the width of an annulus centered at $r$, $p_T(r - \Delta r / 2, r + \Delta r / 2)$ is the scalar \pt sum of all charged particles falling in that annulus, and $A_{\mathrm{annulus}}(r, \Delta r) = \pi[(r + \Delta r / 2)^2 - (r - \Delta r / 2)^2]$ is the annulus area.
Because the sum is averaged over jets, the observable is invariant under overall normalization changes and cannot be interpreted as a cross section.
Note also that a given jet contributes some \pt density to each bin in the radius $r$.
Accounting for these correlations is difficult with binned unfolding methods, but they are tracked naturally in the unbinned \Omnifold measurement since a single event weight is applied to the jets contribution to each bin.
This makes this observable doubly difficult to measure without a full-phase-space measurement.
First each jet definition would need to be calibrated independently as mentioned above, and second the correlations between bins would need to be understood and characterized.
The full-phase-space measurement elegantly avoids both of these issues.

The \pt density is calculated using all charged particles, even those that are not clustered in the jets, allowing the density to be non-zero outside of the jet radius.
The jet selection is $p_T^{\mathrm{jet}} \in [120,\,150]$~GeV and $|y^{\mathrm{jet}}| < 1$, a narrow window just above the peak of the leading-jet \pt spectrum.
\Cref{fig:zjets-results-shape-akt} shows $\langle \rho(r) \rangle$ for jets clustered with the anti-$k_t$ algorithm~\cite{Cacciari:2008gp} with radius parameter $R = 0.4$, together with its uncertainty budget.
The \mgpy and \sherpa generators reproduce the overall shape of $\langle \rho(r) \rangle$ reasonably well, though residual differences at the $5$--$10\%$ level are visible both in the core and in the tails.
The data profile is sharply peaked at small $r$ and falls off with larger annulus radius, with a pronounced dip near $r \approx R$.
This dip is an artifact of the anti-$k_t$ clustering algorithm which centers the jet axis on the highest \pt jet constituent rather than some average over the entire radiation pattern, possibly preventing particles near the jet boundary from being included in the jet.

The unbinned measurement can provide a test of this interpretation by repeating the measurement with different jet clustering algorithms.
\Cref{fig:zjets-results-shape-algos} compares $\langle \rho(r) \rangle$ computed with the anti-$k_t$, $k_t$~\cite{Ellis:1993tq}, and Cambridge--Aachen (C/A)~\cite{Dokshitzer:1997in} algorithms, all with a radius parameter $R = 0.4$.
The dip at $r \approx R$ is much less pronounced for $k_t$ and C/A jets, confirming that the feature is driven by the clustering algorithm rather than by the data.
Producing the measurements in \Cref{fig:zjets-results-shape-algos} would be very difficult with conventional unfolding methods, but requires only a light data analysis built on top of the unbinned spectra given this full-phase-space measurement.
Further exploration of jet shape observables, including many exotic observables that have never been measured in data, should be explored with these data in the future~\cite{Ba:2023hix,Gambhir:2024ndc}.

\FloatBarrier

\subsection{Energy Correlator Observables}
\label{sec:zjets-results-eec}

\begin{figure}[tb]
    \centering
    \includegraphics[width=0.65\linewidth, alt={Measurement of the two-point energy-energy correlator EEC as a function of the angular separation observable z, measured across all pairs of charged particles in the event.}]{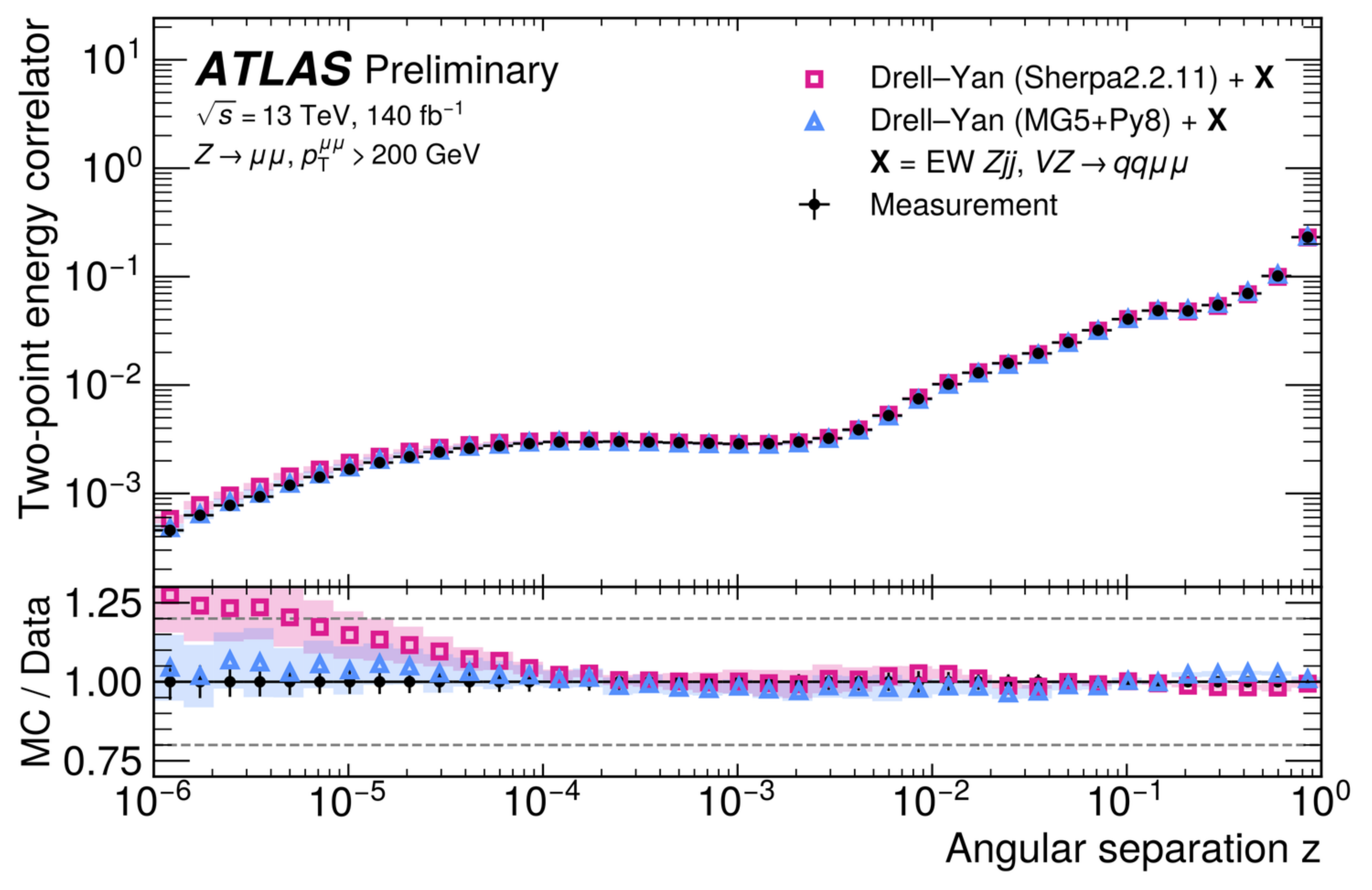}
    \caption{Unfolded measurement of the two-point energy-energy correlator EEC as a function of the dimensionless angular separation observable $z$, computed from all pairs of charged particles in the event. The \Omnifold measurement (black points) is shown together with the truth-level predictions of the \mgpy (blue) and \sherpa (pink) samples. Error bars on the measurement show the total uncertainty. Shaded boxes on the generator predictions show the theory uncertainty. Three scaling regimes are visible: linear at large $z$ (uncorrelated radiation), plateau at intermediate $z$ (perturbative QCD), and linear at small $z$ (free hadrons below $\Lambda_{\mathrm{QCD}}$).}
    \label{fig:zjets-results-eec}
\end{figure}

\begin{figure}[tb]
    \centering
    \includegraphics[width=0.65\linewidth, alt={Covariance matrix for the unfolded EEC measurement, showing the bin-to-bin correlations of the total measurement uncertainty as a function of the angular separation observable z.}]{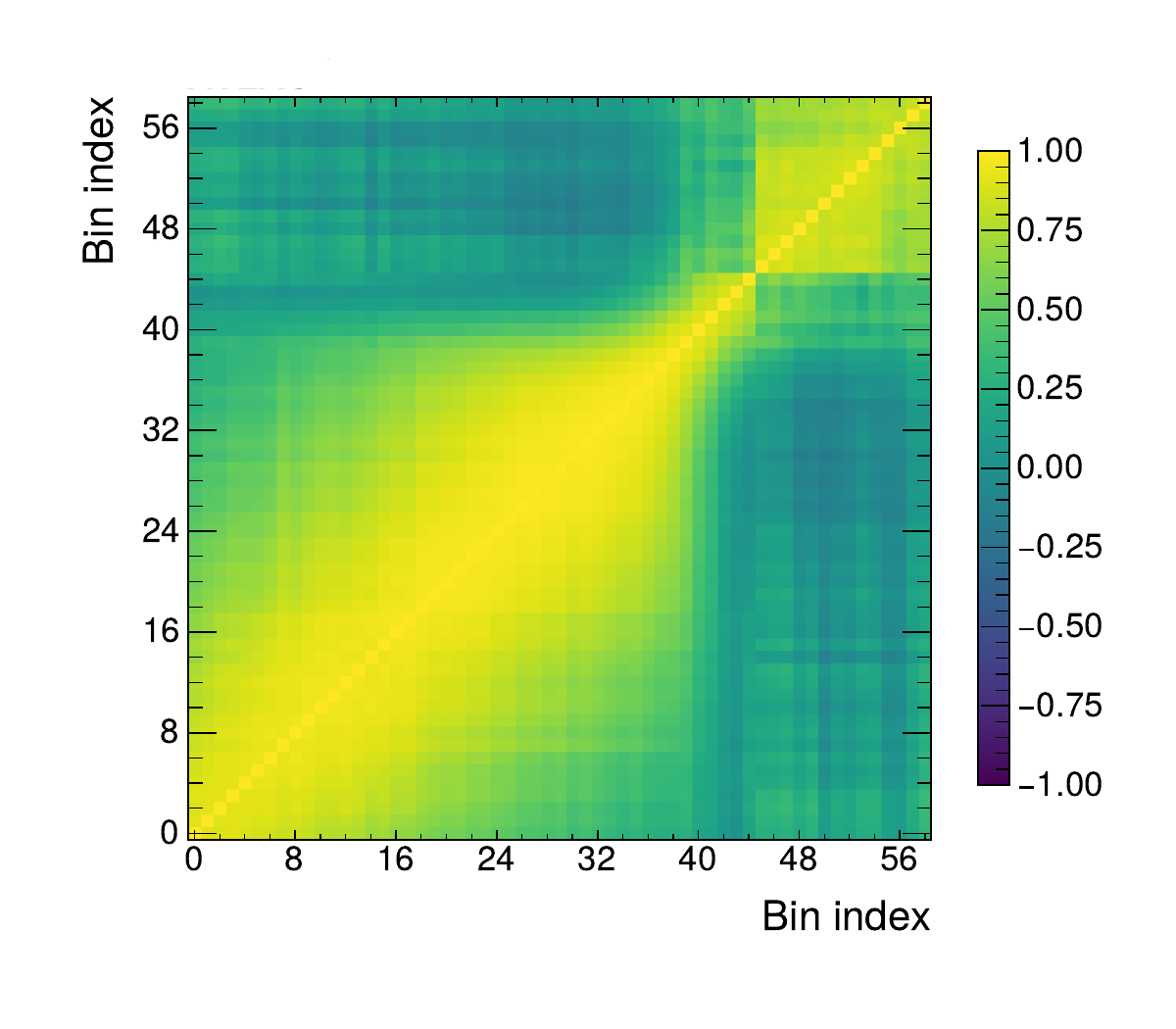}
    \caption{Covariance matrix of the unfolded EEC measurement shown in \cref{fig:zjets-results-eec}. The matrix encodes the bin-to-bin correlations of the total uncertainty as a function of the dimensionless angular separation observable $z$.}
    \label{fig:zjets-results-eec-cov}
\end{figure}

Energy correlator observables were introduced in \Cref{sec:eec}.
All existing experimental measurements of the two-point energy correlator, summarized in \Cref{tab:eec-measurements}, use binned unfolding methods to correct for the detector response.
This is a difficult observable to unfold with binned methods for two reasons.
First, both the opening angle $\theta_{ij}$ and the energy weight $E_i E_j$ in \Cref{eqn:eec} are subject to detector effects and receive corrections from the unfolding.
This means that at least a two-dimensional unfolding in both of these quantities must be performed.
Further many measurements of the EEC are performed in bins of overall jet \pt in order to extract a universal hadronization scale.
This requires a three-dimensional unfolding, which is at the limit of what can be done with binned methods~\cite{CMS:2024mlf}.
Second, even if binned methods can be made to work in two or three dimensions they do not account for correlations in the bin counts resulting from each particle in an event entering the histogram many times.
As with the jet shape observables in \Cref{sec:zjets-results-shape}, the \Omnifold measurement handles this effect naturally since the correlations are preserved by the common event weight.

\Cref{fig:zjets-results-eec} shows the EEC measured using the full-phase-space measurement, with the sum running over all pairs of charged particles in the event.
This is the first measurement of this observable in proton--proton collisions, since all existing measurements either use jets as input objects rather than individual particles, or measure the EEC within individual jets rather than on the event as a whole.
The practical distinction between measuring the EEC with all particles rather than within jets is that the measurement is not truncated when the angular distance between charged particle pairs grows larger than the jet radius.
The measurement achieves $5\%$ accuracy or better across the full range of $z$ shown in the figure, which corresponds to opening angles from $0.002$ to $0.64$ radians.
Three scaling regimes are visible.
At large $z$ ($z > 10^{-2}$), the EEC scales approximately linearly with $z$.
This implies uncorrelated and roughly evenly distributed radiation, which is expected at opening angles outside of the collinear parton shower.
At intermediate $z$, between roughly $10^{-4}$ and $3 \times 10^{-3}$, the EEC flattens into a nearly constant plateau.
This is the perturbative QCD scaling regime in which the two-point correlation is set by the interactions of quarks and gluons above the confinement scale $\Lambda_{\mathrm{QCD}} \sim 200$~MeV.
In previous measurements of the EEC performed inside jets, this scaling regime was cut off by the jet radius.
The presence of this scaling regime is direct evidence for the existence of parton showers in proton--proton collisions.
At small $z < 10^{-4}$, the linear scaling is restored, reflecting the non-interacting final-state hadrons which are produced once the partons confine below $\Lambda_{\mathrm{QCD}}$.

The crossover region between $z \sim 10^{-5}$ and $z \sim 10^{-4}$ is a purely non-perturbative effect which is clearly present in the data but is very difficult to understand within the framework of QCD.
The \mgpy and \sherpa predictions are generally able to reproduce the three-regime structure and show percent level accuracy above $z \sim 10^{-4}$.
Below the hadronization regime at $z < 10^{-4}$, the \sherpa generator substantially over-predicts the correlation.
A similar effect was observed in Ref.~\cite{CMS:2024mlf}.
The \mgpy generator describes this regime to within several percent.
The covariance matrix of the measurement, shown in \cref{fig:zjets-results-eec-cov}, encodes the bin-to-bin correlations of the total uncertainty and is required for any quantitative use of the measurement, such as an extraction of $\alpha_s$.
This matrix would be difficult to extract from a measurement of this observable relying on binned unfolding methods, but is easily available from the full-phase-space measurement.

A further benefit of the full-phase-space measurement is that higher order ($N > 2$) energy correlator observables are also measured.
Though these applications are not shown here, this would be a relatively straightforward extension.
In particular the ratio of the E3C to the E2C is very sensitive to $\alpha_s$ and could be used to extract this parameter from jet substructure.
Higher point correlators than the E3C have never been measured in data, so this application of this measurement would be very novel and valuable.
This application is completely supported by the public data release and so should be explored soon after publication of the measurement.

\FloatBarrier

\subsection{Intrinsic Dimensionality of Jets}
\label{sec:zjets-results-id}

The concept of intrinsic dimensionality and the NNID estimator was introduced in \Cref{sec:id}.
Given intrinsic dimensionality is a property of a dataset as a whole rather than a single event or jet, it is impossible to construct a response matrix that can be used to run binned unfolding methods.
This is part of the reason this class of observables have never been measured in particle physics data.
However unbinned measurements circumvent this problem since the result is a dataset rather than a histogram, and this dataset can be used as input to any intrinsic dimension estimator.
In principle this could have been done using the 24-dimensional cross section produced by the Multifold measurement, but this dimensionality would be unphysical since the inputs would be summary observables of the underlying data rather than the data themselves.
Calculating a physical intrinsic dimensionality requires that the data are used as input, and this requires that the four-momenta of individual particles are unfolded.
This is exactly the setup delivered by the full-phase-space measurement.
This section presents the first measurement of the intrinsic dimensionality of a particle physics dataset using the nearest-neighbor intrinsic dimension estimator (NNID)~\cite{NIPS2004_74934548} evaluated on a sample of jets drawn from the unfolded charged final state.

The procedure for calculating the NNID starts with clustering jets with the anti-$k_T$ algorithm and a radius parameter of 1.0 using the particle-level MC samples needed to build the measurement.
The choice of jet definition is irrelevant and could be changed at will.
It would also be possible to calculate the NNID using the entire event as input rather than the leading jet, but careful consideration would need to be put into the correct equivalent to the pre-processing steps described below.
Jets are then taken from the \pt range $[330, 370]$~GeV, which roughly corresponds to initiating parton \pt in the range $[500, 550]$~GeV when accounting for the missing neutral component.
Next, jets are pre-processed by applying a rotation about the beam axis and Lorentz boost along the beam axis such that the jet is centered in the rapidity--azimuth plane.
The jets are then rotated about the jet axis such that the principal component of the radiation pattern points in the positive $\phi$ direction in the $y$--$\phi$ plane.
These pre-processing steps remove known symmetries of jets.
Then the EMD (introduced in \Cref{sec:id}) between every pair of jets in the sample must be calculated.
In practice this step is compute intensive and requires writing an $N \times N$ array of floating point numbers to disk, where $N$ is the number of jets in the dataset.
To make the calculation manageable, $50,000$ jets are randomly sampled from the set of jets in the \pt selection for each dataset.
After calculation of the EMDs, the EMD ratios are calculated and used as input to a one-dimensional maximization of the likelihood in \Cref{eqn:likelihood}, given a choice of the unphysical index $i$.
The result is an estimate of the intrinsic dimensionality $d$, along with the median EMD between all jets in the dataset and their $i$th nearest neighbor, which is a physical scale that can be used for plotting.
These calculations must be repeated for each weight variation needed to assess the systematic uncertainties.

Unlike the previous three applications of this measurement, the NNID estimator is entirely unbinned.
It is a point estimate provided by the data given a choice of the unphysical index $i$.
This makes the $\chi^2$ tests of \Cref{sec:zjets-validation-chi2} reduce to noting that the total uncertainty on the pseudodata measurement covers the method bias for each choice of $i$.
It was checked that this test passes.

\begin{figure}[tb]
    \centering
    \includegraphics[width=0.48\linewidth, alt={Measurement of the nearest-neighbor intrinsic dimension NNID of anti-kt R=1 jets as a function of the median EMD in the low-EMD regime, highlighting the fine-structure differences between data and generator predictions.}]{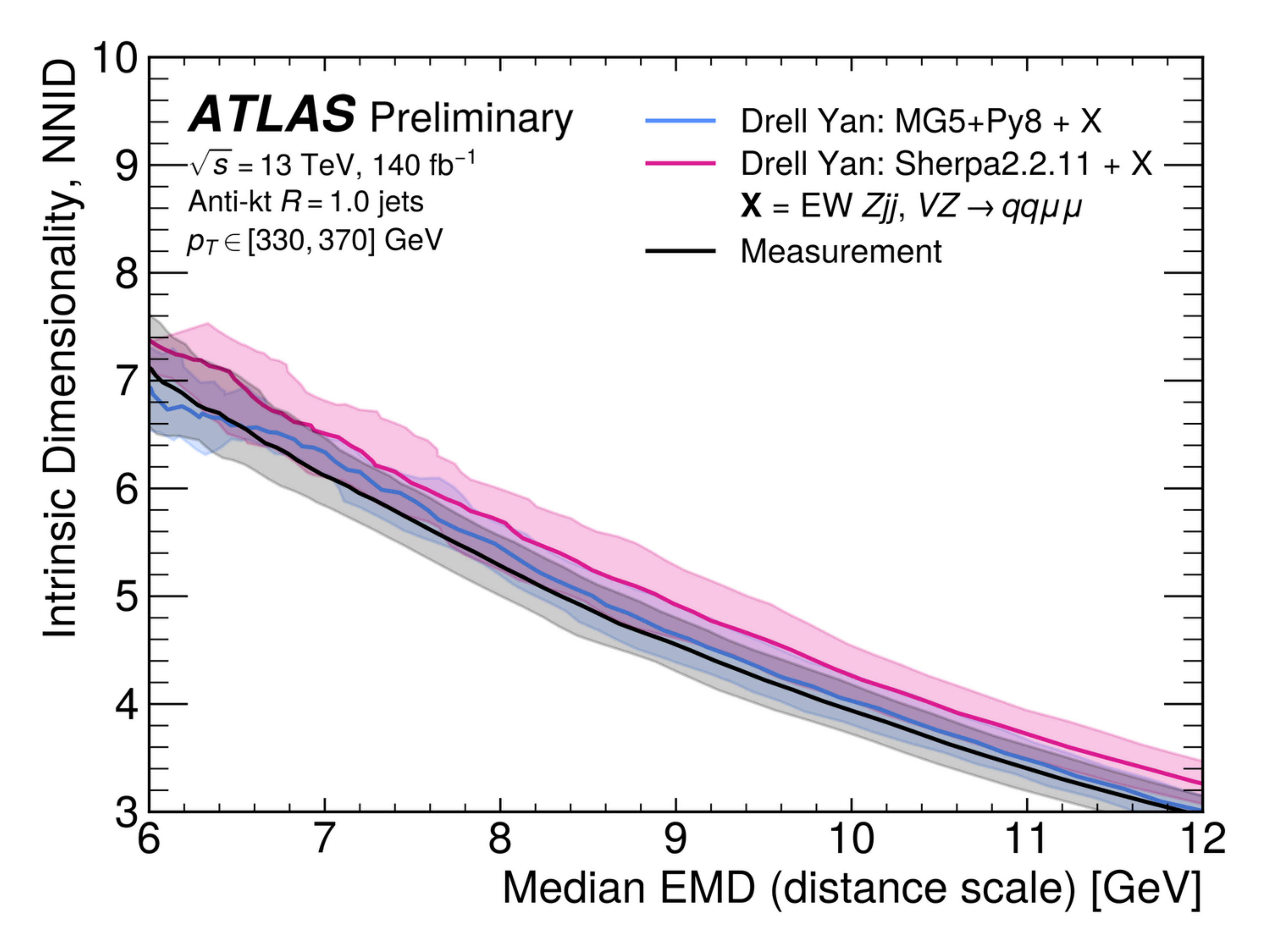}
    \includegraphics[width=0.48\linewidth, alt={Measurement of the nearest-neighbor intrinsic dimension NNID of anti-kt R=1 jets as a function of the median EMD in the high-EMD regime, where the NNID approaches its asymptotic value.}]{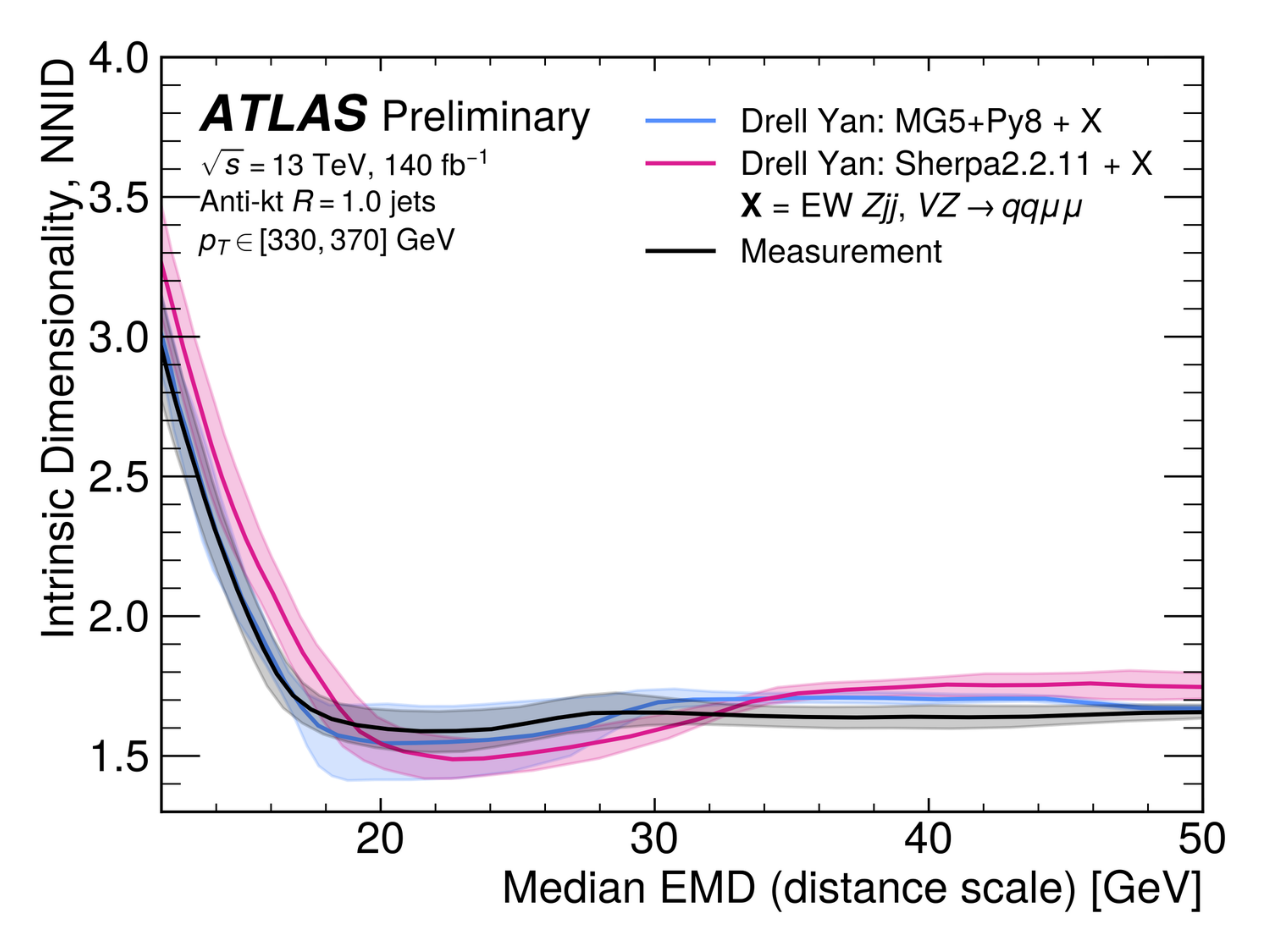}
    \caption{Unfolded measurement of the nearest-neighbor intrinsic dimension (NNID) of a sample of leading anti-$k_t$ $R = 1.0$ jets with $p_T \in [330, 370]$~GeV and $|\eta| < 2.5$, plotted as a function of the weighted median energy mover's distance between a jet and its $i$-th nearest neighbor. Both the NNID and the median EMD are functions of the un-physical neighbor index $i$, which is scanned logarithmically. The two panels show the same curve in two complementary ranges: (left) the low-median-EMD regime highlighting the fine-structure differences between the measurement and generator predictions, and (right) the high-median-EMD regime in which the NNID approaches its asymptotic value. Shaded bands show the total uncertainty on both the measurement and the generator predictions.}
    \label{fig:zjets-results-nnid}
\end{figure}

\Cref{fig:zjets-results-nnid} shows the result in two ranges of the median EMD.
The left panel shows the NNID versus median EMD curve at low values of the median EMD.
The generators reproduce the data within uncertainties, but tend to slightly overestimate the NNID.
The right panel shows the same curve at higher median EMD values.
The dip in NNID between $20$ and $30$~GeV is less pronounced in the data than in the generators, as is the increase in NNID above $30$~GeV.
However overall the generators again agree with the data within uncertainties.

The take-away from this application is that jet data, both provided by MC generators and measured in nature, have no more than several intrinsic dimensions.
Naive representations of jets in terms of the four-momenta of the jet constituents typically require many tens of observables.
\Cref{fig:zjets-results-nnid} demonstrates that this representation is highly redundant.
Even at very small distance scales which resolve fine differences in the substructure between jets, the intrinsic dimensionality is still below 10.
This is direct evidence that the manifold hypothesis discussed in \Cref{sec:id} holds true for particle physics data, in part explaining why machine learning methods are effective in this domain.
Physically speaking, jet data are made to cluster on a lower-dimensional manifold through many constraints imposed by symmetries, conservation laws, and QCD dynamics.
This fact is well-understood from many years of particle physics research, but it is interesting to see this born out in a purely empirical fact about the structure of the data.

As mentioned in \Cref{sec:id}, intrinsic dimension observables are of interest here not because of a deep link to the underlying QCD theory but because they establish basic facts about jet data.
Possible links between this class of observables and the QCD theory are interesting, but have never been explored in the theoretical literature, perhaps in part because measurement of these observables was not possible without unbinned and high-dimensional unfolding methods.
Ideally this measurement also serves as some motivation for the theory community to explore these observables in the future.

\section{Discussion}
\label{sec:zjets-discussion}

This chapter has presented a full-phase-space measurement of $Z(\to\mu\mu)$+jets production in proton--proton collisions at $\sqrt{s} = 13$~TeV using the full Run~2 ATLAS dataset.
The measurement is the highest-dimensional cross section measurement ever performed, reaching up to 843 simultaneous dimensions and a mean of 150 dimensions per event.
It supersedes the previous round Multifold measurement~\cite{ATLAS:2024xxl}, extending from a fixed set of 24 observables to the phase space of all charged particles.
The result is best understood as a dataset rather than a histogram, which can be projected onto any observable of interest without repeating the measurement.
Four physics use cases were presented in \Cref{sec:zjets-results}, but these represent only a small fraction of the possible applications.

To support further analyses, the unbinned spectra are planned to be made publicly available alongside a codebase that documents best practices and reproduces the four use cases presented here\footnote{The equivalent documentation for the Multifold analysis is available at \href{https://gitlab.cern.ch/atlas-physics/public/sm-z-jets-omnifold-2024}{https://gitlab.cern.ch/atlas-physics/public/sm-z-jets-omnifold-2024}}.
The documentation contains a set of usage recommendations, two of which are worth highlighting.
First, the particle-level MC sample used for unfolding must provide sufficient support in any phase space region of interest.
In practice this is enforced by requiring that at least 5,000 effective events are contained in each bin of any histogram constructed from the particle-level data.
Second, users must validate the measurement in their desired binning and phase space by running the pseudodata measurement described in \Cref{sec:zjets-pseudodata} and computing a $\chi^2$ test statistic and p-value as was done for the 26 observables in \Cref{sec:zjets-validation-chi2}.
This validation step is essential because the unfolded spectra contain truth-level information, such as the truth hadron fractions discussed in \Cref{sec:zjets-uncert-unfold}, that is not constrained at detector level.
Observables sensitive to such hidden variables will fail this check, indicating that the observables cannot be constrained with this measurement.

There are many applications of this measurement that can and should be pursued in the future.
A few are worth mentioning here.
The EEC observable measured in \Cref{sec:zjets-results-eec} provides a direct avenue for extracting the strong coupling $\alpha_s$ from the perturbative QCD scaling regime, and extensions to higher-point energy correlators are straightforward given the unbinned representation of the final state.
The underlying event results of \Cref{sec:zjets-results-ue} can be used to tune conventional event generators, or to fit machine learning based hadronization models which assume precisely this kind of unbinned measurement~\cite{Assi:2025gog,Bierlich:2023fmh,Bierlich:2023zzd,Bierlich:2024xzg,Chan:2023ume,Ghosh:2022zdz,Ilten:2022jfm}.
The $Z$ boson kinematic distributions, including the signed $\Delta\phi$ observable, can be used for EFT fits to anomalous couplings.
Because the measurement is fully differential in the hadronic recoil, it can also constrain backgrounds in other measurements and searches, for example electroweak $Zjj$ production is a significant background in many Higgs boson measurements.
The Lund jet plane~\cite{Dreyer:2020brq} can be measured directly from the unfolded charged particles, including the second and tertiary jet planes which have never been measured experimentally.
Finally, quark/gluon tagging applied to the charged particle jets would allow differences in substructure between quark- and gluon-initiated jets to be extracted from data.
One limitation of the measurement is that all observables are constructed from charged particles only, so direct comparison to perturbative QCD calculations is complicated by the absence of the neutral component.
Significant theoretical progress has been made in this direction through the track function formalism~\cite{Chang:2013rca,Chang:2013iba}, which provides a systematic framework for performing perturbative calculations on charged-particle observables.

This measurement also demonstrates that full-phase-space cross section measurements are technically feasible at a hadron collider, so extension of these methods to measurements of other processes is a natural next step.
Dijet production, diboson production, top quark pair production, Higgs processes, and multi-lepton final states are all promising targets.
Measurements of these processes would be very useful for global EFT fits~\cite{CMS:2023xyc,CMS:2025ugn,Heimel:2024drk}, which currently rely on a patchwork of binned measurements that are typically single or double differential at best.
Highly differential cross section measurements are useful for breaking degeneracies between EFT operators in likelihood fits, so highly differential measurements produced through unbinned and high-dimensional unfolding would be ideal inputs for EFT fits performed by phenomenologists.
Two practical limitations of unbinned and high-dimensional unfolding methods will need to be overcome to perform these future measurements.
The acceptance effects in the $Z$+jets phase space discussed in \Cref{sec:zjets-omnifold} are small enough to be negligible, but they become significant when the event selection involves jets, which is typically done for studying many of the processes mentioned above.
Methods for handling acceptance effects in unbinned unfolding have been proposed~\cite{Butter:2025mek,Canelli:2025ybb} but have not yet been applied in a real analysis.
The computational cost of the measurement is also substantial.
Fine-tuning roughly 25 thousand transformer networks required approximately 400,000 A100 GPU-hours.
While this is large in absolute terms, it is a fraction of the resources consumed by frontier language model training, and significant optimizations are possible.
As GPU resources become more available and methodologies mature, computational cost should become less of a bottleneck.

The HL-LHC era will produce datasets of unprecedented size and quality that will be useful many years into the future.
A central goal of particle physics over the next two decades should be to allow wide and open access to these data, and preserve them for future use.
Unbinned and high-dimensional cross section measurements are a tool that address many of the challenges, in addition to offering the ability to measure new observables and improve our understanding of LHC data at present.
They should become standard practice in the HL-LHC physics program.

\FloatBarrier

%% file: ch6_lo_feynman_diagrams.tex

\begin{subfigure}[b]{0.46\textwidth}
  \centering
  \begin{tikzpicture}
    \begin{feynman}
      \vertex (iq)  at (0.0,  1.0) {\(q\)};
      \vertex (iqb) at (0.0, -1.0) {\(\bar{q}'\)};
      \vertex (v1)  at (2.0,  1.0);
      \vertex (v2)  at (2.0, -1.0);
      \vertex (fV)  at (4.0,  1.0) {\(V\)};
      \vertex (fg)  at (4.0, -1.0) {\(g\)};
      \diagram* {
        (iq)  -- [fermion]      (v1),
        (iqb) -- [anti fermion] (v2),
        (v1)  -- [boson]        (fV),
        (v1)  -- [fermion]      (v2),
        (v2)  -- [gluon]        (fg),
      };
    \end{feynman}
  \end{tikzpicture}
  \caption*{(a)}
\end{subfigure}
\hfill
\begin{subfigure}[b]{0.46\textwidth}
  \centering
  \begin{tikzpicture}
    \begin{feynman}
      \vertex (iq)  at (0.0,  1.0) {\(q\)};
      \vertex (iqb) at (0.0, -1.0) {\(\bar{q}'\)};
      \vertex (v1)  at (2.0,  1.0);
      \vertex (v2)  at (2.0, -1.0);
      \vertex (fV)  at (4.0, -1.0) {\(V\)};
      \vertex (fg)  at (4.0,  1.0) {\(g\)};
      \diagram* {
        (iq)  -- [fermion]      (v1),
        (iqb) -- [anti fermion] (v2),
        (v2)  -- [boson]        (fV),
        (v1)  -- [fermion]      (v2),
        (v1)  -- [gluon]        (fg),
      };
    \end{feynman}
  \end{tikzpicture}
  \caption*{(b)}
\end{subfigure}

\vspace{2em}

\begin{subfigure}[b]{0.46\textwidth}
  \centering
  \begin{tikzpicture}
    \begin{feynman}
      \vertex (iq)  at (0.0,  1.0) {\(q\)};
      \vertex (ig)  at (0.0, -1.0) {\(g\)};
      \vertex (v1)  at (1.33,  0.0);
      \vertex (v2)  at (2.66,  0.0);
      \vertex (fV)  at (4.0,  1.0) {\(V\)};
      \vertex (fq)  at (4.0, -1.0) {\(q\)};
      \diagram* {
        (iq)  -- [fermion] (v1),
        (ig)  -- [gluon]   (v1),
        (v1)  -- [fermion] (v2),
        (v2)  -- [boson]   (fV),
        (v2)  -- [fermion] (fq),
      };
    \end{feynman}
  \end{tikzpicture}
  \caption*{(c)}
\end{subfigure}
\hfill
\begin{subfigure}[b]{0.46\textwidth}
  \centering
  \begin{tikzpicture}
    \begin{feynman}
      \vertex (iq)  at (0.0,  1.0) {\(q\)};
      \vertex (ig)  at (0.0, -1.0) {\(g\)};
      \vertex (v1)  at (2.0,  1.0);
      \vertex (v2)  at (2.0,  -1.0);
      \vertex (fV)  at (4.0,  1.0) {\(V\)};
      \vertex (fq)  at (4.0, -1.0) {\(q\)};
      \diagram* {
        (iq)  -- [fermion] (v1),
        (ig)  -- [gluon]   (v2),
        (v1)  -- [fermion] (v2),
        (v1)  -- [boson]   (fV),
        (v2)  -- [fermion] (fq),
      };
    \end{feynman}
  \end{tikzpicture}
  \caption*{(d)}
\end{subfigure}

%% file: tab_ch6_closure_pvalues.tex

\begin{tabular}{l r r r r r}
\toprule
Observable & DoF & \multicolumn{2}{c}{Omnifold} & \multicolumn{2}{c}{Multifold} \\
 &  & $\chi^2$ & $p$ & $\chi^2$ & $p$ \\
\midrule
HT\_tracks & 6 & 7.501 & \cellcolor{green}0.2770 &  &  \\
Ntracks & 5 & 3.851 & \cellcolor{green}0.5711 &  &  \\
Ntracks\_trackj1 & 6 & 7.162 & \cellcolor{green}0.3061 & 7.942 & \cellcolor{green}0.2424 \\
Ntracks\_trackj2 & 5 & 4.726 & \cellcolor{green}0.4502 & 2.680 & \cellcolor{green}0.7492 \\
eta\_l1 & 14 & 16.081 & \cellcolor{green}0.3084 & 14.663 & \cellcolor{green}0.4016 \\
eta\_l2 & 14 & 6.450 & \cellcolor{green}0.9538 & 12.691 & \cellcolor{green}0.5510 \\
m\_trackj1 & 6 & 7.658 & \cellcolor{green}0.2642 & 9.041 & \cellcolor{green}0.1713 \\
m\_trackj2 & 4 & 9.228 & \cellcolor{green}0.0556 & 5.696 & \cellcolor{green}0.2230 \\
pT\_l1 & 6 & 4.345 & \cellcolor{green}0.6301 & 8.864 & \cellcolor{green}0.1814 \\
pT\_l2 & 6 & 10.145 & \cellcolor{green}0.1187 & 8.256 & \cellcolor{green}0.2199 \\
pT\_ll & 5 & 2.032 & \cellcolor{green}0.8447 & 5.049 & \cellcolor{green}0.4100 \\
pT\_trackj1 & 6 & 9.086 & \cellcolor{green}0.1688 & 7.677 & \cellcolor{green}0.2627 \\
pT\_trackj2 & 4 & 2.122 & \cellcolor{green}0.7133 & 1.736 & \cellcolor{green}0.7842 \\
phi\_l1 & 16 & 10.357 & \cellcolor{green}0.8473 & 19.040 & \cellcolor{green}0.2666 \\
phi\_l2 & 16 & 15.519 & \cellcolor{green}0.4870 & 12.876 & \cellcolor{green}0.6818 \\
phi\_trackj1 & 16 & 6.346 & \cellcolor{green}0.9839 & 4.727 & \cellcolor{green}0.9970 \\
phi\_trackj2 & 16 & 7.780 & \cellcolor{green}0.9552 & 6.530 & \cellcolor{green}0.9813 \\
tau1\_trackj1 & 7 & 2.209 & \cellcolor{green}0.9474 & 1.755 & \cellcolor{green}0.9722 \\
tau1\_trackj2 & 5 & 6.970 & \cellcolor{green}0.2229 & 3.920 & \cellcolor{green}0.5610 \\
tau2\_trackj1 & 7 & 11.012 & \cellcolor{green}0.1381 & 8.815 & \cellcolor{green}0.2662 \\
tau2\_trackj2 & 5 & 5.404 & \cellcolor{green}0.3686 & 5.935 & \cellcolor{green}0.3126 \\
tau3\_trackj1 & 4 & 1.726 & \cellcolor{green}0.7860 & 1.062 & \cellcolor{green}0.9002 \\
tau3\_trackj2 & 4 & 1.921 & \cellcolor{green}0.7503 & 5.481 & \cellcolor{green}0.2414 \\
y\_ll & 14 & 19.397 & \cellcolor{green}0.1503 & 18.329 & \cellcolor{green}0.1922 \\
y\_trackj1 & 18 & 9.523 & \cellcolor{green}0.9464 & 7.702 & \cellcolor{green}0.9827 \\
y\_trackj2 & 18 & 10.815 & \cellcolor{green}0.9020 & 7.227 & \cellcolor{green}0.9881 \\
\bottomrule
\end{tabular}

%% file: chapter7.tex
\chapter{Conclusion}
\label{ch:conclusion}

\section{Deep Learning and LHC Physics}
\label{sec:conclusion-dl-lhc}

Machine learning is nothing new to LHC physics, but the deep learning revolution kicked off by AlexNet has produced a large suite of new tools that have transformed how LHC data is processed into physics results.
The research covered in this thesis has been a small part of this transformation.
\Cref{ch:tagging} presented a study of top quark tagging using low-level constituent inputs, which established that neural networks trained directly on low-level jet constituent kinematic information outperform those trained on expert-engineered summary features.
This result mirrors the lessons learned from the AlexNet moment in the computer vision field.
The study also provided a systematic uncertainty quantification framework, which showed that larger, more expressive taggers extract more information from simulation, including information that is imperfectly modeled, so raw performance gains in simulation do not trivially translate into performance gains on data.
This effect is visible in the most recent ATLAS and CMS flavor tagging results~\cite{ATLAS:2025dkv,Sarkar:2024vjz}, and managing it will be a challenge for some applications as jet classification models continue to scale.

\Cref{ch:unfolding} surveyed the recent developments in unbinned and high-dimensional unfolding methods, including diffusion- and flow-matching-based generative approaches to modeling posteriors.
It then introduced the \Omnifold algorithm in detail, which is as of this writing the only deep-learning-based unfolding method to be applied to data.
\Cref{ch:zjets} then detailed one such application of \Omnifold to data: a full charged particle phase space measurement of $Z(\to\mu\mu)$+jets production in proton--proton collisions at the LHC.
This result is the highest-dimensional cross section ever measured, with a maximum of 843 simultaneous dimensions.
The measurement is represented as a reweighted particle-level dataset that can be projected and used to constrain an enormous variety of derived observables.
Four particular derived observables were used as examples, three of which would be very difficult or impossible to measure with traditional binned unfolding techniques, proving that deep-learning-based unfolding methods enable entirely new kinds of measurements.
I believe these new capabilities will produce a shift in how the community understands the purpose of cross section measurements.
Instead of targeting specific observables one measurement at a time, the focus will shift to making general-purpose measurements of various final states and prioritizing downstream applications.
This will allow for flexible use of the data, possibly many years into the future after the LHC is done running and the experimental collaborations no longer exist.

Jet tagging and unfolding were the deep learning applications covered in this thesis, but there are many more examples in LHC physics.
Anomaly detection, simulation-based inference, fast detector simulation, and event reconstruction have all seen substantial activity.
There are few parts of LHC physics that deep learning does not touch in some way.
Despite this progress, no major experimental discovery, for example new particles or dynamics beyond the Standard Model, has been produced.
LHC physics is in many ways still organized around the same measurement and search strategies it used before the deep learning era.
Individual pieces of the data analysis chain have been replaced or substantially improved by deep learning methods, but the overall structure of how physics results are produced has not yet changed.
I believe that the unbinned and high-dimensional measurements described in \Cref{ch:unfolding,ch:zjets} represent genuine progress in how particle physics measurements are performed, and that the order-of-magnitude improvements in jet classification documented in \Cref{ch:tagging} are producing real gains in the sensitivity of searches for rare processes like di-Higgs production, but all of this has not been enough to deliver meaningful progress on the open problems in \Cref{sec:open_problems}.
The sections that follow discuss the two directions I find most promising for changing this.

\section{Higgs and di-Higgs with Large-scale Flavor Tagging}
\label{sec:conclusion-dihiggs}

The Higgs boson is rightfully the current focus of the LHC physics program.
Given that $H \to b\bar{b}$ accounts for roughly 58\% of Higgs decays, flavor tagging is a very important ingredient of many Higgs studies.
This is especially true in searches for di-Higgs production and efforts to constrain the Higgs self-coupling parameter $\lambda$.
This parameter is not a free parameter of the Standard Model.
The Standard Model provides a prediction for $\lambda$ in terms of the other free parameters, and this prediction can be tested by measuring the di-Higgs production cross section.
The self-coupling $\lambda$ determines the shape of the Higgs potential, and any departure from the Standard Model prediction would shift that shape in ways that have significant consequences for many of the open problems listed in \Cref{sec:open_problems}, including the hierarchy problem and the matter-antimatter asymmetry.

The most sensitive di-Higgs production channels ($HH \rightarrow b\bar{b}\gamma\gamma$, $HH \rightarrow b\bar{b}\tau\tau$, etc.) all involve one of the Higgs bosons decaying to a bottom quark pair to increase the cross section.
As discussed in \Cref{sec:tagging-future}, the order-of-magnitude gains in $b$-jet rejection achieved by the current generation of transformer-based flavor taggers~\cite{ATLAS:2025dkv,Sarkar:2024vjz}, and the further gains projected from scaling these models to larger architectures trained on billions of jets~\cite{ATLAS:2026vyw,Vigl:2026ppx,Uslu:2026ywh}, make an HL-LHC observation of di-Higgs production and a constraint on the Higgs self-coupling realistic.
This would be the flagship physics result of the HL-LHC, and continued work on scaling jet classification methods strikes me as essential to making it a reality.

Such a di-Higgs observation could have two outcomes.
If the measured self-coupling is consistent with the Standard Model prediction, it will be an extraordinary confirmation of the theory, and a troubling result that leaves the open problems listed in \Cref{sec:open_problems} exactly as open as they are today.
The direction forward for particle physics in that case is unclear.
We would have a theory that fits all data with no clear next experimental milestone that could point the way forward.
If the self-coupling deviates from the Standard Model prediction, we will have direct evidence that the Higgs boson of nature is not exactly the Higgs boson of the Standard Model, possibly pointing toward explanations of the hierarchy problem, the matter-antimatter asymmetry, or both.
In either case, a di-Higgs observation will be an enormous achievement.
Deep learning, including the flavor tagging results of \Cref{ch:tagging}, will be one of the many innovations that make it possible.

\section{Agentic AI and Physics Research}
\label{sec:conclusion-agentic}

Almost all research on deep learning applications in particle physics has used it as a tool for processing data.
Recently the emergence of agentic AI systems has made it possible to use deep learning to perform the research process itself, rather than simply process data.
Agentic systems can plan multi-step analyses, execute code, search literature, and iterate on their own outputs.
As of May of 2026, such systems can already reproduce existing particle physics analyses with high fidelity~\cite{Birk:2026zpd} and produce reasonable, if imperfect, results on open-ended data analysis tasks~\cite{Moreno:2026mqk}.
If model capability stopped improving today, I think this new capability would already be transformative.

Within the context of particle physics data analysis, the single most constrained resource in experimental collaborations is human time.
The cross section measurement of \Cref{ch:zjets} is only possible because of the dedicated effort of hundreds of researchers who operated the ATLAS detector during run 2, designed analysis software, calibrated analysis objects, and set systematic uncertainties on track measurements.
Maintaining and improving the complicated pipeline that turns raw collision data into physics results is unglamorous but extremely important work.
Typically, the methodology of such work is well understood, but software implementation can be difficult and frustrating.
Furthermore, many of the working groups responsible for these tasks are critically understaffed in ATLAS, making the downstream physics analyses suboptimal.
Agentic AI has the potential to automate these tasks and eliminate the human time bottleneck.
Particle physics experiments could become dramatically more productive if agentic AI can follow up on all possible improvements and deliver a level of optimality that is impossible to support with the current human time constraints.

Additionally, I believe the LHC datasets contain far more physics than human researchers have thus far been able to extract from them.
To give a few examples, there are large regions of unexplored two-body resonance parameter space~\cite{Craig:2016rqv,Kim:2019rhy}, there are still gaps in parameter space for many popular BSM models, and many interesting deep-learning-based data analysis techniques have never been utilized.
The LHC datasets support all of this physics, but there are insufficient human researchers willing to write the necessary software.
This is especially true for research directions that are only partially complete (e.g. the parameter space for some BSM model is only partially excluded) but have been around for long enough that the community has moved on to newer ideas.
Agentic AI can fill these gaps, without the time and career constraints that limit human scientists.

The current paradigm of fundamental physics research features a division of labor between humans and AI systems, where the AI handles the computational ``backend'' of research while the human handles the ideation, high-level implementation, and communication of results.
This is often called ``centaur science'', in reference to the mythical creature that is half human and half horse.
In my opinion, this is a remarkably exciting moment to be a human researcher.
Agentic AI has liberated us from the need to spend the majority of our working hours developing and maintaining software, shifting the bottleneck from implementation to creativity.
I predict that much more elaborate and inventive data analyses will become standard in the near-term future.
To give a concrete example, given the computational complexity of the intrinsic dimension measurement in \Cref{sec:zjets-results-id}, I could not have produced those results in the timeframe of my Ph.D. without agentic coding tools.
I was responsible for the ideation and high-level methodology, but the necessary software development and optimization was entirely performed by agentic tools.

My expectation is that the current balance between human researchers and AI will not last long.
Particle physics data analysis tasks are, at their core, software engineering and logical inference problems, both areas in which AI agents have shown rapid improvement.
As of August of 2026, AI systems are still not as good as a competent human researcher at extracting physics from data, but the gap is not very wide and the systems will only get better in the future.
Over the past decade, betting against all but the most aggressive predictions about AI capability has been a losing proposition.
I think it is likely that within a few years we will have AI systems that can take a physics question, identify the relevant dataset, design and execute the analysis, interpret the result, and propose the next step.
Setting academic labor market implications aside, I believe such a system could produce extremely rapid progress.
A few thoughts on what this may look like, and the conclusion of the thesis, are offered in the next section.

\section{Will AI Deliver the Next Discovery?}
\label{sec:conclusion-discovery}

My prediction is that AI-powered data analysis will allow us to essentially exhaust the physics potential of all available experimental data within the next decade.
I think we are currently operating far below this level, with the potential to make much more accurate claims about the Standard Model and $\Lambda$~CDM cosmology available in our already-collected data.
In particle physics specifically, the painstaking accumulation of results from the first three runs of the LHC has brought us closer to exhausting the potential of the data than any previous moment, and this thesis has been a small contribution to that effort.
However, I suspect the remaining gap is wide, and AI will close this gap to near zero in the coming years.

As with the di-Higgs observation milestone more narrowly, there could be one of two possible outcomes.
The first possibility is that the Standard Model is confirmed to be the correct effective theory of the electroweak scale, including a measurement of the Higgs self-coupling consistent with the SM prediction.
This would be a remarkable scientific achievement, but would also be deeply unsatisfying since the Standard Model is known to be incomplete.
In this case, the field of particle physics would be forced to look toward collecting new and different types of experimental data in the hope that some of them would provide clues to answer the open questions of \Cref{sec:open_problems}.
Neutrino experiments, cosmological probes, cosmic ray observatories, or a next-generation collider would be obvious avenues.

The second possibility is that AI-powered analysis of the HL-LHC data surfaces a deviation from Standard Model predictions.
If this happens, we may produce real answers to some of the open questions in particle physics.
Such a discovery would hopefully make obvious what future experimental efforts to prioritize, but it is likely that the availability of experimental data will always remain a bottleneck.
I think it is unlikely that an AI system will simply return the correct Lagrangian of the universe without the many years of difficult experimentation that science has always required.

I am personally excited about both of these outcomes, and the prospect of knowing which one will occur in a matter of years rather than decades.
Hopefully the work described in this thesis will be part of the foundation on which these largely AI-driven data analyses will be built.
I am optimistic about the probability of the next decade producing tangible progress in particle physics, which by all accounts is badly needed.
Developing and deploying the tooling of modern LHC data analysis has been a fulfilling and fruitful research direction, and I am excited about the potential of AI to accelerate this research.
The day-to-day activity of doing particle physics research is currently changing faster than at any point in the history of the field, and this is typically a promising sign that change is on the horizon.
My hope is that all of this change will teach us something new.

\FloatBarrier